\documentclass[11pt,a4]{article}
\usepackage{amsmath}
\usepackage{amssymb}
\usepackage{graphicx}
\usepackage{subfigure}
\usepackage{listings}
\usepackage{multicol}
\usepackage{float}
\usepackage[usenames]{color}
\usepackage{enumerate}
\usepackage{xspace}

\newcommand{\di}{\ensuremath{{\rm d}}}
\newcommand{\tuple}[1]{\ensuremath{{\bf #1}}}
\newcommand{\T}{\ensuremath{\tuple{T}}}
\newcommand{\R}{\ensuremath{\tuple{R}}}
\newcommand{\B}{\ensuremath{\tuple{B}}}
\newcommand{\D}{\ensuremath{\tuple{D}}}
\newcommand{\M}{\ensuremath{\mathcal{M}}}
\renewcommand{\P}{\ensuremath{\mathcal{P}}}
\newcommand{\U}{\ensuremath{\tuple{U}}}

\newcommand{\pval}{\ensuremath{p{\mbox{-value}}}\xspace}
\newcommand{\pvals}{\ensuremath{p{\mbox{-values}}}\xspace}
\newcommand{\ipb}{pb\ensuremath{^{-1}}\xspace}

\usepackage[pdftex,
      colorlinks=true,
      urlcolor=blue,      
      filecolor=blue,      
      linkcolor=blue,       
      citecolor=blue,
      pdfauthor={Georgios Choudalakis},
      pdfpagemode=None,
      bookmarksopen=true]{hyperref}

\definecolor{MyBkgColor}{rgb}{0.99,0.99,0.90}
\definecolor{commentcolor}{rgb}{0.5,0.3,0.93}

\begin{document}

\title{Fully Bayesian Unfolding}
\author{Georgios Choudalakis\thanks{gchouda@alum.mit.edu} \\ {\it \small University of Chicago, Enrico Fermi Institute,} \\ {\it \small 5640 South Ellis Avenue. Chicago, IL 60637} \\ \it{ \small CERN office: B40-R C32.}}
\date{First version: January 23, 2012.\\ Last revision: \today}

\maketitle
\begin{abstract}
Bayesian inference is applied directly to the problem of unfolding.
The outcome is a posterior probability density for the spectrum before smearing, defined in the multi-dimensional space of all possible spectra.
Regularization consists in choosing a non-constant prior.
Despite some similarity, the {\it fully bayesian unfolding} (FBU) method, presented here, should not be confused with D'Agostini's iterative method \cite{dAgostini}.
\end{abstract}
\tableofcontents

\begin{multicols}{2}

\section{Introduction}

Unfolding, in the context of this document, is the non-parametric inference of a binned histogram.  An overview of unfolding, geared towards High Energy Physics (HEP), is in Ref.~\cite{cowanUnfolding}, and Ref.~\cite{choudalakisUnfolding} contains details on published examples of unfolding from ATLAS.

The motivation for this study is to understand unfolding from first principles.  This sheds light on characteristics common to other unfolding methods too.

The fully bayesian unfolding (FBU) offers a complete solution to unfolding, which can be expressed analytically through Bayes' theorem (Sec.~\ref{sec:formulation}) and computed numerically (Sec.~\ref{sec:sampling}).  When dealing with Poisson-distributed data, the answer of FBU is valid even in regions of low statistics where Poisson probability is not approximated well by a Gaussian, or when the answer is known to respect boundaries, e.g.\ Poisson expectation values be positive, or when regularization conditions distort the posterior's shape.  Such examples are shown.  Here, FBU is formulated for Poisson-distributed data, but an obvious modification in Eq.~\ref{eq:likelihood} would enable it to use non-Poisson data.

FBU differs from D'Agostini's iterative unfolding \cite{dAgostini, dAgostini2}, despite both using Bayes' theorem.  In FBU the answer is not an estimator and its covariance matrix, but a posterior probability density defined in the space of possible spectra. FBU does not involve iterations, thus does not depend on a convergence criterion\footnote{The so-called ``convergence criterion'' in iterative unfolding actually is there to \emph{prevent} convergence, but only after a reasonable number of iterations.  If the goal was convergence, then extra iterations would only make it better.}, nor on the first point of an iterative procedure, which in \cite{dAgostini} is named ``prior''.  If more than one answers are equally likely, as can happen when the reconstructed spectrum has fewer bins than the inferred one, then FBU reveals all of them, while iterative unfolding converges towards some of the possible solutions.  Regularization (see Sec.~\ref{sec:regularization}) is not done by interrupting iterations, but by choosing a prior which favors certain characteristics, such as smoothness.  Thus, FBU offers intuition and full control of the regularizing condition, which makes the answer easy to interpret.

FBU differs significantly also from SVD unfolding \cite{svd}.  In FBU the migrations matrix (defined in Sec.~\ref{sec:nomenclature}) is not distorted by singular value decomposition, therefore FBU assumes the intended migrations model.  The answer of FBU is not an estimator plus covariance matrix, but a probability density function which does not have to be Gaussian, which is important especially in bins with small Poisson event counts.  FBU does not involve matrix inversion and computation of eigenvalues, which makes it more stable numerically.  Finally, SVD imposes curvature regularization (see $S_2(\T)$ in Sec.~\ref{sec:regularization}), while FBU offers the freedom to use different regularization choices.  This freedom becomes necessary when the correct answer actually has large curvature, or when the answer has only two bins, thus curvature is not even defined.

\subsection{Nomenclature}
\label{sec:nomenclature}

The following definitions are used:

\begin{description}

\item[Spectrum:] A binned histogram showing the distribution of entries (``events'') in some observable quantity, $m$.

\item[Smearing:] Any stochastic effect which results in classifying (or ``reconstructing'') events in a wrong bin, i.e., in a bin other than where they would be if the true value of $m$ was always reconstructed.

\item[Truth:] A ``truth-level'' (or ``truth'') spectrum, contains in each bin ($t \in \{1,2,\dots,N_t\}$) a number $T_t \in \mathbb{R}$, which is the number of events \emph{expected} to be \emph{produced} in that bin, before reconstruction.  A truth spectrum is represented by a $N_t$-tuple $\T = (T_1,T_2,\dots,T_{N_t})$, corresponding to a point in a $N_t$-dimensional space.  Two symbols are reserved for two special $\T$-points: The truth-level spectrum from which the data (see below) actually originate, i.e., nature's truth-level spectrum is $\hat{\T}$.  The truth-level spectrum followed by the MC events that populate the migrations matrix (see below) is $\tilde{\T}$.  In the ideal case where MC reproduces reality, then $\tilde{\T} = \hat{\T}$.
  
\item[Reco:] A ``reco-level'' (or ``reco'') spectrum, contains in each bin ($r \in \{1,2,\dots,N_r\}$) a number $R_{r} \in \mathbb{R}$, which is the number of events \emph{expected} to be \emph{reconstructed} in that bin.  A reco spectrum is represented by a $N_r$-tuple $\R = (R_1,R_2,\dots,R_{N_r})$.
  
\item[Data:]  A spectrum where each bin ($r\in \{1,2,\dots,N_r\}$) contains a number $D_r \in \mathbb{N}$, which is the number of events \emph{observed} in that bin, after smearing obviously.  A data spectrum is represented by a $N_r$-tuple $\D = (D_1,D_2,\dots D_{N_r})$.   Without loss of generality, and since this is the most common use case in HEP, it is assumed that $D_r$ follows a Poisson distribution of mean $R_r$.
  
\item[Migrations matrix:] A matrix $\M$ whose element $\M_{tr}$ is the joint probability $P(t,r)$ of an event to be produced in the truth-level bin $t$ and reconstructed in the reco-level bin $r$.

\item[Response matrix:] A matrix $\P$ whose element $\P_{tr}$ is the conditional probability $P(r|t)$ for an event that was produced in the truth-level bin $t$ to be reconstructed in the reco-level bin $r$.

\item[Unfolded:] An unfolded spectrum helps visualize, at least in a limited way, the result of FBU.  Each bin ($t \in \{1,2,\dots,N_t\}$) contains a number $U_t \in \mathbb{R}$, which is the estimated value of the actual truth-level $\hat{T}_t$.  The error bars in each bin show the shortest interval where $\hat{T}_t$ is inferred to be with probability\footnote{Since FBU uses Bayes' theorem, these error bars define an interval that integrates 68\% probability, or ``credibility''.  This percentage should not be confused with frequentist coverage, and the bayesian credibility intervals should not be confused with confidence intervals.} 68\%.  Let this interval be denoted by $[U_t^\ulcorner,U_t^\urcorner]$.   More details in Sec.~\ref{sec:marginalizing}.  An unfolded spectrum is represented by a $N_t$-tuple $\U = (U_1,U_2,\dots,U_{N_t})$.
  
\end{description}

\section{Formulation}
\label{sec:formulation}

In unfolding, the question is:
\begin{quote}
``Given the data ($\D$) and the migrations model, what was the actual truth-level spectrum ($\hat{\T}$)?''
\end{quote}
This is a quintessentially bayesian question, since the true value of an unknown is asked.
It is answered by Bayes' theorem:
\begin{equation}
\label{eq:bayesWithM}
p(\T | \D,\M) = L(\D | \T,\M) \cdot \frac{\pi(\T,\M)}{\rm Norm.\ Const.}
\end{equation}
If the migrations model, represented here by $\M$, is not uncertain, it can be omitted to simplify the notation.  So,
\begin{equation}
\label{eq:bayes}
p(\T | \D) \propto L(\D | \T) \cdot \pi(\T).
\end{equation}

To each point $\T$ in the space of truth spectra corresponds some probability density, $p(\T|\D)$, that $\T$ be the correct truth-level spectrum $\hat{\T}$.  This probability density depends (i) on the observation ($\D$), (ii) on the smearing model encoded in $\M$, (iii) on the probability model followed by the data, e.g., Poisson, and (iv) on the prior probability density $\pi(\T)$.

Assuming that the data follow Poisson statistics, the likelihood is:
\begin{equation}
\label{eq:likelihood}
L(\D|\T) = \prod_{r=1}^{N_r} \frac{R_r^{D_r}}{D_r!} e^{-R_r},
\end{equation}
where
\begin{equation}
\label{eq:R}
R_r = \sum_{t=1}^{N_t} T_t\cdot P(r|t).
\end{equation}
$P(r|t)$ is the probability to be reconstructed in bin $r$, given that the truth-level bin (before smearing) is $t$.  This is extracted from $\M$:
\begin{equation}
  \label{eq:P}
  P(r|t) = \frac{P(t,r)}{P(t)} = \frac{\M_{tr}}{\epsilon_t^{-1}\sum_{k=1}^{N_r} \M_{tk}},
\end{equation}
where $\epsilon_t$ is the \emph{efficiency} of row $t$ of $\M$.  This is the probability that an event produced in truth-level bin $t$ will be reconstructed in one of the $N_r$ reco-level bins included in $\M$.  So,
\begin{equation}
\label{eq:eff}
\epsilon_t = \frac{\sum_{r=1}^{N_r}P(t,r)}{P(t)} = \frac{\sum_{r=1}^{N_r}M_{tr}}{P(t)}.
\end{equation}

Often the data include contamination from background processes, such as noise, or ``fakes'' (to use an example from HEP), namely events which are in \D\ but don't originate from $\hat{\T}$ at truth-level.  The total background has to be taken into account in the likelihood.  The only change, in this case, is that Eq.~\ref{eq:R} should become
\begin{equation}
\label{eq:RwithB}
R_r = B_r + \sum_{t=1}^{N_t} T_t\cdot P(r|t), 
\end{equation}
where $B_r$ is the expected number of background events in bin $r$.  In matrix notation,
\begin{equation}
\label{eq:RwithBmatrix}
\R = \B + \P^T\T. 
\end{equation}

This lays out the fundamentals of FBU.  The solution is written down analytically in Eq.~\ref{eq:bayes}, and much of the rest of this document deals with its computation.

\section{Conceptualization}
\label{sec:conceptualization}

A pervasive problem in unfolding is that the maximum likelihood estimator (MLE) of the truth-level spectrum, which is unbiased and given by $(\P^T)^{-1}\D$ when $\P$ is invertible, suffers from great variance.  Due to this, the MLE often looks unnatural, with bin contents that vary wildly in alternating directions from bin to bin, like saw teeth.  

{\em Regularization} is introduced in order to suppress this variance.  The classical description of regularization is given beautifully by Cowan in \cite{cowanUnfolding}: The MLE is the unbiased estimator with the smallest possible variance, even though the latter is huge.  In order to avoid this variance it is necessary to introduce bias to the estimator.

To introduce bias, instead of maximizing $L(\D|\T)$, the estimator is obtained by maximizing: 
\begin{equation}
\label{eq:classicReg}
\tilde{L}(\T) \equiv L(\D|\T)\cdot e^{-\alpha \cdot S(\T)},
\end{equation}
with $\alpha$ being a positive regularization parameter, and $S(\T)$ an arbitrary regularization function that increases with some undesired property, such as ``non-smoothness''. The resulting estimator is a compromise between high likelihood, which is desirable, and large $S(\T)$, which is not. 

The mode of $\tilde{L}(\T)$ is obviously not the MLE.  It is, however, the mode of the FBU posterior $p(\T|\D)$, since $\tilde{L}(\T) \propto p(\T|\D)$.  The latter becomes obvious if the prior in Eq.~\ref{eq:bayes} is rewritten as
\begin{equation}
\label{eq:priorMinusAlpha}
\pi(\T) = e^{-\alpha\cdot S(\T)}.
\end{equation}
Even without regularization, i.e., when $S(\T)$ is constant, or $\alpha=0$, the classical estimator is still the mode of the posterior, which then coincides with the MLE.  (See \cite{cowanBook}, Sec.~6.13).

The classical estimator is always (with or without regularization), the mode of the FBU posterior $p(\T|\D)$, after assuming the prior which corresponds to the same regularization (Eq.~\ref{eq:priorMinusAlpha}).

Without regularization, i.e., with a constant $\pi(\T)$, what classically is described as large variance of the MLE appears in FBU as large spread in $p(\T|\D)$, which is  $\propto L(\D|\T)$.  So, with FBU it is obvious why regularization may be desired: if the prior is constant the posterior may be too wide, therefore very different \T's may be almost equally likely.  The variance of the classical estimator that maximizes $\tilde{L}(\T)$, and the spread of $p(\T|\D)$ in FBU, are the two faces of the same coin.  

The regularization term $e^{-\alpha\cdot S(\T)}$ in Eq.~\ref{eq:classicReg} reduces the variance of the classical estimator, just like the prior $\pi(\T) = e^{-\alpha\cdot S(\T)}$ reduces the spread of $p(\T|\D)$ in FBU.

The term $e^{-\alpha\cdot S(\T)}$ in Eq.~\ref{eq:classicReg} is, classically, just a way to introduce bias to the classical estimator of $\hat{\T}$.  But at the same time it is the prior in the corresponding FBU.  The classical estimator can be understood as a half-done FBU, where, instead of the full $p(\T|\D)$, only its mode is computed.  Furthermore, instead of computing directly the probability of each \T\ to be the actual $\hat{\T}$, the classical procedure estimates the variance of the mode of $p(\T|\D)$ in pseudo-experiments where $\D$ is substituted with pseudo-data sampled from $\T_{MLE}$ (since $\hat{\T}$ is unknown).  The initial question, however, was not how much the mode of the posterior would vary in pseudo-experiments, but how likely each \T\ was to be $\hat{\T}$.  The variance of the posterior and the variance of its mode in pseudo-experiments are clearly related, but are not the same thing.

Seeing regularization as a choice of prior $\pi(\T)$ is natural.  Even classical regularization is nothing but an a-priori belief in the smoothness (or other property) of $\hat{\T}$.  That's why bias was not introduced by some absurd manipulation\footnote{With the exception of SVD, where distorting the response matrix seems physically unmotivated.}, but through a physically motivated $S(\T)$.  The prior is conceptually the host of any such prior beliefs.

\section{Generation of MC events}
\label{sec:generation}

\begin{figure}[H]
\centering
\subfigure[]{
\includegraphics[width=0.9\columnwidth]{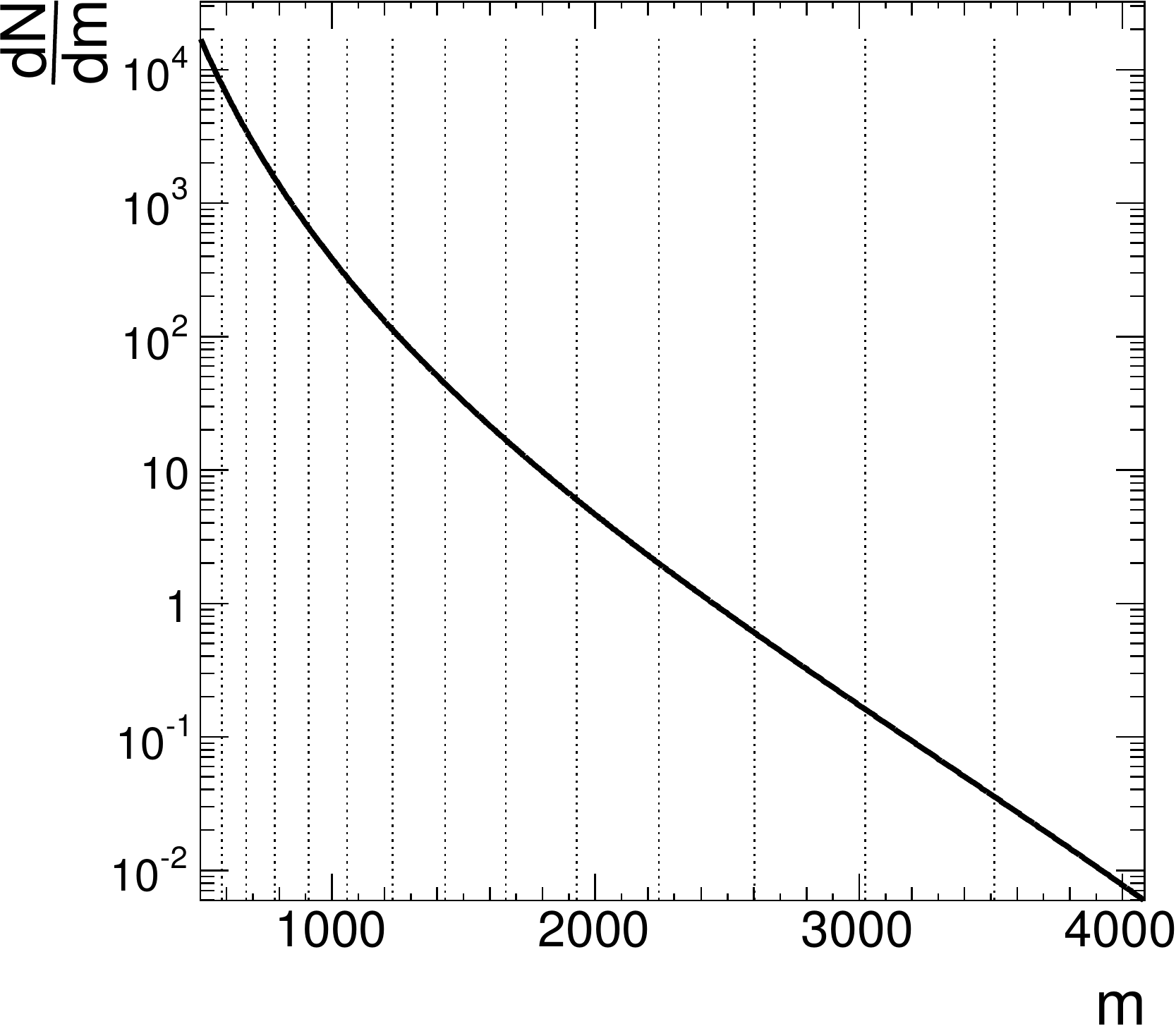}
\label{fig:truthSpectrum}}
\subfigure[]{
\includegraphics[width=0.9\columnwidth]{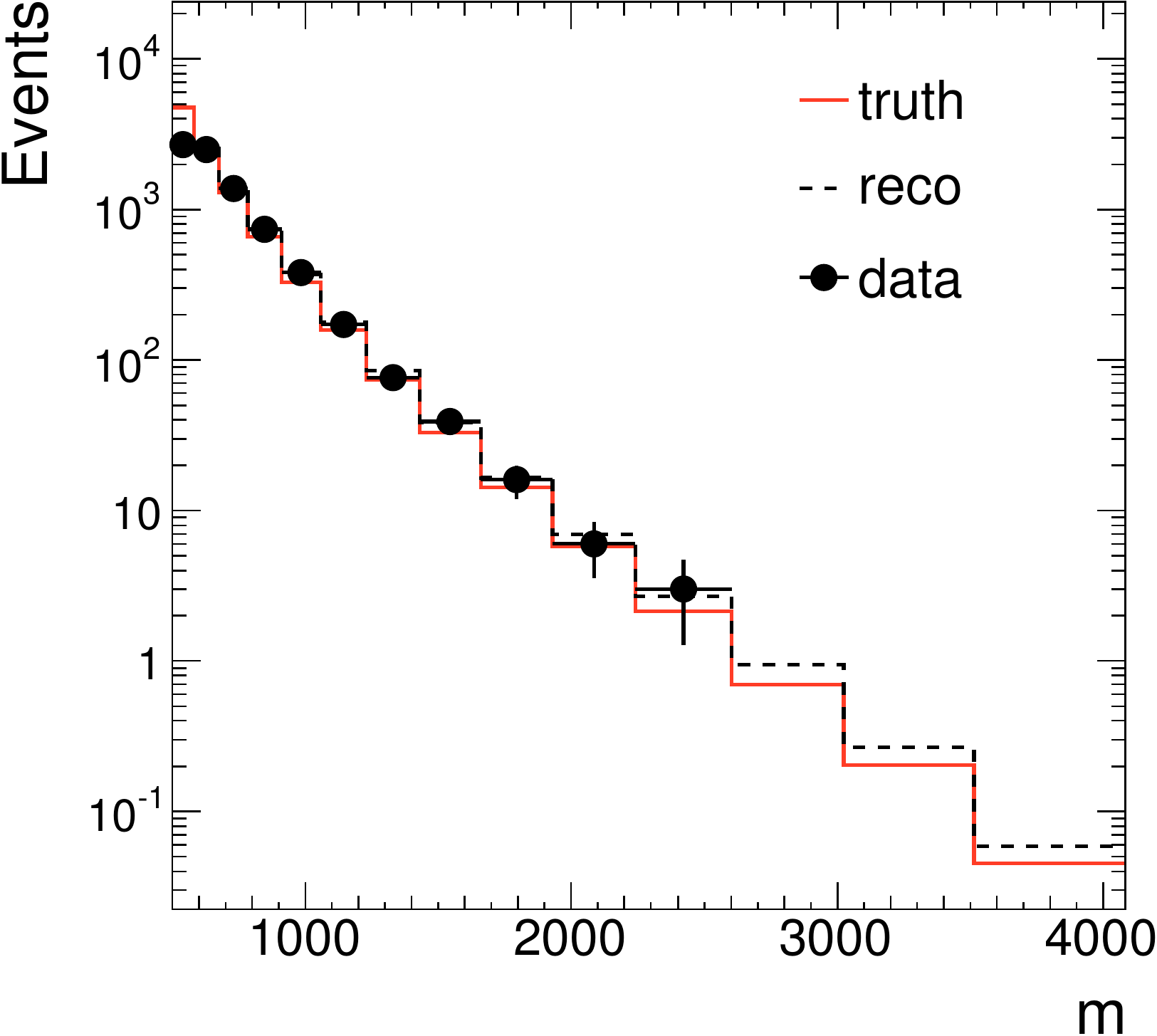}
\label{fig:truthAndReco}}
\caption{(a): The distribution, in observable $m$, that MC events follow (Eq.~\ref{eq:fgen}). The dotted lines indicate the delimiters of $m$ bins. (b): An example of generated MC events.  The red solid line shows the distribution of $10^7$ MC events, each of weight $10^{-3}$, before smearing.   The dashed black line shows their distribution after smearing.  The markers represent observed data, which are random numbers following a Poisson distribution of mean given by the reco-level spectrum.}
\end{figure}

To investigate FBU, Monte Carlo (MC) events are generated and smeared.  They are generated following
\begin{equation}
  \label{eq:fgen}
  \frac{\di N}{\di m} = \left(1-\frac{m}{7000}\right)^6 / \left(\frac{m}{7000}\right)^{4.8},
\end{equation}
shown in Fig.~\ref{fig:truthSpectrum}, which is inspired from the distribution of dijet masses in proton collisions at energy $\sqrt{s} = 7000$~GeV.  

The observable $m$ is binned.  The delimiters of 14 truth-level $m$ bins are set at
\begin{equation}
\label{eq:bins}
m_n = 500\cdot e^{n\cdot 0.15}, \text{\ for\ } 0\le n\le14.
\end{equation}
For simplicity, the same bins are defined in the reco-level $m$, although the formulation of Sec.~\ref{sec:formulation} is so general that the truth and reco bins don't need to share any common properties.  This will be demonstrated in Sec.~\ref{sec:example6}.

Each MC event, generated at $m=m_{\rm truth}$, is smeared and reconstructed at $m=m_{\rm reco}$, where
\begin{equation}
m_{\rm reco} = m_{\rm truth} + \delta m,
\end{equation}
where $\delta m$ is a random variable following a Gaussian distribution of mean 0 and standard deviation 
\begin{equation}
\label{eq:sigma}
\sigma(m_{\rm truth}) = m_{\rm truth} \left( \frac{a}{\sqrt{m_{\rm truth}}} + b \right),
\end{equation}
which is a simplified parametrization of the energy resolution of a calorimeter.

\begin{figure}[H]
\centering
\subfigure[]{
\includegraphics[width=0.8\columnwidth]{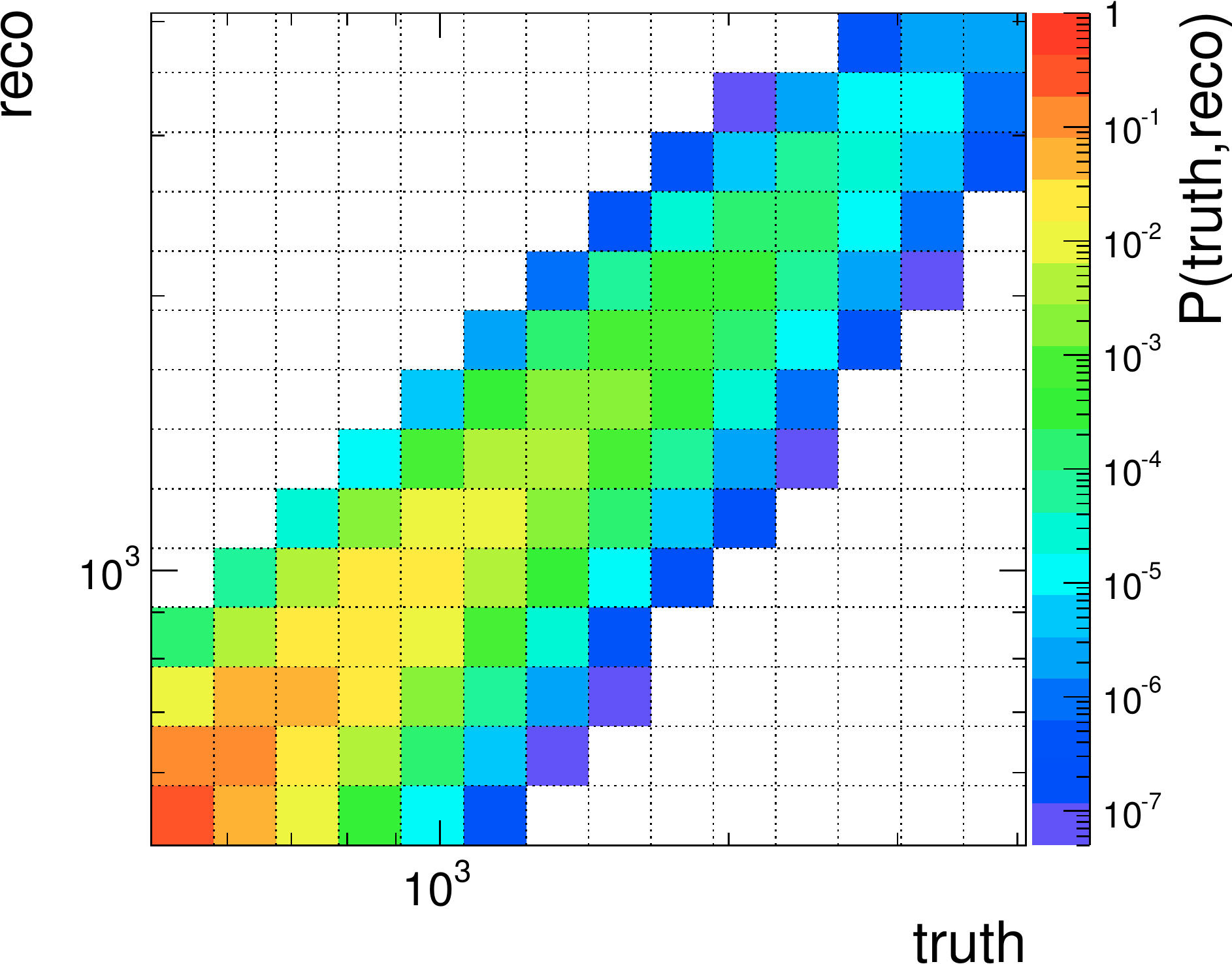} 
\label{fig:mm1}}
\subfigure[]{
\includegraphics[width=0.8\columnwidth]{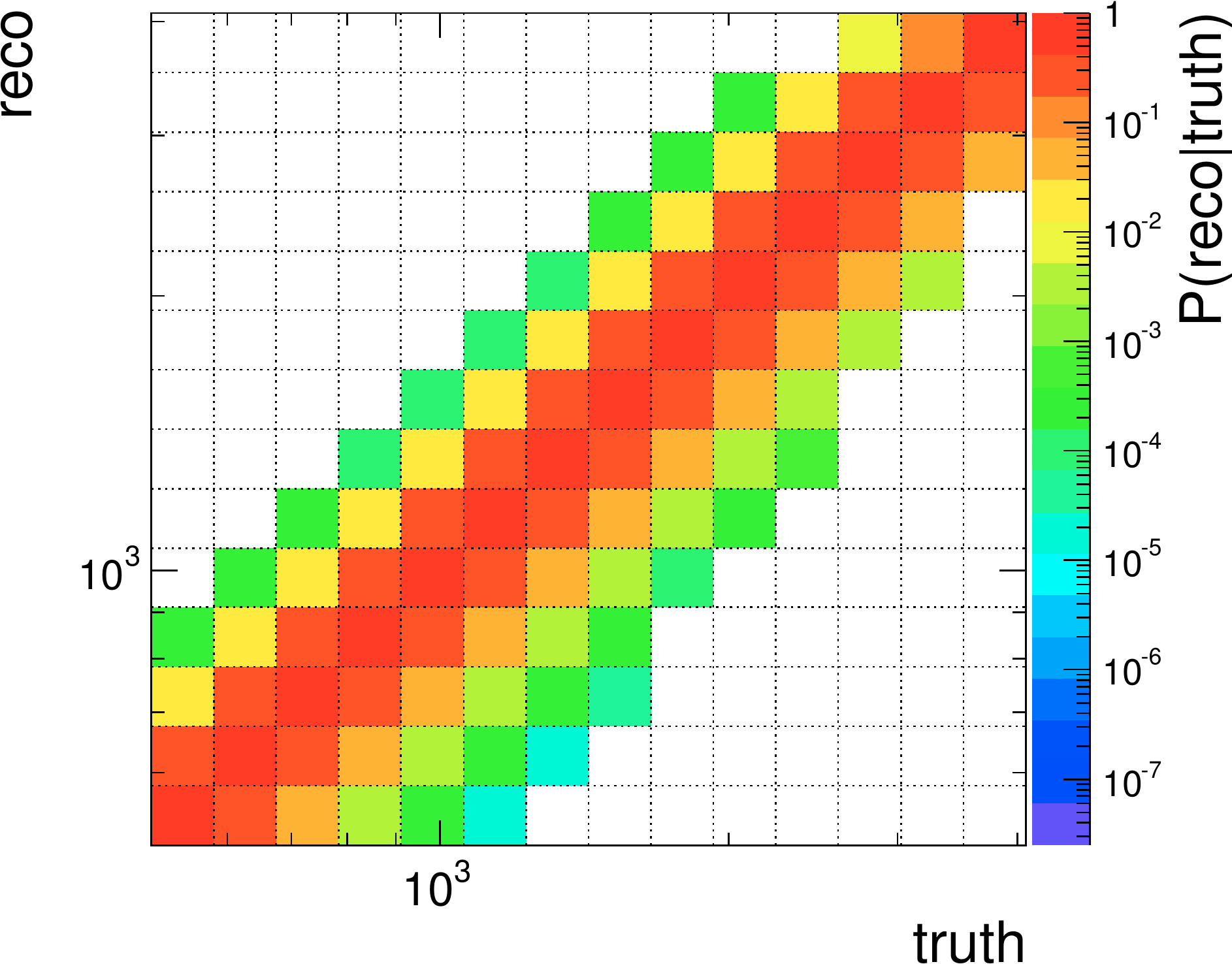} 
\label{fig:mmc1}}
\subfigure[]{\includegraphics[width=0.8\columnwidth]{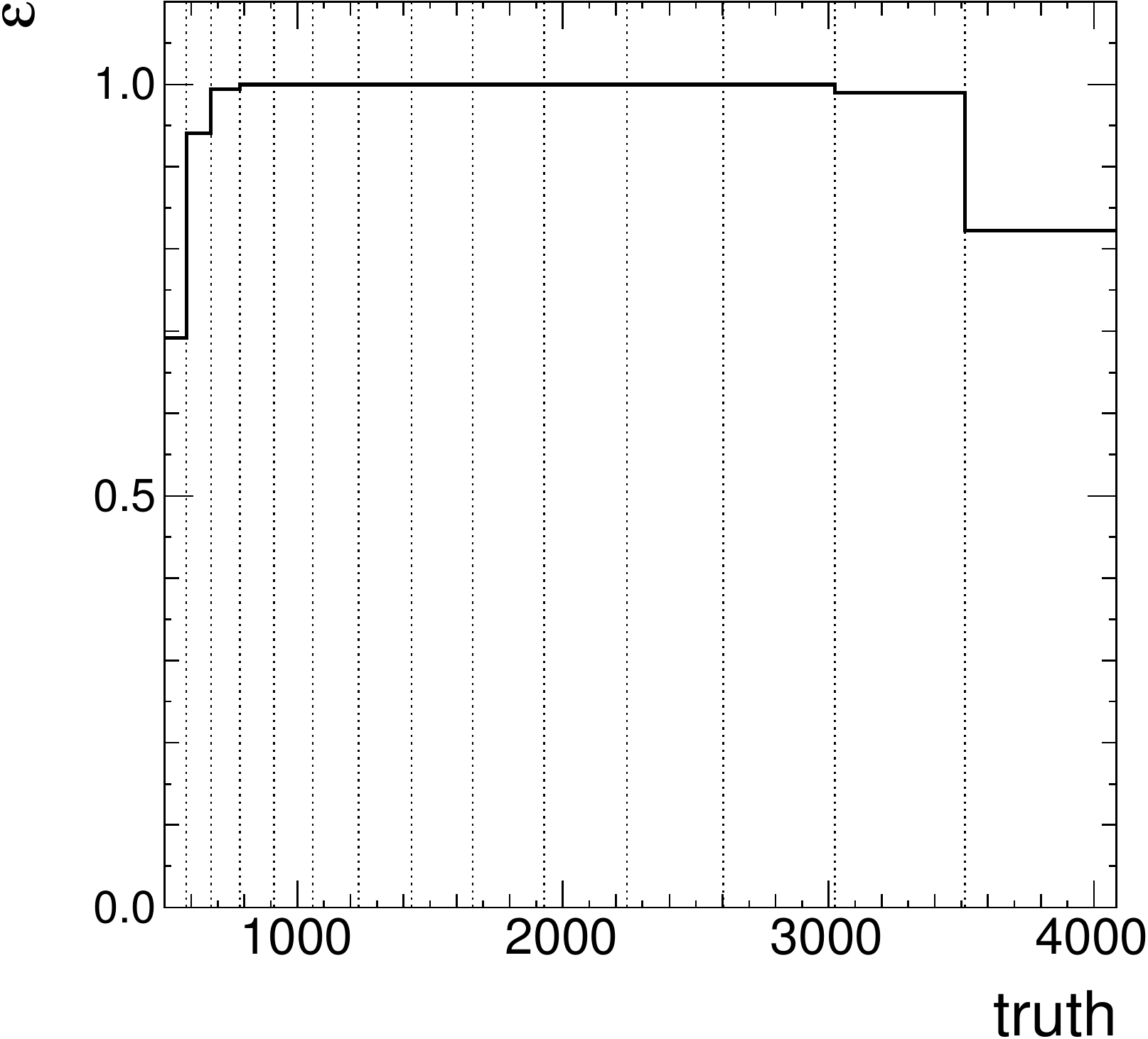} \label{fig:eff1}}
\caption{(a) The migrations matrix, $P({\rm truth},{\rm reco})$, for smearing defined by Eq.~\ref{eq:sigma} with $a=0.5$ and $b=0.1$.  (b) The matrix of $P(r|t)$ for the same smearing. (c) The migrations matrix efficiency of Eq.~\ref{eq:eff} for the same smearing.  The dotted lines indicate the bin delimiters in truth-level and reco-level $m$. \label{fig:genMat}}
\end{figure}

An example of the resulting $\M$, for $a=0.5$ and $b=0.1$, is in Fig.~\ref{fig:mm1}.  The corresponding efficiencies (Eq.~\ref{eq:eff}), are in Fig.~\ref{fig:eff1}.  Fig.~\ref{fig:mmc1} shows the corresponding response matrix $\P$ (see Sec.~\ref{sec:nomenclature} and Eq.~\ref{eq:P}).
Fig.~\ref{fig:truthAndReco} shows the truth, reco, and data spectra.

The migrations matrix \M\ in Fig.~\ref{fig:mm1} is ideal because (i) it reflects the smearing which indeed operated on the MC events, and (ii) it reflects the correct truth-level spectrum, because $\tilde{\T} = \hat{\T}$, by construction.  
In real analyses, $\M$ is not known to be ideal: (i)  the assumed migrations model might not be realistic, which should be treated as a source of systematic uncertainty, and (ii) the MC events used to compute $\M$ may not follow $\hat{\T}$.  Most notably, $\tilde{\T} \neq \hat{\T}$  when an exotic process generates unexpected events.  This scenario is examined in Sec.~\ref{sec:bumpUnknown}.

\section{Living in $N_t$ dimensions}

A technical difficulty with FBU is that $\pi(\T)$, $L(\D|\T)$, and $p(\T|\D)$, are defined in $N_t$ dimensions.  Defining $\pi(\T)$ in $N_t$ dimensions is not difficult, if limited to simple priors.   On the other hand, sampling $L(\D|\T)\cdot\pi(\T)$ in many dimensions is challenging, because it usually is zero at most $\T$-points, and it is difficult to locate the region where it is non-zero.  However, significant progress has been possible through the development of Markov Chain Monte Carlo (MCMC) sampling algorithms.

\subsection{Defining priors}
\label{sec:priors}

The simplest $\pi(\T)$ is a constant (``flat'') prior.  In the case of Poisson-distributed data, $T_t$ has to be positive in all $t$ bins, therefore the simplest $\pi(\T)$ is a $N_t$-dimensional ``step'' function:
\begin{equation}
\pi(\T) \propto \begin{cases}
  1& \text{if\ } T_t > 0 \ \forall t\in[1,N_t] \\
  0& \text{\ otherwise.}
\end{cases}
\end{equation}

For reasons related to sampling (Sec.~\ref{sec:sampling}), it is unpractical to allow non-zero $\pi(\T)$ in the whole, or even half of the $N_t$-dimensional space. Only a region of finite volume should be allowed, by setting $\pi(\T)=0$ outside of it.  This leads to a ``box'' prior, which is constant within some finite $N_t$-dimensional rectangle, or \emph{hyper-box}, extending from $T^\ulcorner_t$ to $T^\urcorner_t$ in dimension $t$:
\begin{equation}
\label{eq:boxPrior}
\pi(\T)\propto \begin{cases}
1& \text{if\ } T_t \in [T^\ulcorner_t , T^\urcorner_t]\ \forall t\in[1,N_t] \\
0 & \text{otherwise.}
\end{cases}
\end{equation}
As long as $T^\ulcorner_t > 0$ for all $t$, the upper edge $T^\urcorner_t$ can be arbitrarily large, so, practically the allowed hyper-box can be large enough to be confident that it contains $\hat{\T}$ and effectively all of $p(\T|\D)$, except for negligible tails that converge exponentially to 0.

With the prior of Eq.~\ref{eq:boxPrior}, nothing is presumed about $\hat{\T}$, except being within a finite volume.  To impose some regularization choice through $S(\T)$ and $\alpha$ (see Sec.~\ref{sec:conceptualization}), one simply uses a non-constant, finite-volume prior:
\begin{equation}
\label{eq:regulPrior}
\pi(\T)\propto \begin{cases}
e^{\alpha S(\T)}& \text{if\ } T_t \in [T^\ulcorner_t , T^\urcorner_t]\ \forall t\in[1,N_t] \\
0 & \text{otherwise.}
\end{cases}
\end{equation}

\subsection{Sampling}
\label{sec:sampling}

Three sampling strategies are implemented to study FBU:
\begin{description}
\item[Grid sampling:] Evaluating $L(\D|\T)\cdot\pi(\T)$ at $n_G$ equally spaced positions along each dimension.  This results in $n_G^{N_t}$ samples, at the nodes of a regular cartesian grid.  The rapidly increasing number of samples is the reason no examples of this sampling method are presented, as this is only practical in cases of $N_t \lesssim 4$.
\item[Uniform sampling:] Sampling $L(\D|\T)\cdot\pi(\T)$ at uniformly distributed random points in $\T$-space.
\item[MCMC:] Using a variation of the Metropolis-Hastings algorithm.  
The first sample of $L(\D|\T)\cdot\pi(\T)$ is taken at $\T = \tilde{\T}$.
The next sample position is proposed randomly, following a uniform distribution within a hyper-box centered at the latest sampled position.  Typically, the hyper-box has edge length $\frac{T_t^\urcorner-T_t^\ulcorner}{100}$ along dimension $t$, though this may be adjusted differently if necessary.  If the proposed point has greater $L(\D|\T)\cdot\pi(\T)$ than the latest sample, it is accepted, and it is registered as the next sample.  Otherwise, the ratio is found between the $L(\D|\T)\cdot\pi(\T)$ at the proposed point and the latest sample, and it is used as the probability to accept the proposed $\T$-point.  The result of this algorithm is a random walk which drifts towards the most likely region in the $\T$-space, and samples it.  Improvements, such as MCMC with adaptive step size, are possible.  However, this basic implementation suffices for the purposes of this study.
\end{description}

\subsection{Marginalizing}
\label{sec:marginalizing}

Each sample is a value of $L(\D|\T)\cdot\pi(\T)$, which is proportional to $p(\T|\D)$, at a given point $\T$ in the $N_t$-dimensional $\T$-space.  The set of samples contains information about the shape of $p(\T|\D)$ in $\T$, which is the essential output of FBU, but is difficult to visualize.  One typically wants the unfolded spectrum (see Sec.~\ref{sec:nomenclature}).

To produce the unfolded spectrum,  one must first compute from $p(\T|\D)$ a set of 1-dimensional marginal posteriors, $p_t(T_t|\D)$, for $t\in[1,\dots,N_t]$.  The shortest interval in $T_t$ that integrates 68\% of $p_t(T_t|\D)$, denoted by $[U_t^\ulcorner,U_t^\urcorner]$, is shown as error bars in bin $t$ of the unfolded spectrum.

There are several options to define the bin content, $U_t$:
\begin{enumerate}[i)]
\item $U_t$ be the most likely value of $T_t$.  It may take too many samples to accurately estimate the mode of $p_t(T_t|\D)$ when $N_t$ is large. \label{item:Umax}
\item $U_t$ be the expectation value of $T_t$. \label{item:Uexp}
\item $U_t$ be in the middle of $[U_t^\ulcorner , U_t^\urcorner]$.  Unlike \ref{item:Umax} and \ref{item:Uexp}, this does not require handling any asymmetric error bars. \label{item:used}
\item $U_t$ be the mean of a Gaussian (or other function) fitted to $p_t(T_t|\D)$, and $[U_t^\ulcorner,U_t^\urcorner]$ reflect the mean $\pm$ standard deviation of this Gaussian.  The assumption of a parametrization for $p_t(T_t|\D)$ is problematic, especially if $\pi(\T)$ is not constant, and fitting is a source of possible complications.
\end{enumerate}
Option \ref{item:used} is used, for its practical advantage, since $U_t$ itself is not as important as $[U_t^\ulcorner , U_t^\urcorner]$.   $U_t$ probably should not be used in any computation by itself, whereas  $[U_t^\ulcorner , U_t^\urcorner]$ could be compared to some theoretical $T_t$, to see if it is favored.  One should remember, after all, that in FBU the real answer is $p(\T|\D)$, and the unfolded spectrum is a mere device to visualize that.

The rest of this section discusses how samples are used to compute marginal posteriors from $p(\T|\D)$.

The marginal probability density in dimension $t$ is
\begin{equation}
p_t(T_t|D) = \iint  p(T|D) \di T_1\dots\di T_{t-1}\di T_{t+1}\dots \di T_{N_t}.
\end{equation}
From this, one defines a finely binned probability distribution, with a parameter $\delta$ controlling the bin size along $T_t$.  The probability integrated in $[T_t-\delta,T_t+\delta]$ is\footnote{Note the different notation: $p_t$ is used for the marginal probability \emph{density}, and $P_t$ for the \emph{total} marginal probability in fine bins of $T_t$.}
\begin{equation}
P_t(T_t|\D) = \int_{T_t-\delta}^{T_t+\delta} p_t(T'_t|\D) \di T'_t.
\end{equation}
Let there be $N_s$ samples, with sample $i\in[1,N_s]$ taken at point $\T_i=(T_{i,1},T_{i,2},\dots,T_{i,N_t})$.  The value of this sample is $w_i = L(\D|\T_i)\cdot \pi(\T_i) \propto p(\T_i|\D)$.  The boolean 
\begin{equation}
  \mathcal{B}(i,t,T_t) \equiv
\begin{cases} 
1&, \text{if\ } T_{i,t} \in [T_t-\delta,T_t+\delta), \\
0&, \text{otherwise}
\end{cases}
\end{equation}
allows to sum only the samples ($i$) for which the $t^{\rm th}$ element of $\T_i$ belongs in the bin which contains the value $T_t$:
\begin{equation}
W(t,T_t) \equiv \sum_{i=1}^{N_s} w_i \mathcal{B}(i,t,T_t).
\end{equation}

If the $N_s$ samples are uniformly distributed in the $\T$-space, then
\begin{equation}
\label{eq:convergence}
\lim_{N_s\to\infty}\frac{W(t,T_t)}{const} = P_t(T_t|\D),
\end{equation}
where $const$ is a normalization constant independent of $T_t$.  

One needs to be careful with MCMC sampling (Sec.~\ref{sec:sampling}), because it does not sample the $\T$-space uniformly.  This invalidates Eq.~\ref{eq:convergence}.  To make it clear, consider that the MCMC random walk has reached an equilibrium where the samples are distributed according to $p(\T|\D)$.  Then, $W(t,T_t)$ would converge towards $P_t(T_t|\D)^2$, and $\sum_{i=1}^{N_s} \mathcal{B}(i,t,T_t)$ would converge towards $P_t(T_t|\D)$ instead.  For this reason, only when the MCMC sampling method is used, $w_i$ is set to 1 for all samples.  However, even this is not enough to ensure that $P_t(T_t|\D)$ is computed correctly, because there is no guarantee that the MCMC has reached equilibrium.  For this reason, MCMC is used in combination with uniform sampling, for which Eq.~\ref{eq:convergence} holds, in the way explained in Sec.~\ref{sec:narrowing}.  

Finally, 2-dimensional marginal posteriors are obtained in a similar way, showing $P_{t_1,t_2}(T_{t_1},T_{t_2}|D)$, to visualize the correlation between the contents of truth-level bins $t_1$ and $t_2$.

\subsection{Volume reduction}
\label{sec:narrowing}

In Sec.~\ref{sec:marginalizing} it is explained that MCMC is not as trustworthy for marginalizing $p(\T|\D)$ as uniform sampling.  The latter, however, is not as efficient when the sampled hyper-box is very large(Sec.~\ref{sec:sampling}).

For this reason, an optional procedure is used to reduce the volume of the a-priori allowed hyper-box (Sec.~\ref{sec:priors}).  The initial hyper-box can be large enough to be confident that it contains $\hat{\T}$.  In such a large volume, the MCMC random walk travels towards the region of large $p(\T|\D)$, and samples it.  Soon MCMC approaches the equilibrium mentioned in Sec.~\ref{sec:marginalizing}, and the marginal posteriors $P_t(T_t|\D)$ are determined with adequate accuracy for the purpose of the next step:  In each dimension $t \in [1,N_t]$, the shortest interval containing 99\% of $P_t(T_t|\D)$ is found.  Then, this interval is used to redefine the boundaries of the allowed hyper-box in each dimension.

When the allowed hyper-box is reduced, and located where $L(\D|\T)\cdot \pi(\T)$ is large, it is easier to use a uniform random sampling.

Note that volume reduction is optional, and its motivation is merely practical.

Also, in practice one can immediately see if the reduced hyper-box is too narrow in some dimension, because the corresponding marginal posterior will drop to 0 abruptly at the boundary of the hyper-box.  It is obviously better to not allow the boundaries of the hyper-box to interfere with the posterior, especially if the chopping is drastic. Fortunately, this is easy to detect and avoid in practice, when the posterior needs to be very precise even in the tails.





\section{Applications without regularization}
\label{sec:applications}

Various examples of FBU follow, applying the devices described above, to understand the behavior of FBU, and hopefully some more general characteristics of unfolding.  No regularization is used yet; that will be the subject of Sec.~\ref{sec:regularization}.

Some of the aspects to investigate are:
\begin{enumerate}
 \item High vs low statistics data.
 \item Low dimensionality ($N_t$) vs large.
 \item Heavy smearing vs little (or no) smearing.
 \item Having $N_r = N_t$ vs $N_r \neq N_t$.
 \item Sampling strategies and how they affect convergence.
 \item Building $\M$ with MC that follows the actual truth spectrum ($\tilde{\T} = \hat{\T}$), vs allowing some unexpected features ($\tilde{\T} \neq \hat{\T}$).
\end{enumerate}

\subsection{No smearing, 2 bins, high statistics}
\label{sec:example1}

\begin{figure}[H]
\centering
\includegraphics[width=0.9\columnwidth]{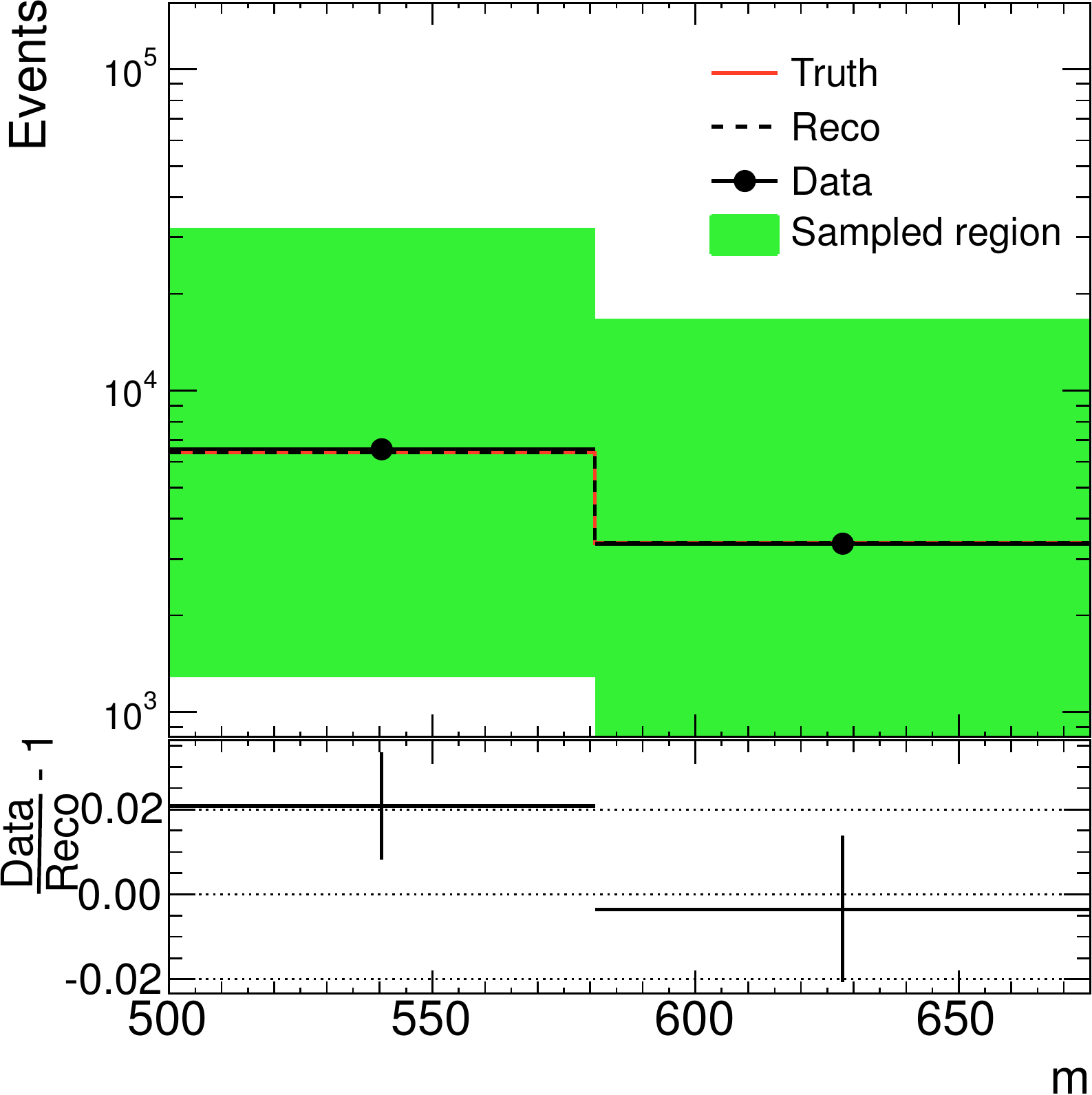}
\caption{Input truth-level, reco-level, and data spectrum of the example in Sec.~\ref{sec:example1}.  Due to assuming no smearing, the truth and the reco spectra coincide.  The green area shows the initial sampled hyper-box (Sec.~\ref{sec:priors}).   \label{fig:recoTruthDataPrior1}}
\end{figure}

\begin{figure}[H]
\centering
\includegraphics[width=0.9\columnwidth]{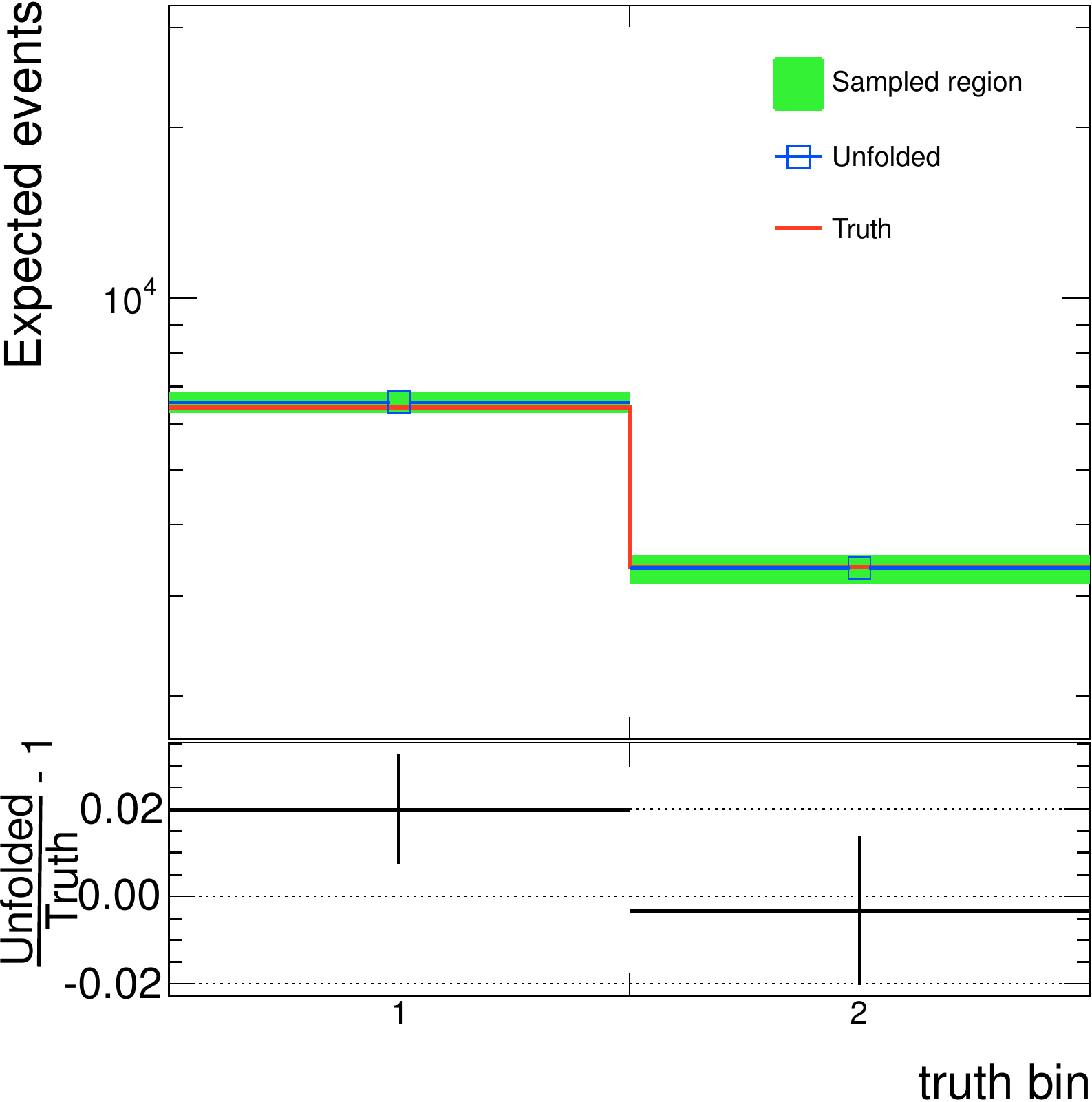}
\caption{The unfolded spectrum, compared to the truth spectrum, in the example of Sec.~\ref{sec:example1}.   \label{fig:unfolded1}}
\end{figure}

The simplest example involves just two reco-level and truth-level bins ($N_r=N_t=2$).  Events are generated in the first 2 bins defined in Eq.~\ref{eq:bins}: $\{[m_0,m_1),[m_1,m_2)\}$.
To have no smearing, $a$ and $b$ in Eq.~\ref{eq:sigma} are set to 0.  The result is the $2\times 2$ diagonal migrations matrix with elements
\begin{equation}
\begin{split}
 P(t=1,r=1)&=\frac{T_1}{T_1+T_2}=0.66,\\
 P(t=2,r=2)&=0.34, \\
 P(1,2)&=P(2,1)=0,
\end{split}
\end{equation}
which has efficiency $\epsilon_1=\epsilon_2=1$.
Fig.~\ref{fig:recoTruthDataPrior1} shows the input data, and the initial sampled region, which is reduced (see Sec.~\ref{sec:narrowing}) into the one shown in Fig.~\ref{fig:unfolded1}.

\begin{figure}[H]
\centering
\includegraphics[width=0.9\columnwidth]{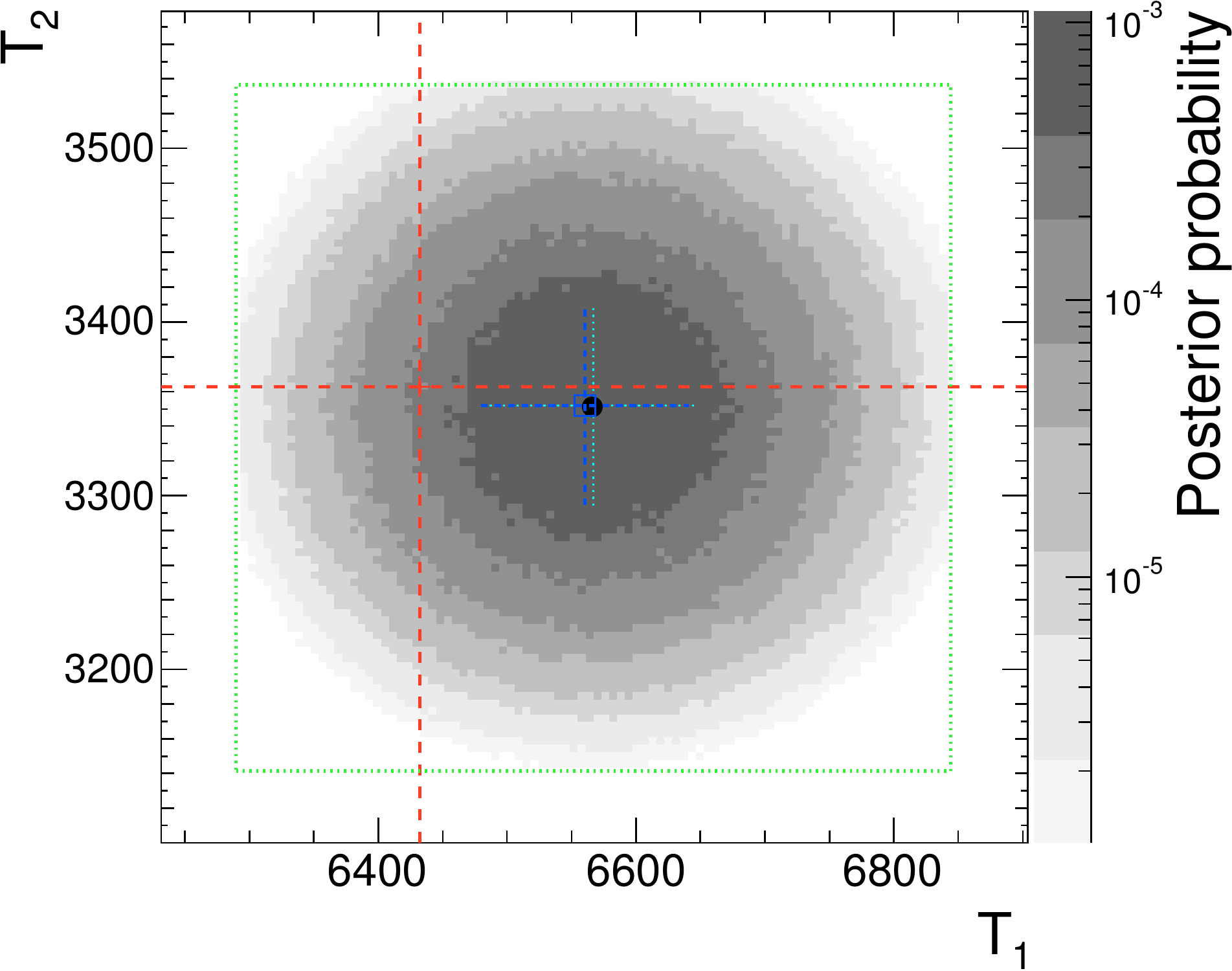}
\caption{For the example of Sec.~\ref{sec:example1} the following are shown: The 2-dimensional $P_{1,2}(T_1,T_2|\tuple{D})$, in gray.  The cyan dotted lines cross at $(\langle T_1 \rangle, \langle T_2 \rangle)$, where $\langle T_t \rangle = \int T_t p(\tuple{T}|\tuple{D}) \di \tuple{T}$, and their half-width is equal to the RMS of $p_{t}(T_t|\tuple{D})$ for $t=\{1, 2\}$.  The green doted box shows the limits of the sampled region.  The red dashed lines cross at the correct values of $(T_1,T_2)$.  The black circle corresponds to the observed data $(D_1,D_2)$.  The blue dashed lines and the empty blue square marker indicate the content of the unfolded spectrum. \label{fig:2Dim1}}
\end{figure}

The full result of FBU, $p(\T|\D)$, is easy to visualize when $N_t=2$, as in Fig.~\ref{fig:2Dim1}.  Since no smearing is assumed, the reco-level spectrum is equal to the truth-level.  The data are obviously somewhat different, due to Poisson fluctuations.   As a result of assuming no migrations, (i) the posterior probability distribution peaks around the observed data, and (ii) $T_1$ and $T_2$ are uncorrelated.   In the bottom inset of Fig.~\ref{fig:unfolded1} it becomes clear that the unfolded spectrum differs from the actual truth-level spectrum as much as the data spectrum differs from the truth-level (and reco-level) spectrum (Fig.~\ref{fig:recoTruthDataPrior1}).

\subsection{No smearing, 2 bins, low statistics}
\label{sec:example2}

\begin{figure}[H]
\centering
\includegraphics[width=0.9\columnwidth]{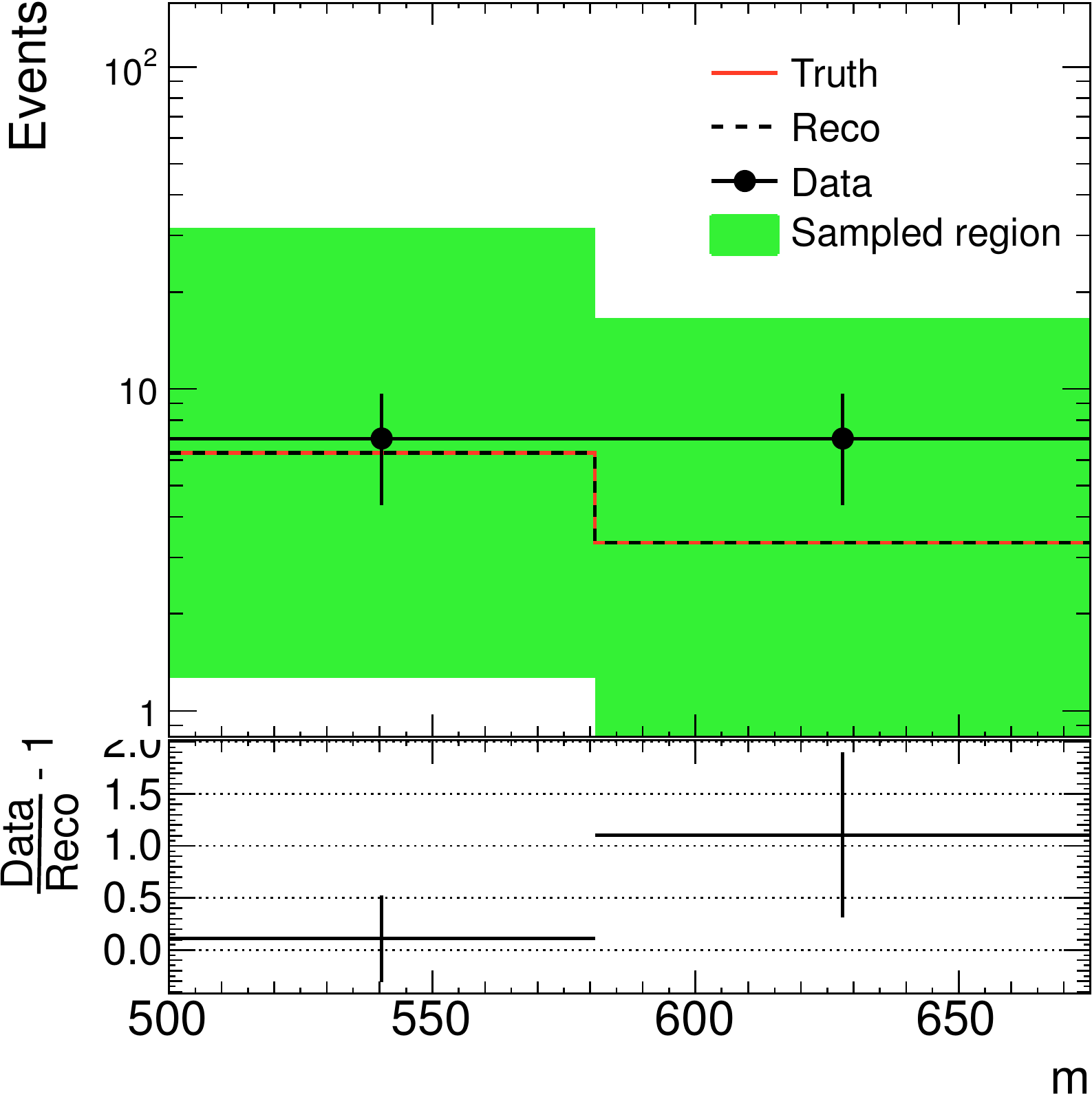}
\caption{Input spectra of the example in Sec.~\ref{sec:example2}.  \label{fig:recoTruthDataPrior2}}
\end{figure}

Keeping all the settings of the example in Sec.~\ref{sec:example1}, with only one difference:  The MC events are generated with 1000 times smaller weight, resulting in the data spectrum of lower statistics in Fig.~\ref{fig:recoTruthDataPrior2}.

\begin{figure}[H]
\centering
\includegraphics[width=0.9\columnwidth]{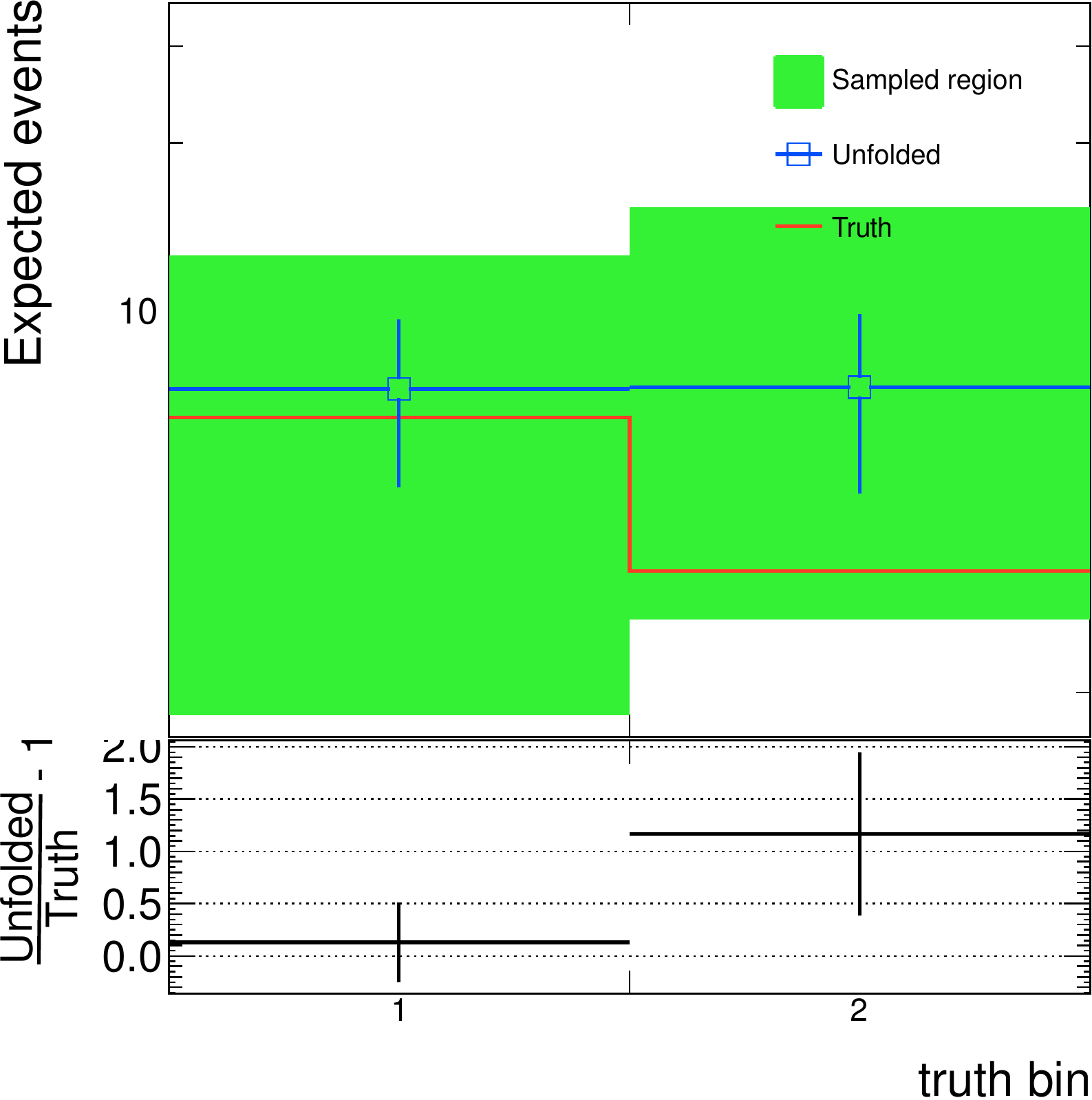}
\caption{The unfolded spectrum, compared to the truth spectrum, in the example of Sec.~\ref{sec:example2}.   \label{fig:unfolded2}}
\end{figure}

The unfolded spectrum, and the volume-reduced sampled region, are shown in Fig.~\ref{fig:unfolded2}.  It remains true, as in Sec.~\ref{sec:example1}, that the unfolded spectrum differs from the actual truth-level spectrum by as much as the data differ from the reco-level (and truth-level) spectrum.

\begin{figure}[H]
\centering
\includegraphics[width=0.9\columnwidth]{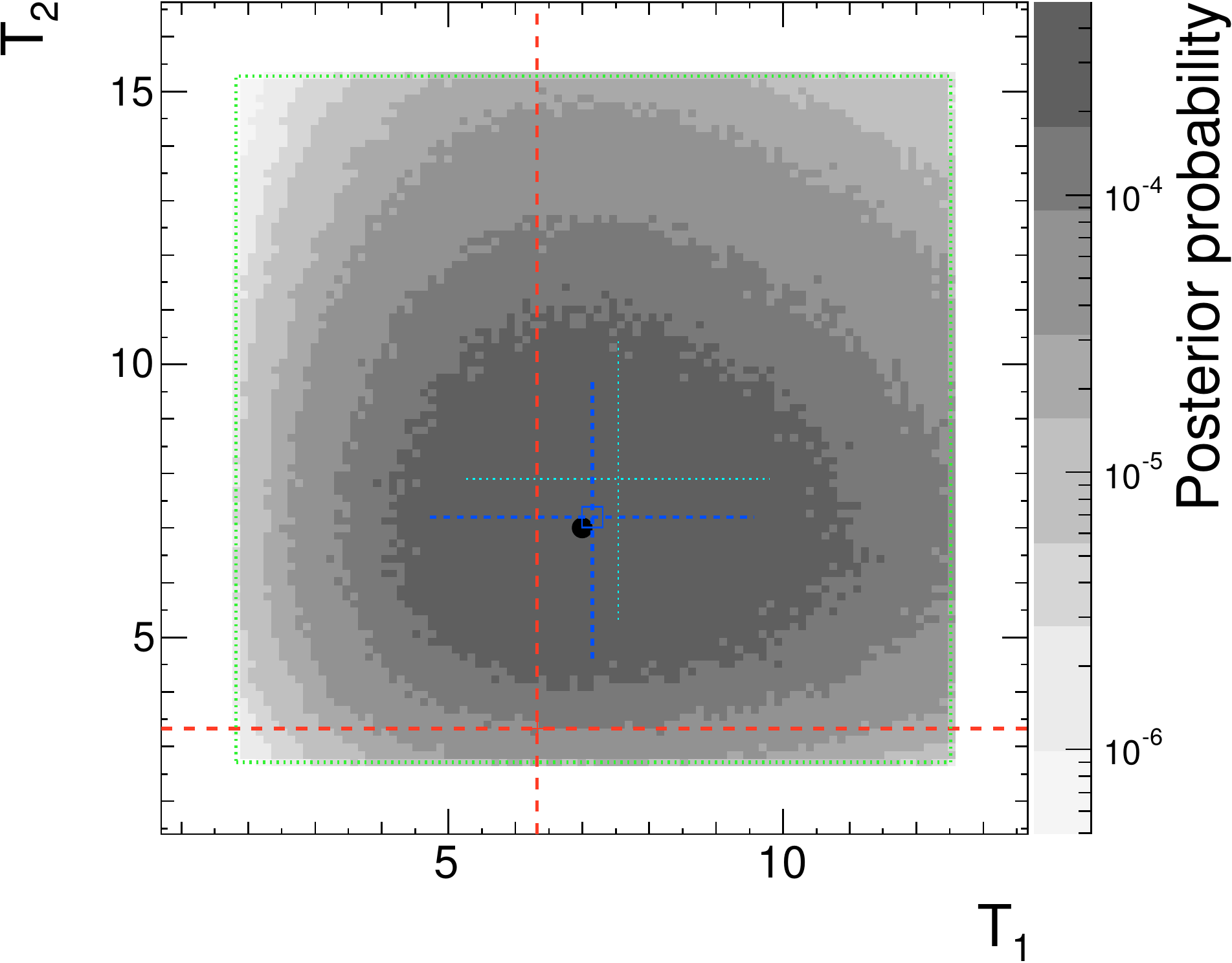}
\caption{Corresponding to Fig.~\ref{fig:2Dim1}, but for the example in Sec.~\ref{sec:example2}.  \label{fig:2Dim2}}
\end{figure}

The full $p(\T|\D)$ is in Fig.~\ref{fig:2Dim2}.  Unlike Fig.~\ref{fig:2Dim1}, where the $p(\T|\D)$ is a nearly perfect 2-dimensional Gaussian, in Fig.~\ref{fig:2Dim2}, due to low data statistics, the Poisson asymmetric shape is visible in both dimensions.  The maximum likelihood, as in Sec.~\ref{sec:example1}, remains at the point $\D=(D_1,D_2)$, but here the difference between the expectation value $E(\T) = (E(T_1),E(T_2))$ and the most likely $\T$ is clear due to the asymmetry of $p(\T|\D)$.

\subsection{More smearing, in 2 bins}
\label{sec:example3}

To show the effect of smearing, various degrees of smearing will be applied, and the corresponding migrations matrices will be used to perform FBU in a spectrum with $N_t=N_r=2$.

\begin{figure}[H]
\centering
\subfigure[]{
  \includegraphics[width=0.5\columnwidth]{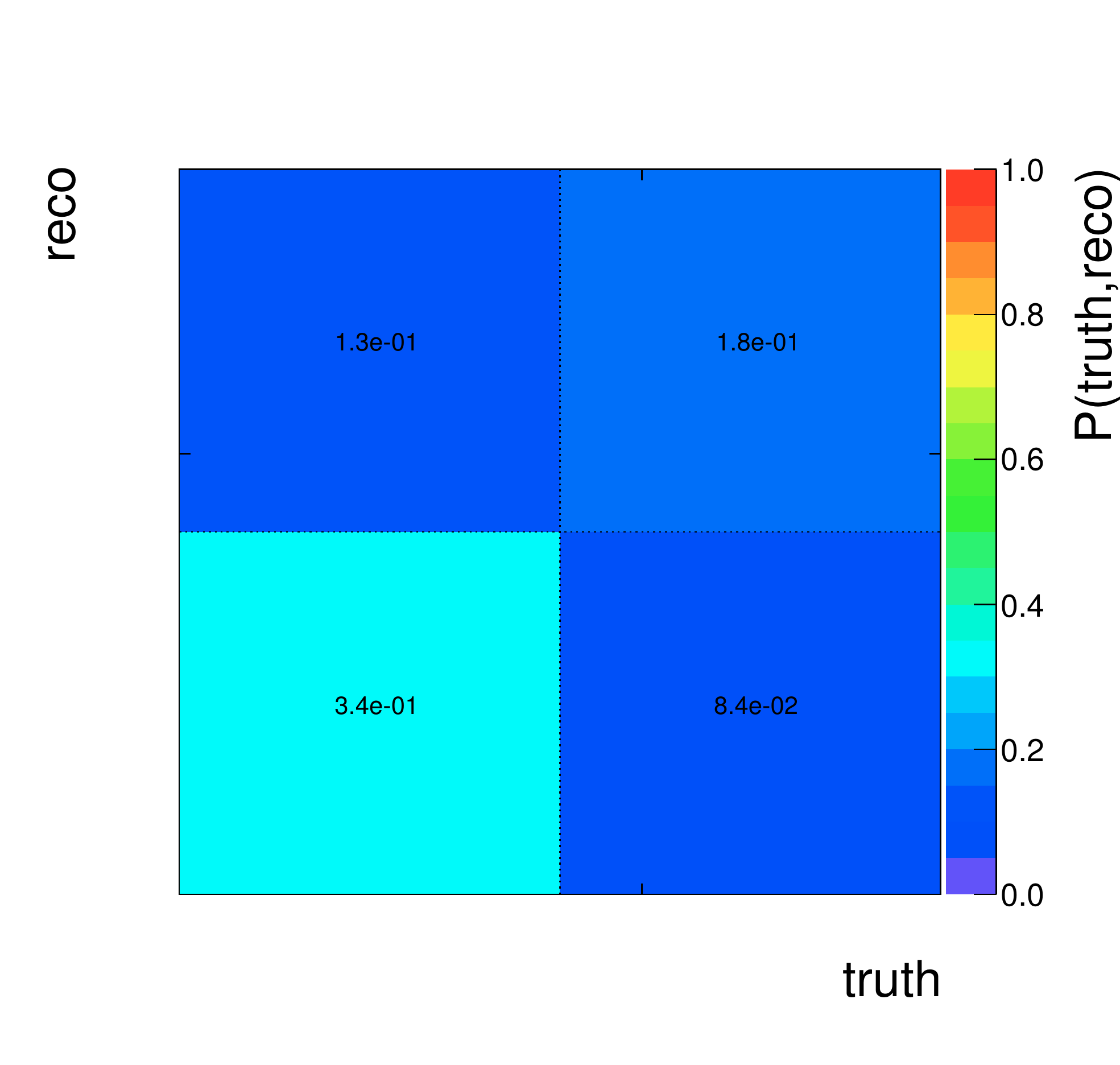}
  \includegraphics[width=0.5\columnwidth]{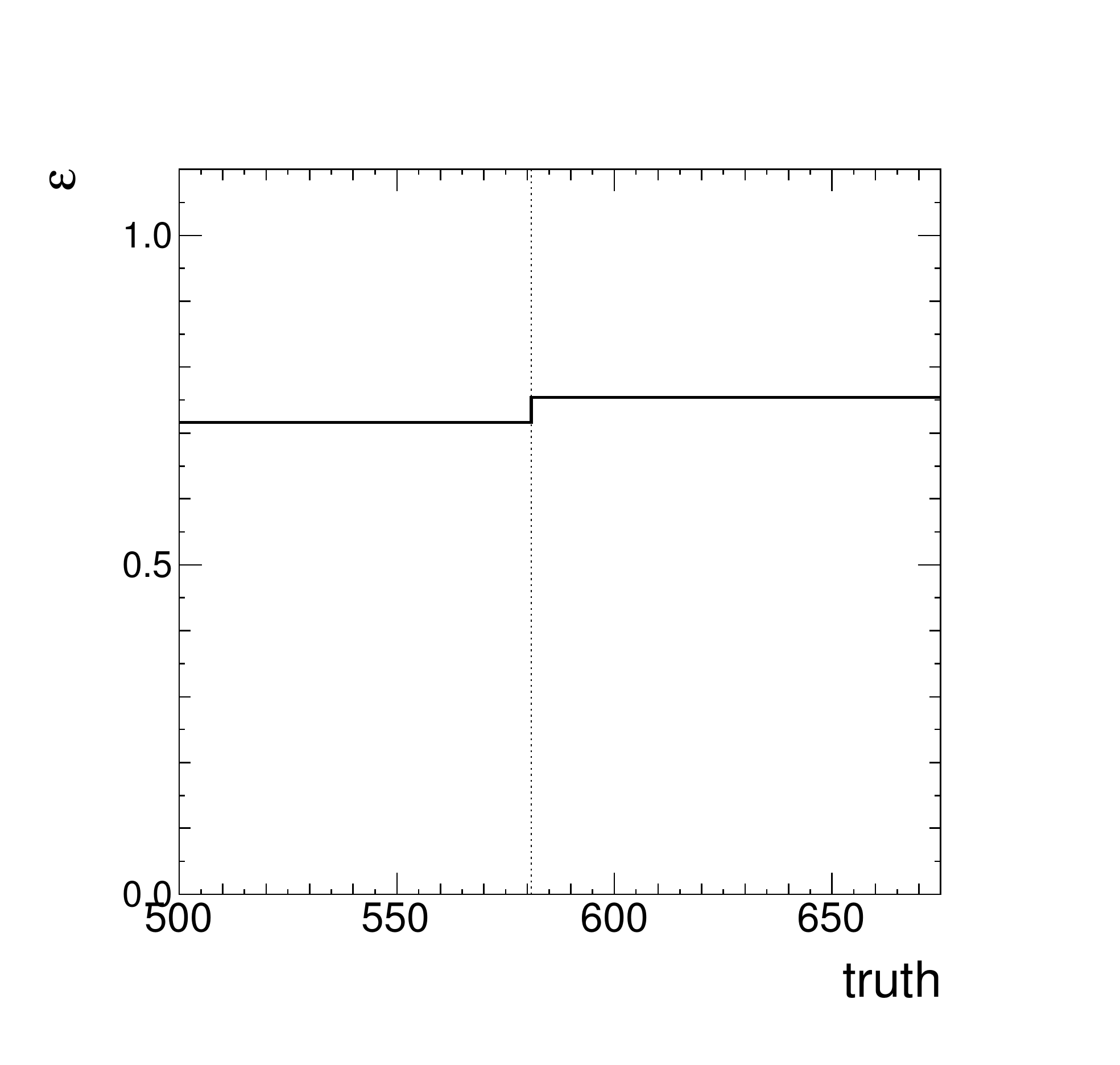}
}\\
\subfigure[]{
  \includegraphics[width=0.5\columnwidth]{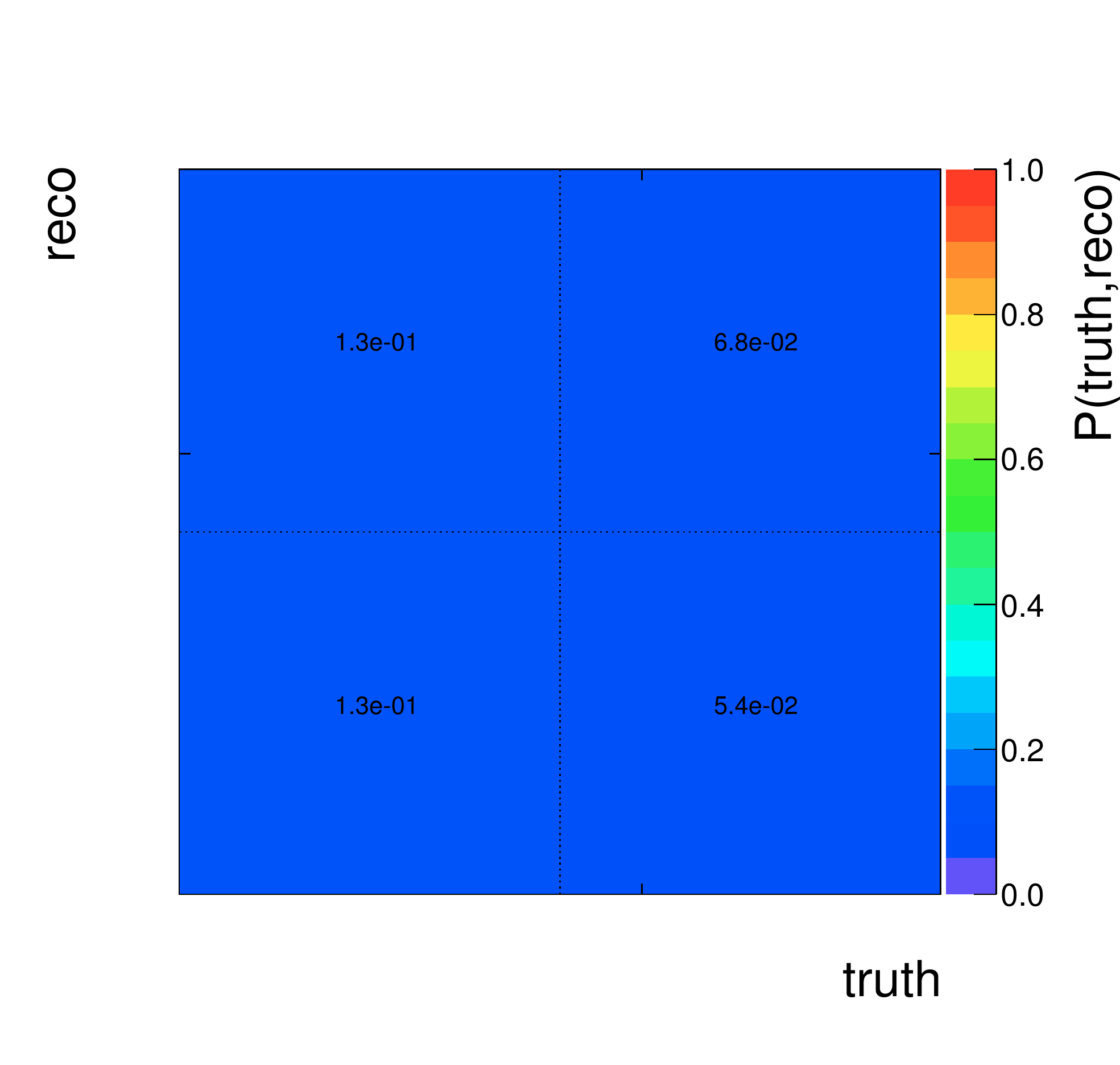}
  \includegraphics[width=0.5\columnwidth]{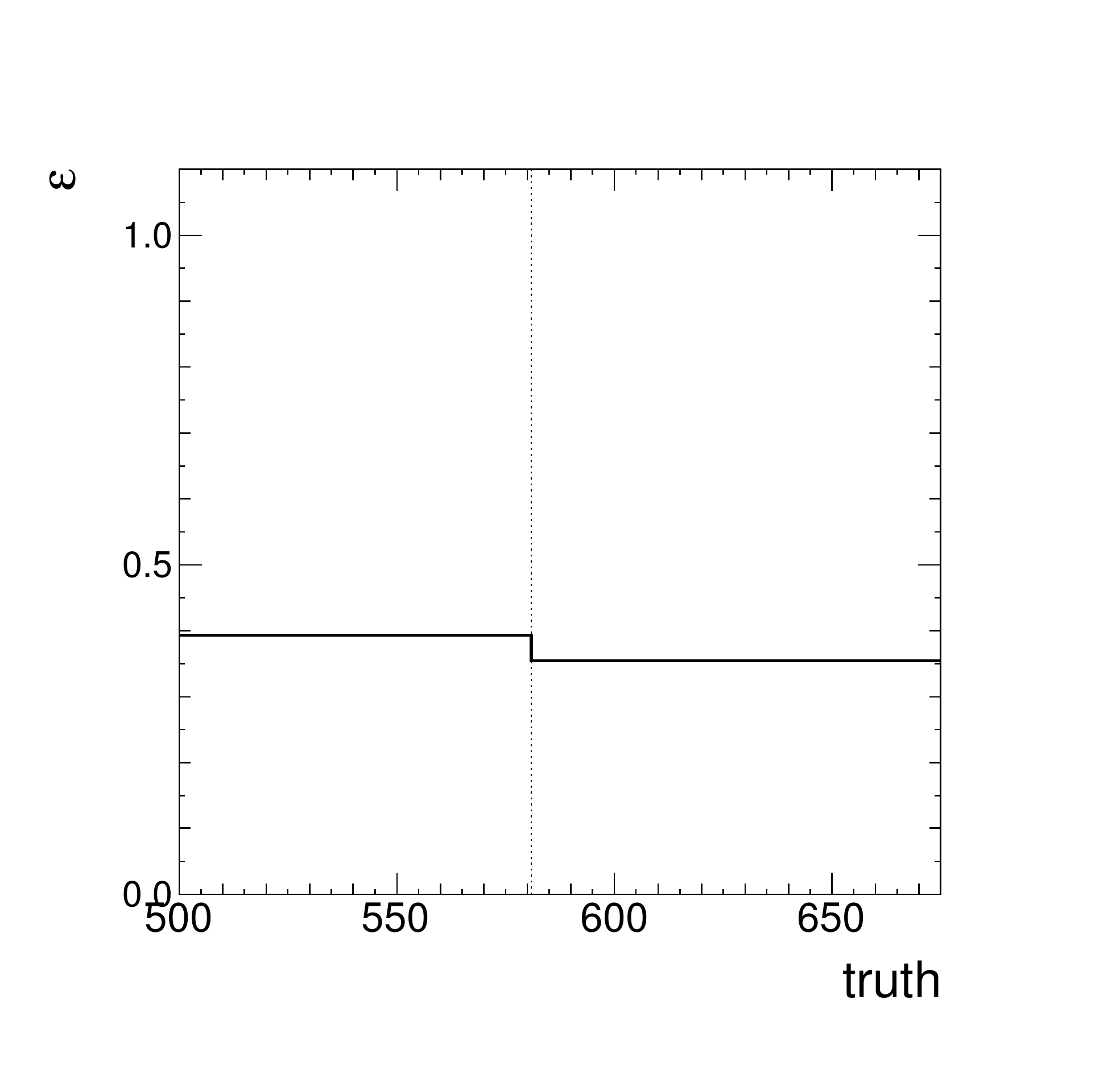}
}\\
\subfigure[]{
  \includegraphics[width=0.5\columnwidth]{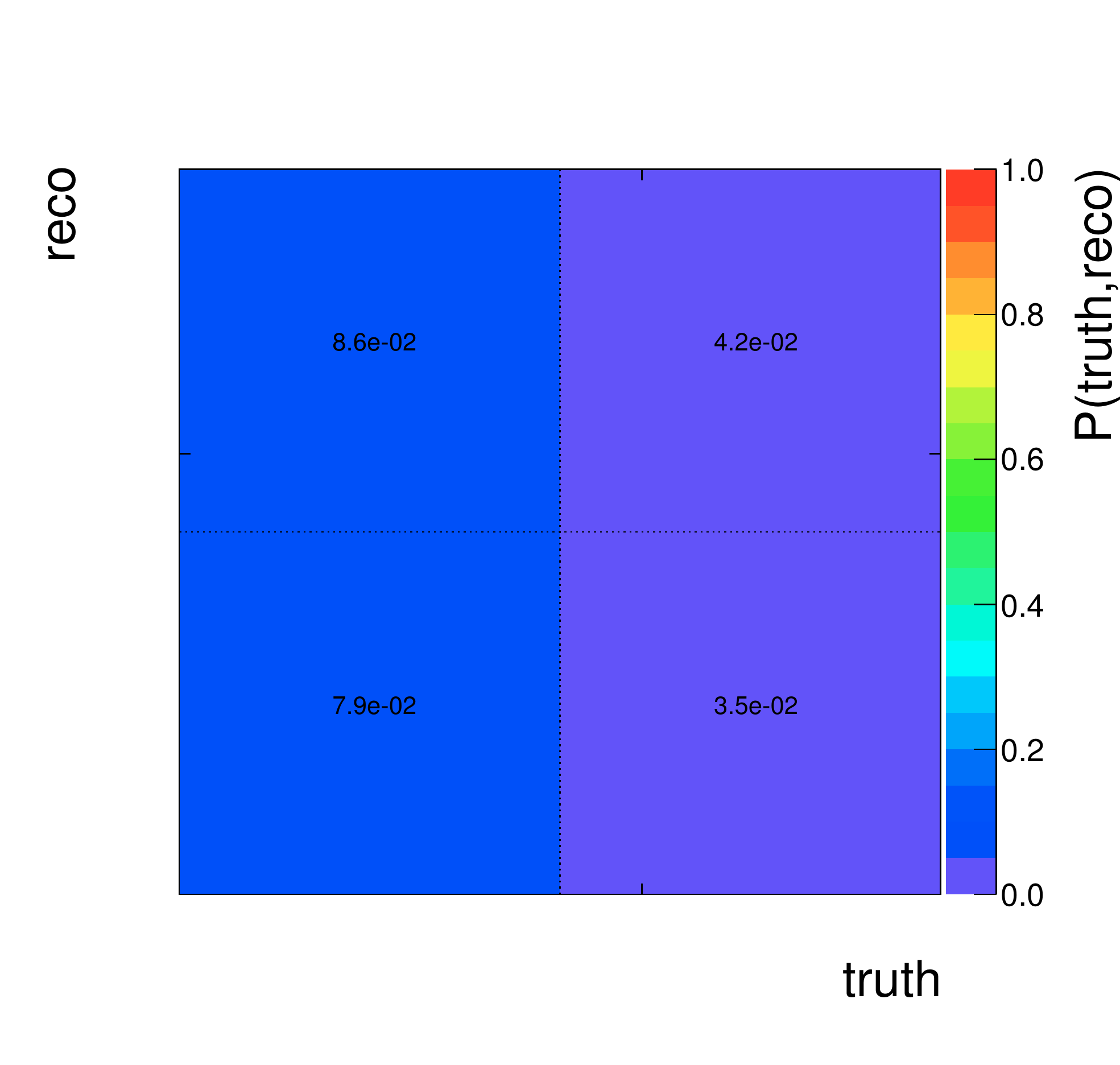}
  \includegraphics[width=0.5\columnwidth]{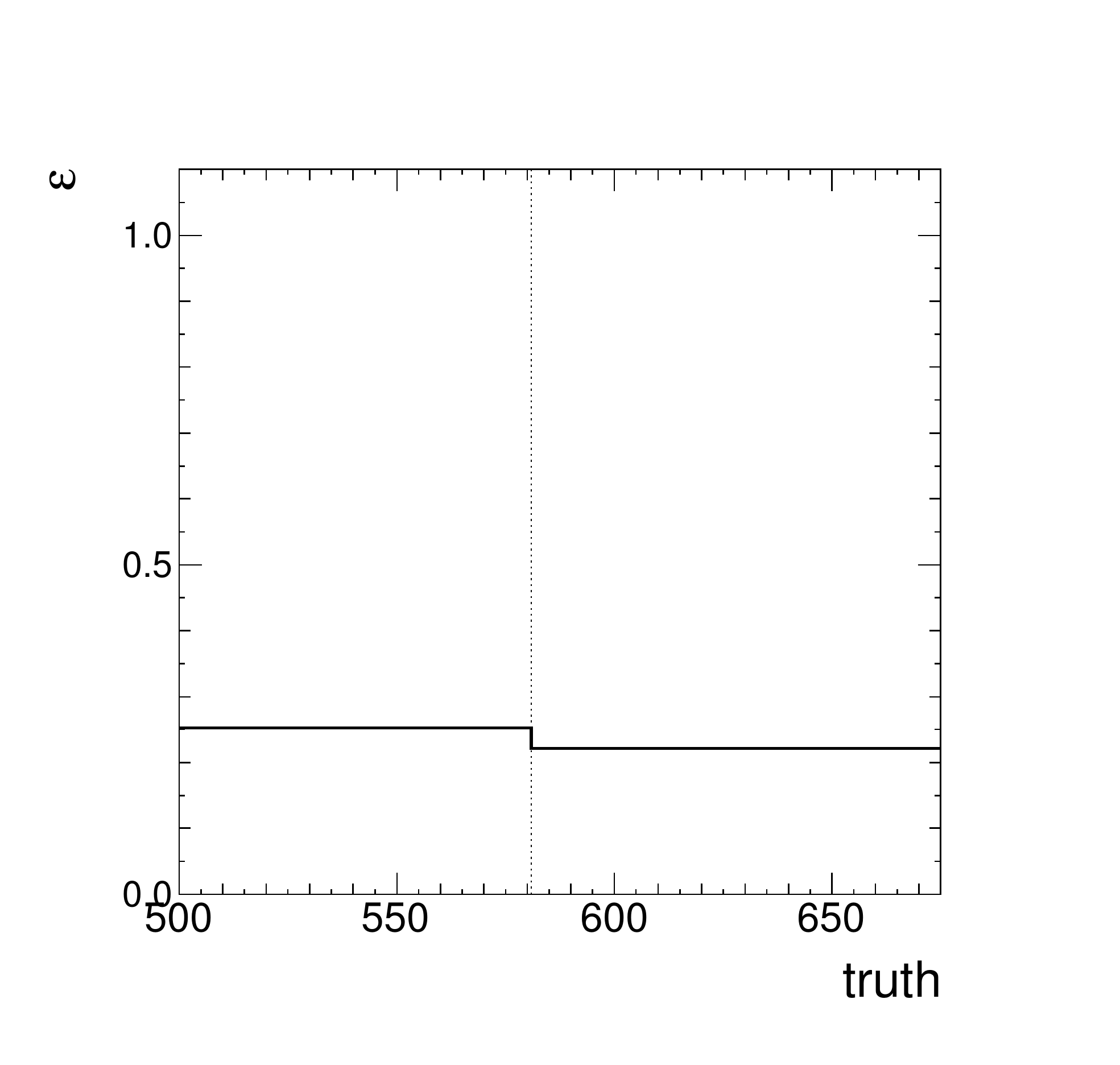}
}\\
\subfigure[]{
  \includegraphics[width=0.5\columnwidth]{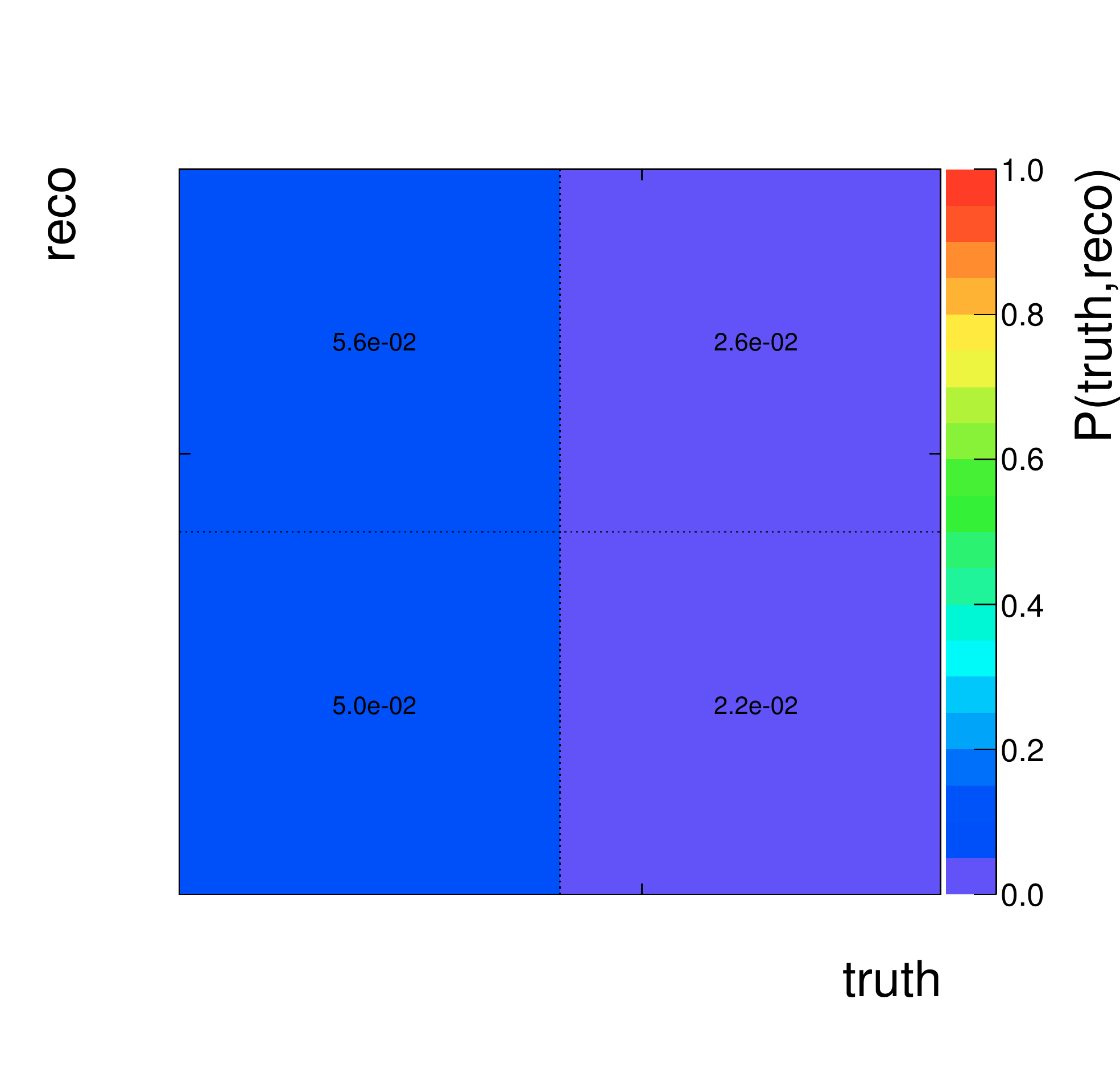}
  \includegraphics[width=0.5\columnwidth]{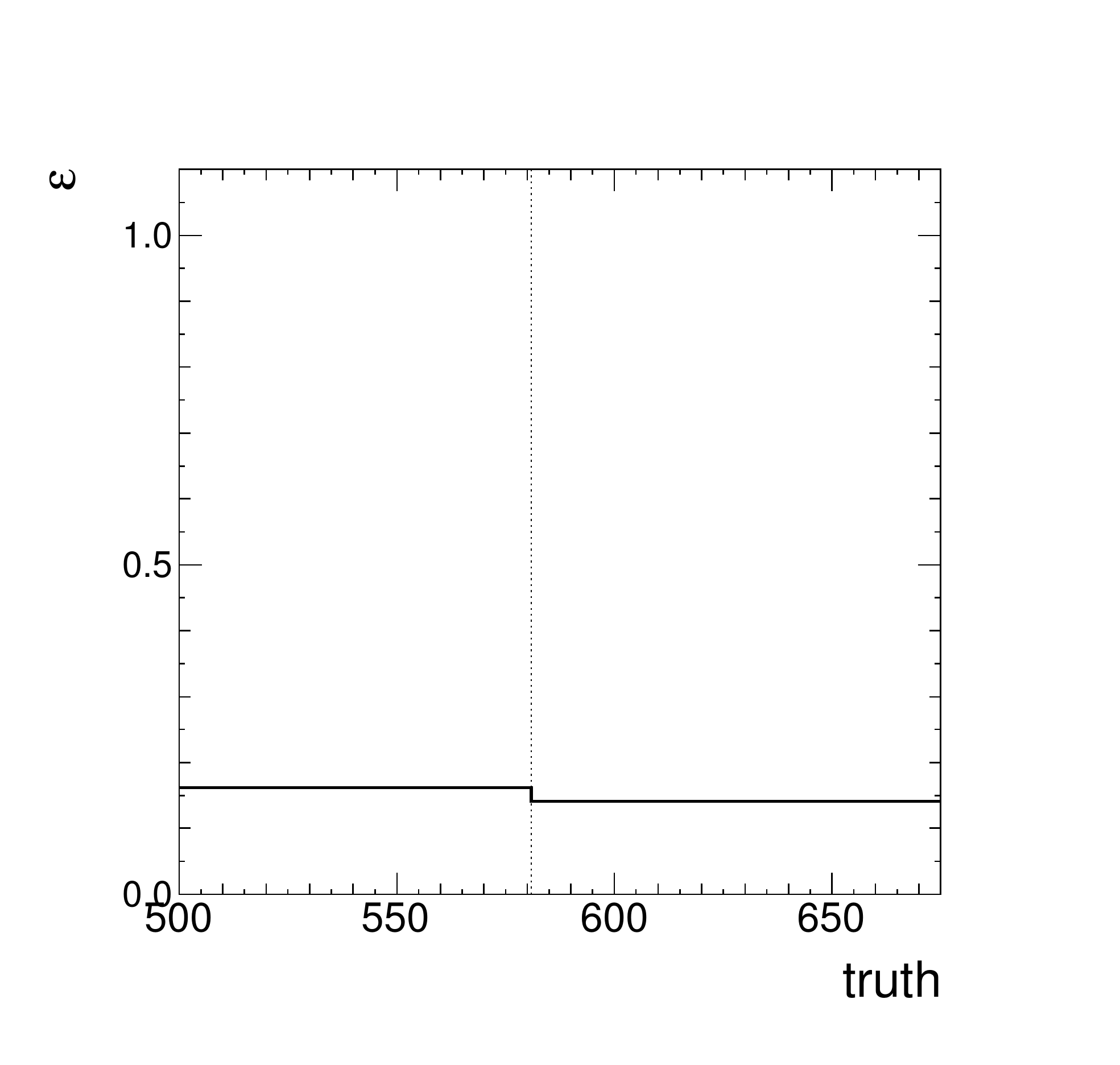}
}
\caption{The migrations matrix (left) and efficiency (right) resulting from setting the smearing parameter $b$ to 0.1 (a), 0.3 (b), 0.5 (c), and 0.8 (d), as in Sec.~\ref{sec:example3}. \label{fig:matrices3}}
\end{figure}

\begin{figure}[H]
\centering
\subfigure[]{
  \includegraphics[width=0.45\columnwidth]{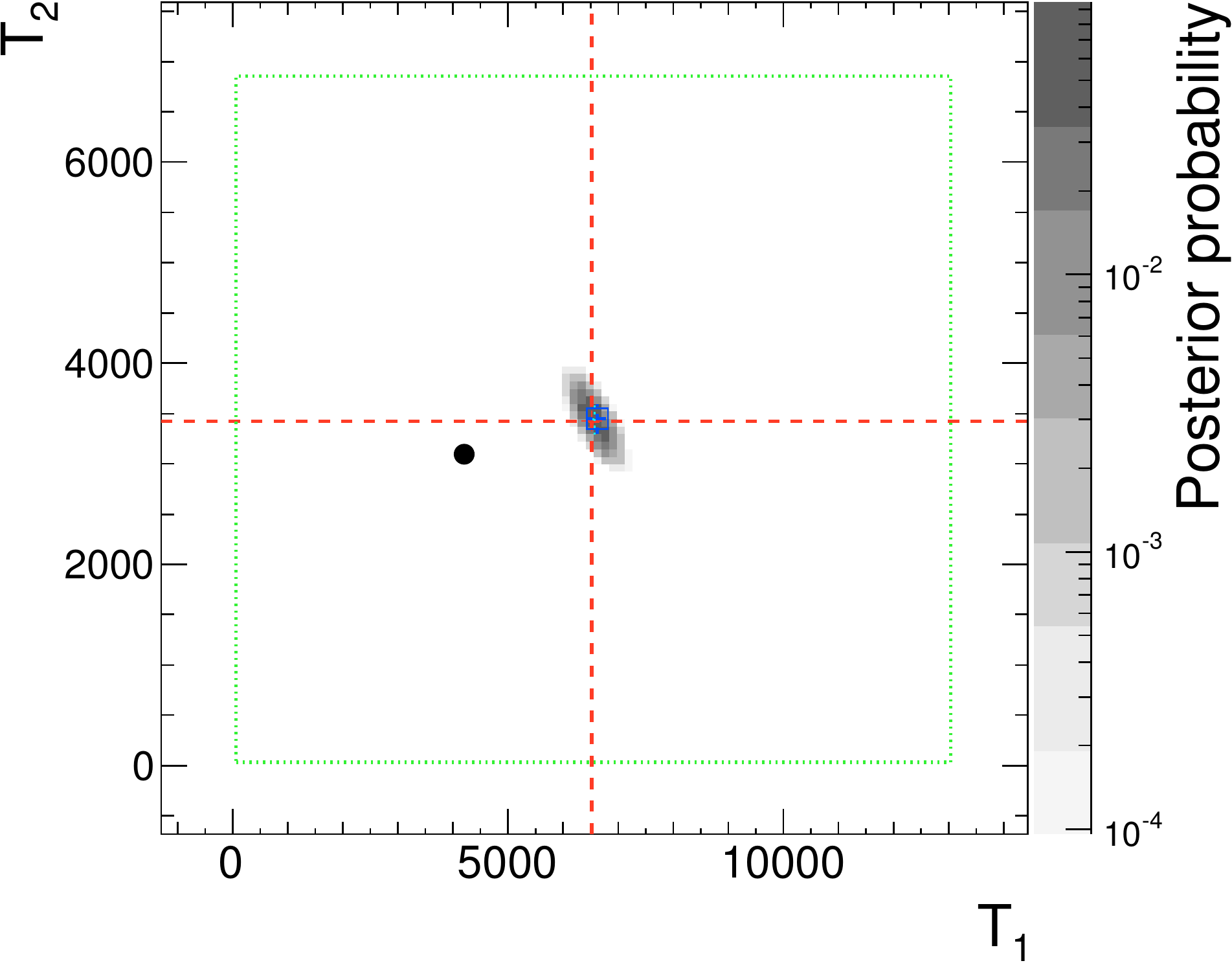}
}
\subfigure[]{
  \includegraphics[width=0.45\columnwidth]{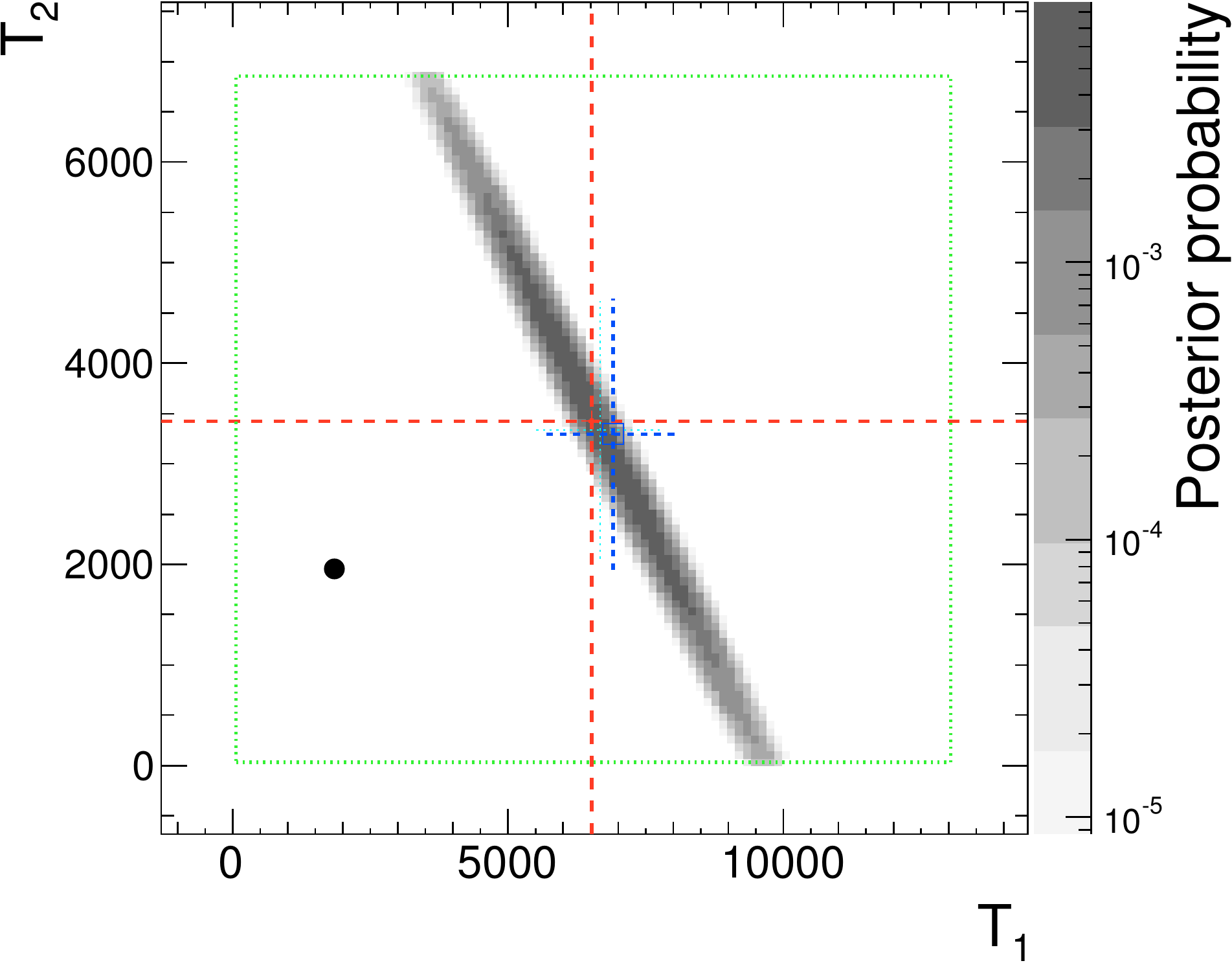}
}\\
\subfigure[]{
  \includegraphics[width=0.45\columnwidth]{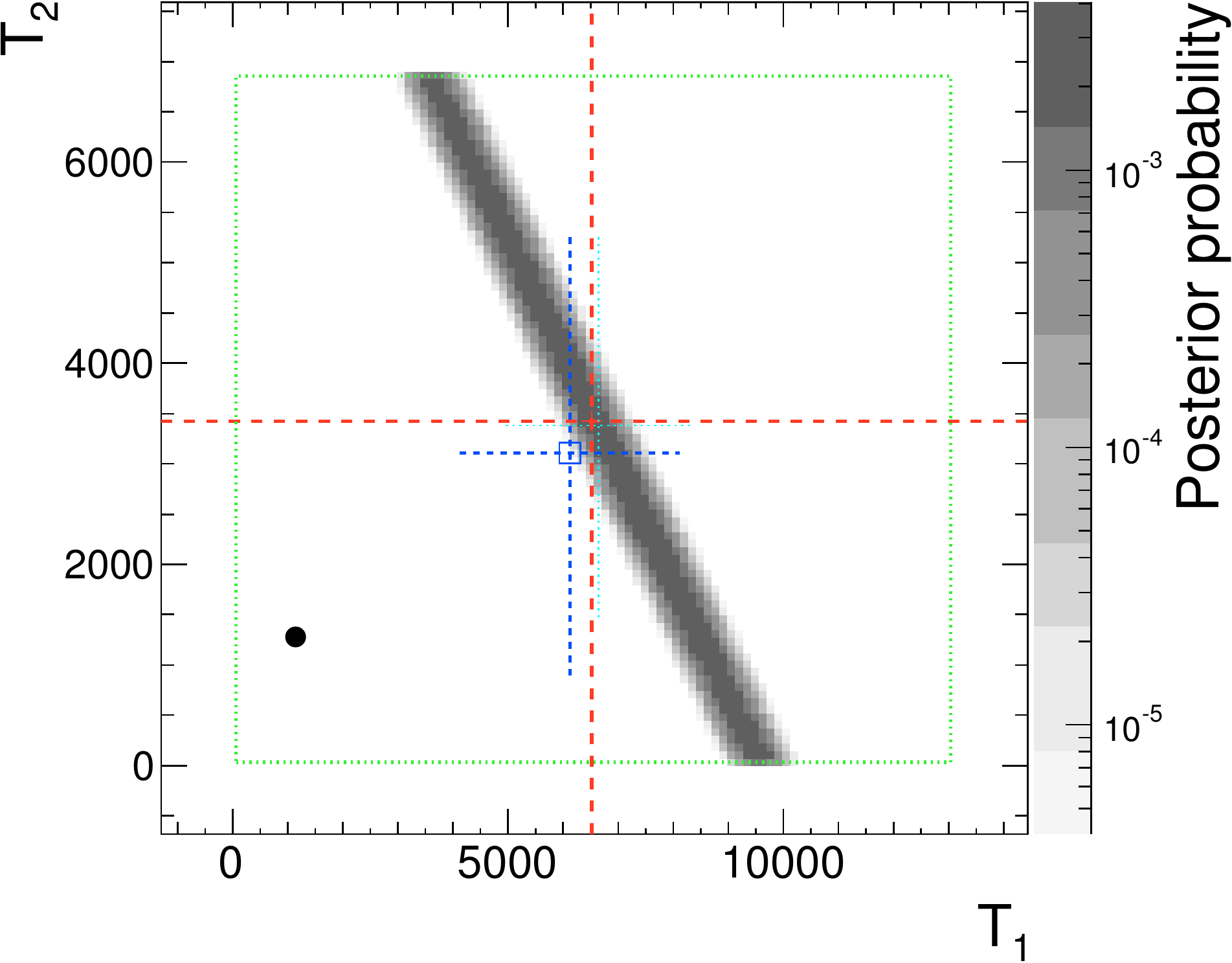}
}
\subfigure[]{
  \includegraphics[width=0.45\columnwidth]{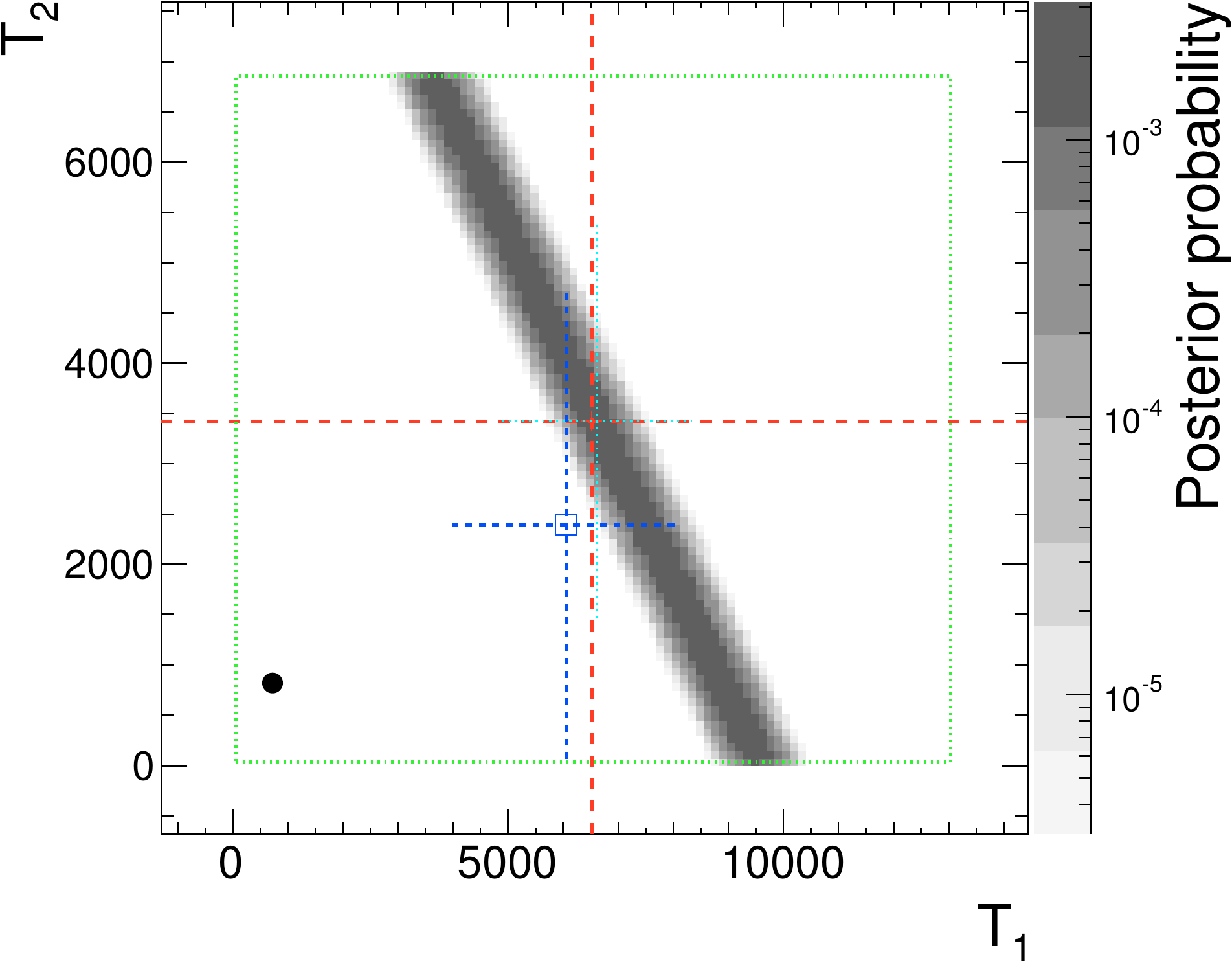}
}
\caption{The inferred $p(\tuple{T}|\tuple{D})$ for the smearing parameter $b$ to set to 0.1 (a), 0.3 (b), 0.5 (c), and 0.8 (d), as in Sec.~\ref{sec:example3}. The appearing markers and lines are explained in Fig.~\ref{fig:2Dim1}. \label{fig:results3}}
\end{figure}

In Eq.~\ref{eq:sigma}, the parameter $a$ is kept at 0, and $b$ is given the values $\{0.1, 0.3, 0.5, 0.8\}$, resulting to the migrations matrices in Fig.~\ref{fig:matrices3}, the input spectra and the inferred $p(\T|\D)$ in Fig.~\ref{fig:results3}.

The correlation between $T_1$ and $T_2$ stays the same, but the spread of the posterior increases quickly with smearing, and unlike the case without smearing, the spread is much greater than the statistical uncertainty of the data.  This is a direct demonstration that, unless some regularizing presumption is imposed through the prior  (see Sec.~\ref{sec:conceptualization}), unfolding can not provide a precise answer.  This is true not only for FBU, but for all unfolding methods; if the likelihood $L(\D|\T)$ is so widely spread, there is no method that can recover the information lost with smearing, unless some external information is utilized, in the form of prior assumptions about the answer.

Imprecise as the answer may be, it is at least accurate, in the sense that the correct $\T$ lies well within the unfolded spectrum error bars, shown with the blue dashed lines in Fig.~\ref{fig:results3}.  This happens under extreme smearing, because the error bars are much larger.   This can be untrue, though, in situations of very little (or no) smearing, as in Fig.~\ref{fig:2Dim1} or Fig.~\ref{fig:2Dim2}, where a merely statistical fluctuation of the data by more than 1 standard deviation in one of the $N_t=2$ dimensions is enough to drag the bulk of $p(\T|\D)$ equivalently far from the correct $\T$, while $p(\T|\D)$ does not spread enough to keep including the correct $\T$ within its 68\% core.

\begin{figure}[H]
\centering
\subfigure[]{
  \includegraphics[width=0.45\columnwidth]{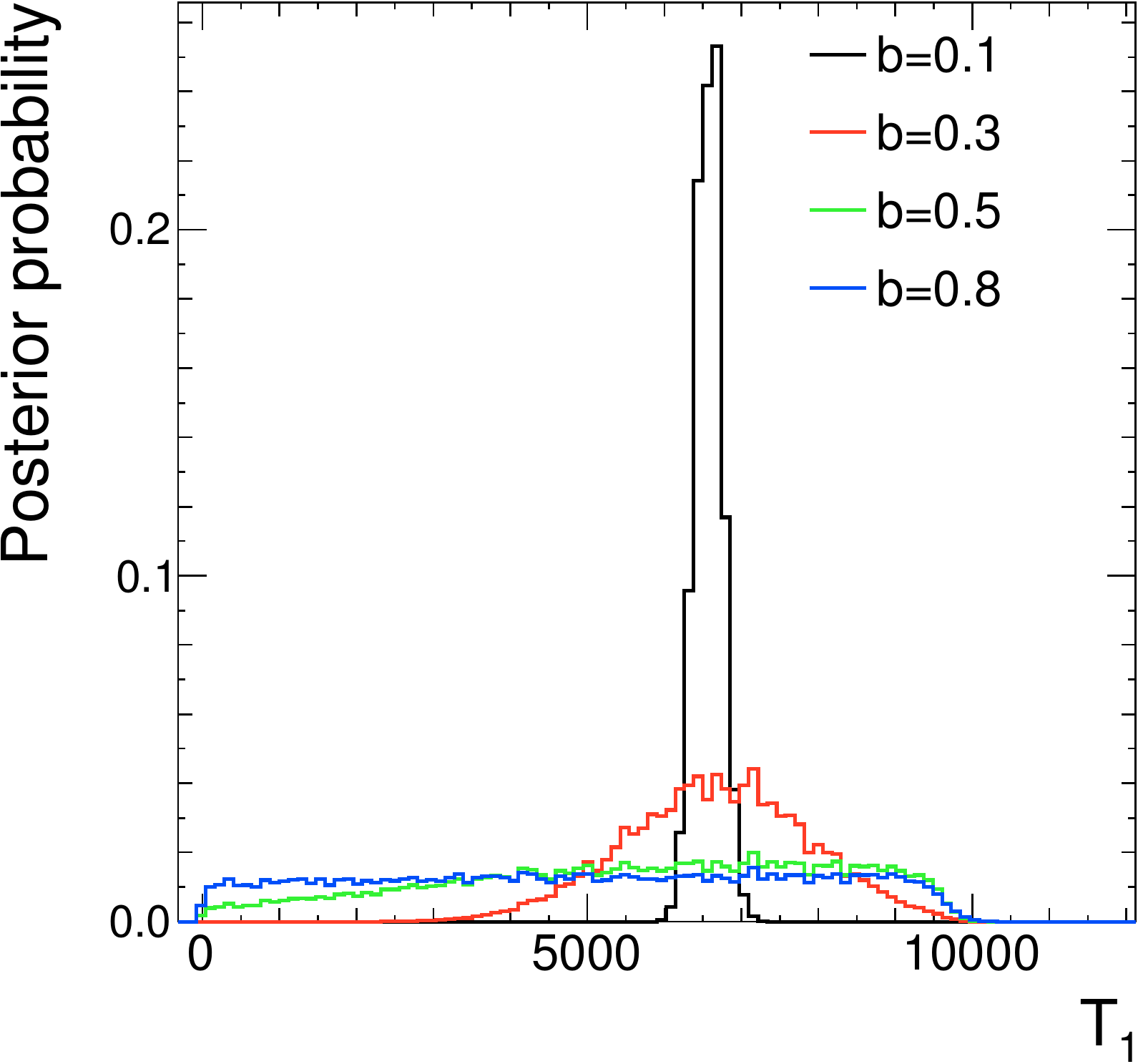}
}
\subfigure[]{
  \includegraphics[width=0.45\columnwidth]{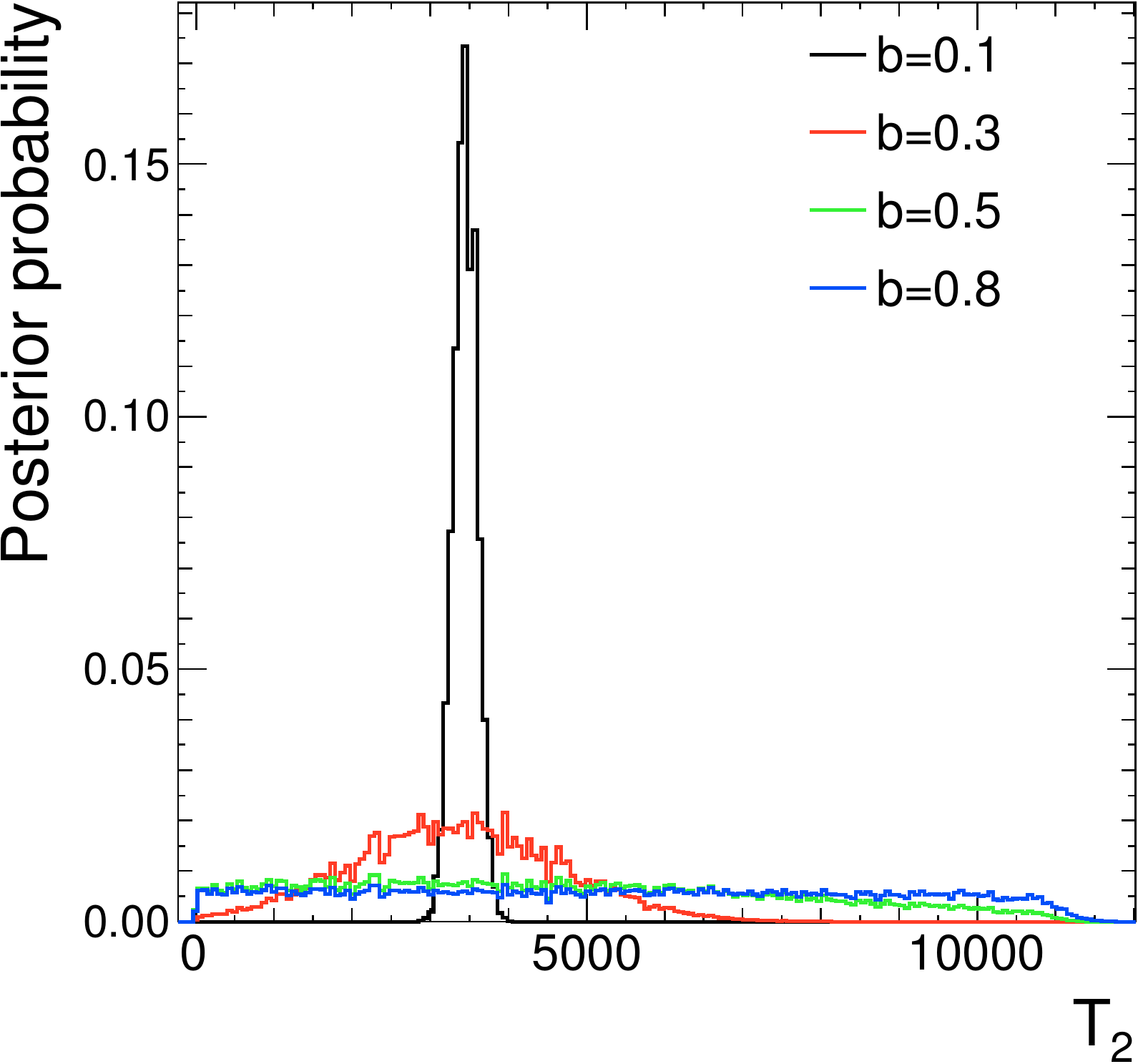}
}
\caption{Marginal distributions $P_1(T_1|\tuple{D})$ (a), and $P_2(T_2|\tuple{D})$ (b), for $b$ set to 0.1 (a), 0.3 (b), 0.5 (c), and 0.8 (d), as in Sec.~\ref{sec:example3}. \label{fig:projections3}}
\end{figure}

Regarding the shape of $p(\T|\D)$, while for little (or no) smearing it resembles a $N_t$-dimensional Gaussian (assuming high event counts), this is not true under heavy smearing.  This happens because $T_t > 0$, for all $t\in[1,N_t]$, so, when smearing increases, $p(\T|\D)$ gets chopped.  This starts happening at different amounts of smearing in each dimension.   Fig.~\ref{fig:projections3} shows this effect, through 1-dimensional marginal distributions.

As a final remark, the same results are obtained by fixing the smearing parameter $b$ to 0, and increasing $a$ instead.  The correlations between ($T_{t_1}$,$T_{t_2}$) pairs stays the same.  Of course, to attain the same amounts of smearing, $a$ has to increase to about 10, due to the large $\sqrt{m}$ denominator in Eq.~\ref{eq:sigma}.

\subsection{$N_t=14$ dimensions}
\label{sec:example4}

A more realistic example, with $N_t=N_r=14$ bins is produced.  The smearing is as described in Sec.~\ref{sec:generation}, i.e., assuming $(a,b)=(0.5,0.1)$.  The migrations matrix in Fig.~\ref{fig:genMat}, and the truth, reco, and data spectrum are in Fig.~\ref{fig:truthAndReco}.  

\begin{figure}[H]
\centering
 \subfigure[]{
   \includegraphics[width=0.45\columnwidth]{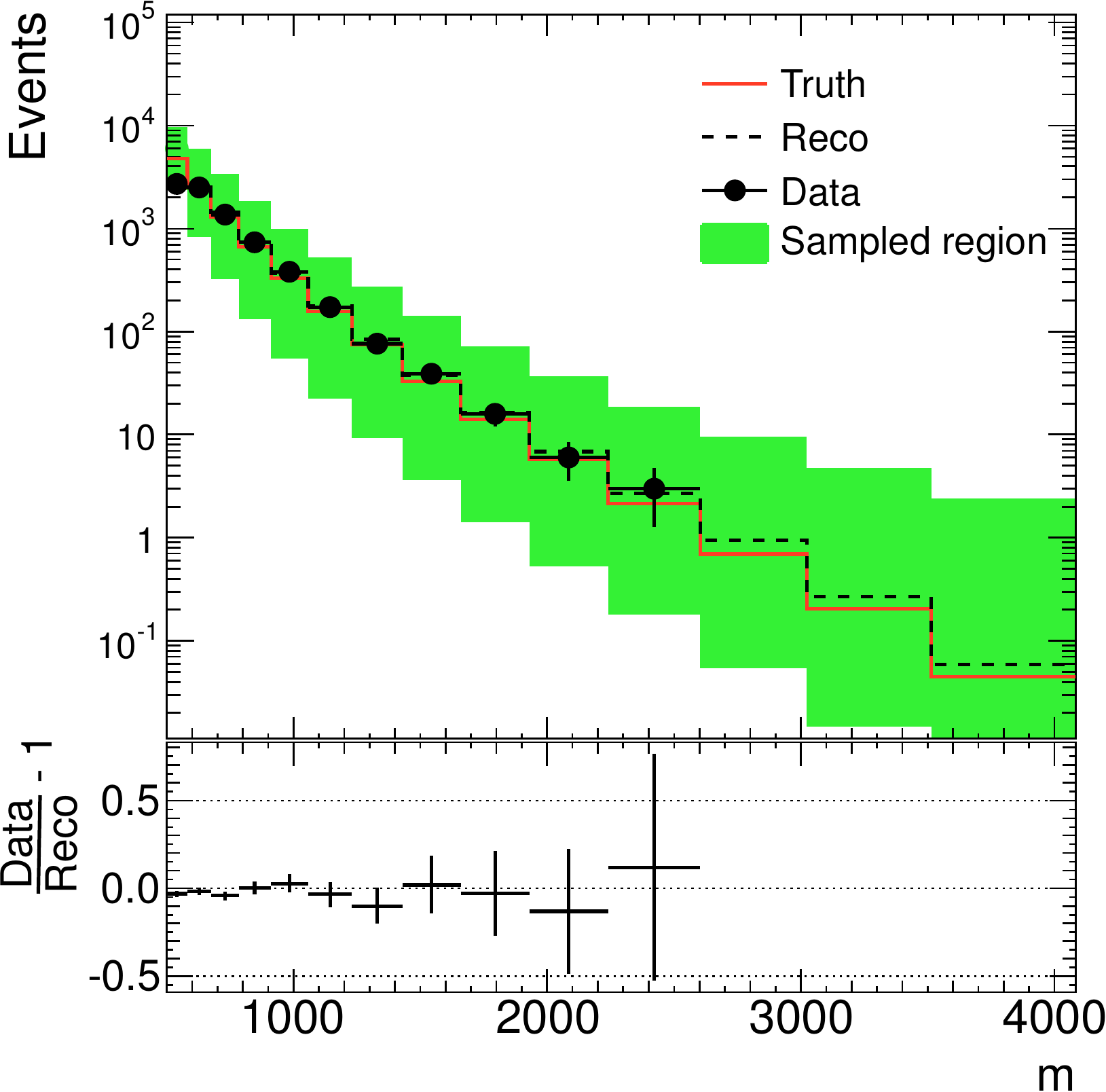}
   \label{fig:inputs4}
  }
  \subfigure[]{
   \includegraphics[width=0.45\columnwidth]{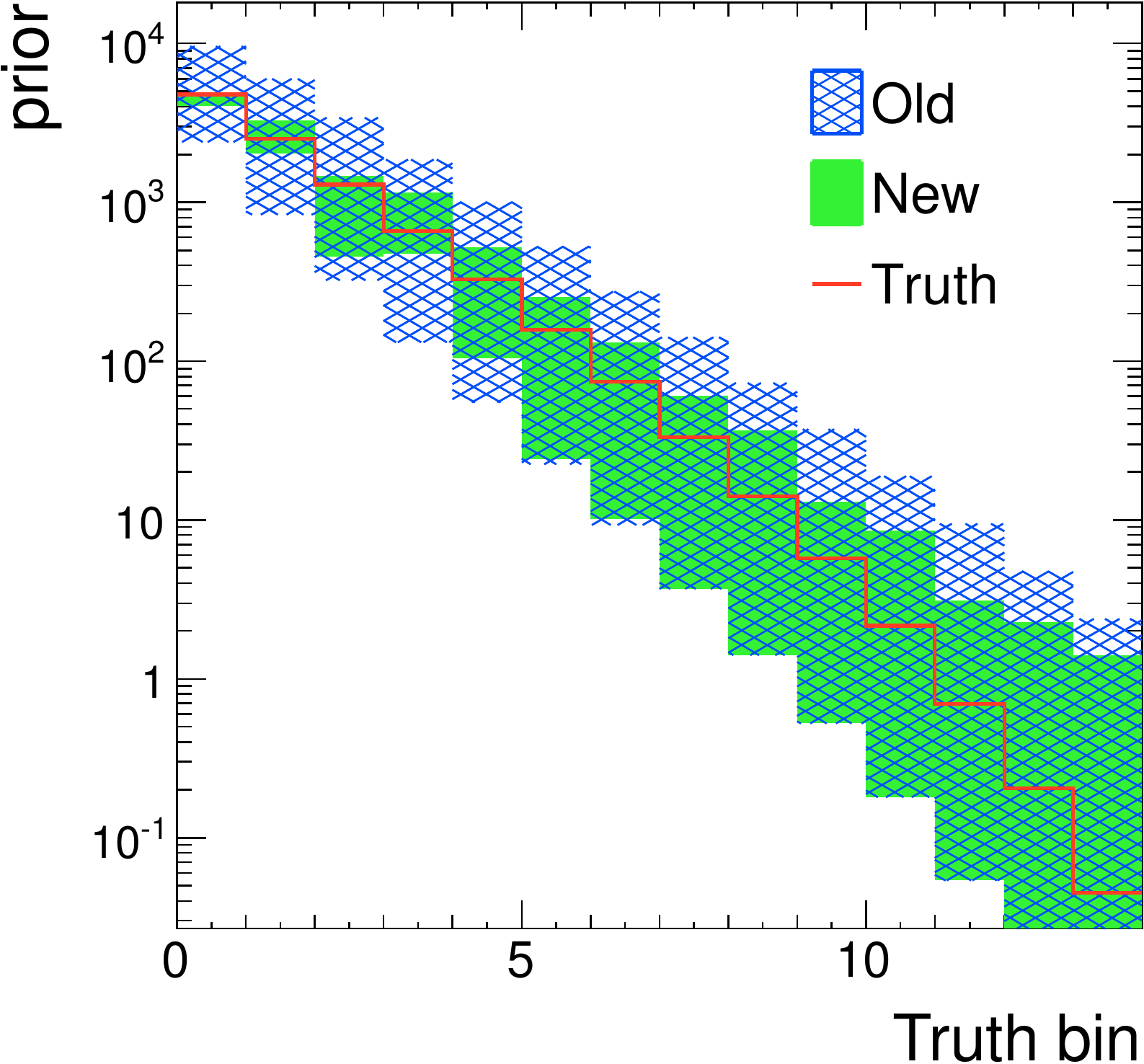}
   \label{fig:priorRedefined4}
 }
\caption{In (a): The truth, reco, and data spectrum of the example in Sec.~\ref{sec:example4}.  These are the same spectra shown in Fig.~\ref{fig:truthAndReco}, except that here the initial sampling region is overlaid, defined by Eq.~\ref{eq:prior4}.  In (b):  The initial sampling region (called ``old'') is shown and compared to the ``new'' hyper-box, which is found by volume reduction, according to Sec.~\ref{sec:narrowing}.  Instead of showing the observable $m$ in the horizontal axis, in (b) the horizontal axis shows simply the index of each bin.
\label{fig:prior4}}
\end{figure}

The initial sampling hyper-box (Sec.~\ref{sec:priors}) is quite large, to minimize its effect on the answer.  Its limits, in dimension $t$, are:
\begin{equation}
\label{eq:prior4}
\begin{split}
T_t^\ulcorner & \equiv \tilde{T}_{t} / (t+1) , t\in[1,N_t] \\
T_1^\urcorner & \equiv \tilde{T}_1\cdot 2 \\
T_t^\urcorner & \equiv  T_{t-1}^\ulcorner , t\in[2,N_t] 
\end{split}
\end{equation}
and, since no extra processes are assumed, $\tilde{\T} = \hat{\T}$ (see Sec.~\ref{sec:nomenclature}).  Fig.~\ref{fig:inputs4} shows this initial sampled hyper-box, which is then reduced (see Sec.~\ref{sec:narrowing}) to the hyper-box shown in Fig.~\ref{fig:priorRedefined4}.

The sampling of the reduced hyper-box proceeds with $10^7$ uniformly distributed samples, which takes less than 30 seconds with an typical modern CPU core.  The final unfolded spectrum would be practically identical even with $10^5$ samples, but $10^7$ are used for aesthetic reasons.

\begin{figure}[H]
\centering
\begin{tabular}{ccccc}
   \includegraphics[width=0.18\columnwidth]{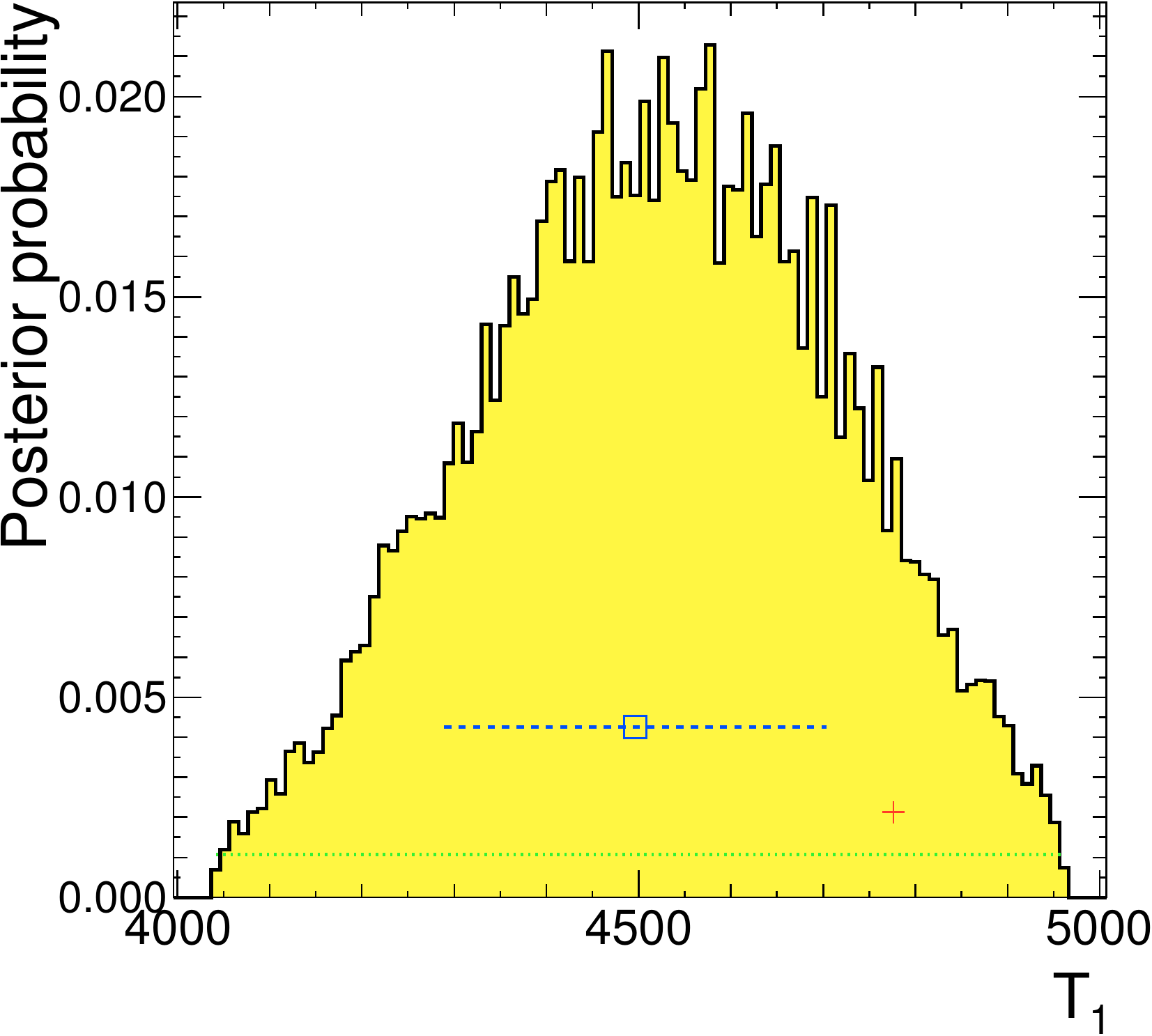} &
   \includegraphics[width=0.18\columnwidth]{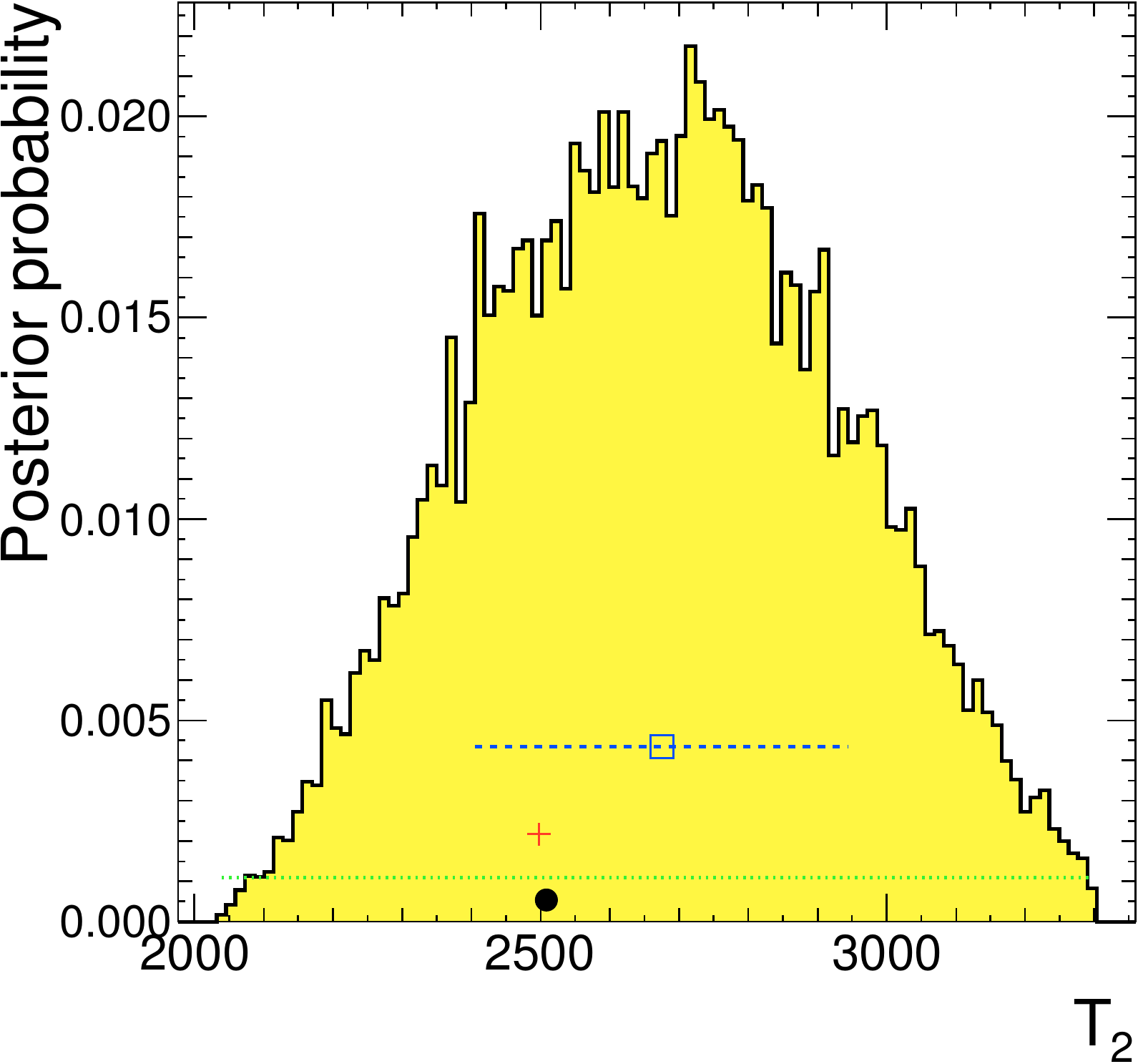} &
   \includegraphics[width=0.18\columnwidth]{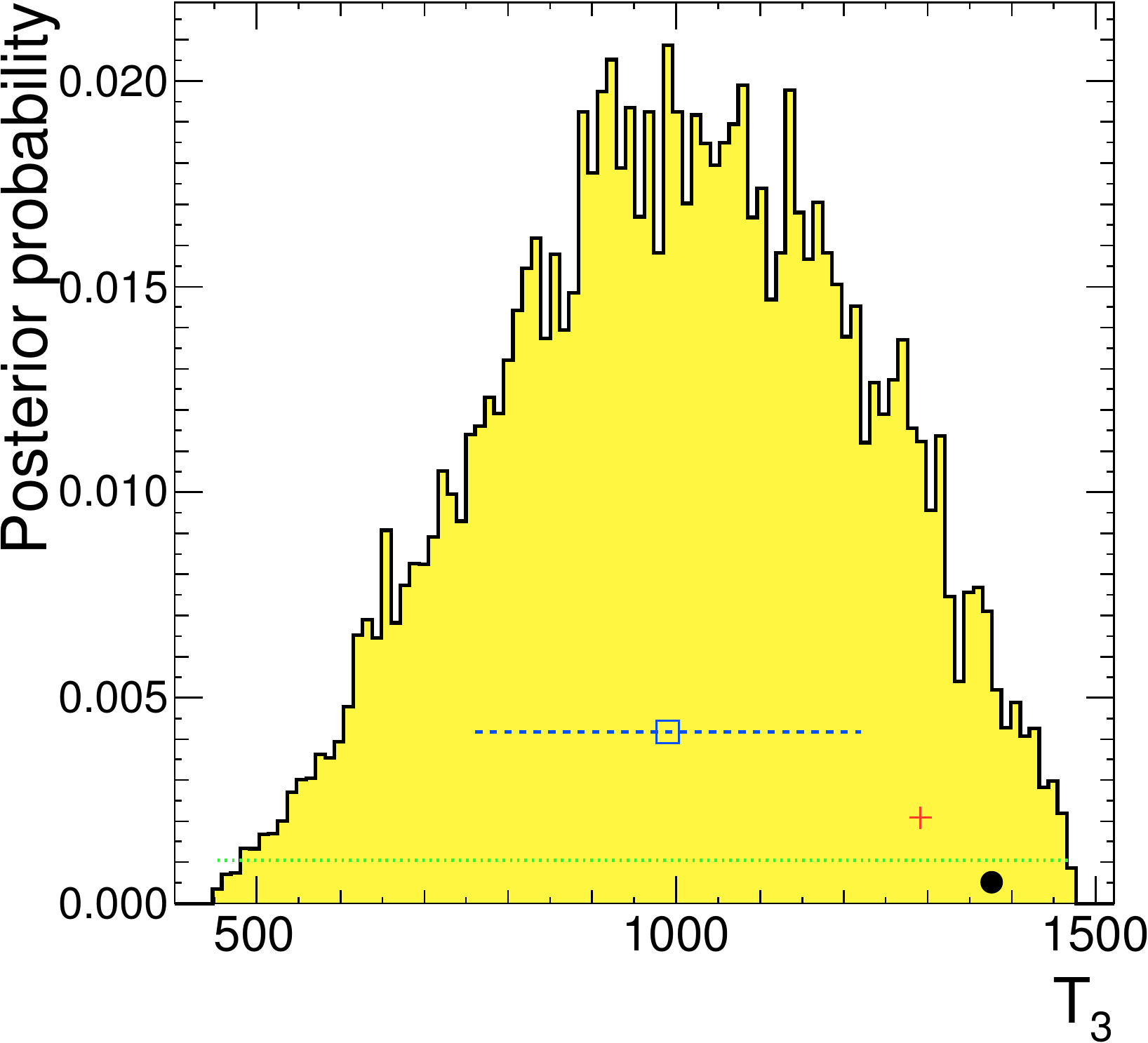} &
   \includegraphics[width=0.18\columnwidth]{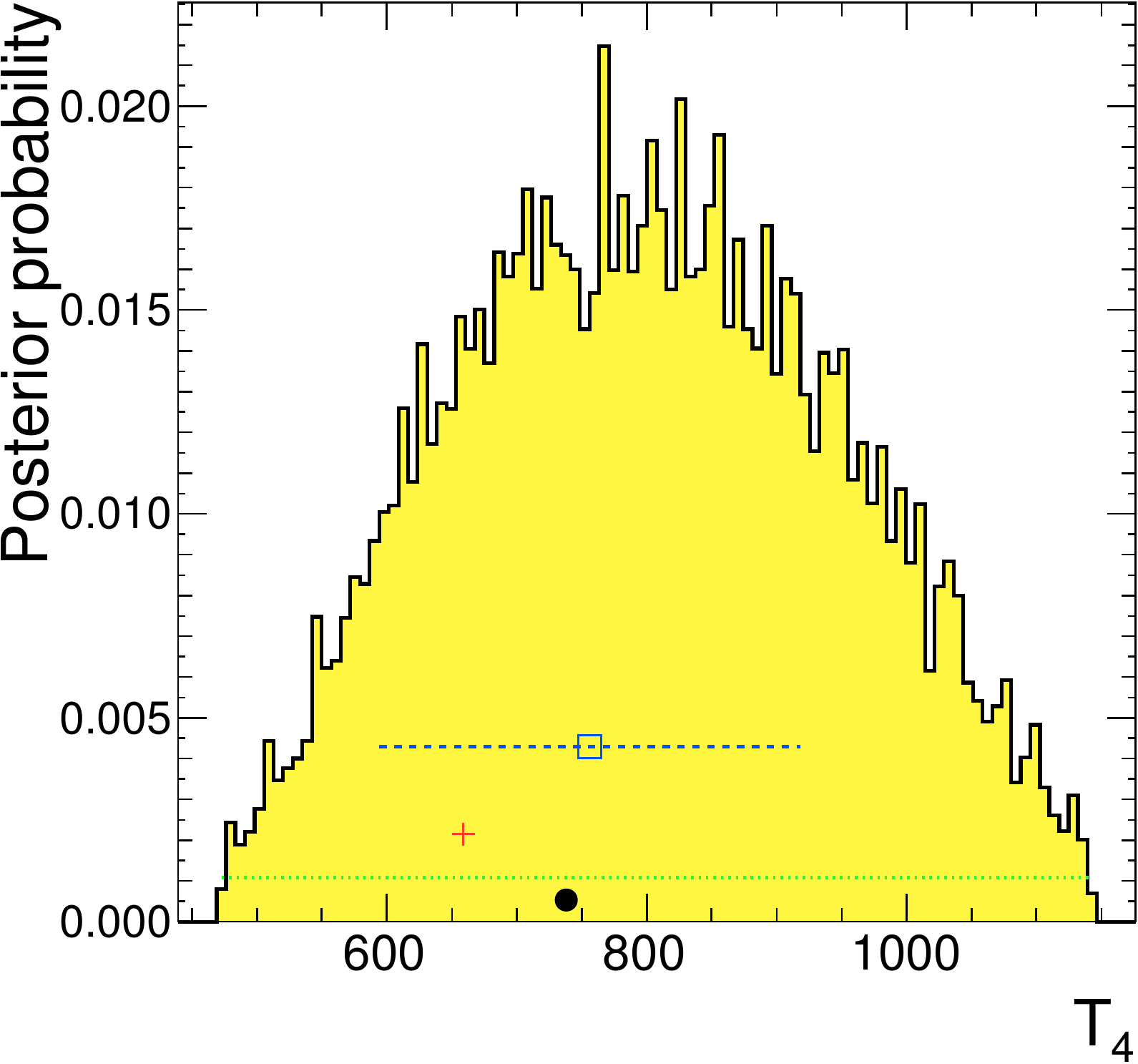} &
  \includegraphics[width=0.18\columnwidth]{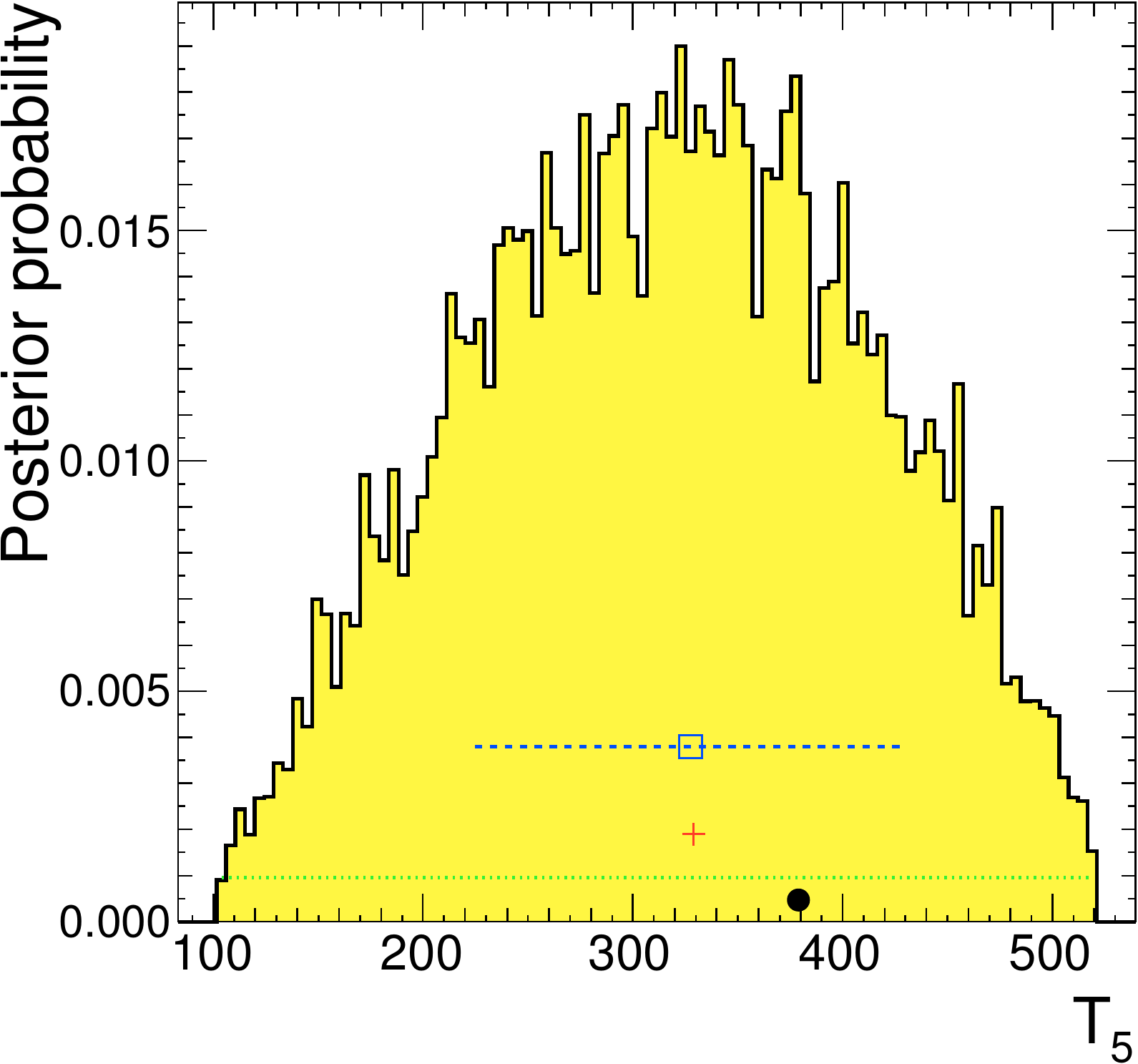} \\

   \includegraphics[width=0.18\columnwidth]{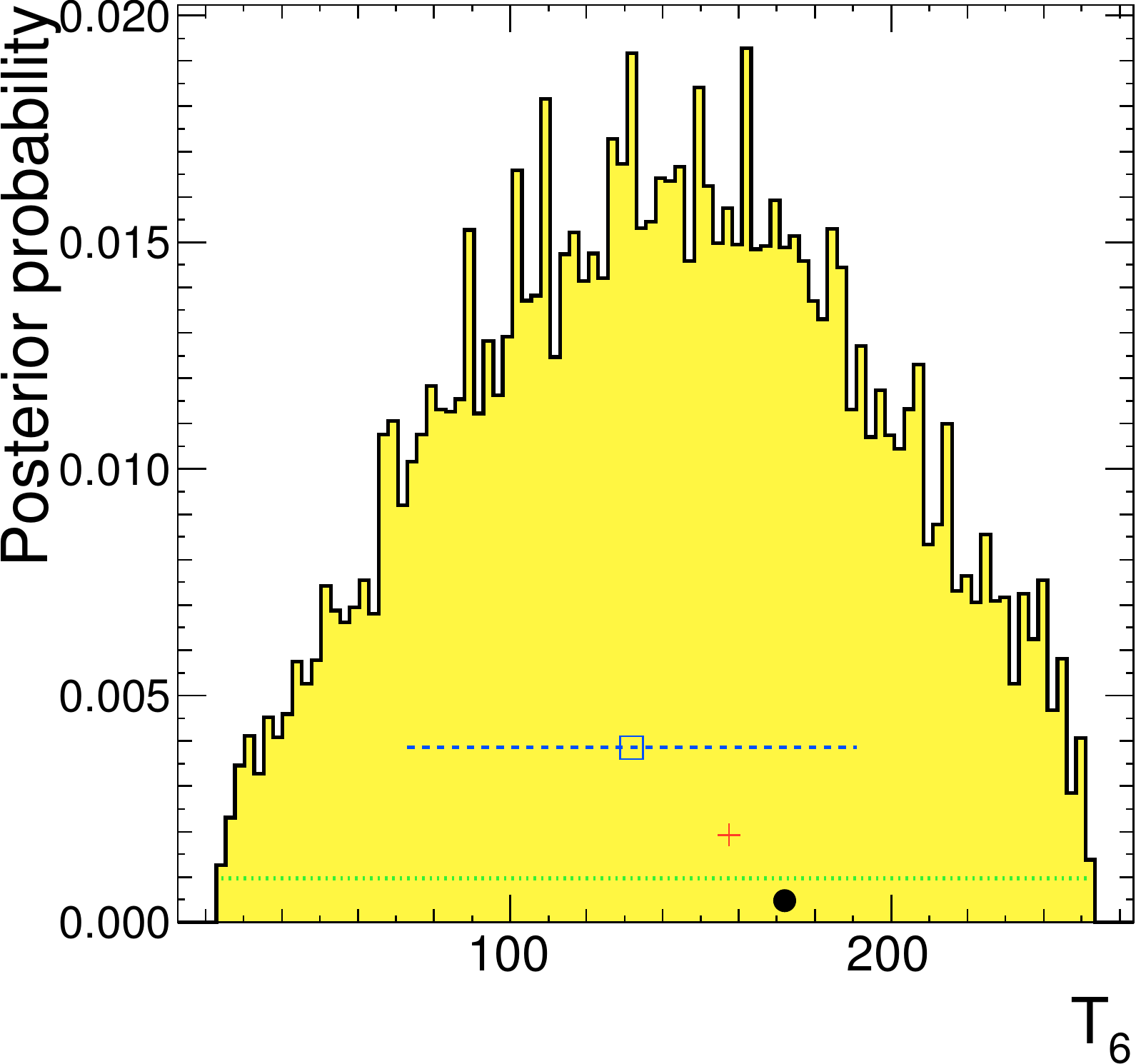} &
   \includegraphics[width=0.18\columnwidth]{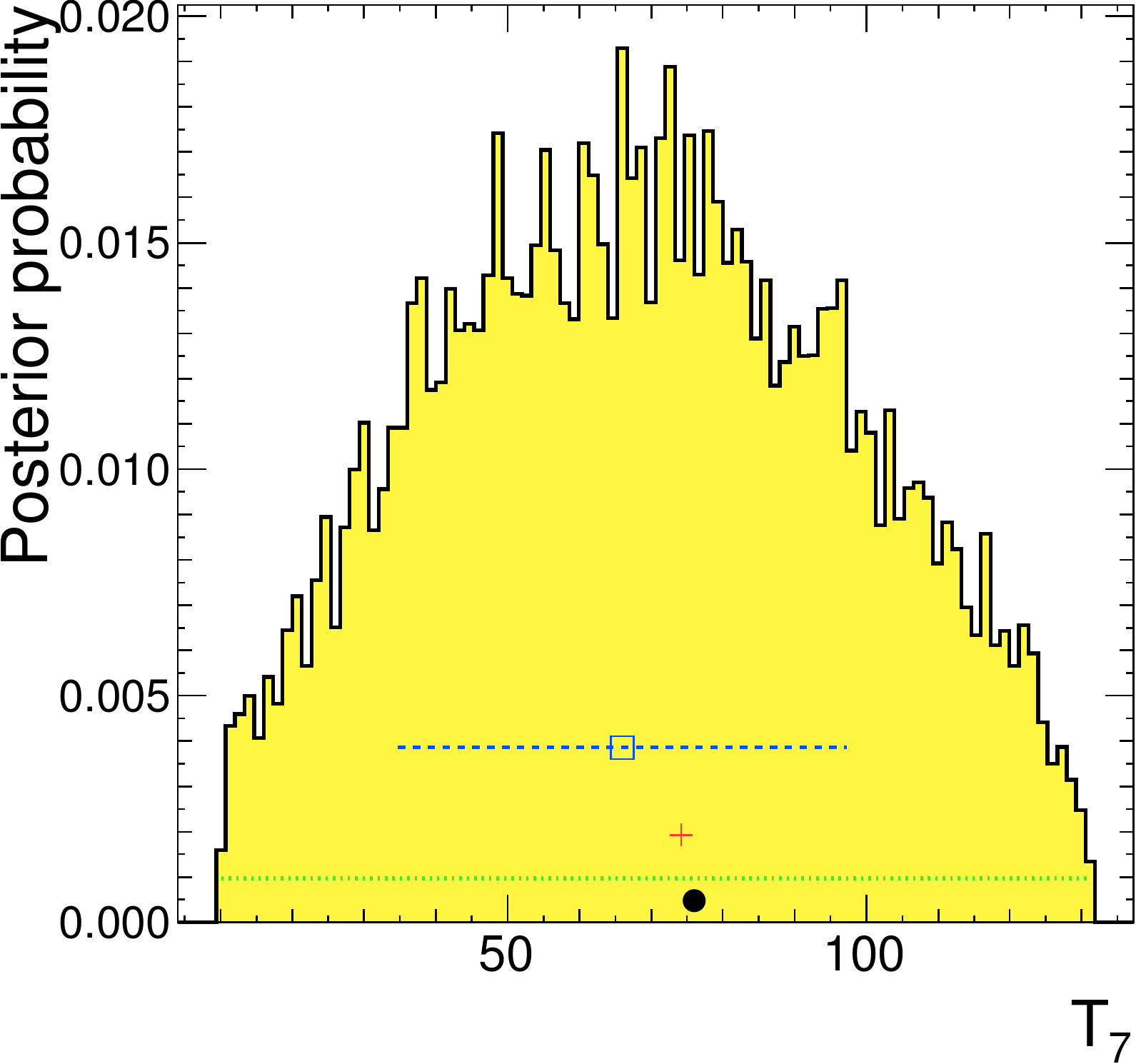} &
   \includegraphics[width=0.18\columnwidth]{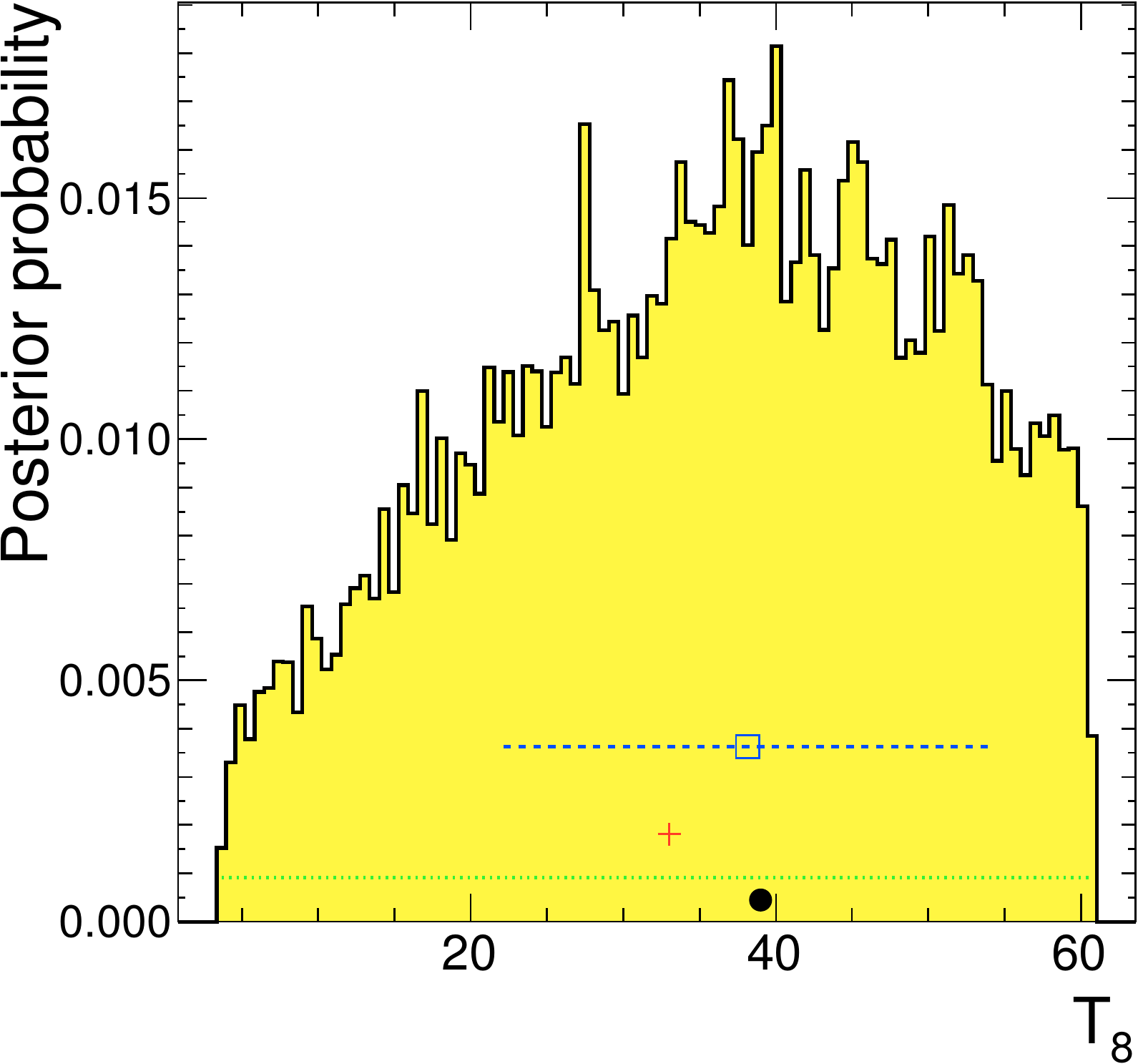} &
   \includegraphics[width=0.18\columnwidth]{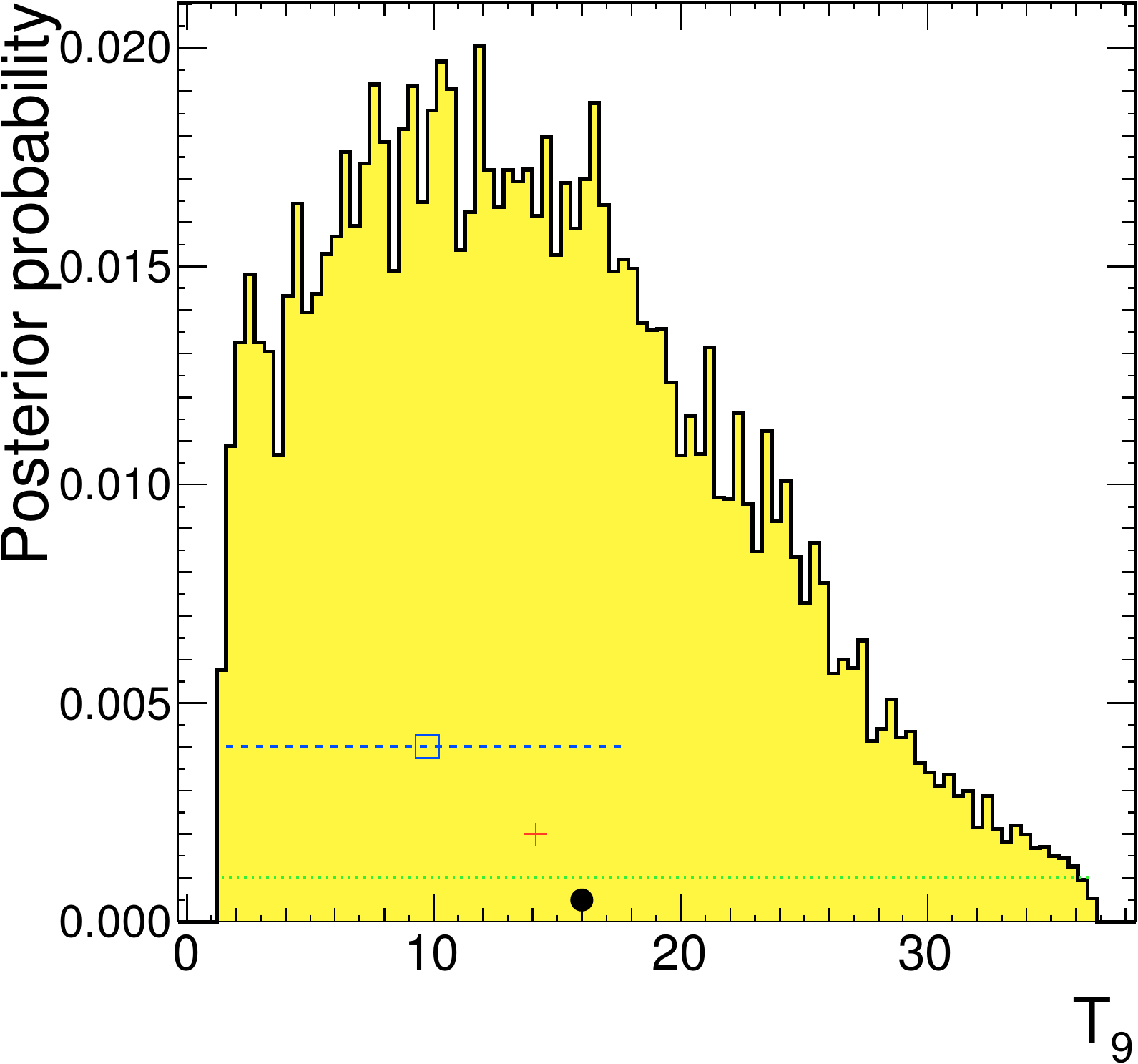} &
   \includegraphics[width=0.18\columnwidth]{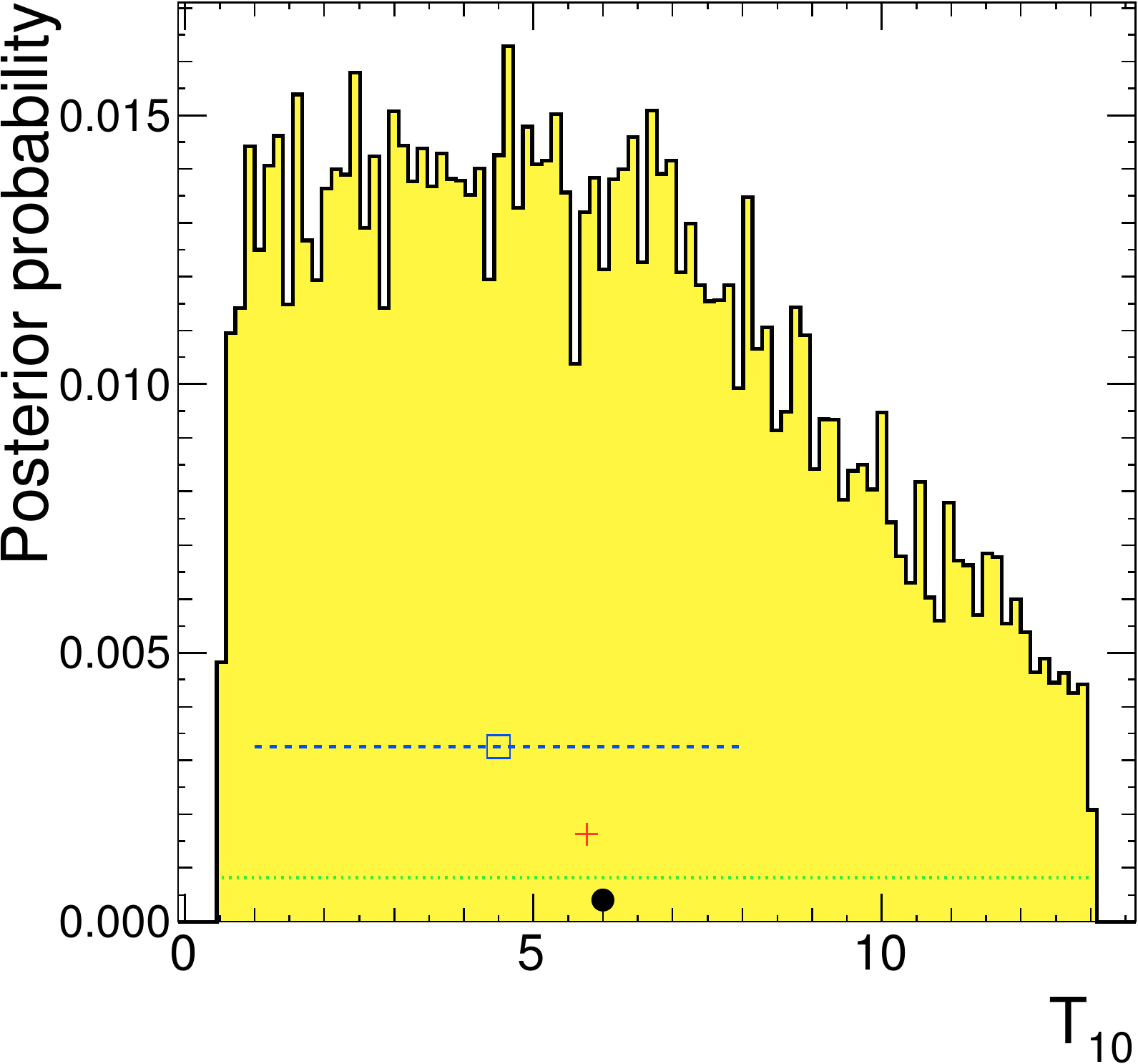} \\

   \includegraphics[width=0.18\columnwidth]{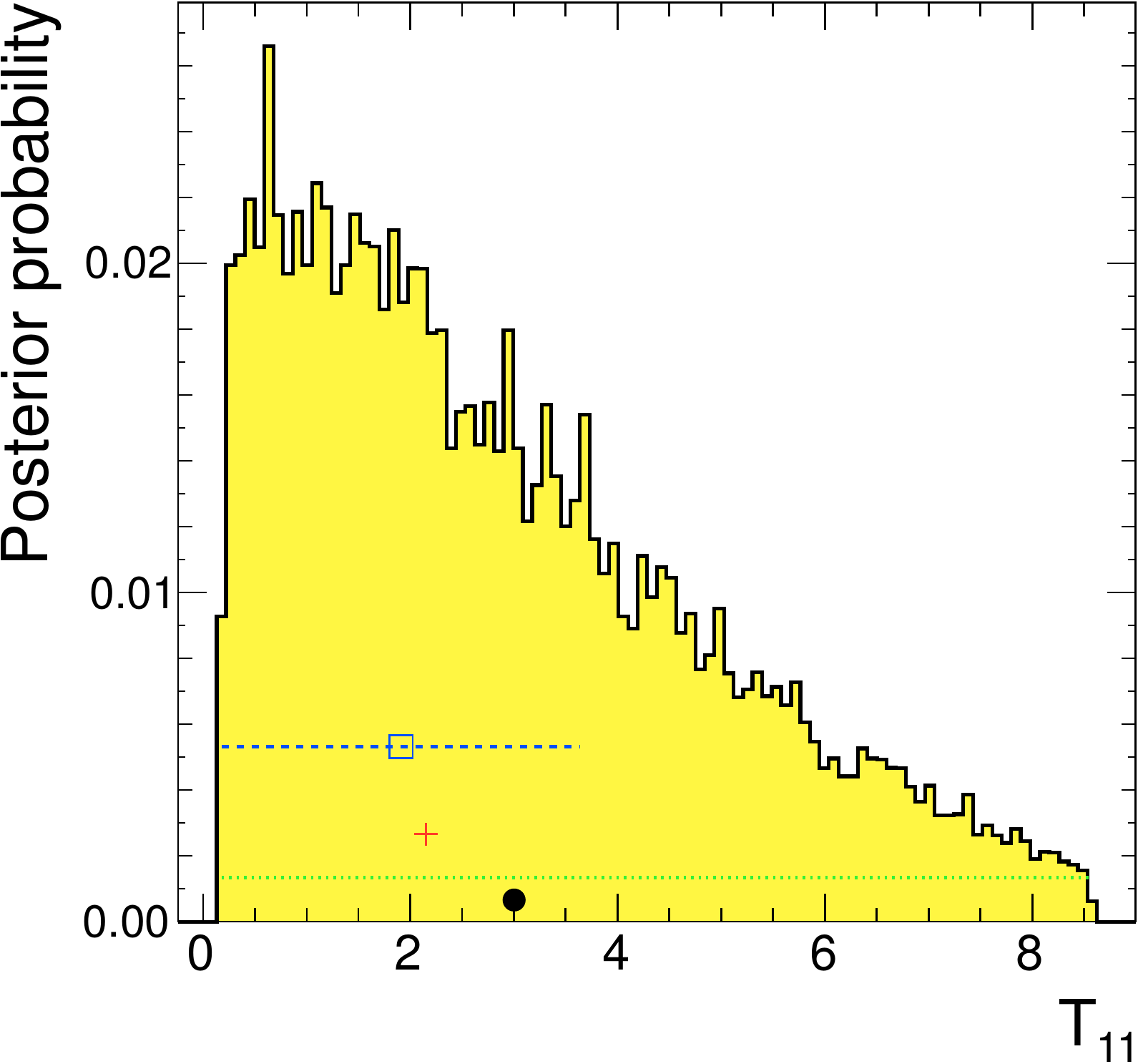} &
   \includegraphics[width=0.18\columnwidth]{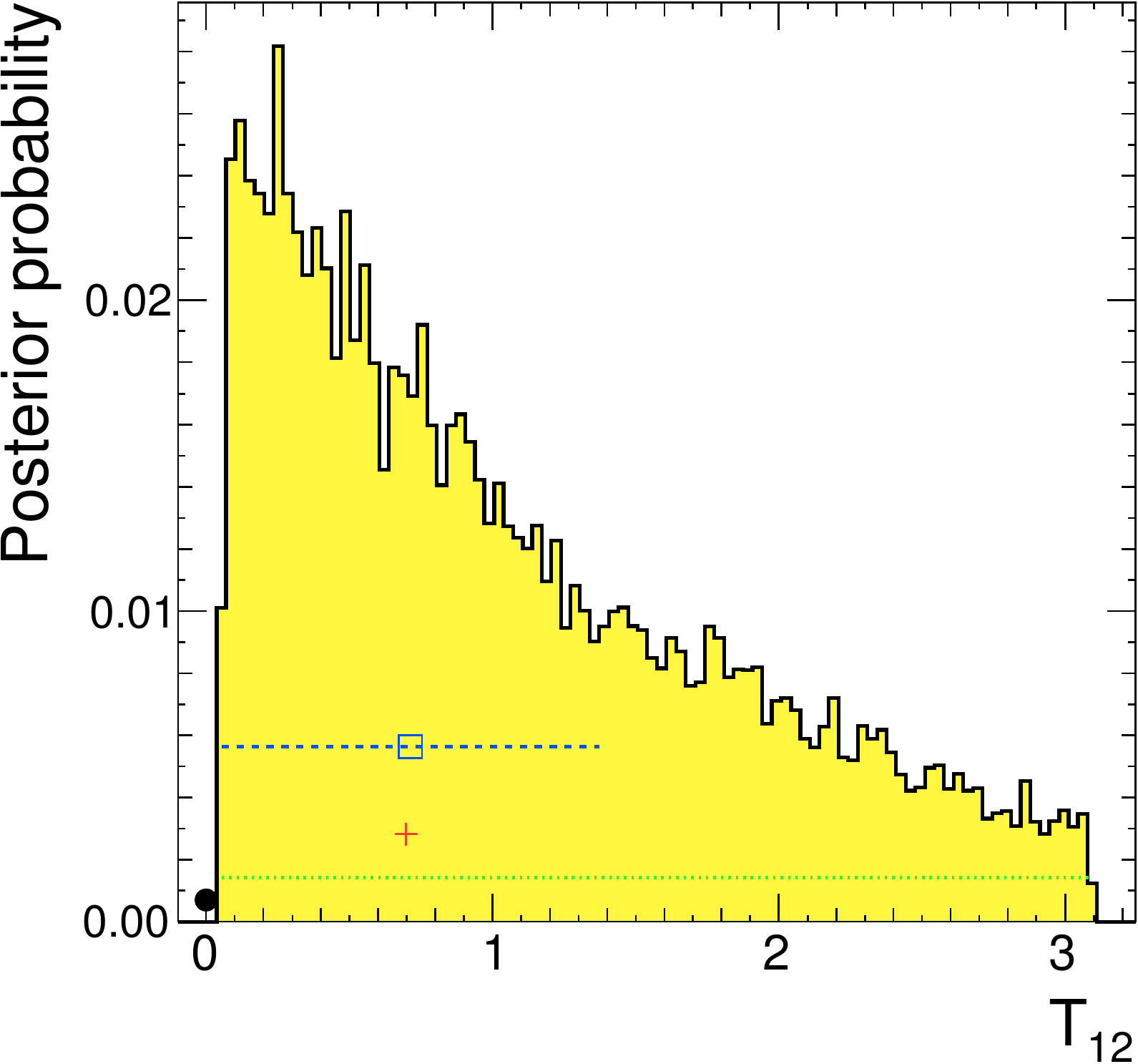} &
   \includegraphics[width=0.18\columnwidth]{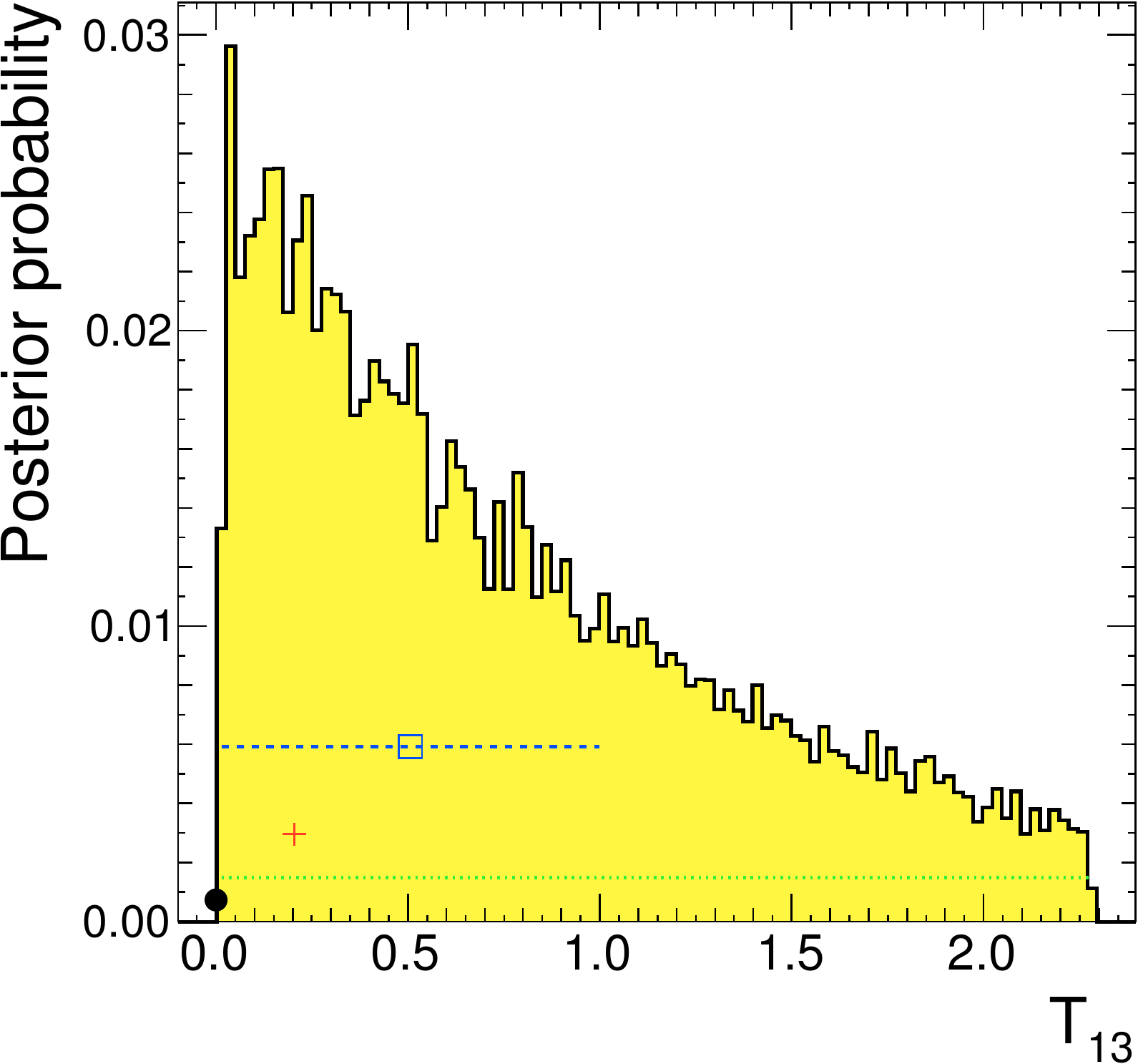} &
   \includegraphics[width=0.18\columnwidth]{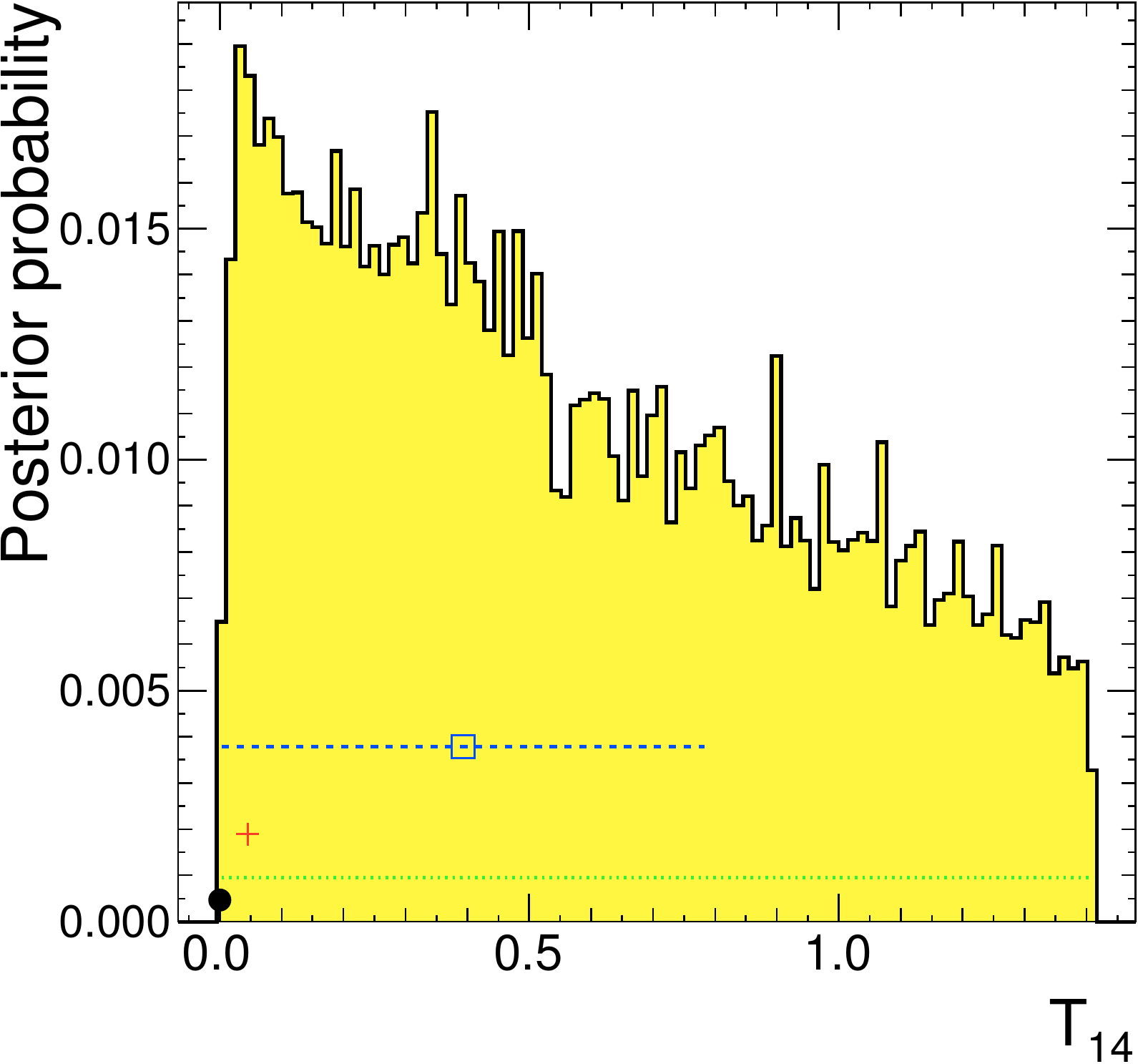} 
 \end{tabular}
 \caption{The 1-dimensional marginal distributions of $p(\tuple{T}|\tuple{D})$ in the example of Sec.~\ref{sec:example4}.  The yellow distribution is $P_t(T_t|\tuple{D})$.  The red cross marker shows the actual truth spectrum content in each bin ($\hat{T}_t$).  The black circle marker shows the observed data in each bin ($D_t$).  The blue dashed line and the blue square marker show the unfolded spectrum contents $[U_t^\ulcorner,U_t^\urcorner]$ and $U_t$.  The green dotted line shows the range in $T_t$ that is included in the sampled hyper-box (after volume reduction).
\label{fig:1Dim4}}
\end{figure}

\begin{figure}[H]
  \centering
  \subfigure[]{
    \includegraphics[width=0.3\columnwidth]{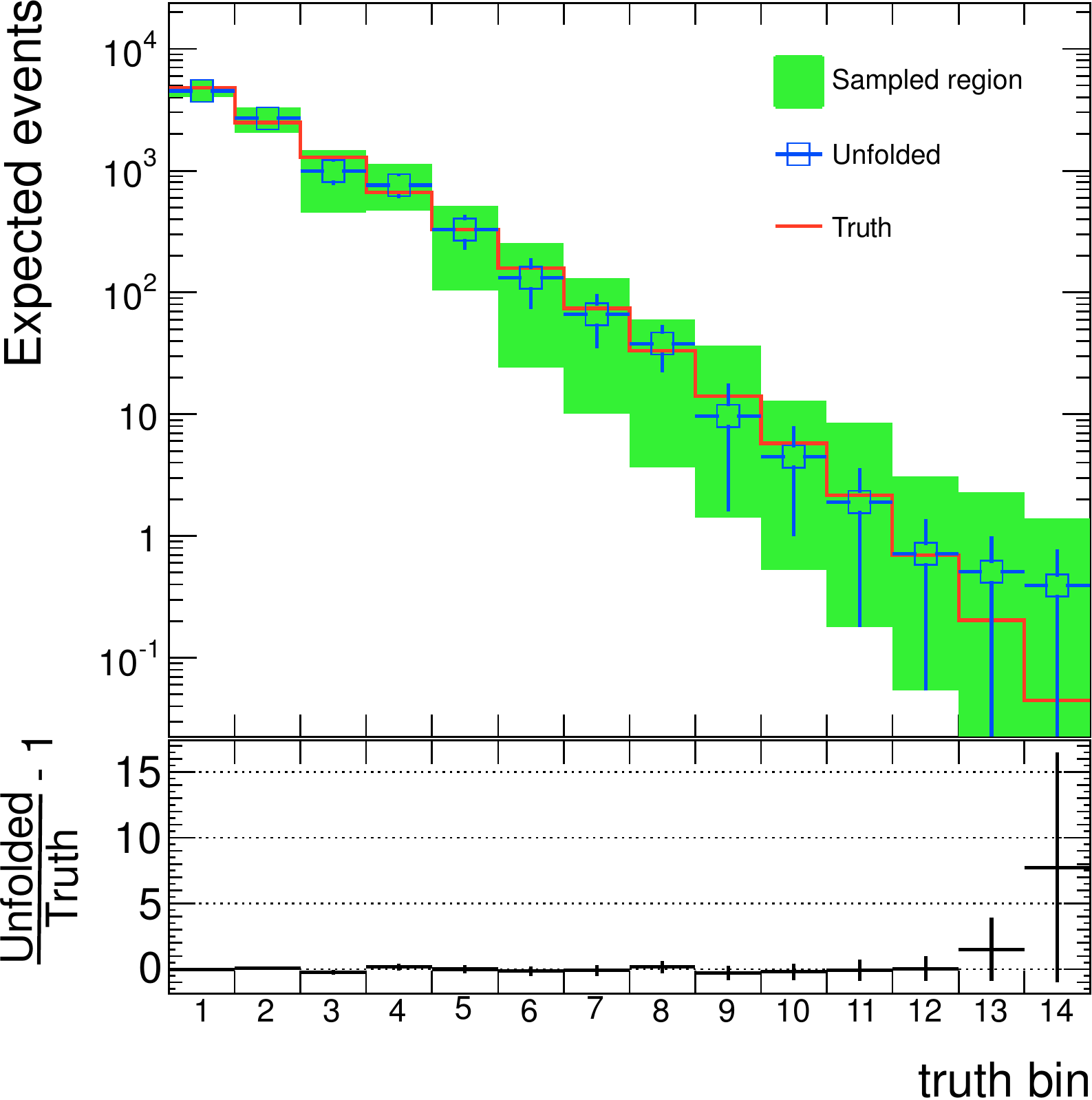}
  }
  \subfigure[]{
    \includegraphics[width=0.3\columnwidth]{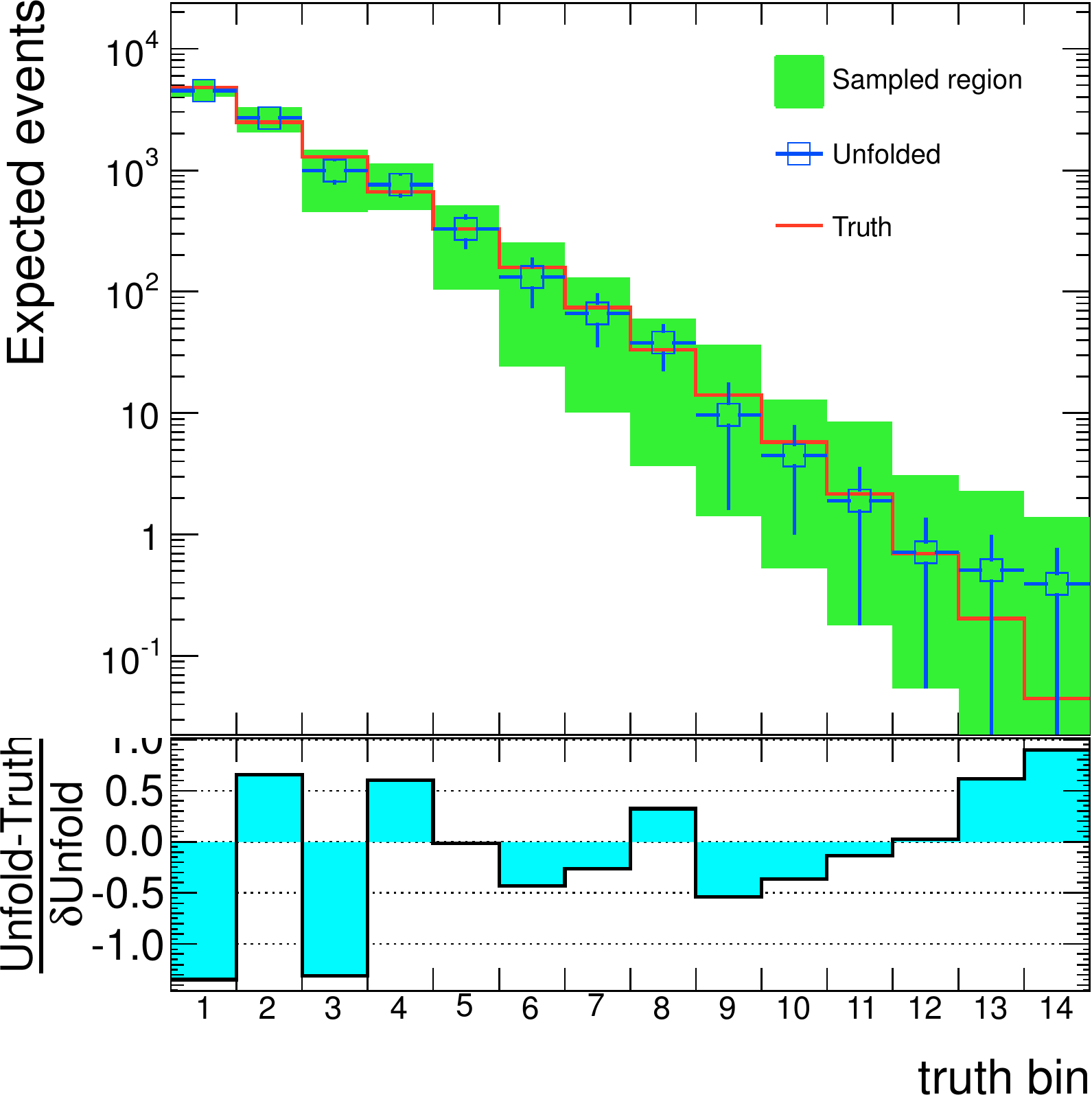}
  }
   \subfigure[]{
     \includegraphics[width=0.3\columnwidth]{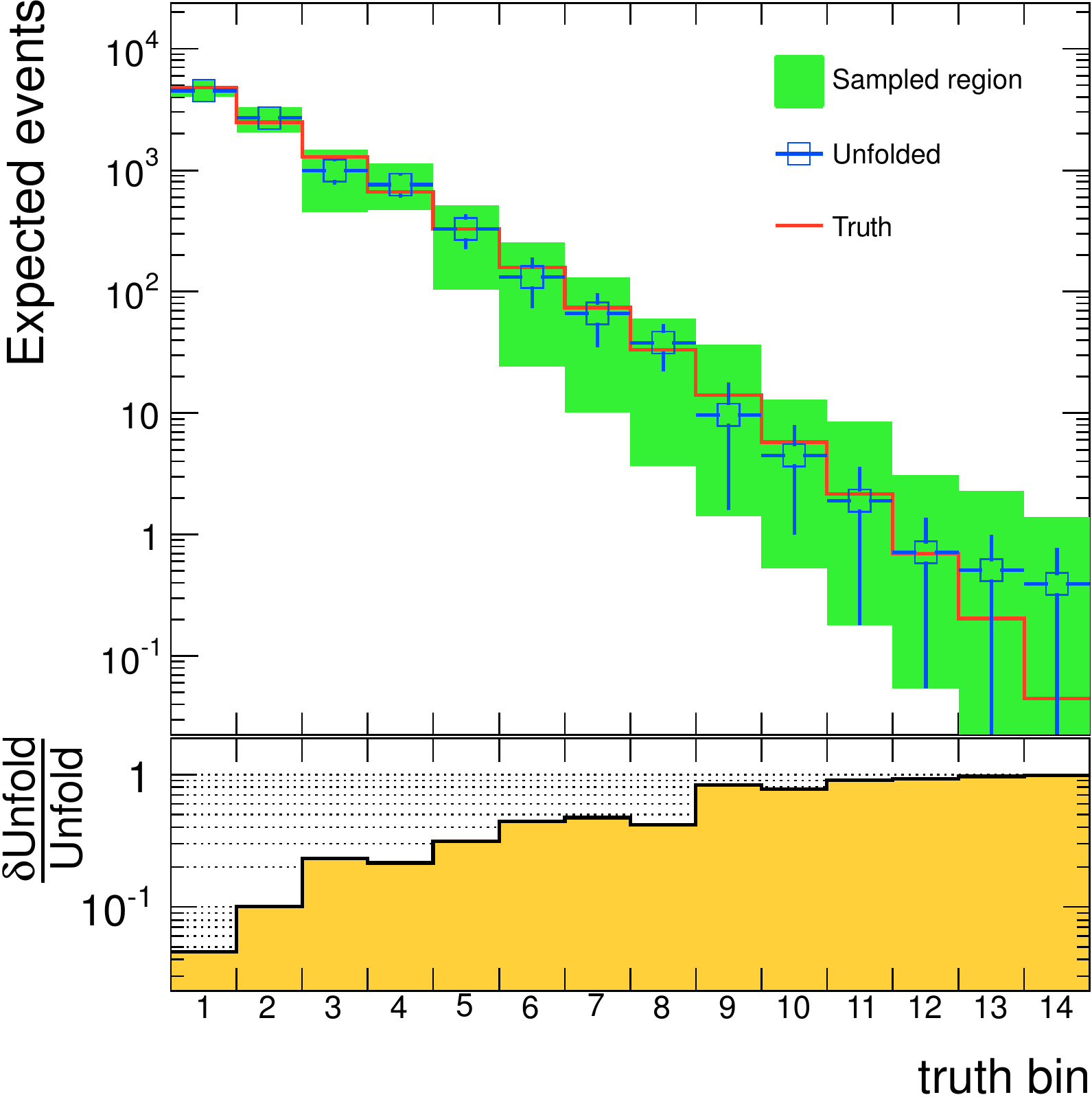}
     \label{fig:unfolded4c}
  }
\caption{The unfolded spectrum of the example in Sec.~\ref{sec:example4}.  The relative difference is plotted in the inset of (a), but due to the large error bars it is hard to see the difference in the first truth bins.  In (b), the difference $U_t-\hat{T}_t$ is divided by the error bars of the unfolded spectrum $\delta \text{Unfold} \equiv \frac{U_t^\urcorner - U_t^\ulcorner}{2}$.  Only bins 1 and 3 have $\hat{T}_t$ outside of $[U_t^\ulcorner,U_t^\urcorner]$.  In (c), the relative error bars of the unfolded spectrum is shown, $\frac{|U_t^\ulcorner - U_t^\urcorner|/2}{U_t} = \frac{|U_t^\ulcorner - U_t^\urcorner|}{U_t^\ulcorner + U_t^\urcorner}$.
\label{fig:unfolded4}}
\end{figure}

The 1-dimensional marginal distributions of $p(\T|\D)$ are shown in Fig.~\ref{fig:1Dim4}, and the unfolded spectrum in Fig.~\ref{fig:unfolded4}.

\begin{figure}[H]
\centering
\begin{tabular}{ccccc}
   \includegraphics[width=0.18\columnwidth]{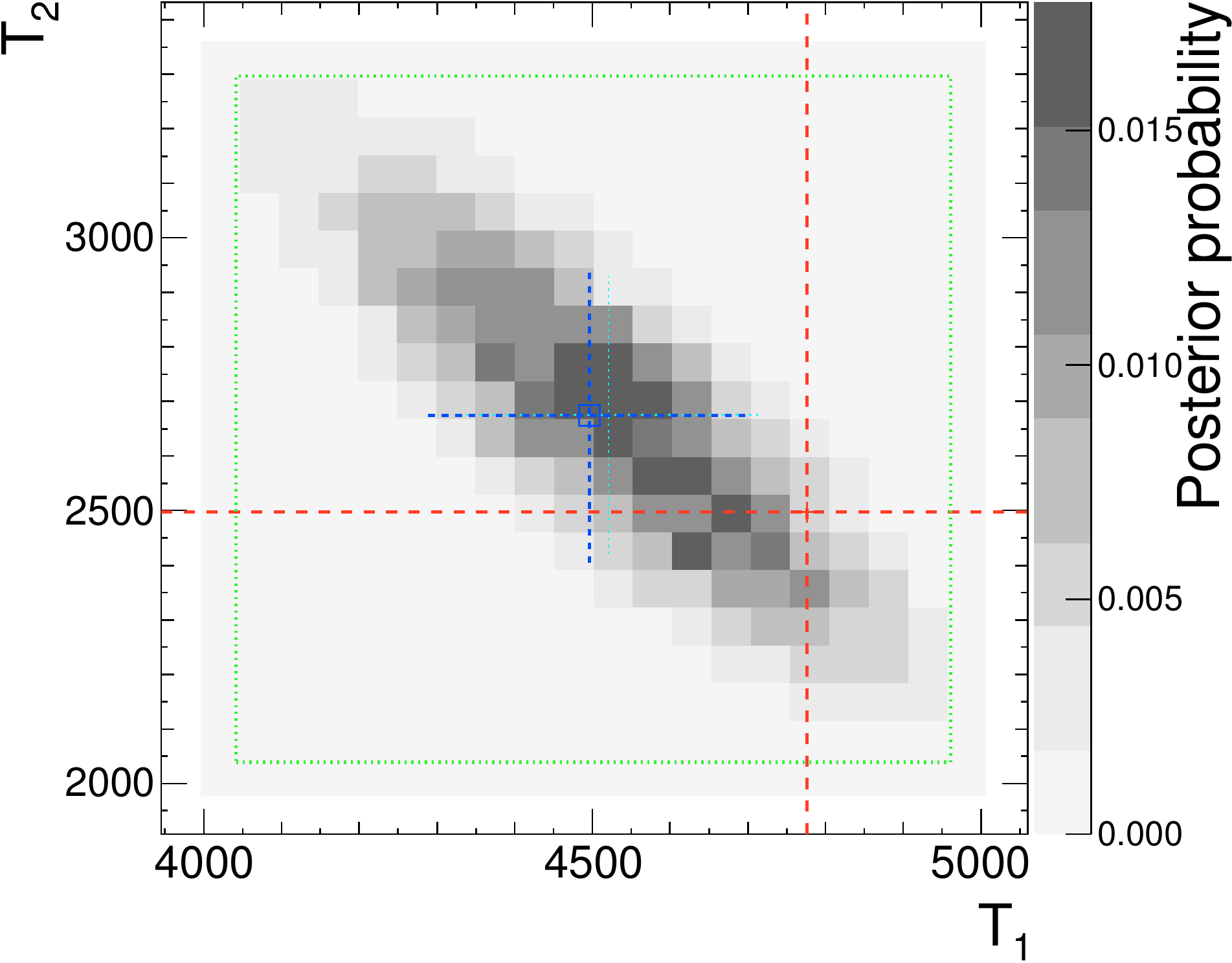} &
   \includegraphics[width=0.18\columnwidth]{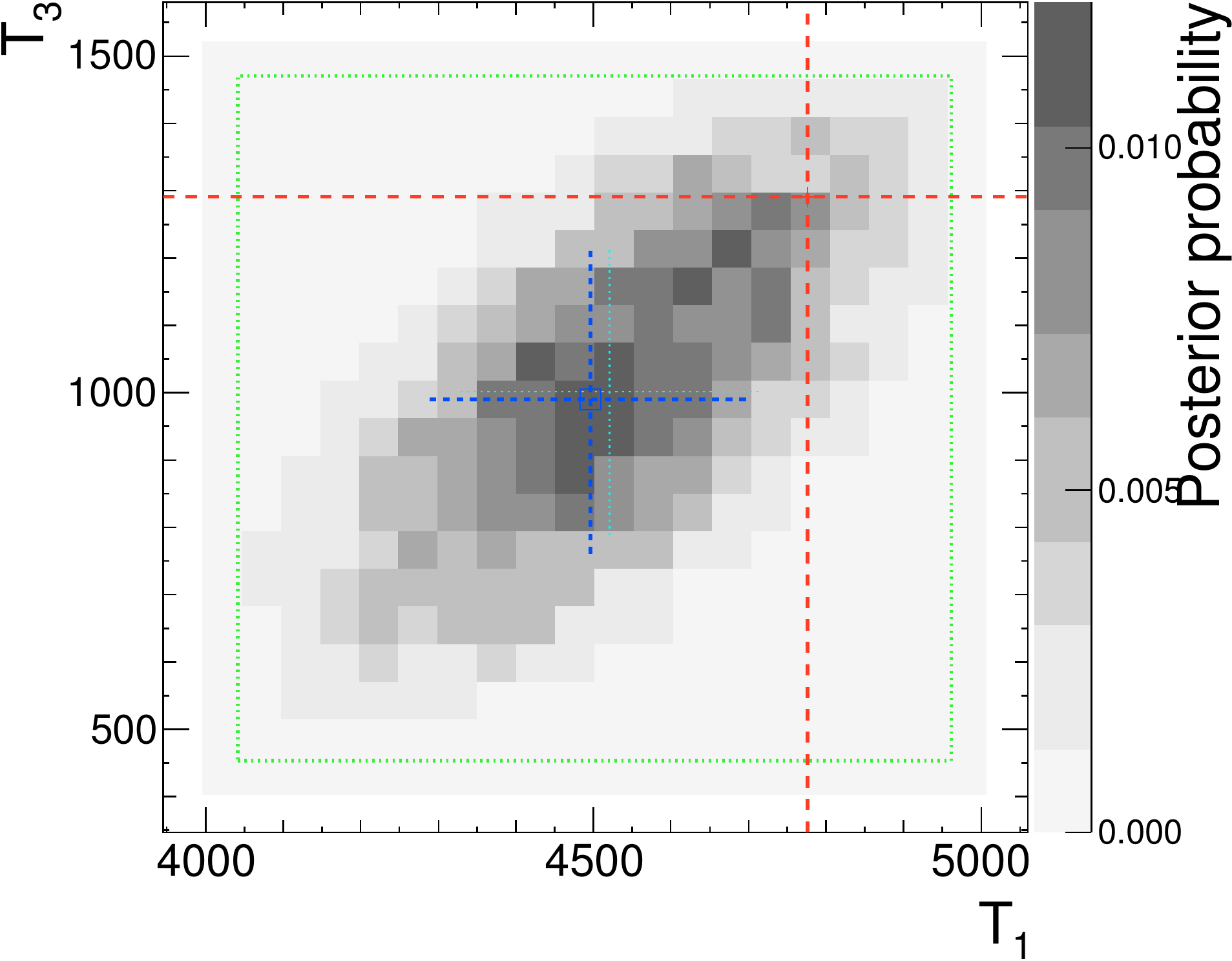} &
   \includegraphics[width=0.18\columnwidth]{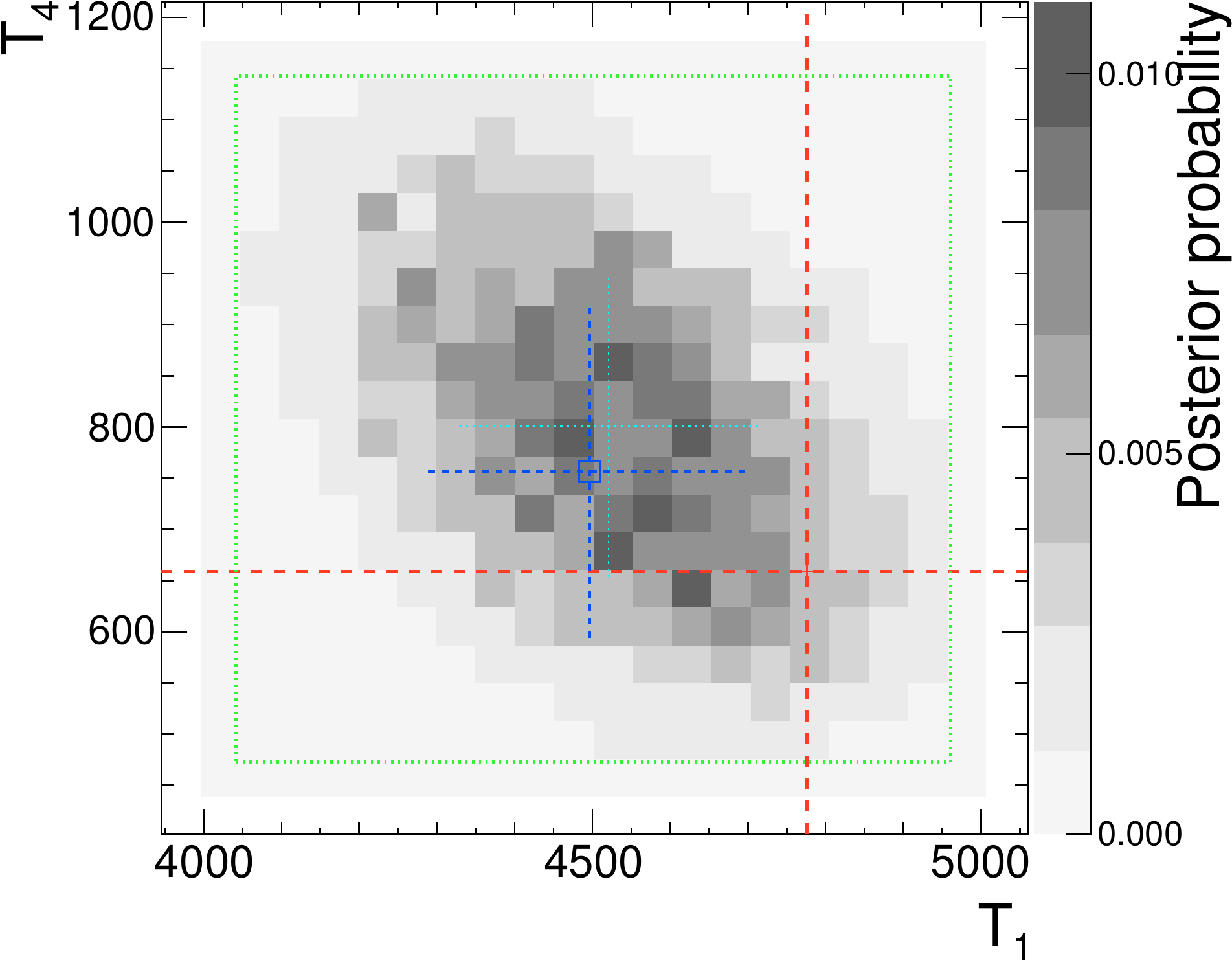} &
   \includegraphics[width=0.18\columnwidth]{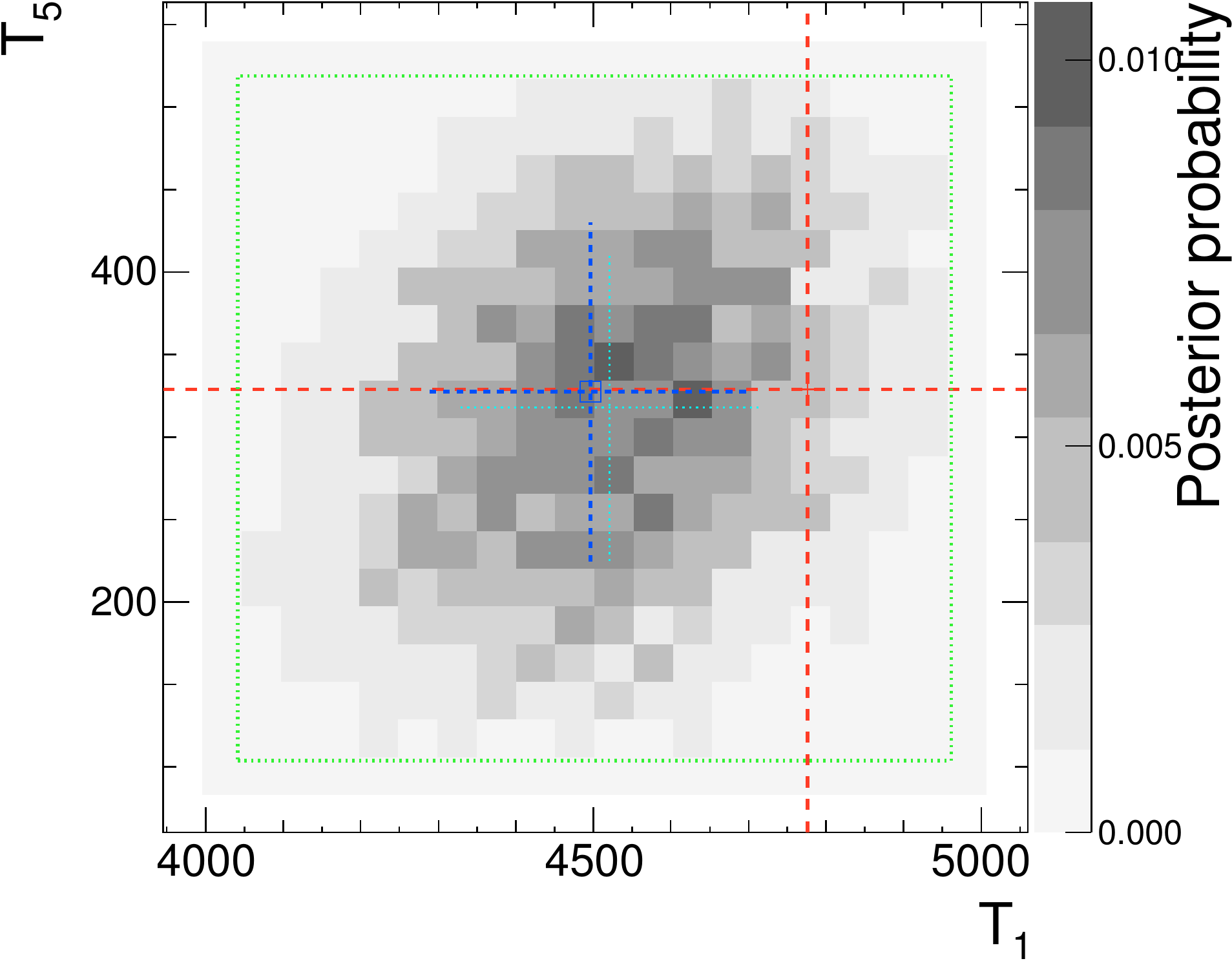} &
   \includegraphics[width=0.18\columnwidth]{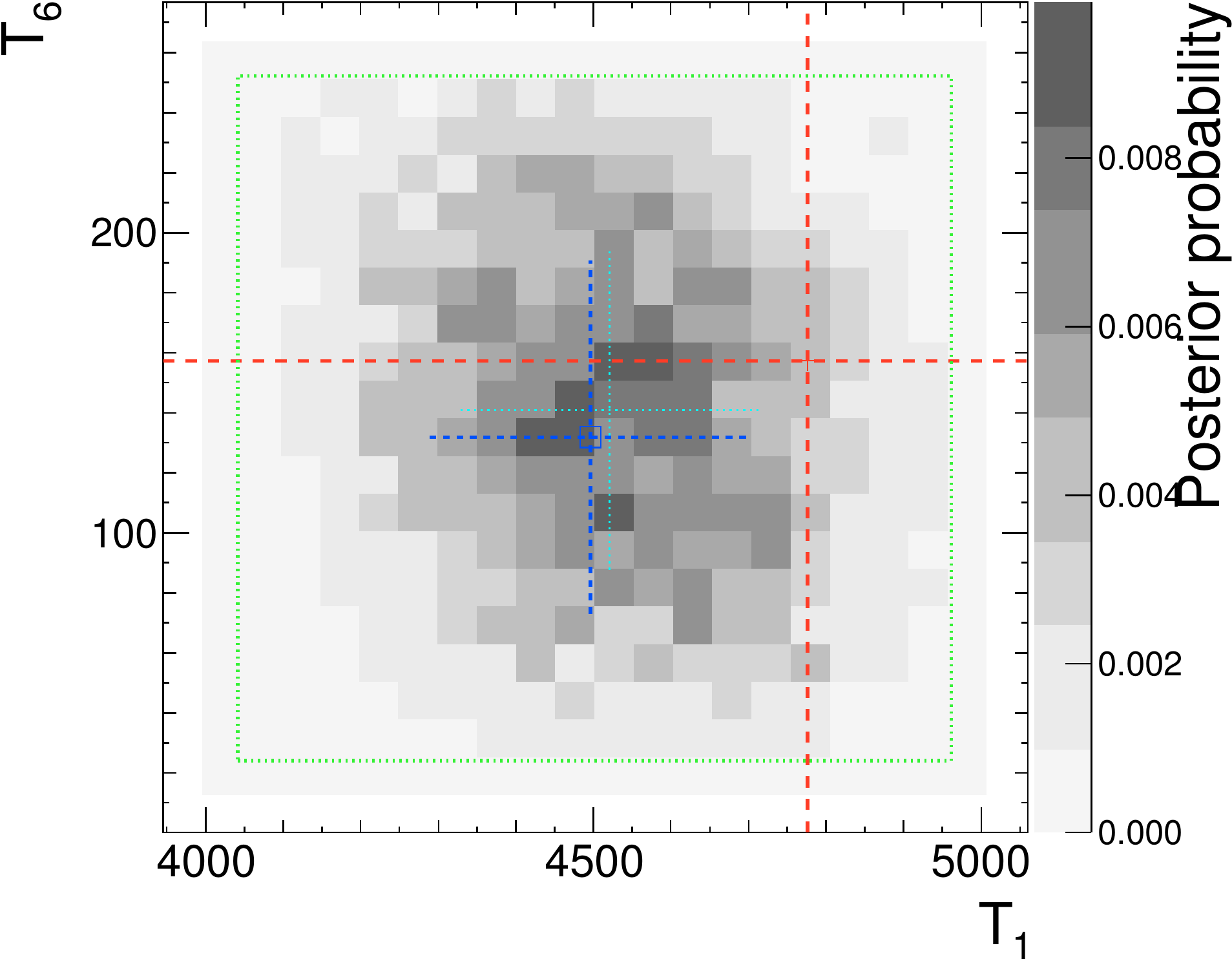} \\

   \includegraphics[width=0.18\columnwidth]{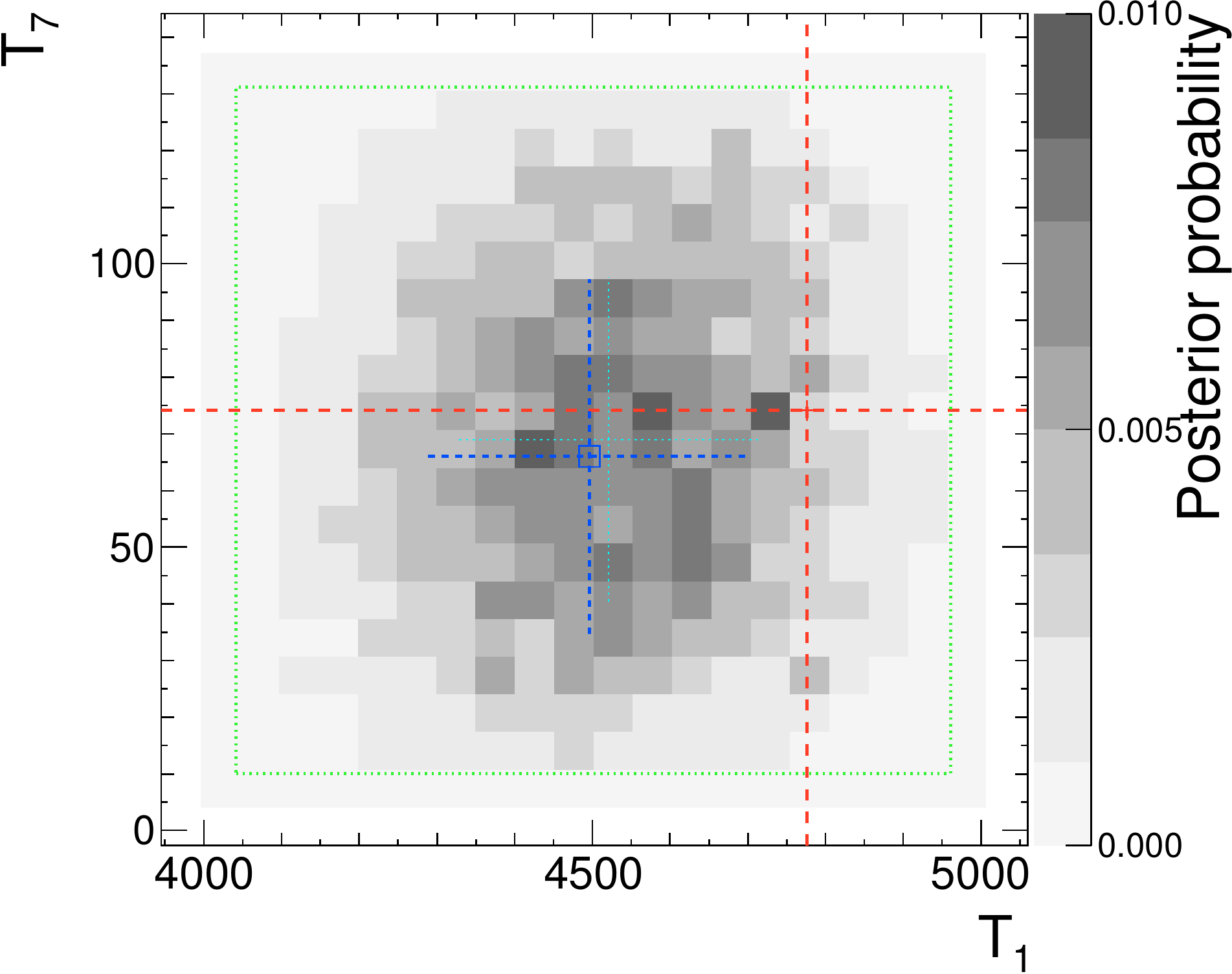} &
   \includegraphics[width=0.18\columnwidth]{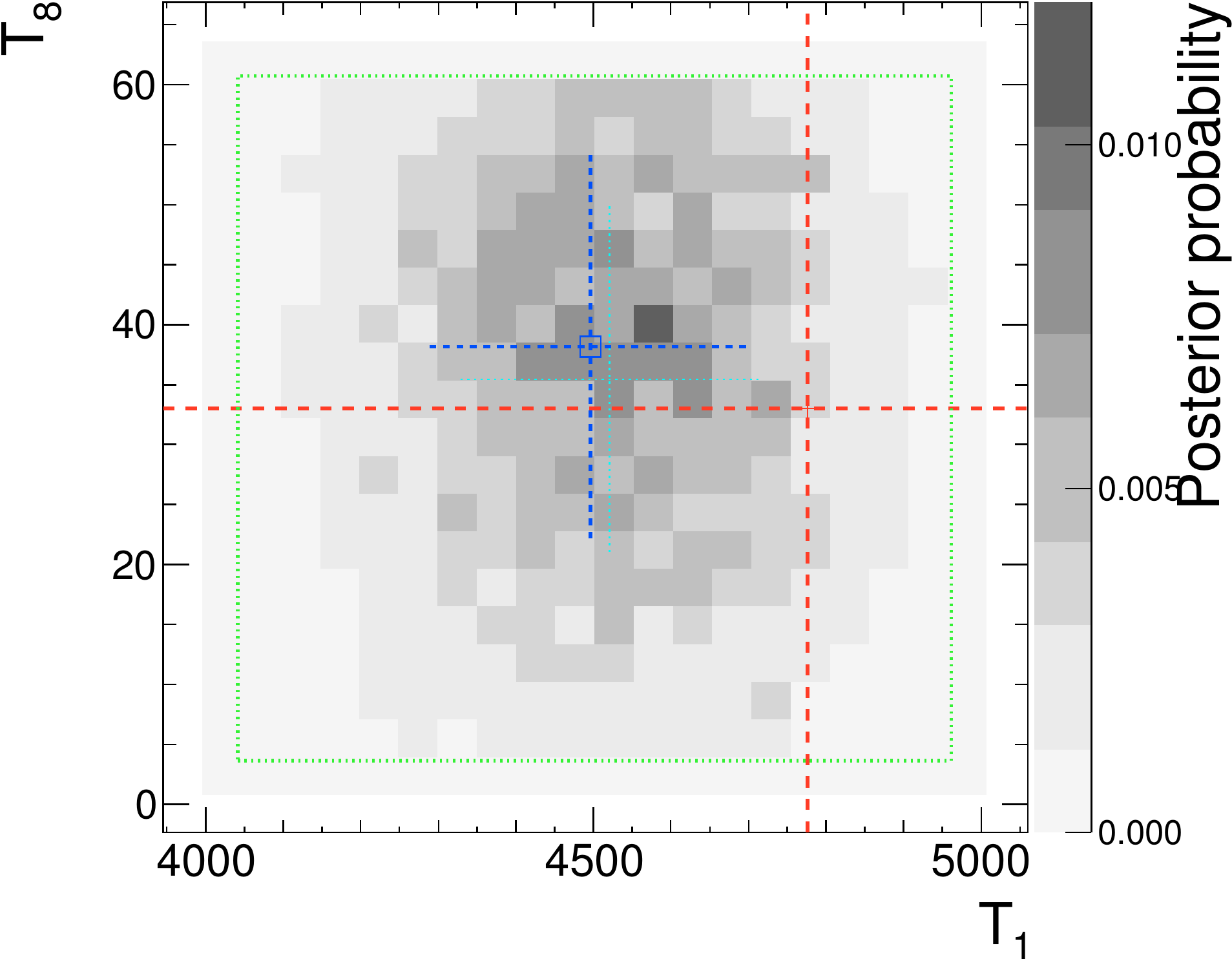} &
   \includegraphics[width=0.18\columnwidth]{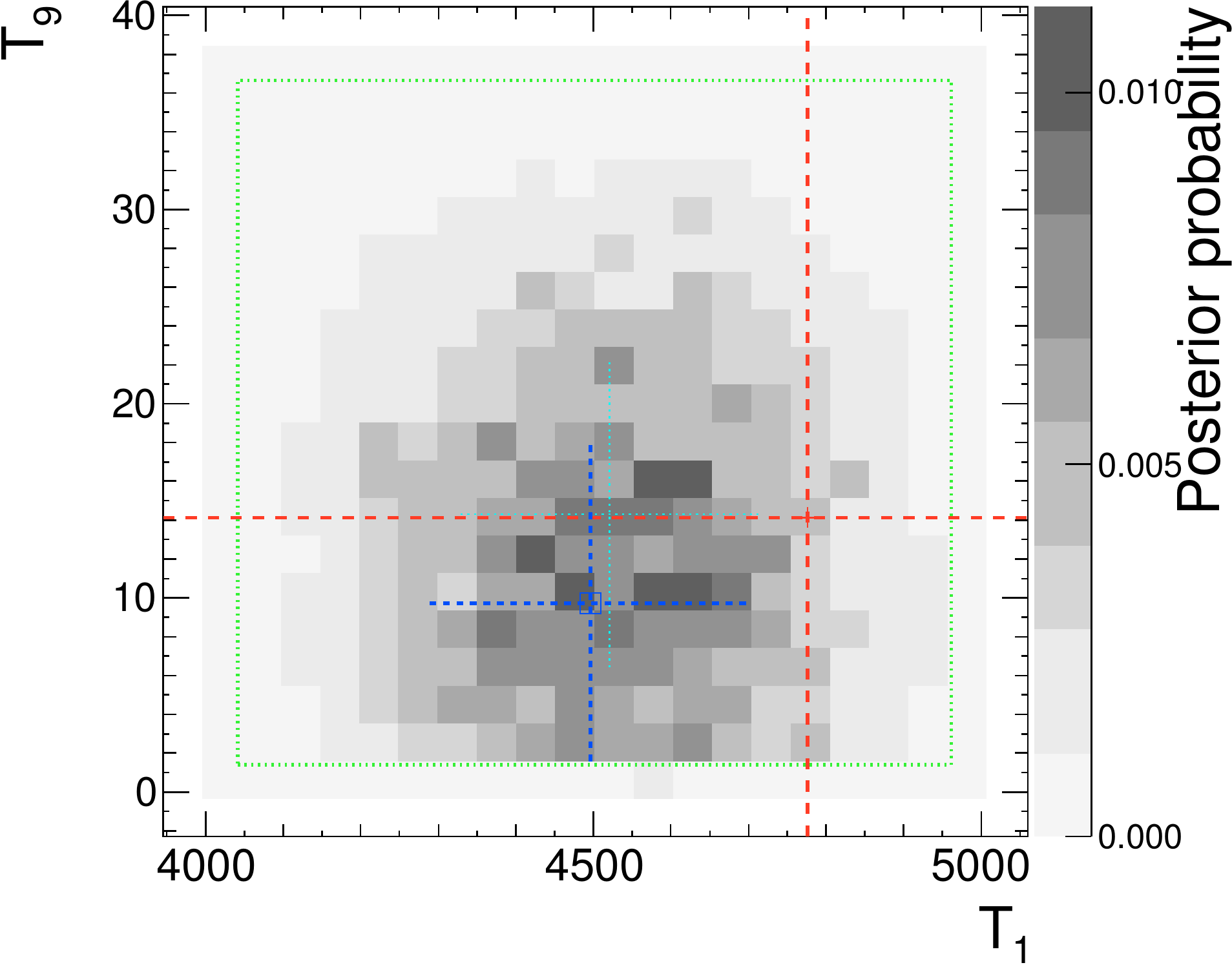} &
   \includegraphics[width=0.18\columnwidth]{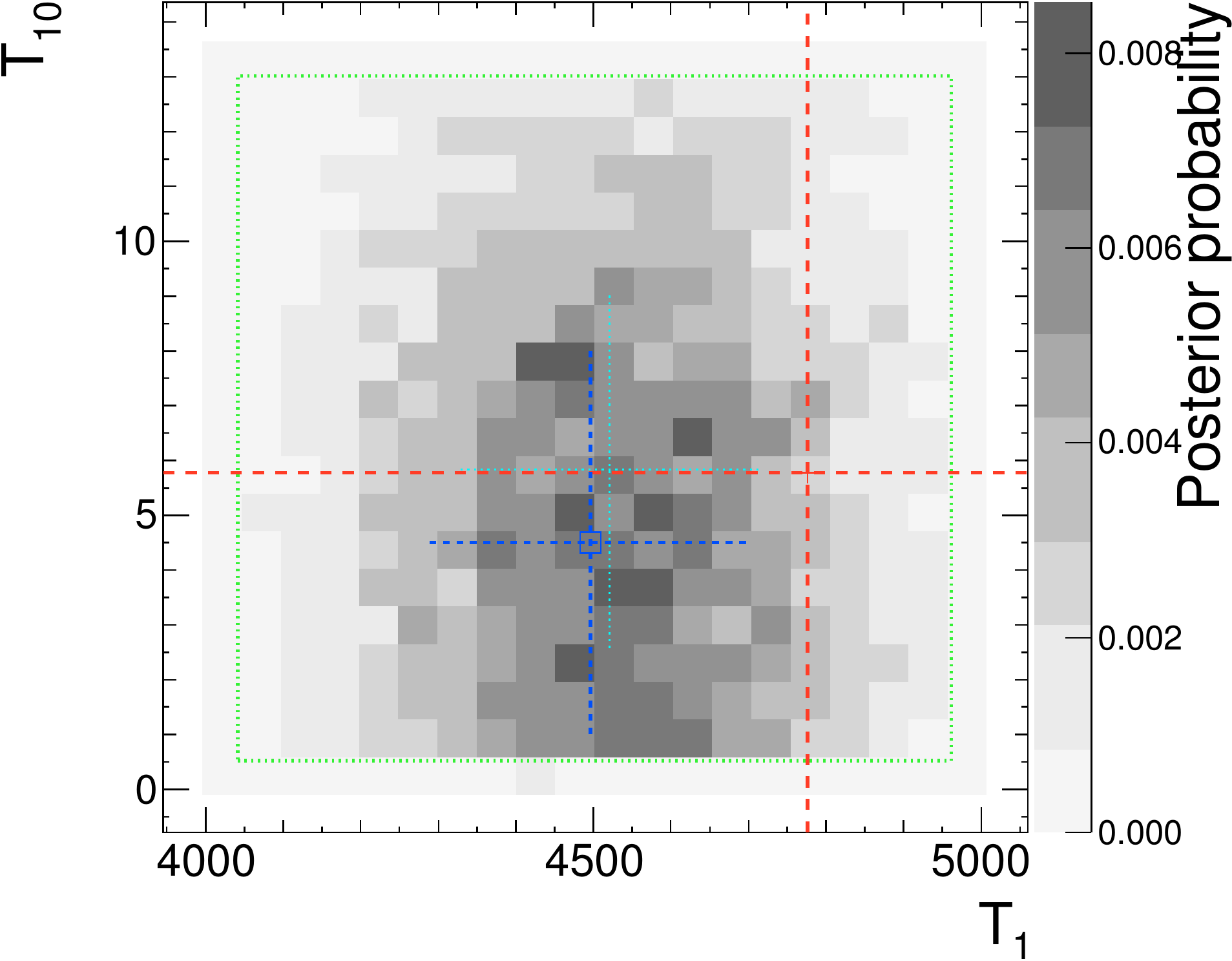} &
   \includegraphics[width=0.18\columnwidth]{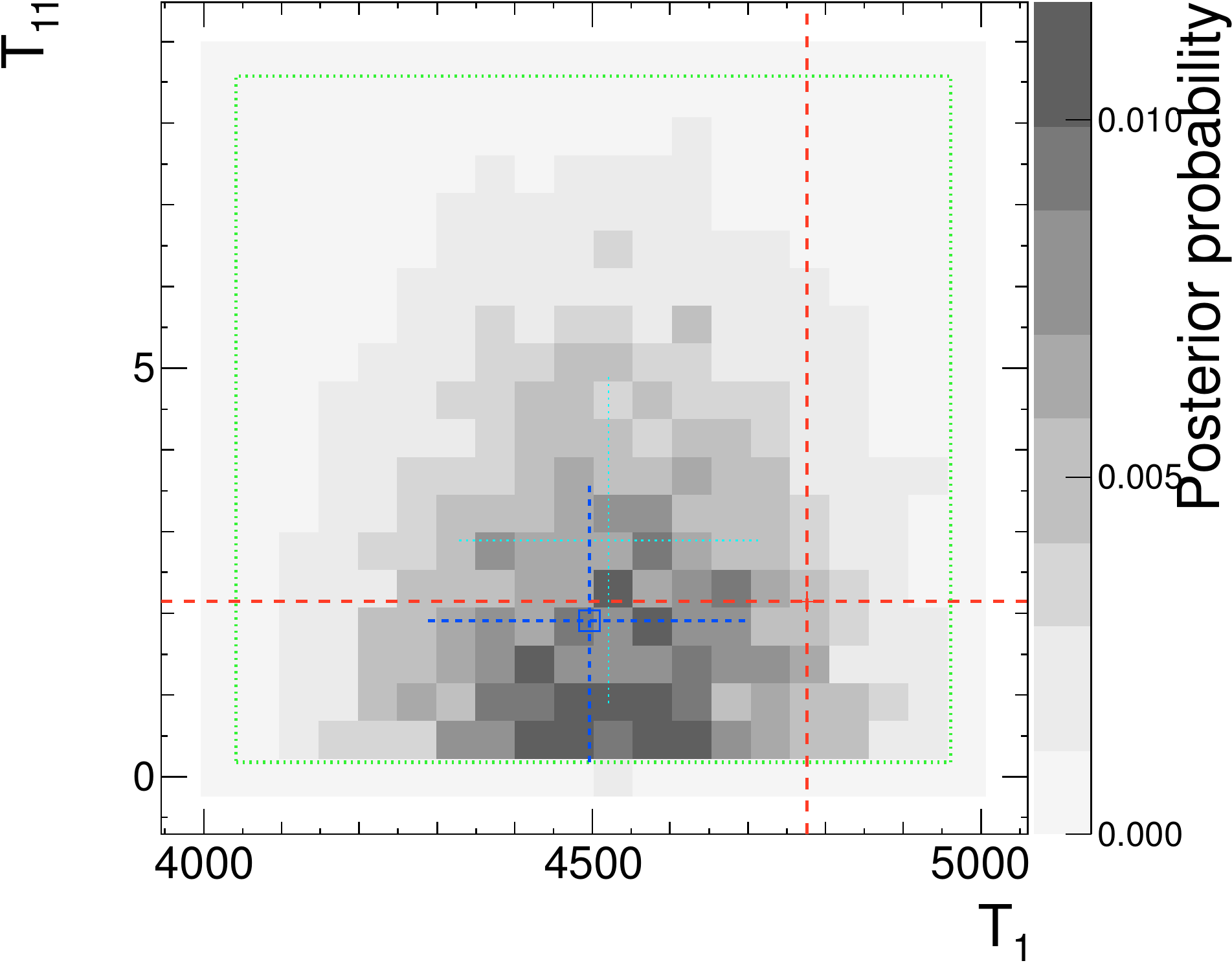} \\

   \includegraphics[width=0.18\columnwidth]{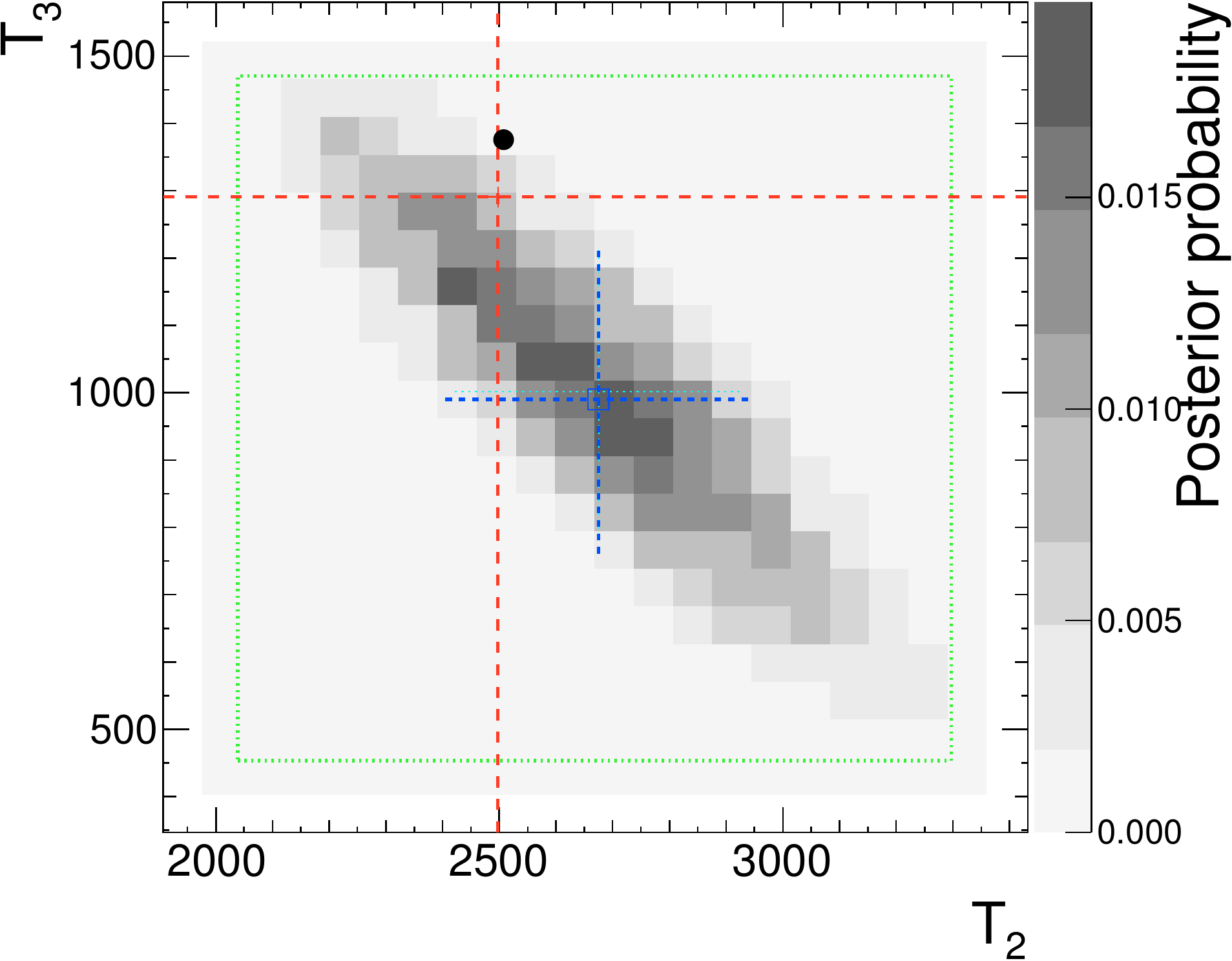} &
   \includegraphics[width=0.18\columnwidth]{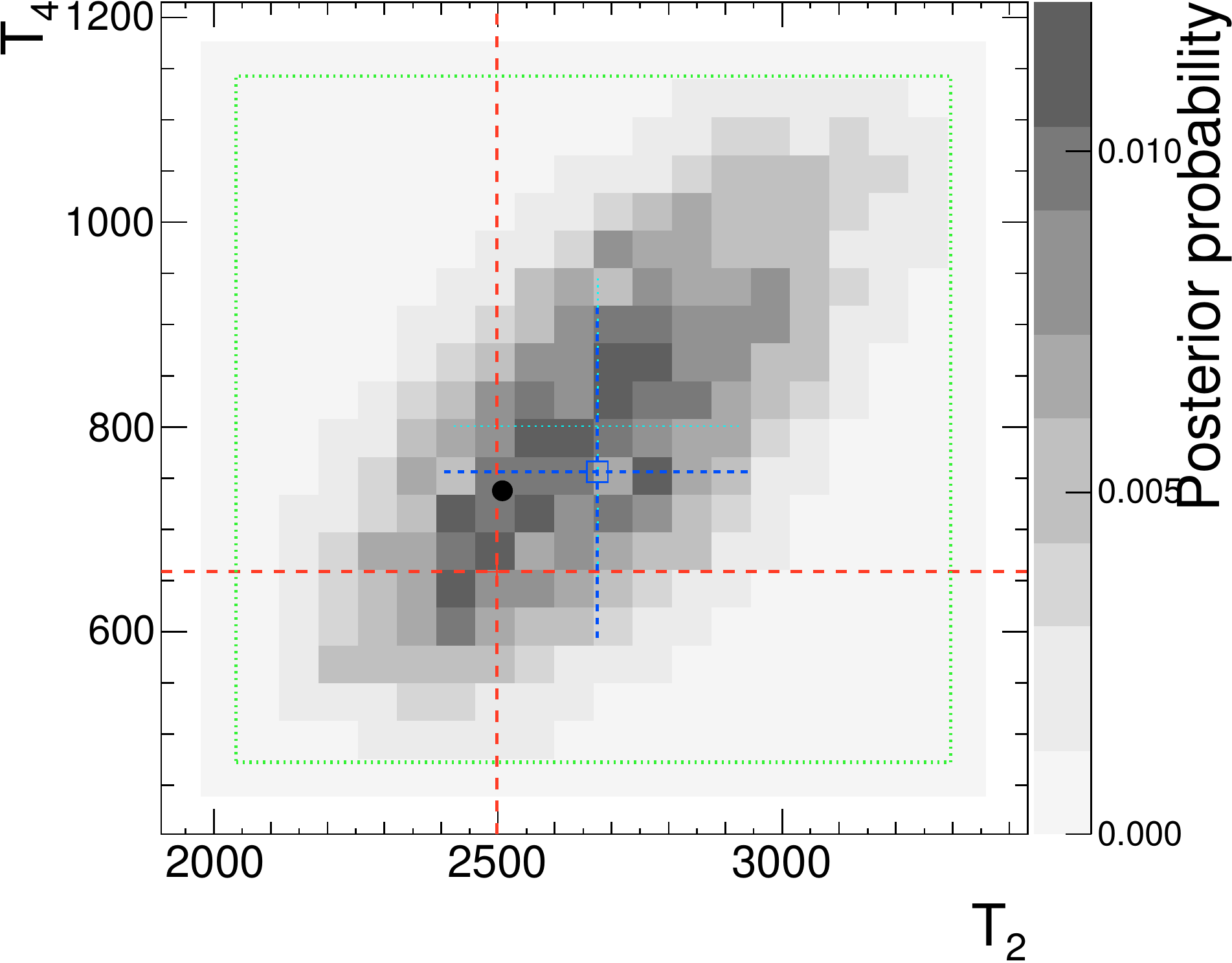} &
   \includegraphics[width=0.18\columnwidth]{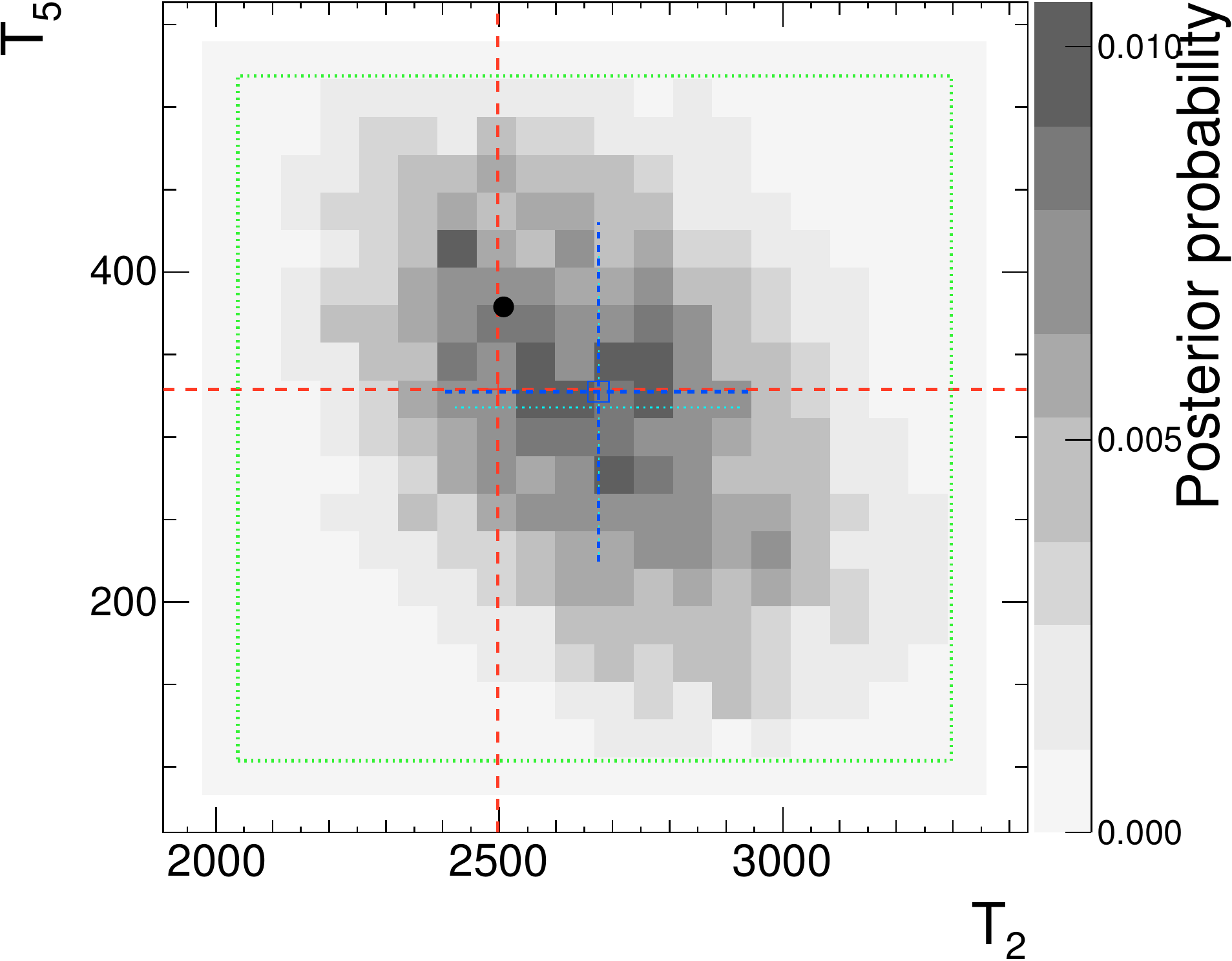} &
   \includegraphics[width=0.18\columnwidth]{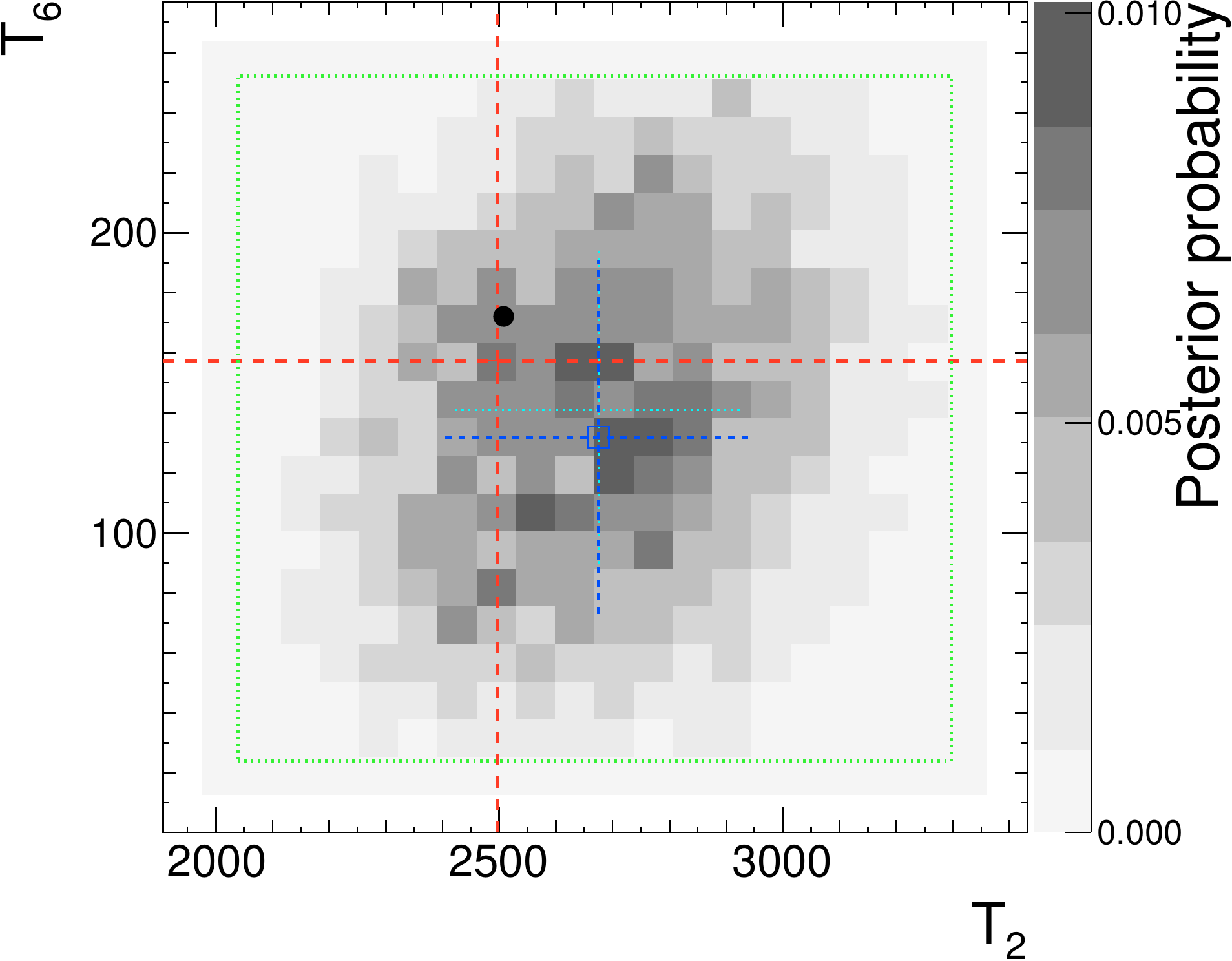} &
   \includegraphics[width=0.18\columnwidth]{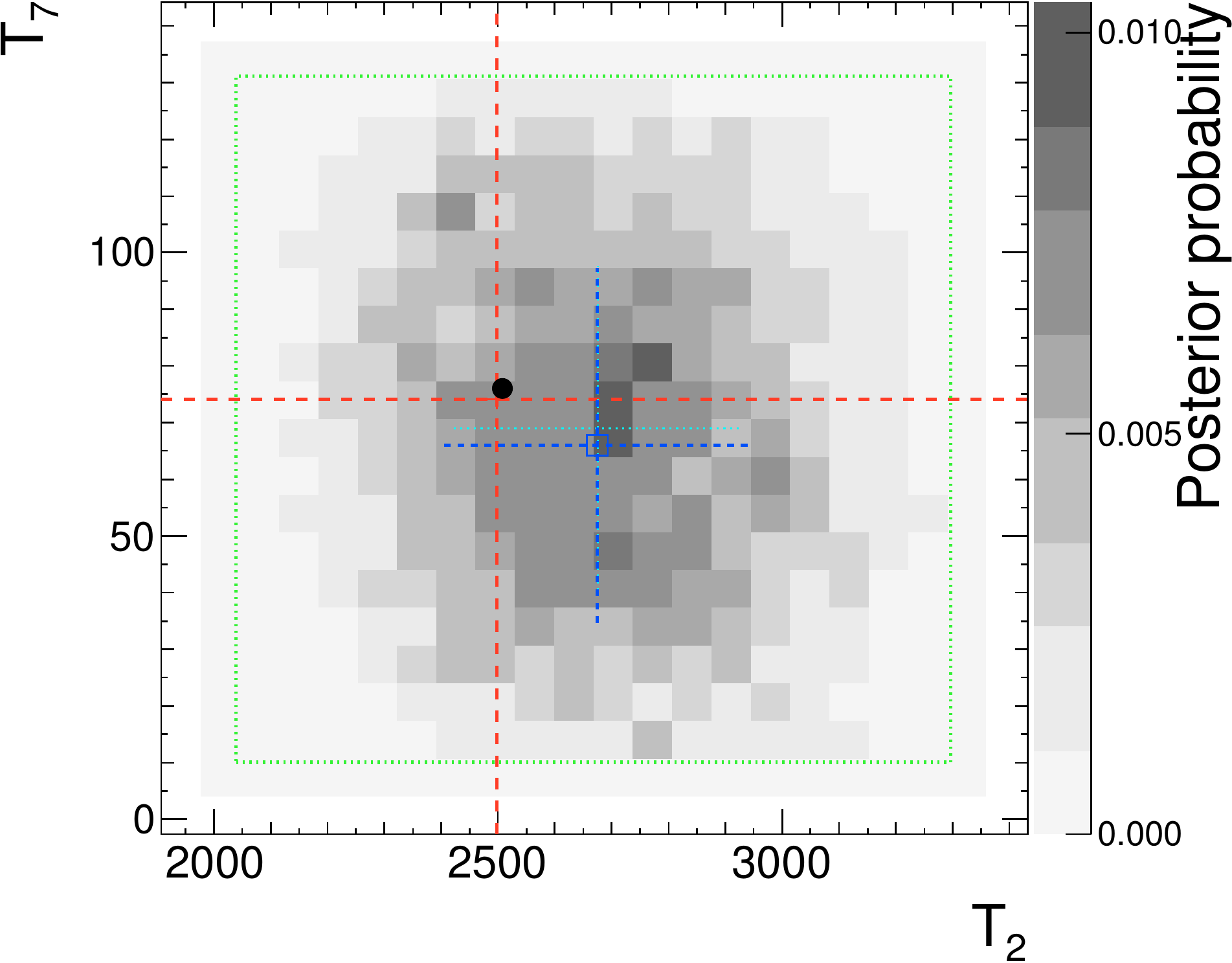} \\

   \includegraphics[width=0.18\columnwidth]{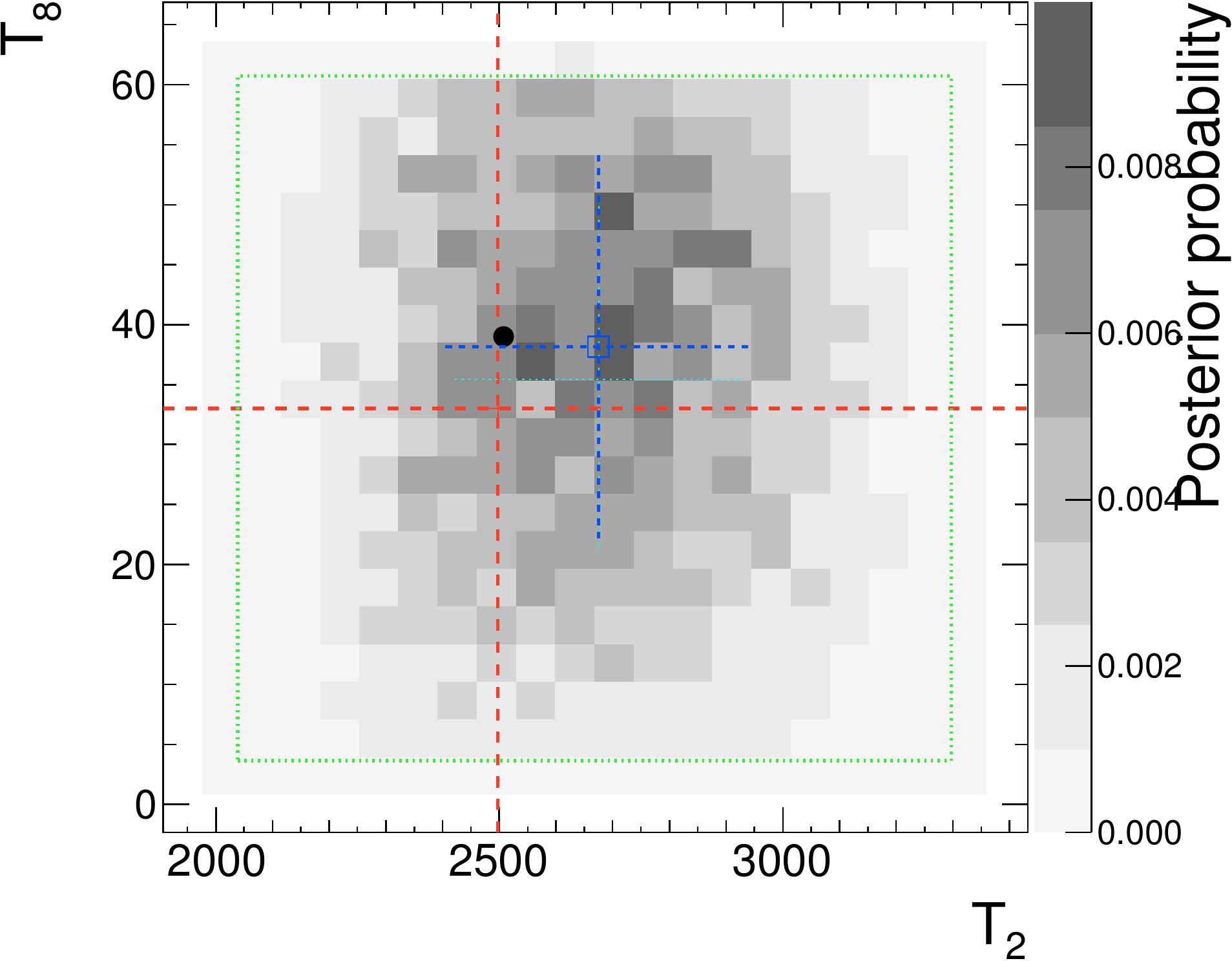} &
   \includegraphics[width=0.18\columnwidth]{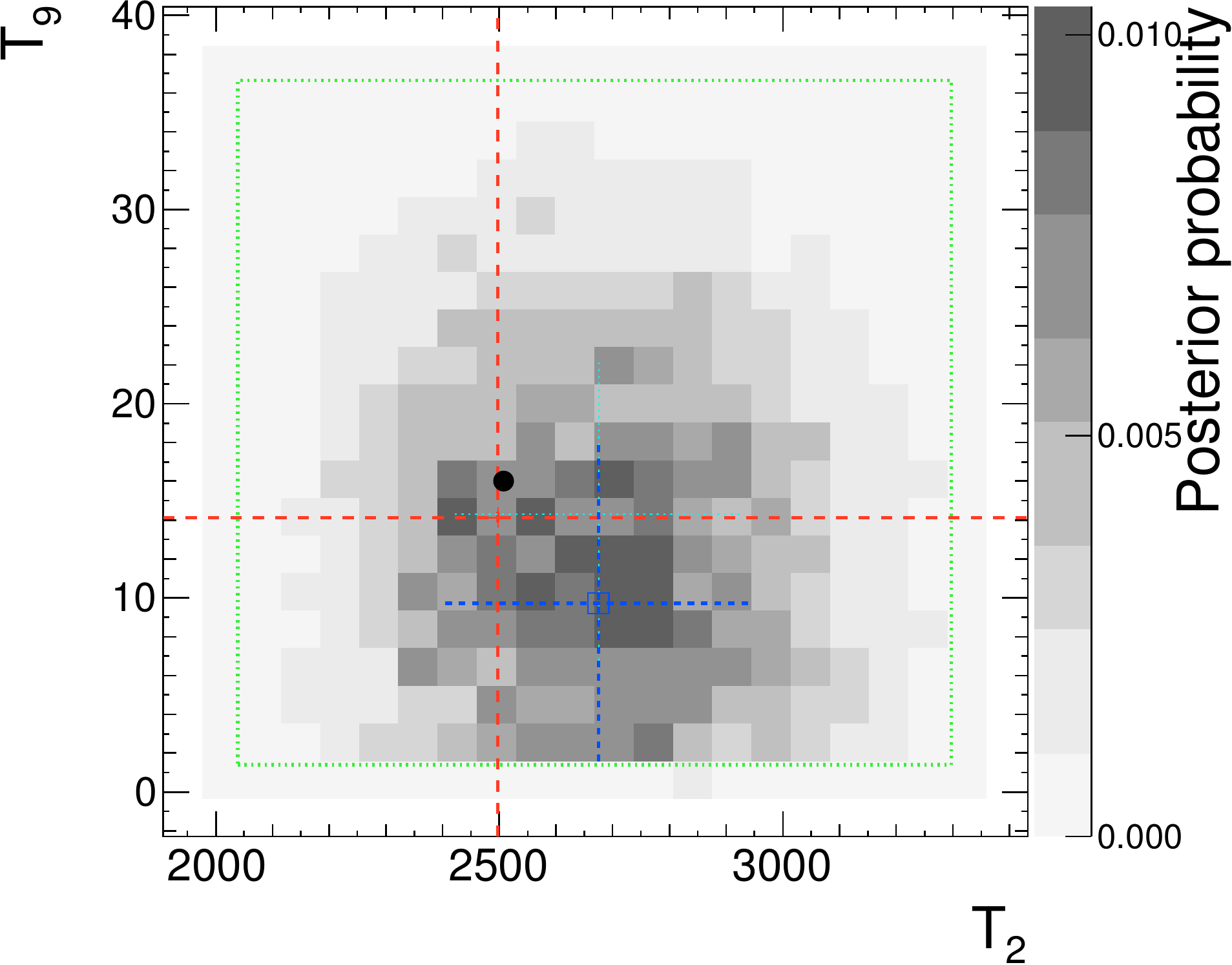} &
   \includegraphics[width=0.18\columnwidth]{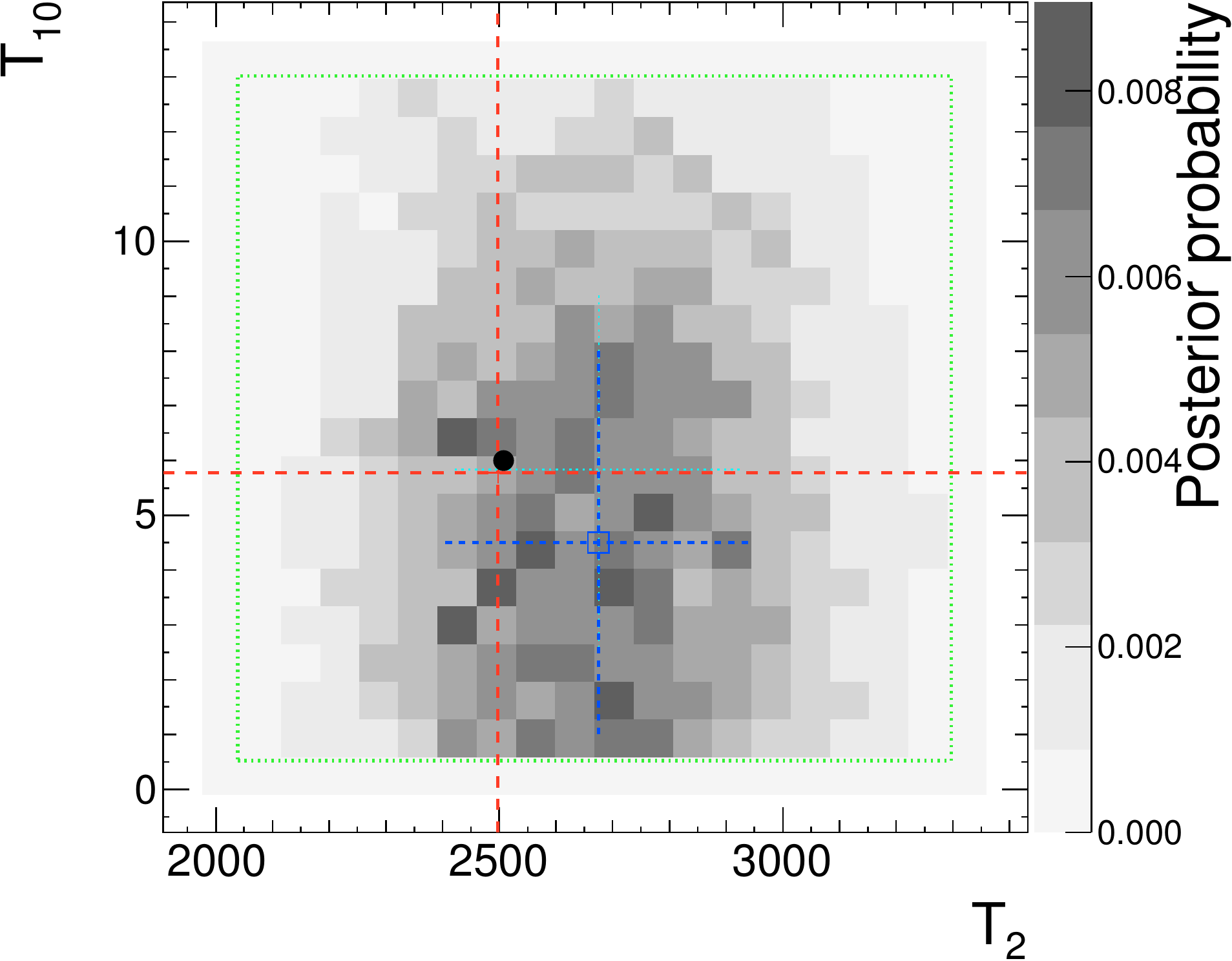} &
   \includegraphics[width=0.18\columnwidth]{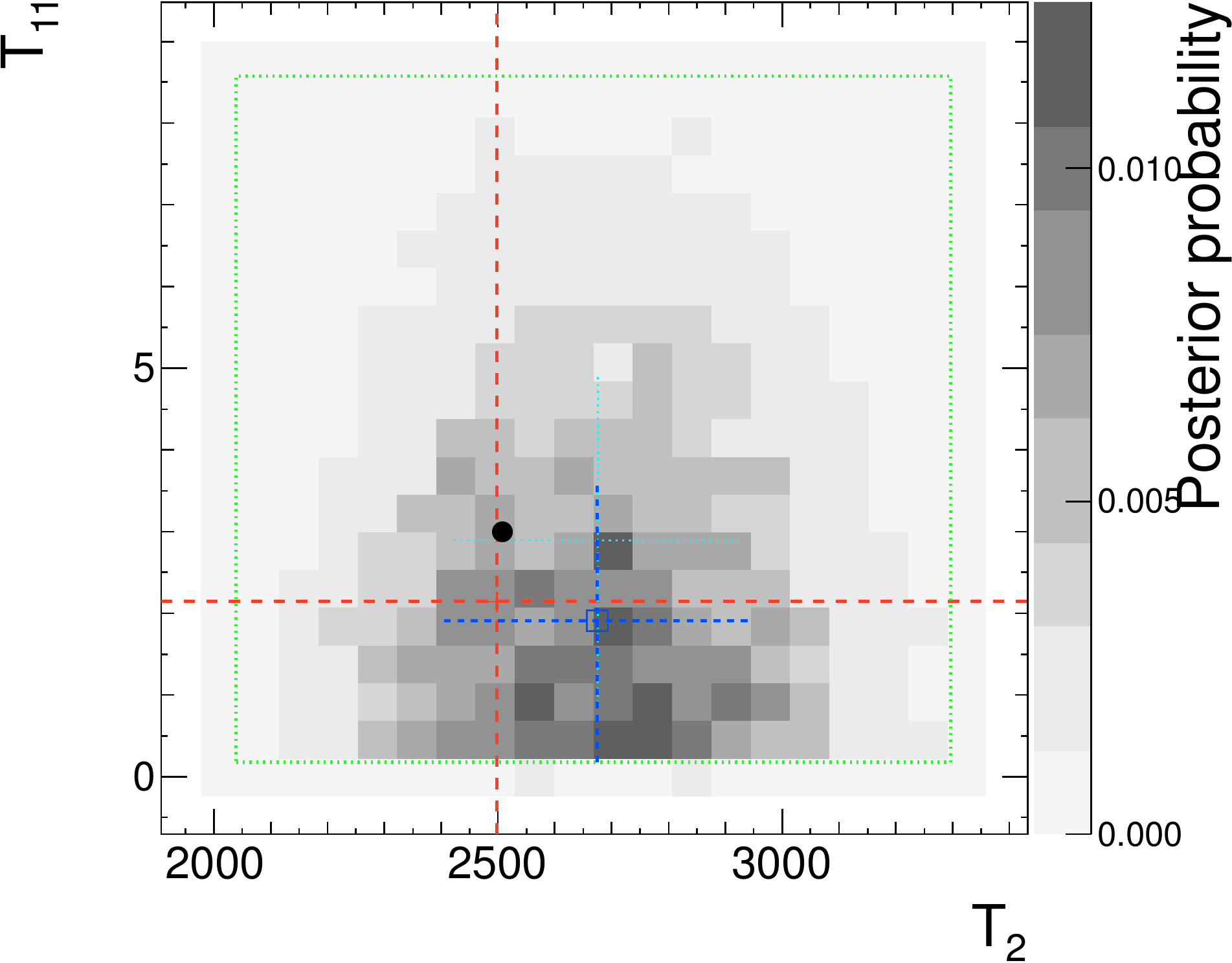} &
   \includegraphics[width=0.18\columnwidth]{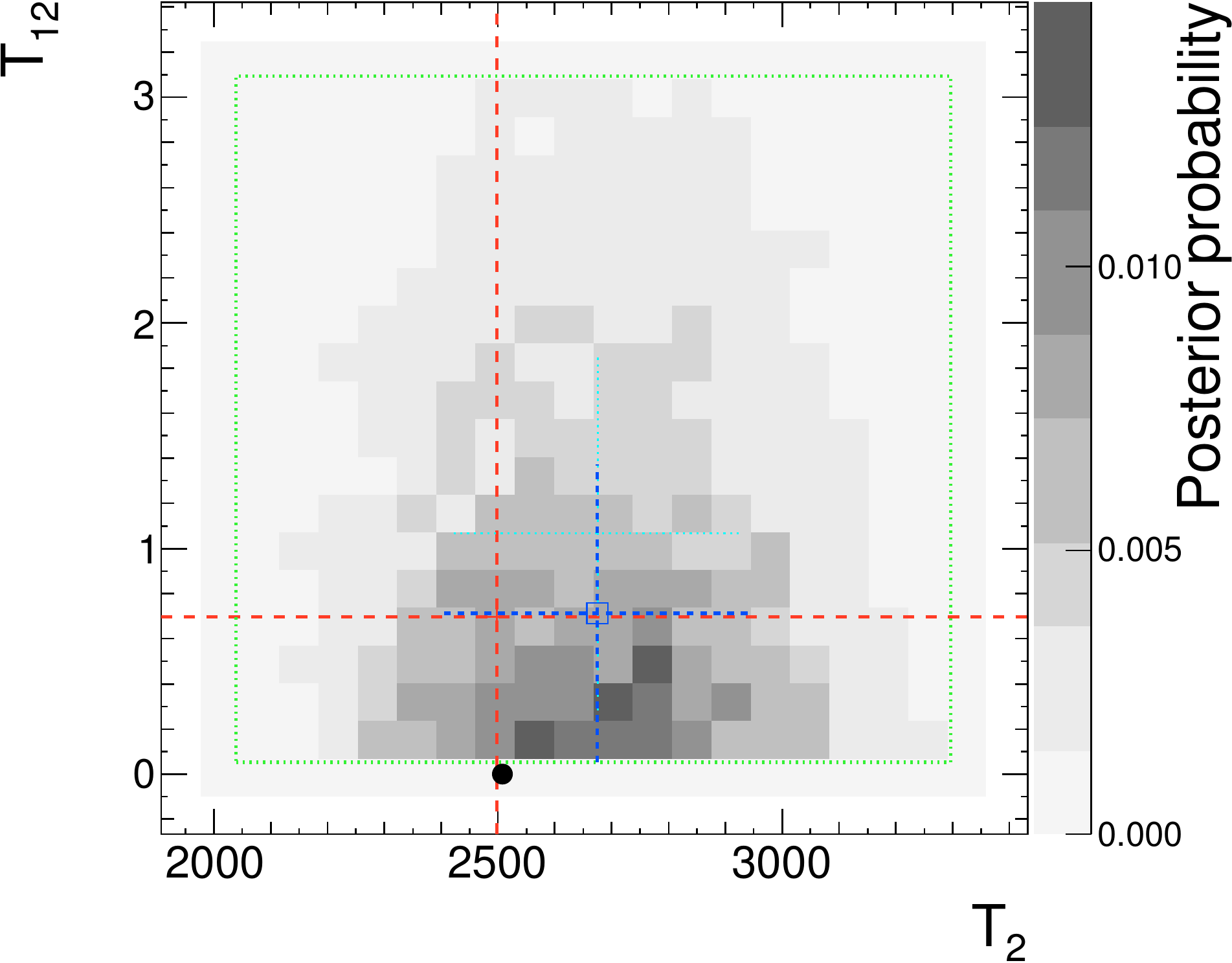} \\

   \includegraphics[width=0.18\columnwidth]{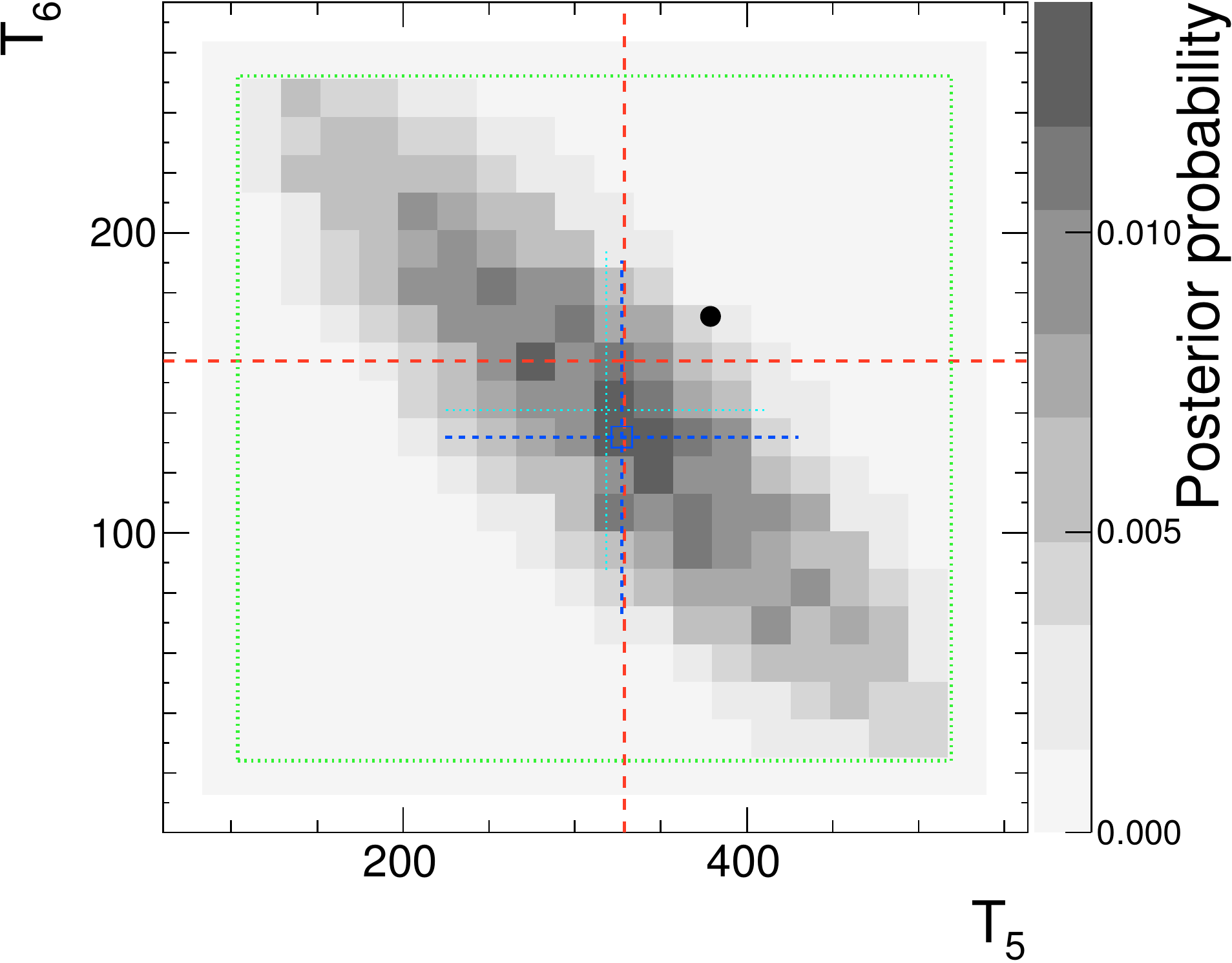} &
   \includegraphics[width=0.18\columnwidth]{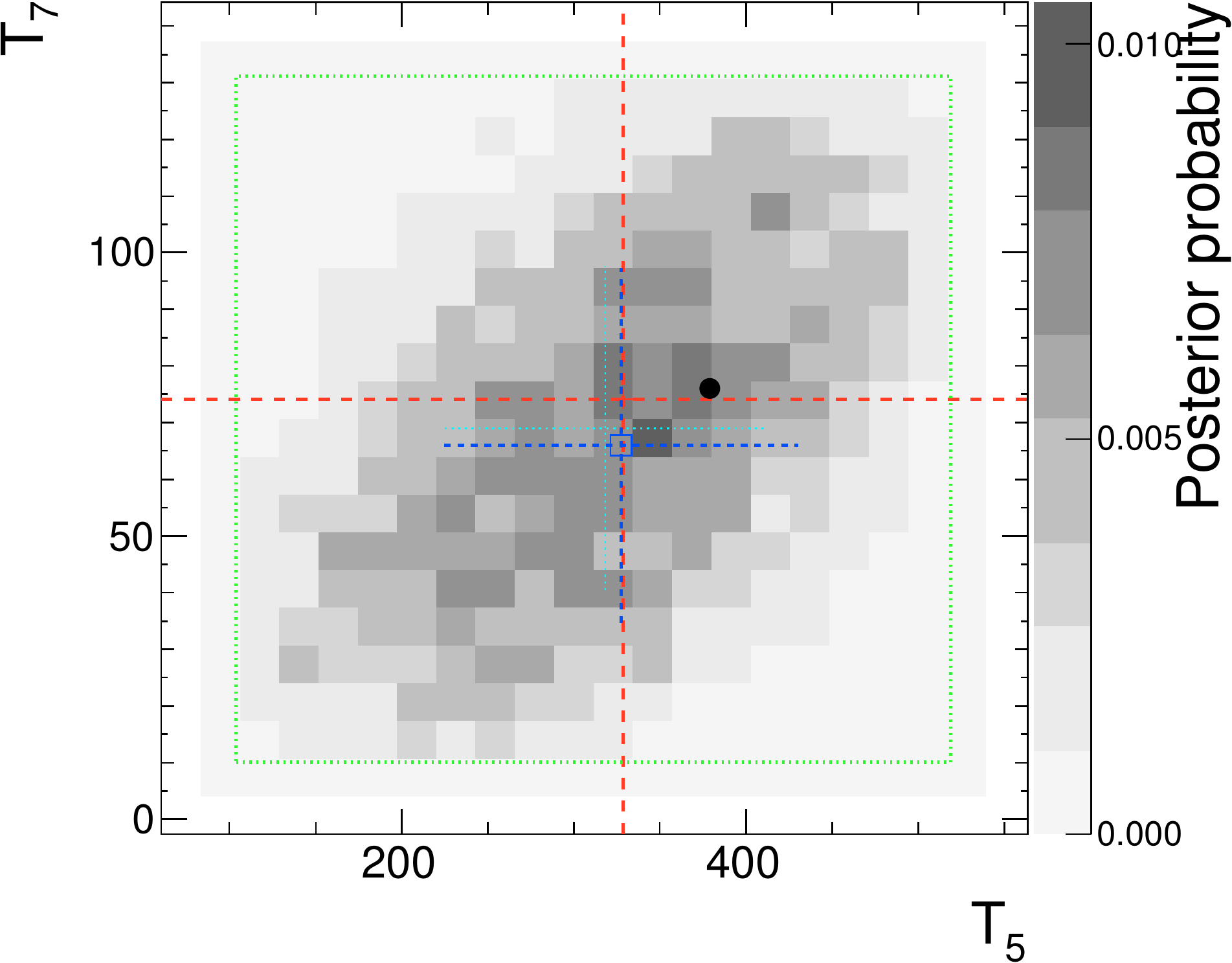} &
   \includegraphics[width=0.18\columnwidth]{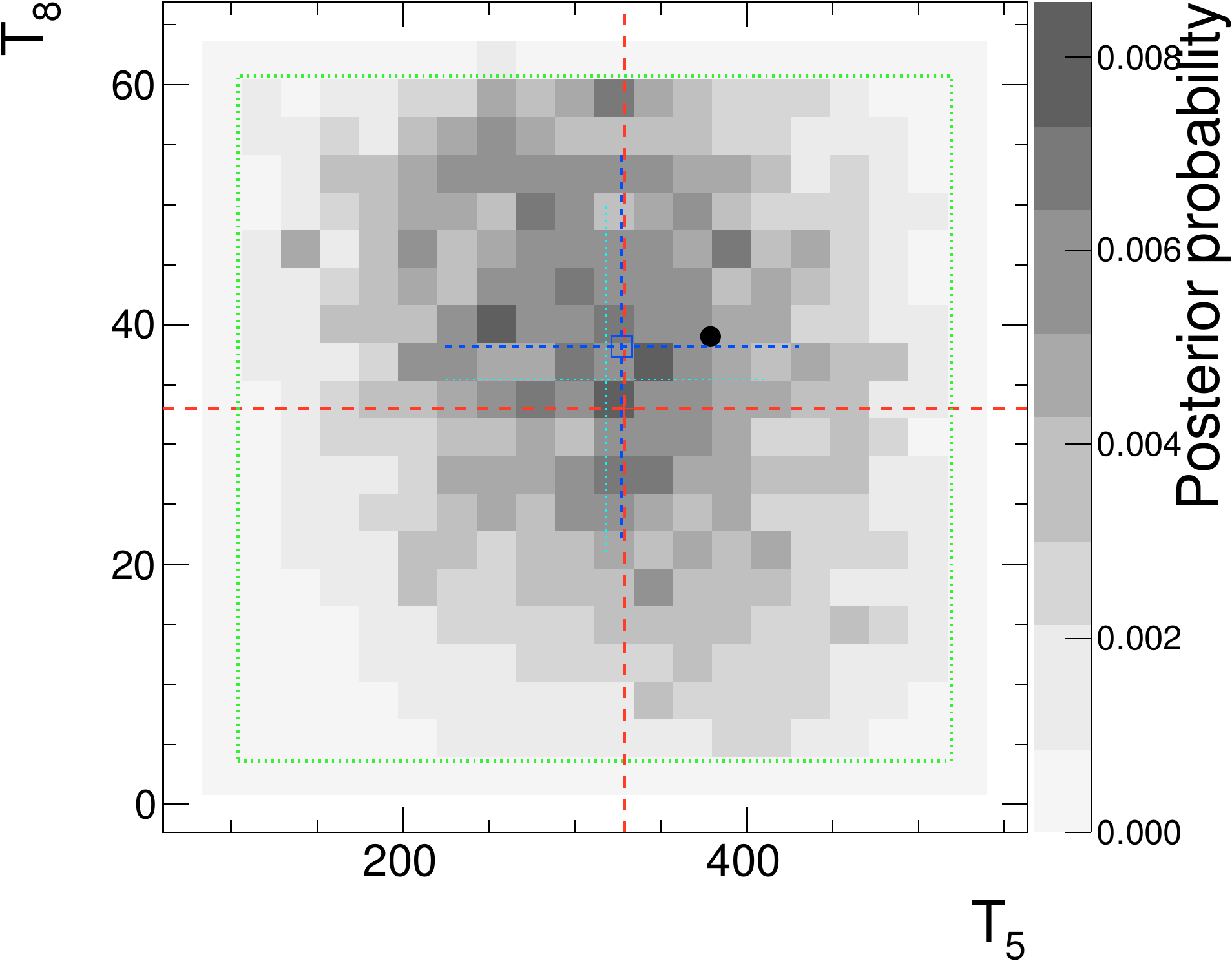} &
   \includegraphics[width=0.18\columnwidth]{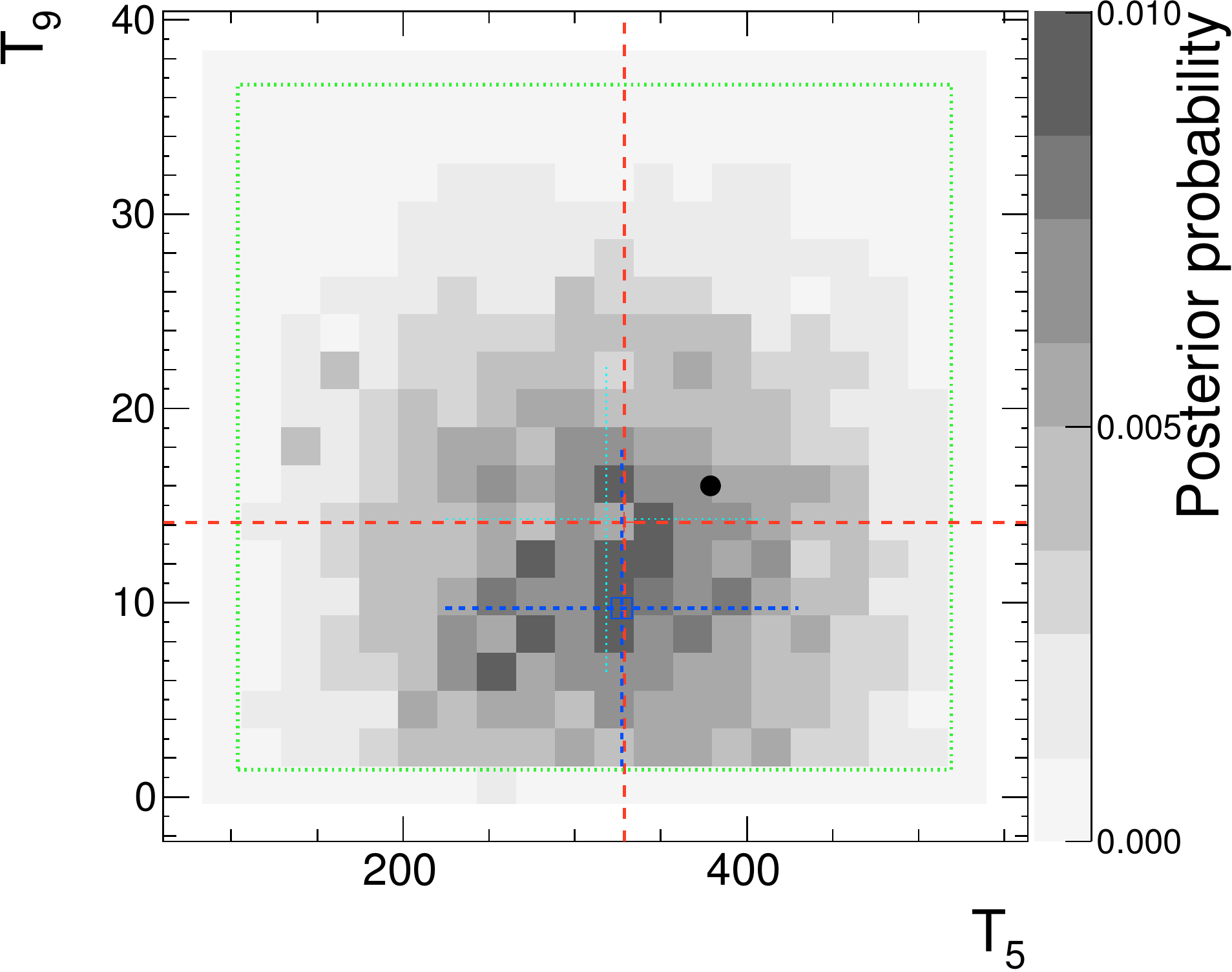} &
   \includegraphics[width=0.18\columnwidth]{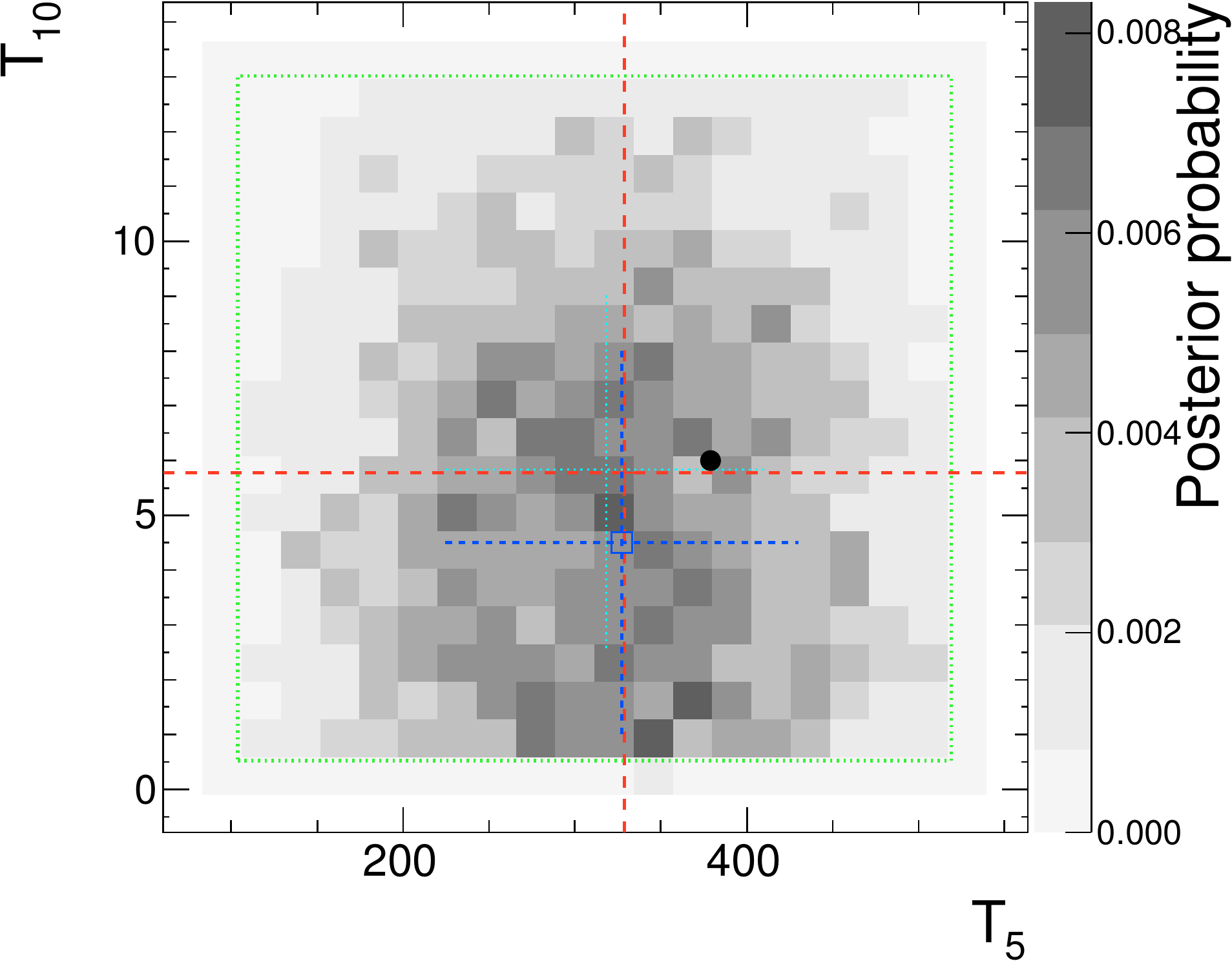} \\

   \includegraphics[width=0.18\columnwidth]{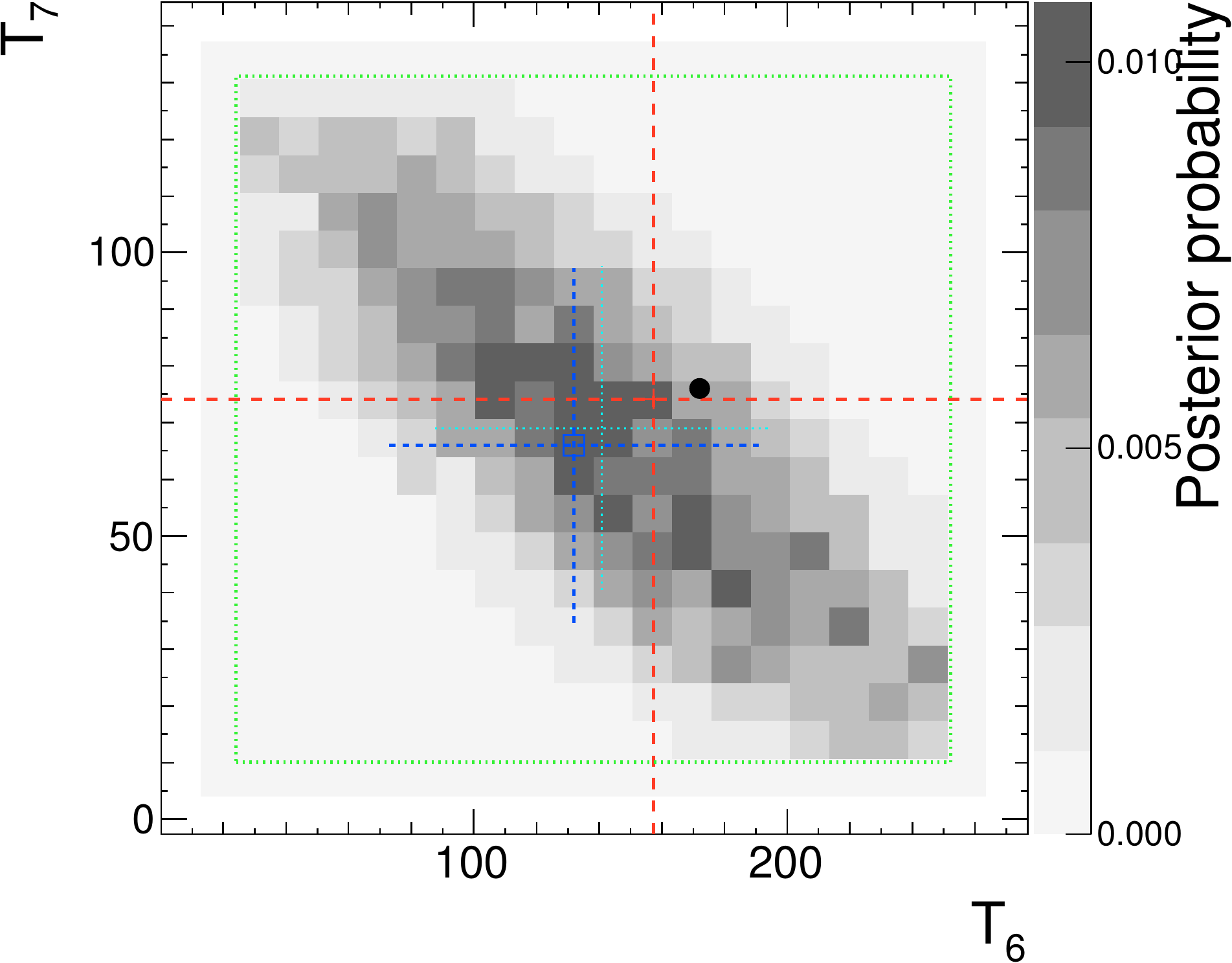} &
   \includegraphics[width=0.18\columnwidth]{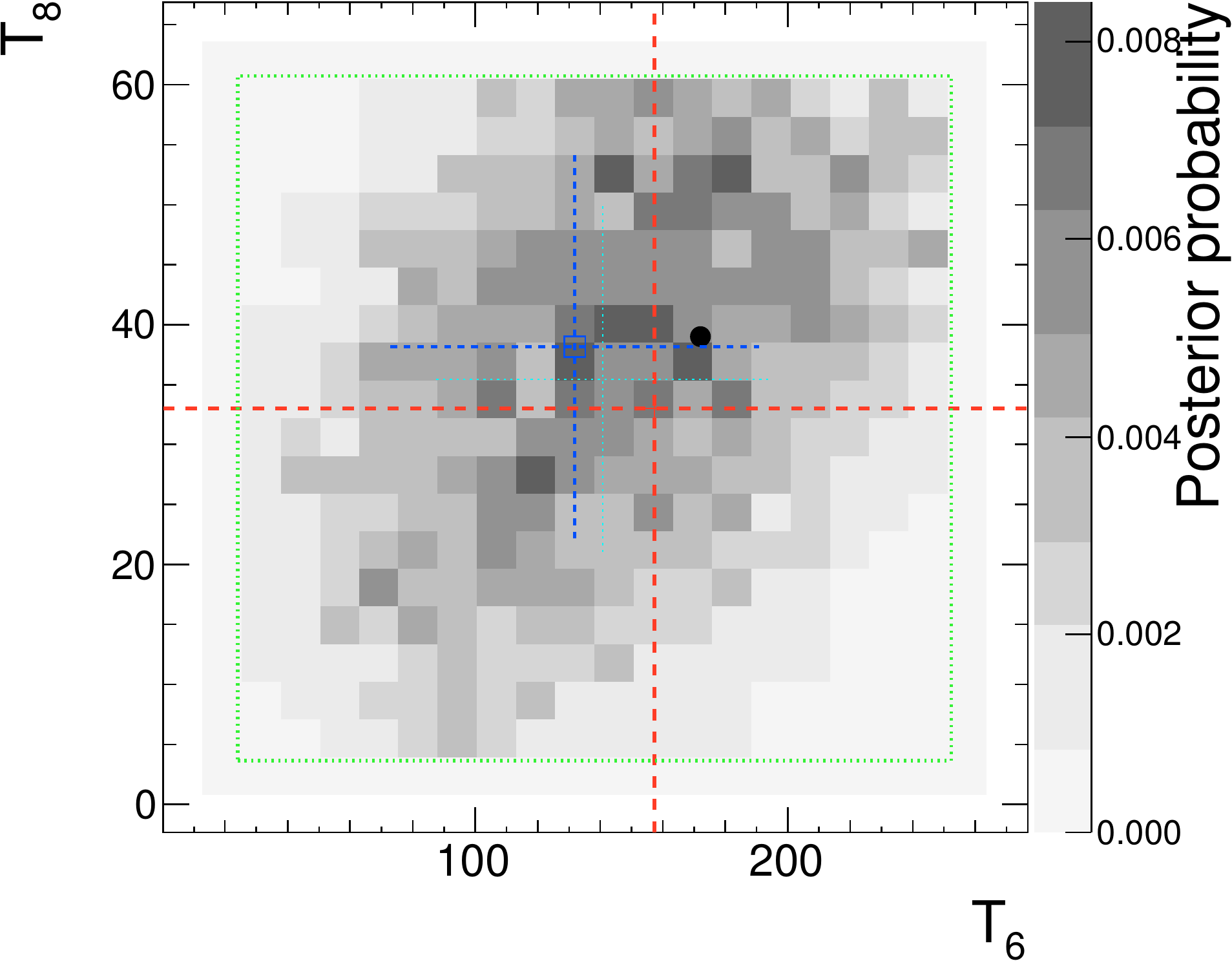} &
   \includegraphics[width=0.18\columnwidth]{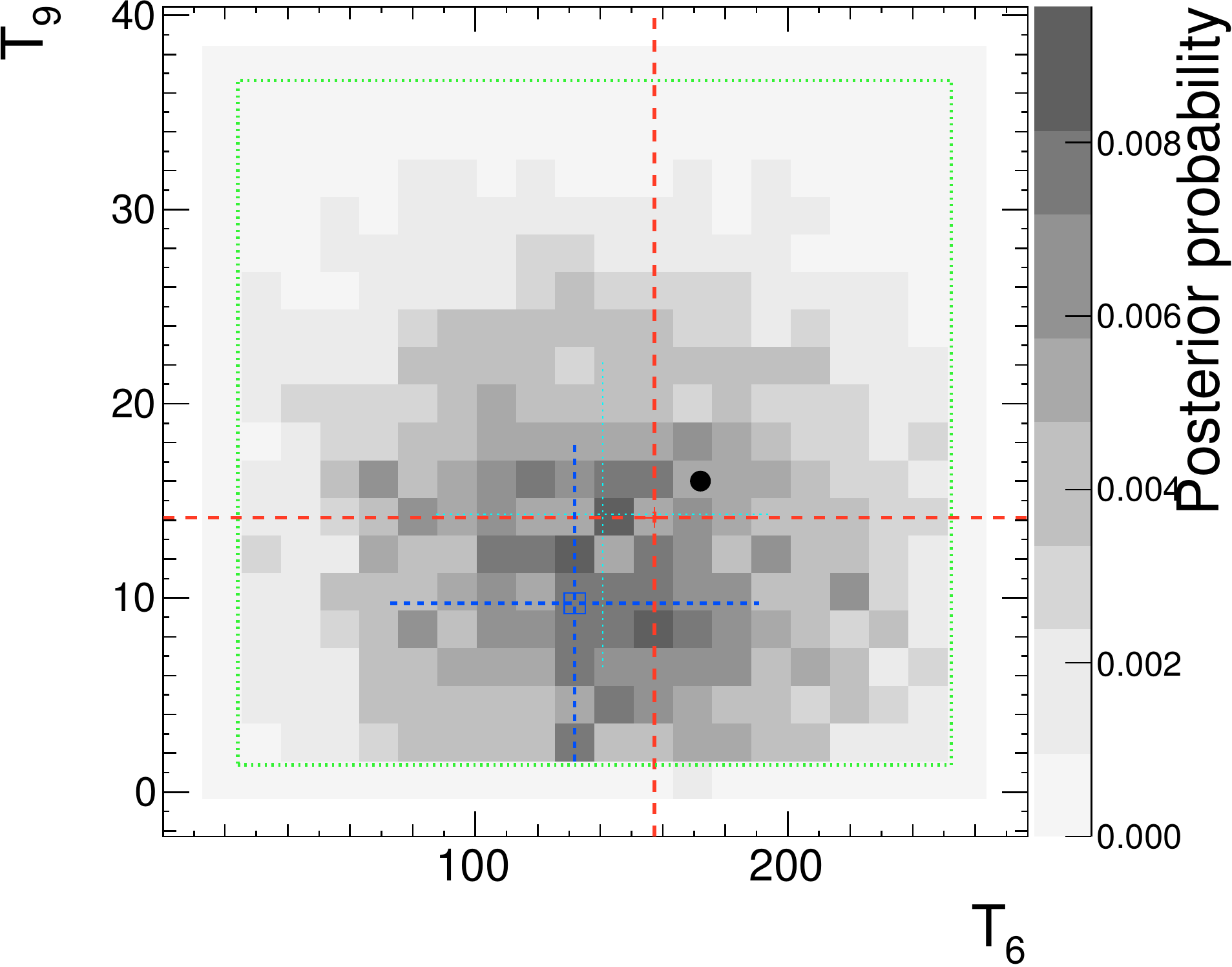} &
   \includegraphics[width=0.18\columnwidth]{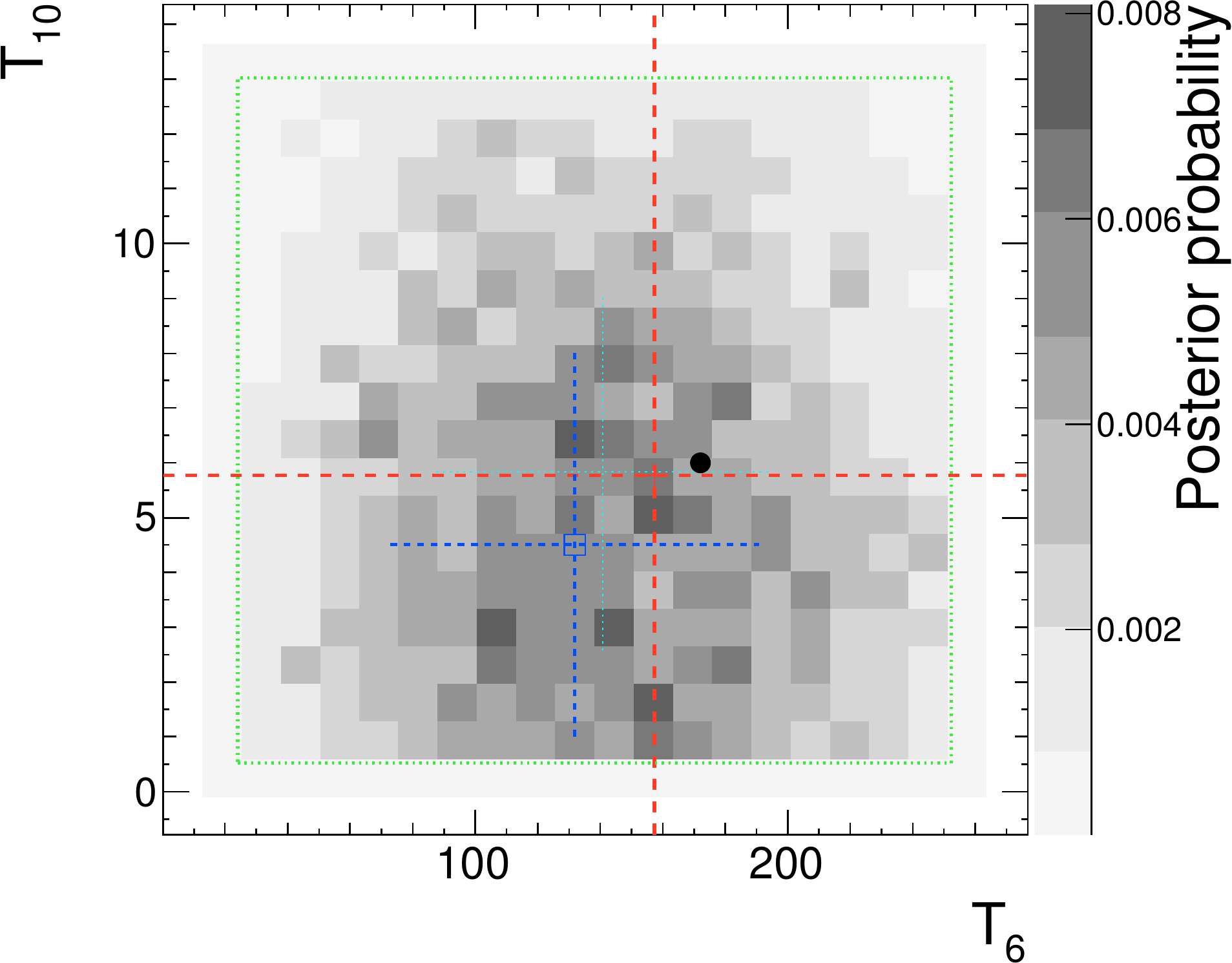} &
   \includegraphics[width=0.18\columnwidth]{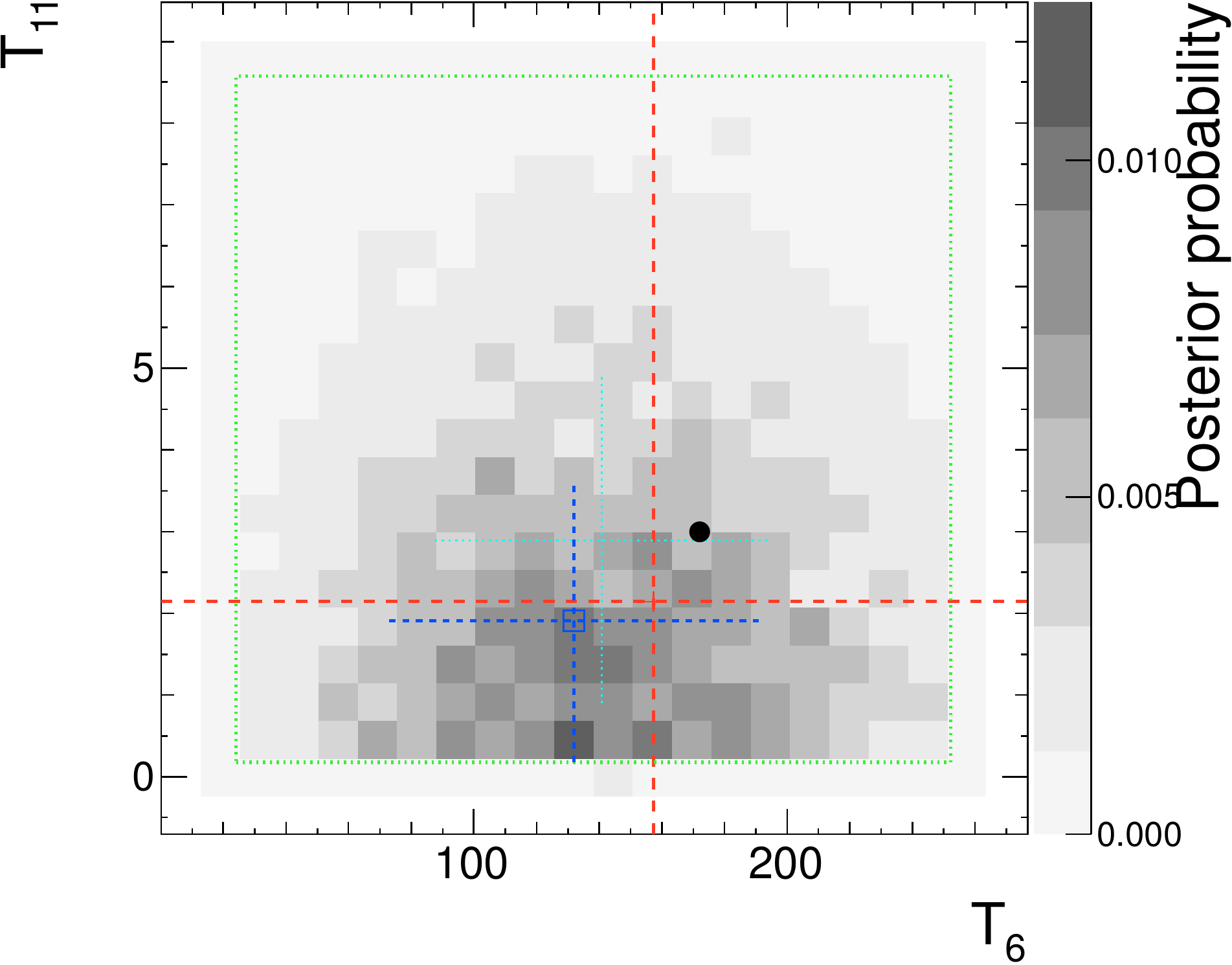} \\

   \includegraphics[width=0.18\columnwidth]{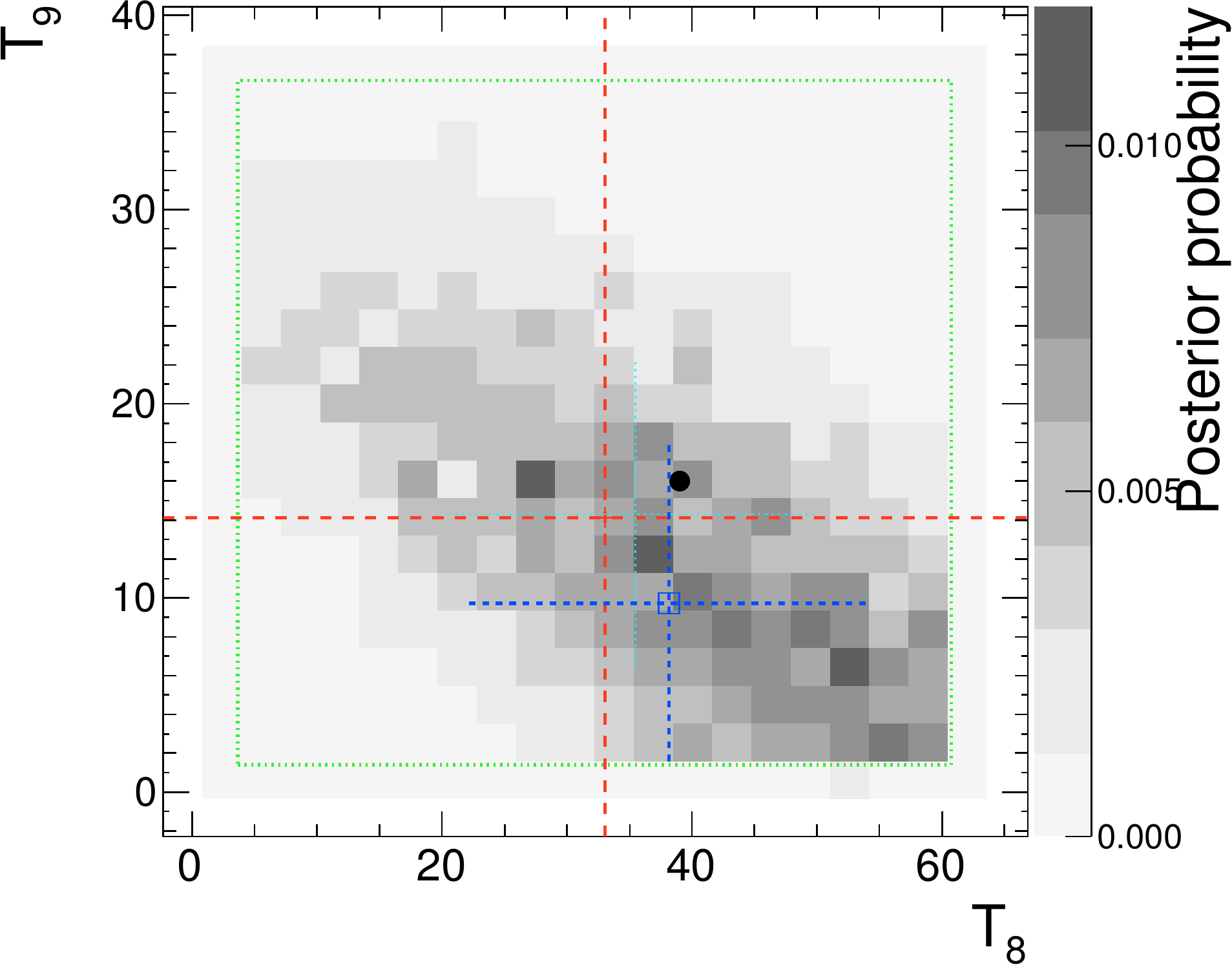} &
   \includegraphics[width=0.18\columnwidth]{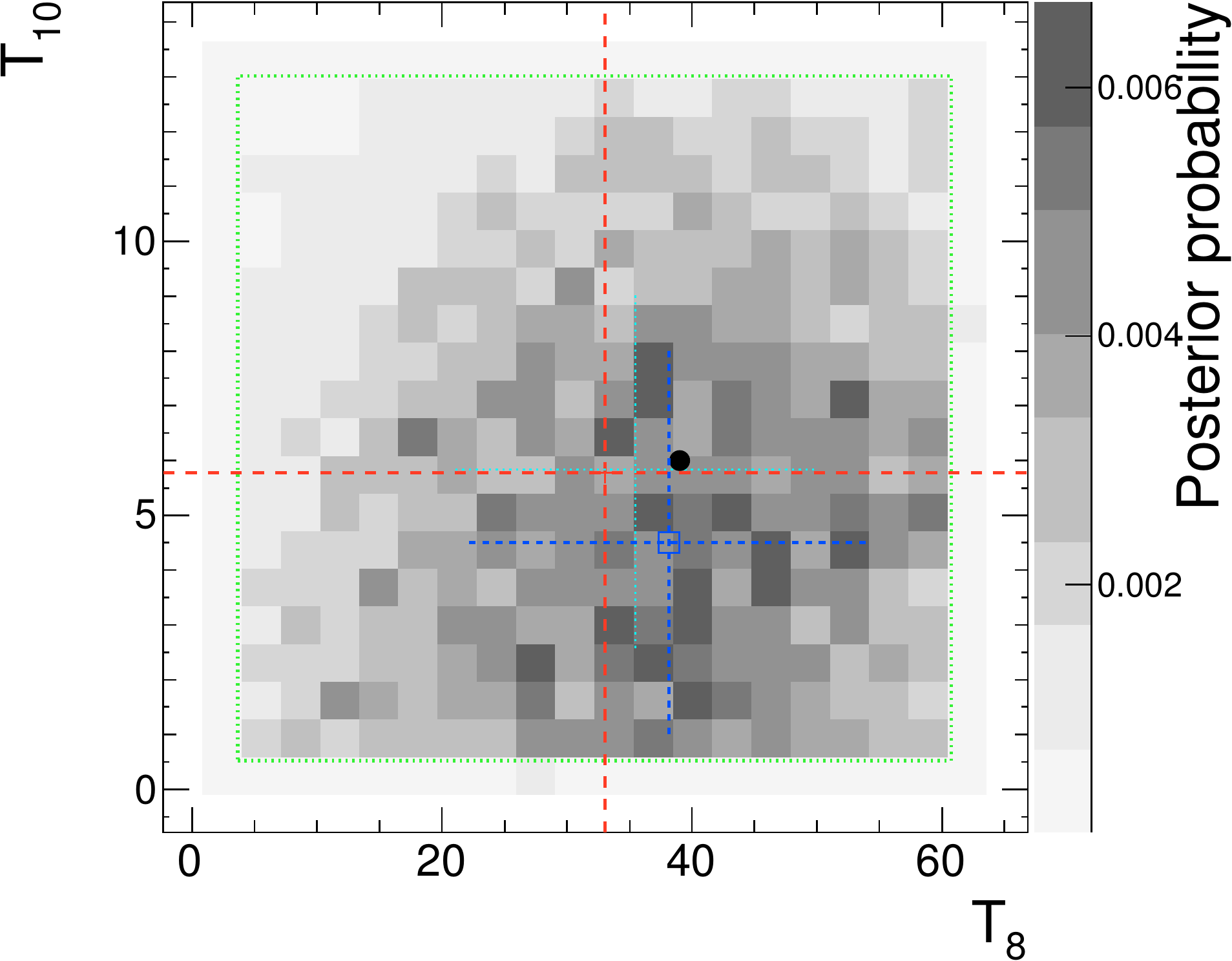} &
   \includegraphics[width=0.18\columnwidth]{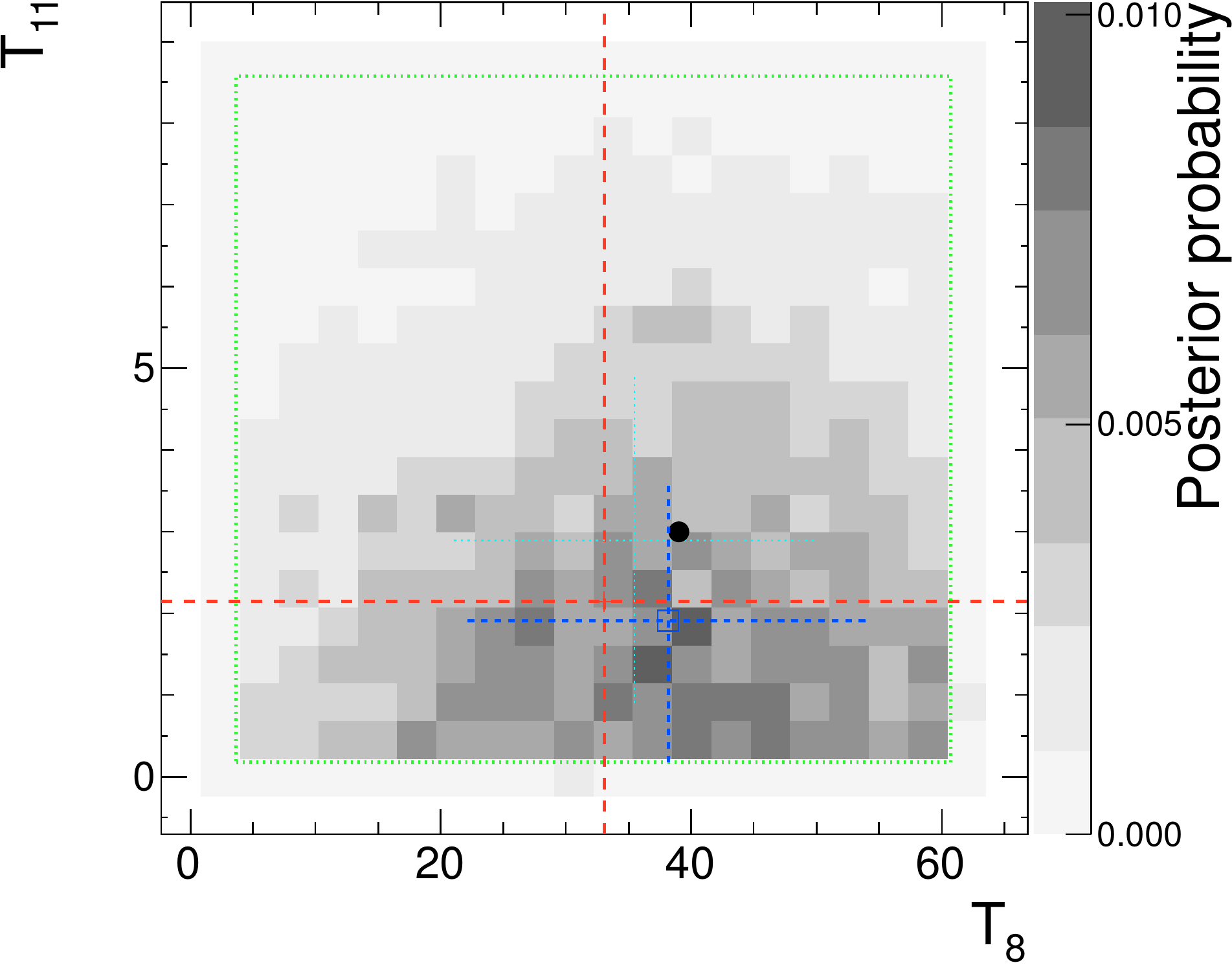} &
   \includegraphics[width=0.18\columnwidth]{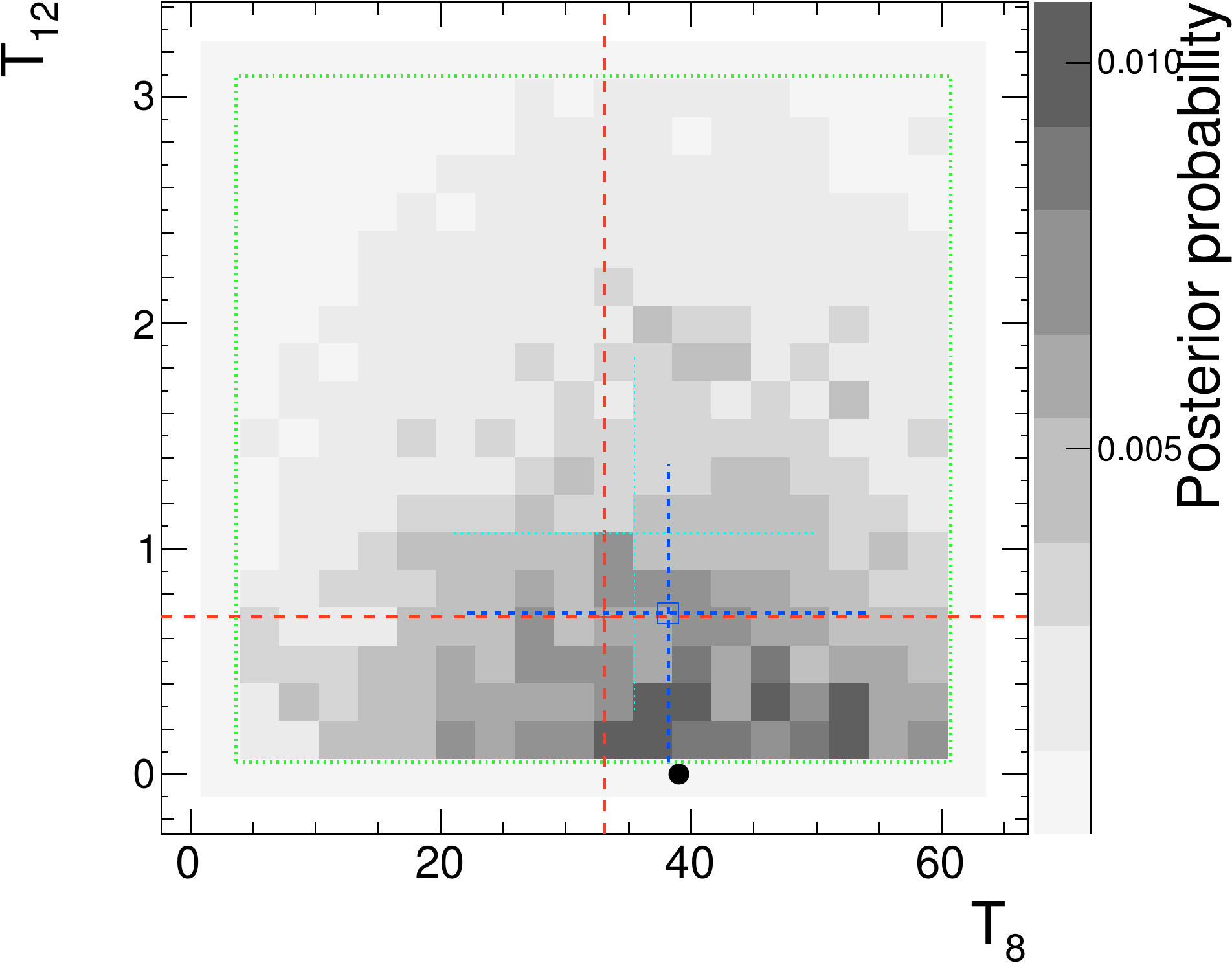} &
   \includegraphics[width=0.18\columnwidth]{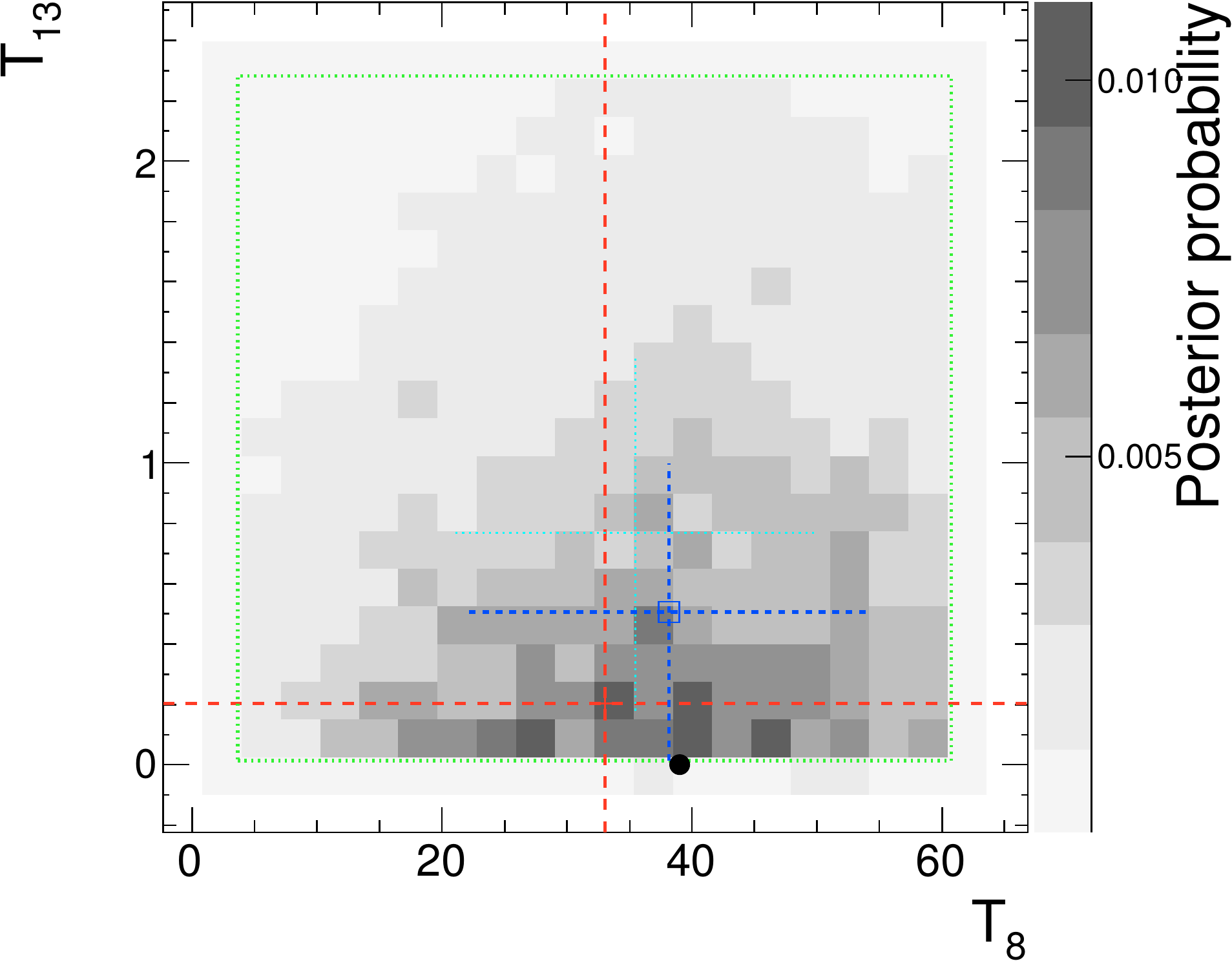} \\

   \includegraphics[width=0.18\columnwidth]{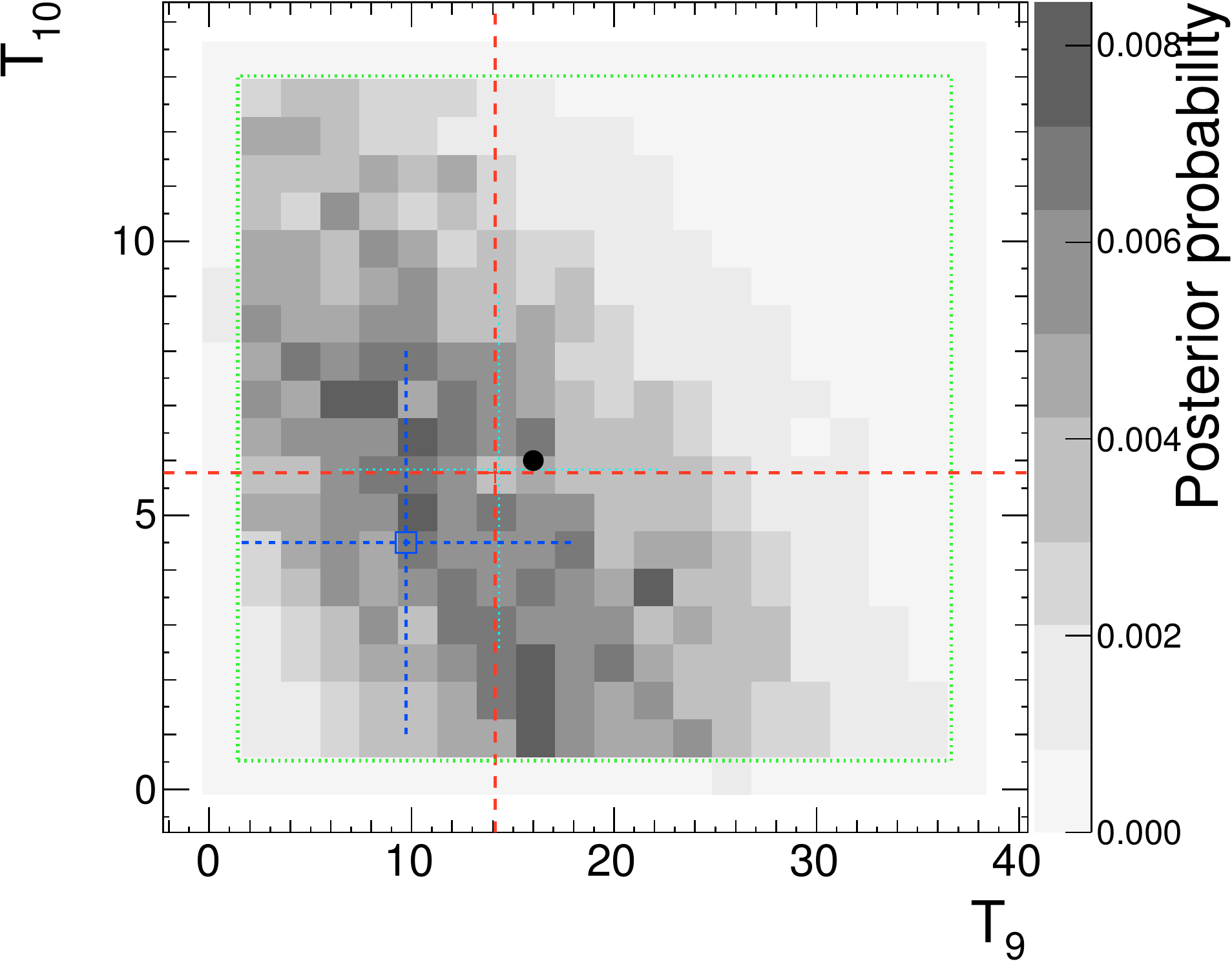} &
   \includegraphics[width=0.18\columnwidth]{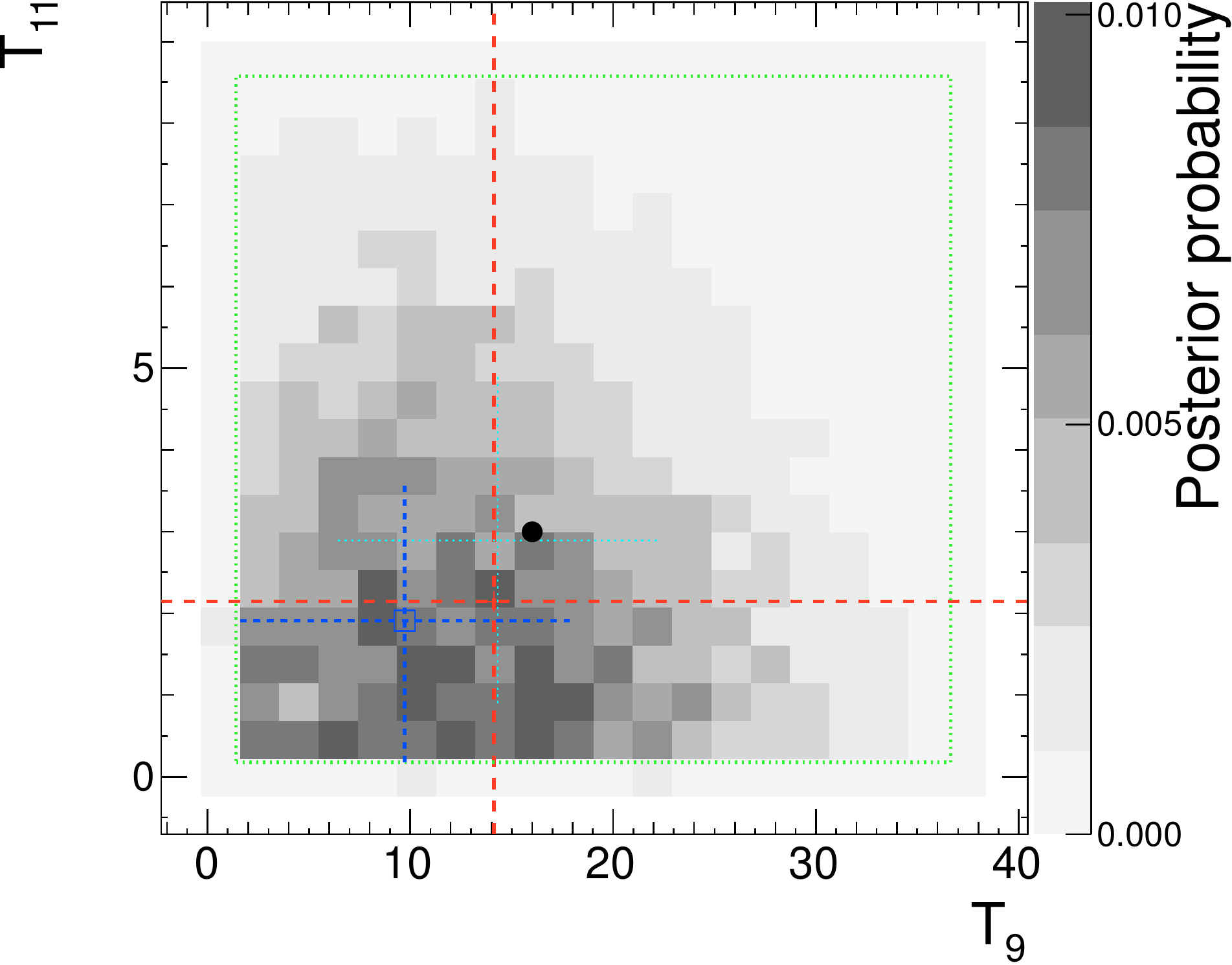} &
   \includegraphics[width=0.18\columnwidth]{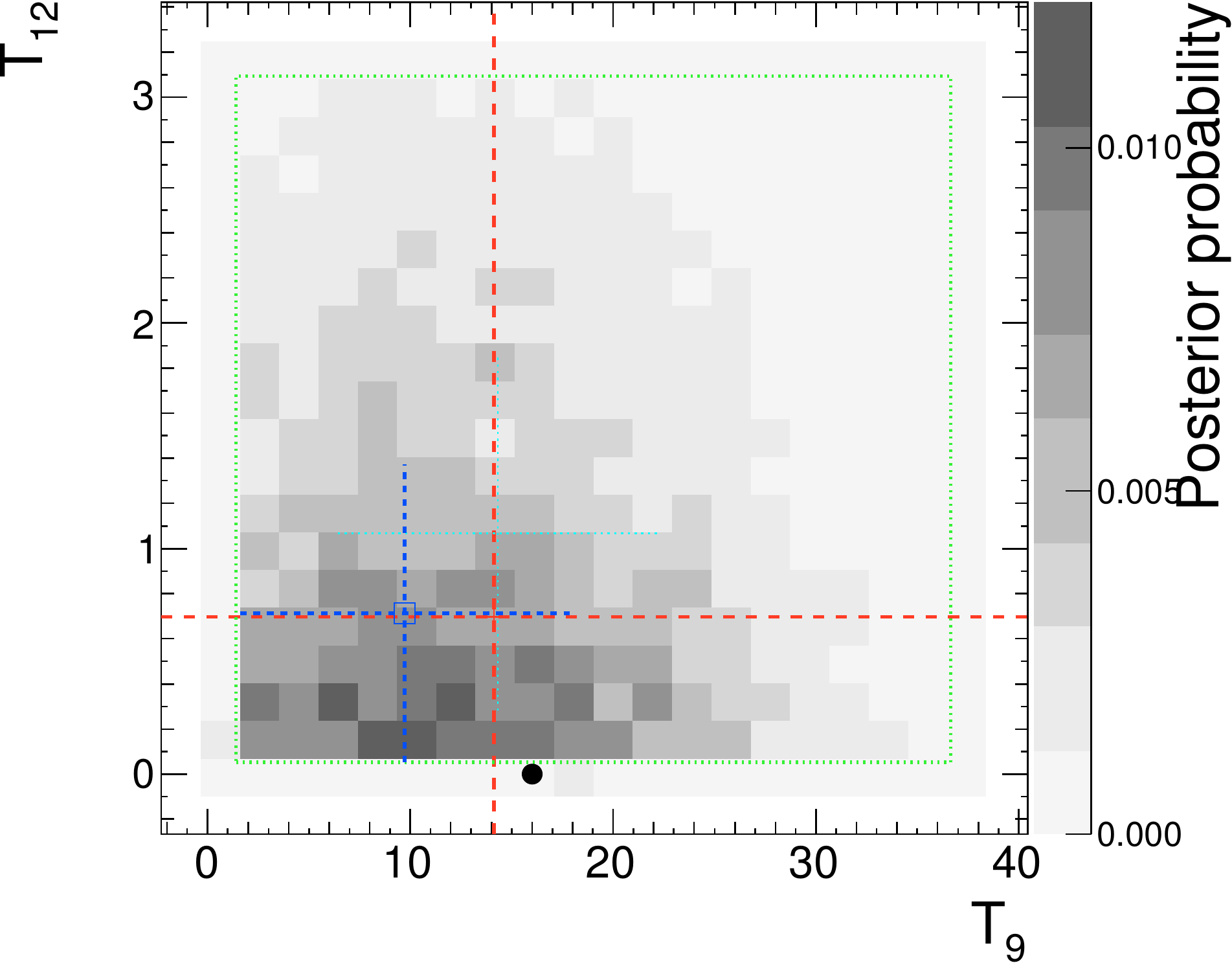} &
   \includegraphics[width=0.18\columnwidth]{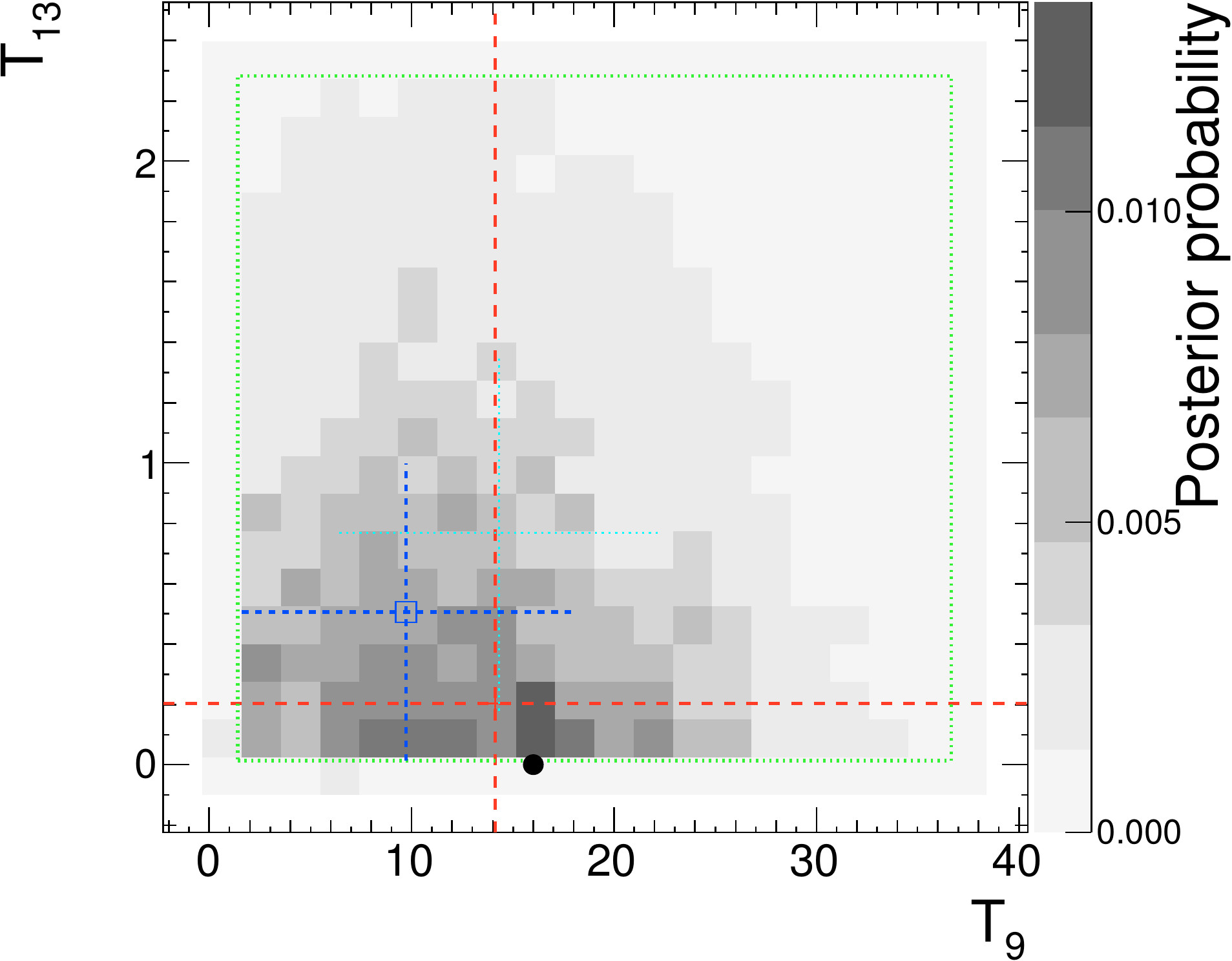} &
   \includegraphics[width=0.18\columnwidth]{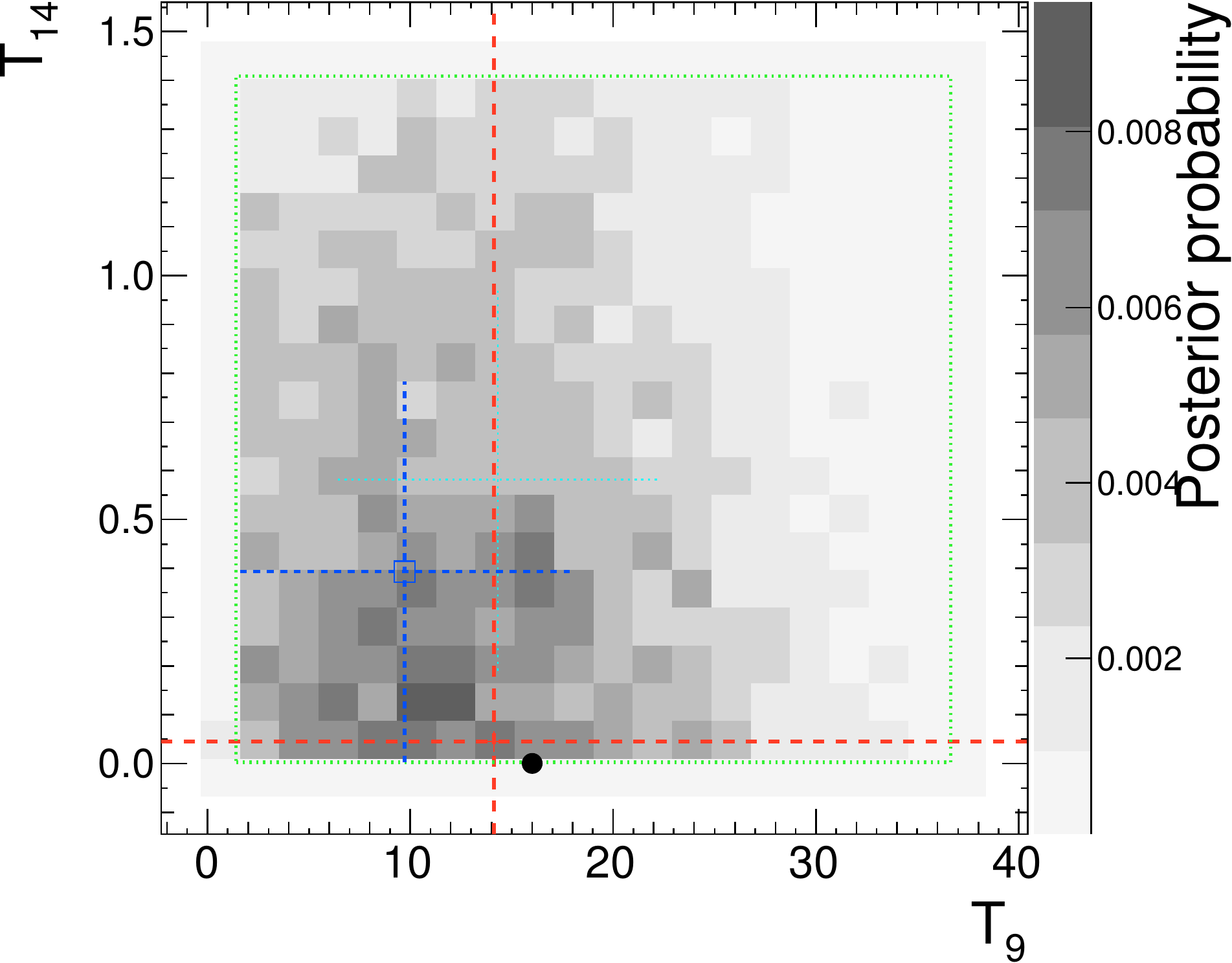} \\
 \end{tabular}
 \caption{Some of the 2-dimensional marginal distributions of $p(\tuple{T}|\tuple{D})$ in the example of Sec.~\ref{sec:example4}.
\label{fig:2Dim4}}
\end{figure}

In 14 dimensions it is not possible to visualize the whole $p(\T|\D)$, but it is helpful to show some of its 91 2-dimensional marginal distributions in Fig.~\ref{fig:2Dim4}.
The smearing is significant, which enhances the (anti)correlation of the pair $(T_t,T_{t+1})$.   This (anti)correlation is weaker between $(T_t,T_{t+2})$, even weaker for $(T_t,T_{t+3})$, etc.  This happens because migrations are more rare between bins that are farther apart.

\subsubsection{MCMC sampling}
\label{sec:example5}

The advantage of MCMC is that fewer of its samples are taken at $\T$ points where $L(\D|\T) \simeq 0$, therefore the sampling is more efficient.  For the reason explained in Sec.~\ref{sec:sampling}, it may cause bias in the marginalization of $p(\T|\D)$.  The goal of this example is to examine this bias in practice.

\begin{figure}[H]
\centering
\begin{tabular}{ccccc}
   \includegraphics[width=0.18\columnwidth]{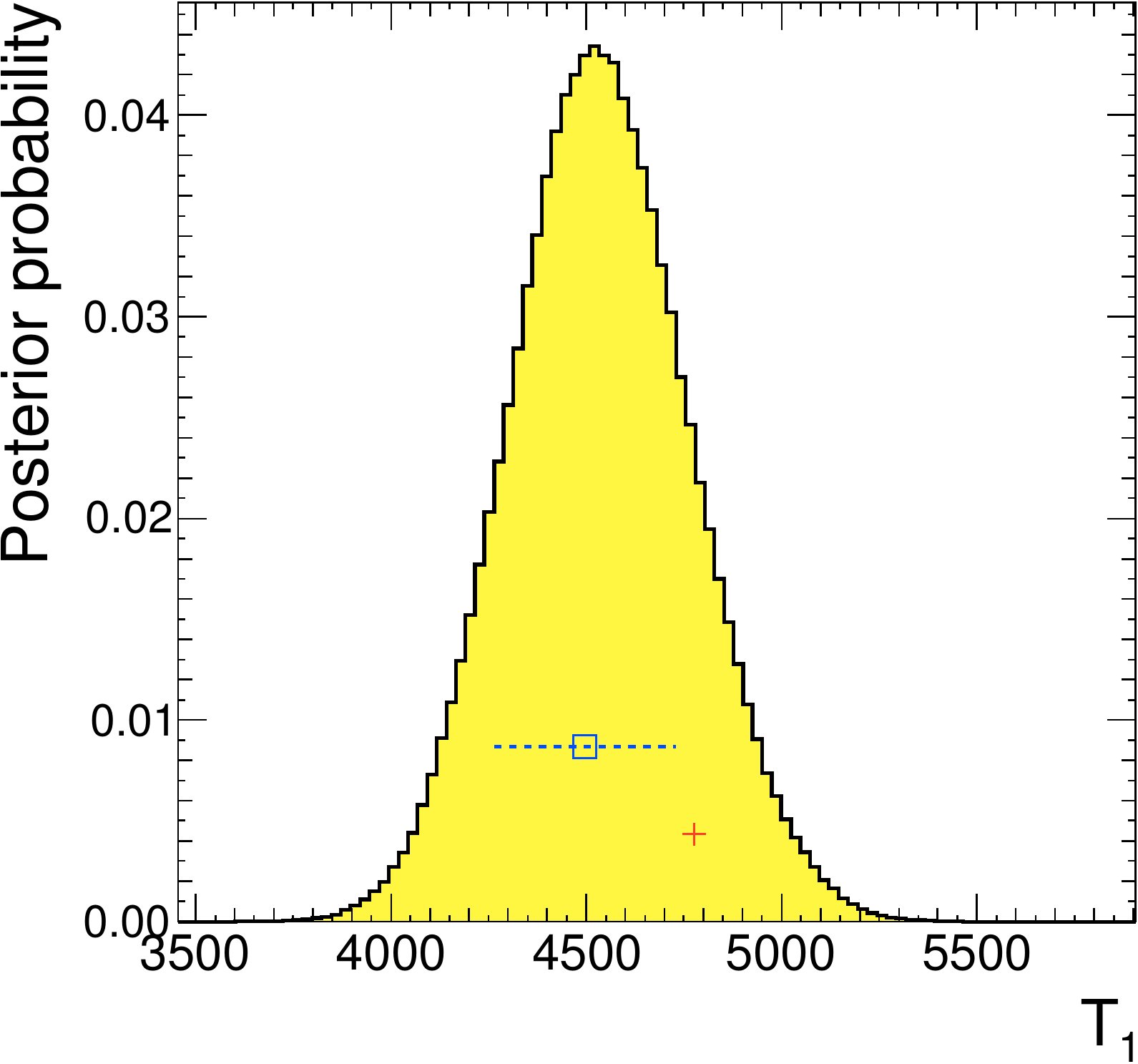} &
   \includegraphics[width=0.18\columnwidth]{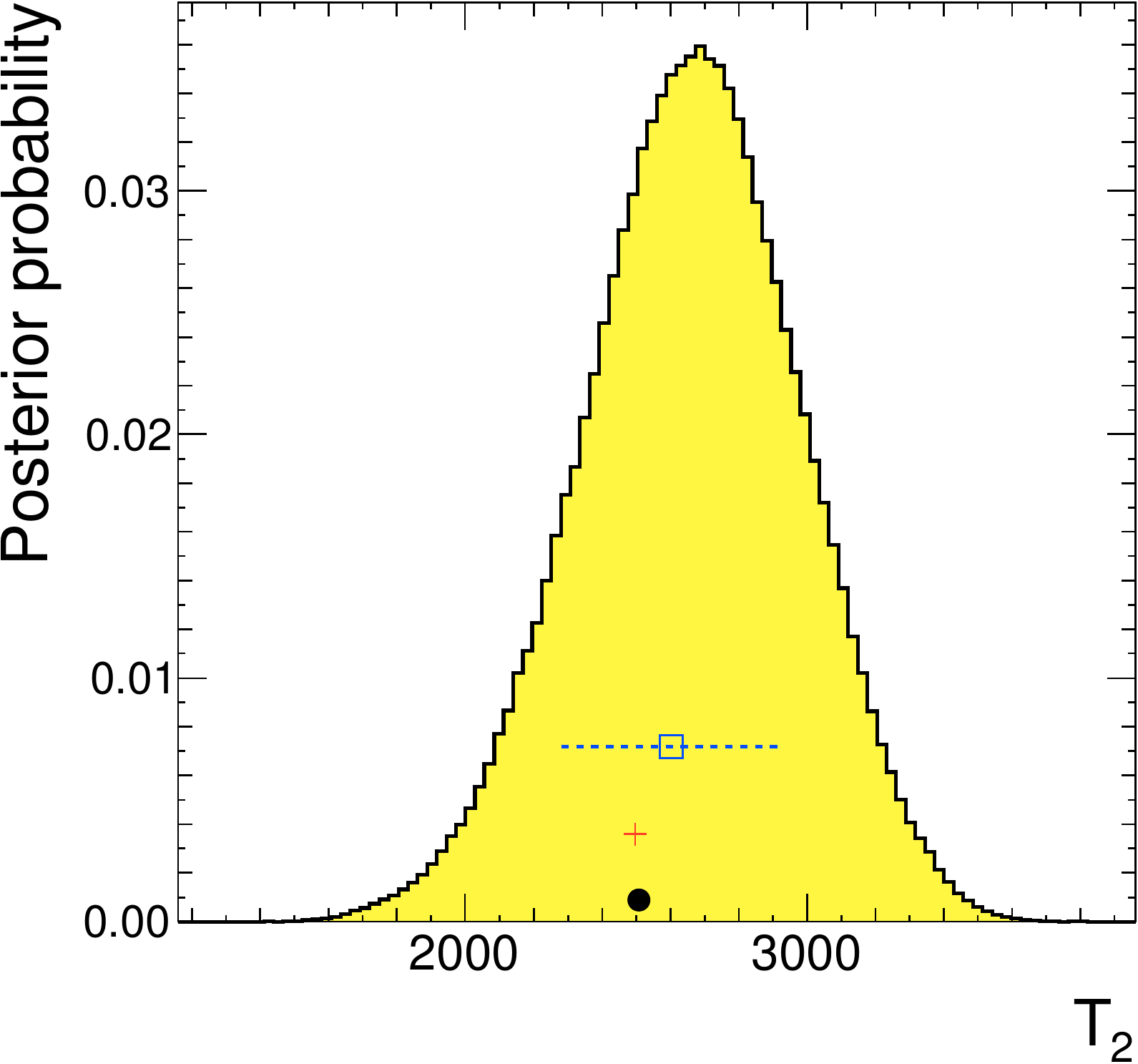} &
   \includegraphics[width=0.18\columnwidth]{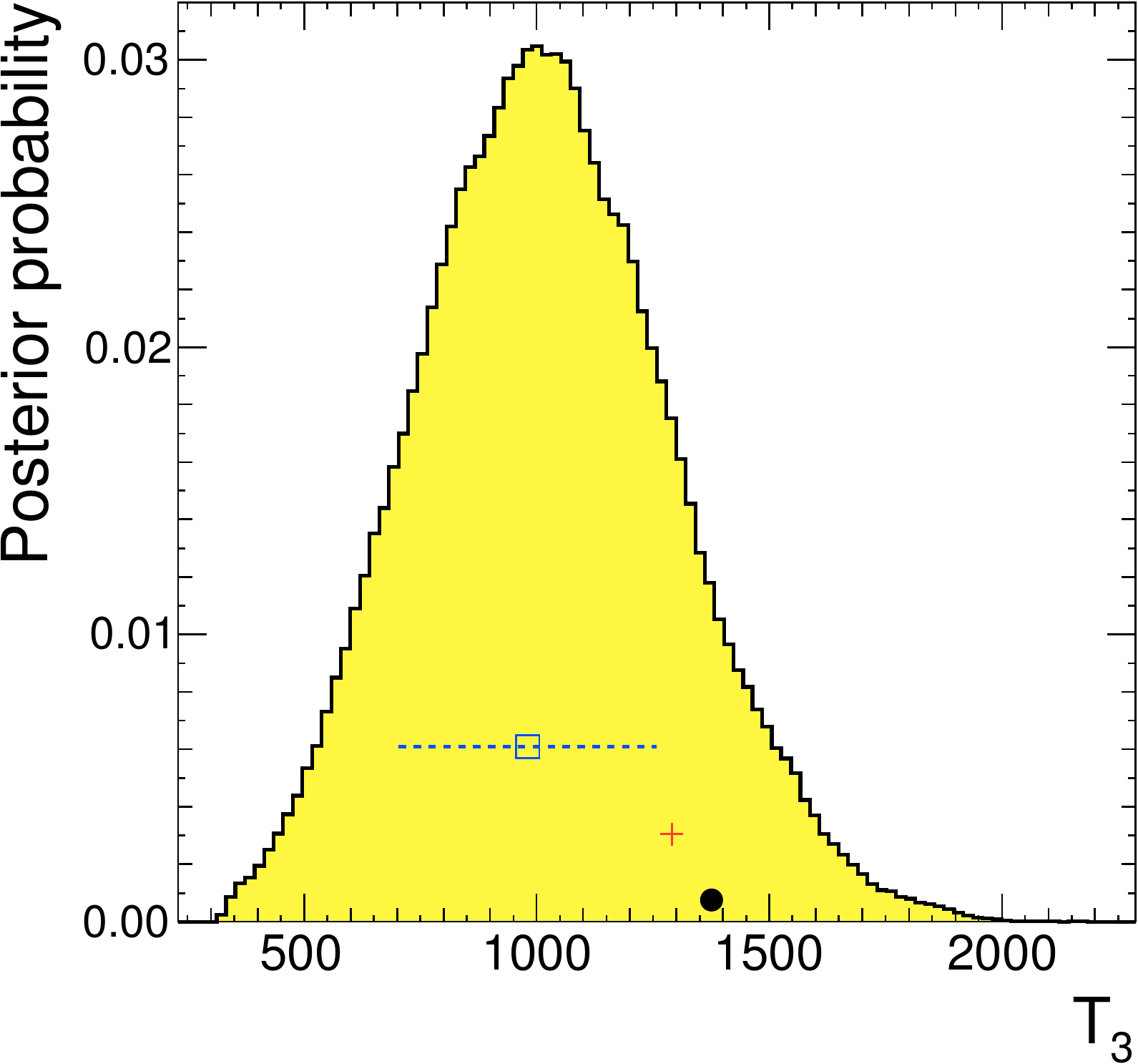} &
   \includegraphics[width=0.18\columnwidth]{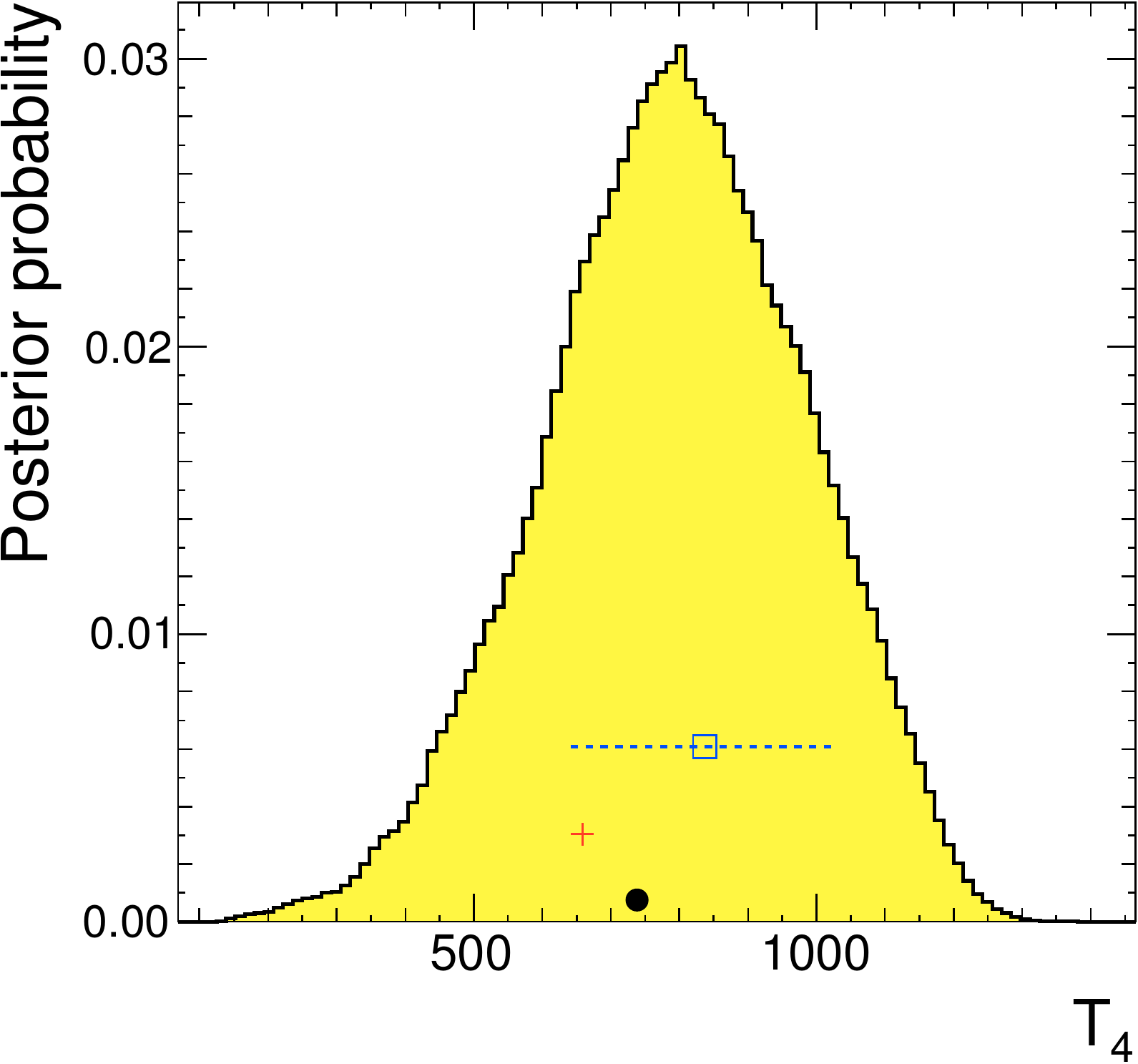} &
   \includegraphics[width=0.18\columnwidth]{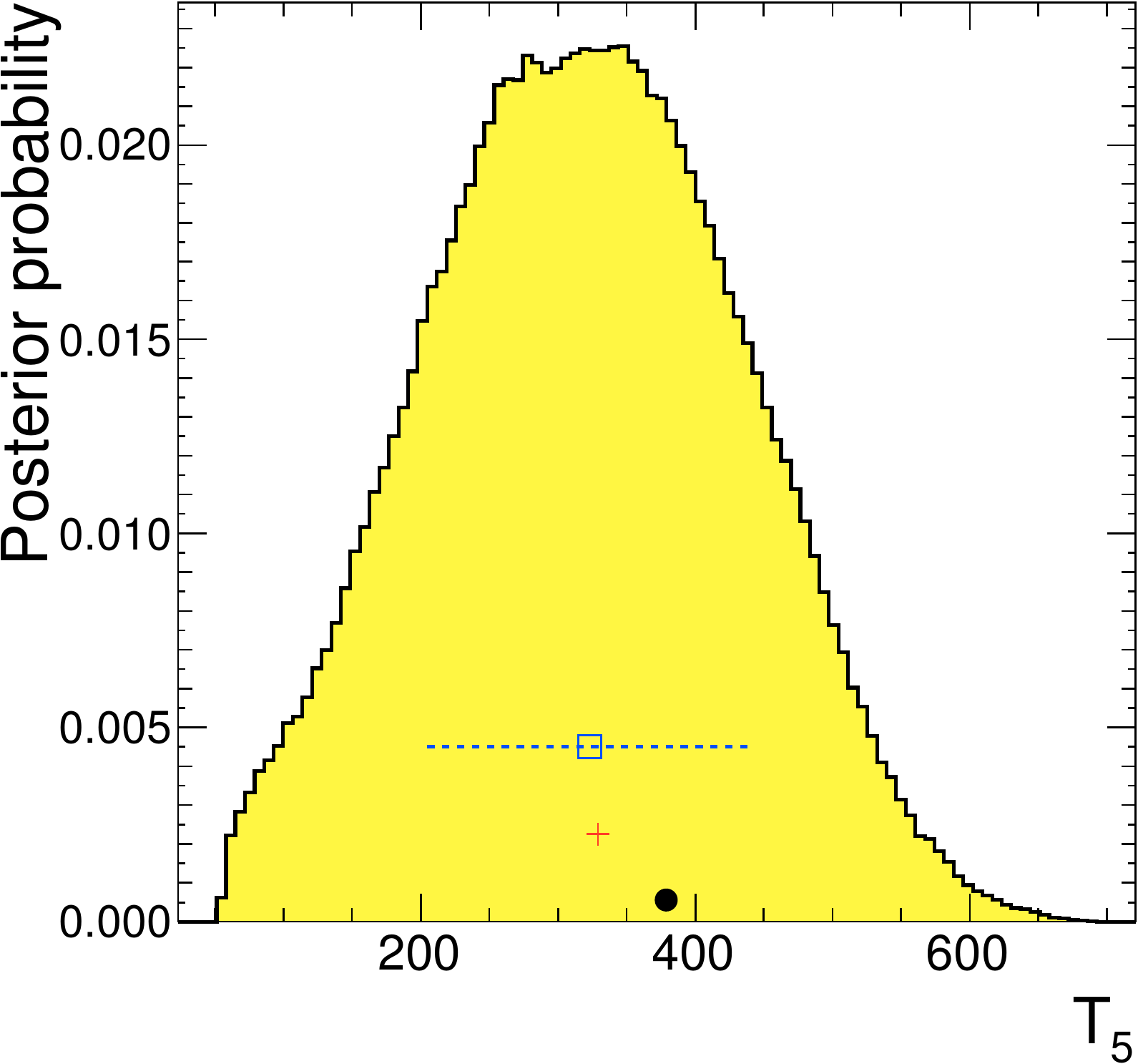} \\

   \includegraphics[width=0.18\columnwidth]{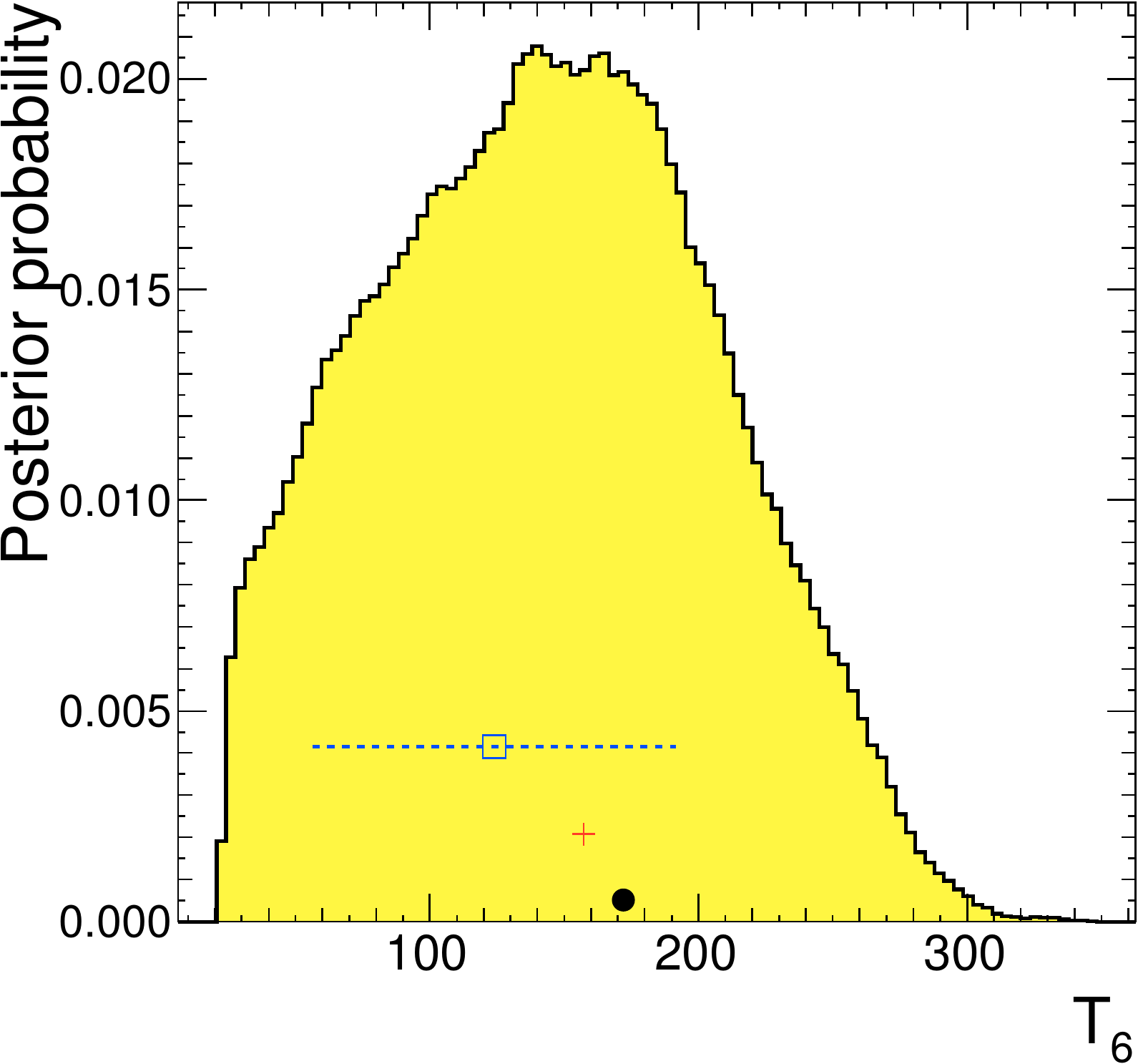} &
   \includegraphics[width=0.18\columnwidth]{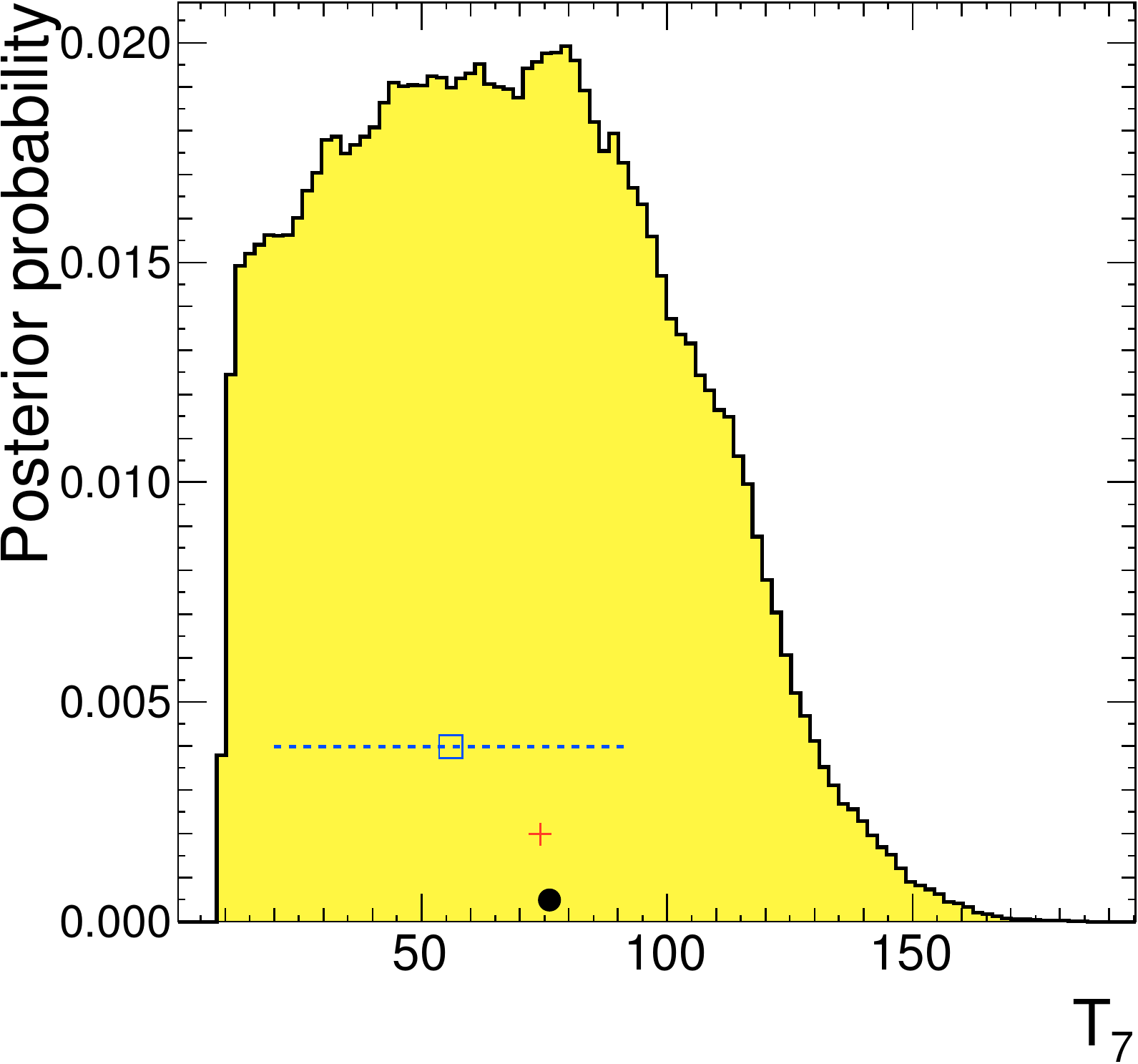} &
   \includegraphics[width=0.18\columnwidth]{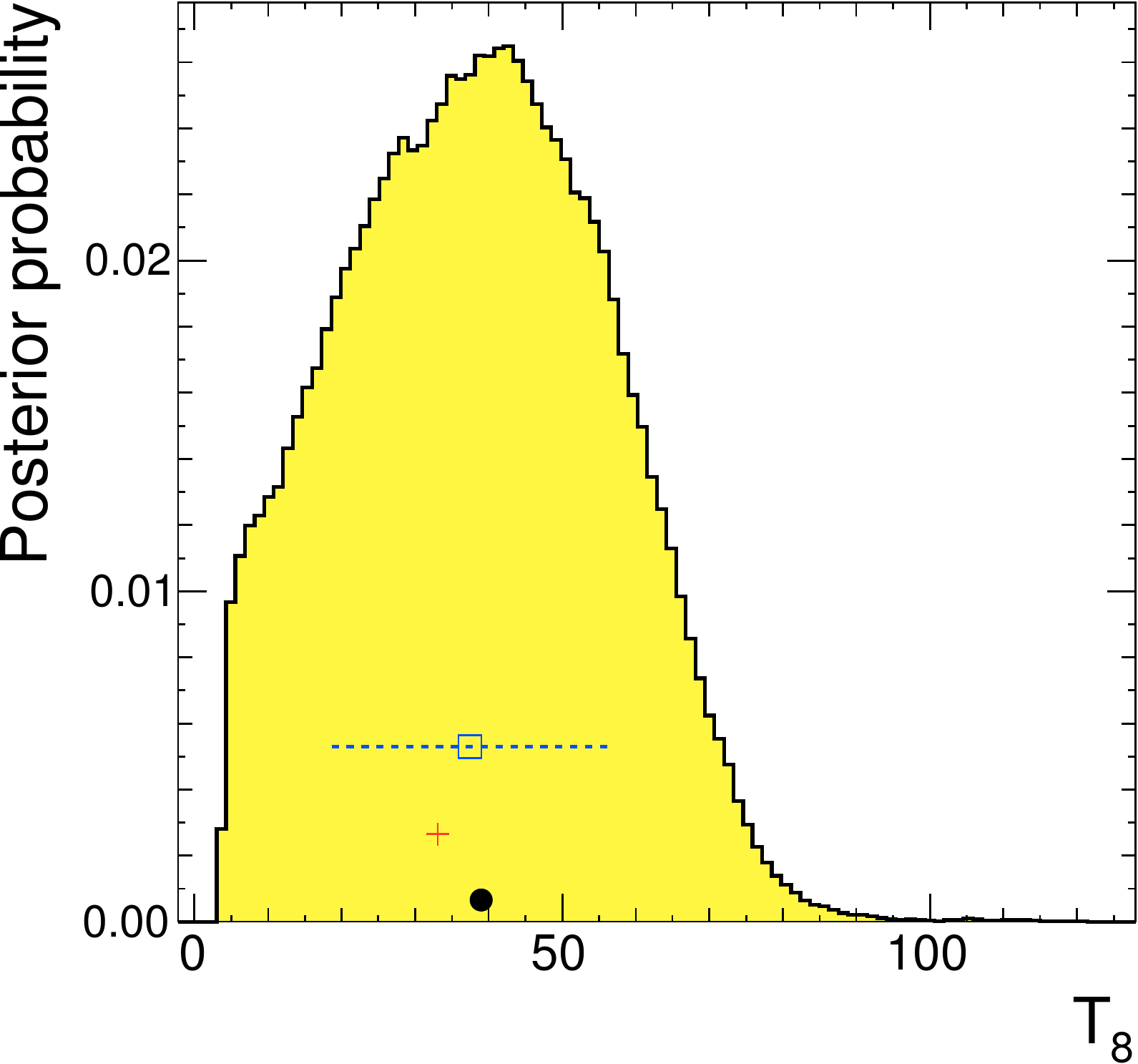} &
   \includegraphics[width=0.18\columnwidth]{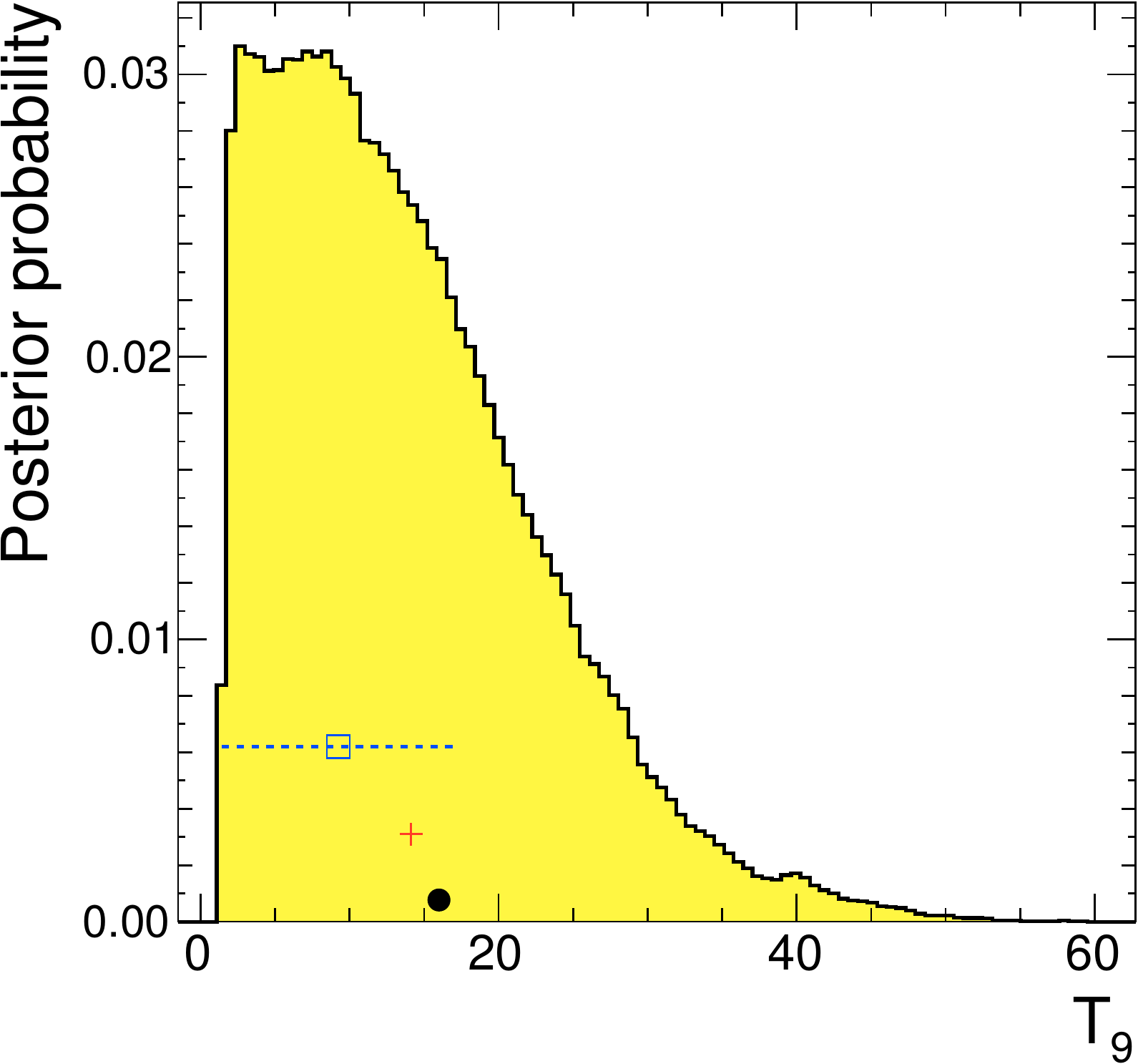} &
   \includegraphics[width=0.18\columnwidth]{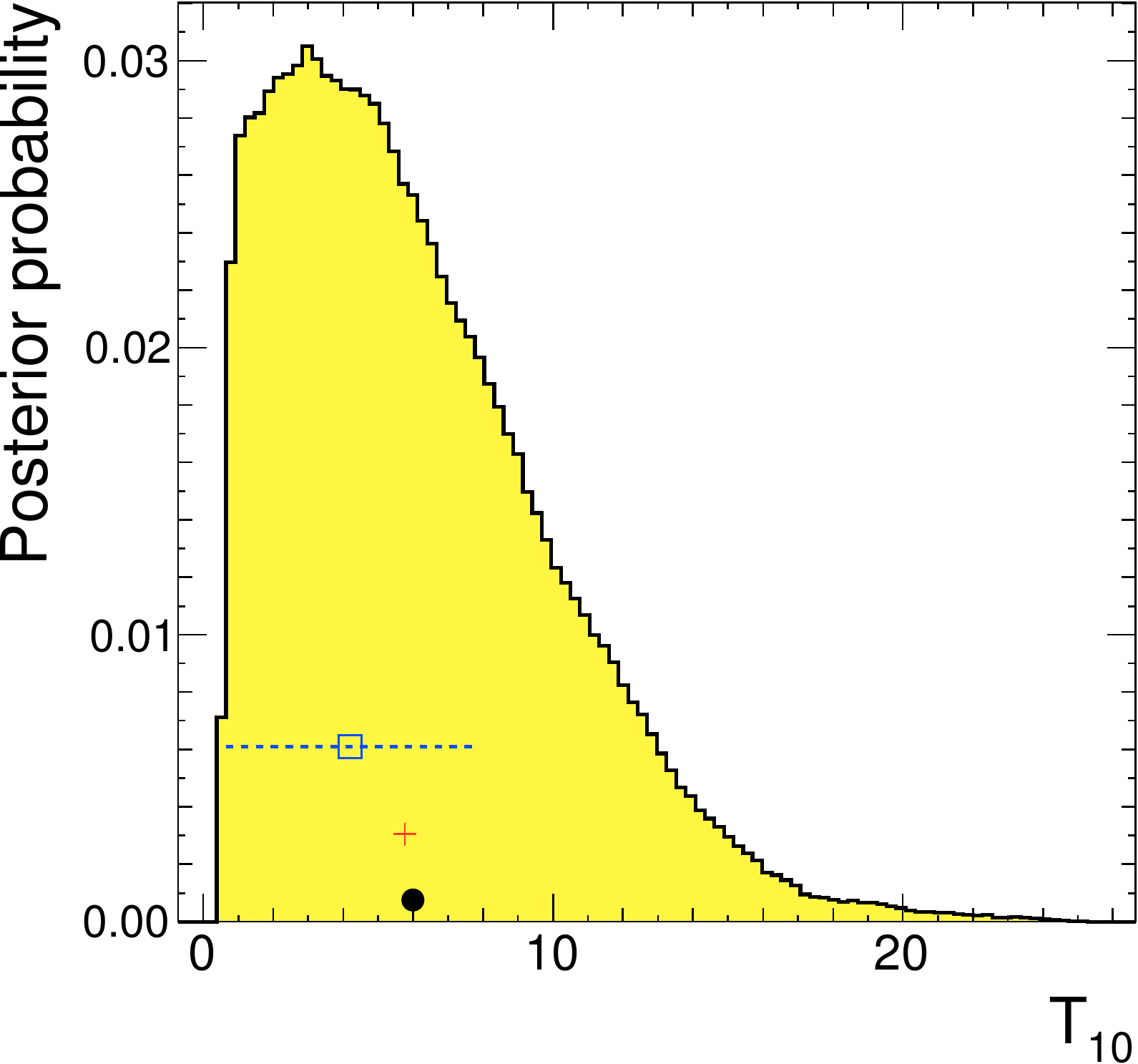} \\

   \includegraphics[width=0.18\columnwidth]{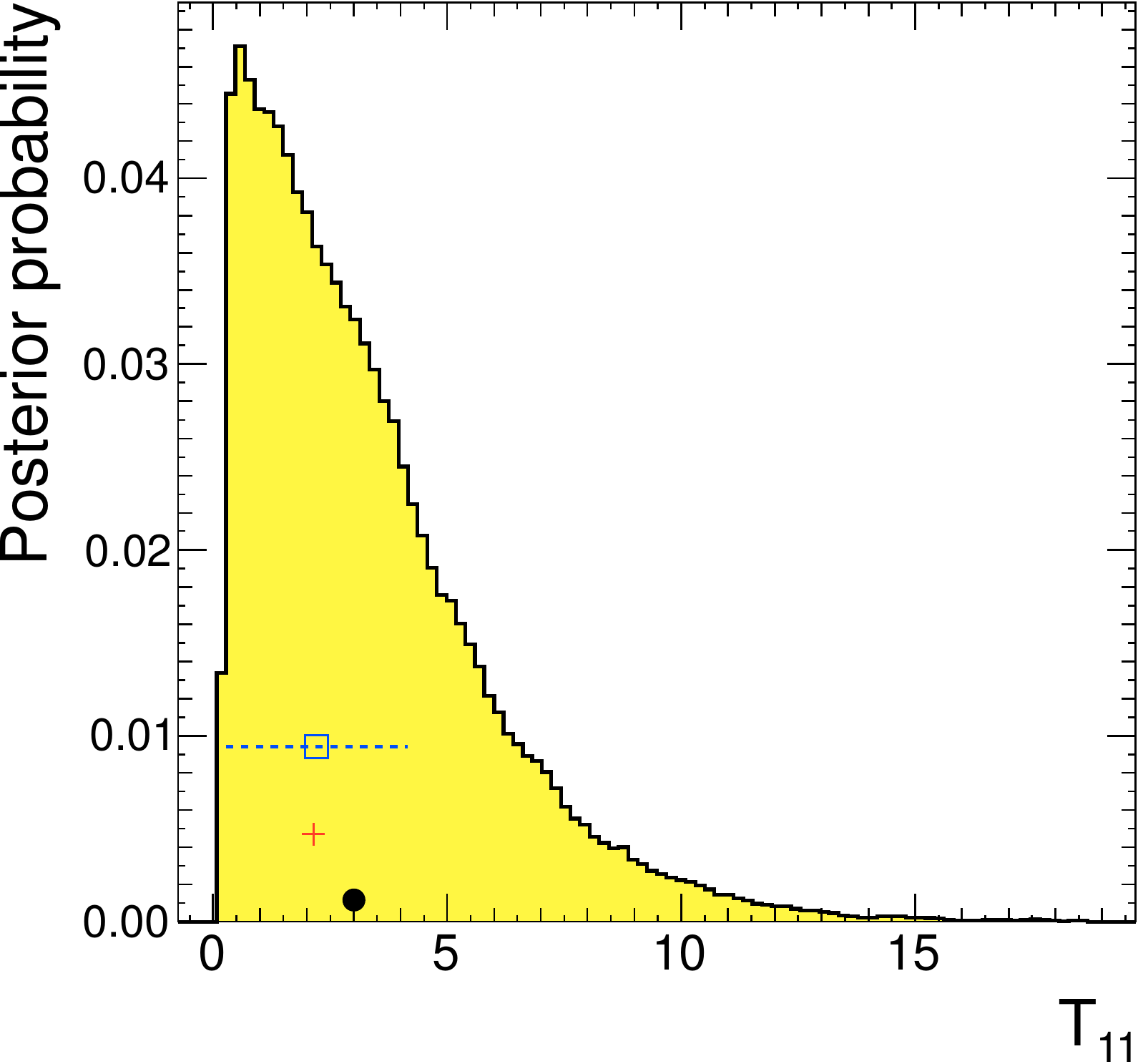} &
   \includegraphics[width=0.18\columnwidth]{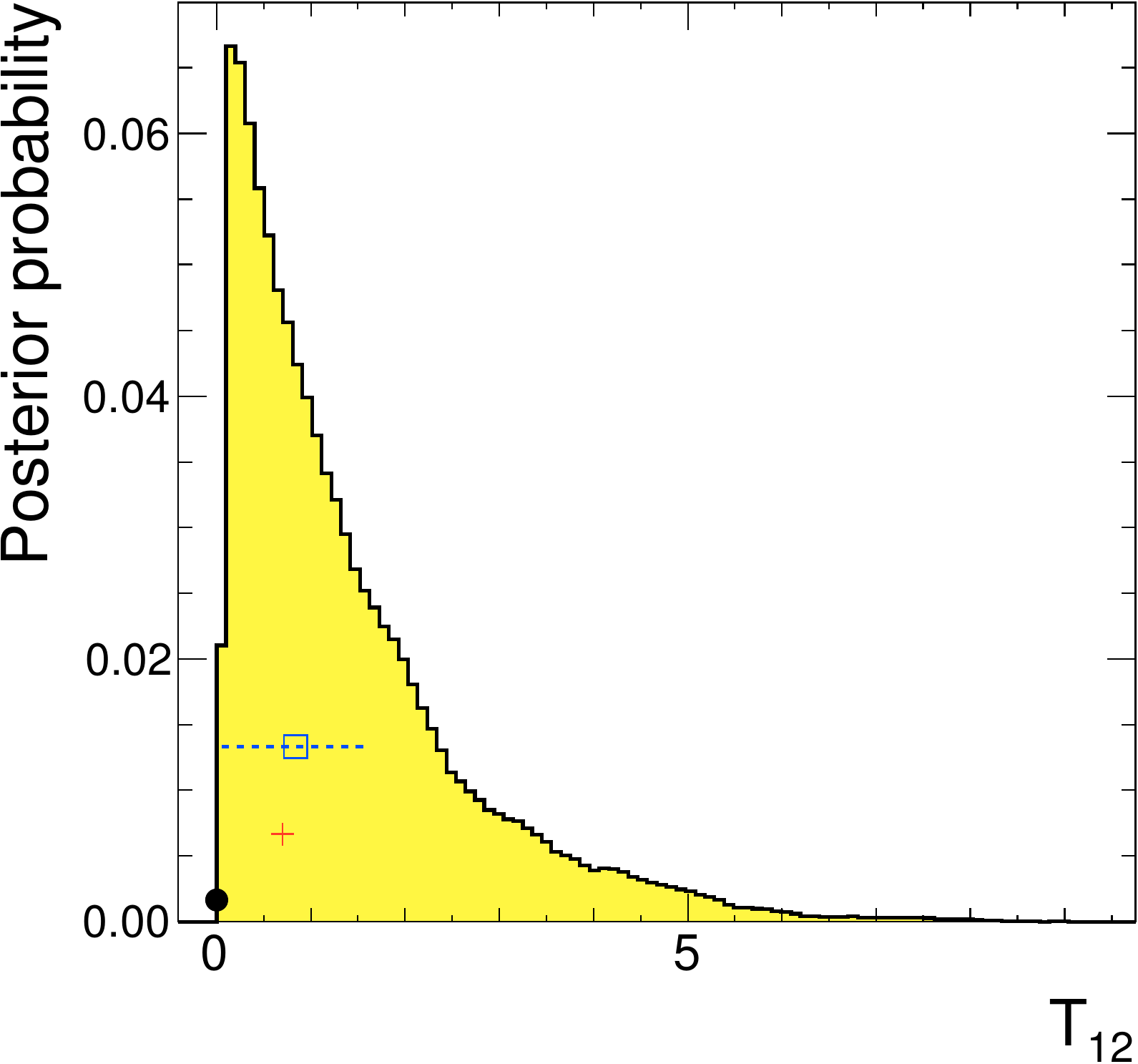} &
   \includegraphics[width=0.18\columnwidth]{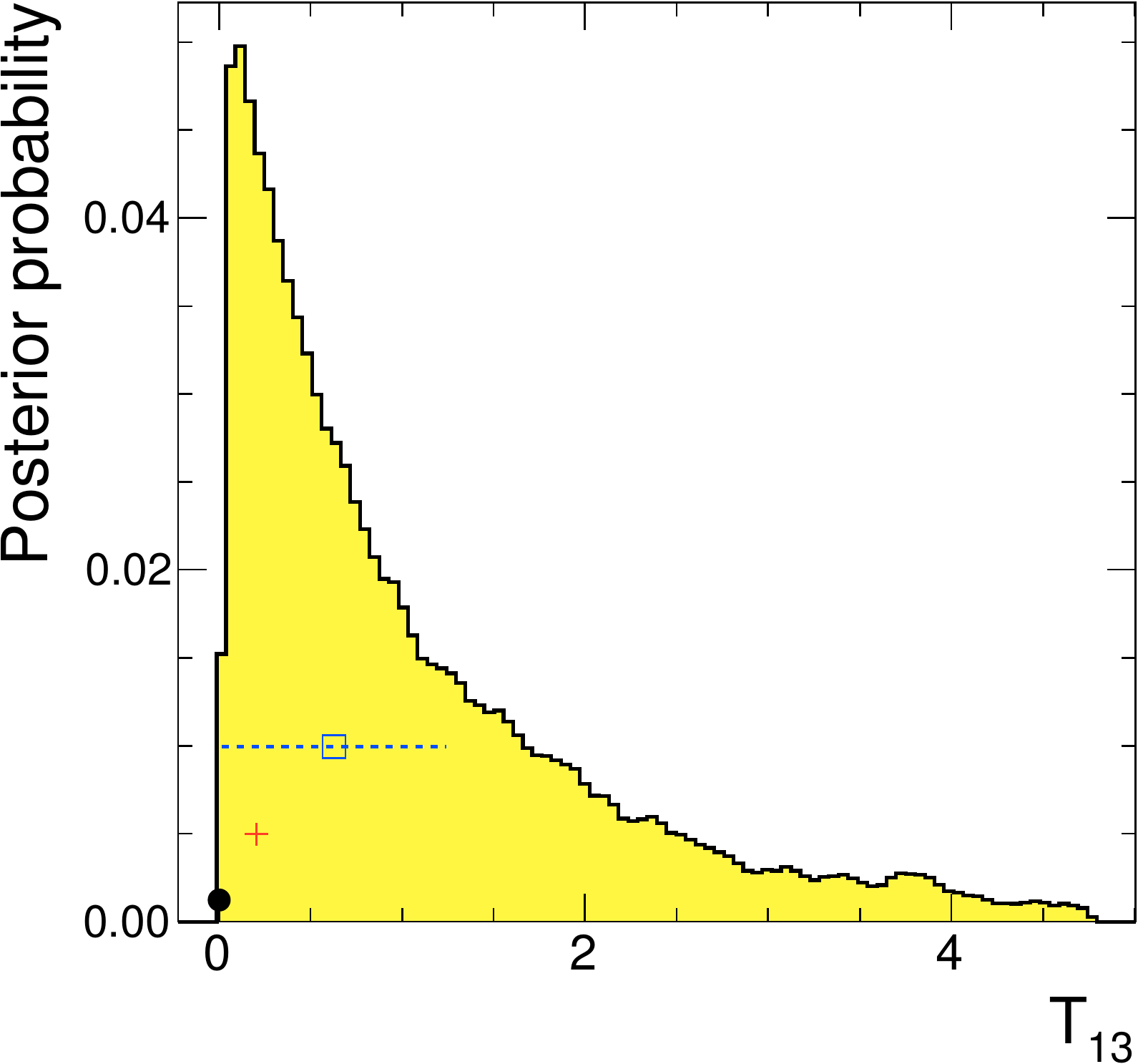} &
   \includegraphics[width=0.18\columnwidth]{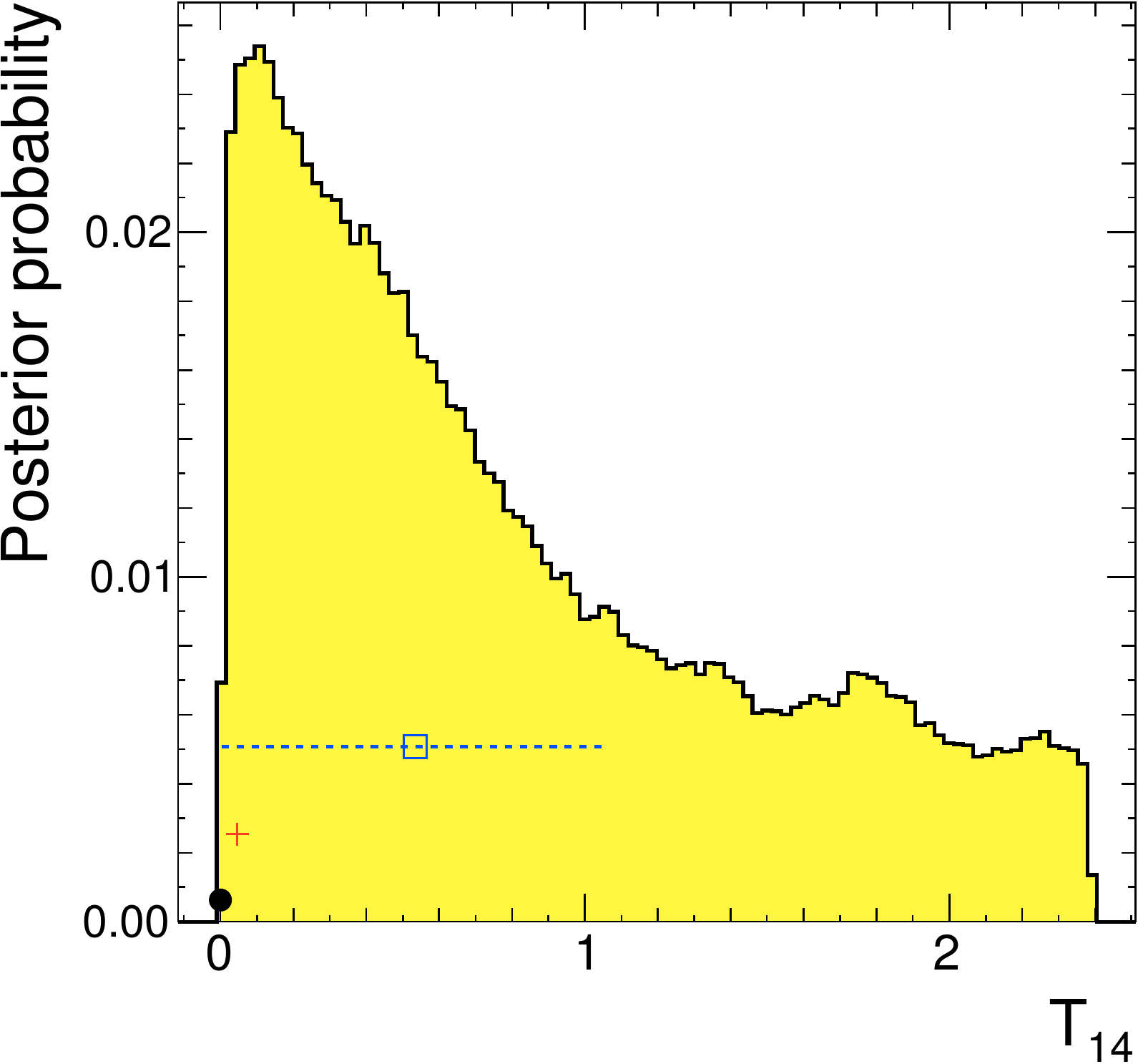} 
 \end{tabular}
 \caption{The 1-dimensional marginal distributions of $p(\tuple{T}|\tuple{D})$ in the example of Sec.~\ref{sec:example5}.
\label{fig:1Dim5}}
\end{figure}

First, the initial sampling hyper-box shown in Fig.~\ref{fig:inputs4} is used.  It is not reduced to a smaller volume, exploiting the ability of MCMC to navigate through large spaces towards the region of interest.  With $10^6$ MCMC samples, the 1-dimensional marginal distributions of Fig.~\ref{fig:1Dim5} are obtained.  

Comparing Fig.~\ref{fig:1Dim5} to \ref{fig:1Dim4}, the statistical fluctuations are much smaller in the former, using MCMC, even though the MCMC samples are $10^6$ instead of $10^7$, as a result of more efficient sampling.  The shape of the distributions in Fig.~\ref{fig:1Dim5}, though, seems to have unnatural anomalies, especially in truth-level bins with small event counts.

The anomalies get worse if the sampling is limited to the reduced volume used in Sec.~\ref{sec:example4}, and shown in Fig.~\ref{fig:priorRedefined4}.  This has to do with the ability of the MCMC algorithm to reach equilibrium, which depends on the sampled volume and on the step size of the MCMC random walk (Sec.~\ref{sec:sampling}).  Improvements are possible, by adjusting the MCMC step size, but they are left for future study.

The unfolded spectra are shown in Fig.~\ref{fig:unfolded5}, as they are found with and without reducing the sampled hyper-box volume.  Qualitatively the results are similar, but not identical.  More interestingly, they are similar (but not identical) to the result using uniform random sampling in the reduced hyper-box, in Sec.~\ref{sec:example4}, shown in Fig.~\ref{fig:unfolded4}.

So, using MCMC instead of uniform random sampling has the advantage of speed and lower statistical fluctuations in the result, but it probably should not be used when emphasis is put on very detailed computations.  Luckily, after volume reduction (Sec.~\ref{sec:narrowing}), uniform sampling is not prohibitively slow, as shown in Sec.~\ref{sec:example4}.  When MCMC is used, to get quick results, it is recommended to inspect the 1-dimensional marginal distributions, and make sure they look reasonably smooth, especially in bins with large $T_t$ values.  If this is not the case, a different step size in MCMC may help.

\subsection{$N_r \neq N_t$}
\label{sec:example6}

The goal of this example is to demonstrate the possibility of applying FBU even when the $N_r$ reco-level bins do not correspond to the $N_t$ truth-level bins.  The computer program written to produce these FBU examples required no modification to tackle this extreme scenario.

\begin{figure}[H]
  \centering
\subfigure[]{
  \includegraphics[height=0.25\columnwidth]{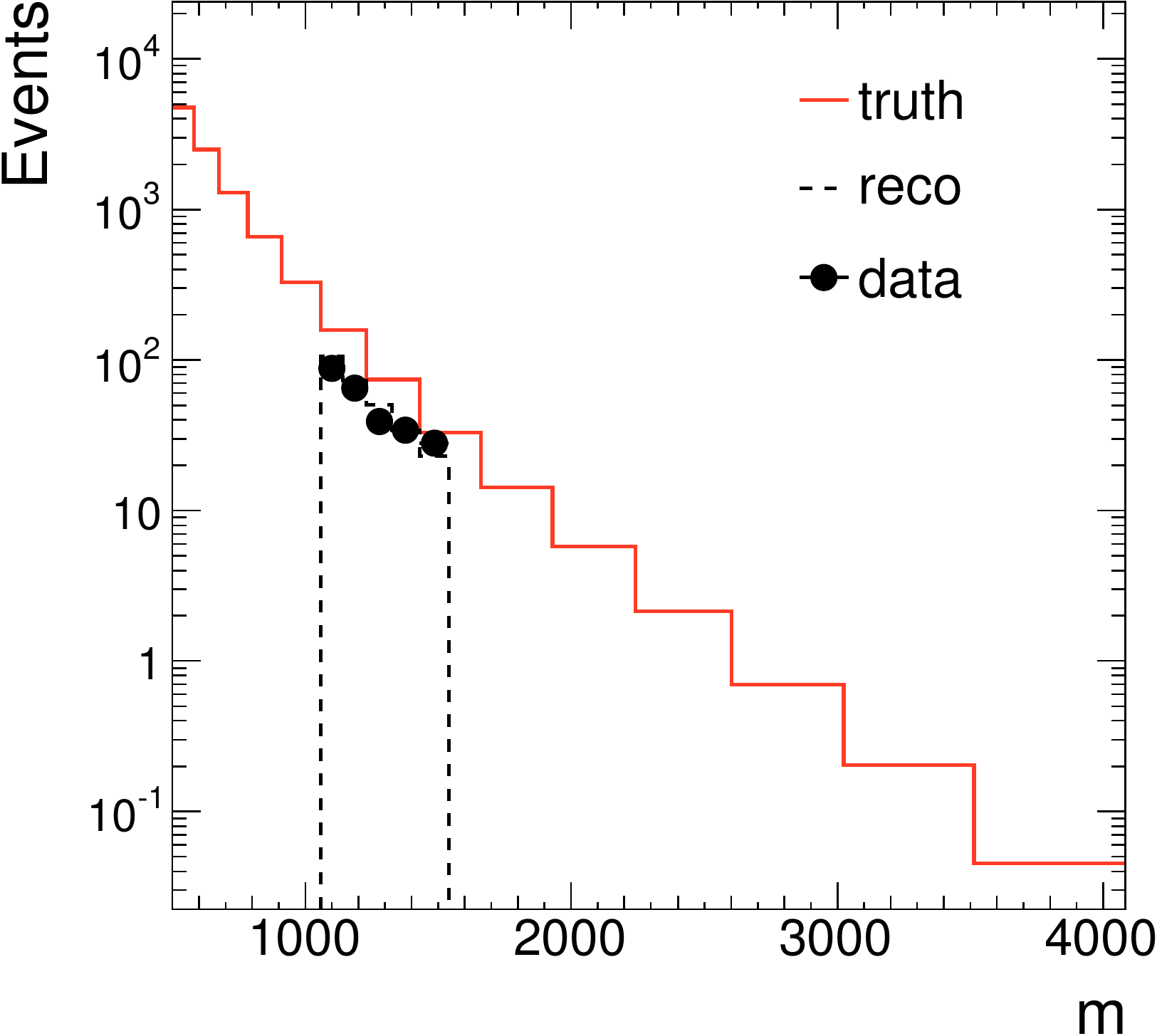}
  \label{fig:truthAndReco6}
}
\subfigure[]{
  \includegraphics[height=0.25\columnwidth]{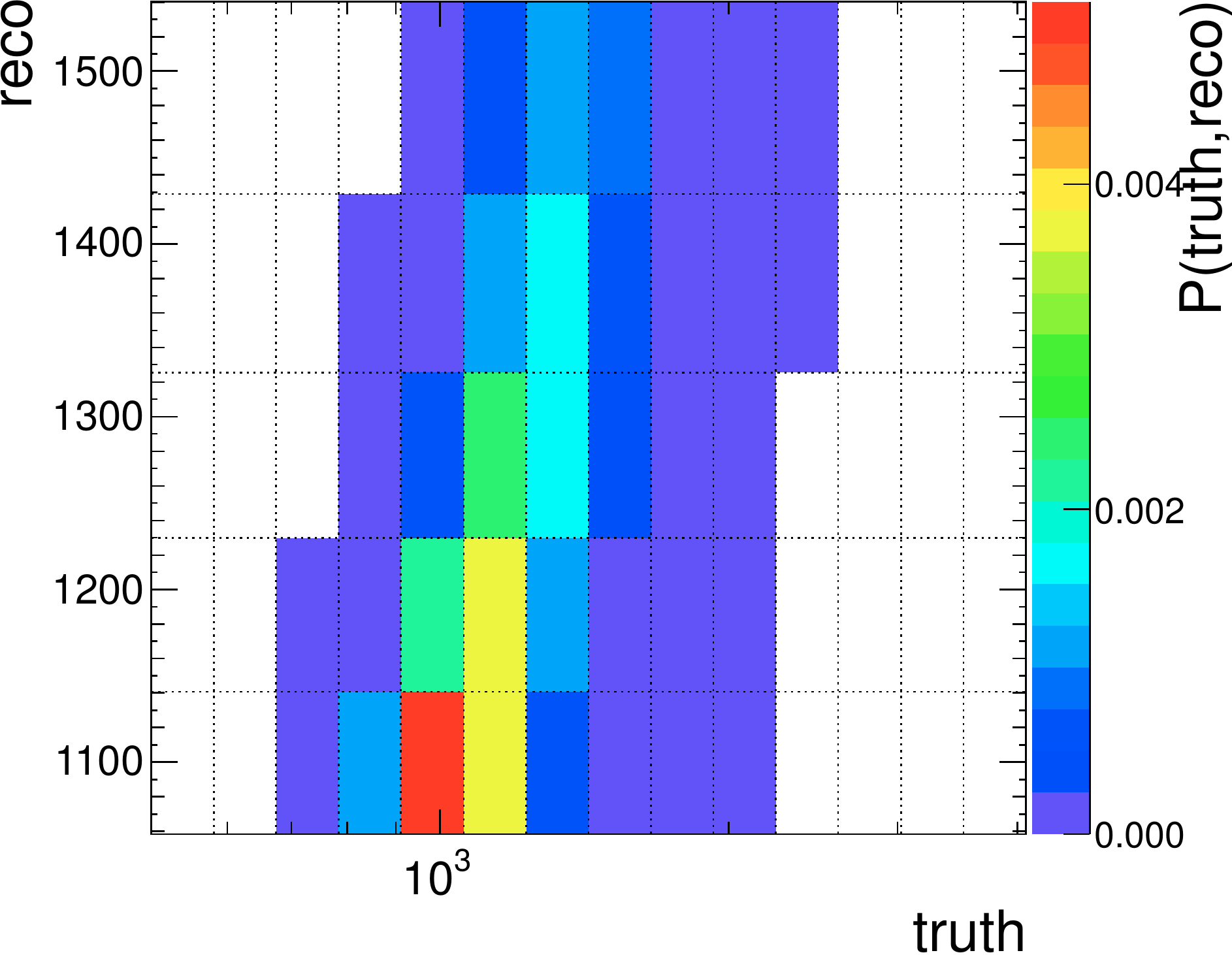}
  \label{fig:migrations6}
}
\subfigure[]{
  \includegraphics[height=0.25\columnwidth]{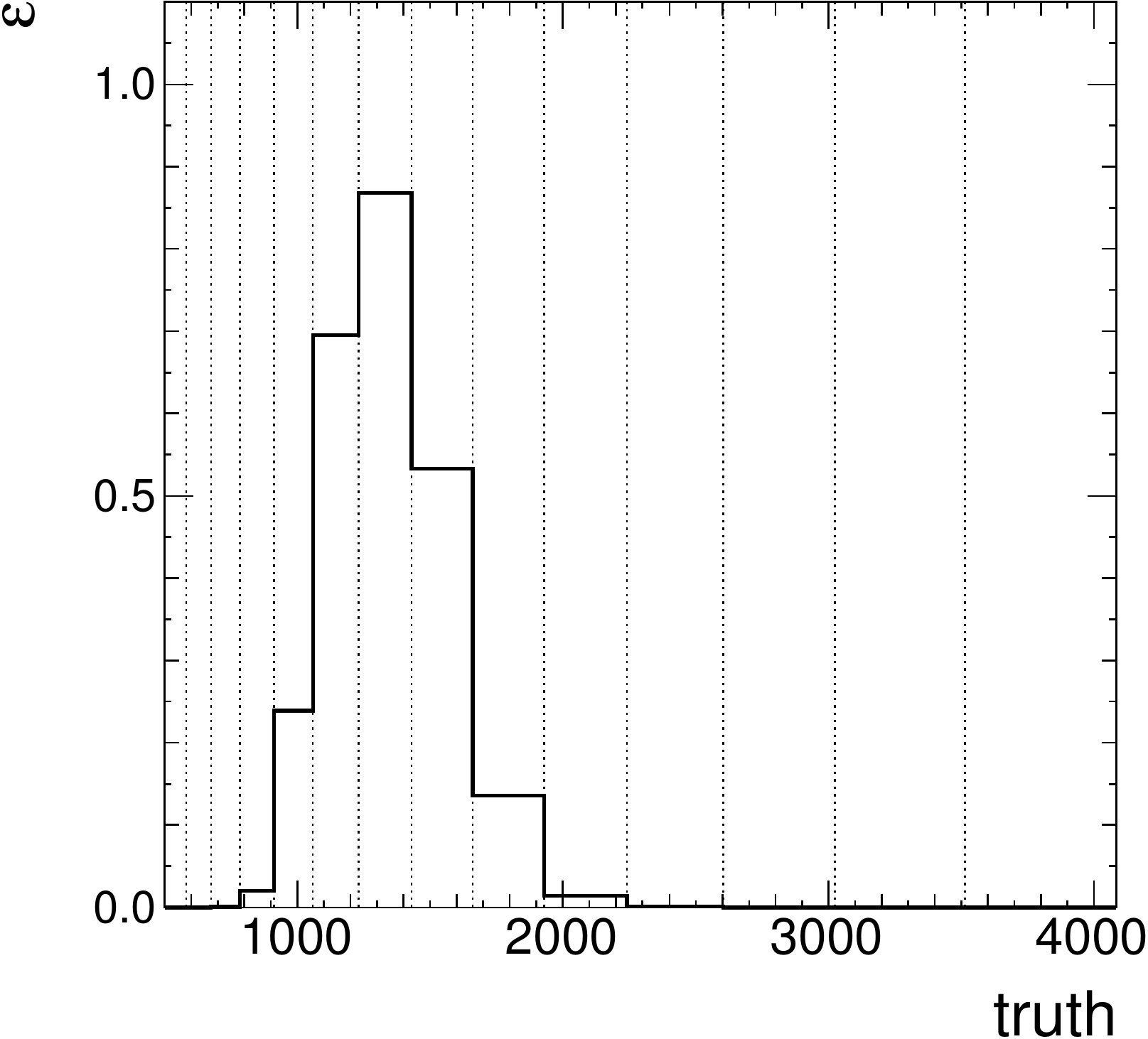}
  \label{fig:eff6}
}
\caption{In (a): The truth-level spectrum generated (red), the corresponding reco-level spectrum (black dashed), and the pseudo-data (black markers) for the example of Sec.~\ref{sec:example6}.  (b): The corresponding migrations matrix, of dimension $N_t\times N_r = 14\times 5$.  (c): The efficiency of the migrations matrix.
}
\end{figure}

The truth-level distribution of MC events in $m$, with $N_t=14$ bins, does not change, but they are reconstructed only in $N_r=5$ narrower bins shown in Fig.~\ref{fig:truthAndReco6}.  Clearly the migrations matrix (Fig.~\ref{fig:migrations6}) is not square, and its efficiency (Fig.~\ref{fig:eff6}) is zero in truth-level bins where smearing is not capable of migrating events into the reconstructed bins.

\begin{figure}[H]
  \centering
 \includegraphics[height=0.5\columnwidth]{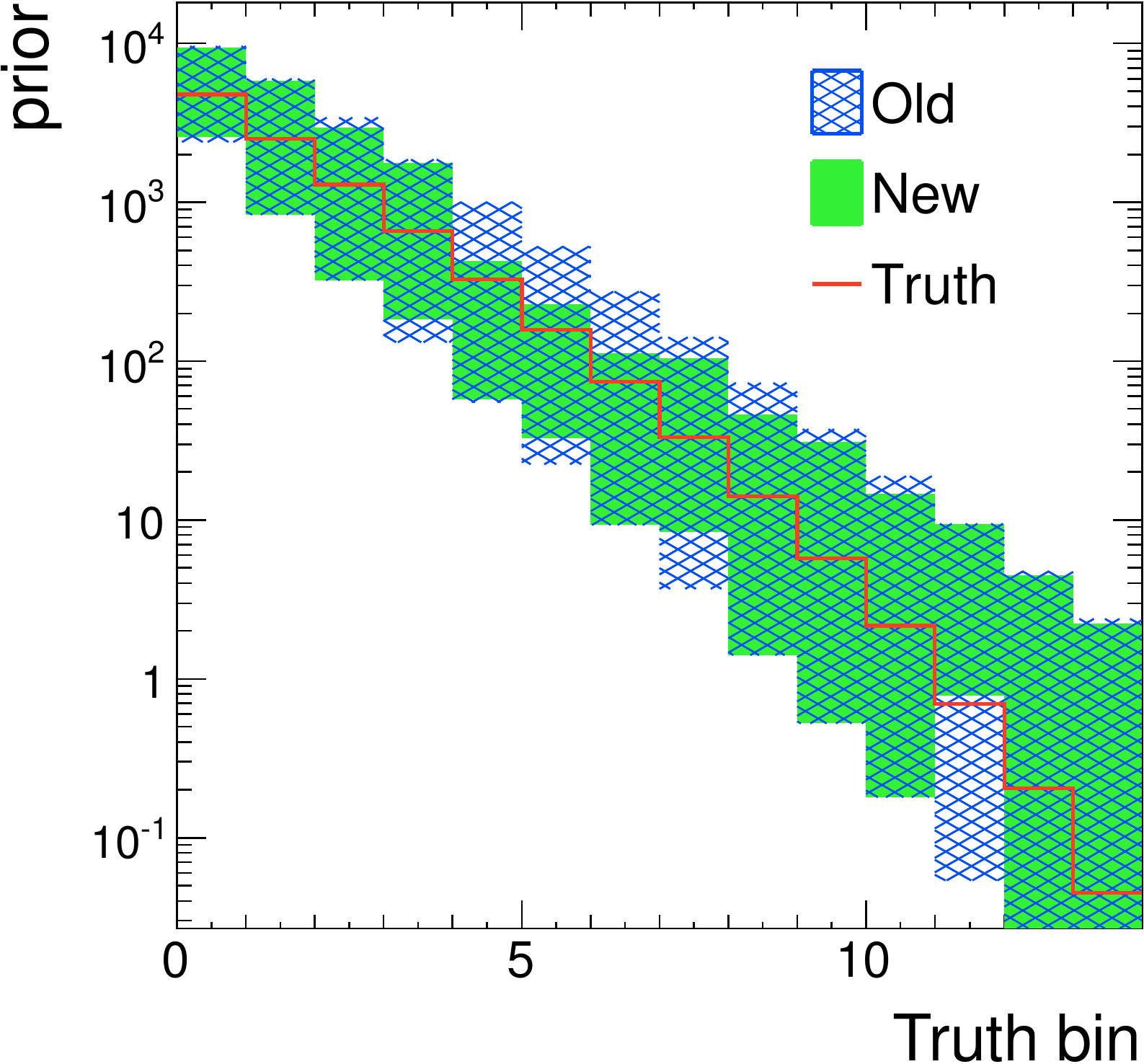}
\caption{Initial sampled hyper-box (``old''), and hyper-box with reduced volume (``new'') corresponding to Sec.~\ref{sec:example6}.
\label{fig:priorRedefined6}
}
\end{figure}

The same initial sampled hyper-box is defined as in Eq.~\ref{eq:prior4}, which is reduced according to Sec.~\ref{sec:narrowing}, with the result shown in Fig.~\ref{fig:priorRedefined6}.  The volume reduction is not great, due to the wide spread of $p_t(T_t|\D)$ for some $T_t$ for which the data provide no constraint.

\begin{figure}[H]
\centering
\begin{tabular}{ccccc}
   \includegraphics[width=0.18\columnwidth]{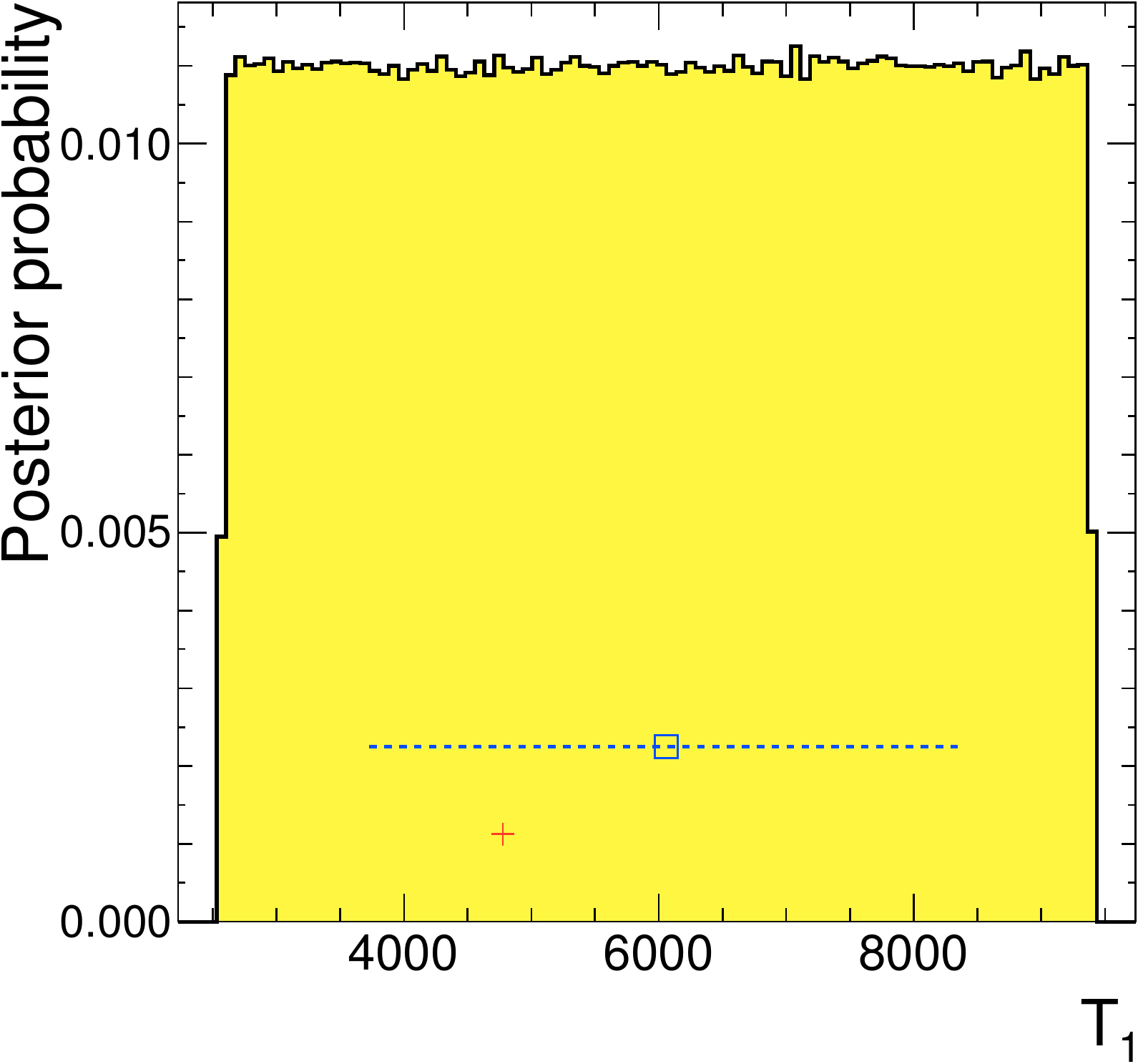} &
   \includegraphics[width=0.18\columnwidth]{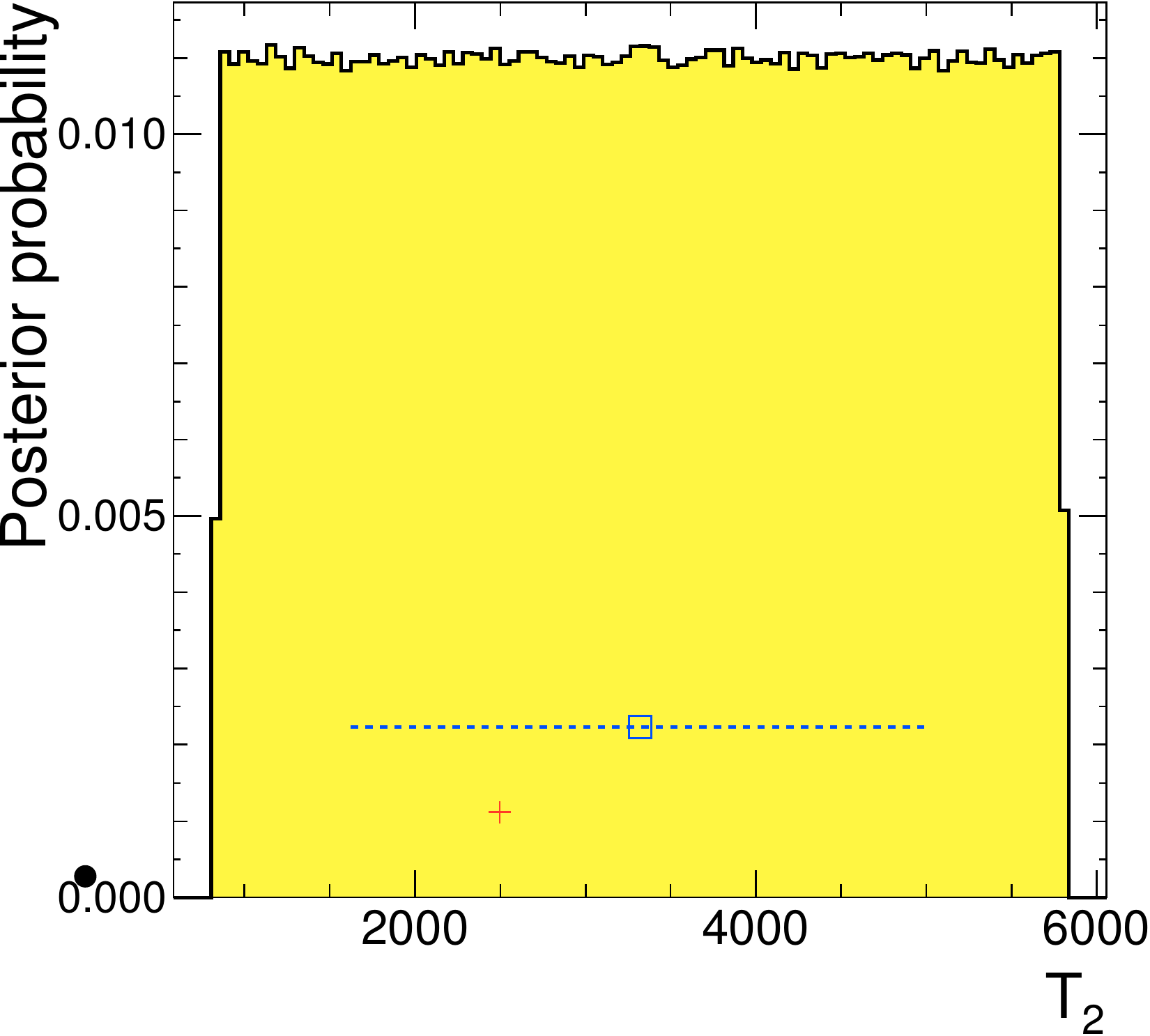} &
   \includegraphics[width=0.18\columnwidth]{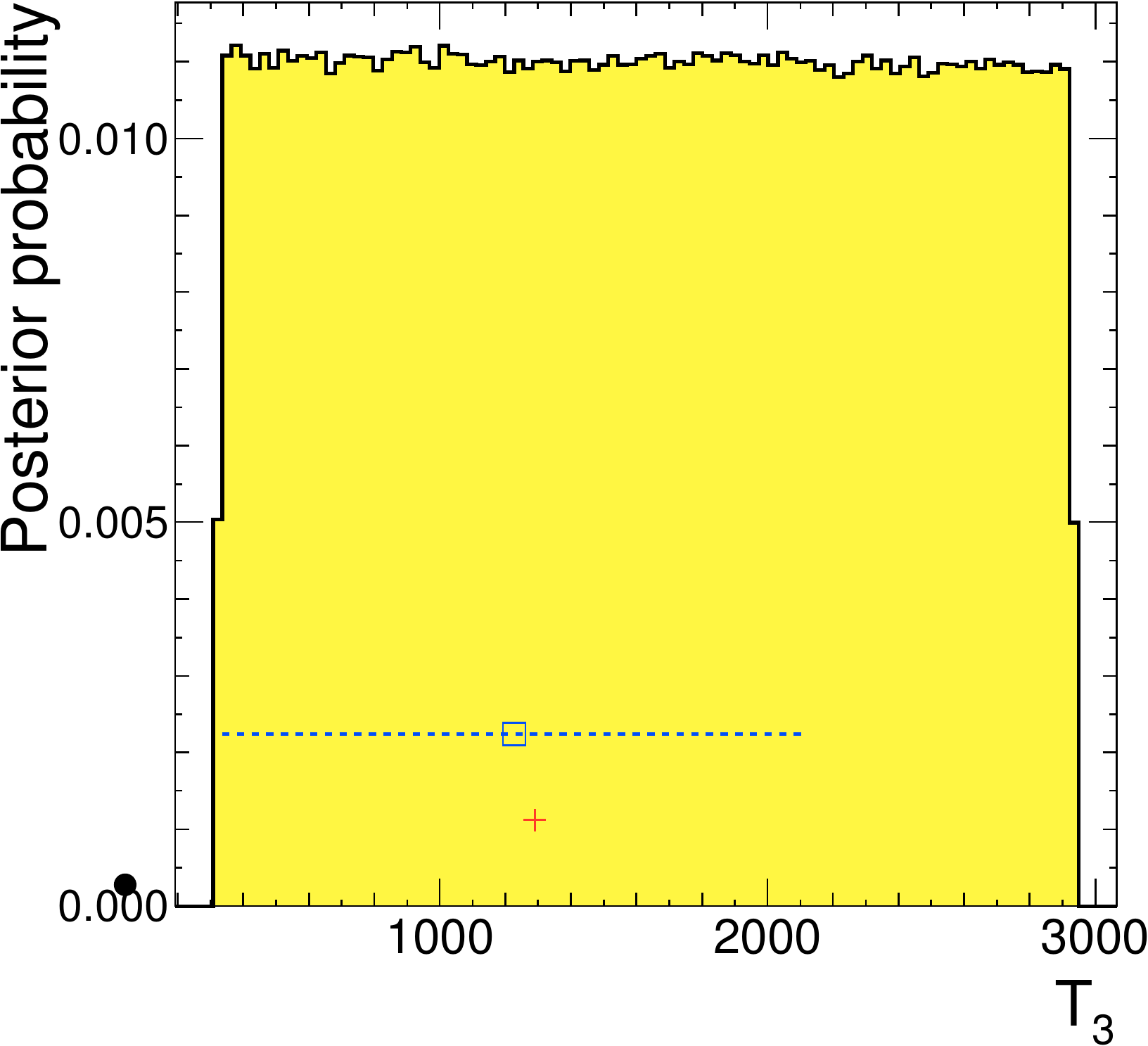} &
   \includegraphics[width=0.18\columnwidth]{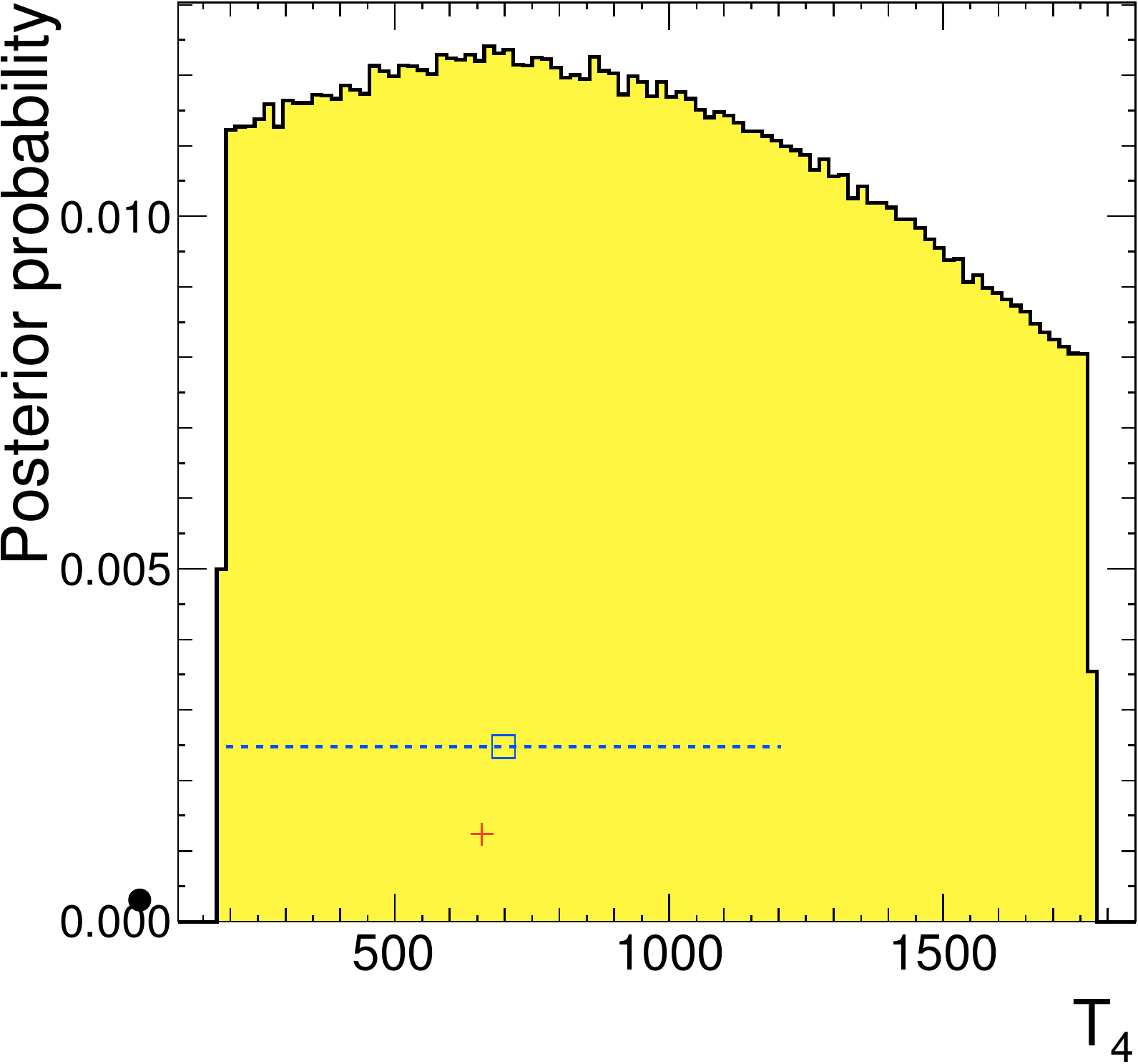} &
   \includegraphics[width=0.18\columnwidth]{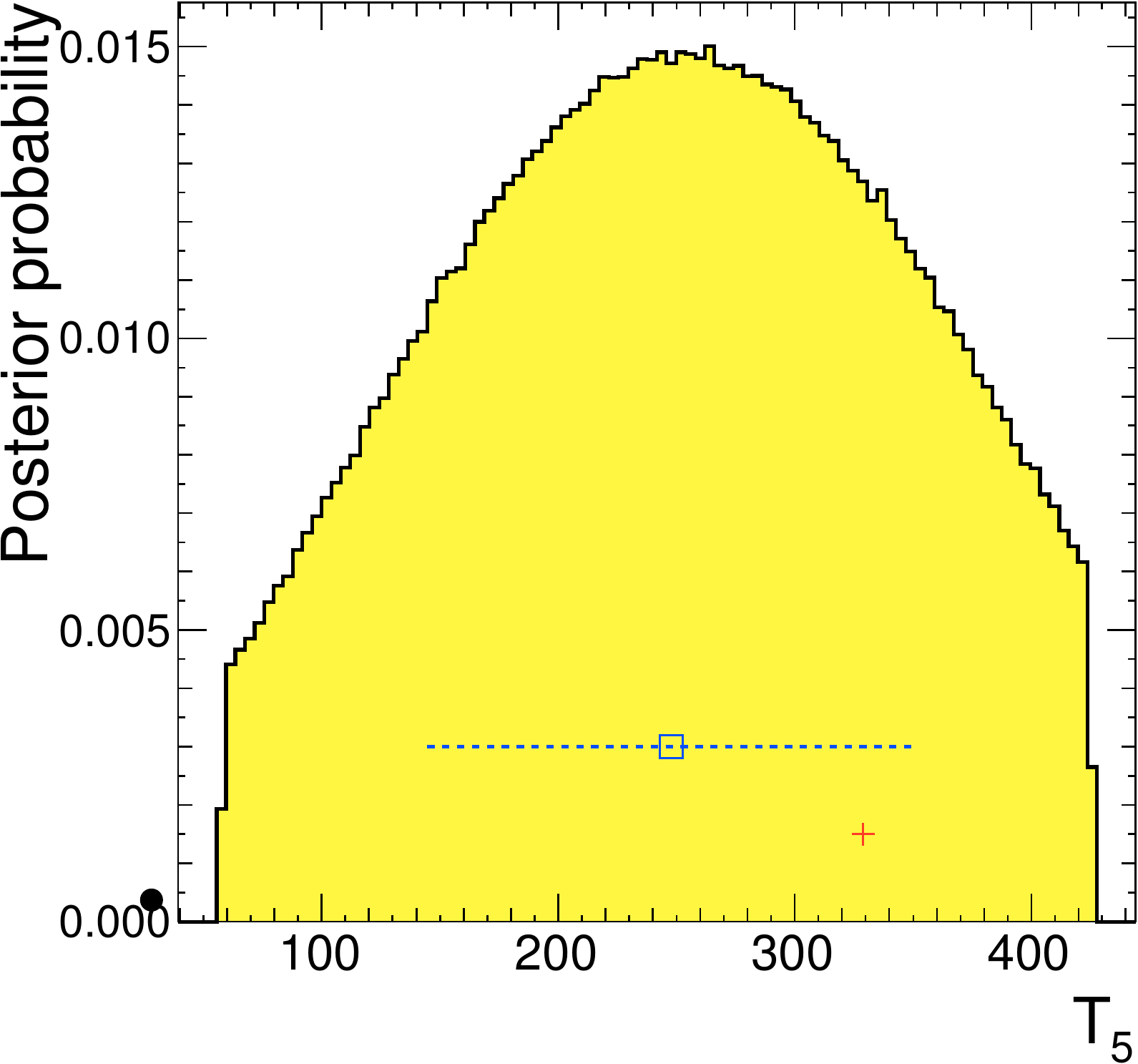} \\

   \includegraphics[width=0.18\columnwidth]{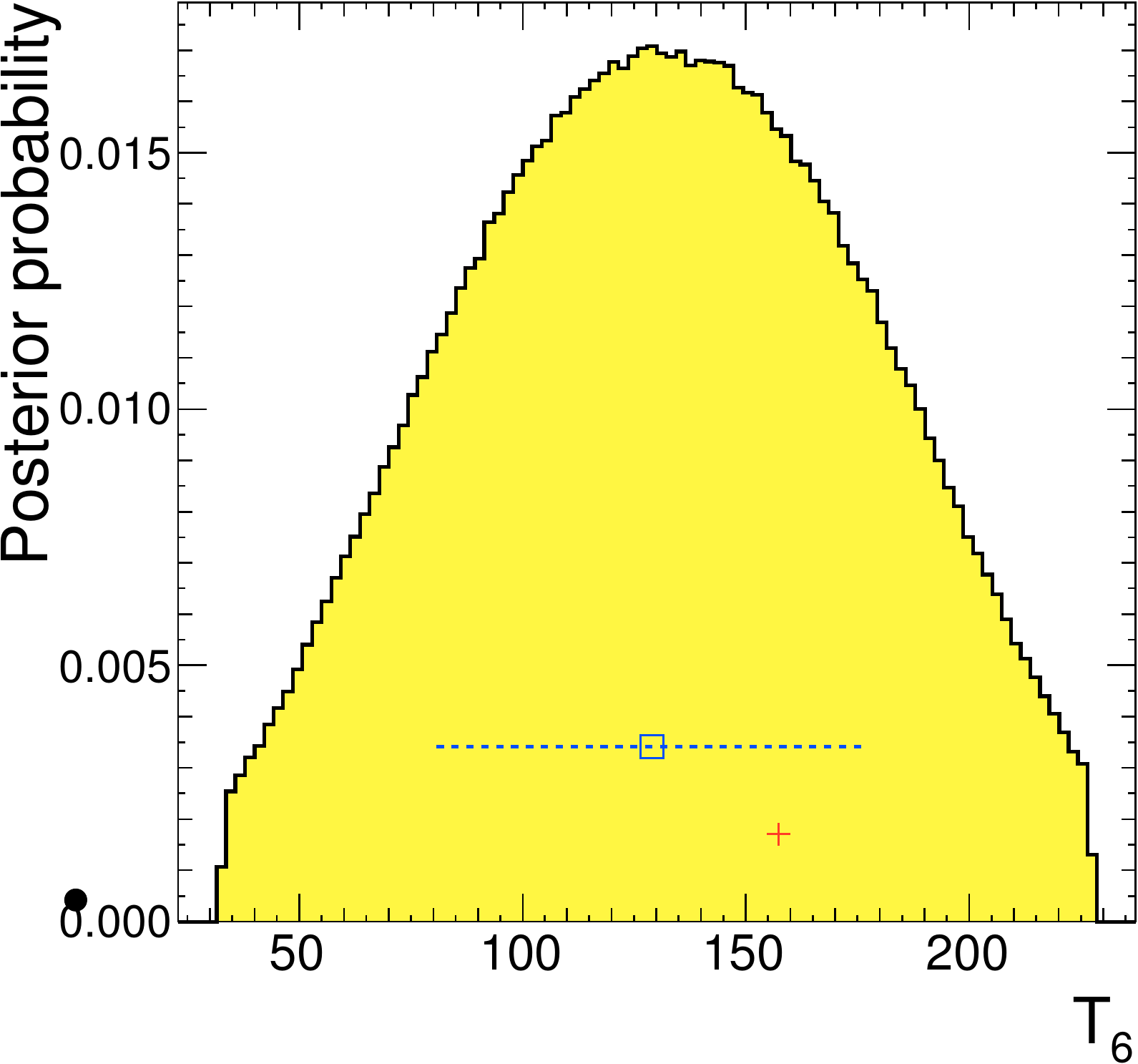} &
   \includegraphics[width=0.18\columnwidth]{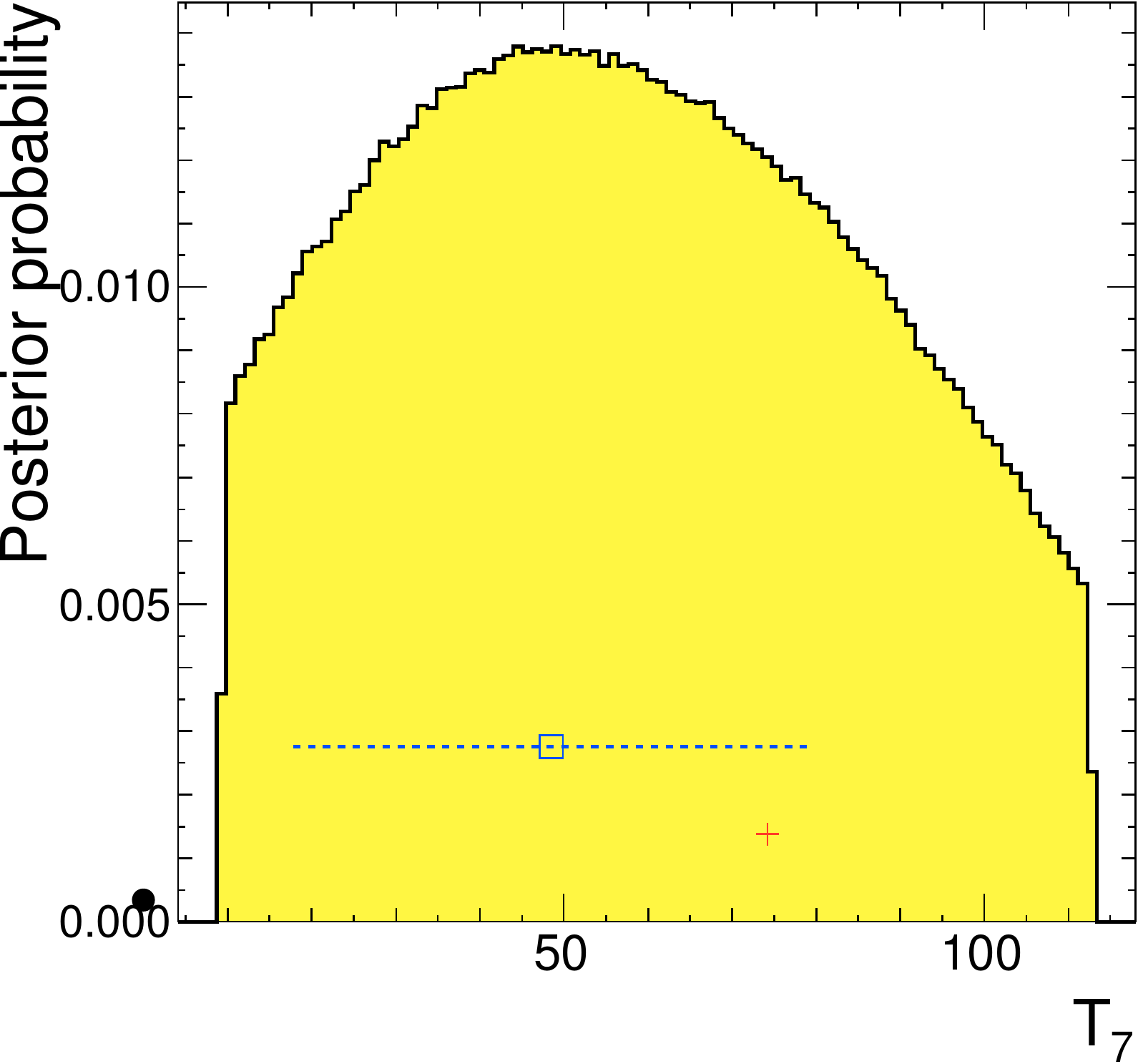} &
   \includegraphics[width=0.18\columnwidth]{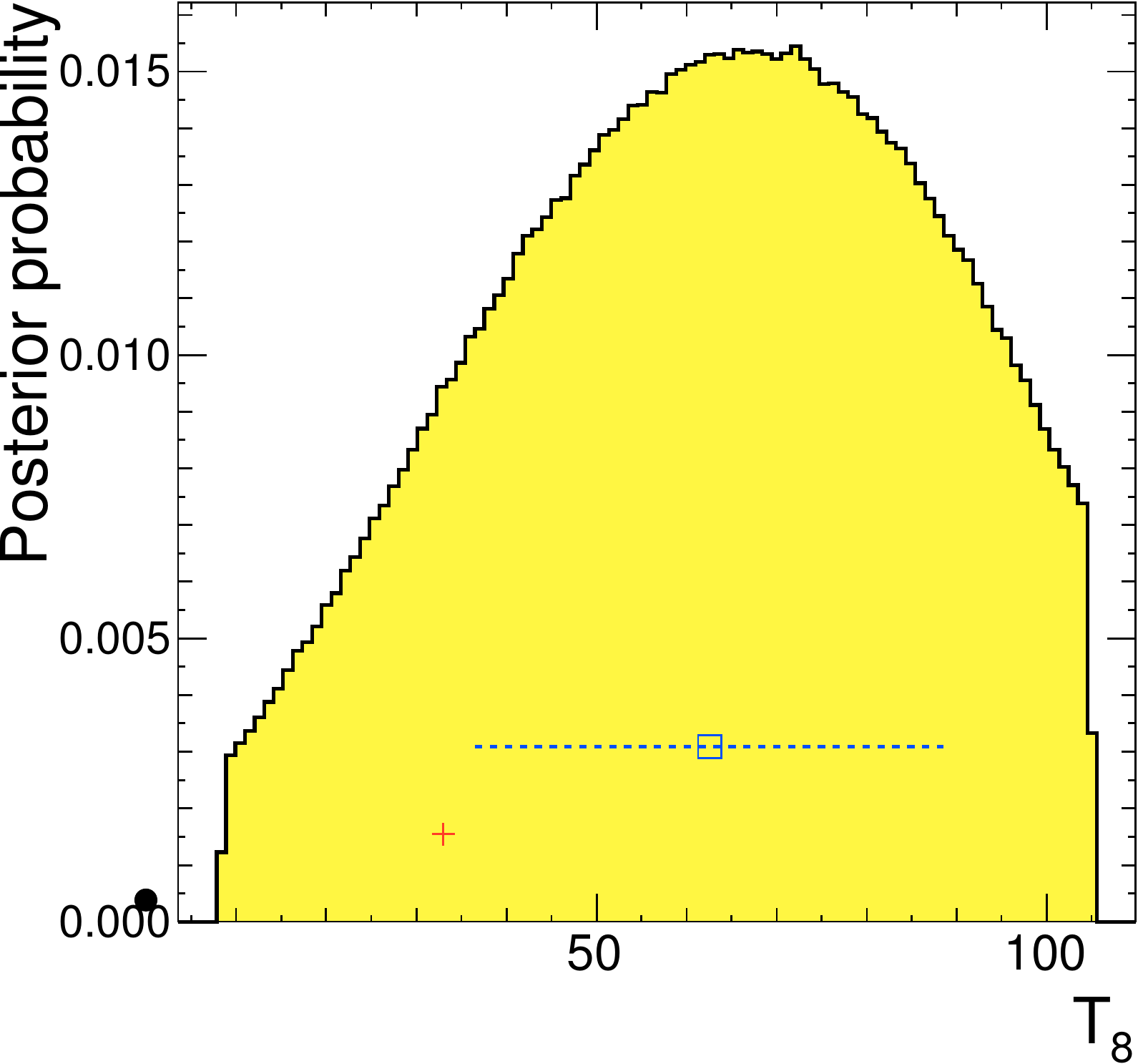} &
   \includegraphics[width=0.18\columnwidth]{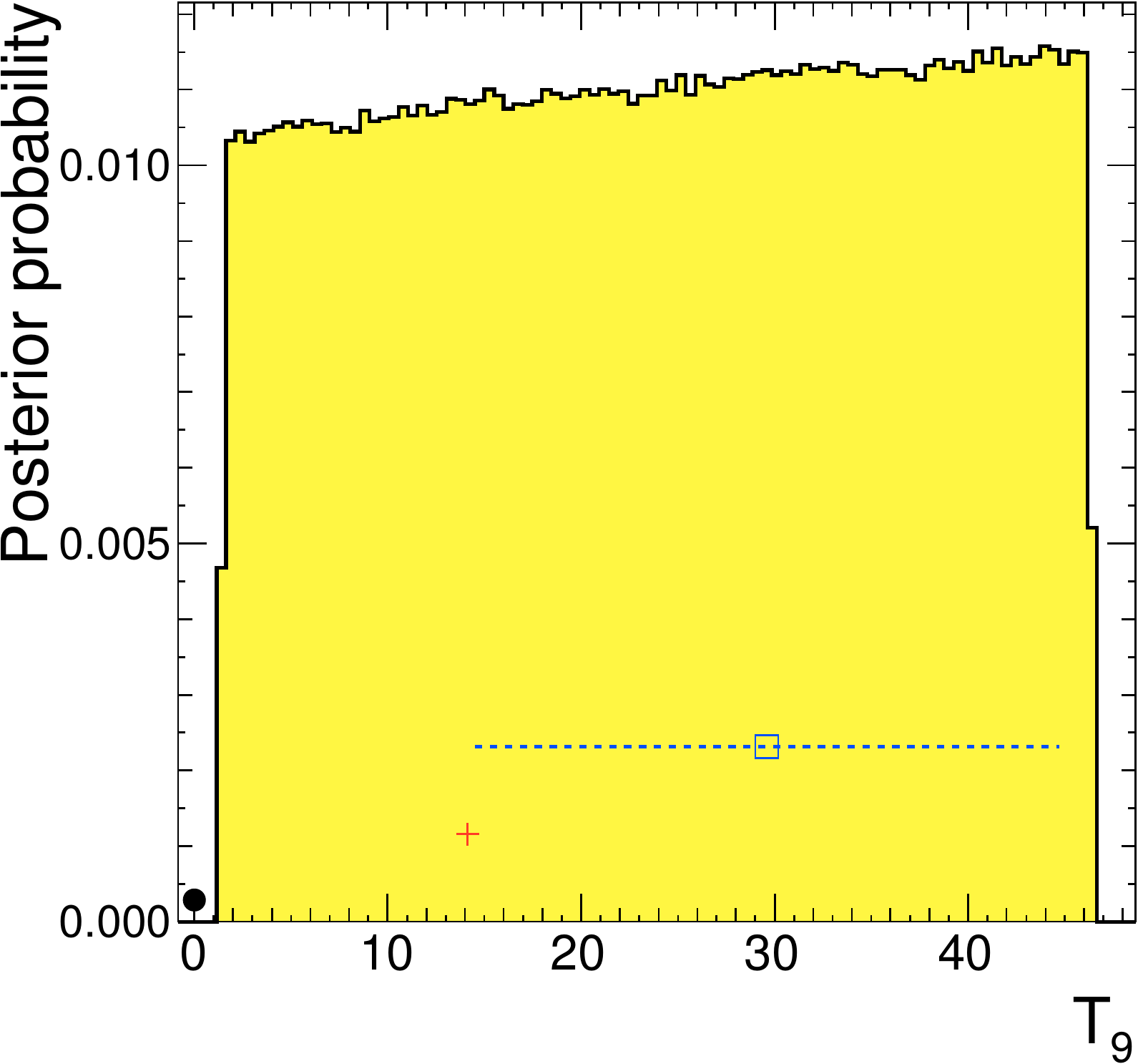} &
   \includegraphics[width=0.18\columnwidth]{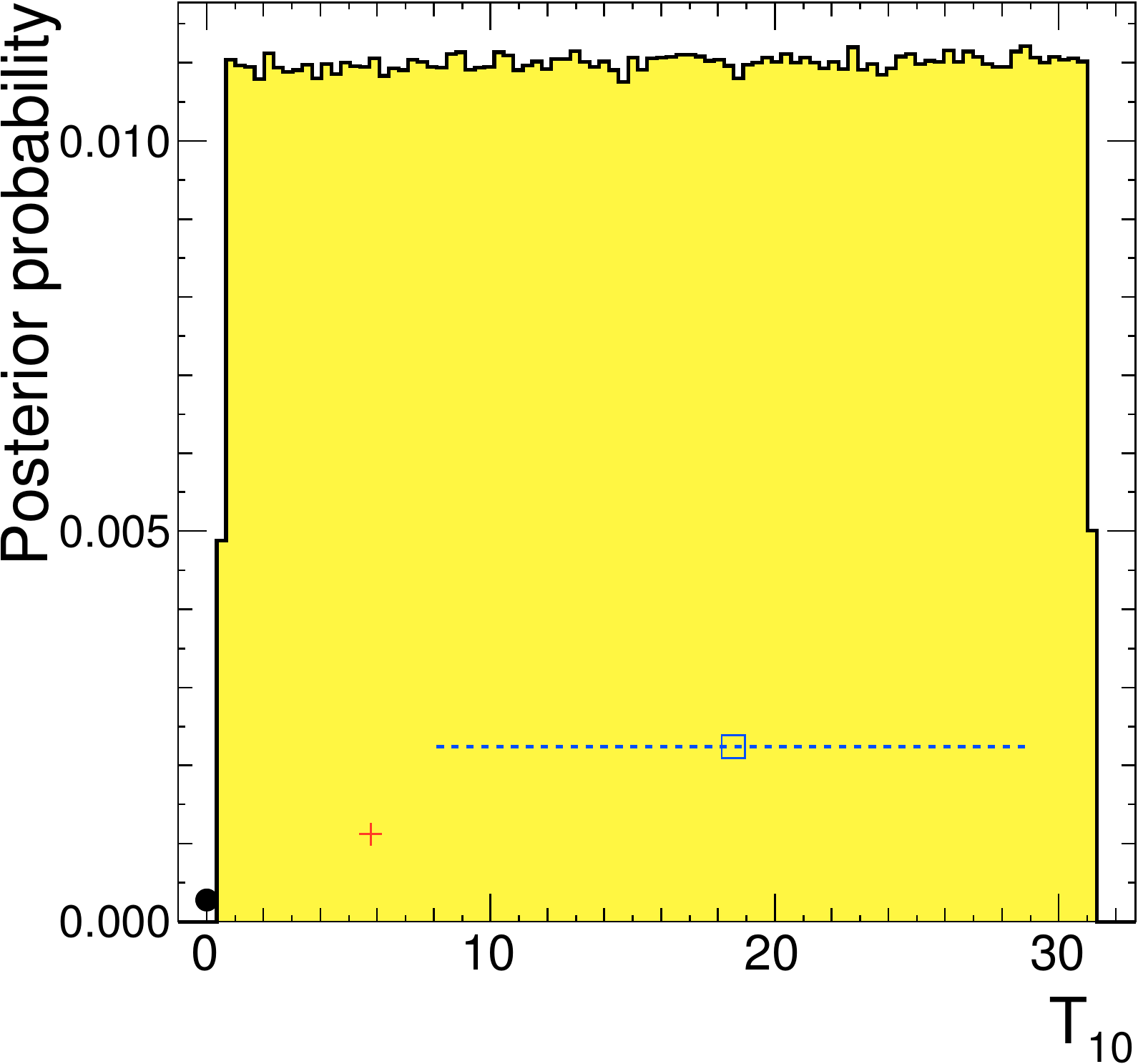} \\

   \includegraphics[width=0.18\columnwidth]{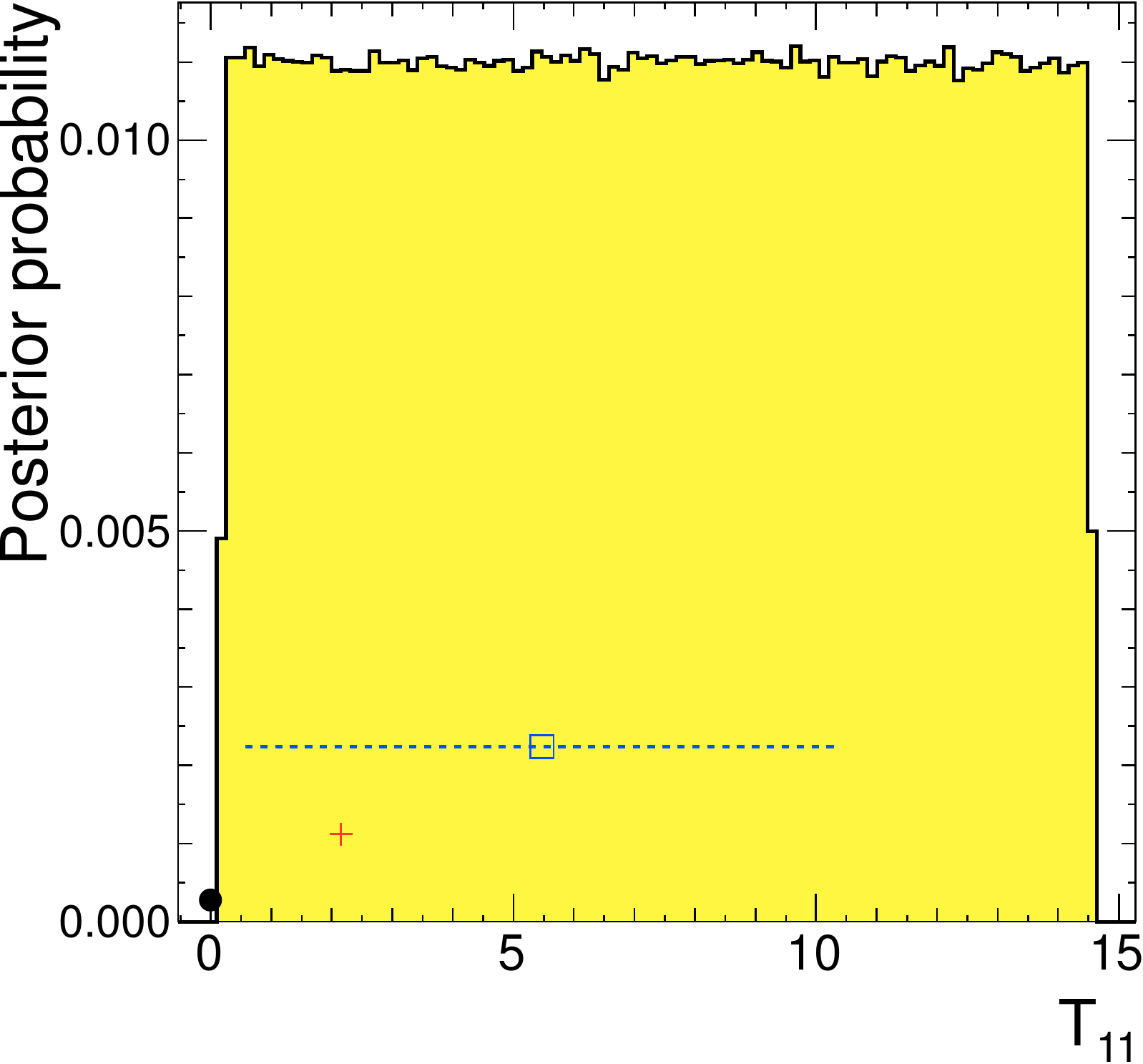} &
   \includegraphics[width=0.18\columnwidth]{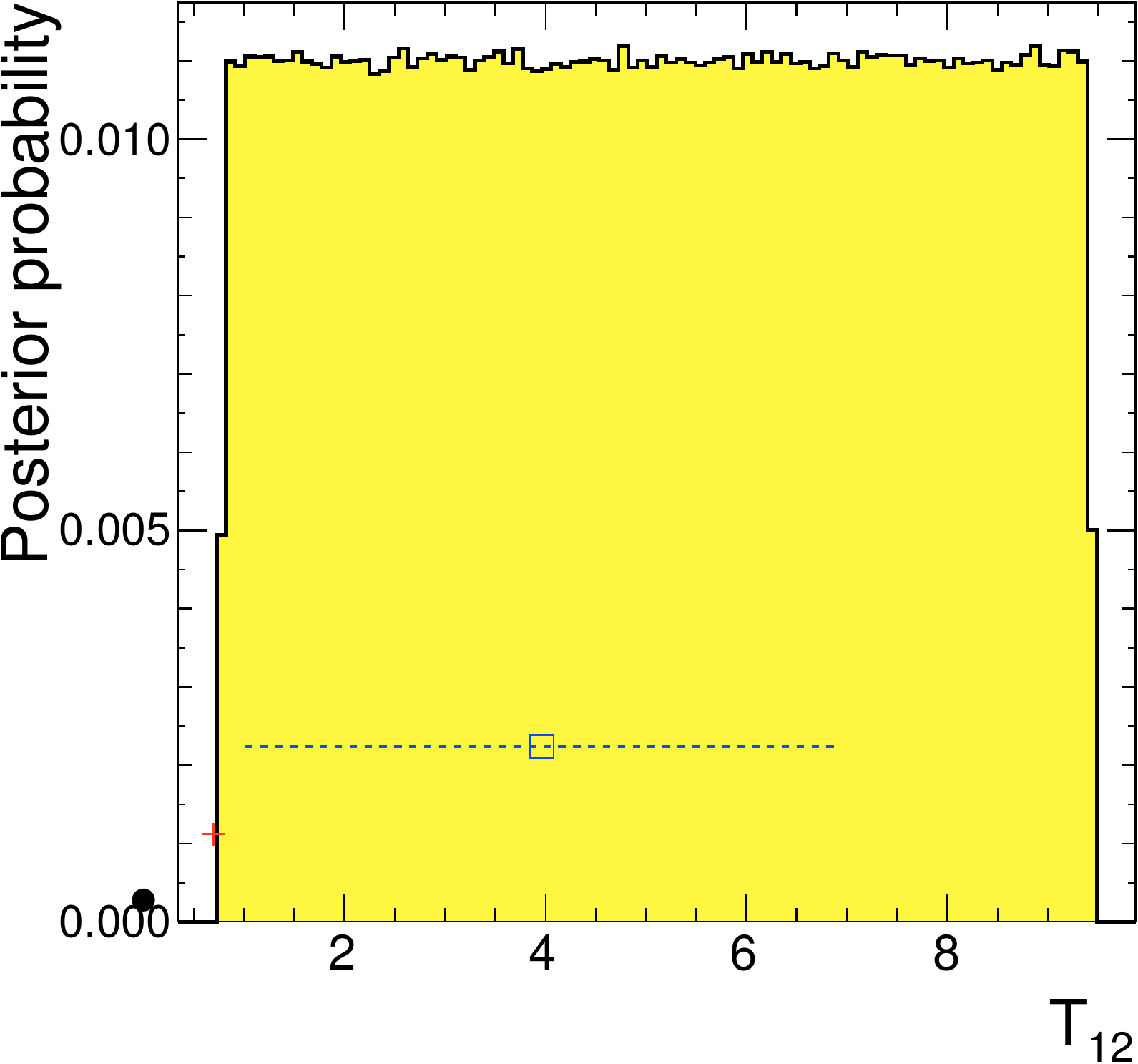} &
   \includegraphics[width=0.18\columnwidth]{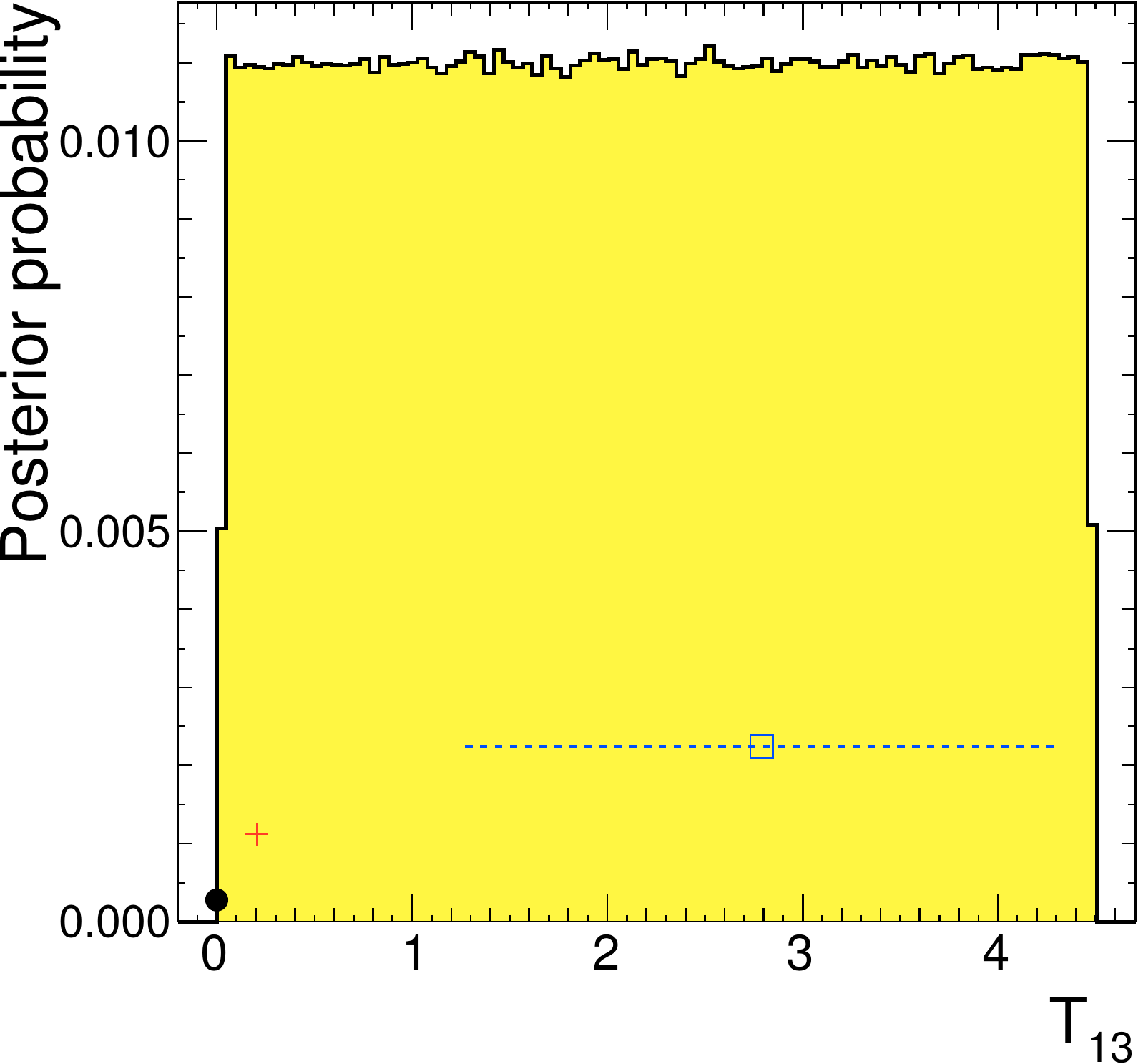} &
   \includegraphics[width=0.18\columnwidth]{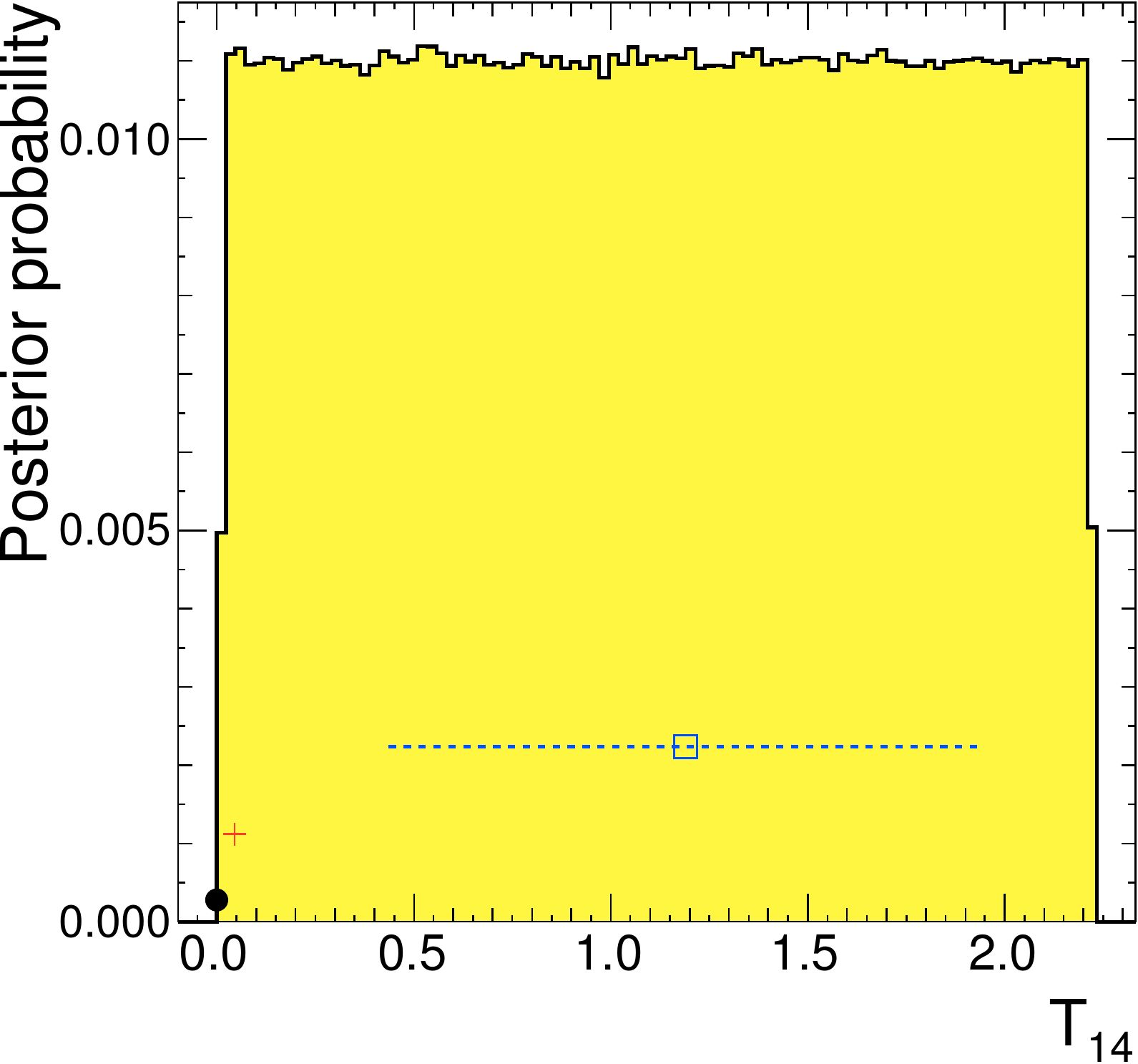} 
 \end{tabular}
 \caption{The 1-dimensional marginal distributions of $p(\tuple{T}|\tuple{D})$ in the example of Sec.~\ref{sec:example6}.  
\label{fig:1Dim6}}
\end{figure}

\begin{figure}[H]
  \centering
  \subfigure[]{
    \includegraphics[width=0.3\columnwidth]{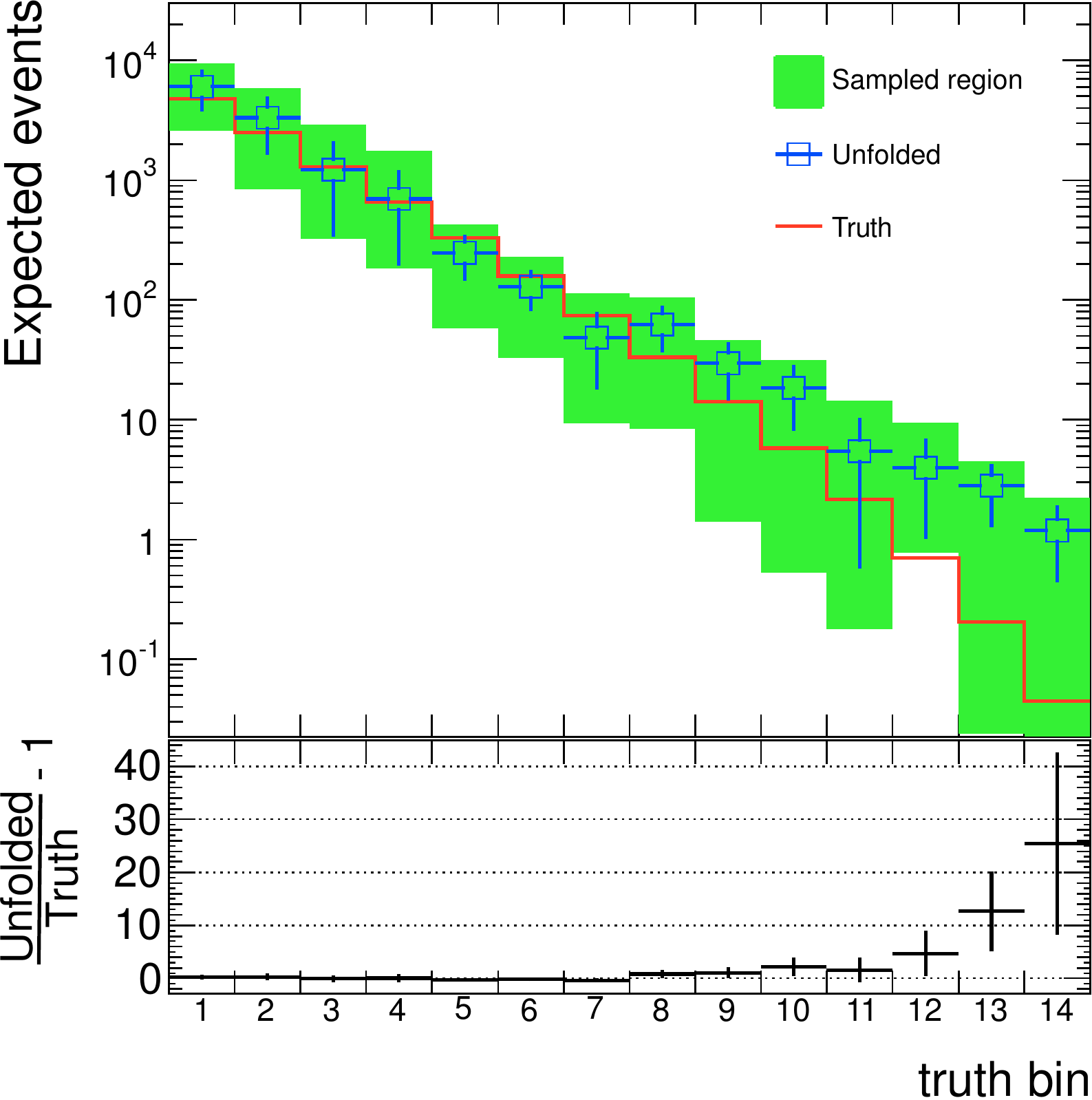}
  }
  \subfigure[]{
    \includegraphics[width=0.3\columnwidth]{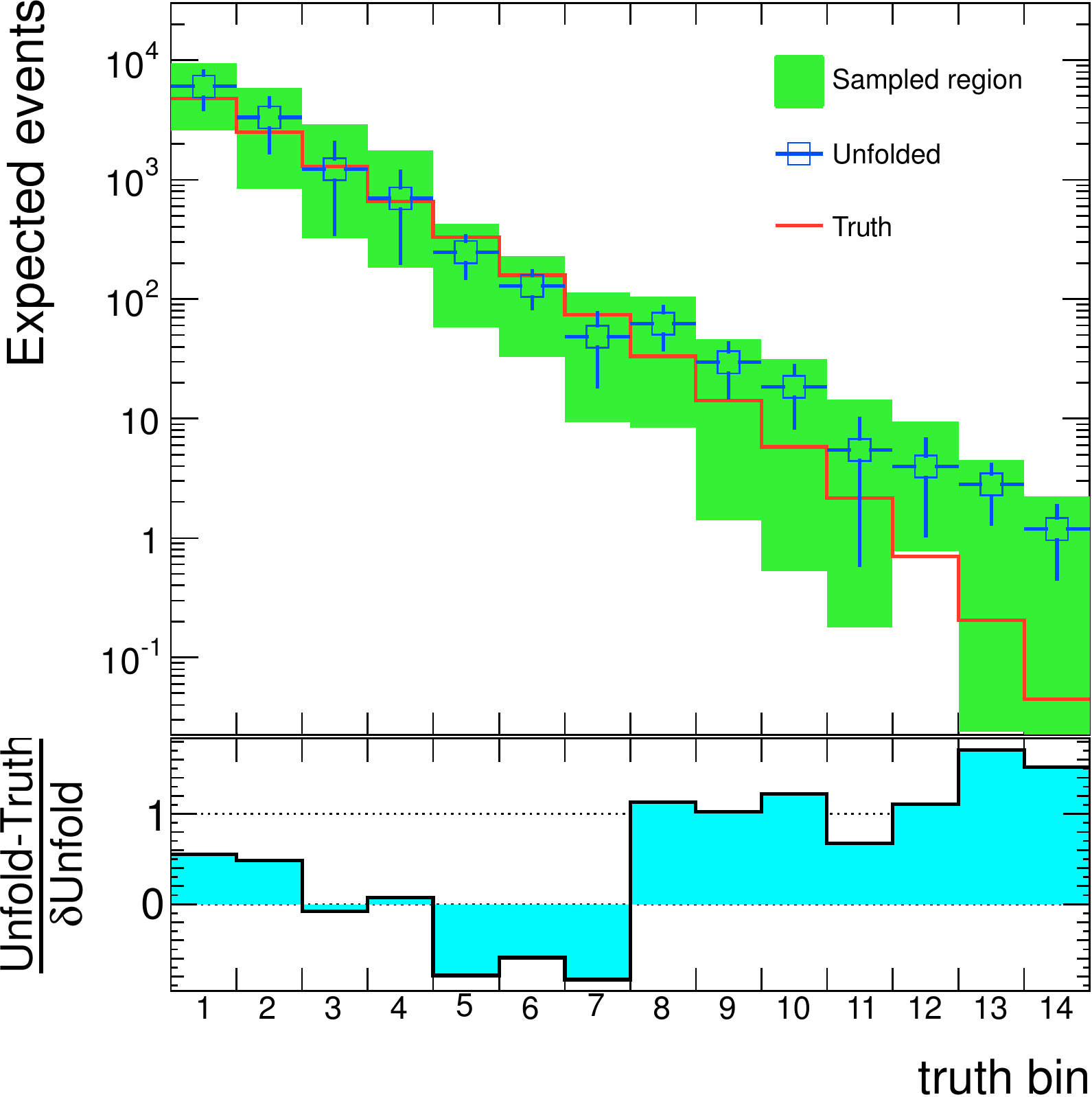}
  }
   \subfigure[]{
    \includegraphics[width=0.3\columnwidth]{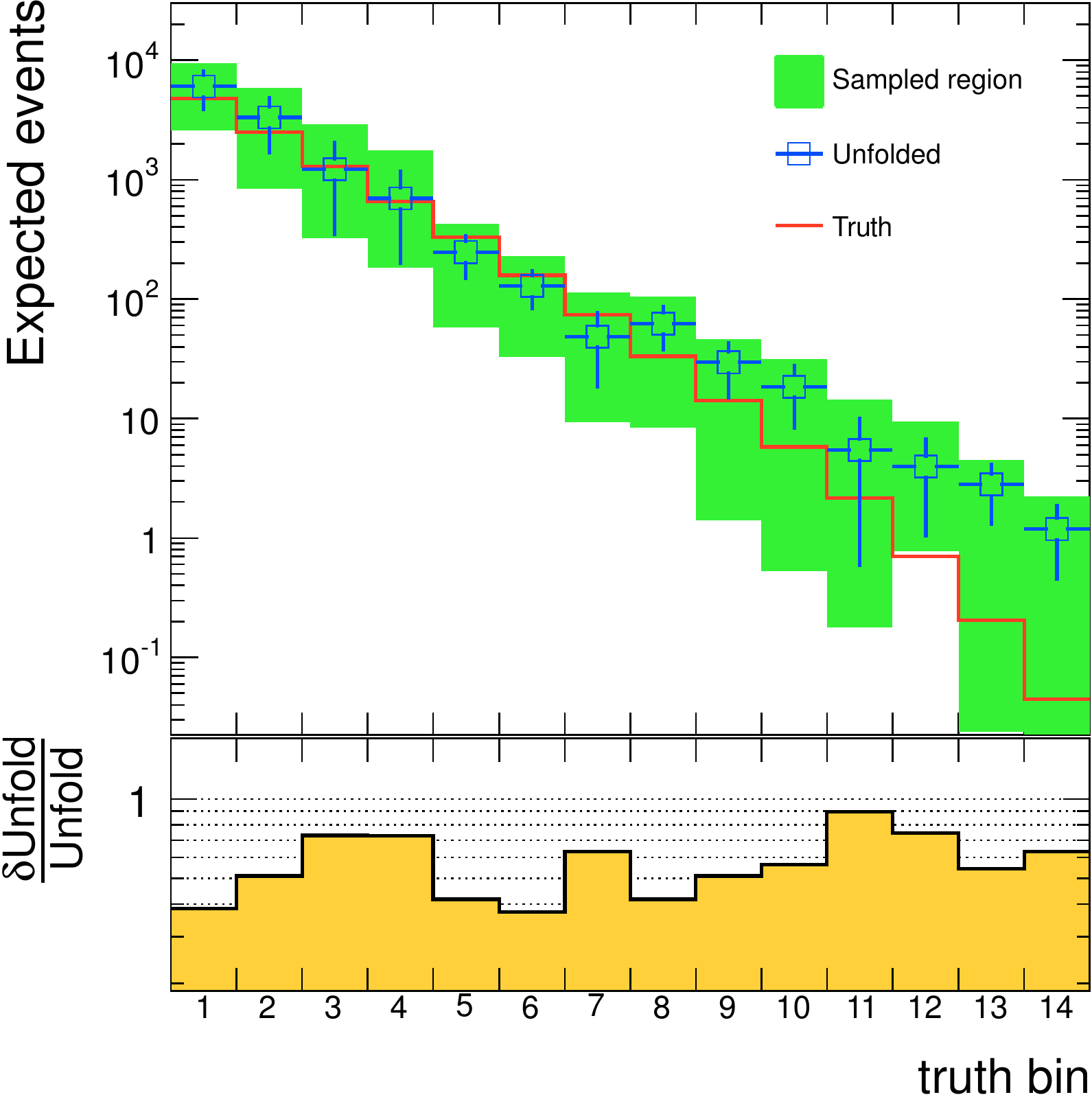}
  }
\caption{The unfolded spectrum of the example in Sec.~\ref{sec:example6}.  As Fig.~\ref{fig:1Dim6} makes clear, $T_t$ bins $t=\{1,2,3,10,11,12,13,14\}$ is unconstrained by the data, which means that the unfolded spectrum content in these bins is a mere reflection of the prior, namely of the sampled region.
\label{fig:unfolded6}
}
\end{figure}

In Fig.~\ref{fig:1Dim6}, all 1-dimensional marginal distributions of $p(\T|\D)$ are shown.
The $T_t$ in bins $t=\{1,2,3,10,11,12,13,14\}$ are undetermined, because the migrations matrix (Fig.~\ref{fig:migrations6}) does not relate these $T_t$ values with any of the reconstructed data.  These are precisely the bins for which $\epsilon_t = 0$ (Fig.~\ref{fig:eff6}).
The resulting unfolded spectrum is in Fig.~\ref{fig:unfolded6}.  In bins that are unconstrained by the data, the unfolded spectrum is a mere reflection of the flat prior, namely, of the arbitrary sampled hyper-box.  So, the only thing known about these truth-level bins is the prior.  This is what one would expect from Bayes' theorem when the data are unrelated to the parameter of interest.  However, an inference is possible about $T_4$ and $T_5$ (Fig.~\ref{fig:1Dim6}), despite the lack of reconstructed bins corresponding directly to the 4$^{\rm th}$ and 5$^{\rm th}$ truth-level bin (Fig.~\ref{fig:truthAndReco6}); they are constrained by the data only through possible migrations into the $m$ bins where data exist.

\subsection{Unfolding a bump}
\label{sec:bump}

It is expectable that unfolding can make a known bump, such as the $Z$ boson mass peak, sharper, as it is before smearing.  In this case, it is known that the bump is generated, so, $\tilde{\T} = \hat{\T}$, and $\M$ contains this bump information.  What if there is an exotic process, unknown to the MC and to $\M$?  Will unfolding then make the bump sharper, or will it conceal it?

The scenario of unfolding an expected bump is investigated in Sec.~\ref{sec:bumpKnown}, and unfolding an unknown bump in Sec.~\ref{sec:bumpUnknown}.

As seen in the inset of Fig.~\ref{fig:unfolded4c}, the precision of unfolding deteriorates quickly in later bins.  This happens partly because the Poisson distribution is wider for smaller mean values (see Sec.~\ref{sec:example1} and \ref{sec:example2}), and partly because of larger migrations, which magnify the impact of these statistical fluctuations (Sec.~\ref{sec:example3}).  To avoid such complications, and to focus on the bump, some changes are made in Sec.~\ref{sec:bumpKnown} and \ref{sec:bumpUnknown}:
\begin{enumerate}[i)]
\item The spectrum is not steeply falling, but constant, with the addition of a bump.
\item The smearing is not taken from Eq.~\ref{eq:sigma}, but a constant $\sigma$ is used, independent of $m$.
\item Instead of the 14 unequal bins of Eq.~\ref{eq:bins}, $N_t=30$ bins are used, to have enough bins to describe the bump.  The $m$ bins span from 500 to 3500, in steps of 100.
\end{enumerate}

\subsubsection{FBU with a known bump}
\label{sec:bumpKnown}

\begin{figure}[H]
  \centering
  \begin{tabular}{ccc}
    \includegraphics[height=0.23\columnwidth]{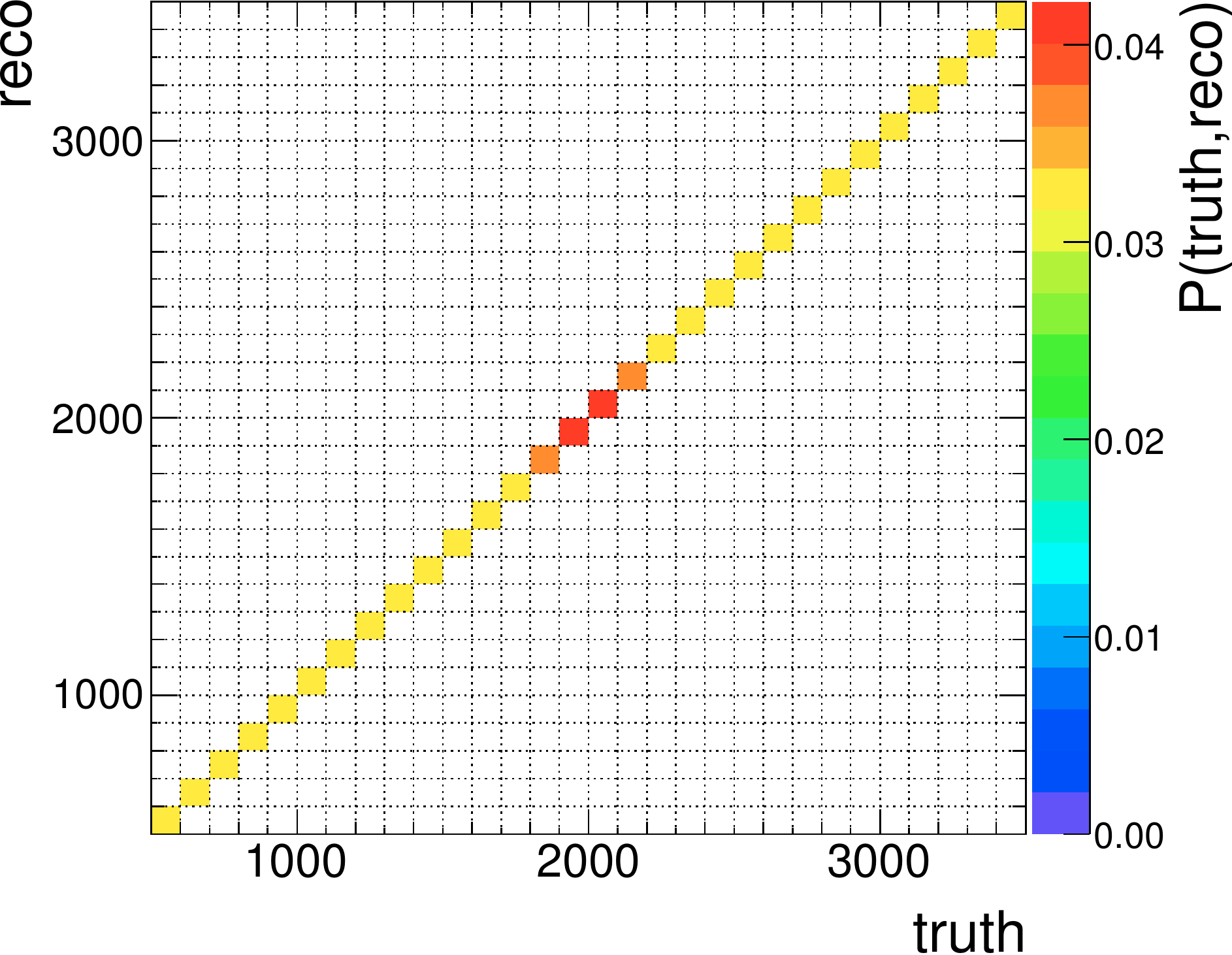} &
    \includegraphics[height=0.23\columnwidth]{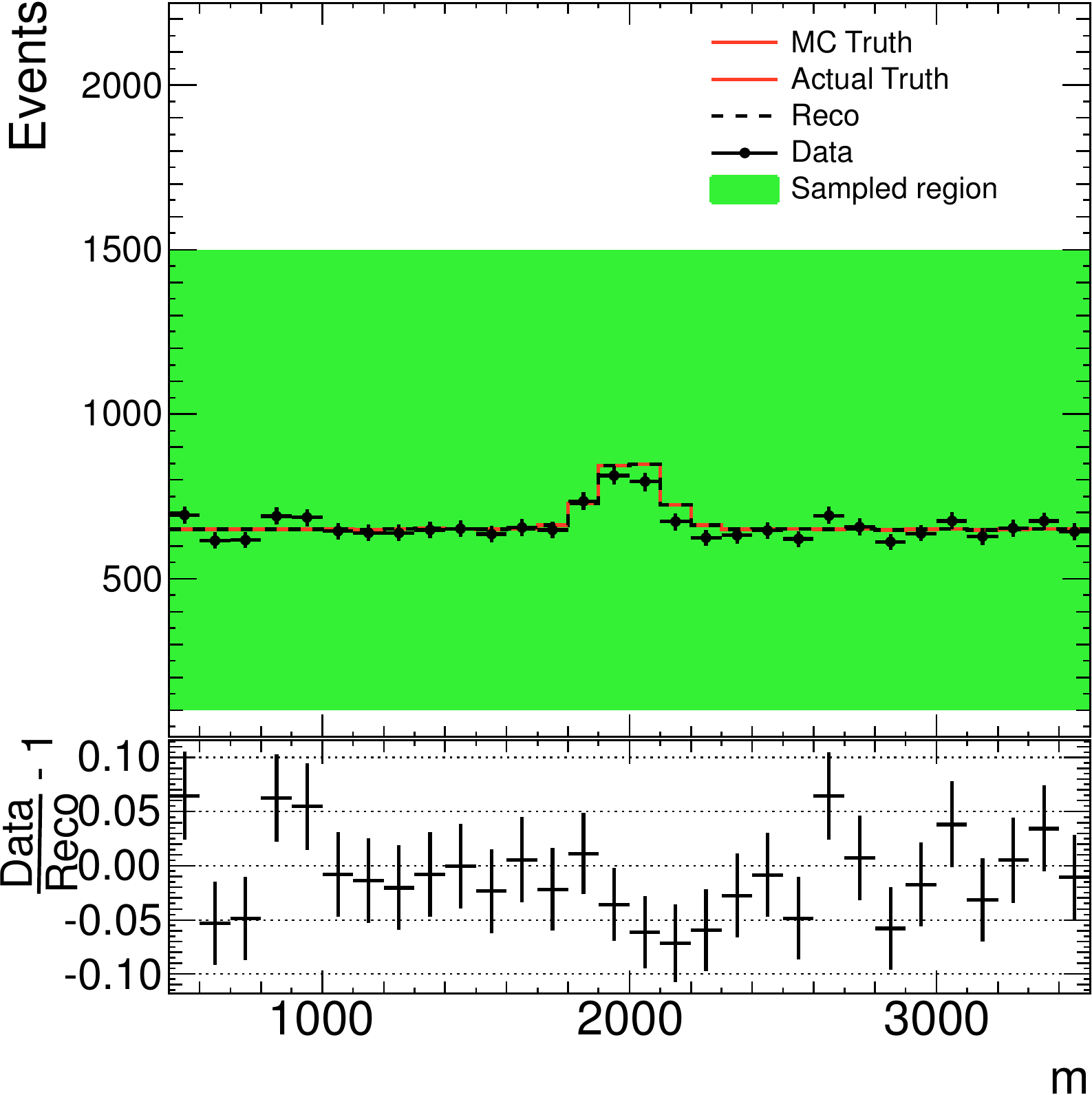} &
    \includegraphics[height=0.23\columnwidth]{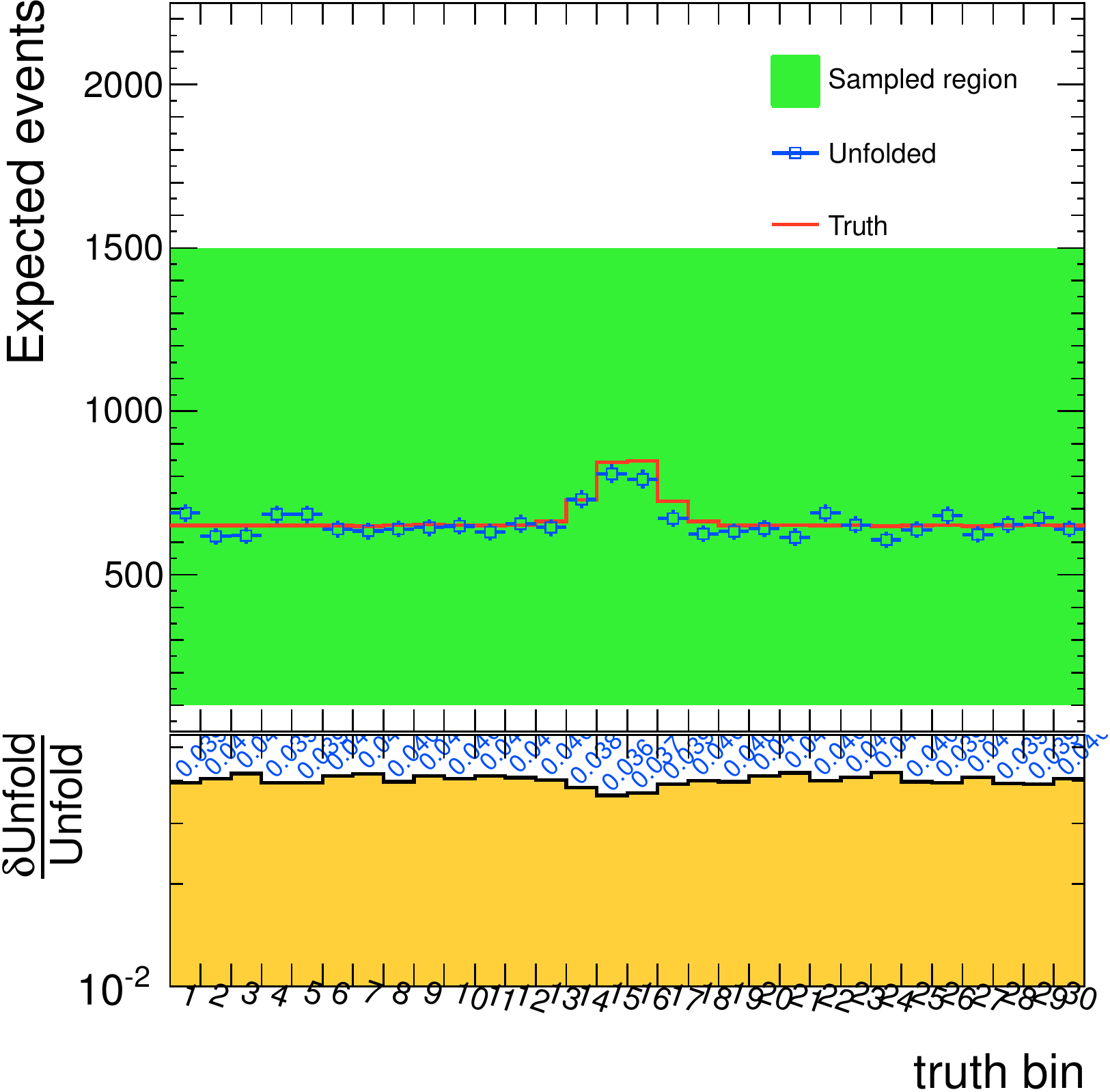} \\
    \includegraphics[height=0.23\columnwidth]{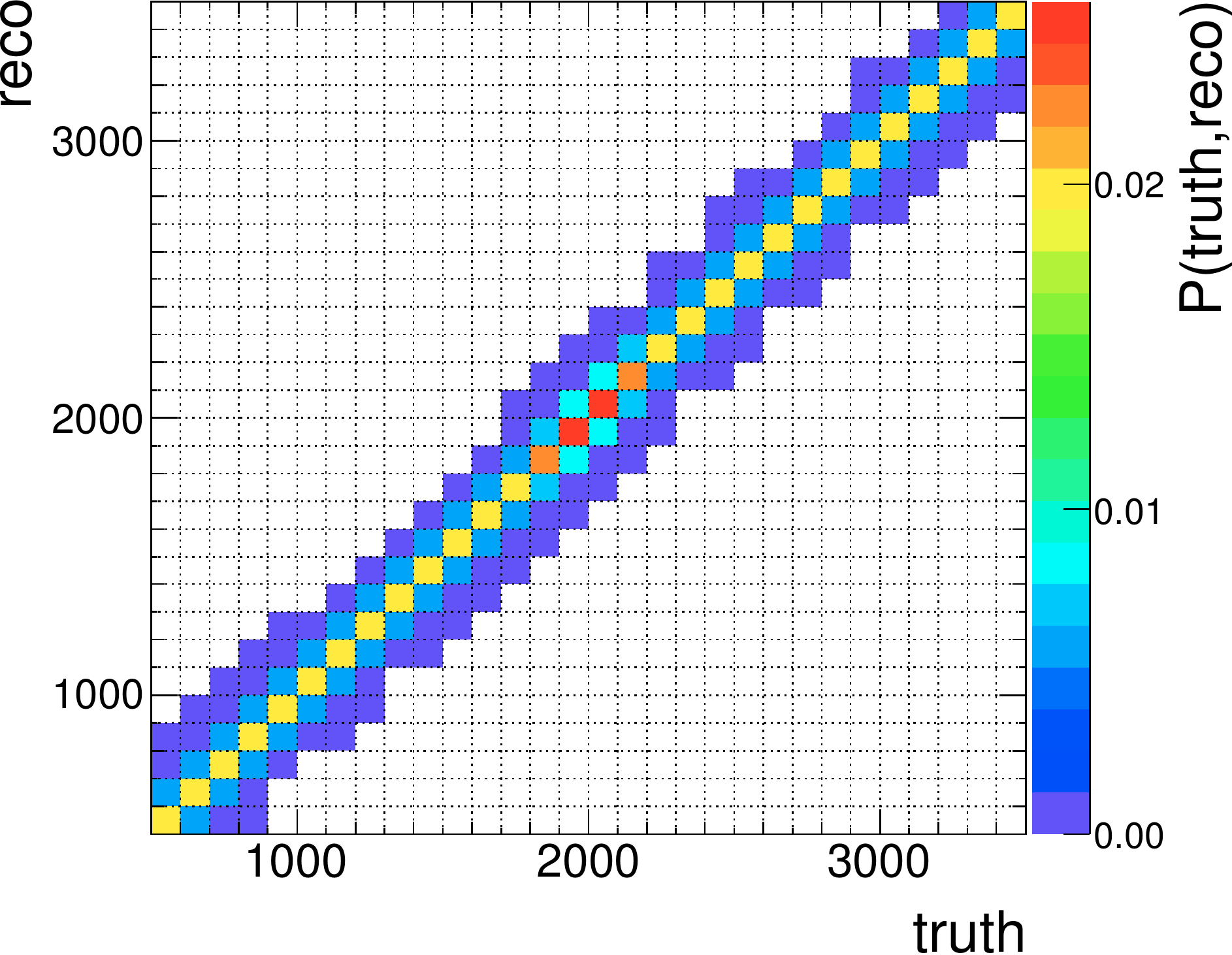} &
    \includegraphics[height=0.23\columnwidth]{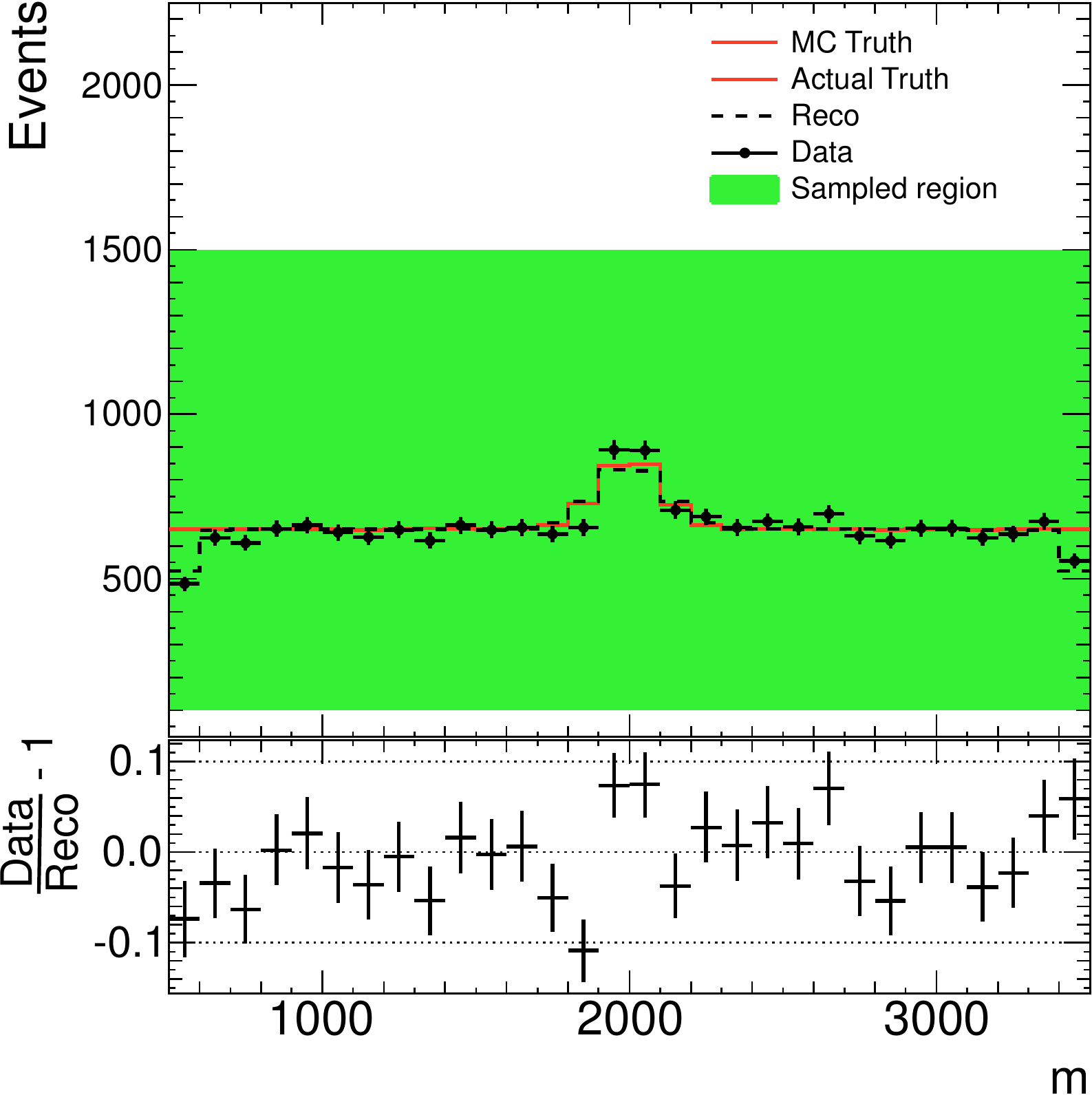} &
    \includegraphics[height=0.23\columnwidth]{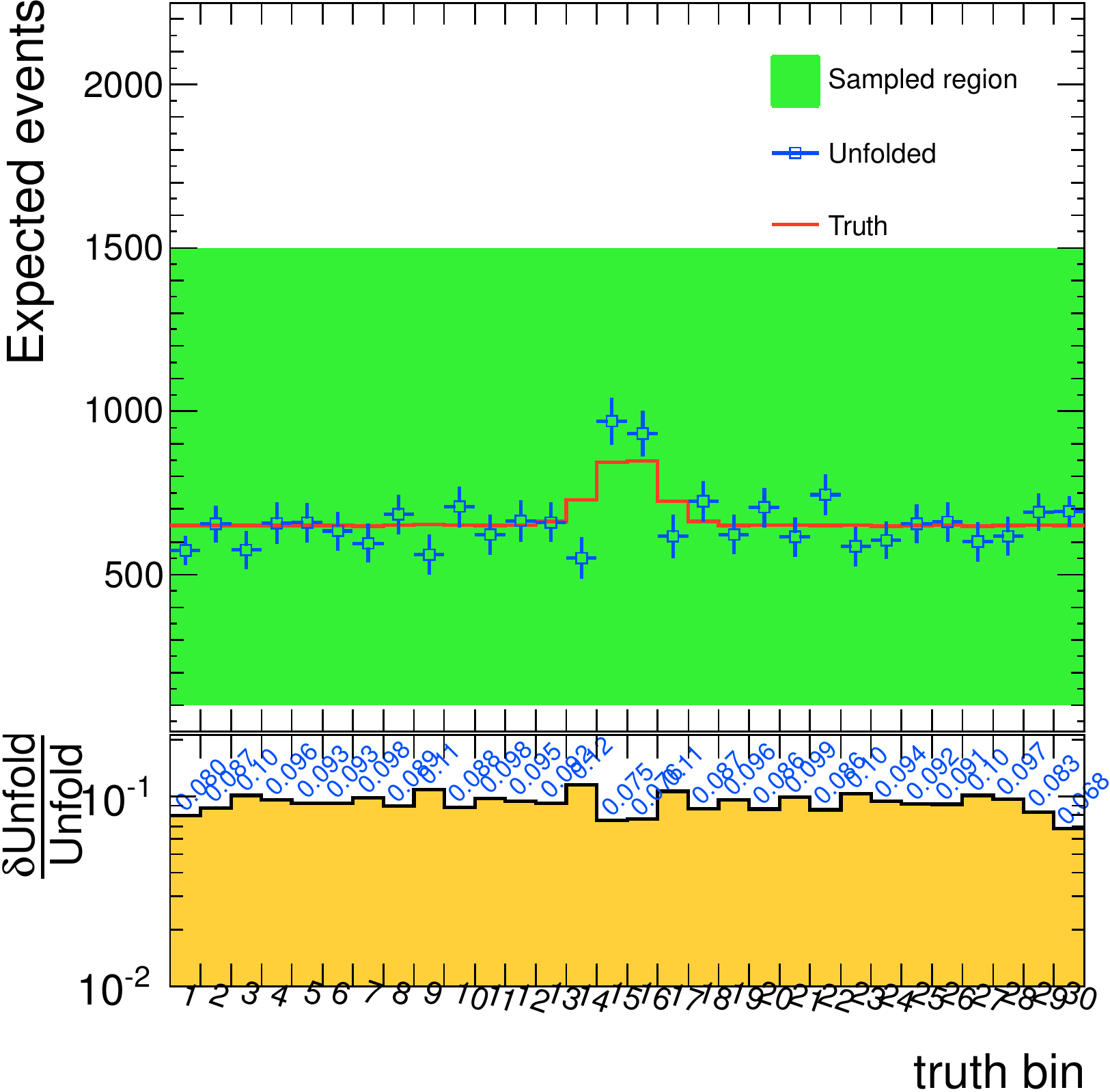} \\
    \includegraphics[height=0.23\columnwidth]{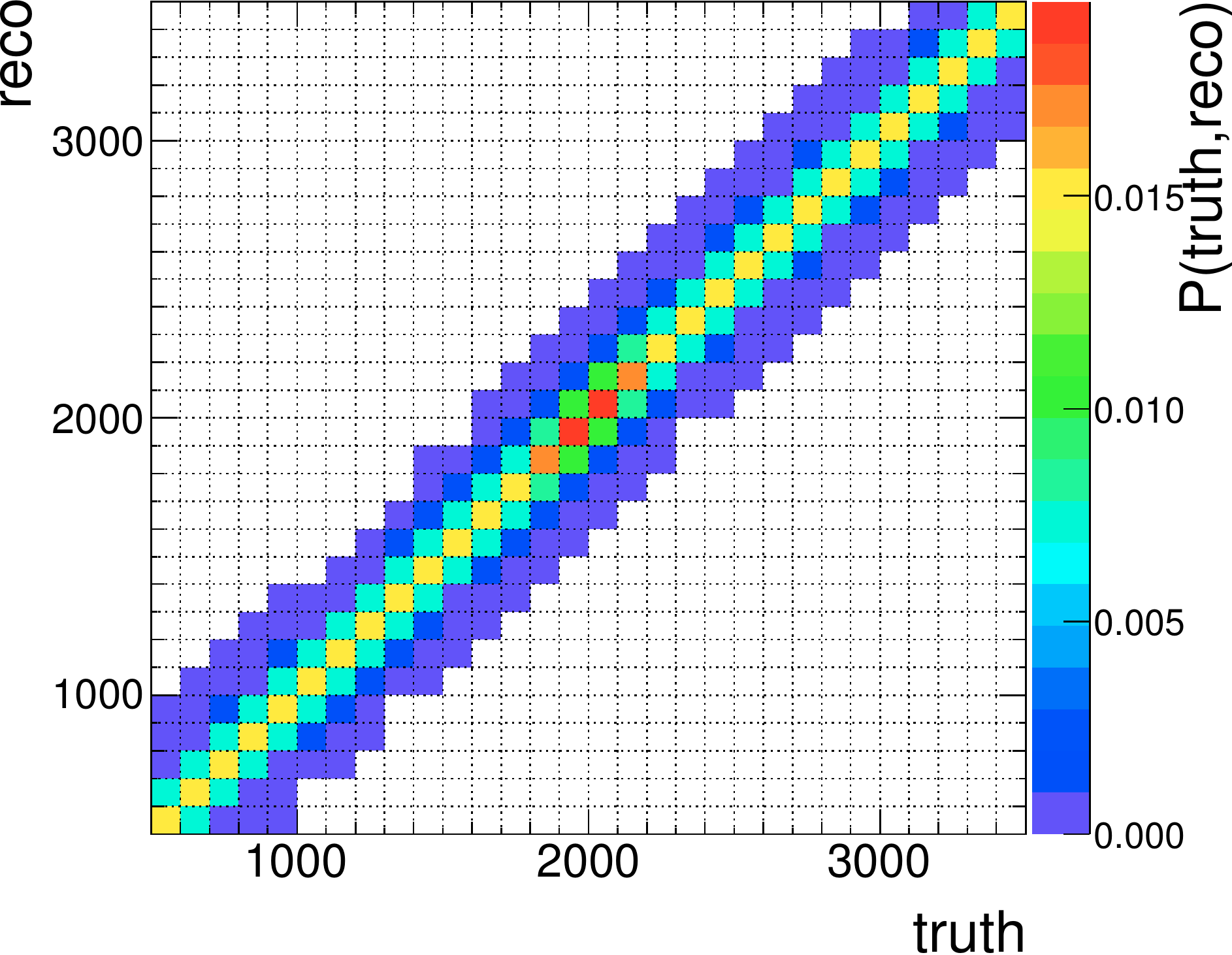} &
    \includegraphics[height=0.23\columnwidth]{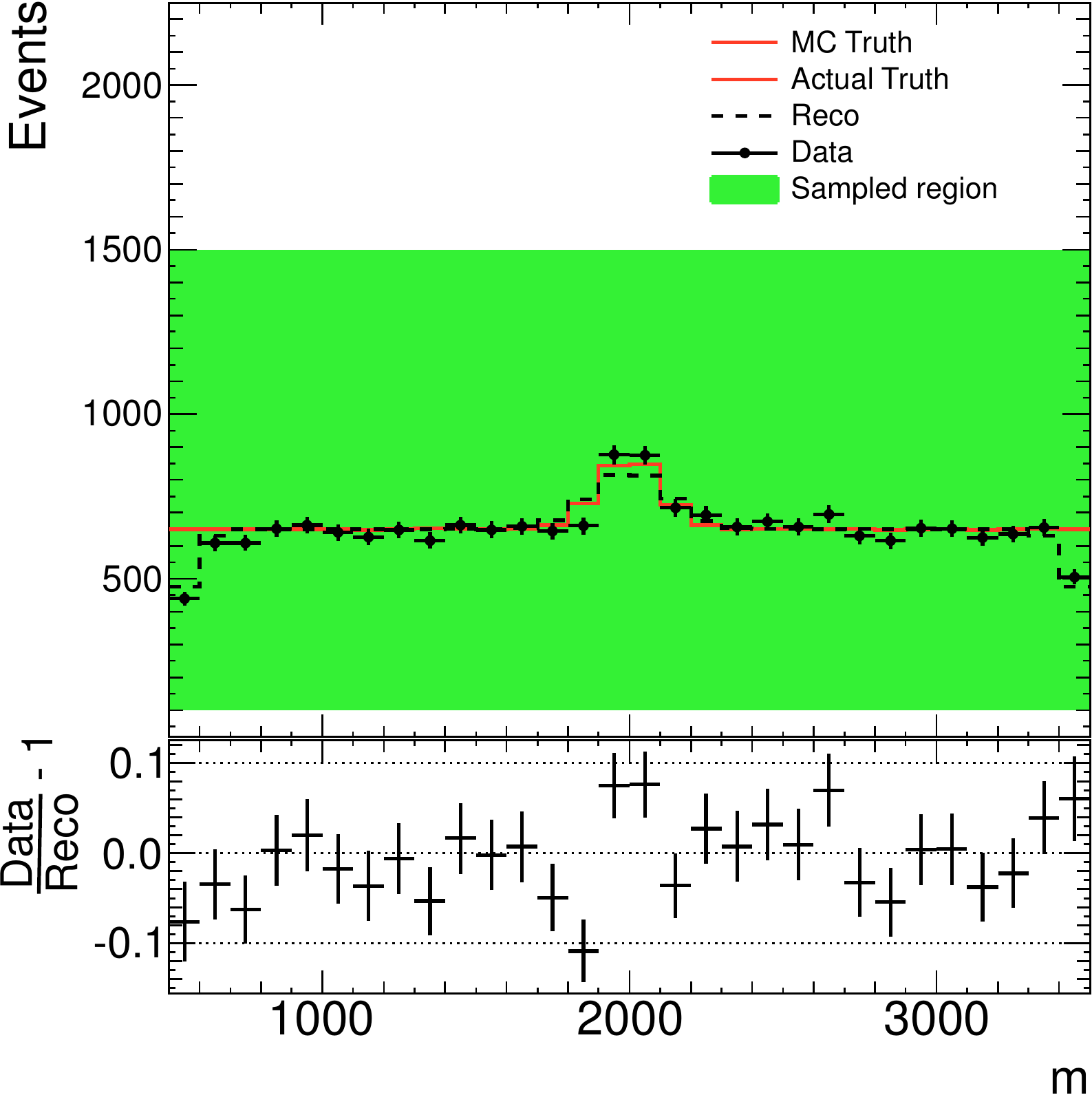} &
    \includegraphics[height=0.23\columnwidth]{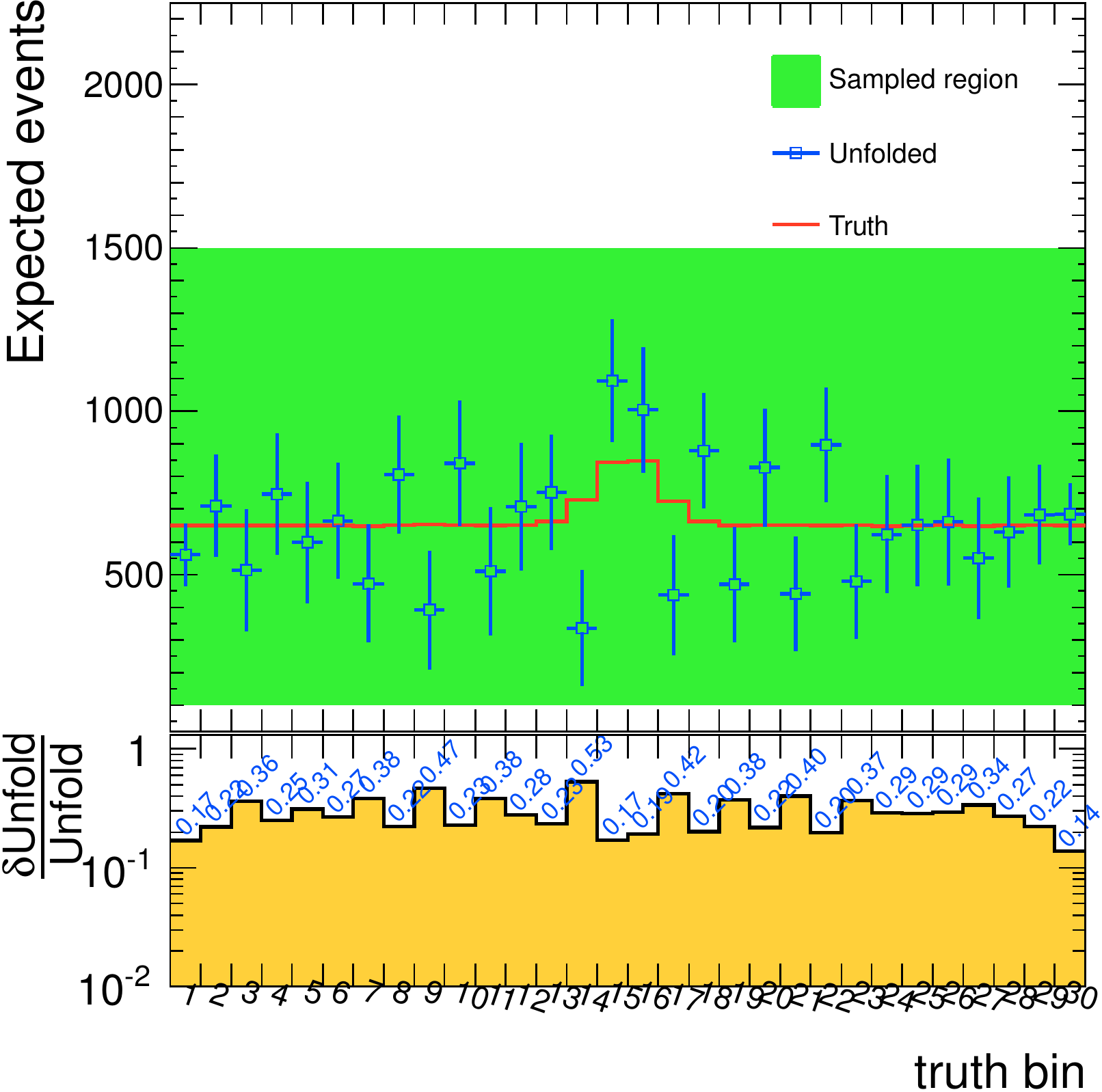} \\
    \includegraphics[height=0.23\columnwidth]{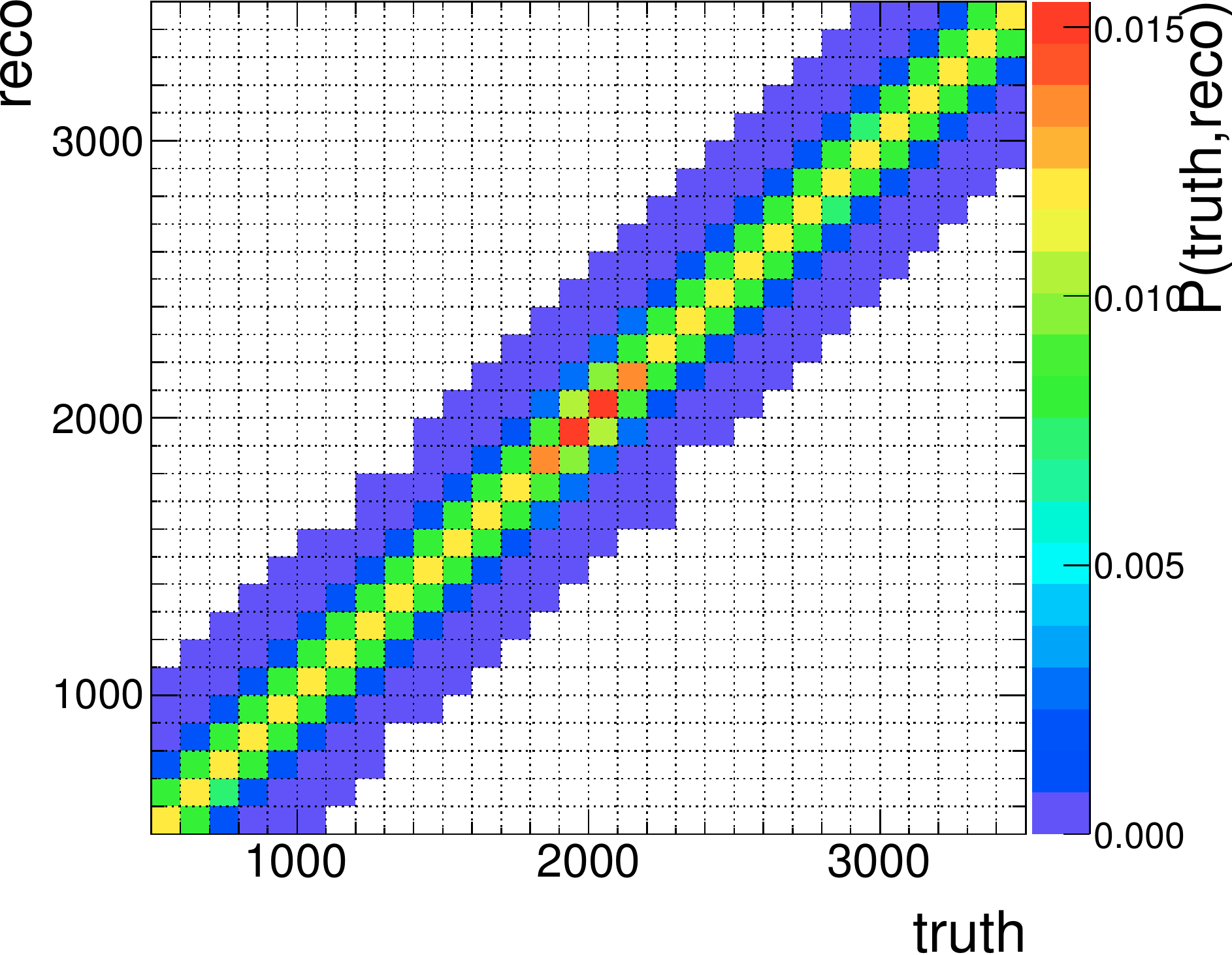} &
    \includegraphics[height=0.23\columnwidth]{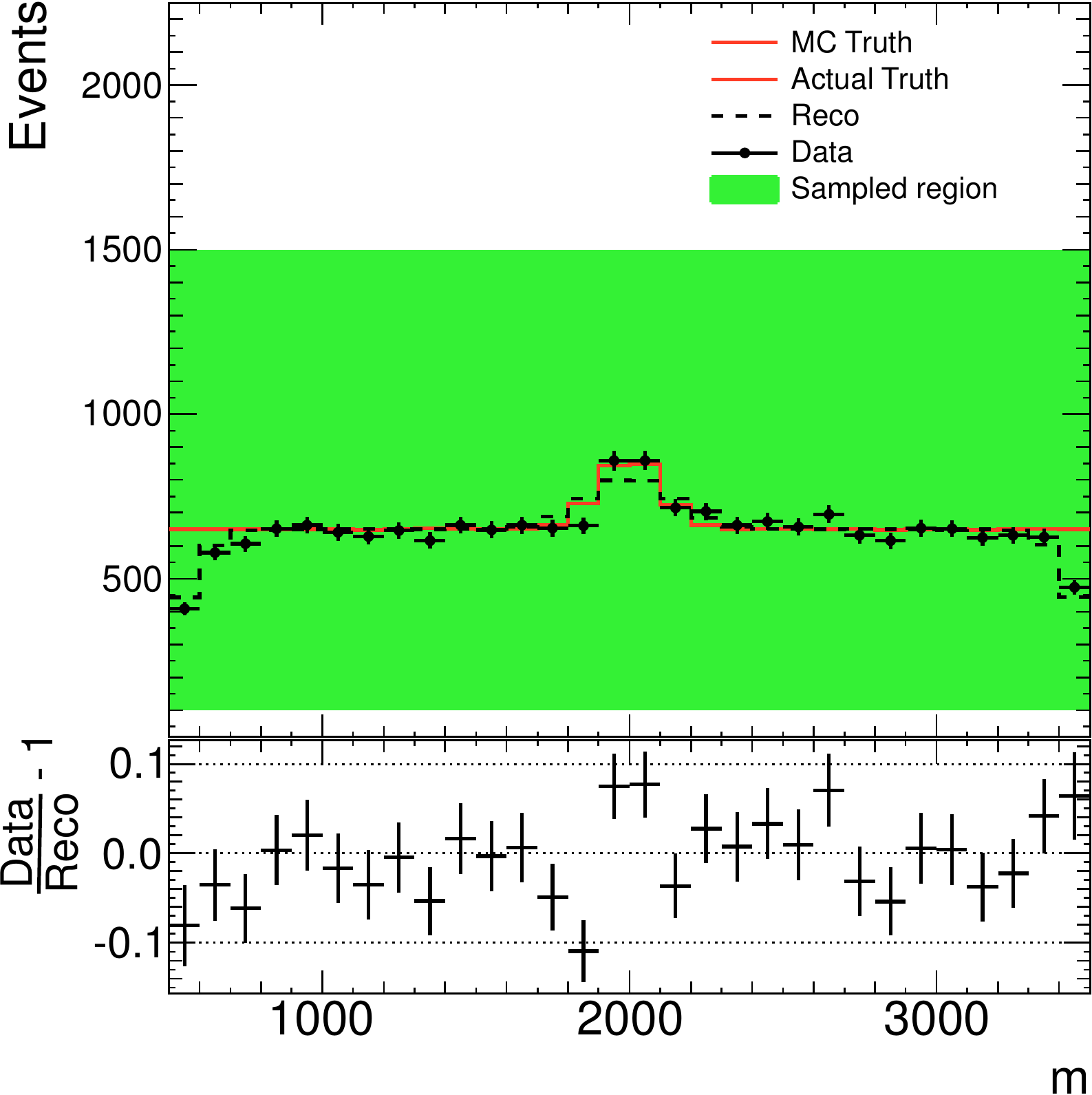} &
    \includegraphics[height=0.23\columnwidth]{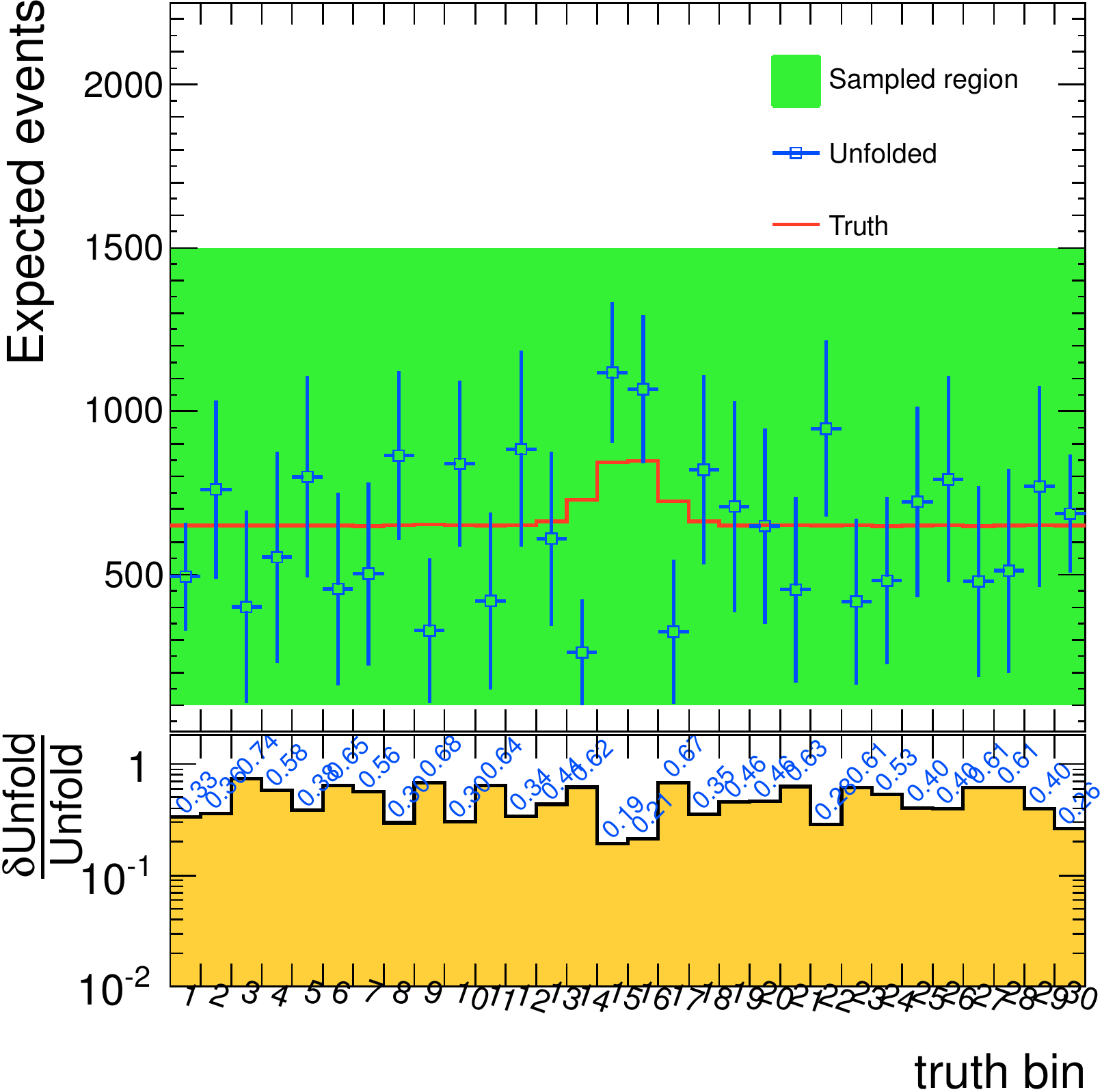} \\
    \includegraphics[height=0.23\columnwidth]{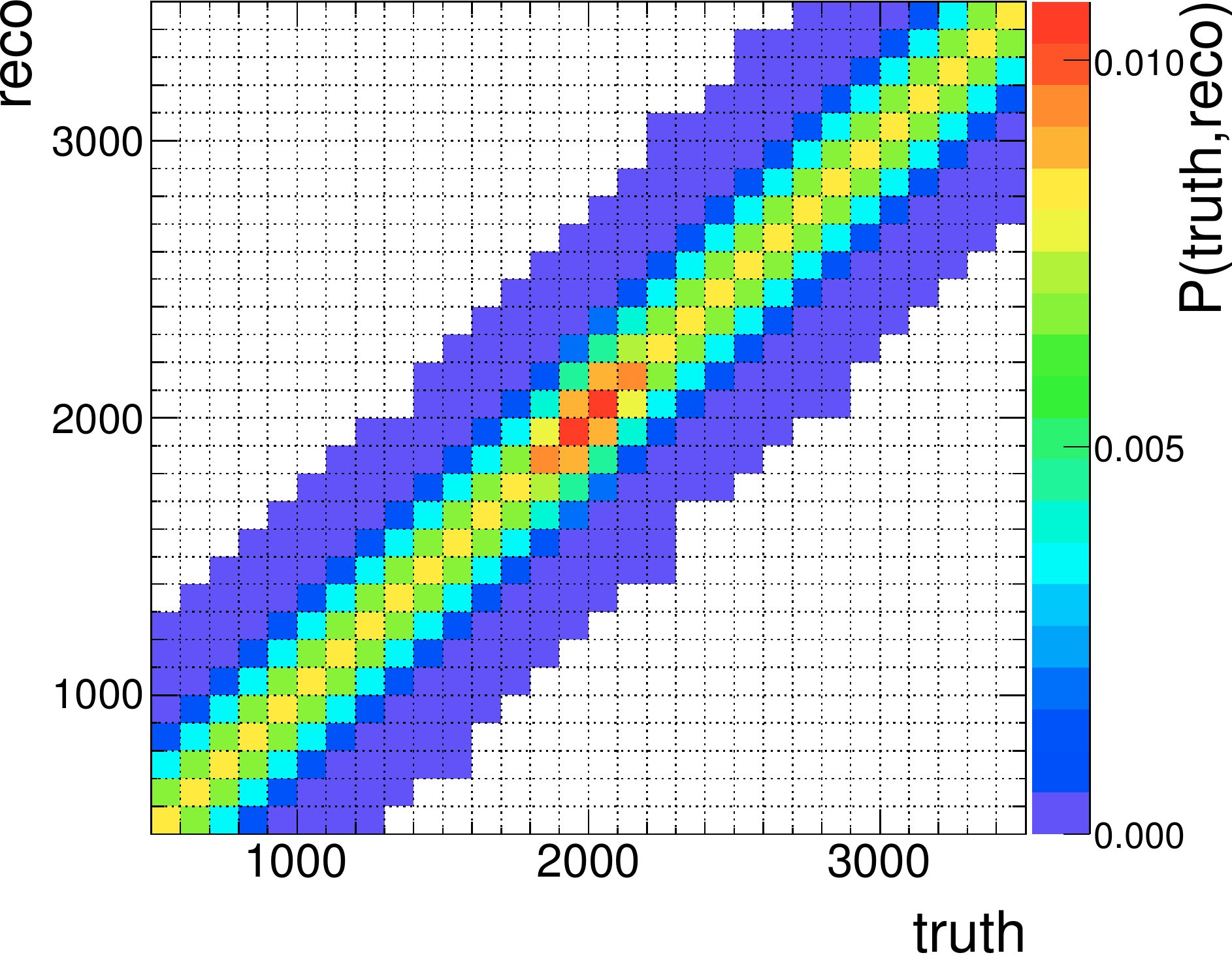} &
    \includegraphics[height=0.23\columnwidth]{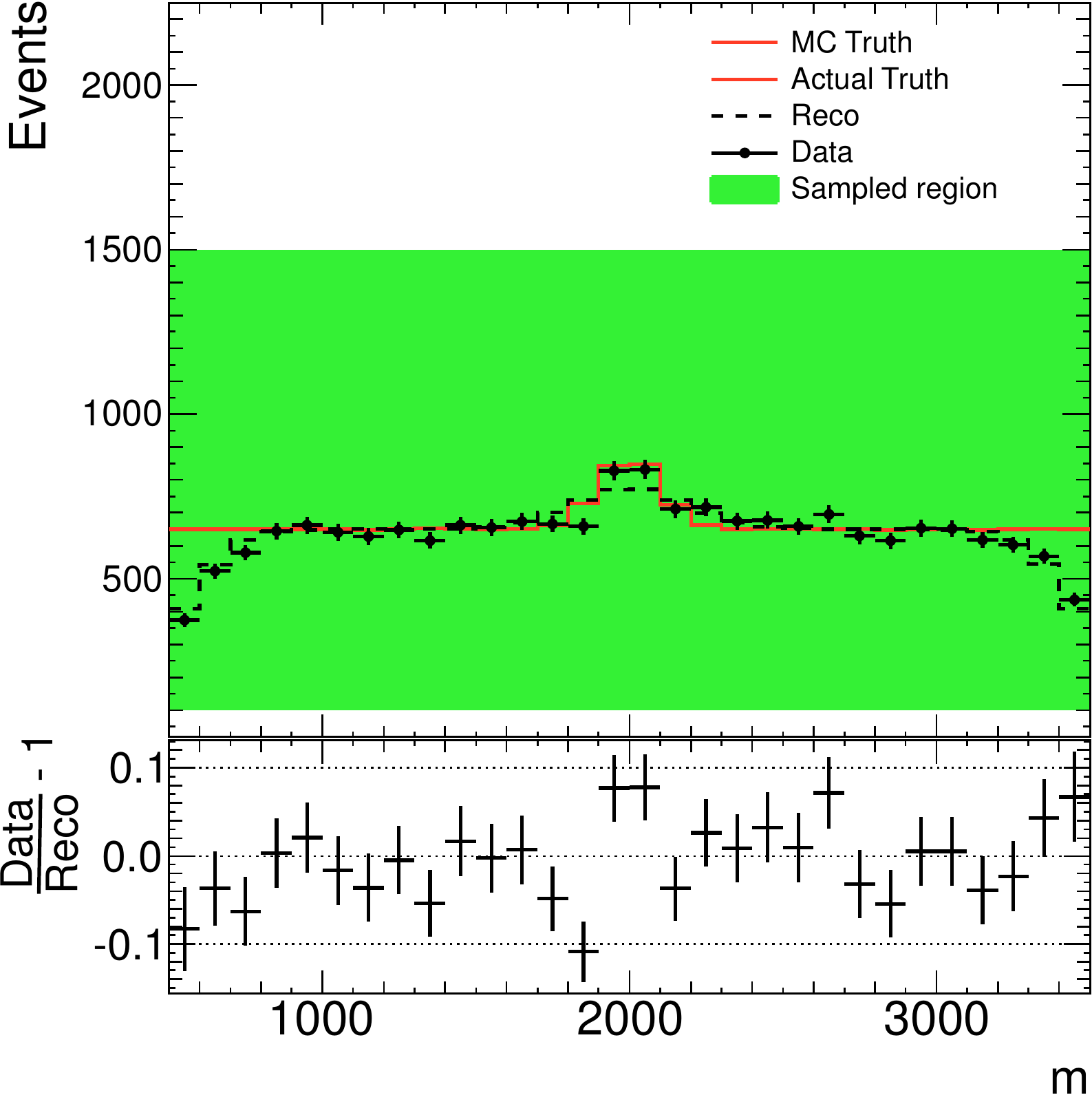} &
    \includegraphics[height=0.23\columnwidth]{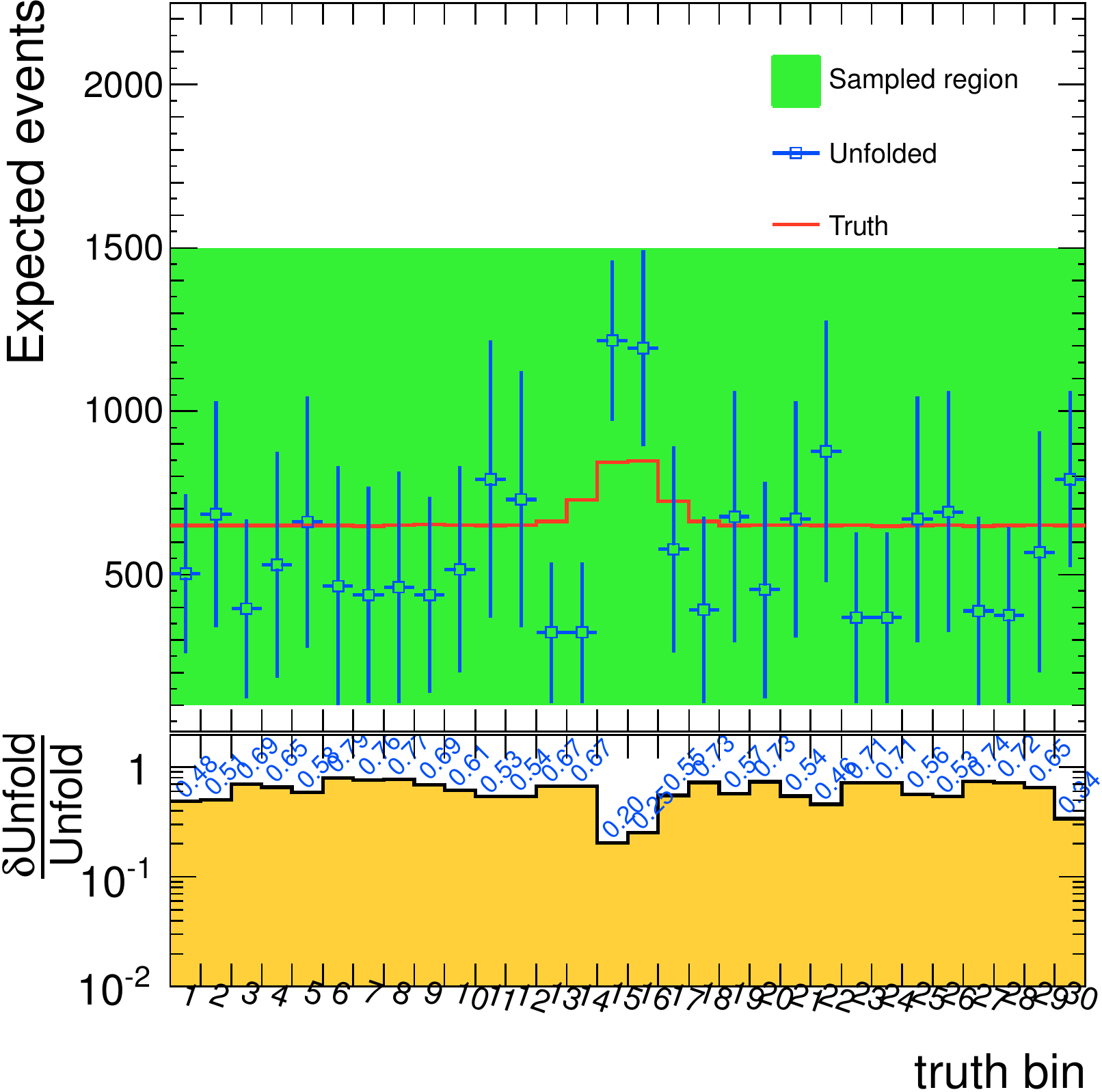} 
 \end{tabular}
\caption{The migrations matrix (left); data, reco and truth spectra (middle), where ``Actual truth'' is $\hat{\T}$, ``MC truth'' is $\tilde{\T}$, and in this case $\tilde{\T} = \hat{\T}$; and truth and unfolded spectra (right).  Each row corresponds to $\sigma=\{0,50,75,100,150\}$.  Details in Sec.~\ref{sec:bumpKnown}.
\label{fig:bumpKnown}}
\end{figure}

Initially, no smearing is assumed ($\sigma=0$), which results in the diagonal $\M$ in the first row of Fig~\ref{fig:bumpKnown}.  The truth-level spectrum contains a profound bump: a Gaussian of mean 2000 and RMS 100.  This bump is reflected in $\M$.   The reconstructed spectrum is identical to the truth-level one, and that along with the observed data and the sampled hyper-box are shown in the middle first row of Fig.~\ref{fig:bumpKnown}.  FBU is performed, without any regularization.  In the interest of speed, MCMC sampling is used, without volume reduction.  The resulting 1-dimensional distributions are very regularly shaped, and smooth, which suggests that this approximation is satisfactory (see Sec.~\ref{sec:example5}).  

The unfolded spectrum is compared to the truth-level spectrum on the right of the first row of Fig.~\ref{fig:bumpKnown}.  The bottom inset shows the relative uncertainty of the unfolded spectrum, which is equal to the data fluctuations.  
The unfolded spectrum is effectively identical to the data, and its relative uncertainty in bin $t$ is about equal to $\sqrt{D_t}/D_t$, which in the sidebands of the bump is about $\frac{1}{\sqrt{650}} \simeq 4\%$, and in the bump, where statistics are greater, it is smaller.
This result is consistent with Sec.~\ref{sec:example1}.

The same FBU procedure is repeated for gradually increasing smearing.  The rows of Fig.~\ref{fig:bumpKnown} correspond to $\sigma=\{0,50,75,100,150\}$ respectively.
The relative uncertainty of the unfolded spectrum ($\frac{|U_t^\ulcorner-U_t^\urcorner|}{U_t^\ulcorner+U_t^\urcorner}$) increases from 4\% to roughly 9\%, 20\%, 40\% and 60\%, with wild fluctuations covarying with $U_t$.   With $\sigma=50$ and uncertainty $\sim$9\% (2nd row of Fig.~\ref{fig:bumpKnown}), the bump is still visible, and it seems slightly sharper than the reco spectrum, though, the error bars of the unfolded spectrum are large enough to make it also consistent with being less sharp than the reco spectrum. So, it is not clear that in this example unfolding made the feature sharper.  If it made it a little sharper, it also made the error bars big enough to cancel this benefit.
When $\sigma=50$, the smearing is not very strong, so, it may be thought that unfolding could demonstrate its benefits more clearly when smearing is stronger.  Unfortunately, with $\sigma \ge 75$ (3rd, 4th, 5th row in Fig.~\ref{fig:bumpKnown}), the bump is hardly discernible in the unfolded answer, because the error bars are about as large as the bump itself.

So, it seems that, if FBU is making the bump in Fig.~\ref{fig:bumpKnown} more sharp, it is simultaneously increasing the uncertainty of the unfolded spectrum so much that the feature is less obvious.  

To see if this behavior is different for sharper truth-level bumps, the previous Gaussian is replaced with one with mean 2050 and RMS 5, so, at truth-level it populates just one bin.  The results are summarized in Fig.~\ref{fig:bumpKnown2}.   Just as in Fig.~\ref{fig:bumpKnown}, smearing with $\sigma \gtrsim 75$ is enough to make the error bars about as big as the feature itself, when no regularization is used.  

\subsubsection{FBU with an unexpected bump} 
\label{sec:bumpUnknown}

\begin{figure}[H]
  \centering
  \begin{tabular}{ccc}
    \includegraphics[height=0.23\columnwidth]{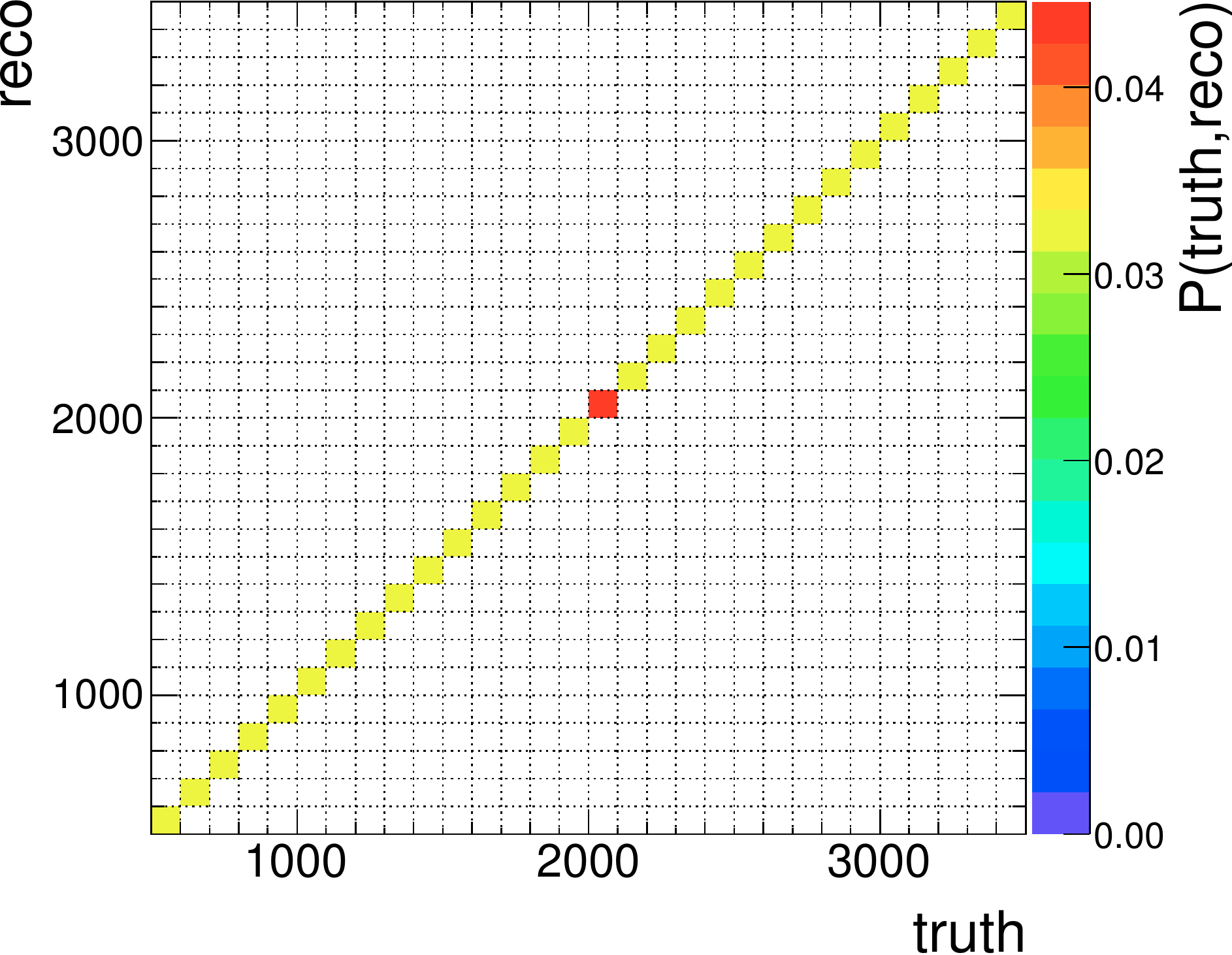} &
    \includegraphics[height=0.23\columnwidth]{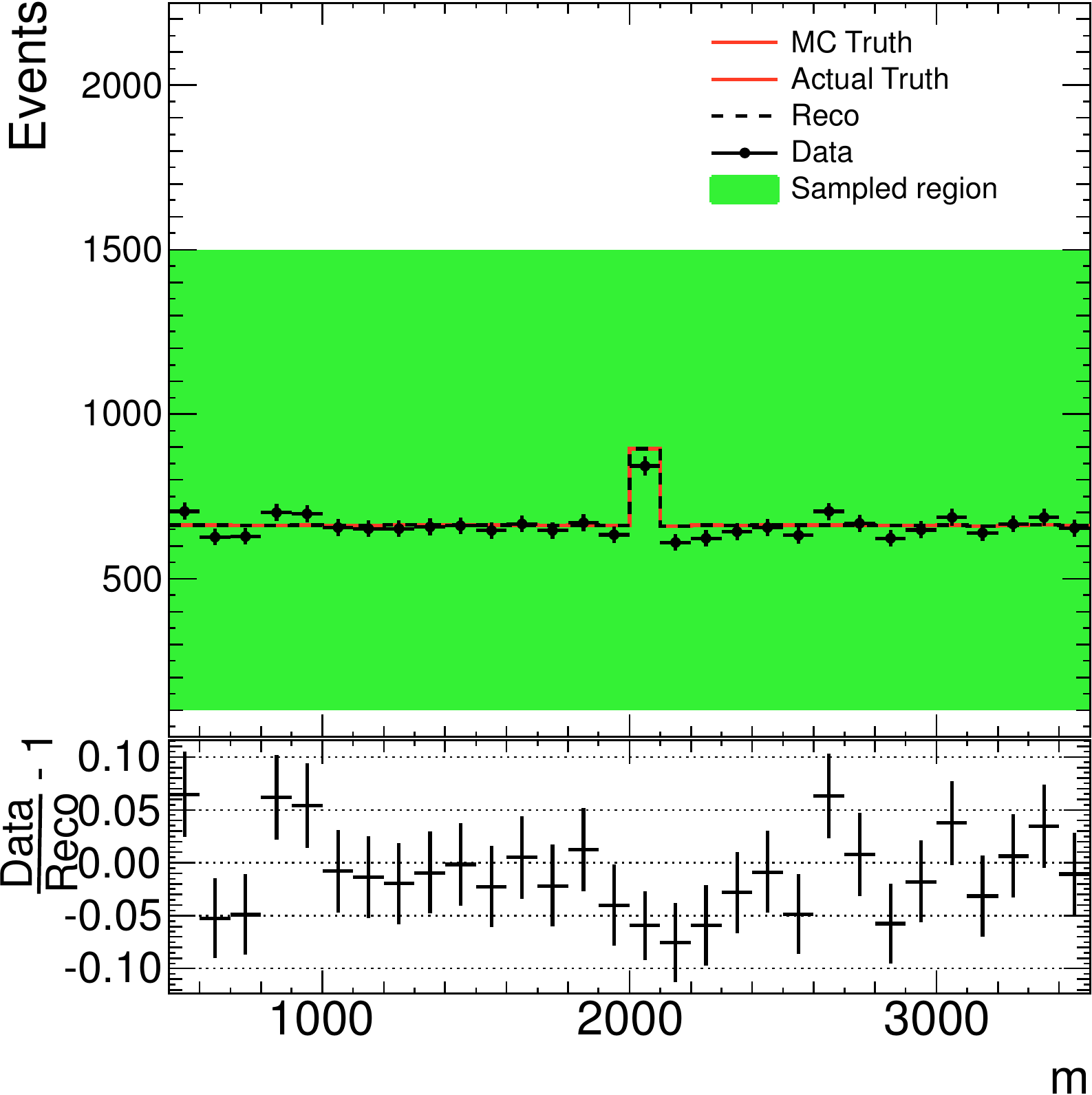} &
    \includegraphics[height=0.23\columnwidth]{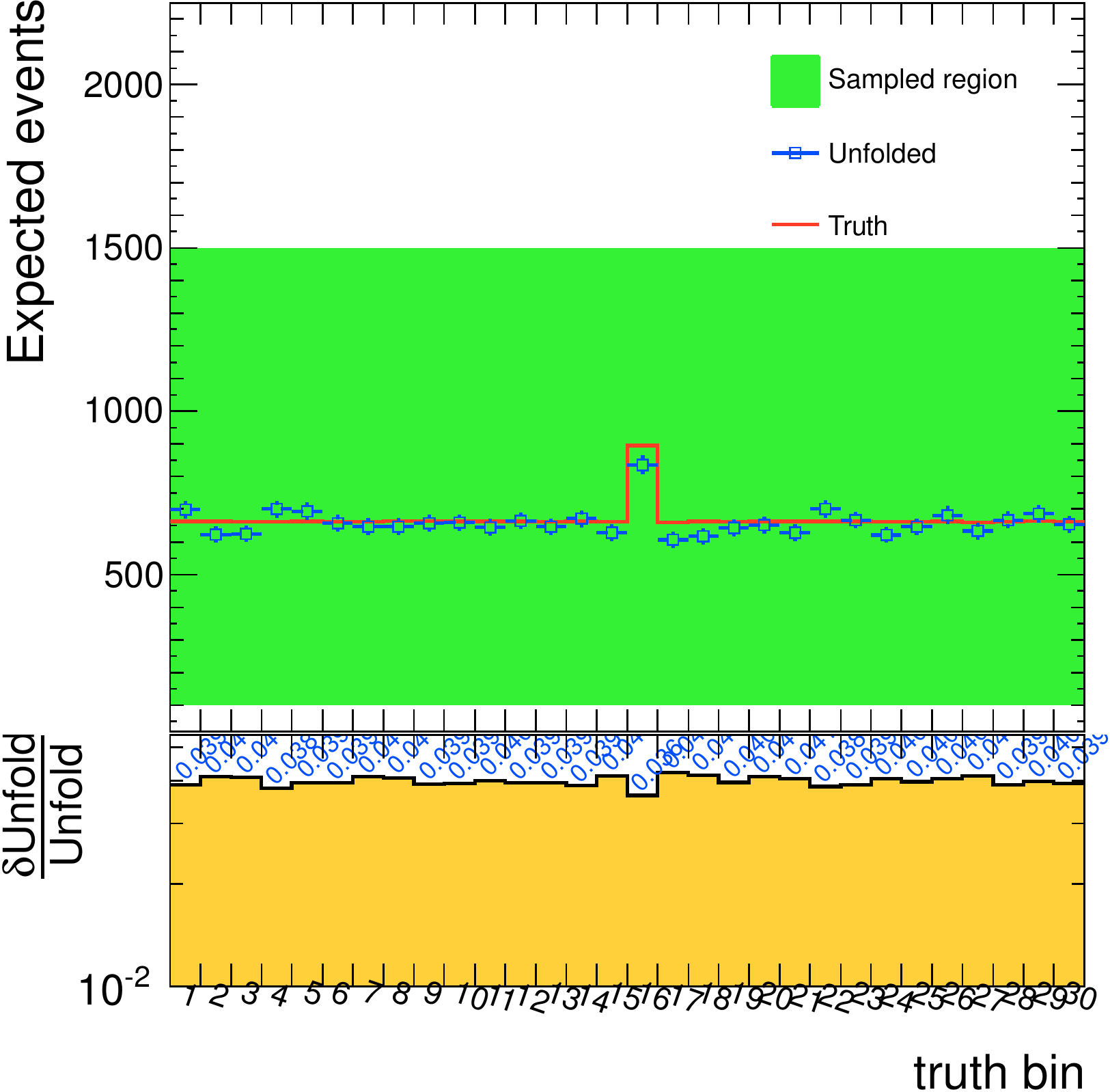} \\
    \includegraphics[height=0.23\columnwidth]{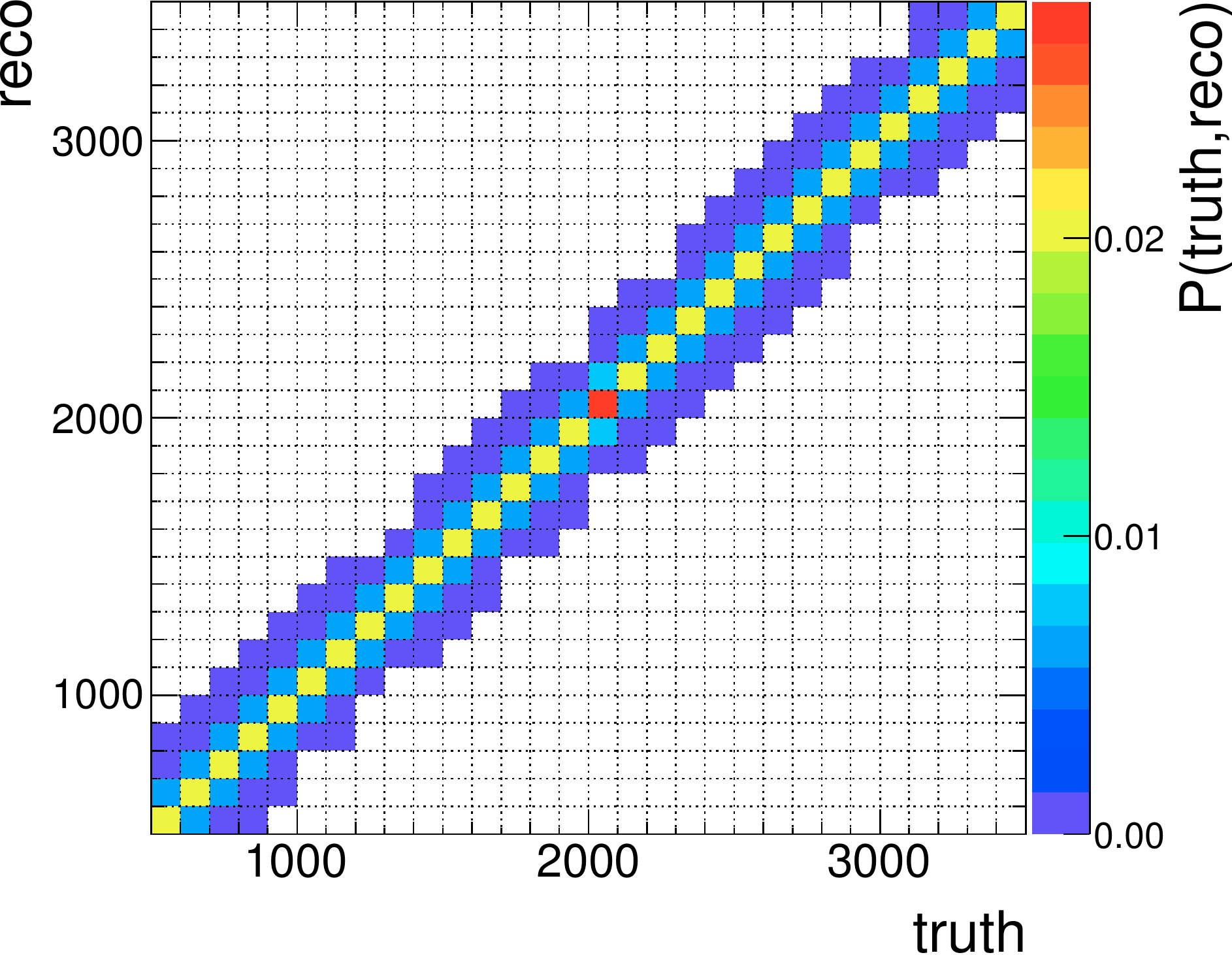} &
    \includegraphics[height=0.23\columnwidth]{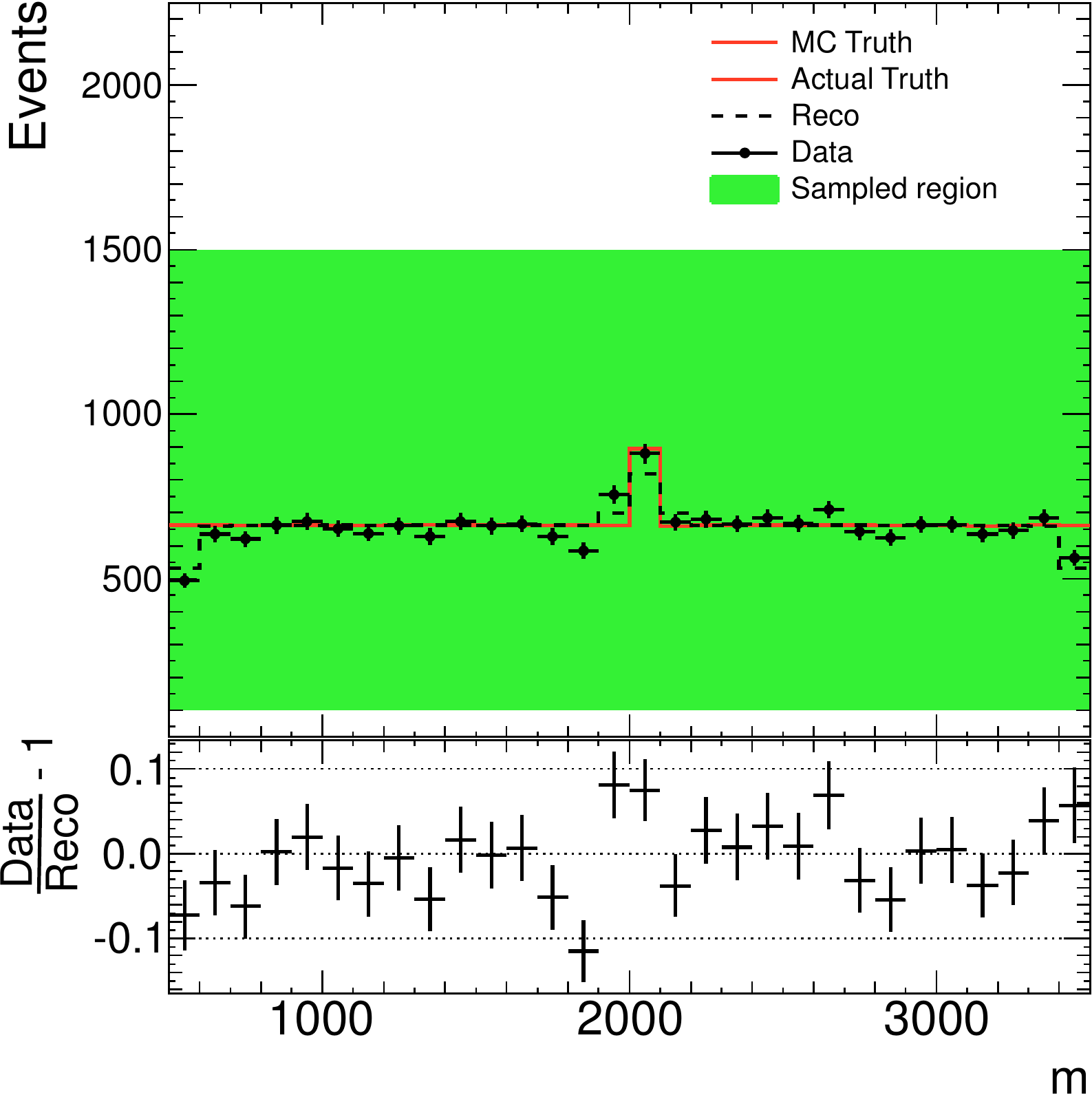} &
    \includegraphics[height=0.23\columnwidth]{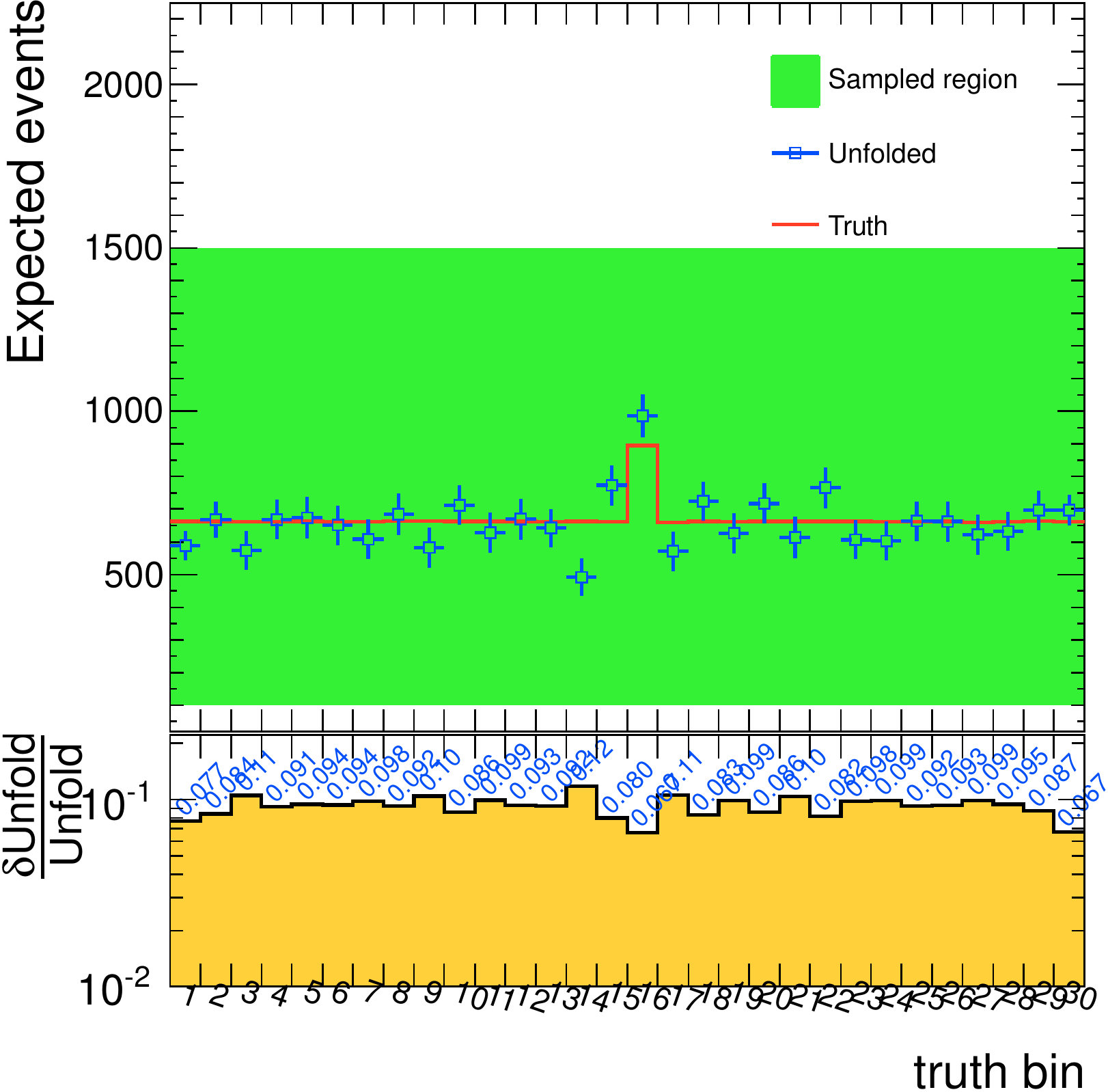} \\
    \includegraphics[height=0.23\columnwidth]{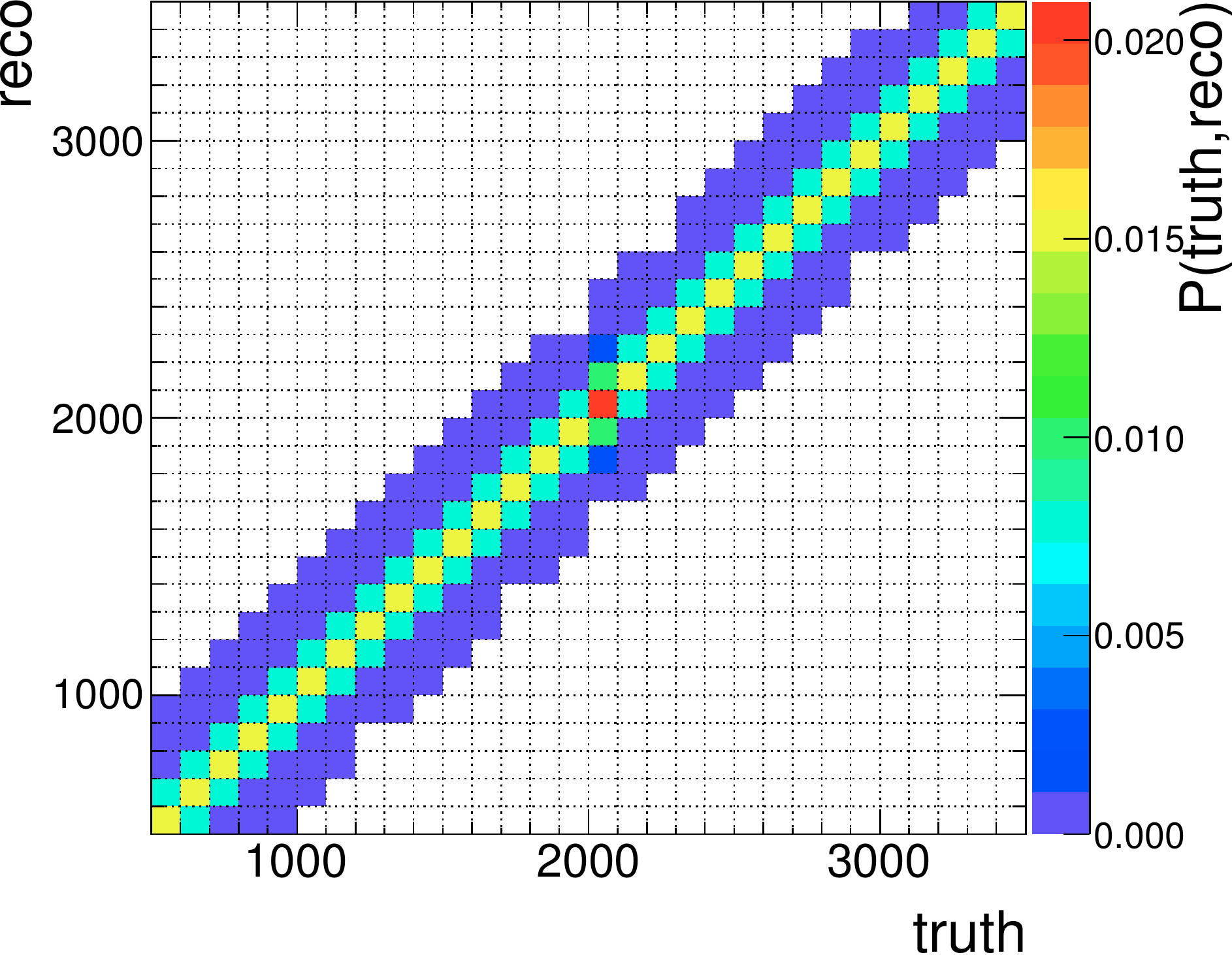} &
    \includegraphics[height=0.23\columnwidth]{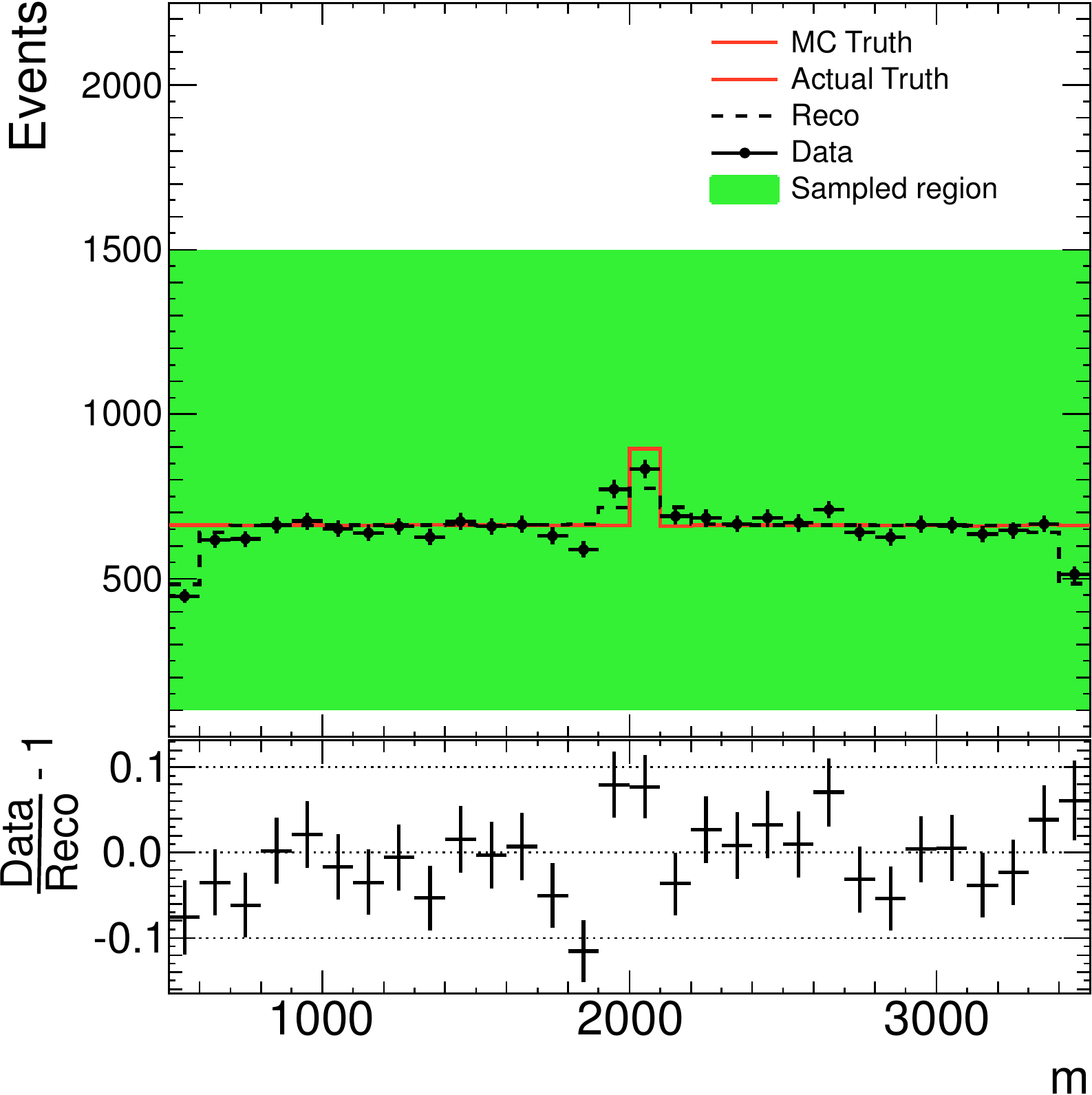} &
    \includegraphics[height=0.23\columnwidth]{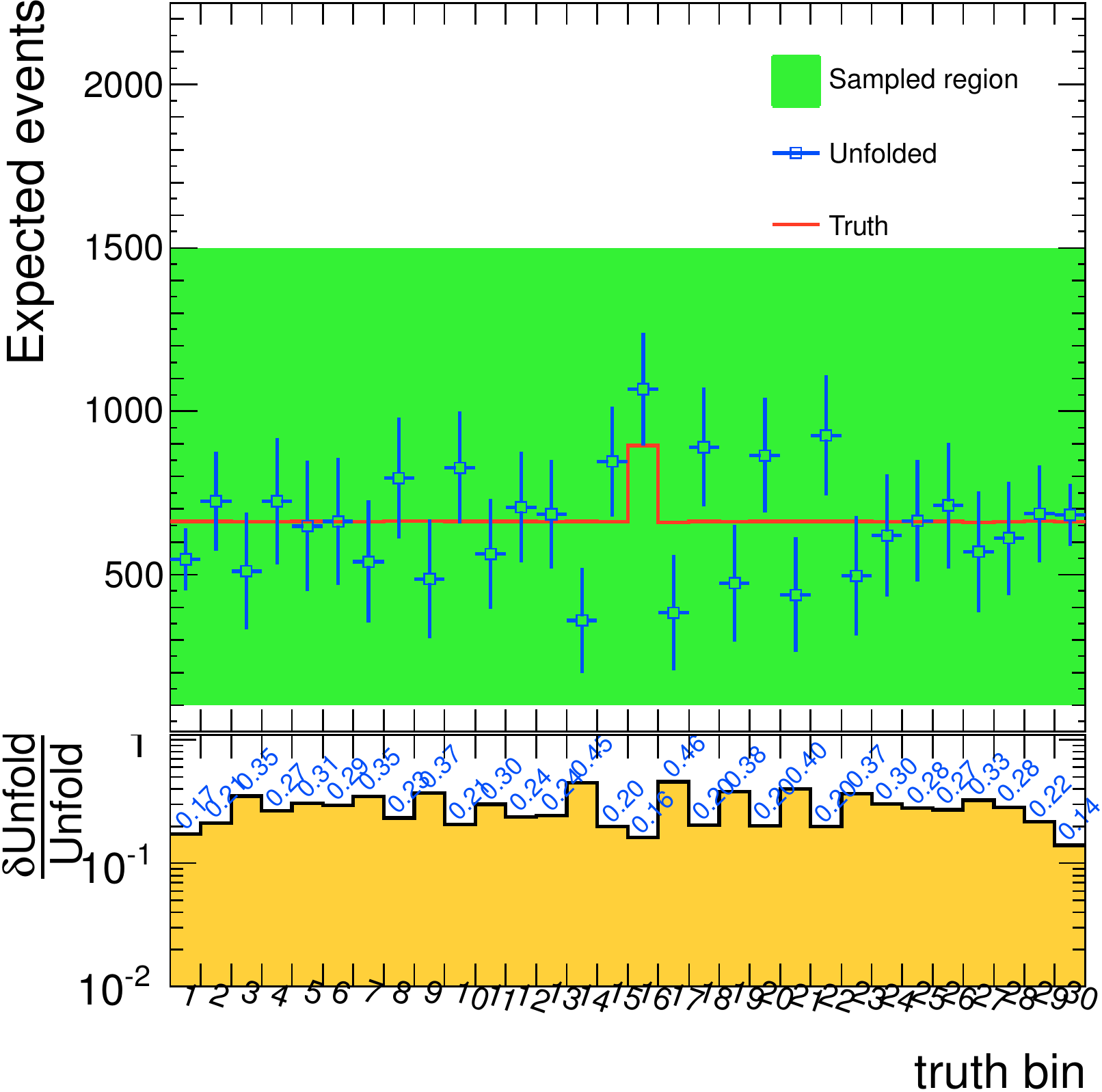} \\
    \includegraphics[height=0.23\columnwidth]{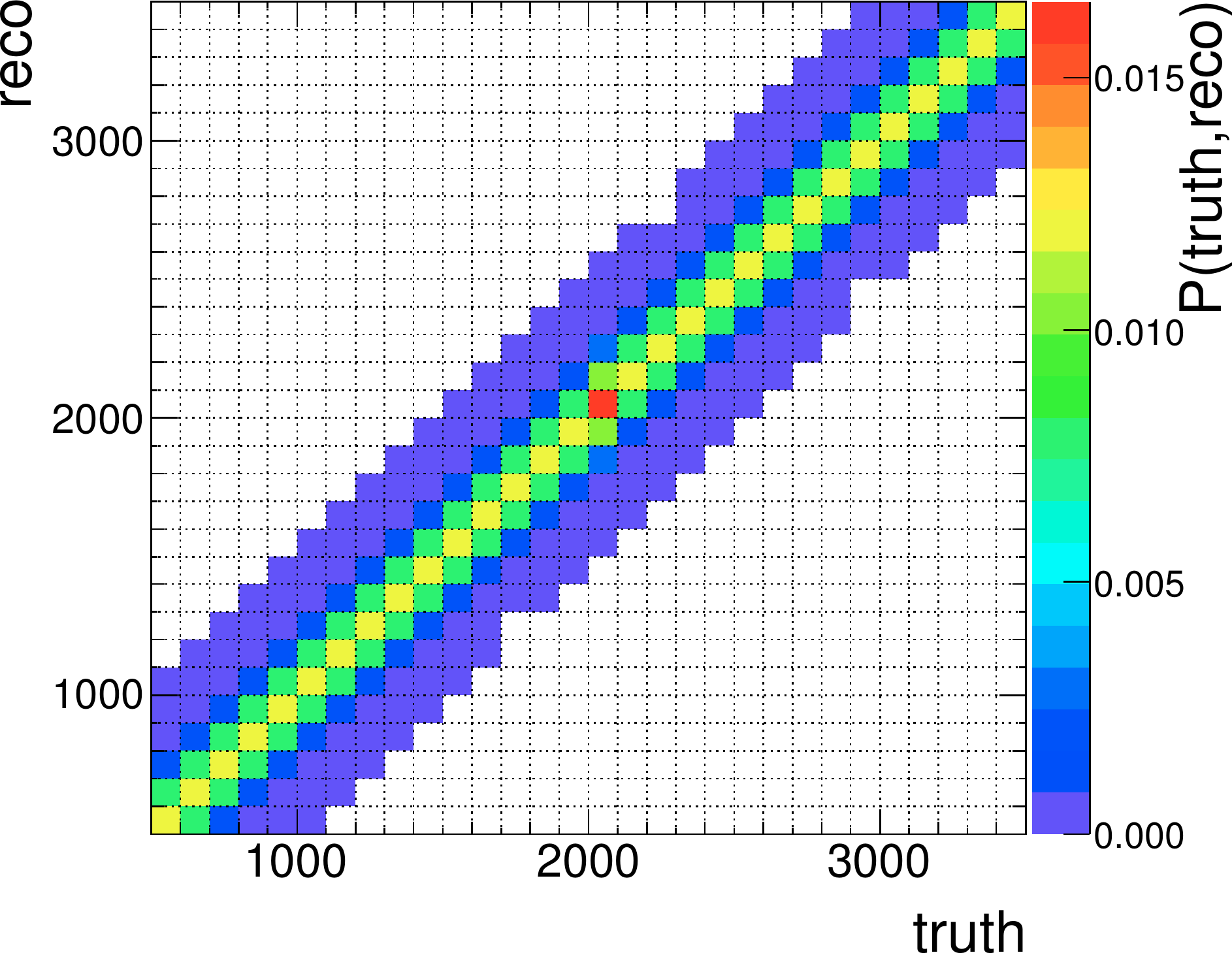} &
    \includegraphics[height=0.23\columnwidth]{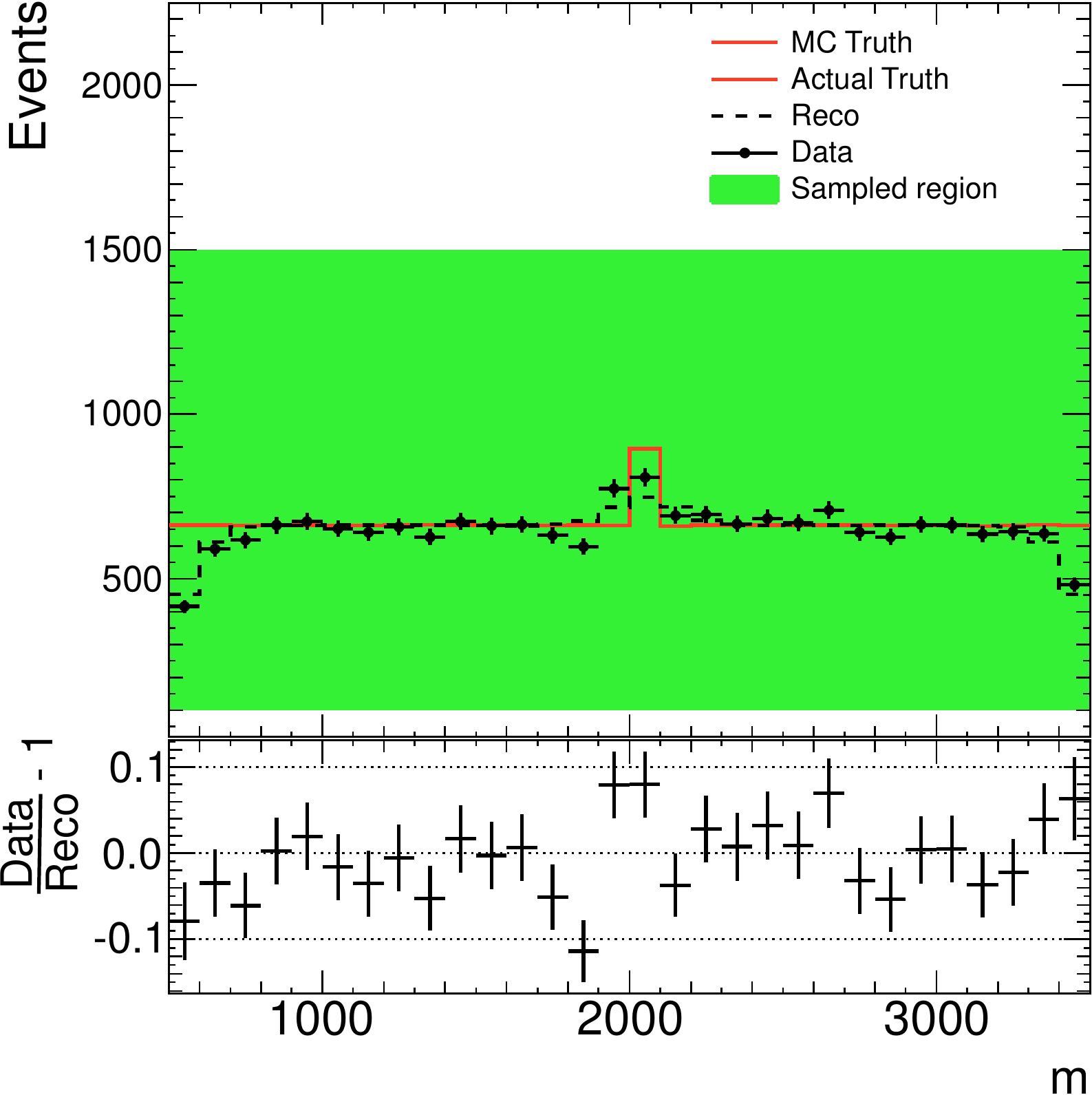} &
    \includegraphics[height=0.23\columnwidth]{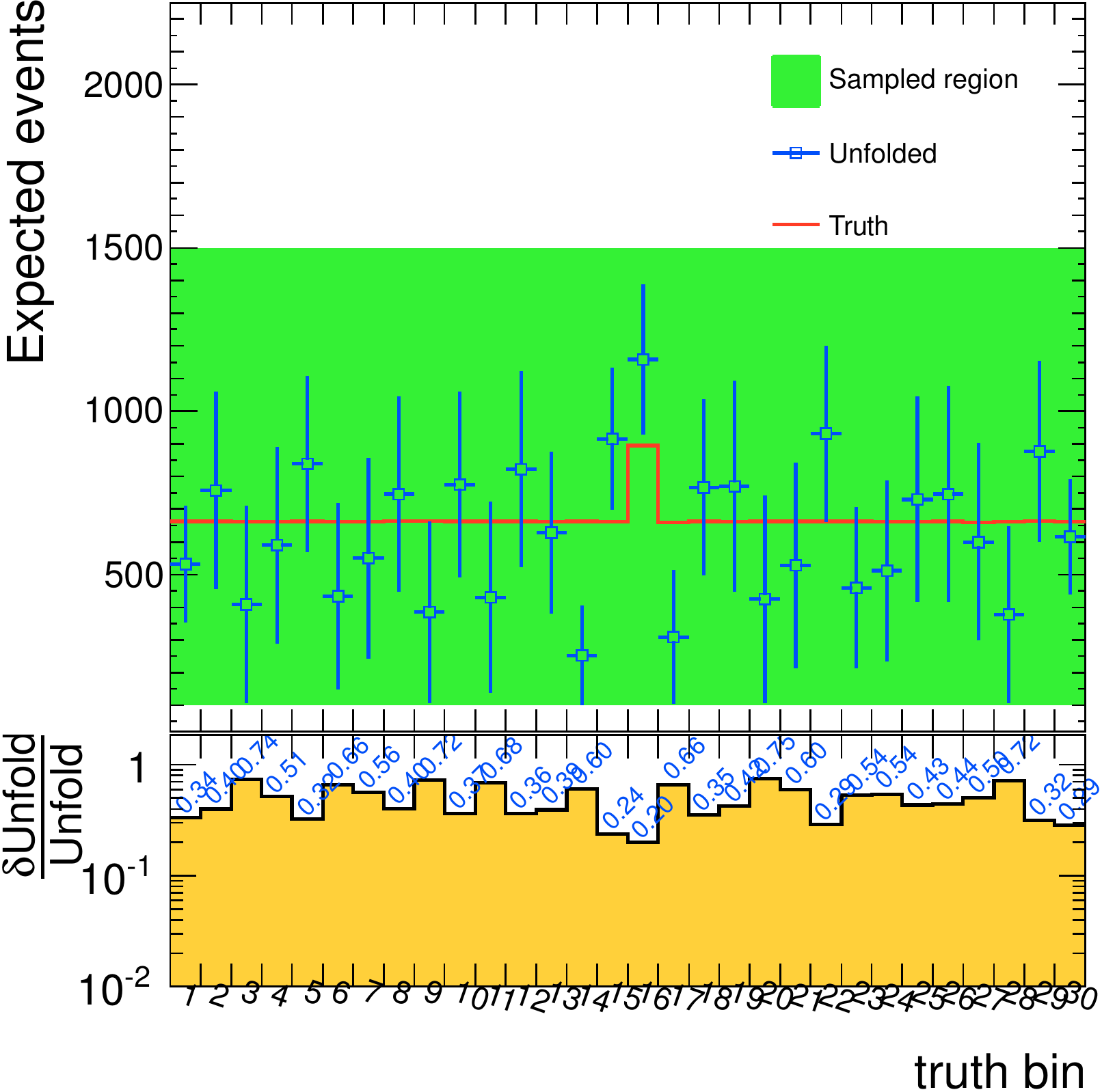} \\
    \includegraphics[height=0.23\columnwidth]{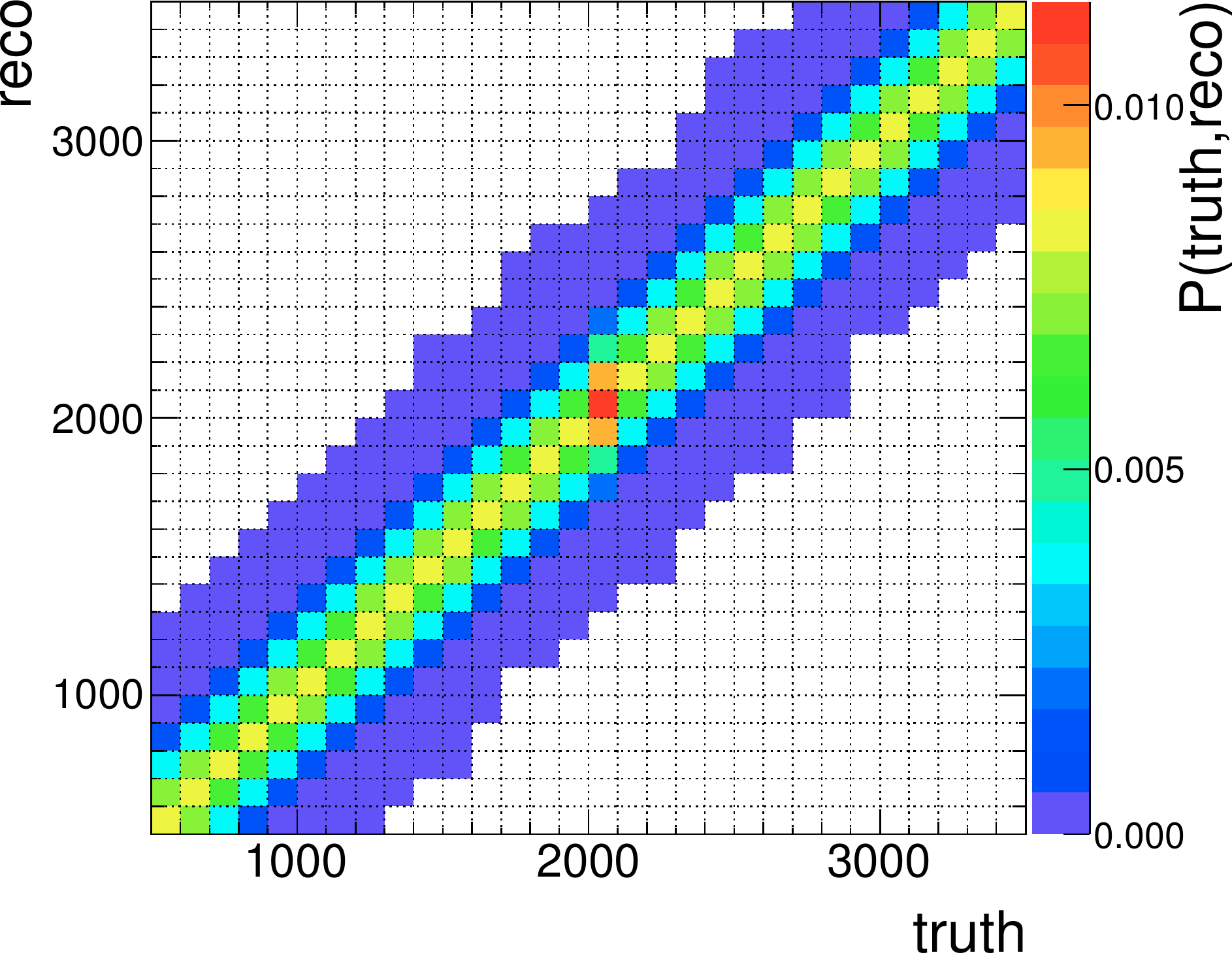} &
    \includegraphics[height=0.23\columnwidth]{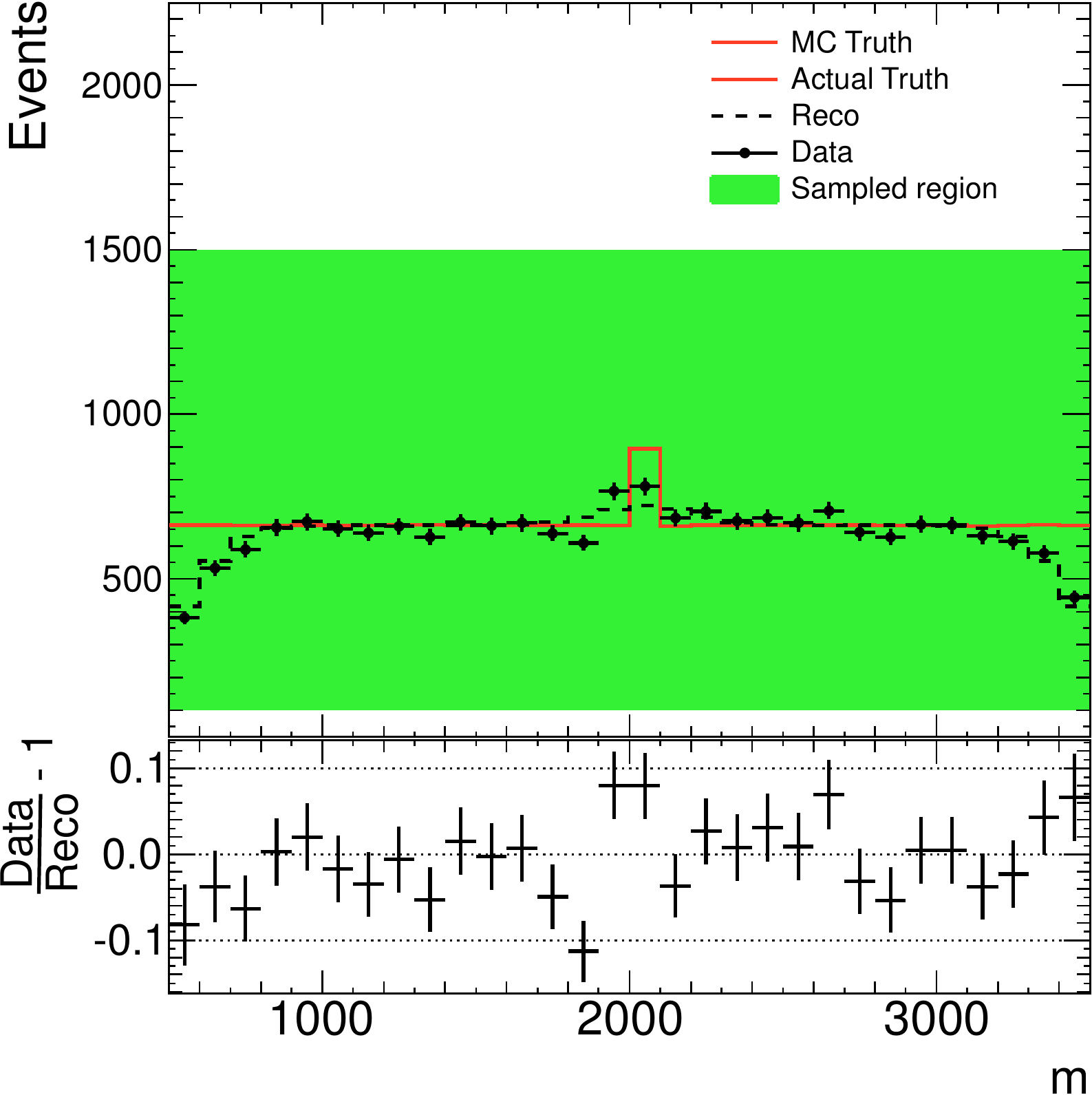} &
    \includegraphics[height=0.23\columnwidth]{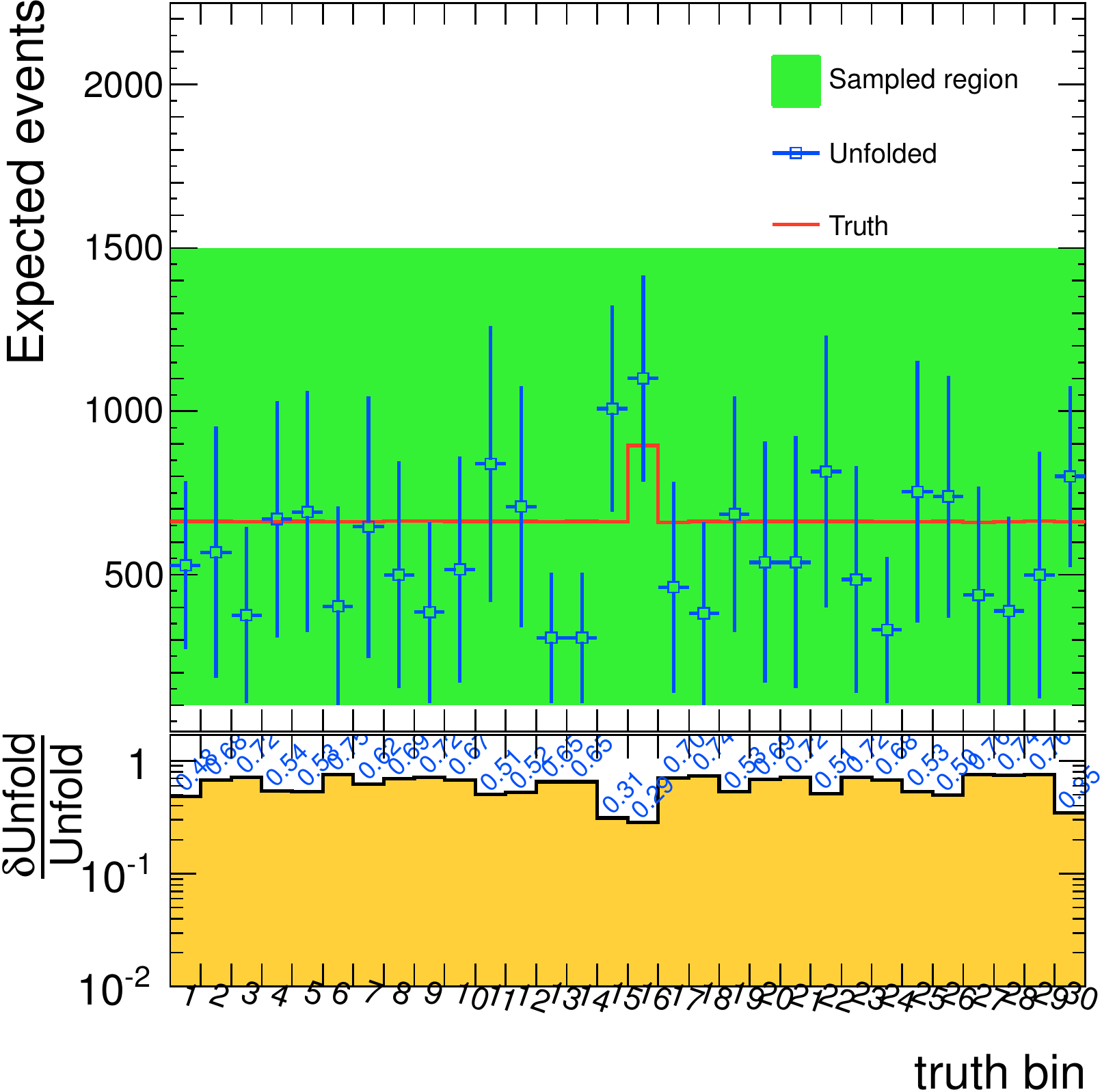} 
 \end{tabular}
\caption{Same as Fig.~\ref{fig:bumpKnown}, but with a narrower truth-level bump.  Details are given in Sec.~\ref{sec:bumpKnown}.
\label{fig:bumpKnown2}}
\end{figure}

A similar truth-level spectrum is generated to the one in Sec.~\ref{sec:bumpKnown}.  MC events are then smeared, to form the expected reconstructed spectrum, which is then allowed to fluctuate according to Poisson, to produce pseudo-data similar to those in Fig.~\ref{fig:bumpKnown} and \ref{fig:bumpKnown2}.  The difference, in this section, is that the MC events, used to compute $\M$, follow a flat truth level spectrum $\tilde{\T}$, namely different from the actual truth spectrum $\hat{\T}$.  

\begin{figure}[H]
  \centering
  \begin{tabular}{ccc}
    \includegraphics[height=0.23\columnwidth]{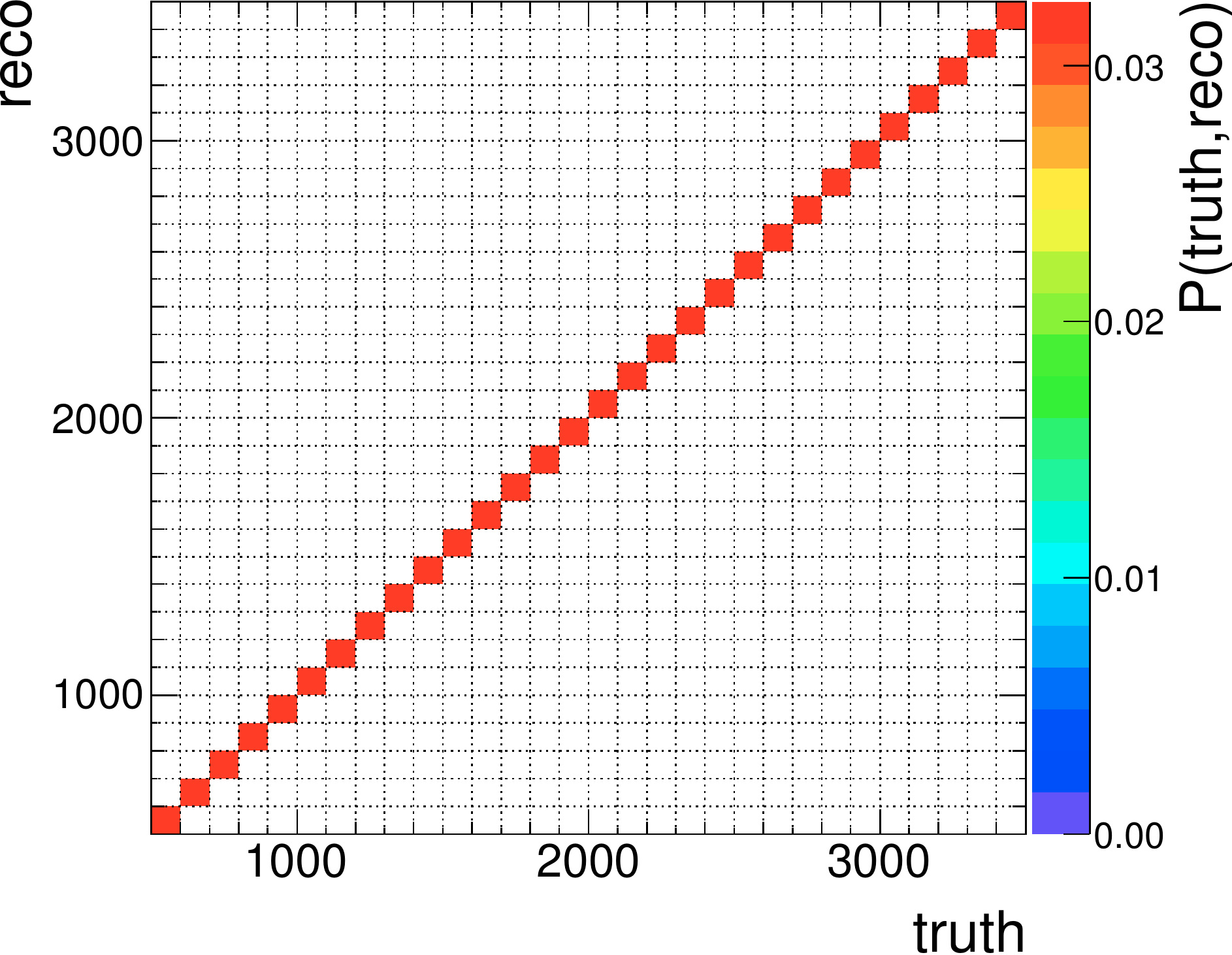} &
    \includegraphics[height=0.23\columnwidth]{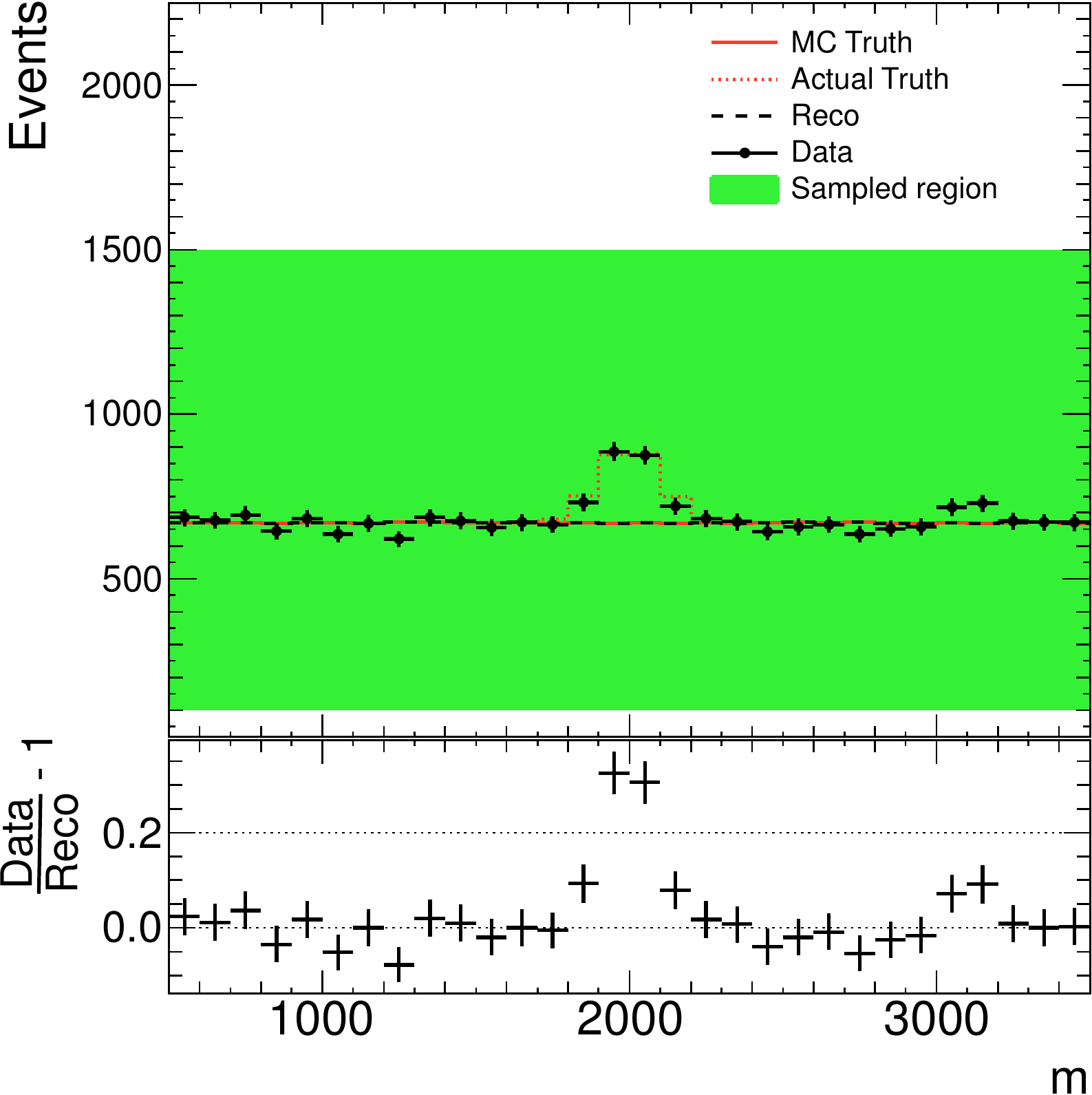} &
    \includegraphics[height=0.23\columnwidth]{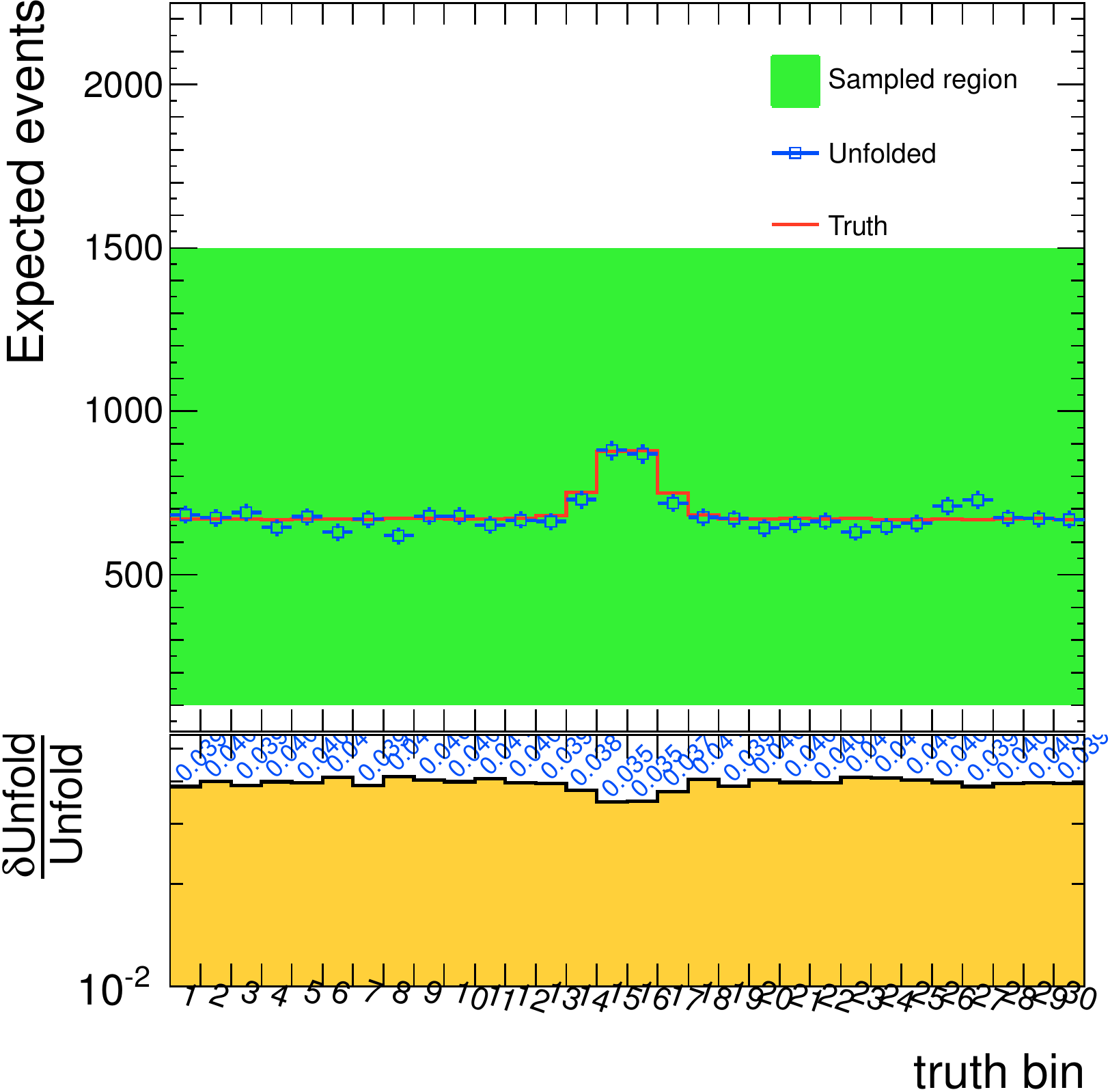} \\
    \includegraphics[height=0.23\columnwidth]{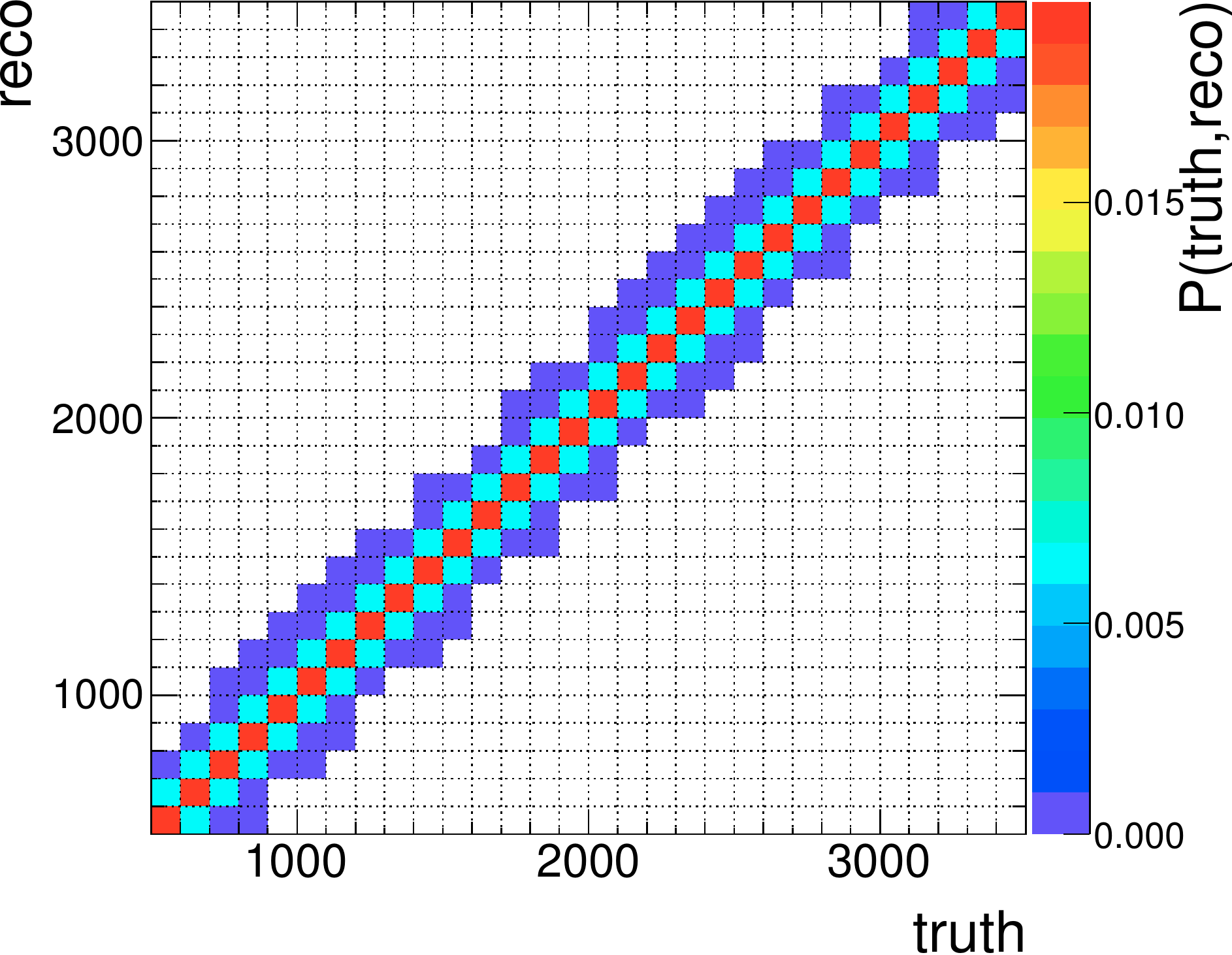} &
    \includegraphics[height=0.23\columnwidth]{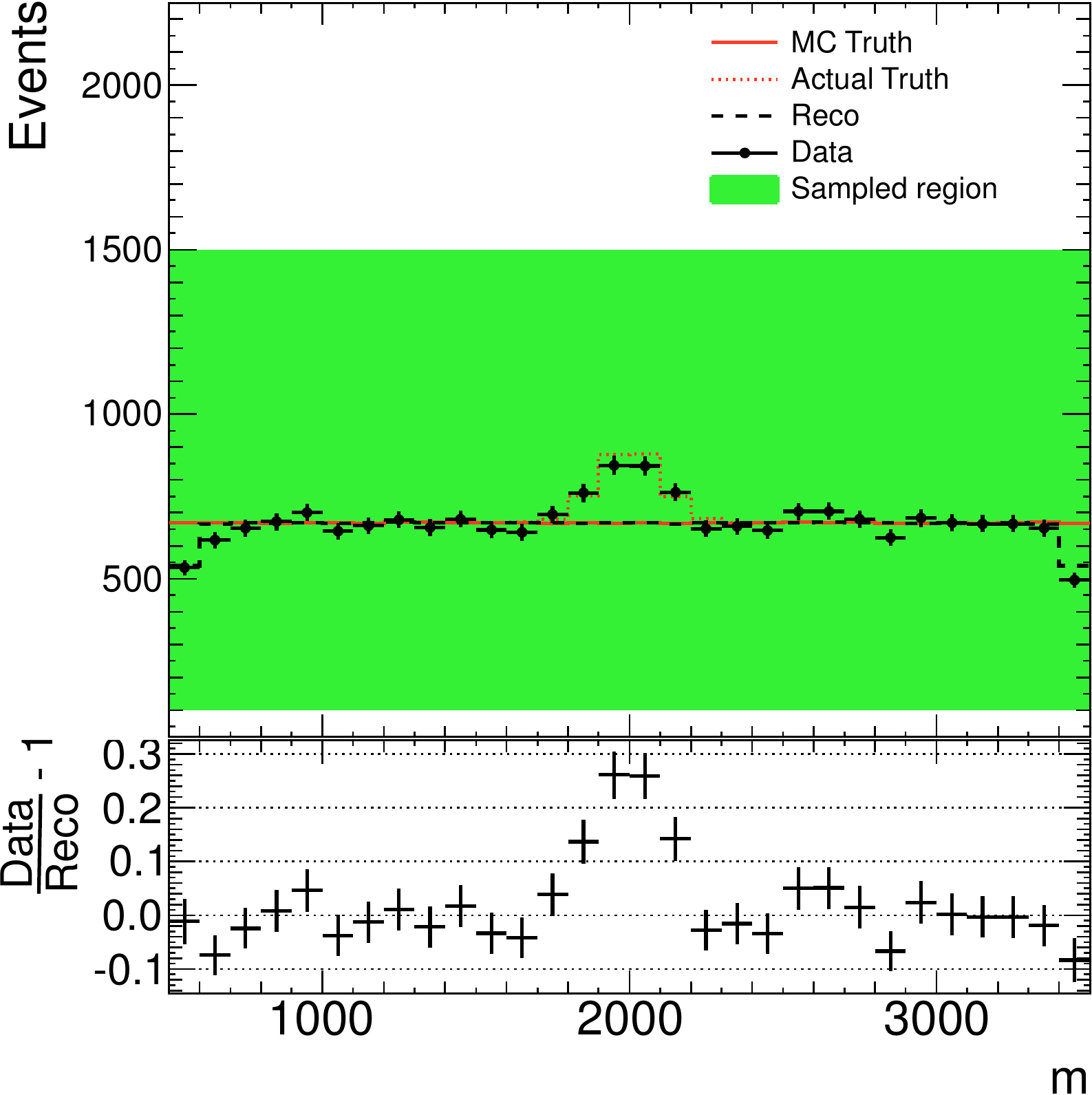} &
    \includegraphics[height=0.23\columnwidth]{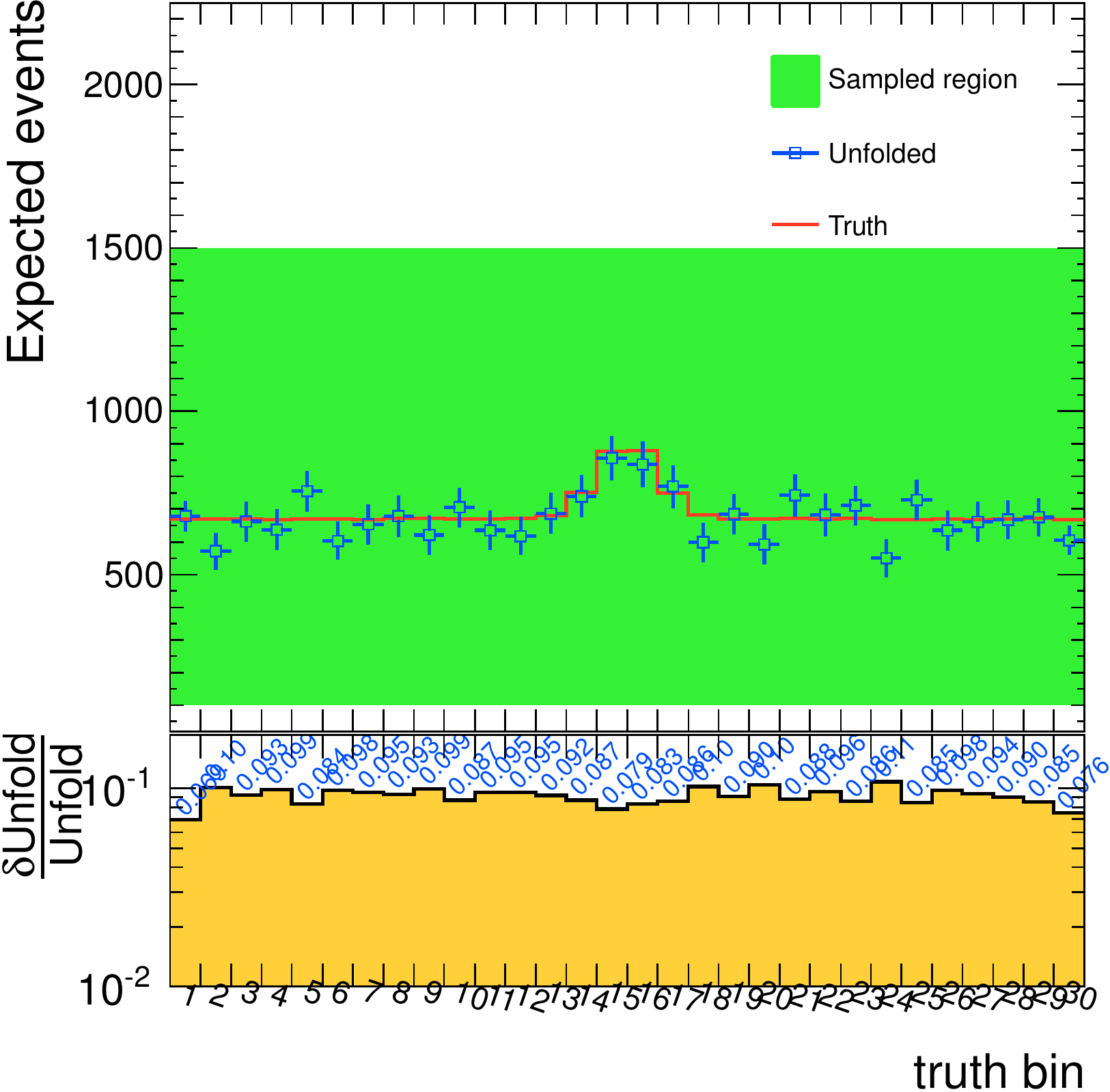} \\
    \includegraphics[height=0.23\columnwidth]{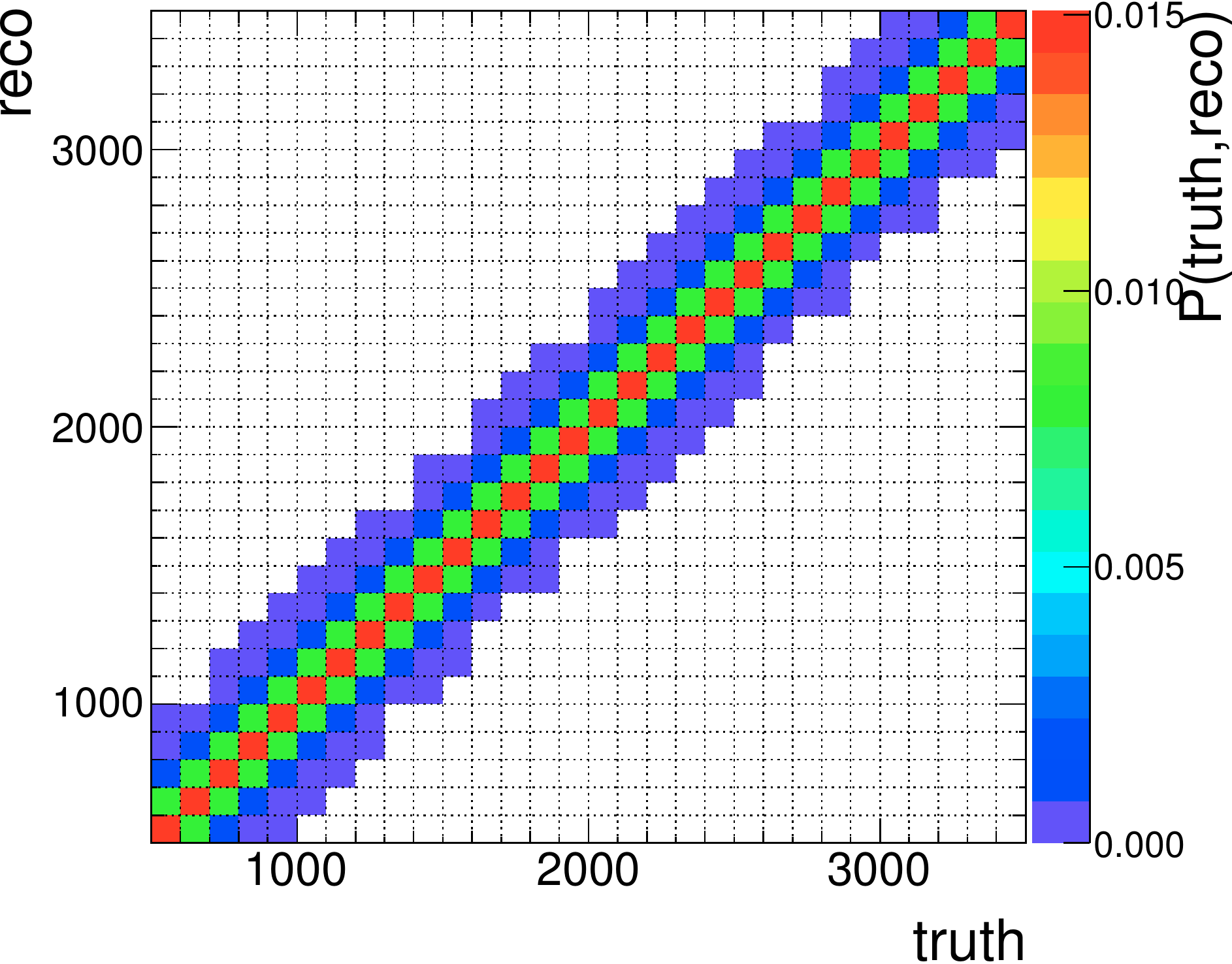} &
    \includegraphics[height=0.23\columnwidth]{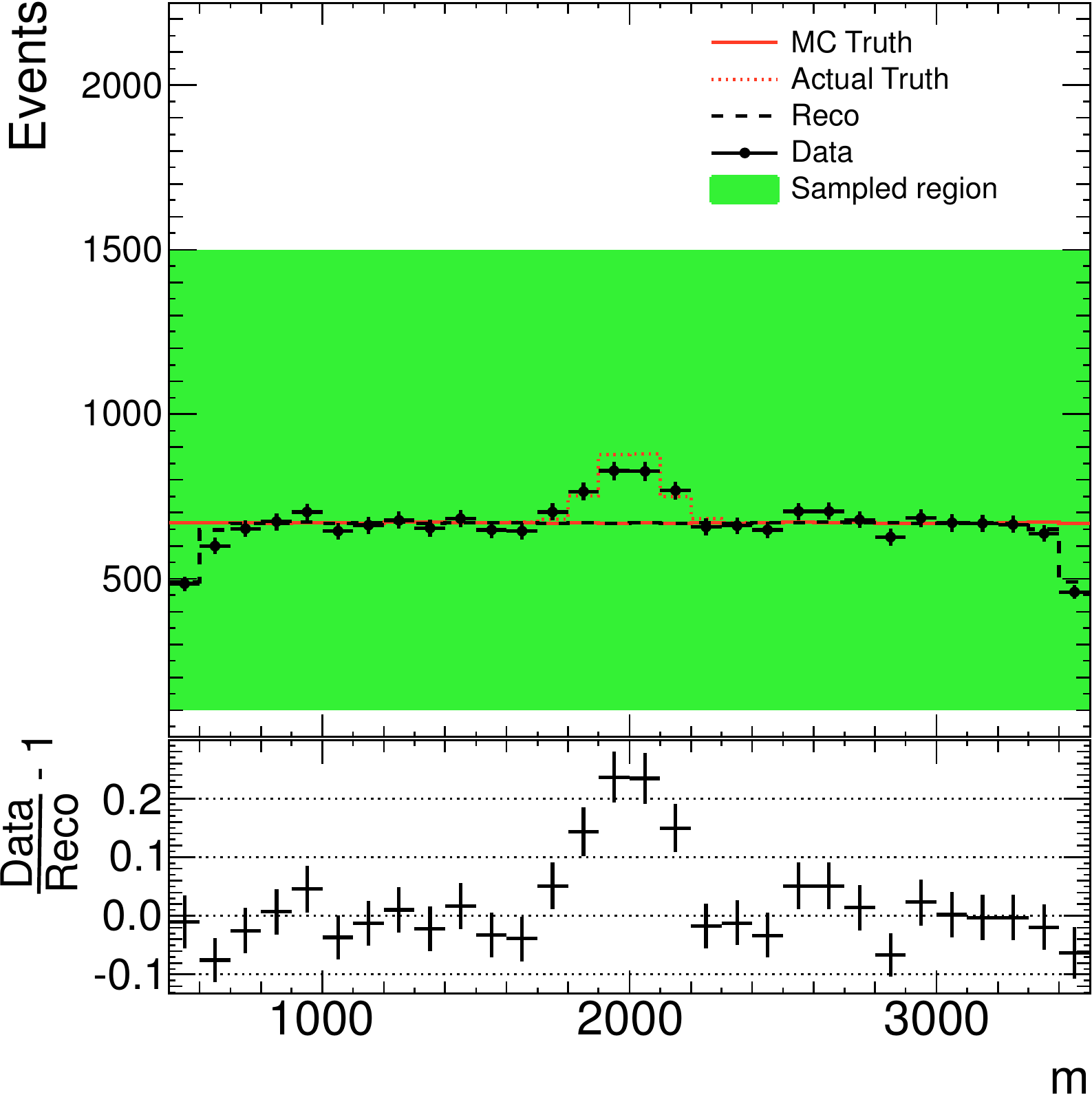} &
    \includegraphics[height=0.23\columnwidth]{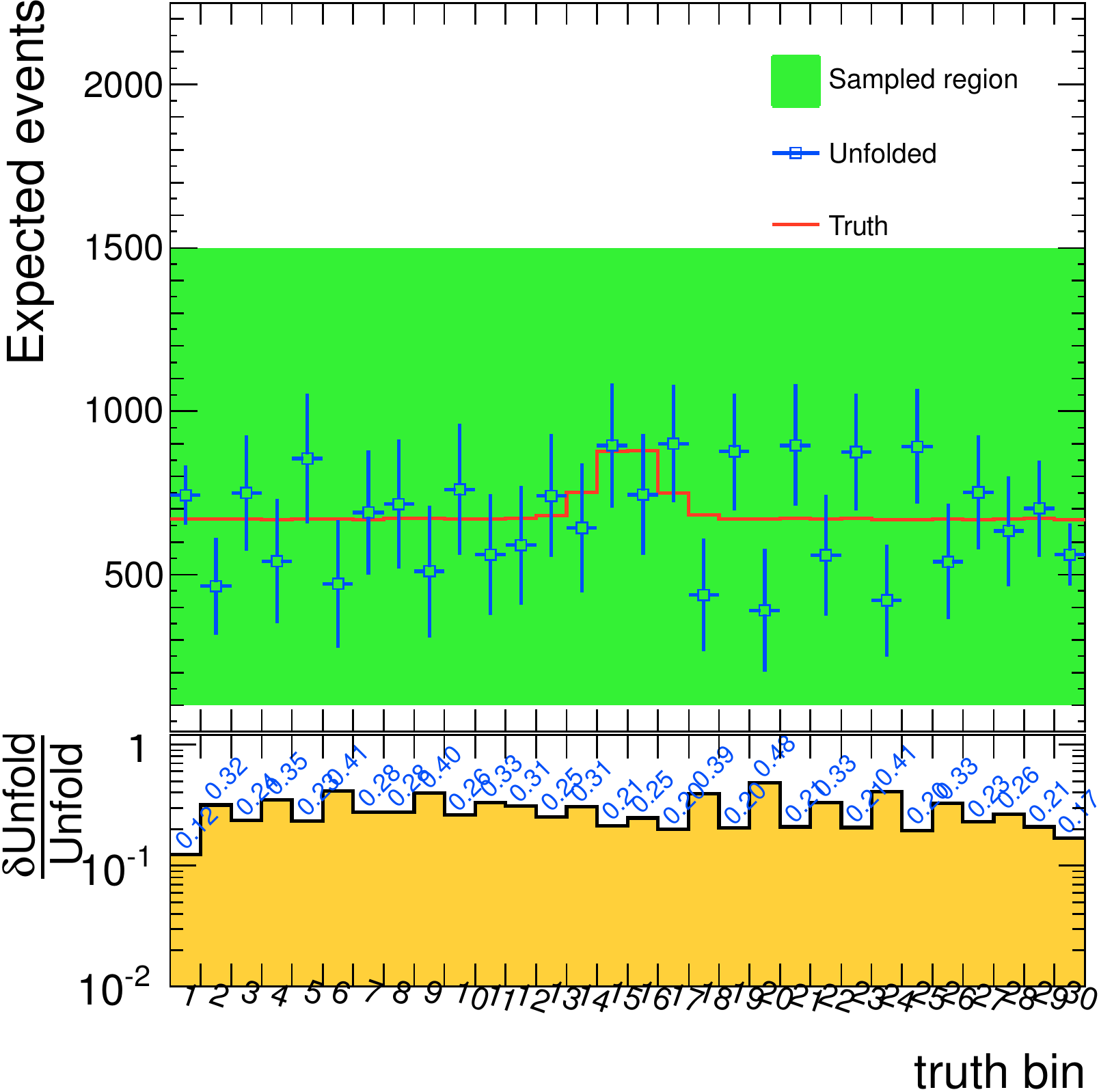} \\
    \includegraphics[height=0.23\columnwidth]{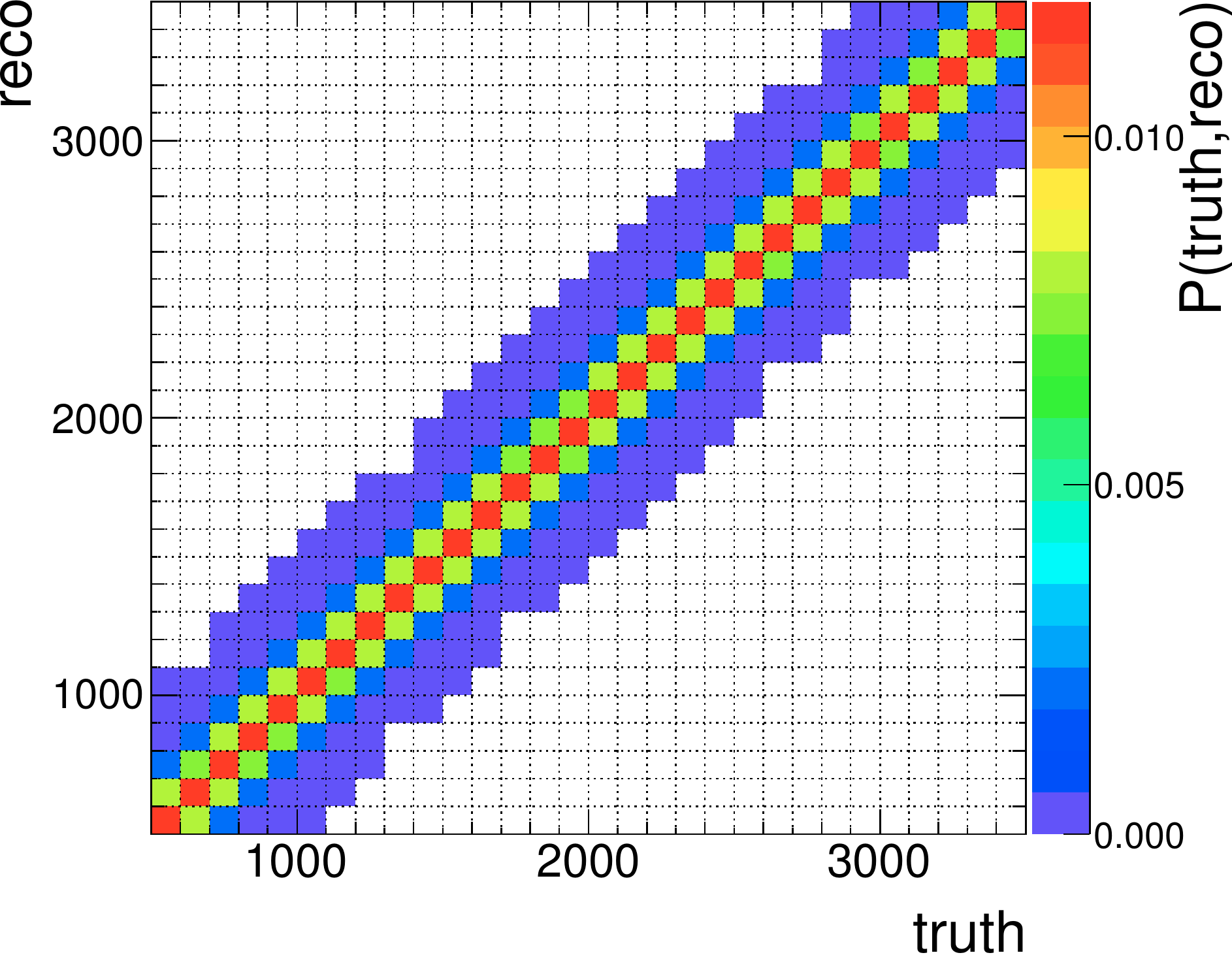} &
    \includegraphics[height=0.23\columnwidth]{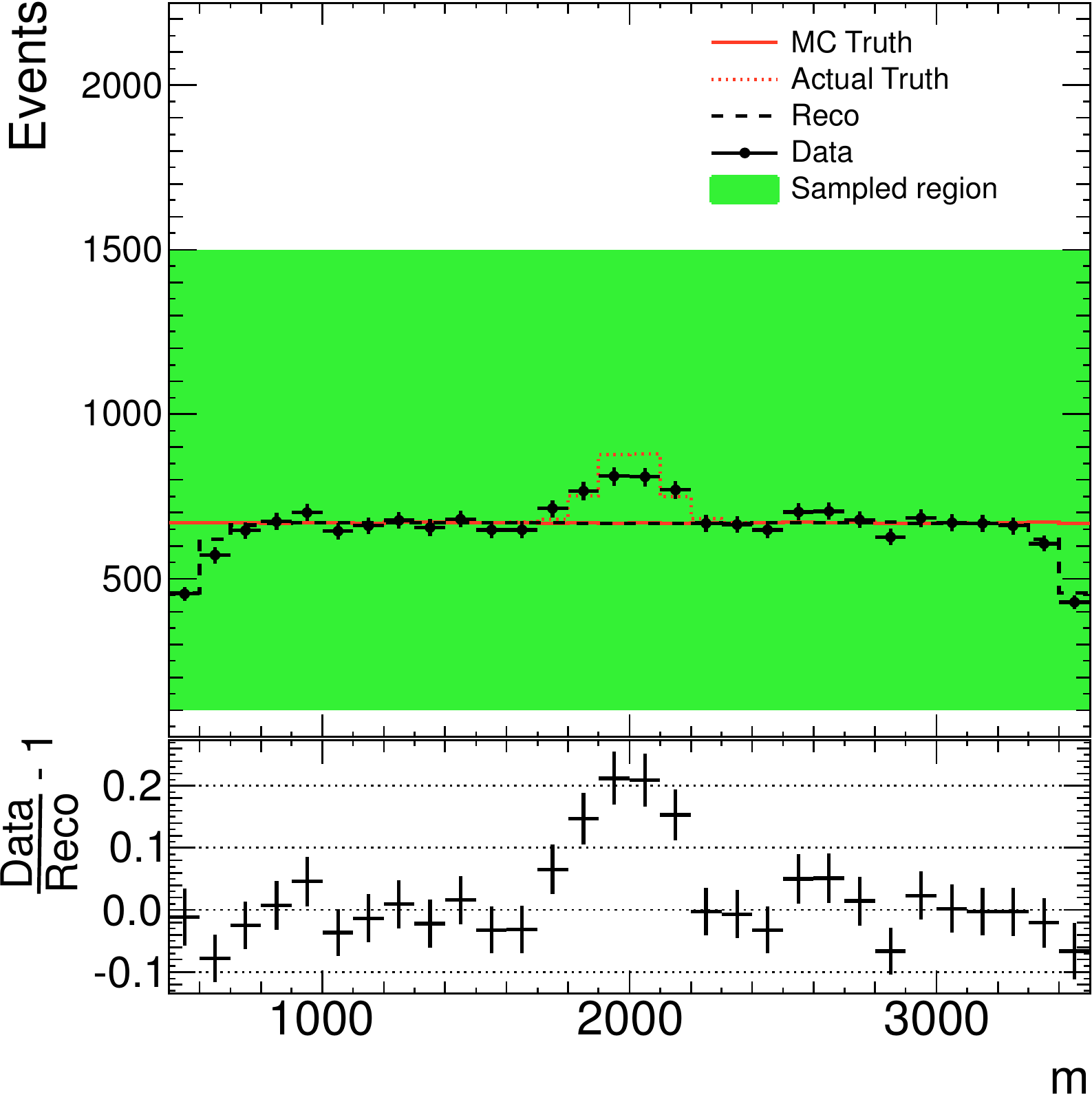} &
    \includegraphics[height=0.23\columnwidth]{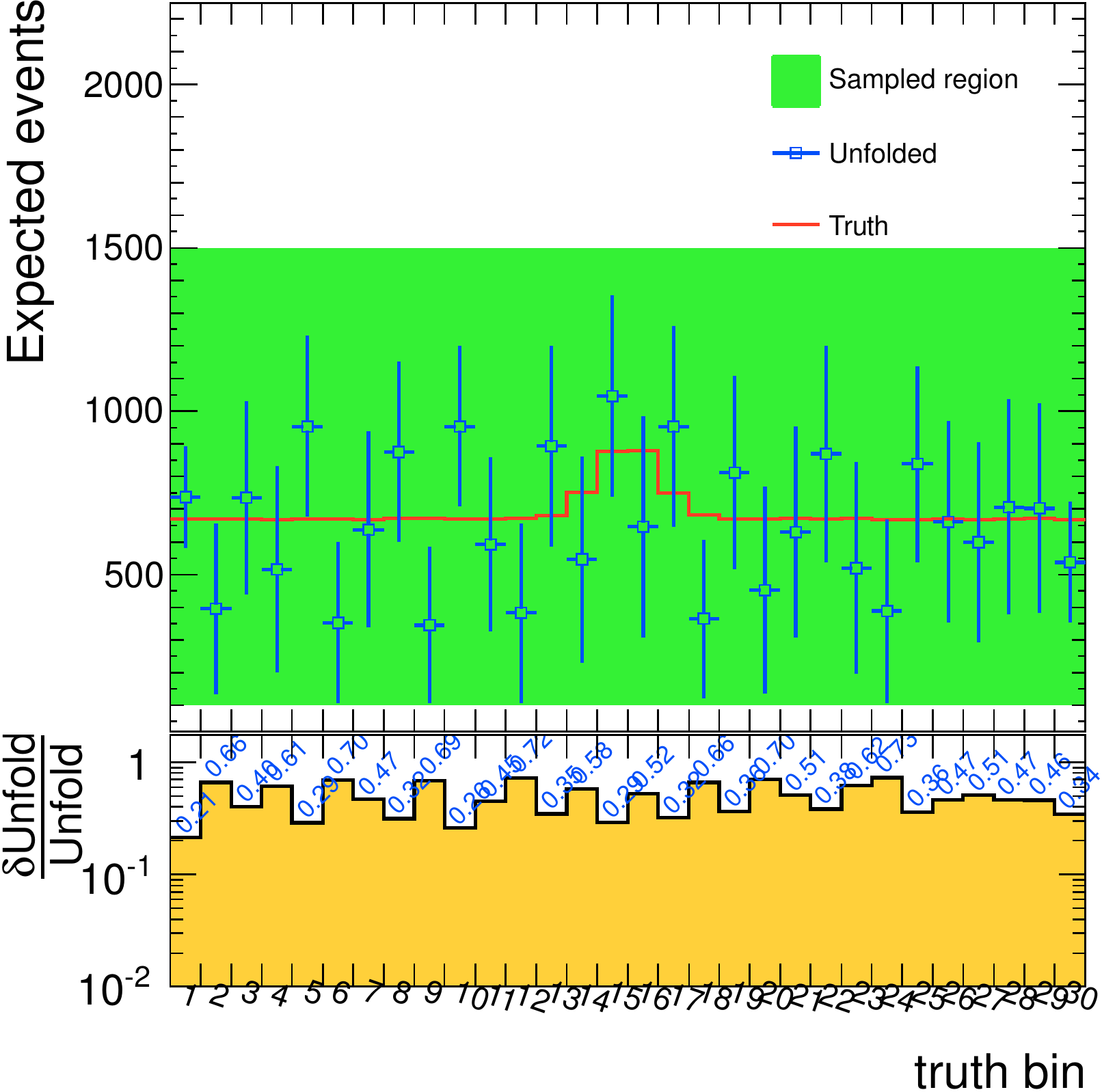} \\
    \includegraphics[height=0.23\columnwidth]{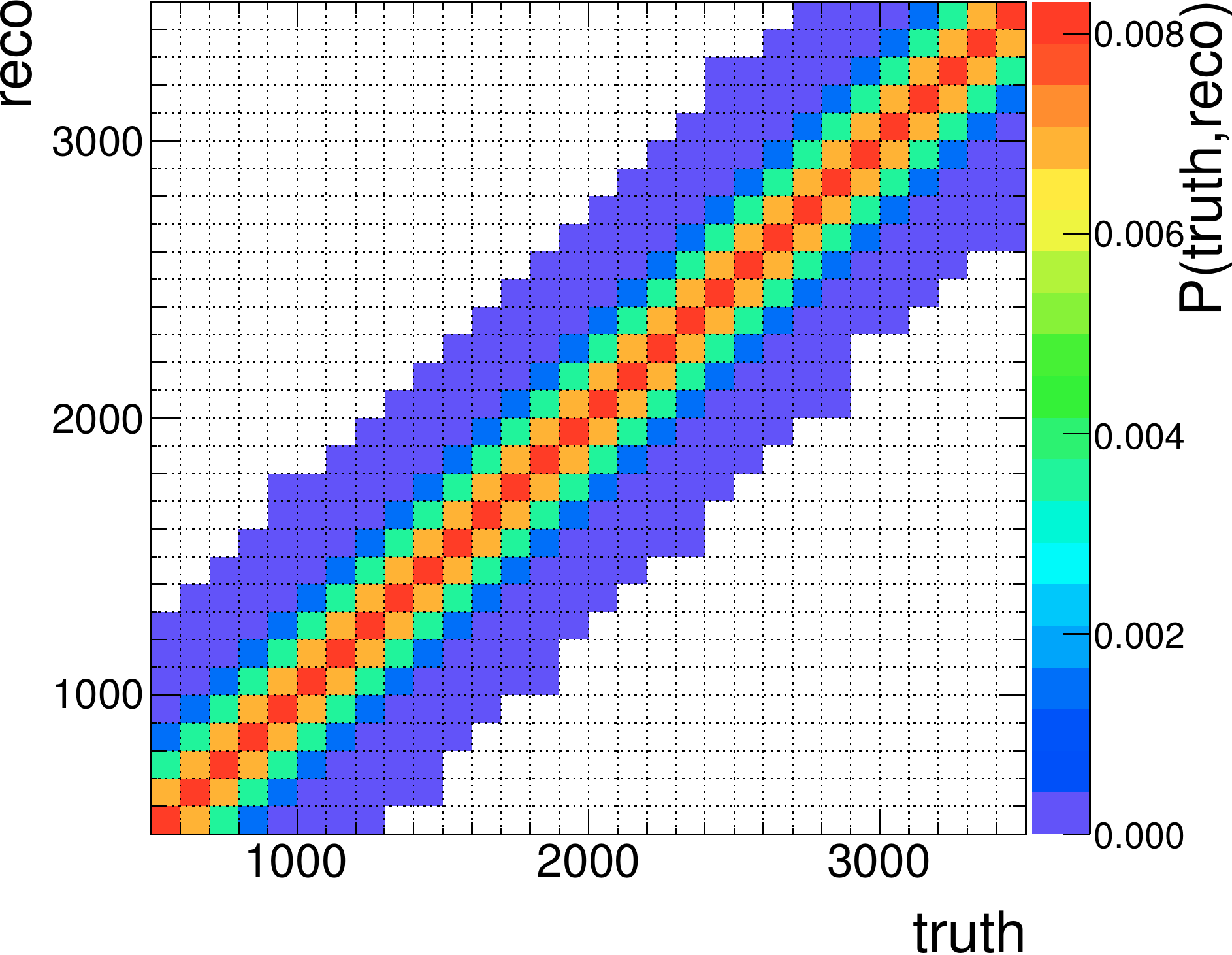} &
    \includegraphics[height=0.23\columnwidth]{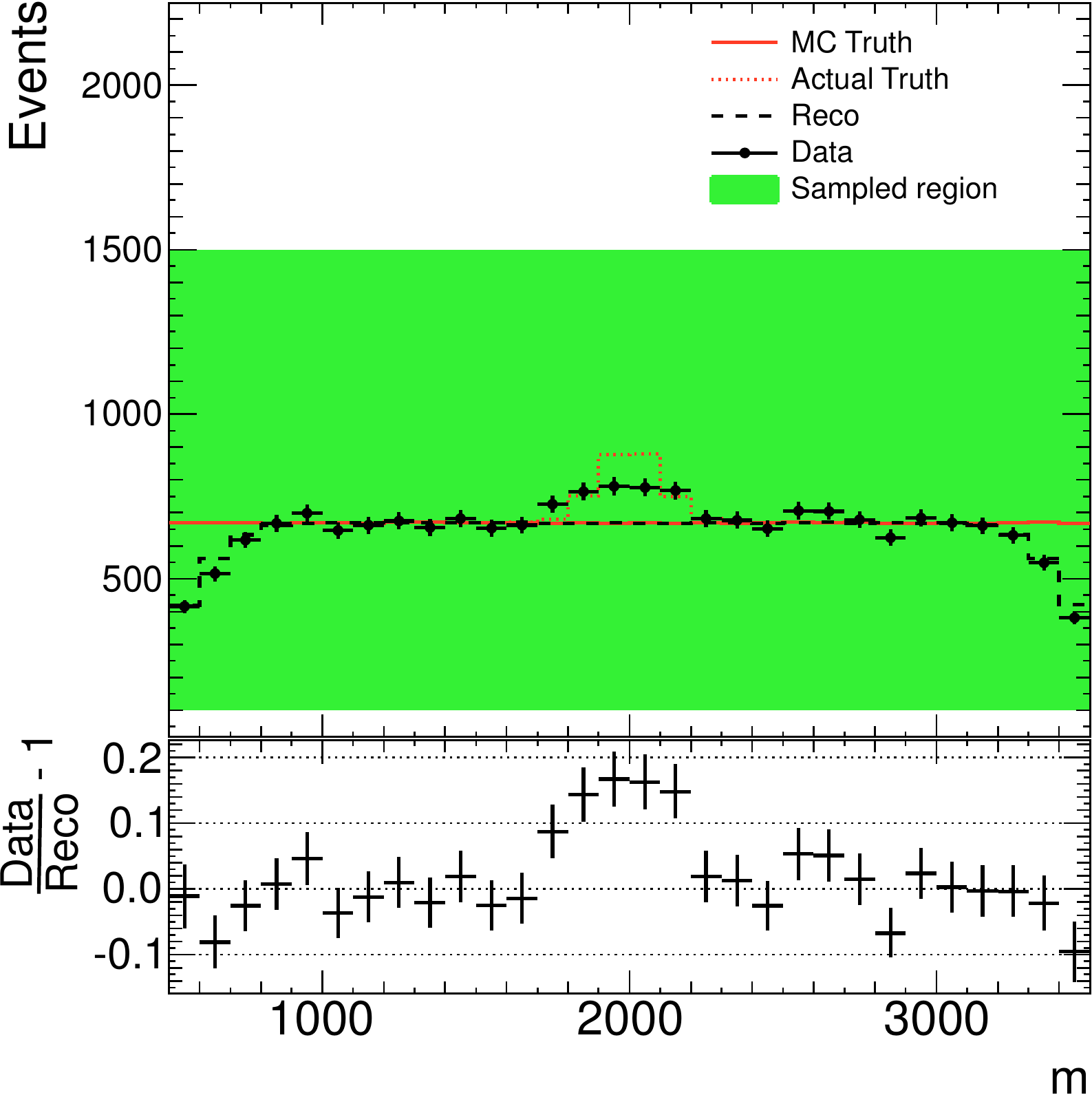} &
    \includegraphics[height=0.23\columnwidth]{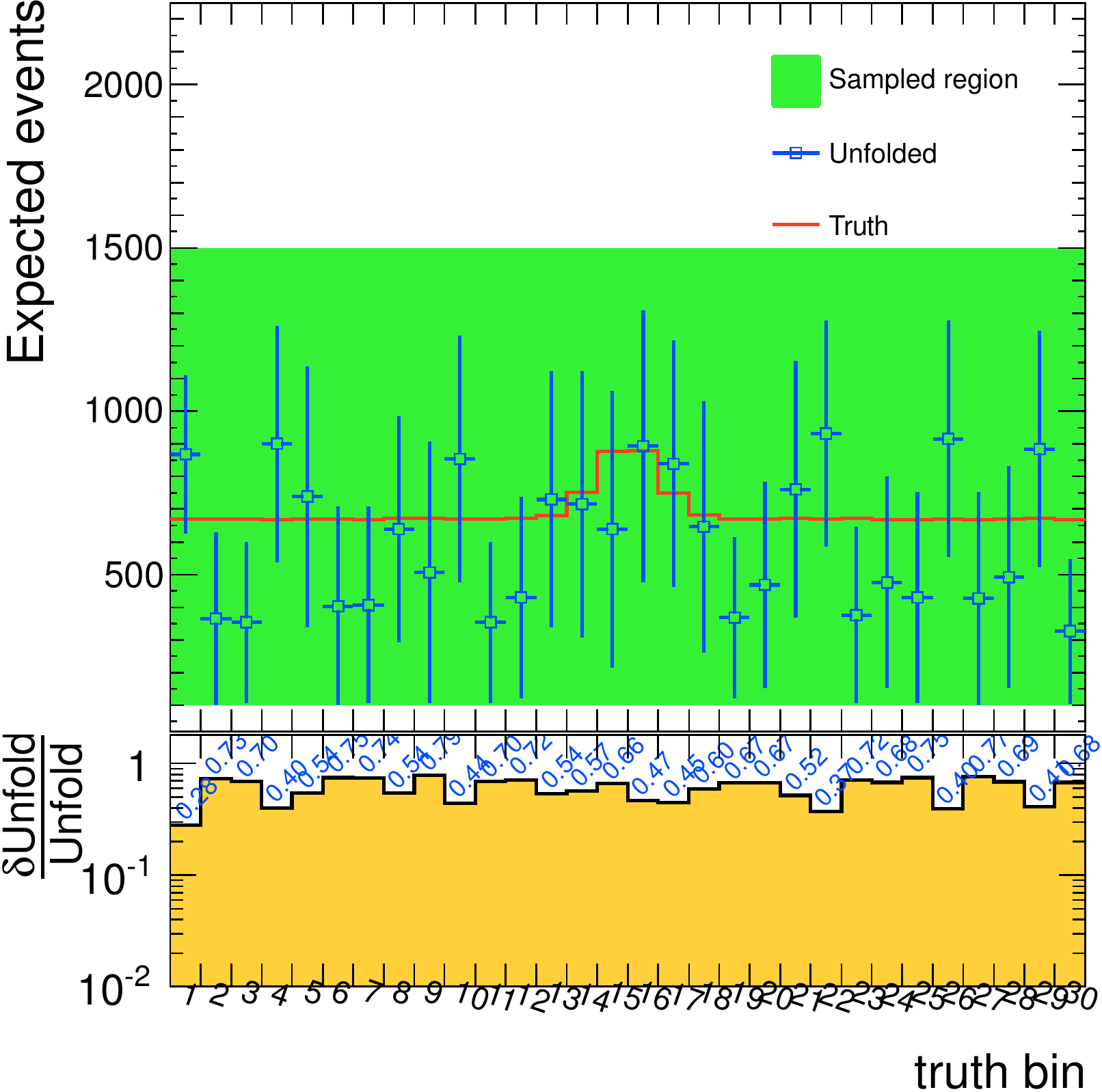} 
 \end{tabular}
\caption{The migrations matrix (left); MC reco, MC truth $\tilde{\T}$, actual truth $\hat{\T}$, and the data, which contain a bump unknown to the MC (middle); and truth ($\hat{\T}$) and unfolded spectra (right).  Each row corresponds to $\sigma=\{0,50,75,100,150\}$.  Details in Sec.~\ref{sec:bumpUnknown}.
\label{fig:bumpUnknown}}
\end{figure}

\begin{figure}[H]
  \centering
  \begin{tabular}{ccc}
    \includegraphics[height=0.23\columnwidth]{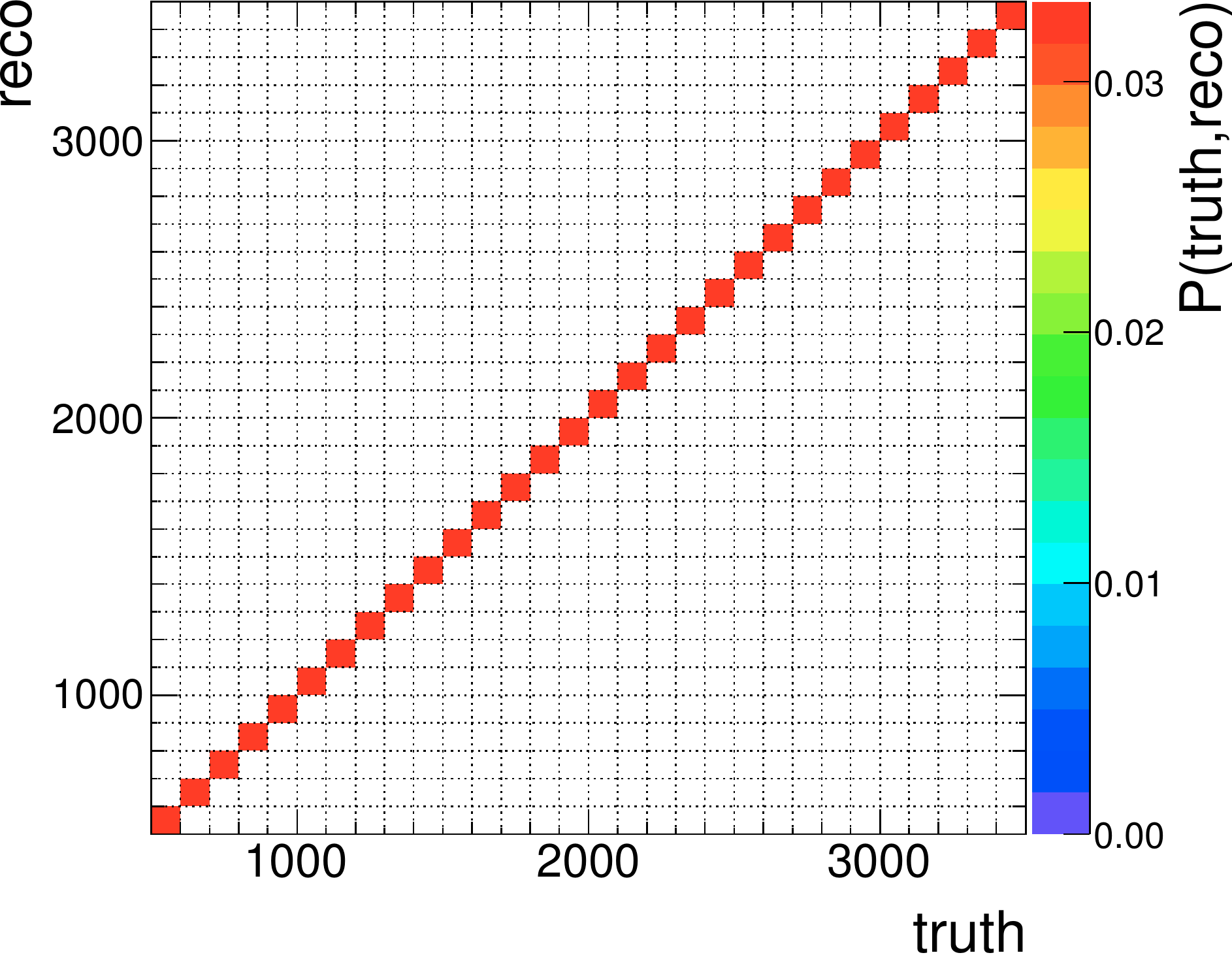} &
    \includegraphics[height=0.23\columnwidth]{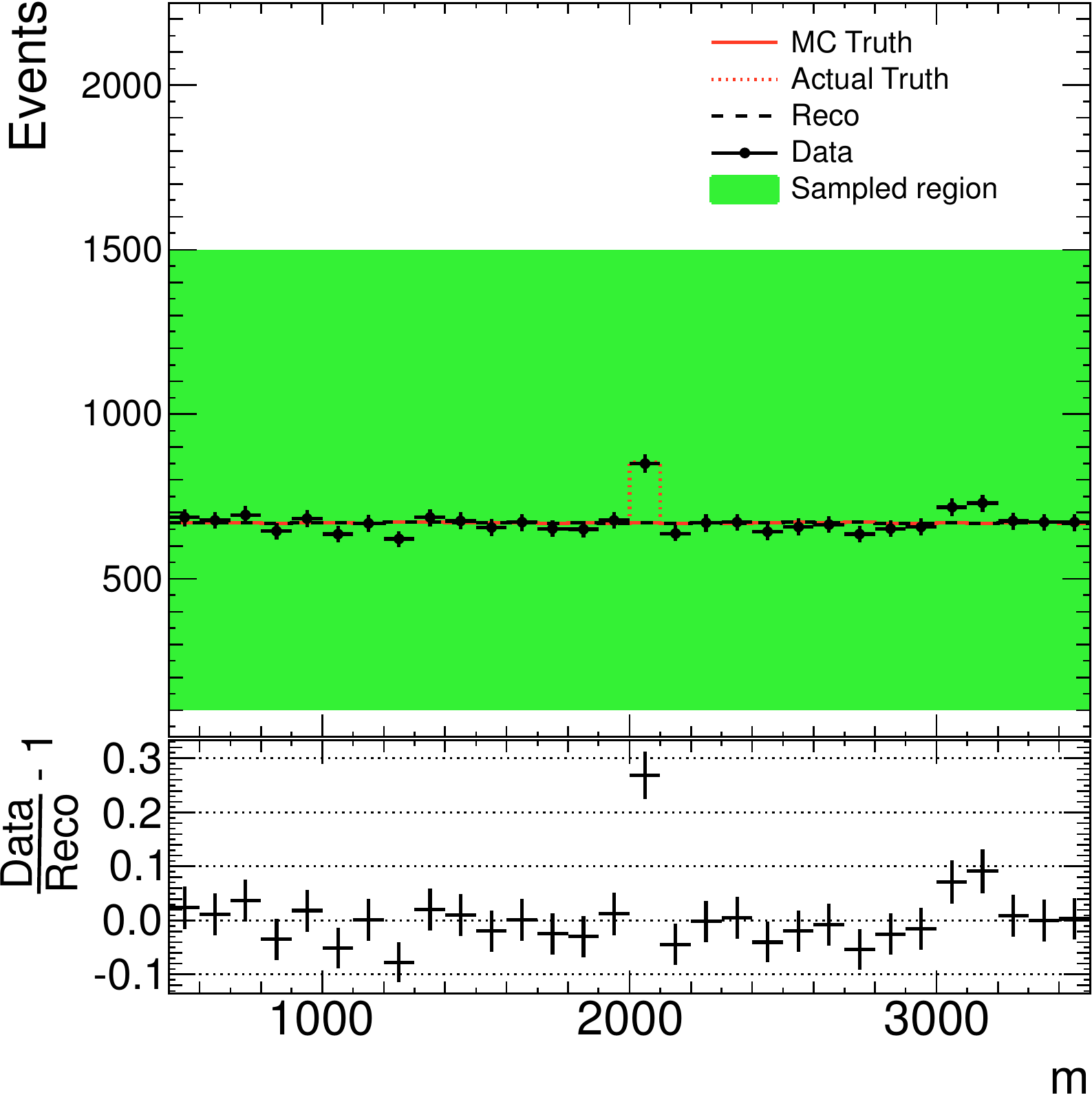} &
    \includegraphics[height=0.23\columnwidth]{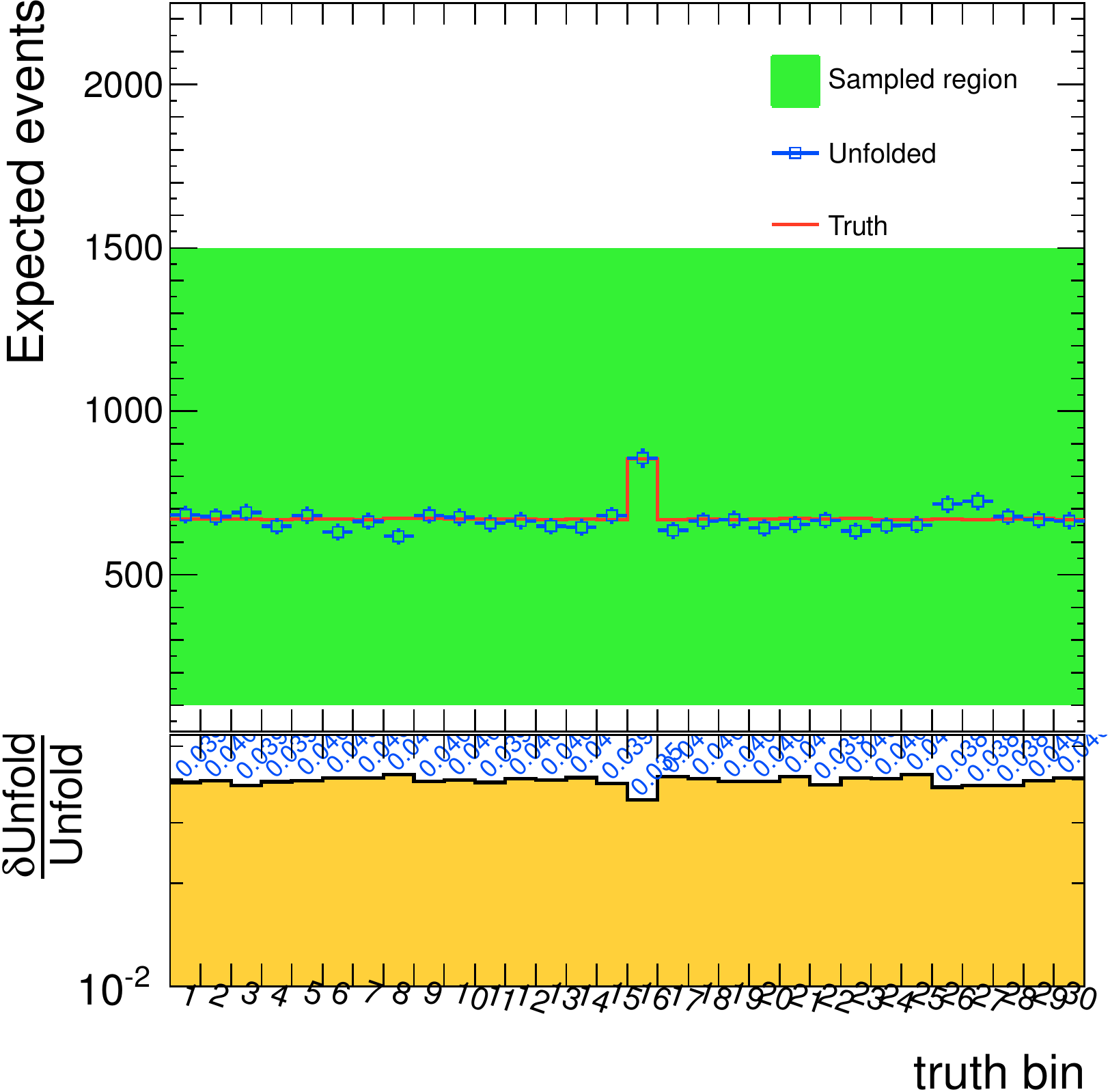} \\
    \includegraphics[height=0.23\columnwidth]{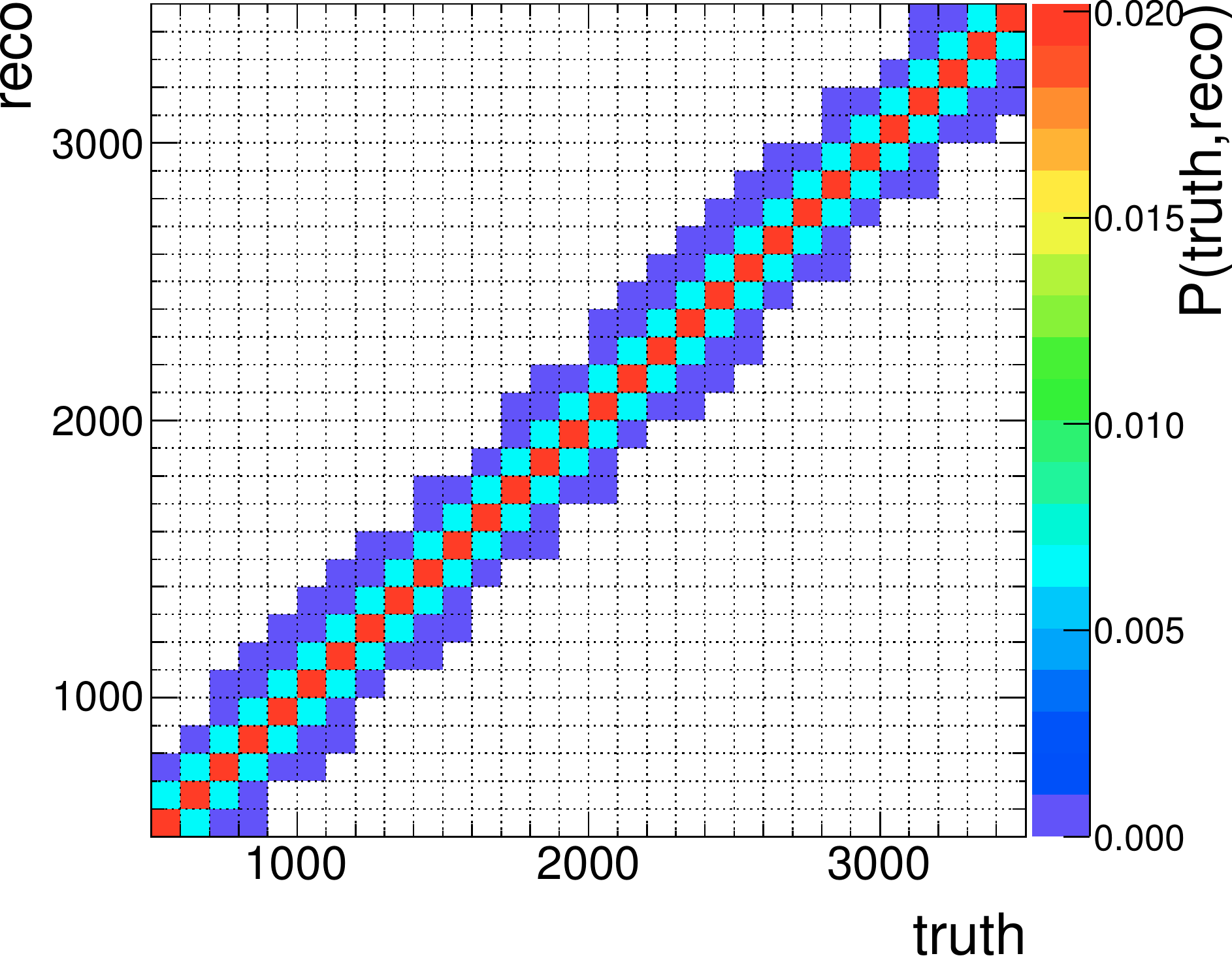} &
    \includegraphics[height=0.23\columnwidth]{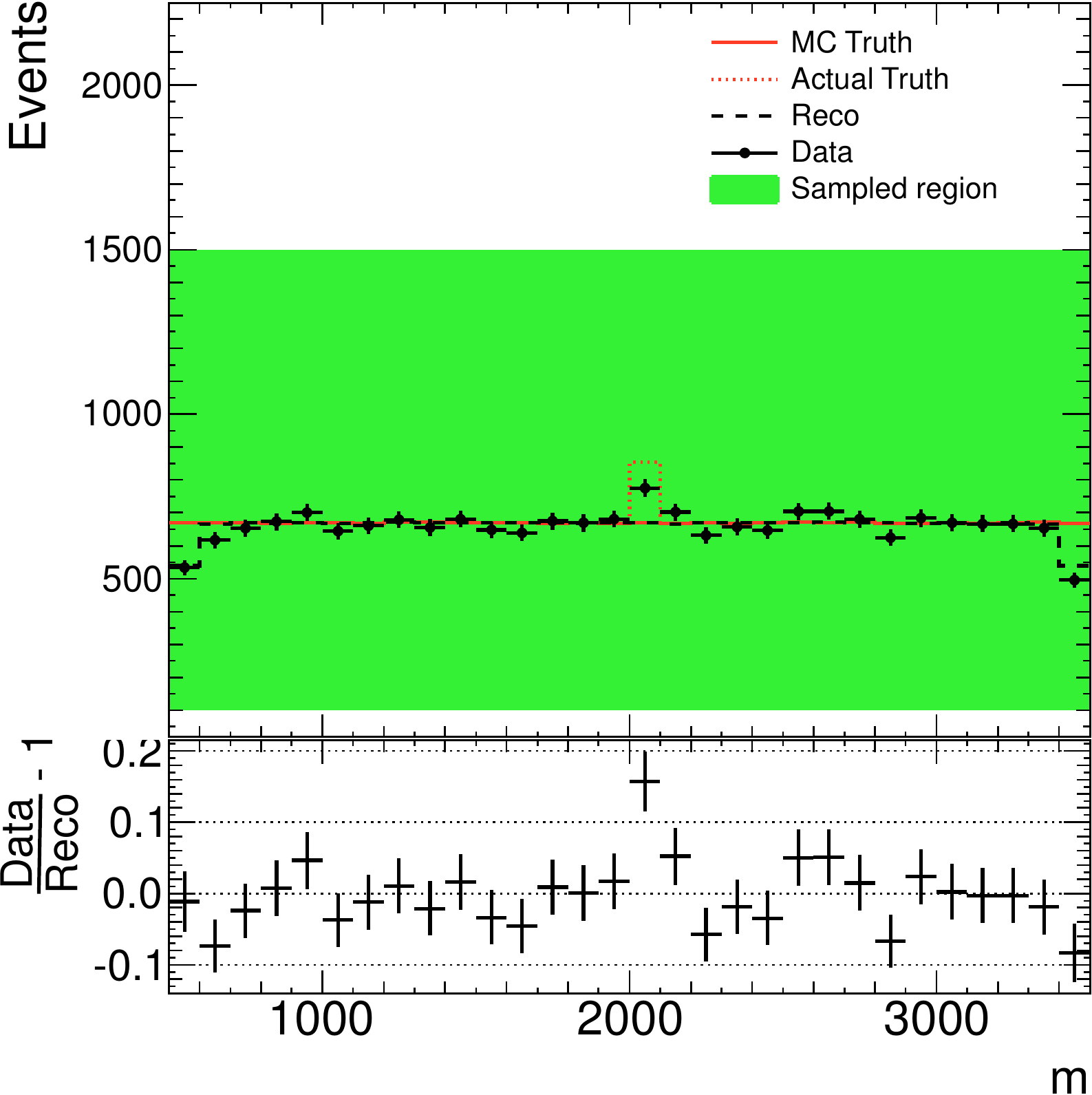} &
    \includegraphics[height=0.23\columnwidth]{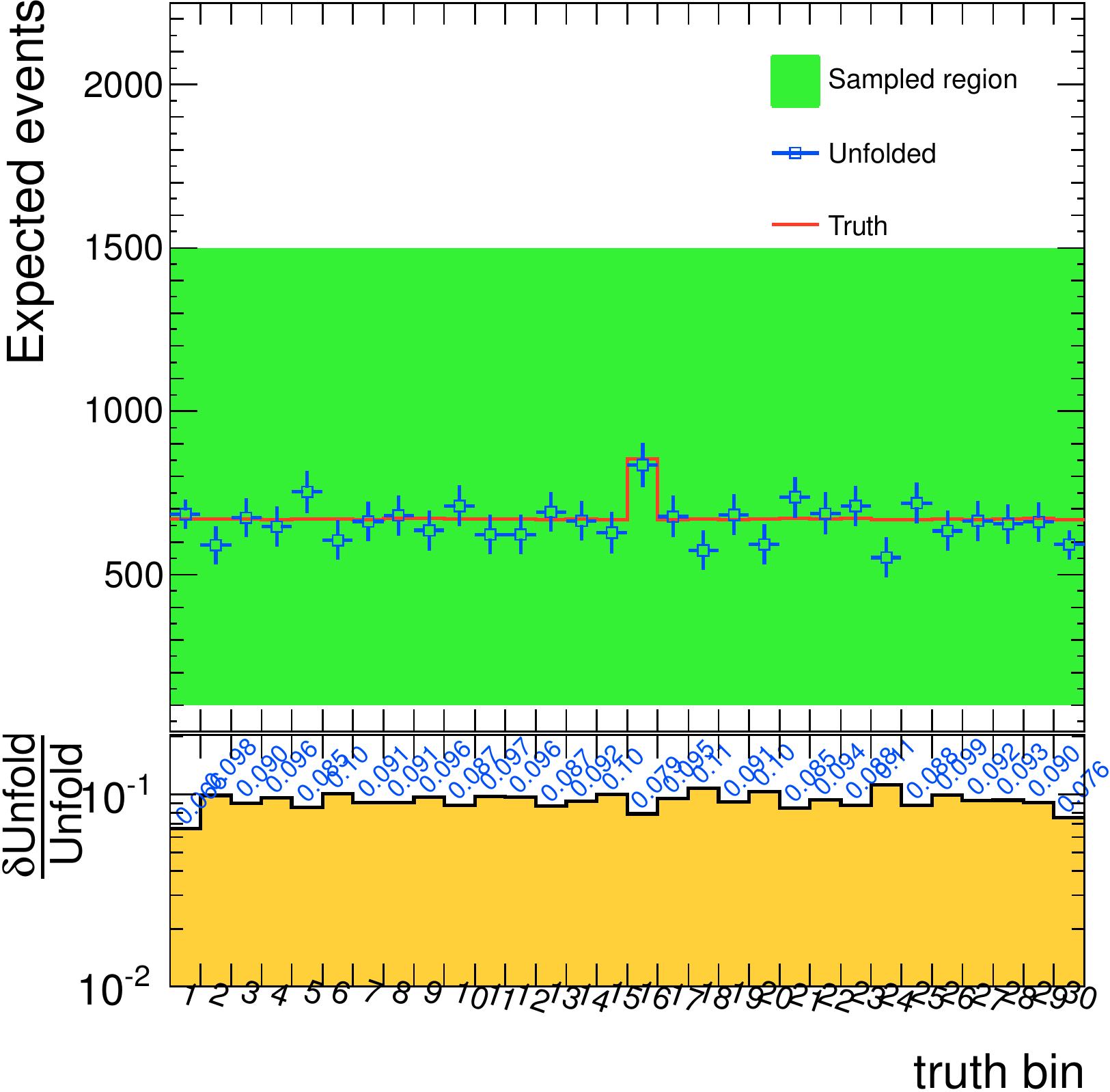} \\
    \includegraphics[height=0.23\columnwidth]{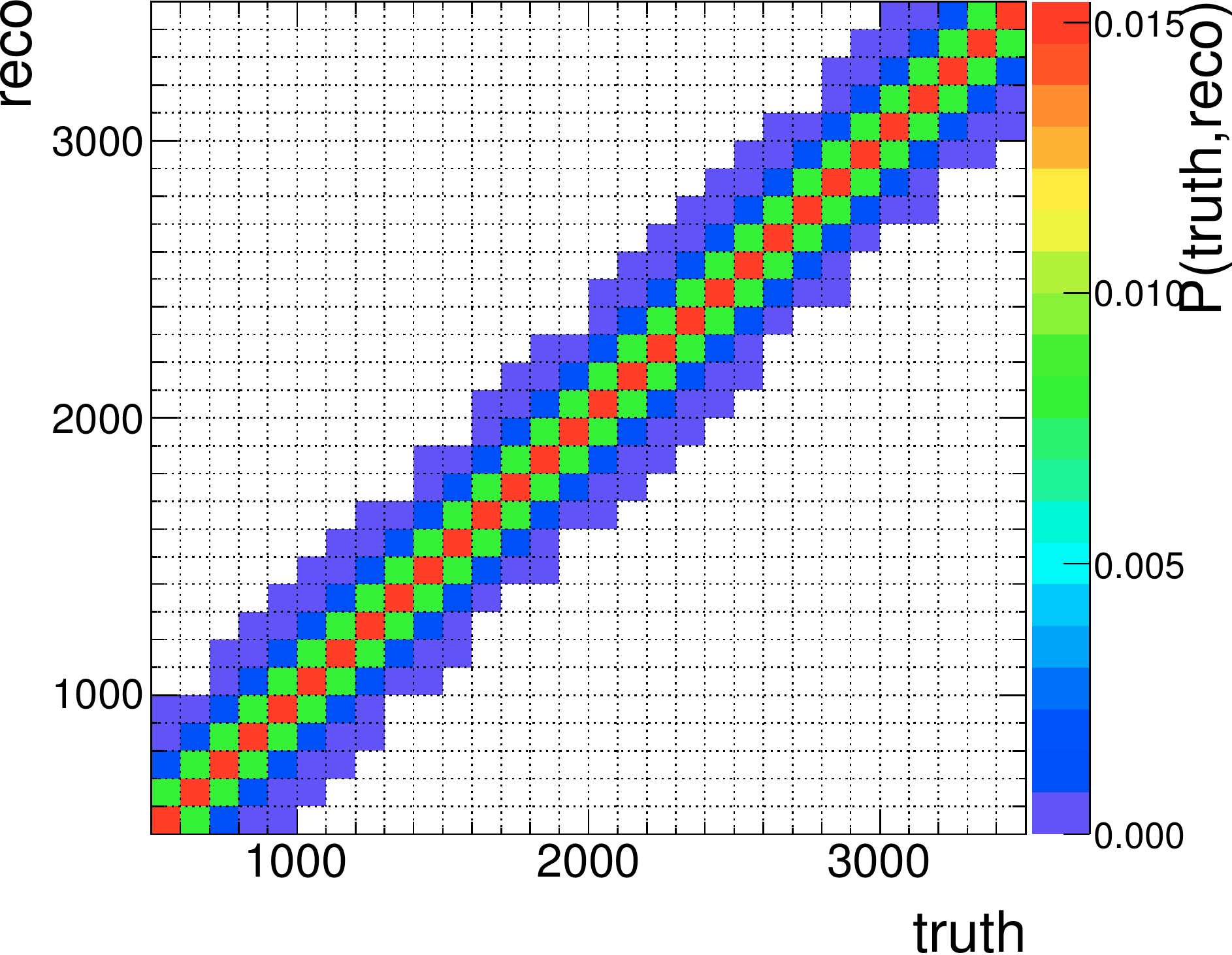} &
    \includegraphics[height=0.23\columnwidth]{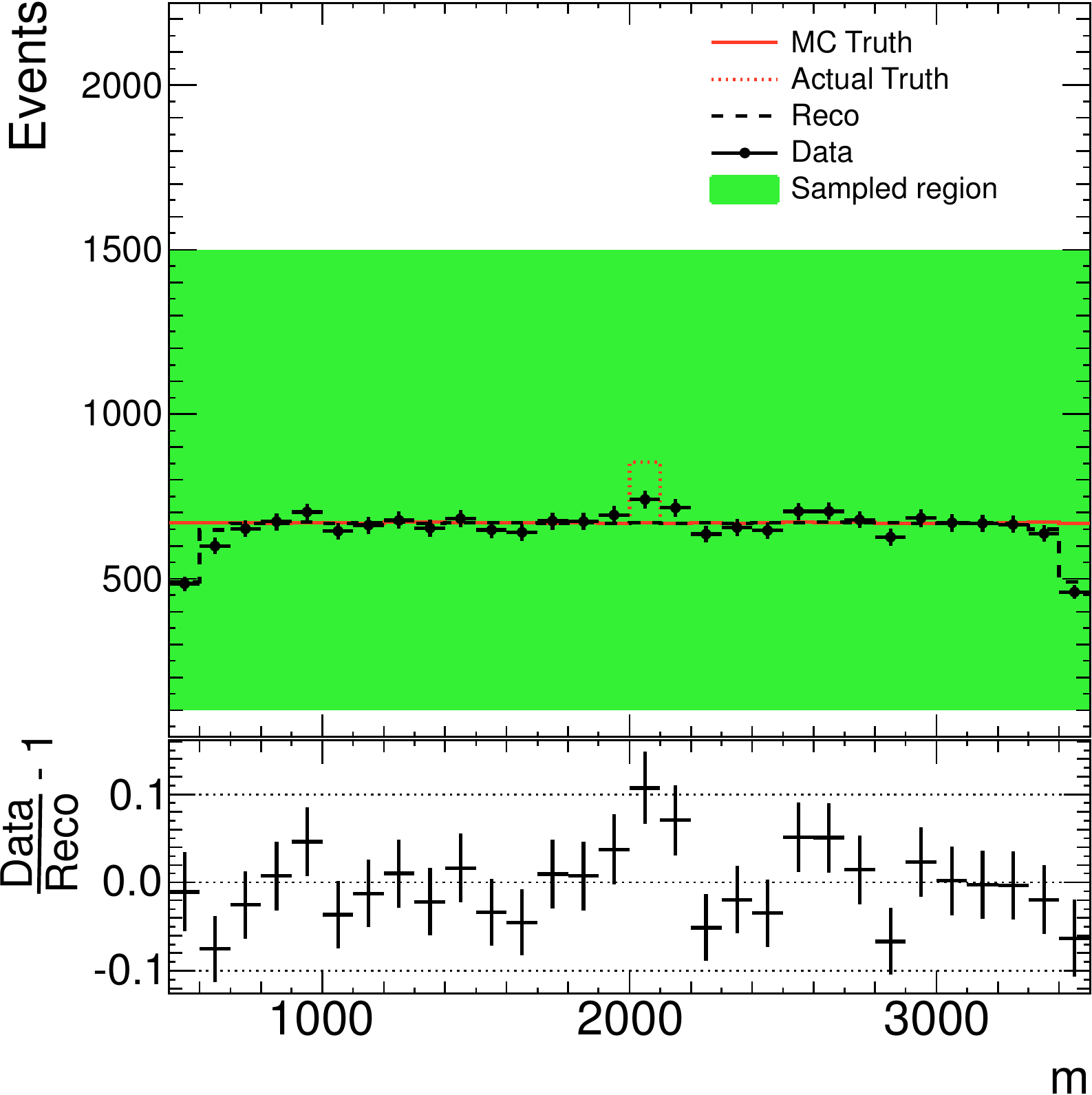} &
    \includegraphics[height=0.23\columnwidth]{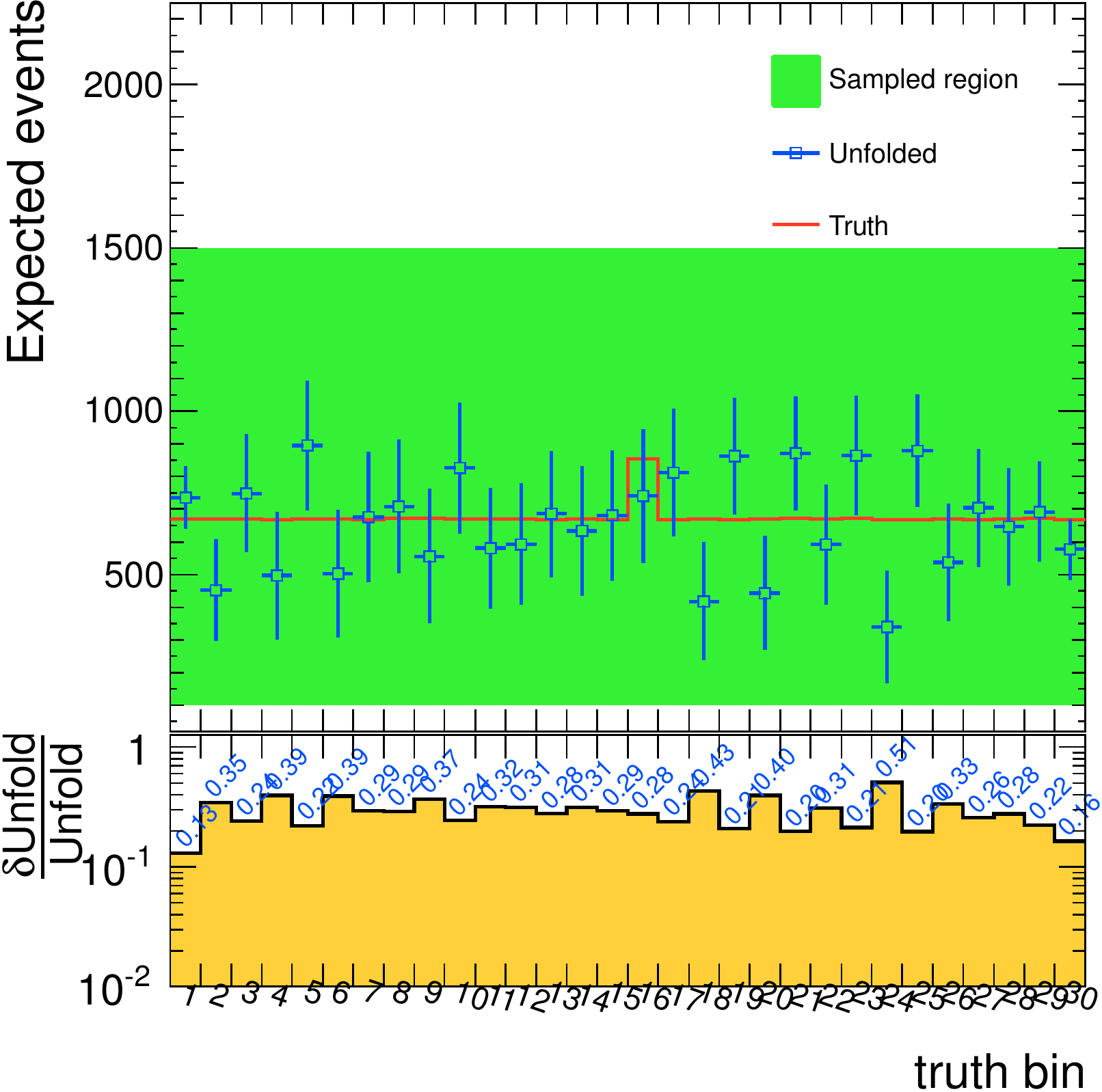} \\
    \includegraphics[height=0.23\columnwidth]{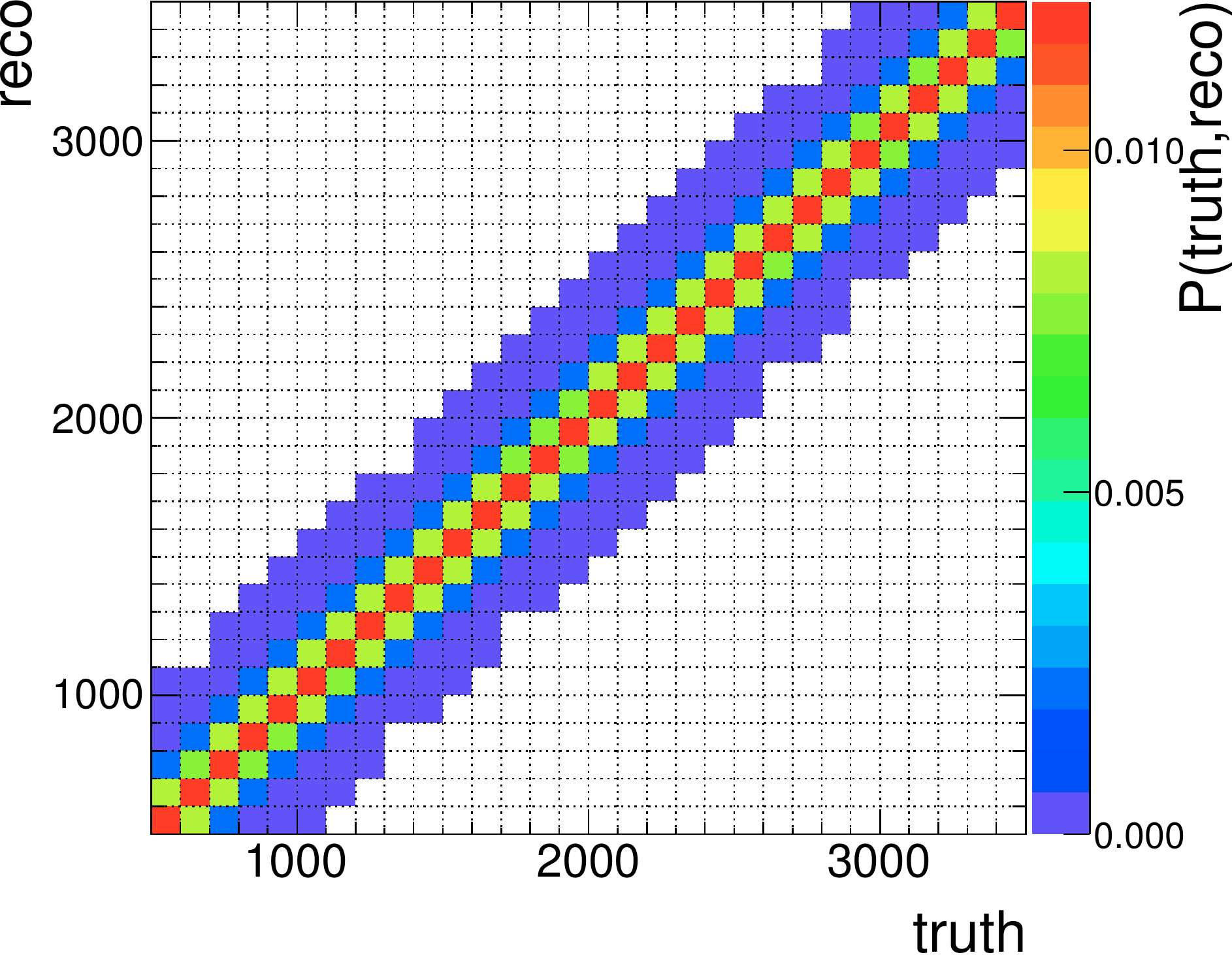} &
    \includegraphics[height=0.23\columnwidth]{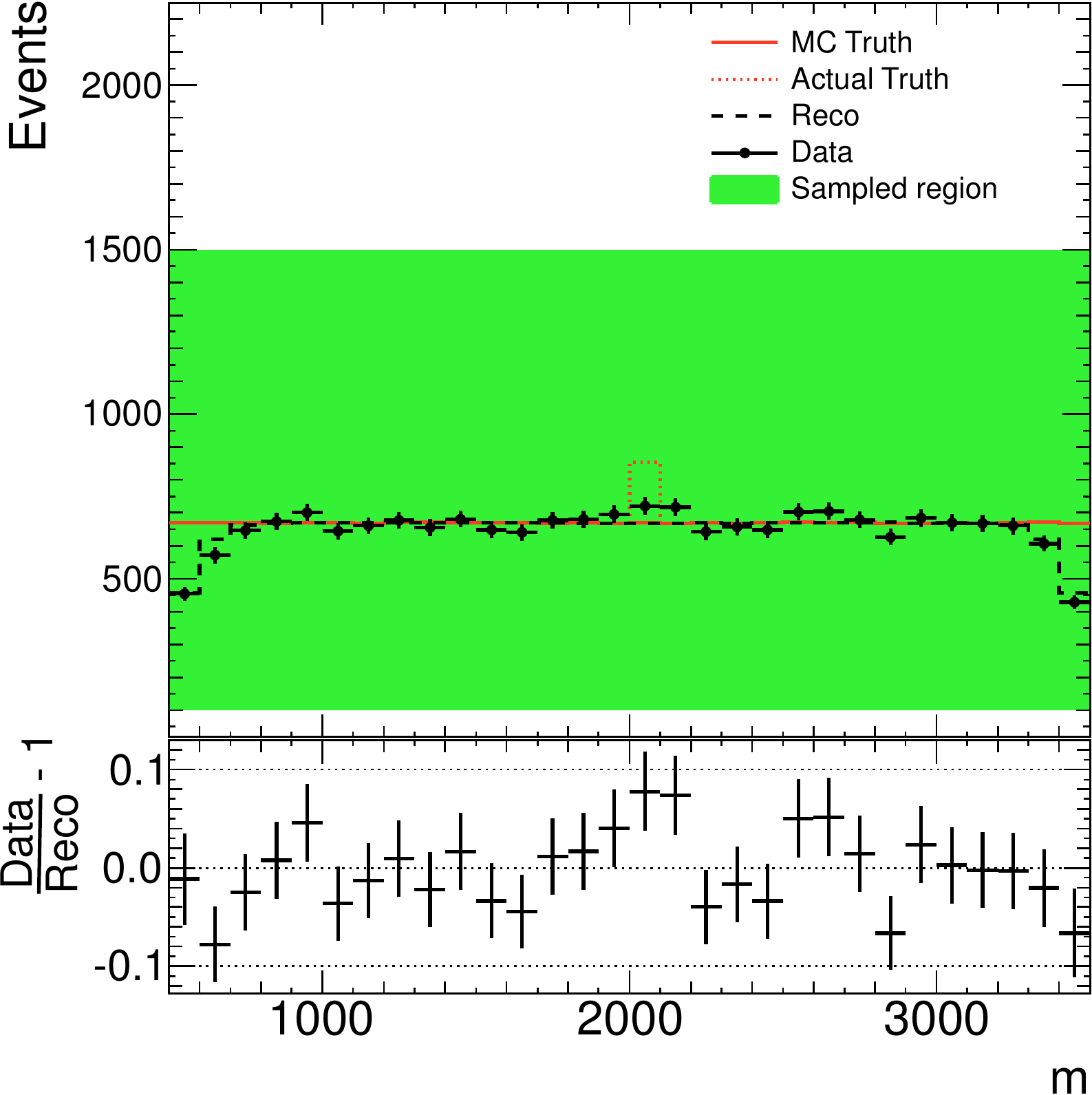} &
    \includegraphics[height=0.23\columnwidth]{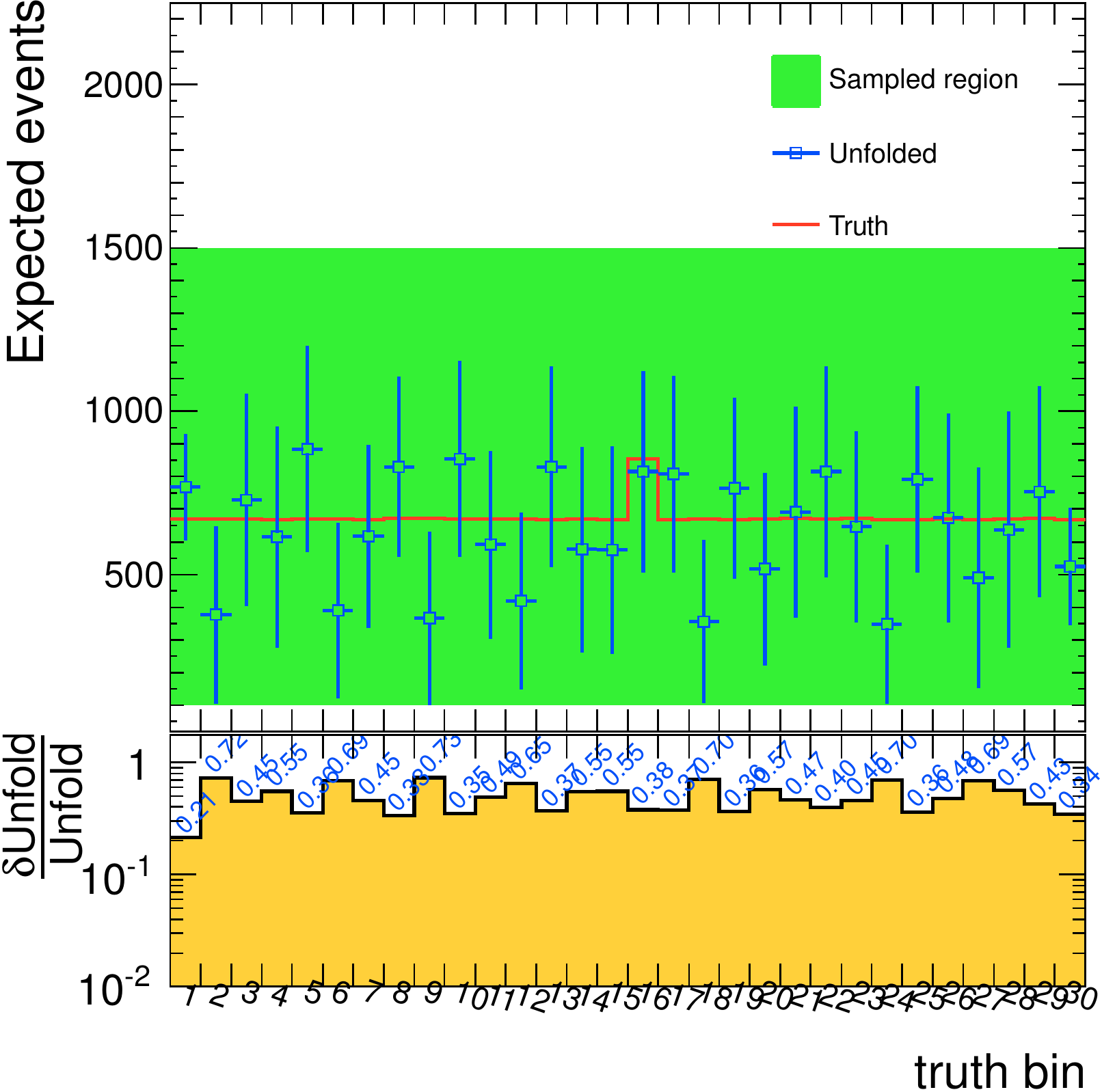} \\
    \includegraphics[height=0.23\columnwidth]{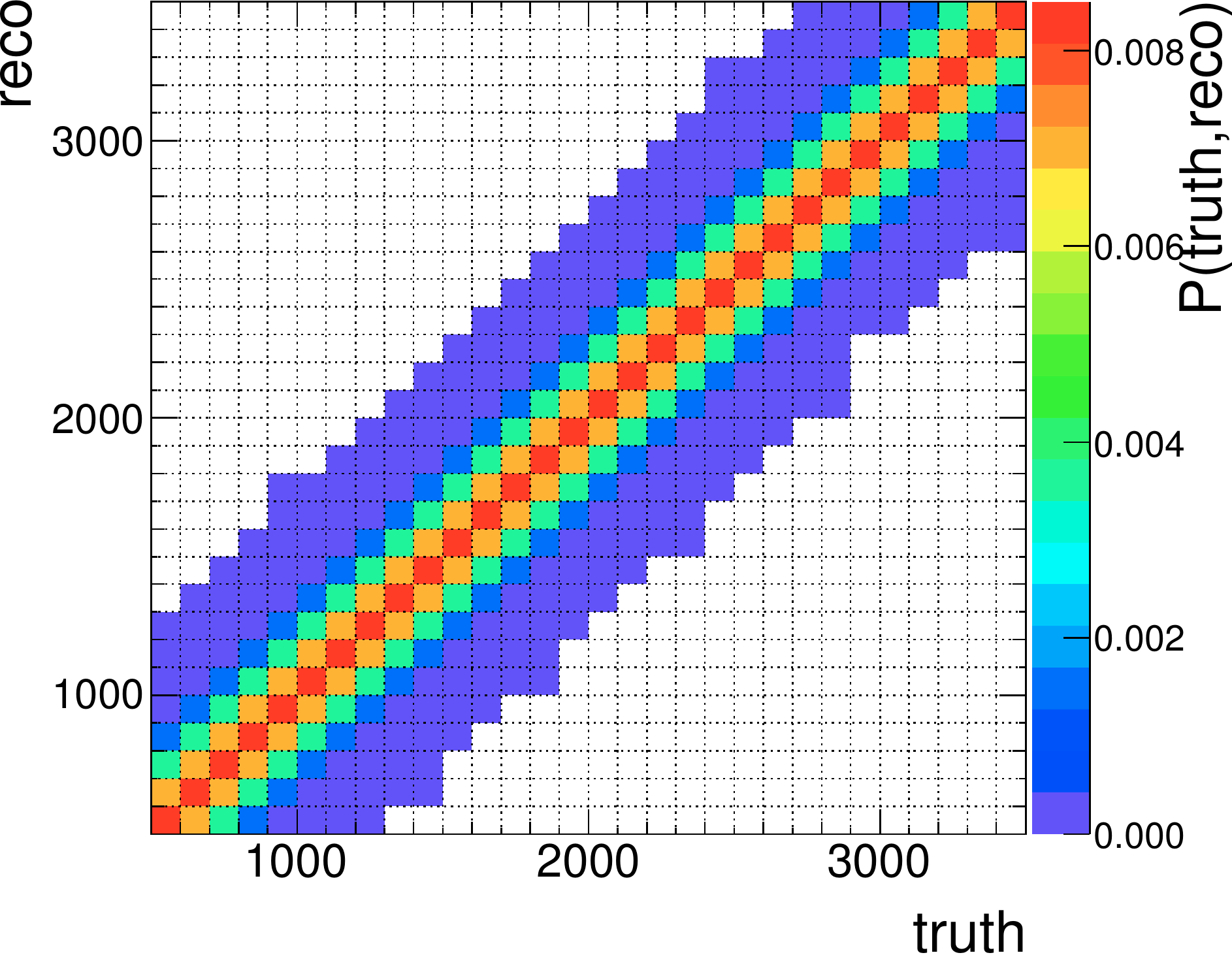} &
    \includegraphics[height=0.23\columnwidth]{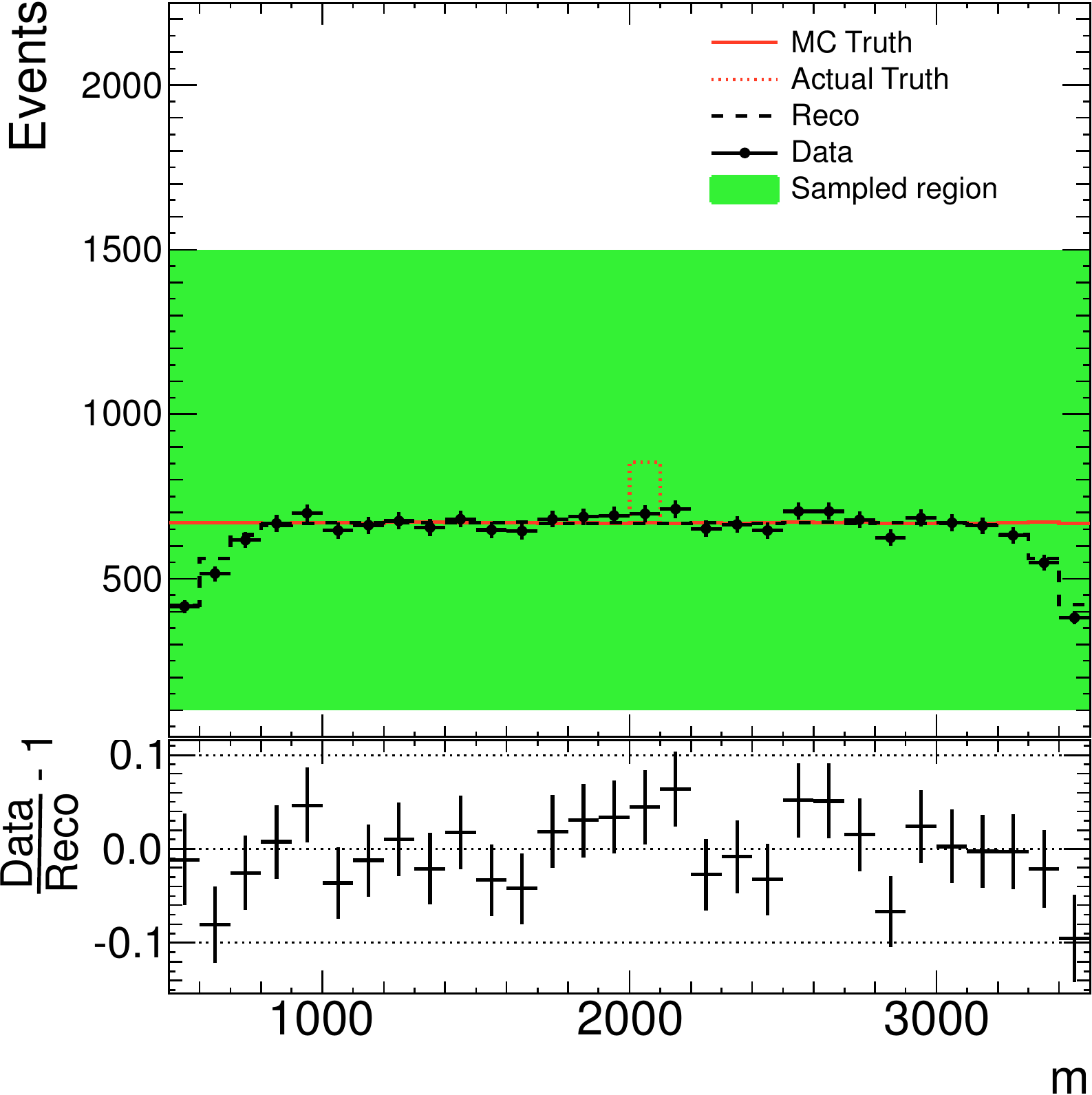} &
    \includegraphics[height=0.23\columnwidth]{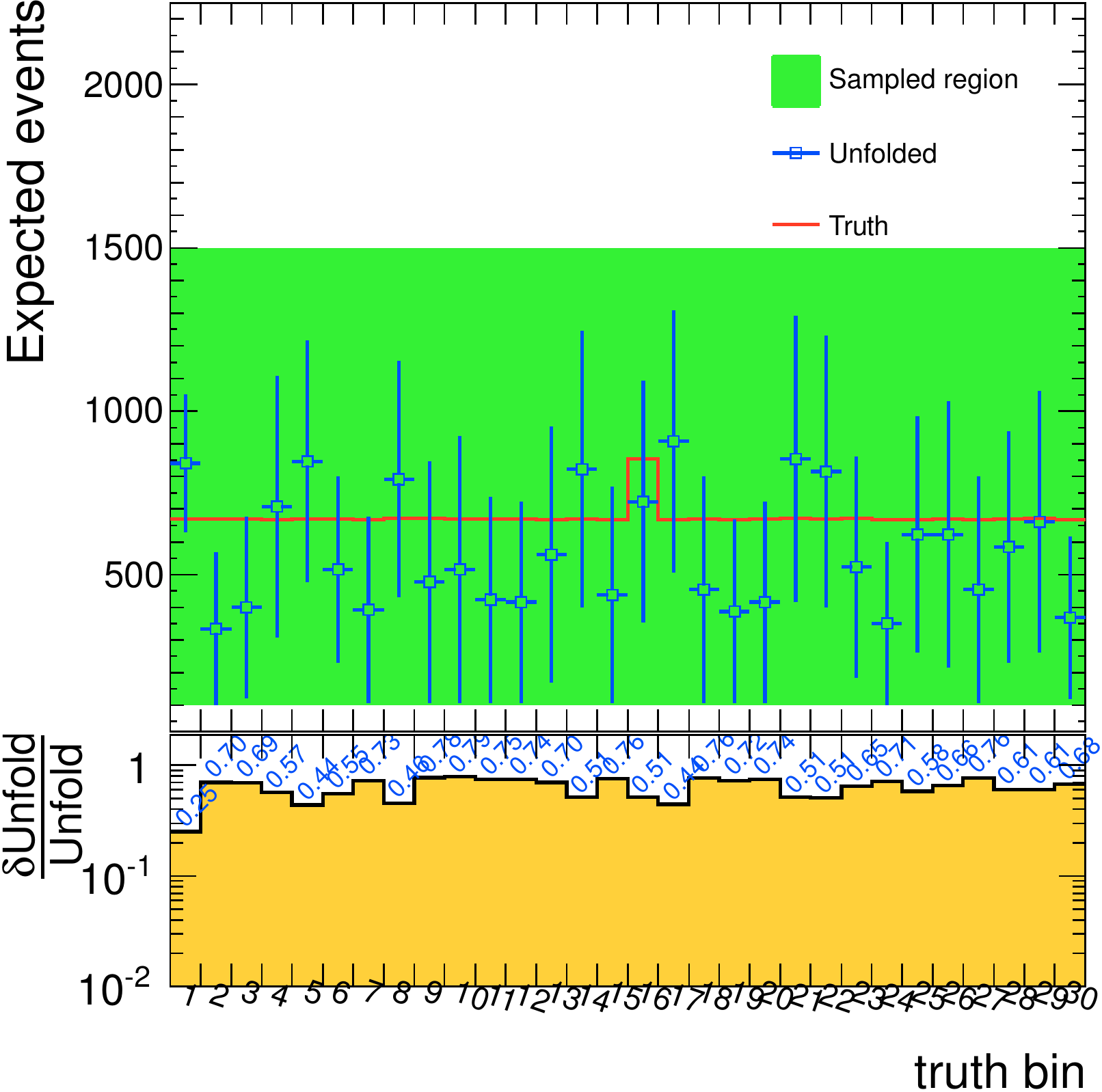} 
 \end{tabular}
\caption{Same as Fig.~\ref{fig:bumpUnknown}, but with a narrower unexpected bump in the data.  Details are given in Sec.~\ref{sec:bumpUnknown}.
\label{fig:bumpUnknown2}}
\end{figure}

Fig.~\ref{fig:bumpUnknown} and \ref{fig:bumpUnknown2} present the cases where the unexpected bump has width 100 and 5 respectively, analogously to Fig.~\ref{fig:bumpKnown} and \ref{fig:bumpKnown2} discussed in Sec.~\ref{sec:bumpKnown}.  The same amounts of smearing are shown in each row of these figures, namely $\sigma=\{0,50,75,100,150\}$.

From Fig.~\ref{fig:bumpUnknown} and \ref{fig:bumpUnknown2}, it seems that the unfolded spectrum maintains traces of the bump.  For a narrow bump, and $\sigma \ge 100$, it seems that the unfolded spectrum is enhanced in bin $t=16$, so, by eye at least, the feature is more distinguishable in the unfolded spectrum than in the data.  However, when smearing is greater, the uncertainty of the unfolded spectrum grows, making it impossible to see the feature.

FBU does not hide this unexpected feature, when performed without regularization, because Eq.~\ref{eq:R} does not use directly the elements of $\M$, $P(t,r)$, but instead the conditional probabilities $P(r|t)$, which are independent of the population in each truth-level bin $t$.  On the other hand, when smearing is significant, and no regularization is used, the posterior $p(\T|\D)$ can be so wide that the feature is obscured.

Later, in Sec.~\ref{sec:regSteepBump}, an example with a much larger unexpected bump on a steeply falling spectrum is given.  There, it will be more visible that the unfolded spectrum has a narrower bump than the reconstructed spectrum, so it undoes the effect of smearing in the bump, but for this to be visible the bump has to be so larger that the posterior's spread (see Fig.~\ref{fig:regGenSteepBumpA}, and the result with $\alpha=0$ in any of Fig.~\ref{fig:unfoldSteepBumpGaus}, \ref{fig:unfoldSteepBumpS3}, \ref{fig:unfoldSteepBumpS2}, or \ref{fig:unfoldSteepBumpS1}).

\section{Regularization}
\label{sec:regularization}

It is demonstrated in Sec.~\ref{sec:applications} that, when $\pi(\T)$ is constant within the sampled hyper-box, and smearing is significant, the posterior $p(\T|\D)$ is too widely spread.   So, the unfolded spectrum has large error bars.  Only if smearing is zero these error bars reduce to the level of $\sqrt{T_t}$.

From observation makes it clear that the posterior's information content is limited by (i) lack of data, and (ii) smearing.  
Just like limited data are an insurmountable limitation, so is smearing.  Information can not be recovered after smearing, just like it can not be made up if the data are limited.  \emph{Unless} information comes in from somewhere beyond the data and beyond what is known about migrations!  This is possible by shaping $\pi(\T)$, or through parametric estimation of $\hat{\T}$, i.e., by fitting a function through $\D$.

FBU makes it easy to try various regularization choices, without changing the formulation of the problem.  Choices are unlimited, but the following few are studied here:

\begin{enumerate}[i)]
\item $S(\T)$ is the entropy \cite{cowanUnfolding}, multiplied by $-1$ for reasons of convention explained below:
\begin{equation}
S_1(\T)\equiv - \left( -\sum_{t=1}^{N_t} \frac{T_t}{\sum T_t} \log \frac{T_t}{\sum T_t} \right).
\end{equation}
\item $S(\T)$ is the curvature (Eq.~39 in Ref~\cite{svd}):
\begin{equation}
S_2(\T)\equiv\sum_{t=2}^{N_t-1} \left( \Delta_{t+1,t} - \Delta_{t,t-1} \right)^2,
\end{equation}
where 
\begin{equation}
  \Delta_{t_1,t_2}\equiv T_{t_1}-T_{t_2}.
\end{equation}

\item $S(\T)$ is a function that sums up the relative variations of the first derivative, taking into account the possibly varying bin sizes:
\begin{equation}
S_3(\T)\equiv \sum_{t=2}^{N_t-1} \frac{|\delta_{t+1,t}-\delta_{t,t-1}|}{|\delta_{t+1,t}+\delta_{t,t-1}|},
\end{equation}
where 
\begin{equation}
  \delta_{t_1,t_2}\equiv \frac{\frac{T_{t_1}}{\mathcal{W}_{t_1}} - \frac{T_{t_2}}{\mathcal{W}_{t_2}}}{\mathcal{C}_{t_1}-\mathcal{C}_{t_2}},
\end{equation}
where 
\begin{equation}
  \mathcal{W}_t \equiv m_{t}-m_{t-1}
\end{equation}
is the width of bin $t$, and
\begin{equation}
  \mathcal{C}_t \equiv \frac{1}{2}(m_{t}+m_{t-1})
\end{equation}
is the center of bin $t$.



\item \label{item:gaussian} The prior is proportional to a multivariate Gaussian ($\pi_G$), without correlations, which disfavors $\T$-points far from $\tilde{T}$ (Sec.~\ref{sec:nomenclature}):
  \begin{equation}
    \pi_G(\T) = \prod_{t=1}^{N_t} \exp{\frac{(T_t - \tilde{T}_t)^2}{2(\tilde{T}_t / \alpha)^2}}.
  \end{equation}
In this case, the parameter $\alpha$ adjusts the RMS of the Gaussian, which is set to $\tilde{T}_t / \alpha$ in dimension $t$.

\end{enumerate}

In all the above cases, except for the Gaussian prior (\ref{item:gaussian}), within the sampled hyper-box the prior is given by Eq.~\ref{eq:priorMinusAlpha}.

When $\alpha= 0$, no regularization applies.  The larger the $\alpha$, the stronger the prior belief that $S(\T)$ must be small (or, in case \ref{item:gaussian}, the stricter the Gaussian constraint).  It is informative to try various values.  The result may not be satisfactory, if the bias introduced is unacceptable and the uncertainty reduction is small.  An exception is the choice \ref{item:gaussian}, where, if $\tilde{\T}$ happens to be the correct truth-level spectrum ($\hat{\T}$), then larger $\alpha$ values reduce both the posterior uncertainty and bias; the posterior is concentrated closer to $\hat{\T}$.  Of course, it is difficult to trust that $\tilde{\T} = \hat{\T}$, because that presumes a perfect MC simulation, and absence of exotic processes.

In the interest of speed, the sampling method in the following regularization examples is MCMC, without volume reduction.  The 1-dimensional marginal distributions of $p(\T|\D)$ are smooth, indicating that this approximation is satisfactory, as shown in Sec.~\ref{sec:example5}.


\subsection{Steeply falling spectrum}
\label{sec:regSteep}

Sections \ref{sec:regSteepNoSmearing} and \ref{sec:regSteepSmearing} assume a steeply falling spectrum with $N_t$=14 bins, as in Sec.~\ref{sec:generation}.  In Sec.~\ref{sec:regSteepNoSmearing} no smearing is assumed, and in \ref{sec:regSteepSmearing} smearing is applied with parameters $(a,b)=(0.5,0.1)$, as in Sec.~\ref{sec:generation}.  Multiple regularization attempts will be made in each case, in an attempt to build intuition.

\subsubsection{Without smearing}
\label{sec:regSteepNoSmearing}

\begin{figure}[H]
  \centering
   \includegraphics[width=0.45\columnwidth]{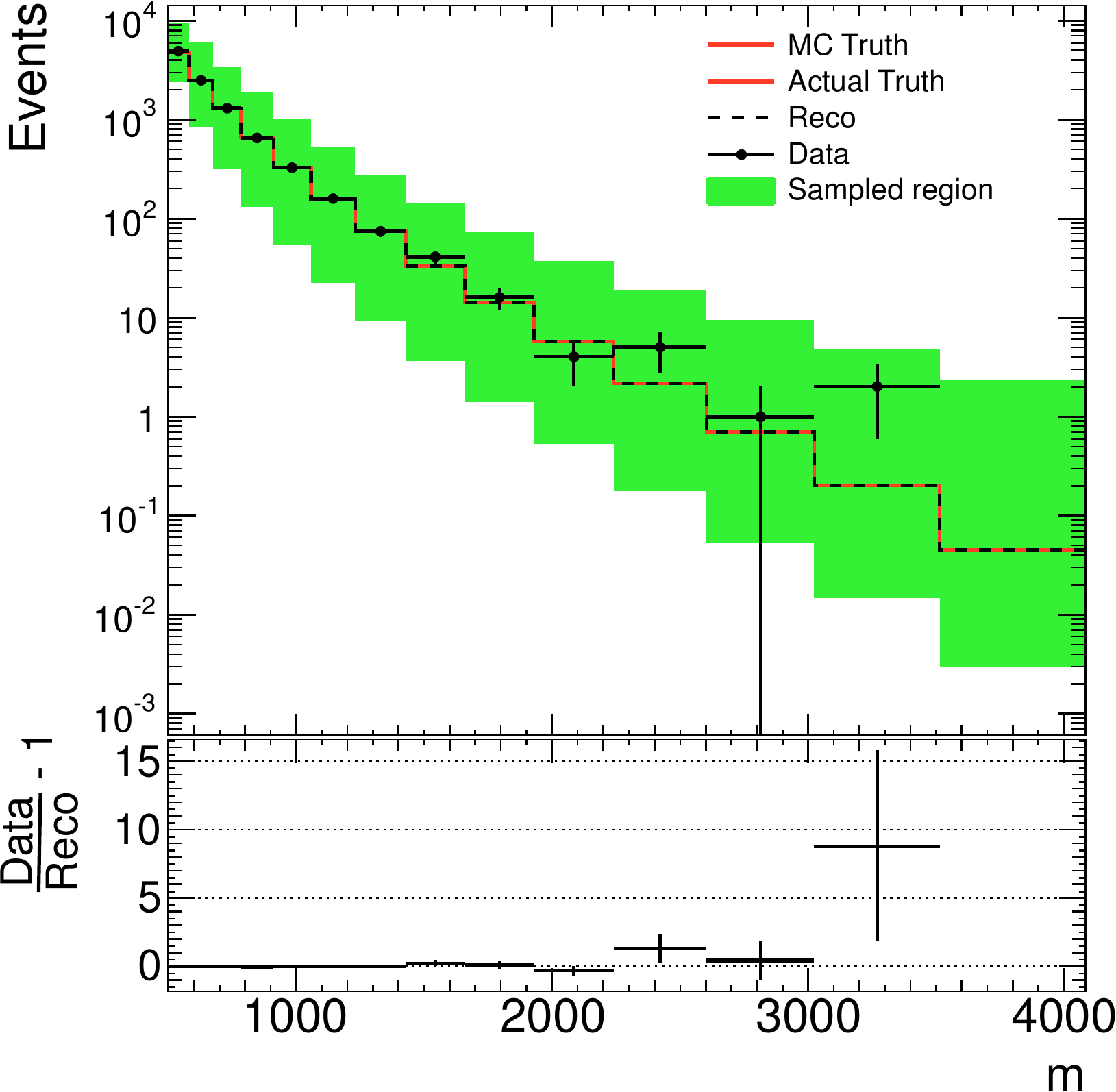}
\caption{The input data, truth-level, and reco-level spectrum used in the examples of Sec.~\ref{sec:regSteepNoSmearing}.   Since no smearing is modeled, the reco spectrum is identical to the truth, and since no unexpected processes are assumed, the actual truth $\hat{\T}$ is identical to the MC truth spectrum $\tilde{\T}$.  The sampled region is shown, which is sampled with MCMC, without need for volume reduction.
\label{fig:spectrumSteepNoSmear}
}
\end{figure}

The spectrum used in this section is in Fig.~\ref{fig:spectrumSteepNoSmear}, where no smearing is assumed.

\begin{figure}[H]
  \centering
  \subfigure[$\alpha=0$]{
    \includegraphics[width=0.3\columnwidth]{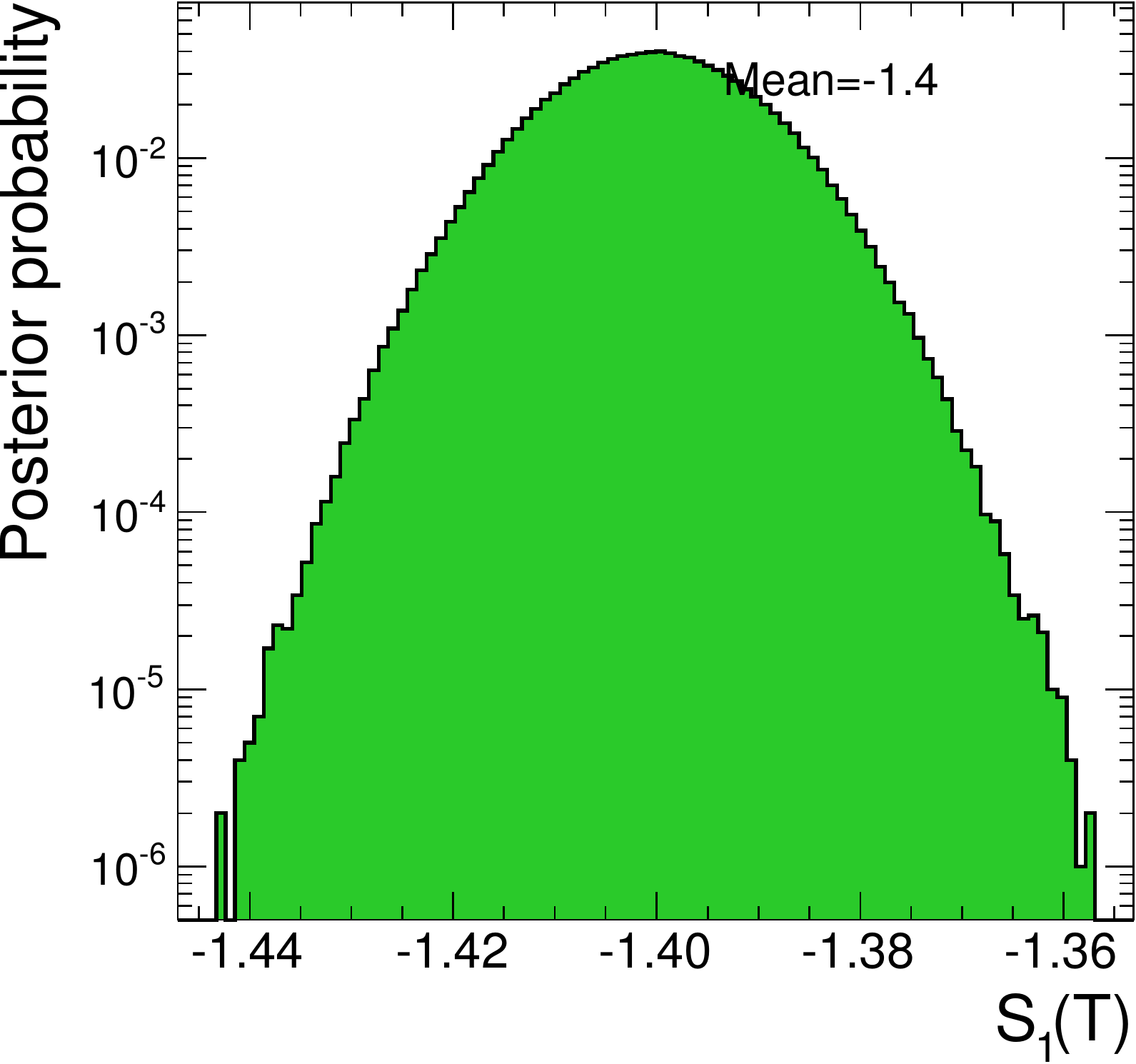}
  }
  \subfigure[$\alpha=1\times 10^3$]{
    \includegraphics[width=0.3\columnwidth]{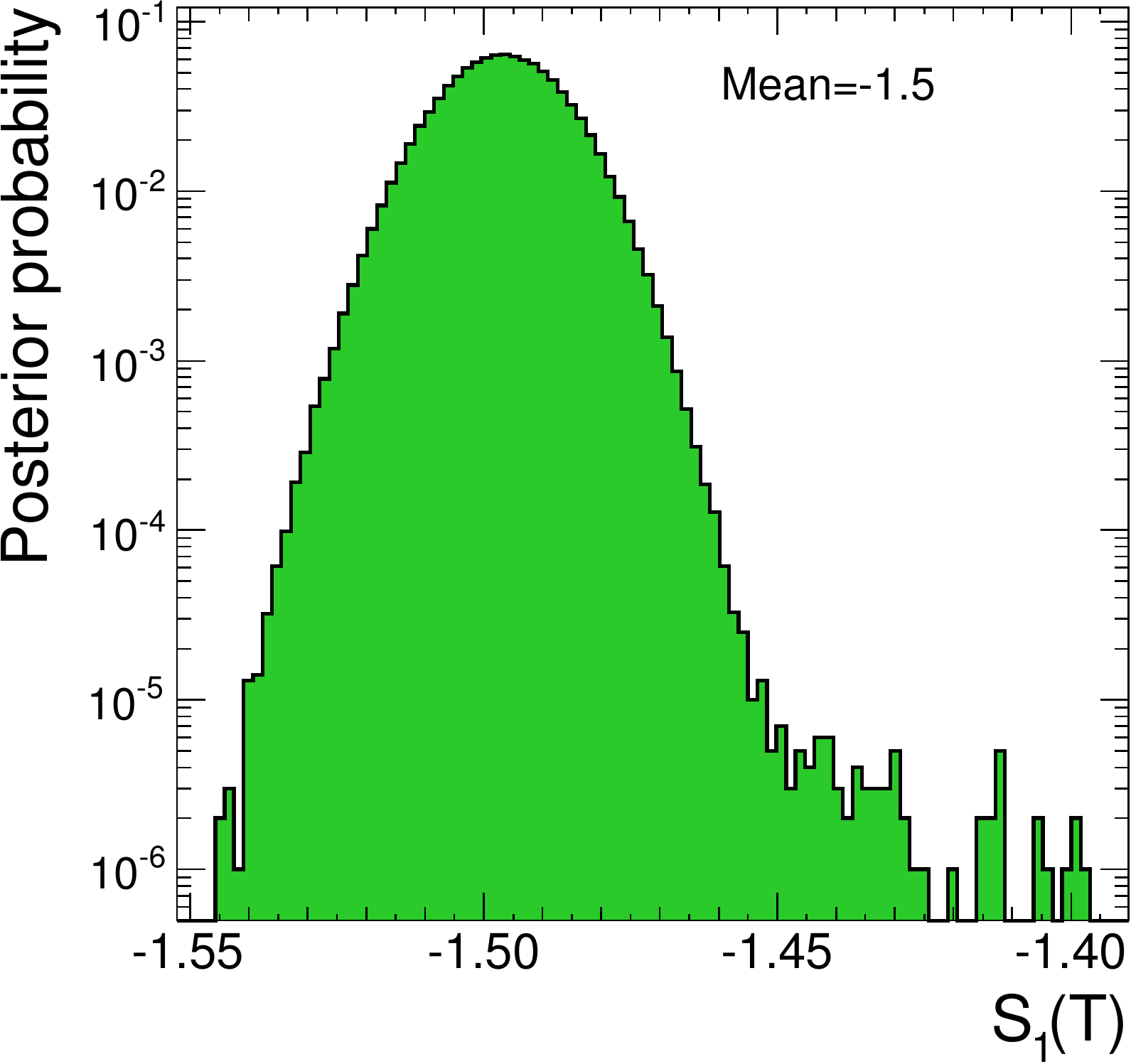}
  }
  \subfigure[$\alpha=3\times 10^3$]{
    \includegraphics[width=0.3\columnwidth]{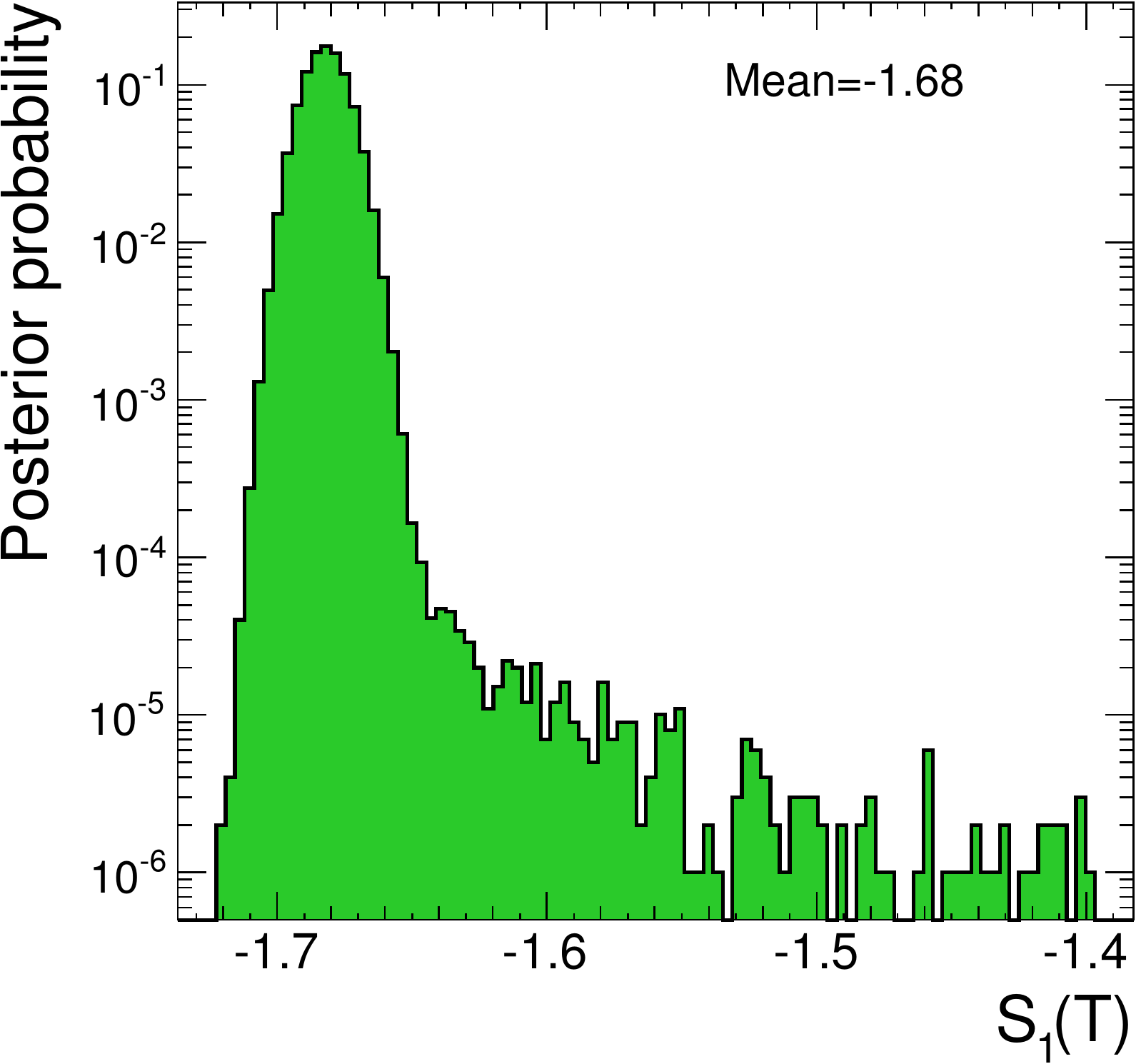}
  }
\caption{The posterior $P(S_1(\tuple{T})|\tuple{D})$, for three different choices of the regularization parameter $\alpha$, corresponding to Sec.~\ref{sec:regSteepNoSmearing}.  For $\alpha=0$, $p(\tuple{T}|\tuple{D})$ is unaffected by regularization.  As the regularization constraint becomes stronger, the posterior $p(\tuple{T}|\tuple{D})$ is ``pushed'' towards $\tuple{T}$-points which give smaller $S_1(\tuple{T})$.  The sampling method is MCMC, and the small tails in (b) and (c) are reflecting the part of the MCMC random walk before it had reached the vicinity of the most likely $\tuple{T}$.
\label{fig:regFuncSteepNoSmearS1}
}
\end{figure}

\begin{figure}[H]
  \centering
  \subfigure[$\alpha=0$]{
    \includegraphics[width=0.3\columnwidth]{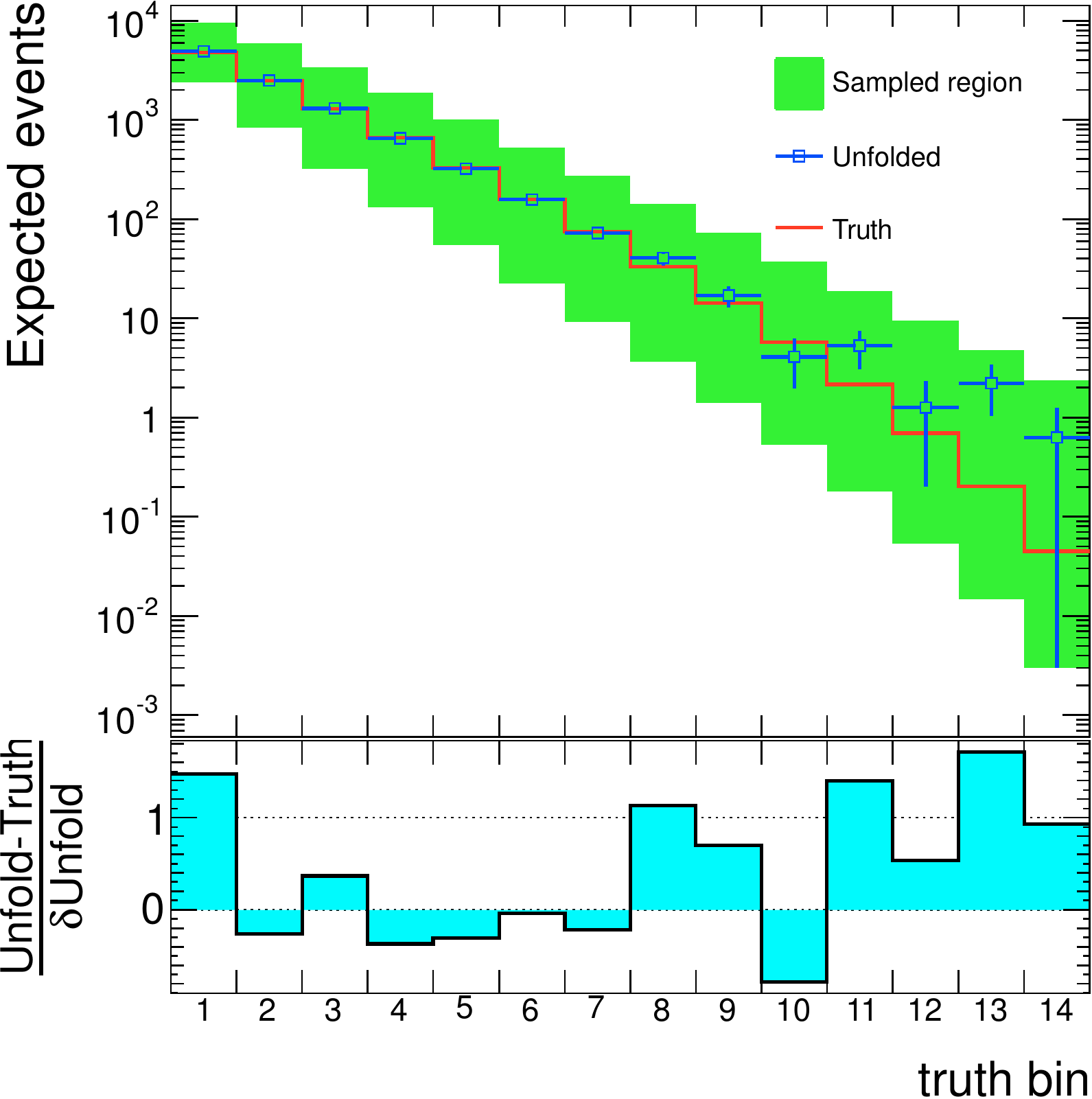}
  }
  \subfigure[$\alpha=1\times 10^3$]{
    \includegraphics[width=0.3\columnwidth]{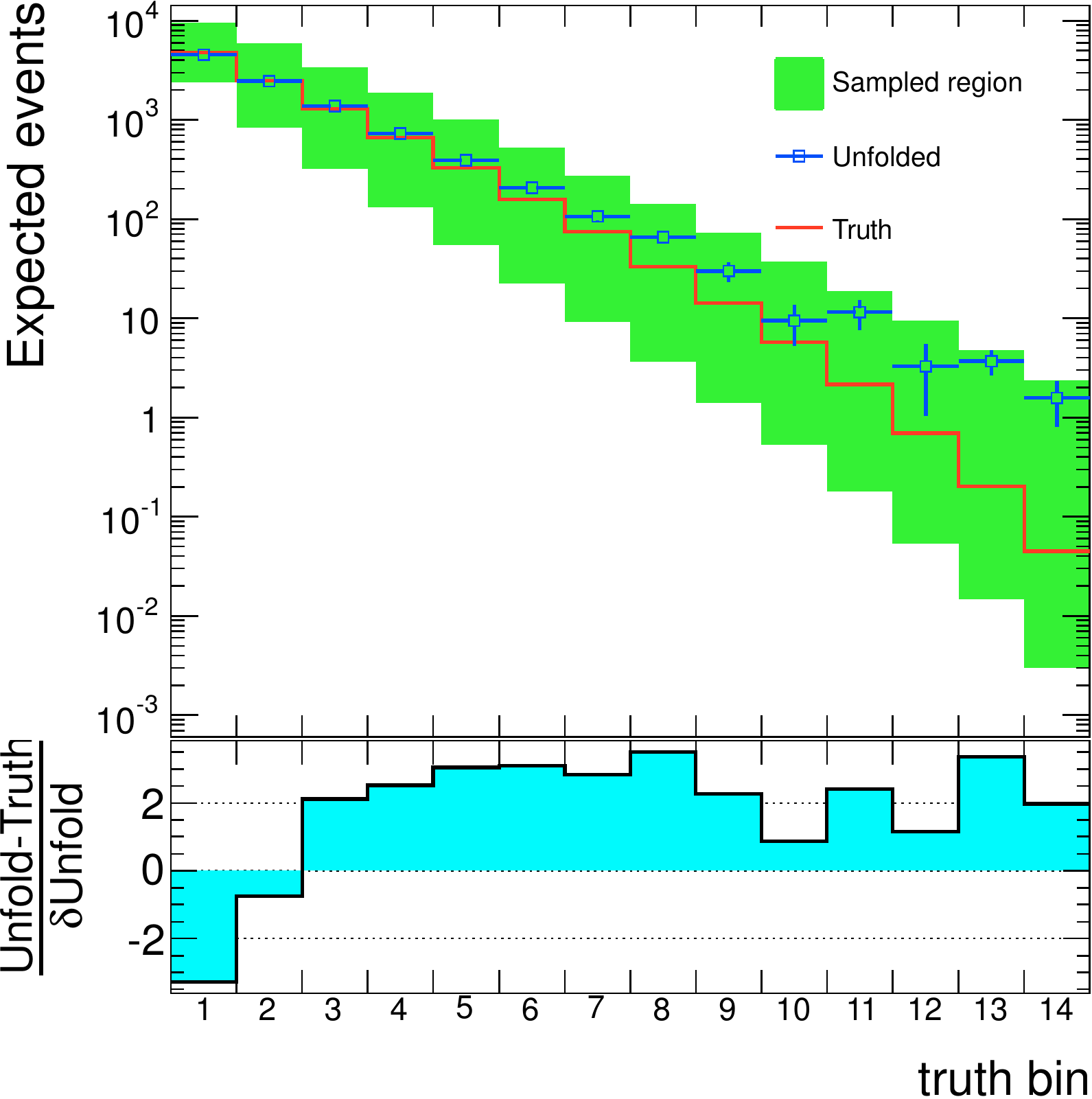}
  }
  \subfigure[$\alpha=3\times 10^3$]{
    \includegraphics[width=0.3\columnwidth]{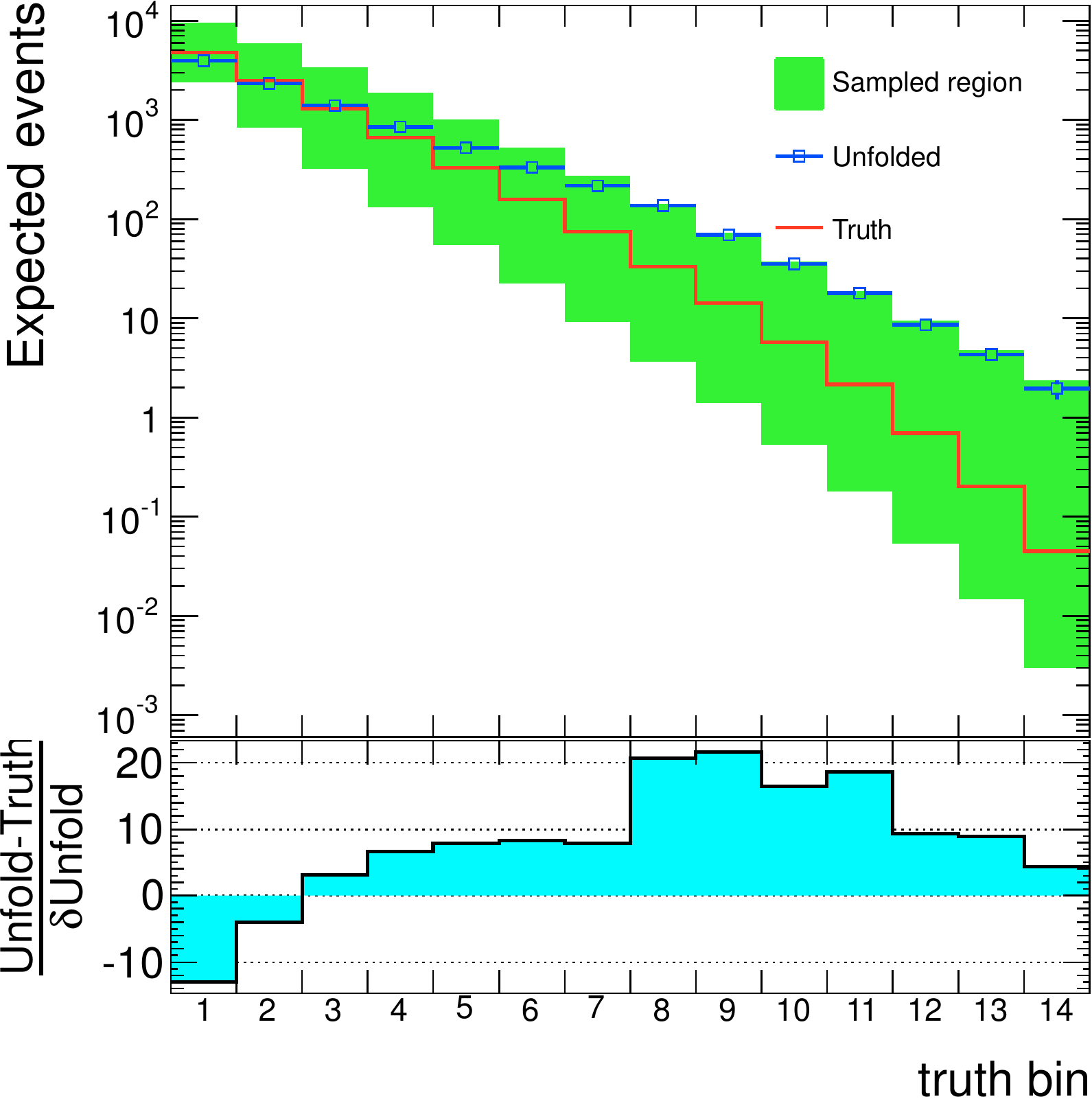}
  }\\
 \subfigure[$\alpha=0$]{
    \includegraphics[width=0.3\columnwidth]{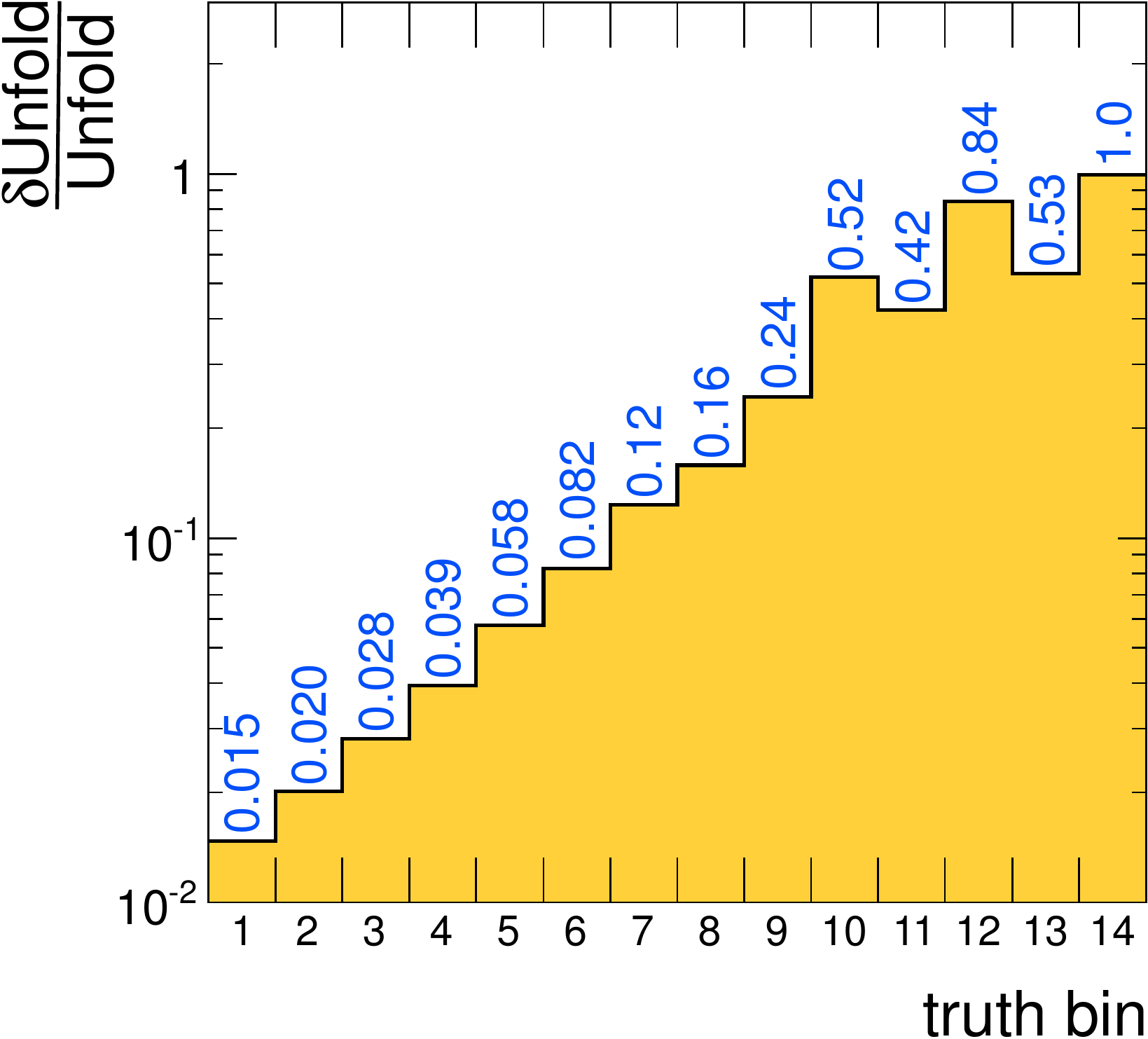}
  }
  \subfigure[$\alpha=1\times 10^3$]{
    \includegraphics[width=0.3\columnwidth]{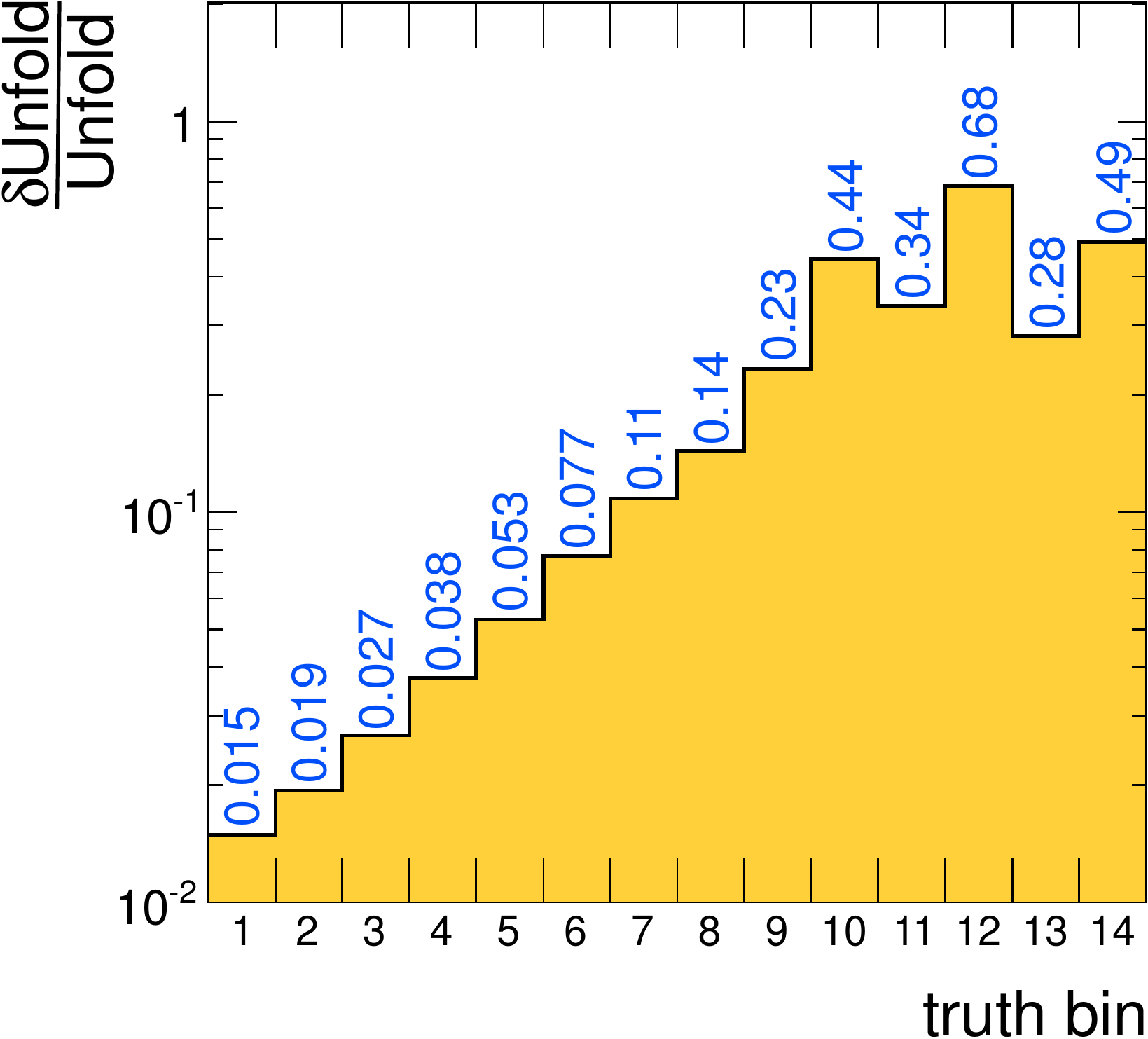}
  }
  \subfigure[$\alpha=3\times 10^3$]{
    \includegraphics[width=0.3\columnwidth]{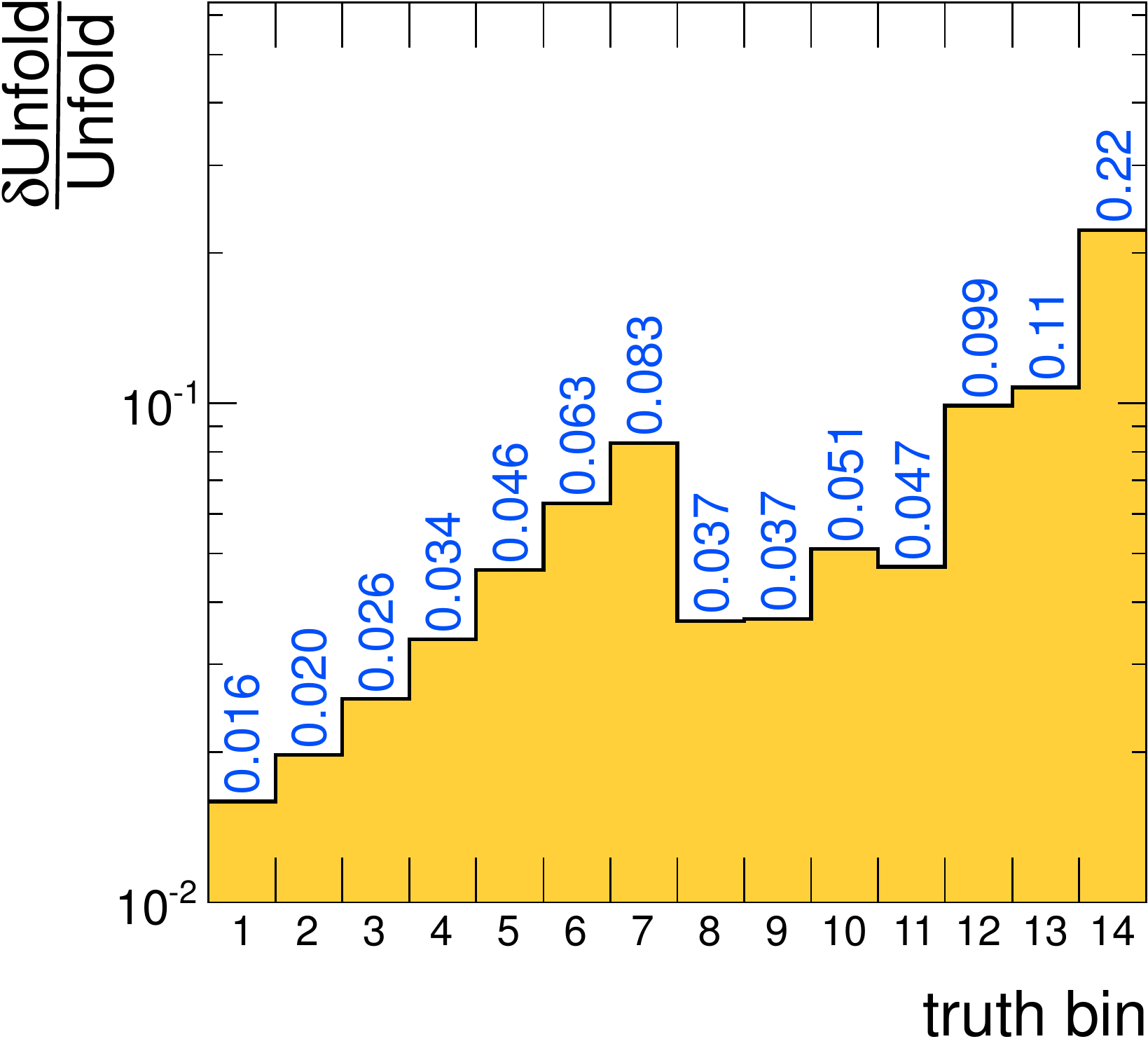}
  }
  \caption{The result of unfolding of Sec.~\ref{sec:regSteepNoSmearing}, with regularization function $S_1$, for three values of $\alpha$.  The upper row (a,b,c) demonstrates that the bias of the unfolded spectrum increases with increasing regularization strength.
   The lower row (d,e,f) shows the relative uncertainty of the unfolded spectrum ($\frac{|U_t^\ulcorner-U_t^\urcorner|}{U_t^\ulcorner+U_t^\urcorner}$), which is expected to reduce with increasing $\alpha$. There is a tiny reduction for $\alpha=10^3$, at the cost of considerable bias.  For $\alpha=3\times 10^3$, the posterior is forced to be more constant, which maximizes entropy, but is limited by the upper edge of the sampled hyper-box, so, the apparent reduction of uncertainty is simply because the posterior is squeezed against that edge (see Fig.~\ref{fig:1DimSteepNoSmear}, 3rd row).
\label{fig:unfoldedSteepNoSmearS1}
}
\end{figure}

The first attempt is to use $S_1$ (Sec.~\ref{sec:regularization}).  The parameter $\alpha$ is varied, from 0 to $3\times 10^3$.  Fig.~\ref{fig:regFuncSteepNoSmearS1} demonstrates that increasing $\alpha$ results in a $p(\T|\D)$ that favors $\T$ points with smaller $S_1(\T)$.   The resulting unfolded spectra are shown in Fig.~\ref{fig:unfoldedSteepNoSmearS1}.  
The posterior becomes highly biased with respect to the actual truth-level spectrum ($\hat{\T}$), and the reduction in its spread is small.\footnote{In some bins, the uncertainty even increases, which could be due to statistical fluctuations in sampling.}  It seems that, when there is no smearing to enhance the spread of the posterior, there is not much uncertainty for regularization to reduce.  In Sec.~\ref{sec:regSteepSmearing}, where smearing is activated, the reduction in posterior spread is significant.

\begin{figure}[H]
  \centering
  \subfigure[$\alpha=0$]{
    \includegraphics[width=0.3\columnwidth]{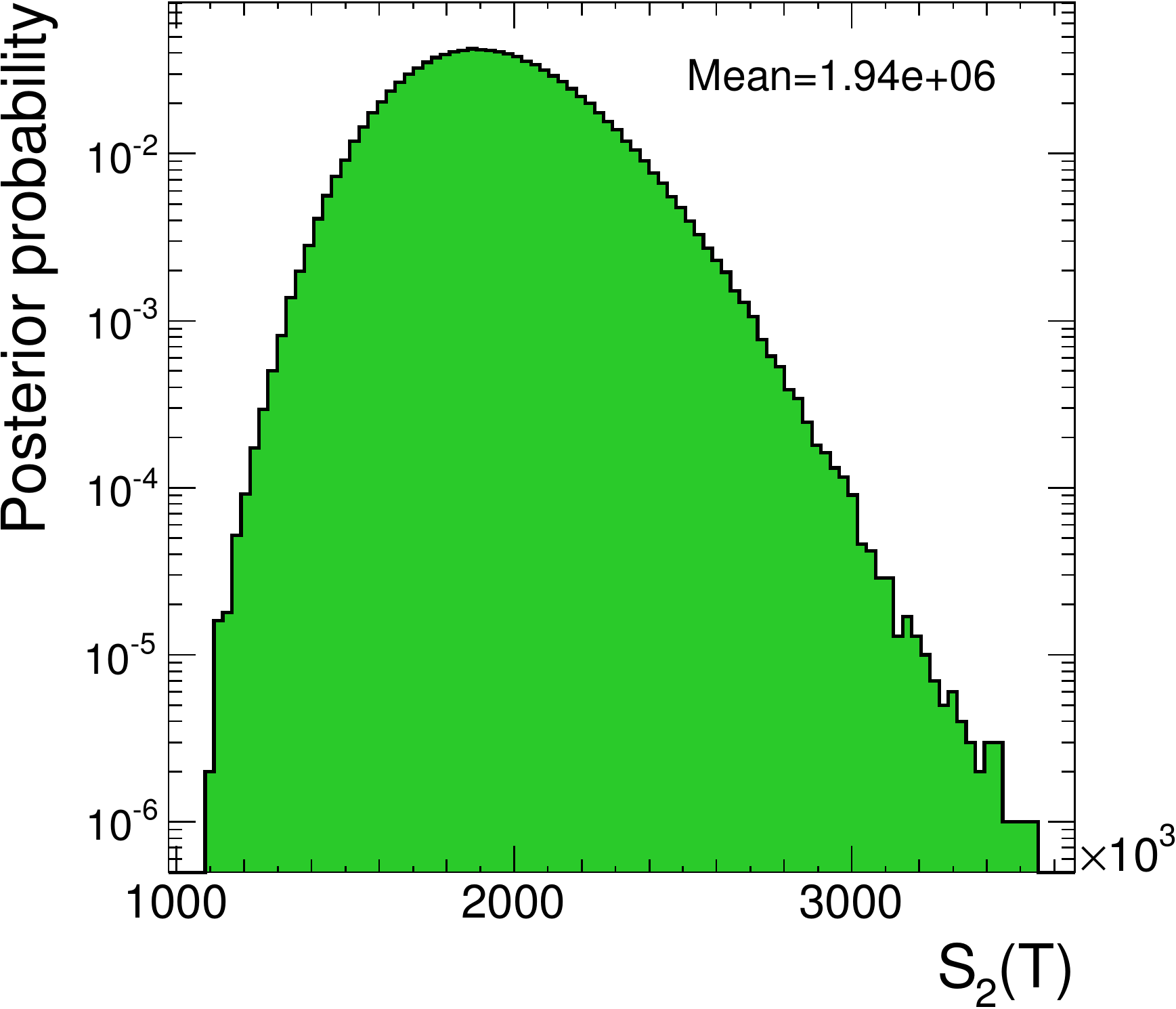}
  }
  \subfigure[$\alpha=3\times 10^{-4}$]{
    \includegraphics[width=0.3\columnwidth]{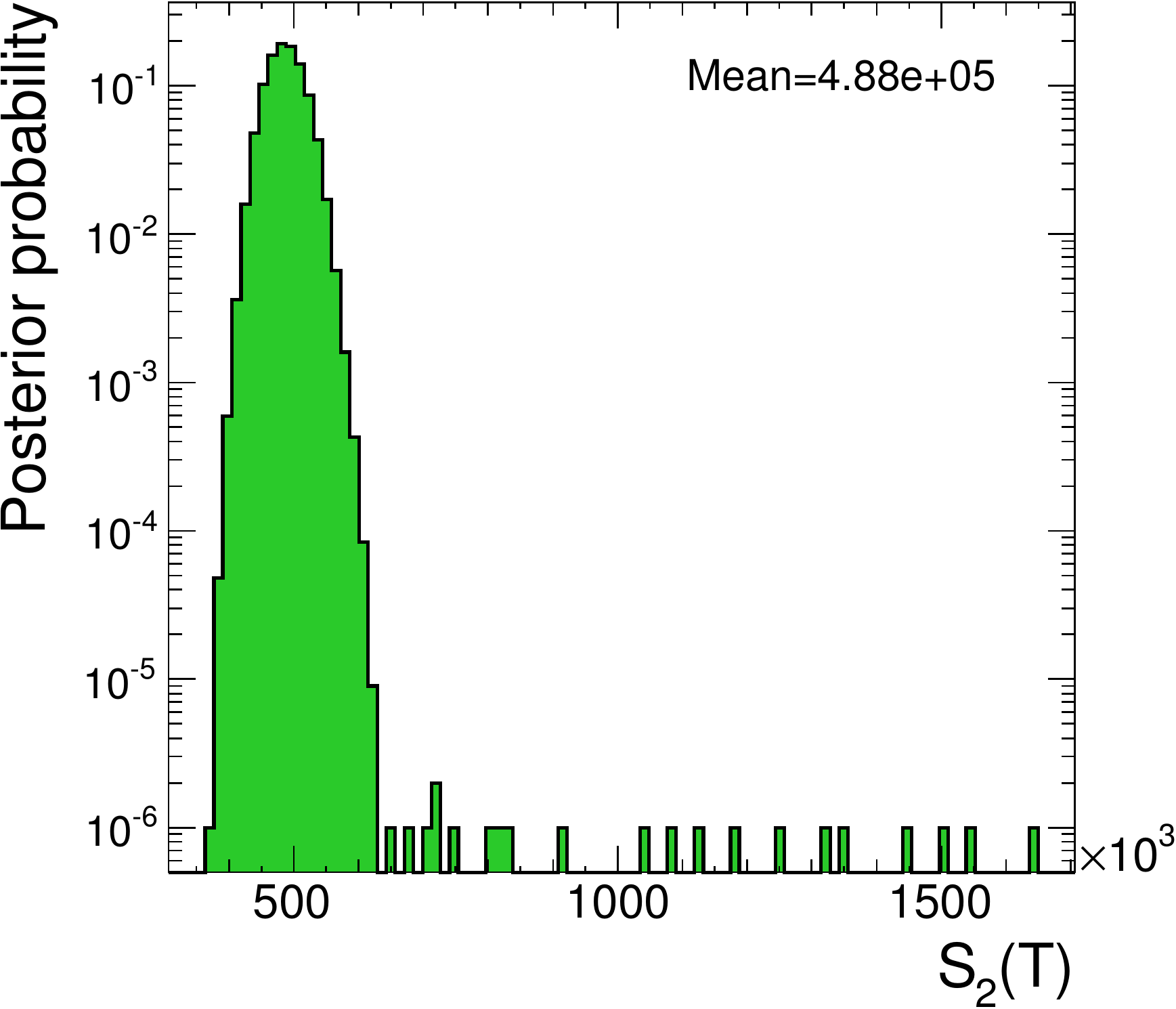}
  }
  \subfigure[$\alpha=6\times 10^{-4}$]{
    \includegraphics[width=0.3\columnwidth]{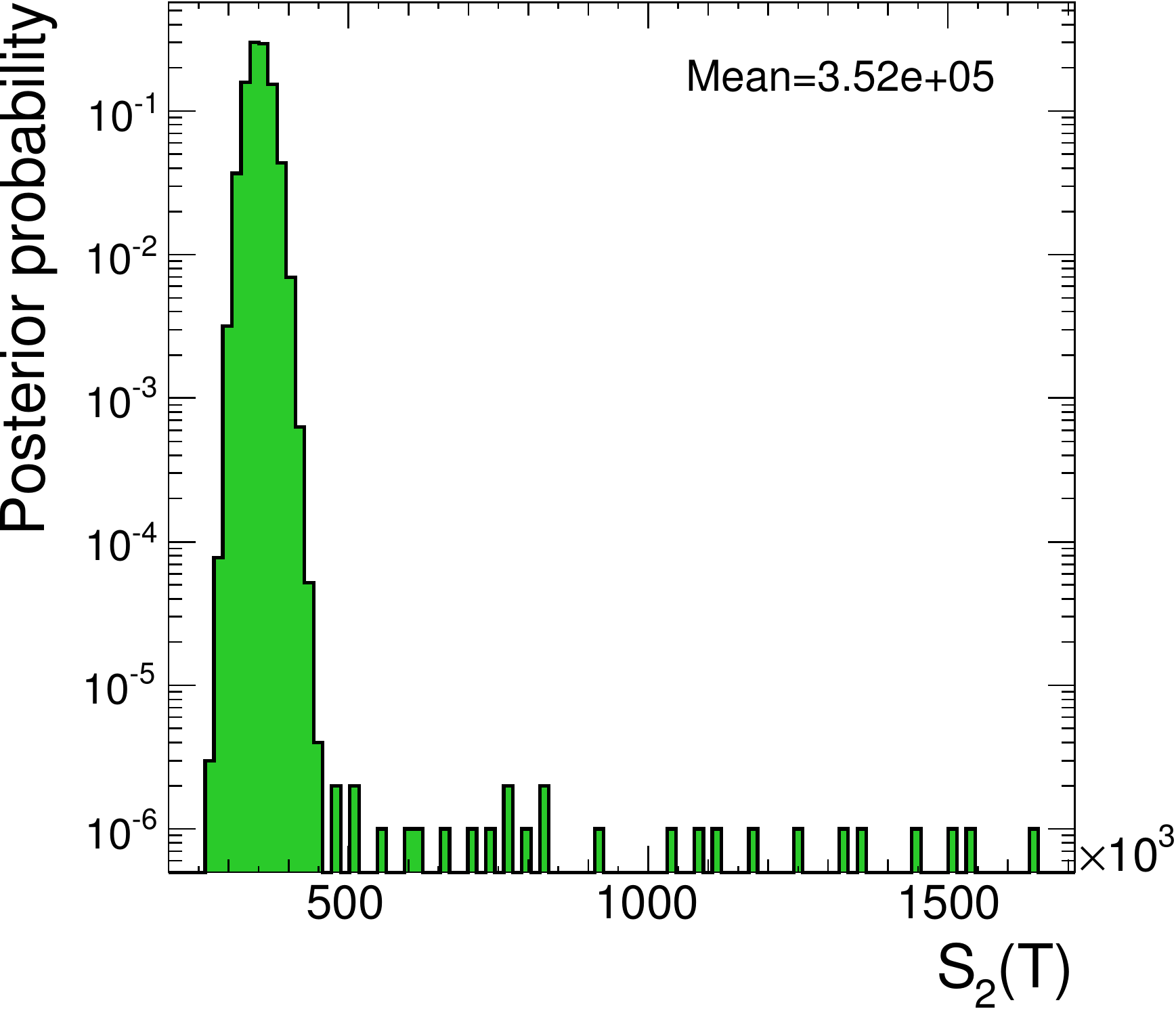}
  }
\caption{The posterior $P(S_2(\tuple{T})|\tuple{D})$, for three different choices of the regularization parameter $\alpha$, corresponding to Sec.~\ref{sec:regSteepNoSmearing}.
\label{fig:regFuncSteepNoSmearS2}
}
\end{figure}

\begin{figure}[H]
  \centering
  \subfigure[$\alpha=0$]{
    \includegraphics[width=0.3\columnwidth]{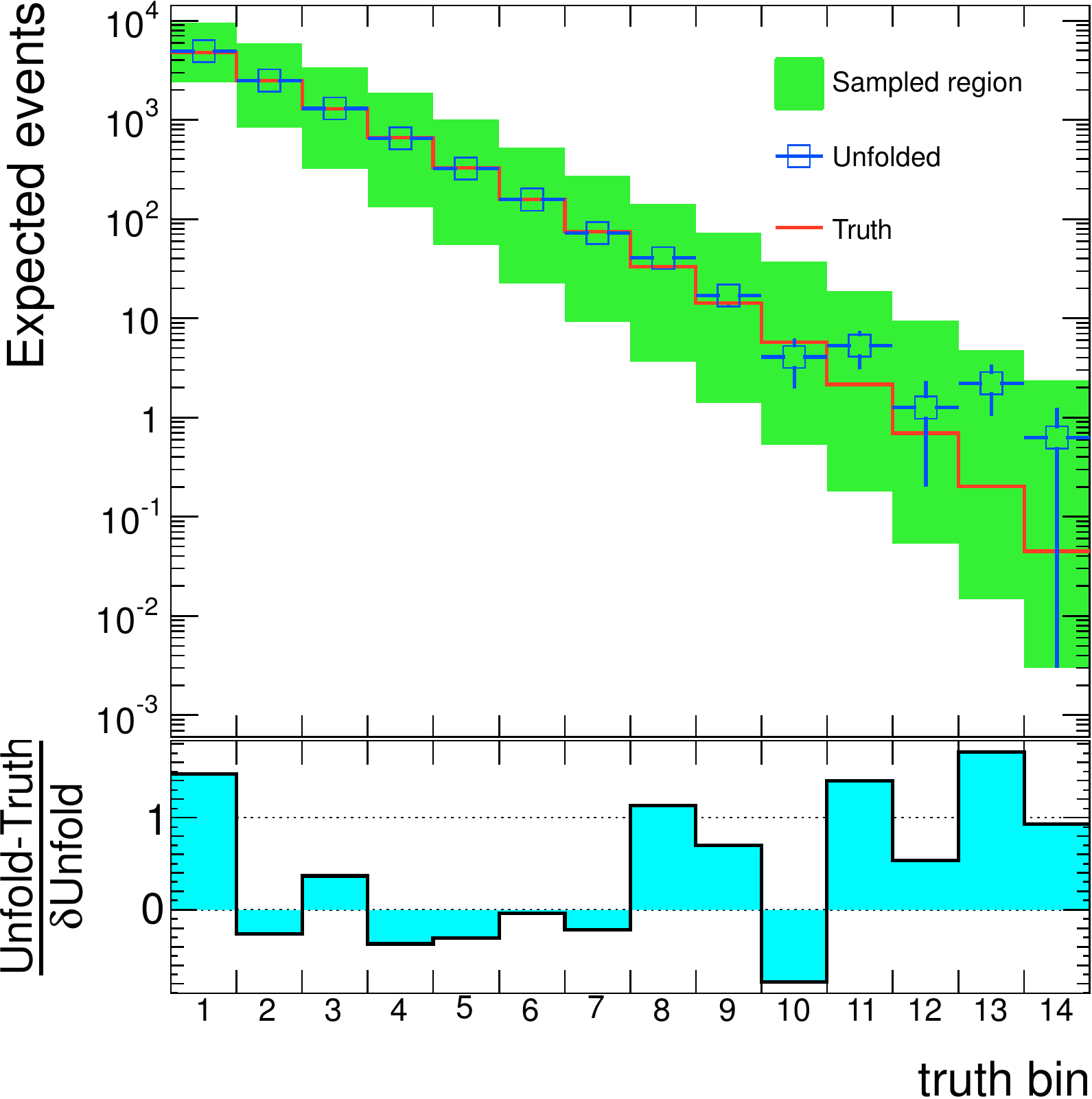}
  }
  \subfigure[$\alpha=3\times 10^{-4}$]{
    \includegraphics[width=0.3\columnwidth]{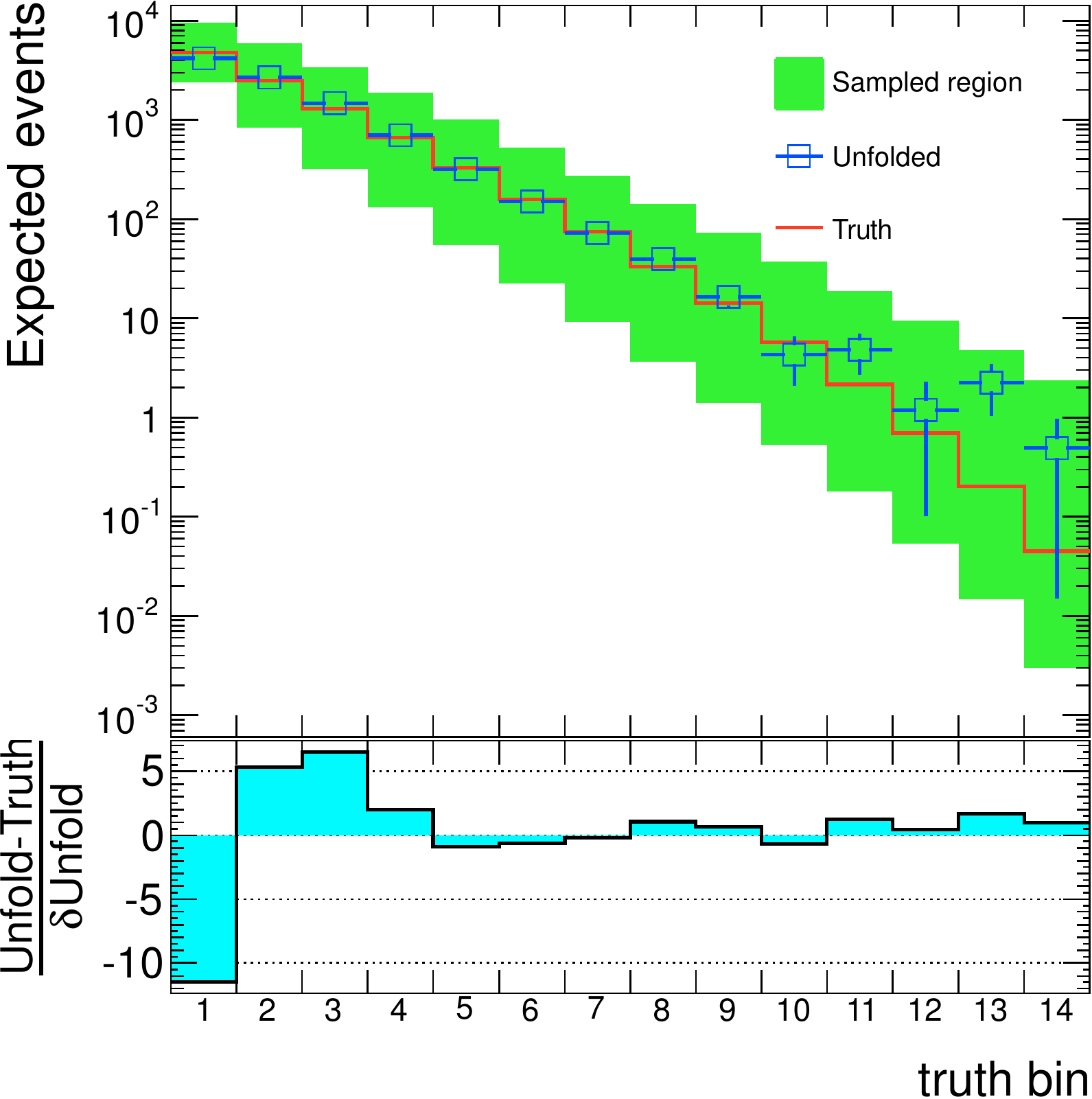}
  }
  \subfigure[$\alpha=6\times 10^{-4}$]{
    \includegraphics[width=0.3\columnwidth]{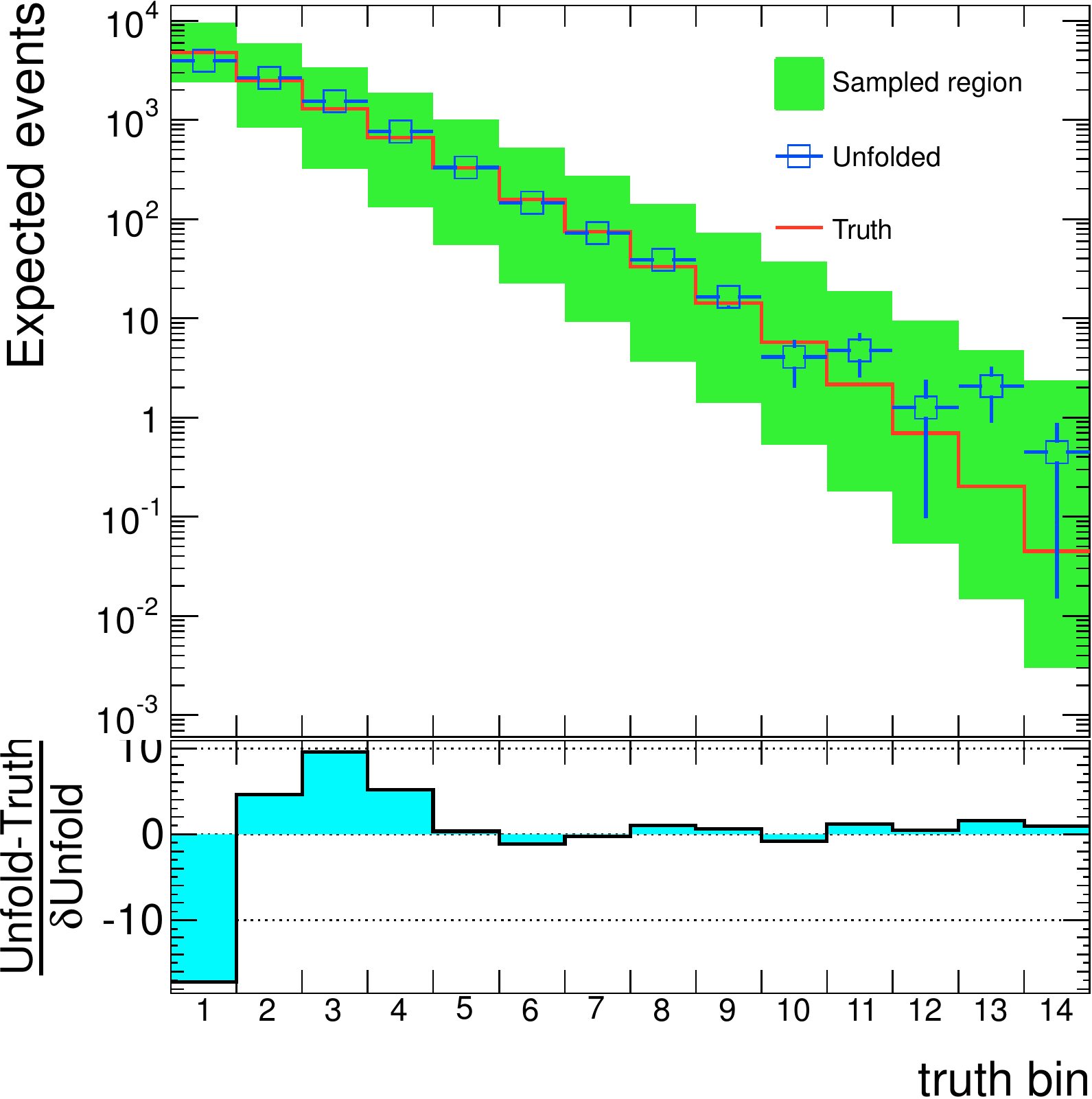}
  }\\
 \subfigure[$\alpha=0$]{
   \includegraphics[width=0.3\columnwidth]{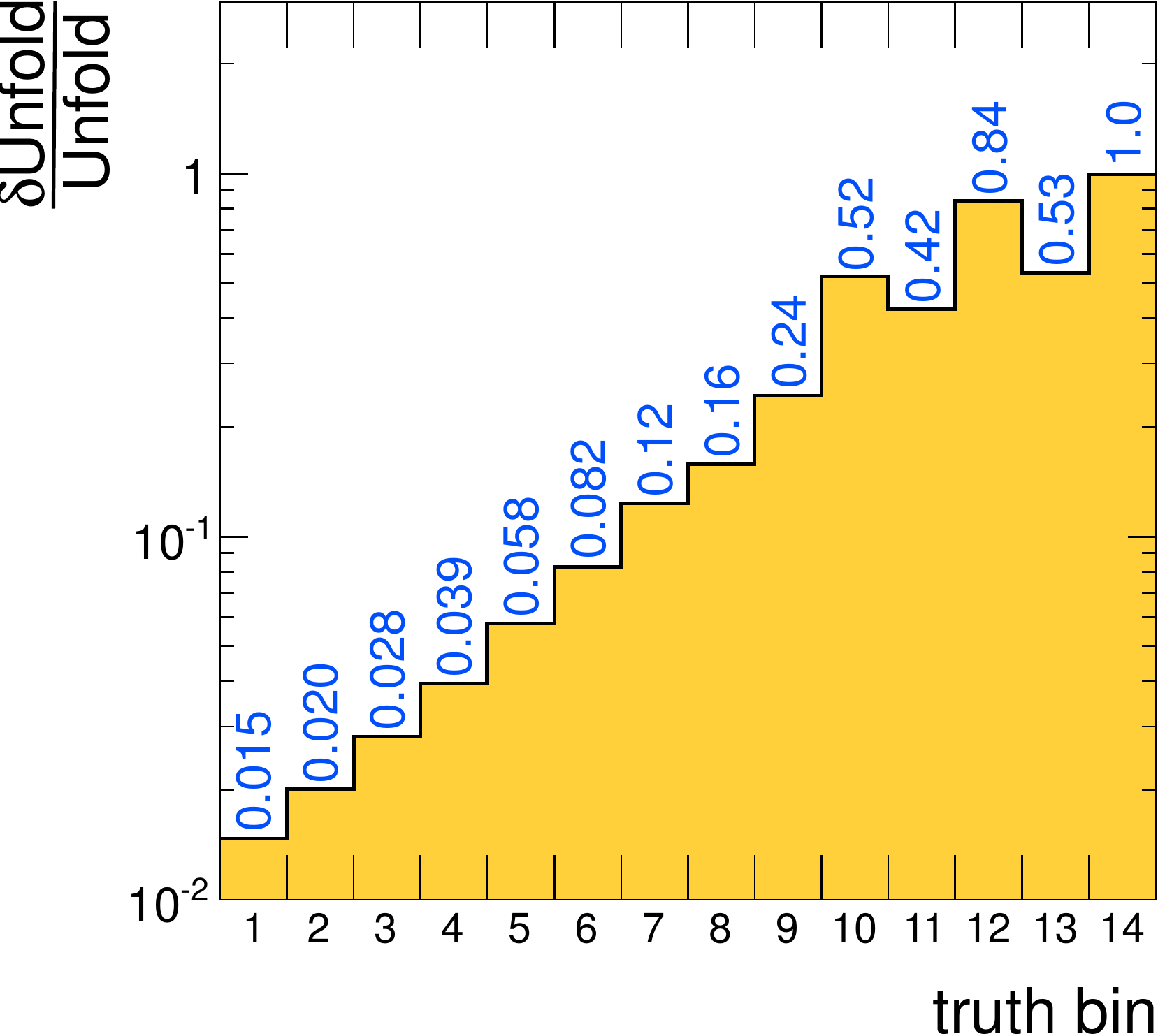}
   \label{fig:unfoldedSteepNoSmearS2d}
  }
  \subfigure[$\alpha=3\times 10^{-4}$]{
    \includegraphics[width=0.3\columnwidth]{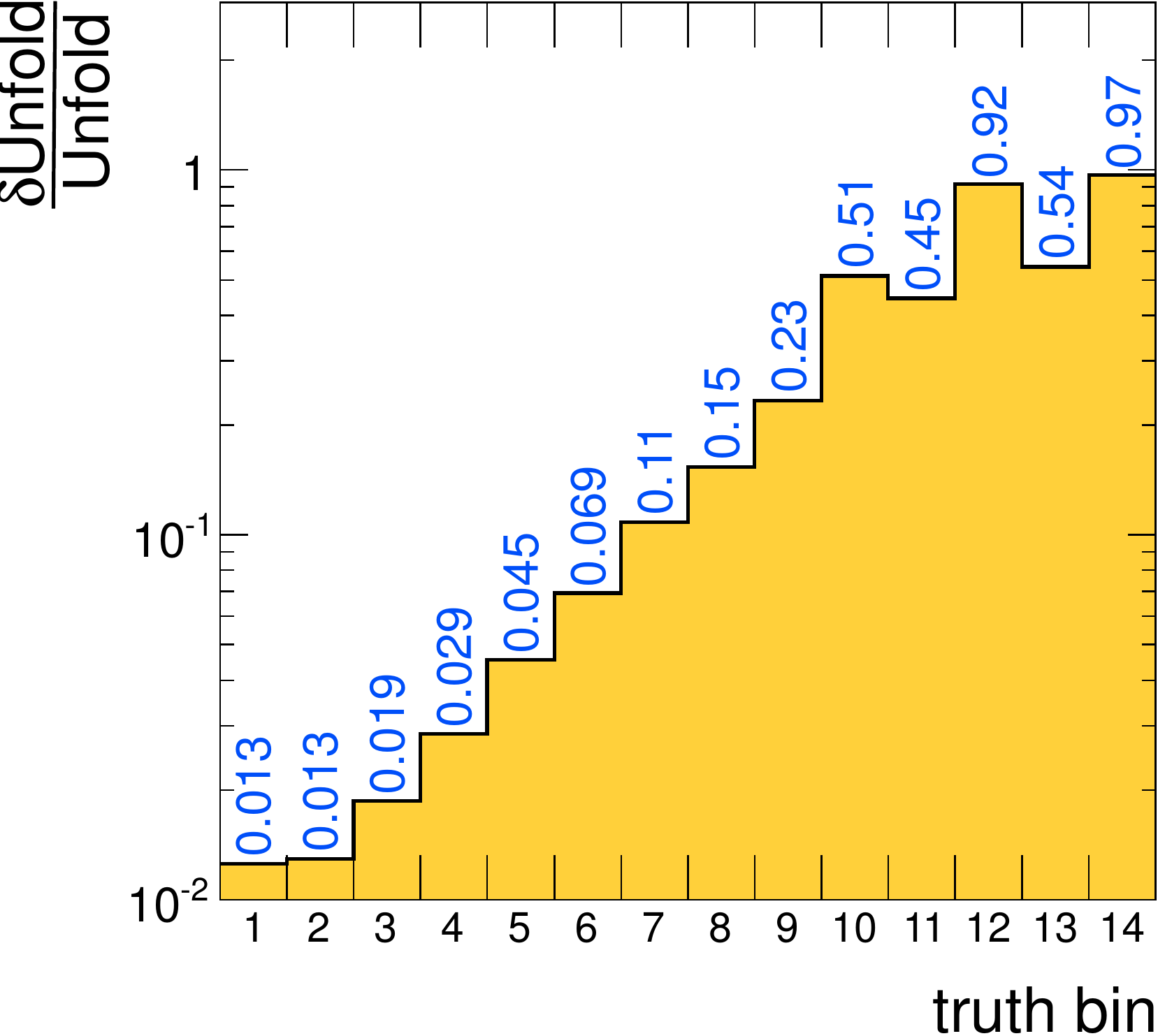}
  }
  \subfigure[$\alpha=6\times 10^{-4}$]{
    \includegraphics[width=0.3\columnwidth]{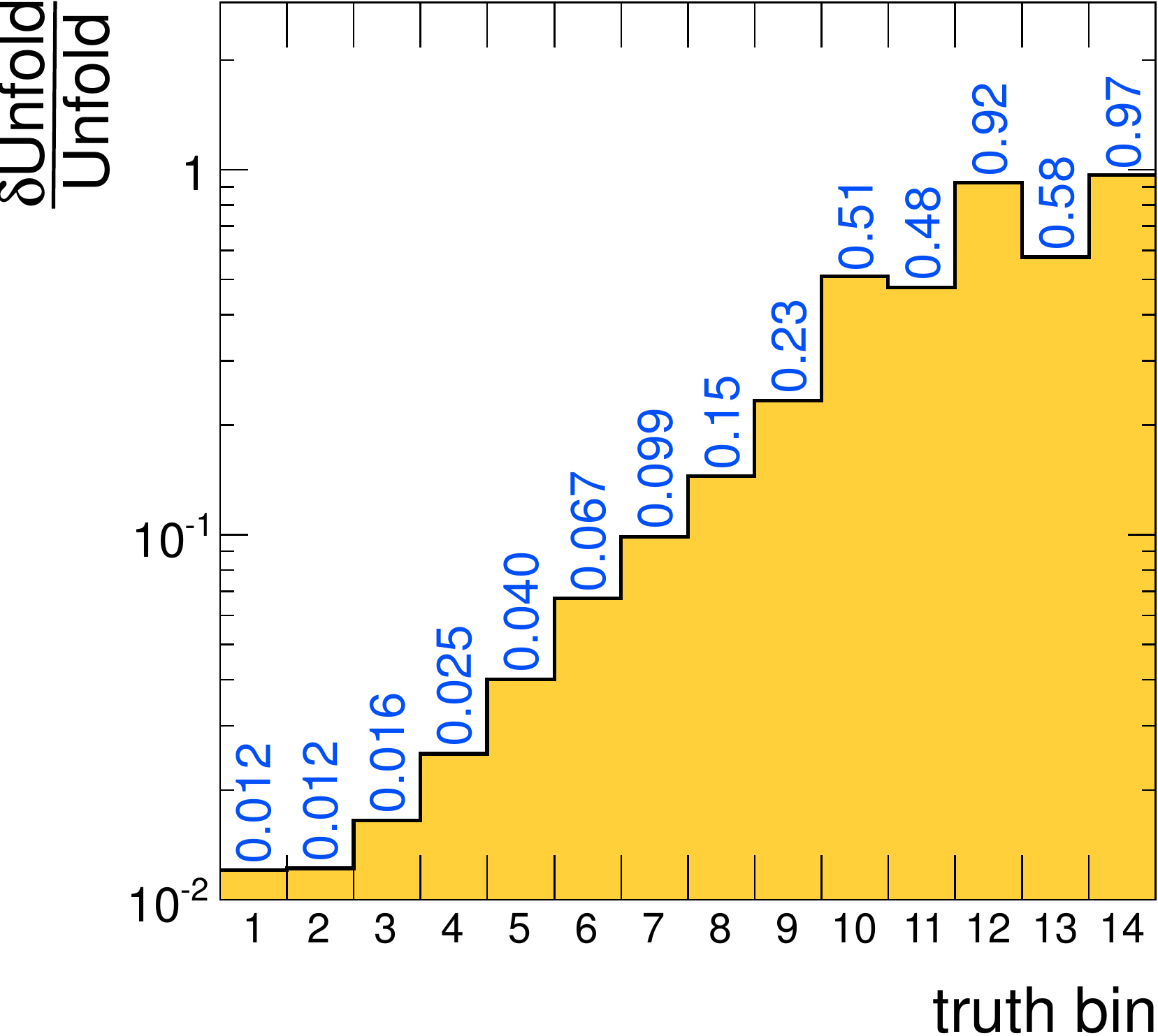}
  }
  \caption{The result of unfolding of Sec.~\ref{sec:regSteepNoSmearing}, with regularization function $S_2$, for three values of $\alpha$.   The first bins are more affected by this regularization.  Very small improvement is observed in the uncertainty of the unfolded spectrum.
\label{fig:unfoldedSteepNoSmearS2}
}
\end{figure}

The next attempt is with $S_2$.  The posterior probability distribution in $S_2$ is in Fig.~\ref{fig:regFuncSteepNoSmearS2}, and the unfolding results in Fig.~\ref{fig:unfoldedSteepNoSmearS2}.  This regularization only affects the first bins, and, as expected, increases their bias and slightly reduces their uncertainty.

\begin{figure}[H]
  \centering
  \subfigure[$\alpha=0$]{
    \includegraphics[width=0.3\columnwidth]{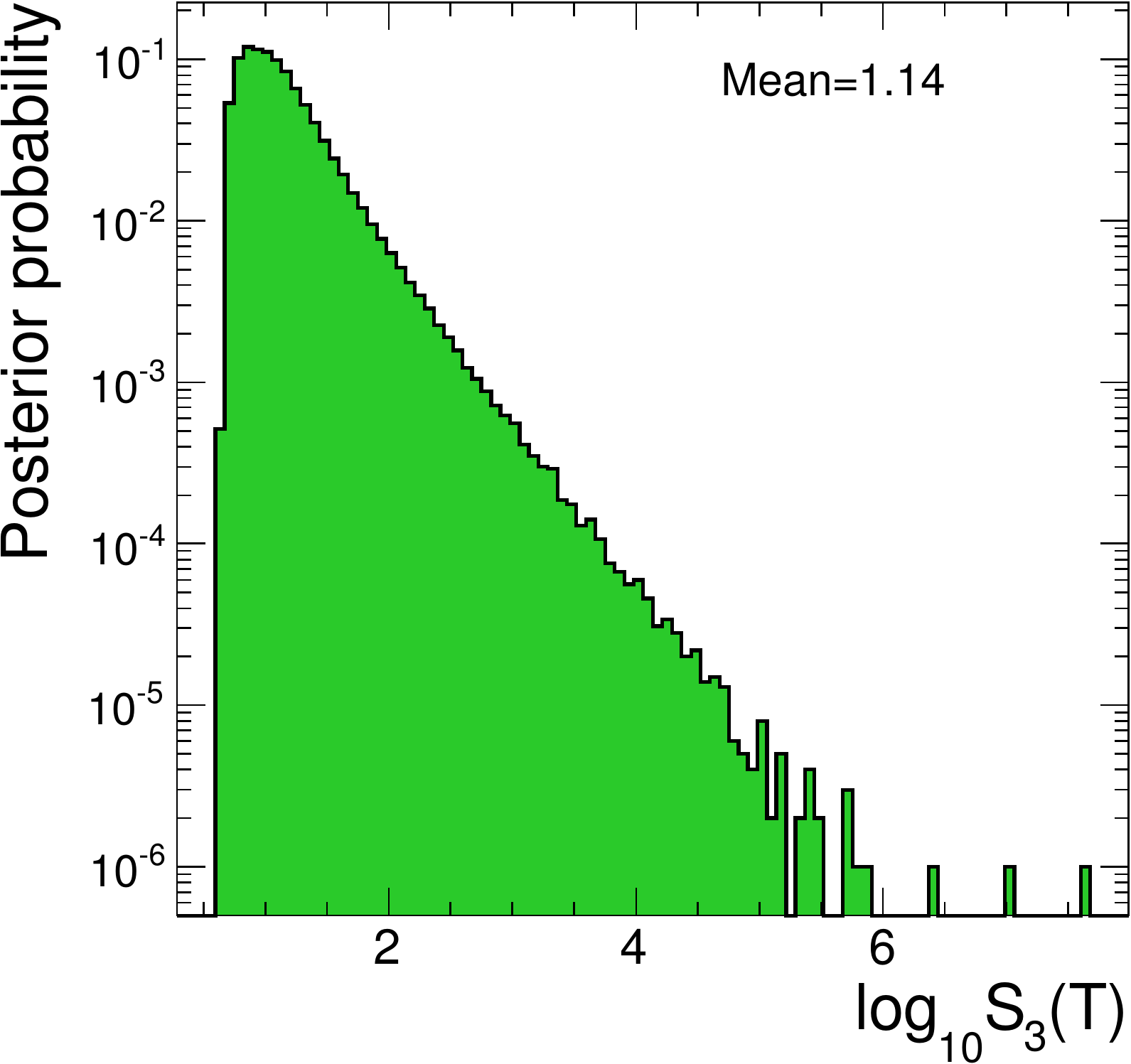}
  }
  \subfigure[$\alpha=20$]{
    \includegraphics[width=0.3\columnwidth]{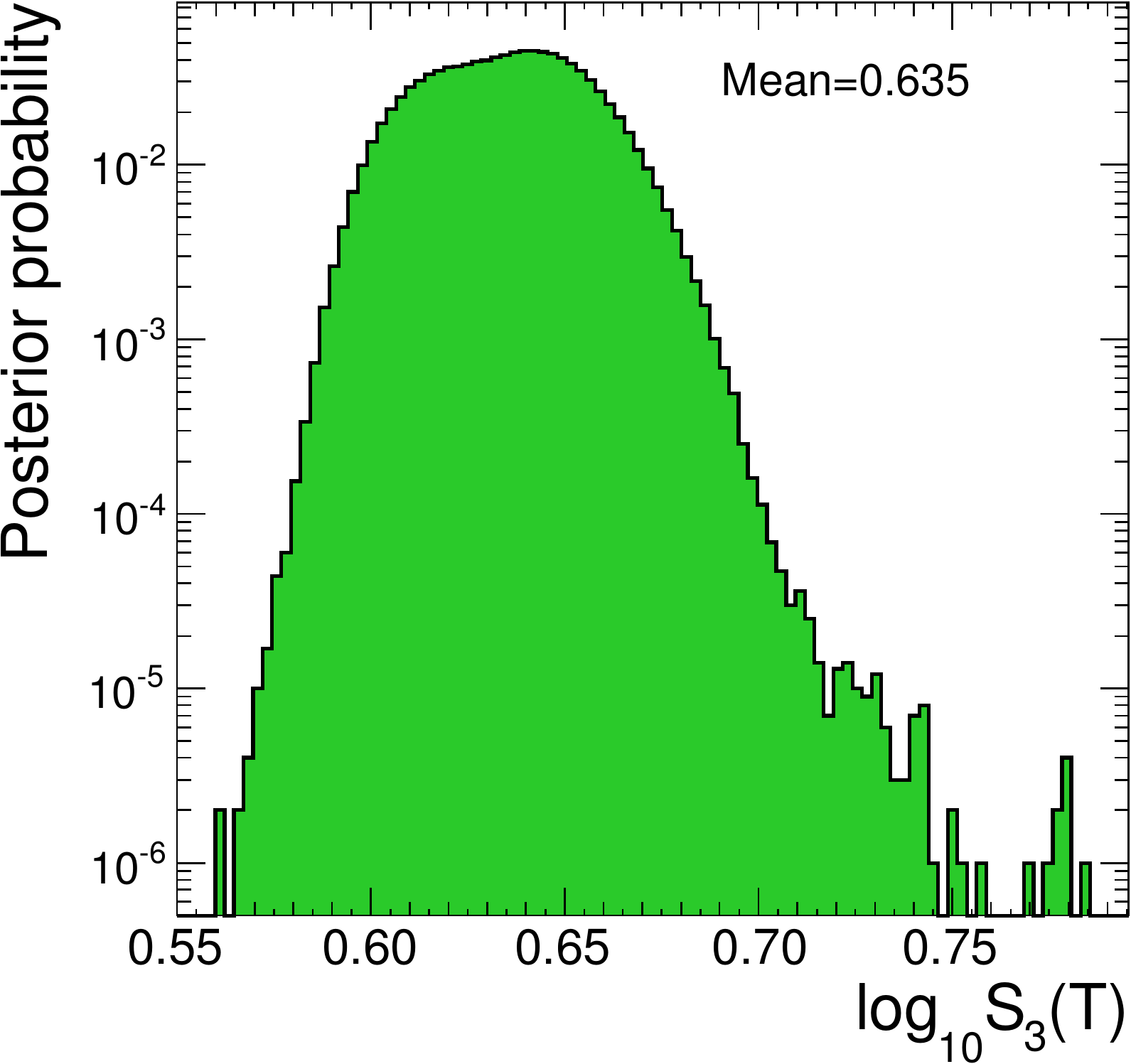}
  }
  \subfigure[$\alpha=40$]{
    \includegraphics[width=0.3\columnwidth]{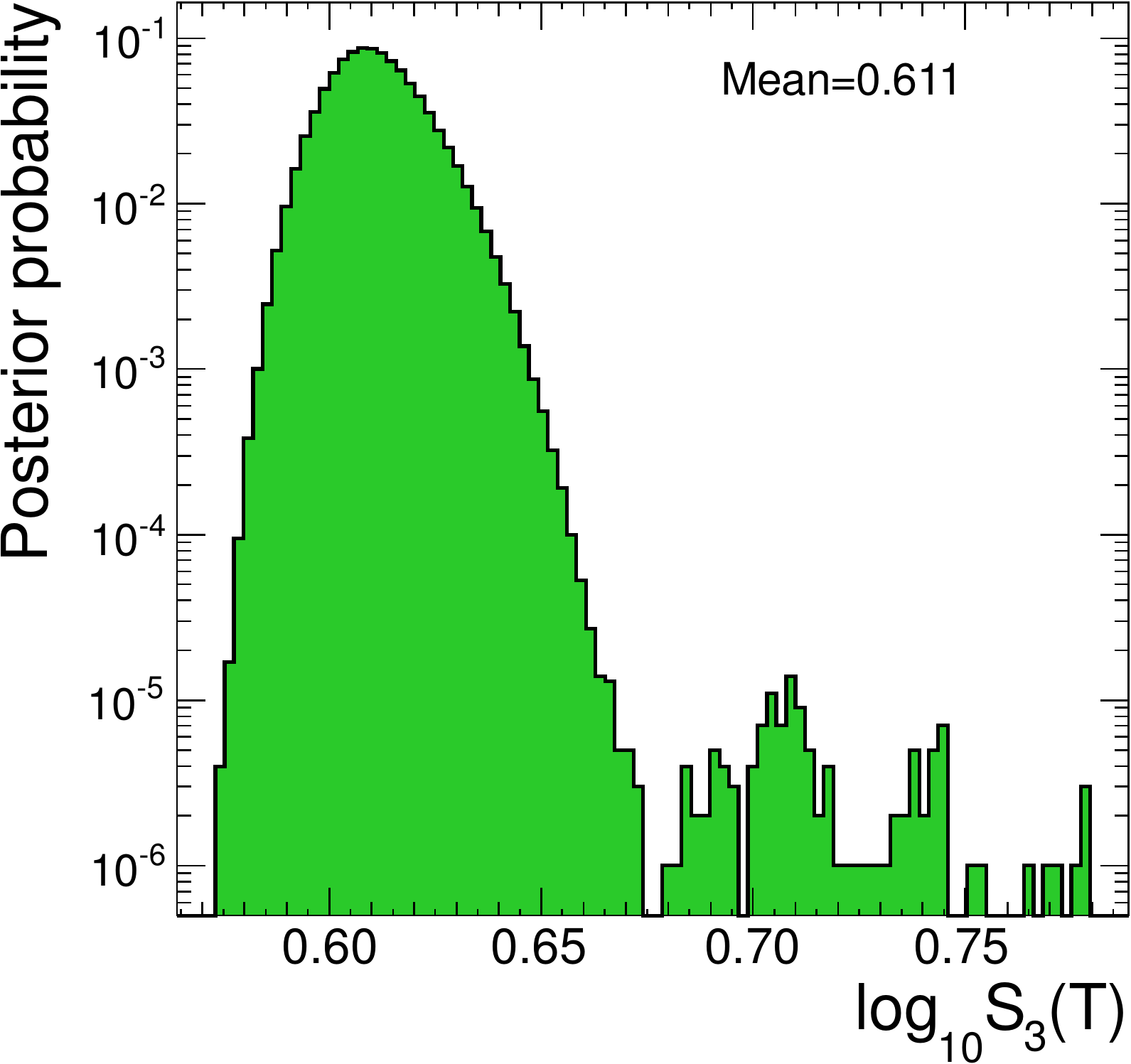}
  }
\caption{The posterior $P(\log_{10} S_3(\tuple{T})|\tuple{D})$, for three different choices of the regularization parameter $\alpha$, corresponding to Sec.~\ref{sec:regSteepNoSmearing}.
\label{fig:regFuncSteepNoSmearS3}
}
\end{figure}

\begin{figure}[H]
  \centering
  \subfigure[$\alpha=0$]{
    \includegraphics[width=0.3\columnwidth]{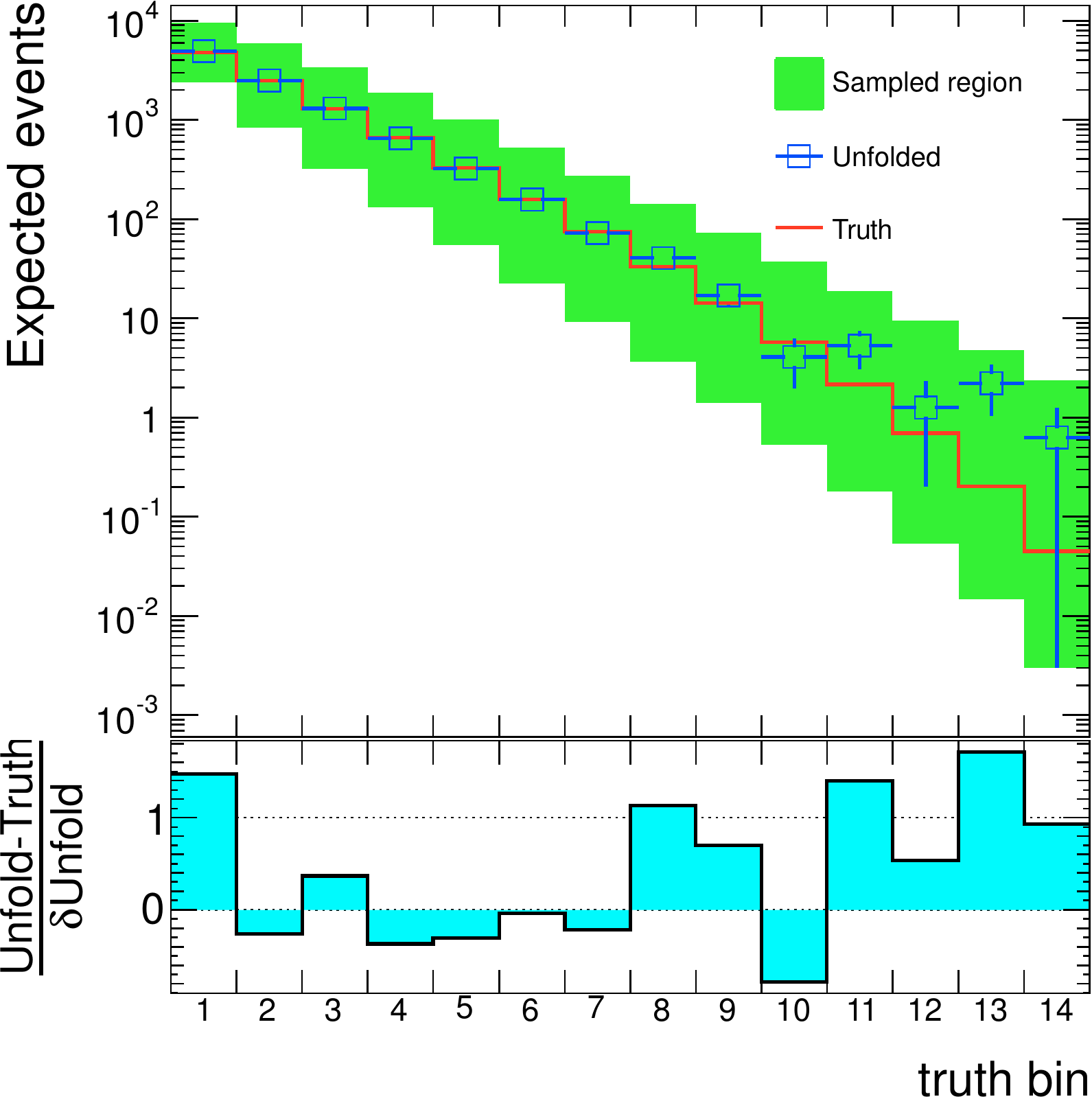}
  }
  \subfigure[$\alpha=20$]{
    \includegraphics[width=0.3\columnwidth]{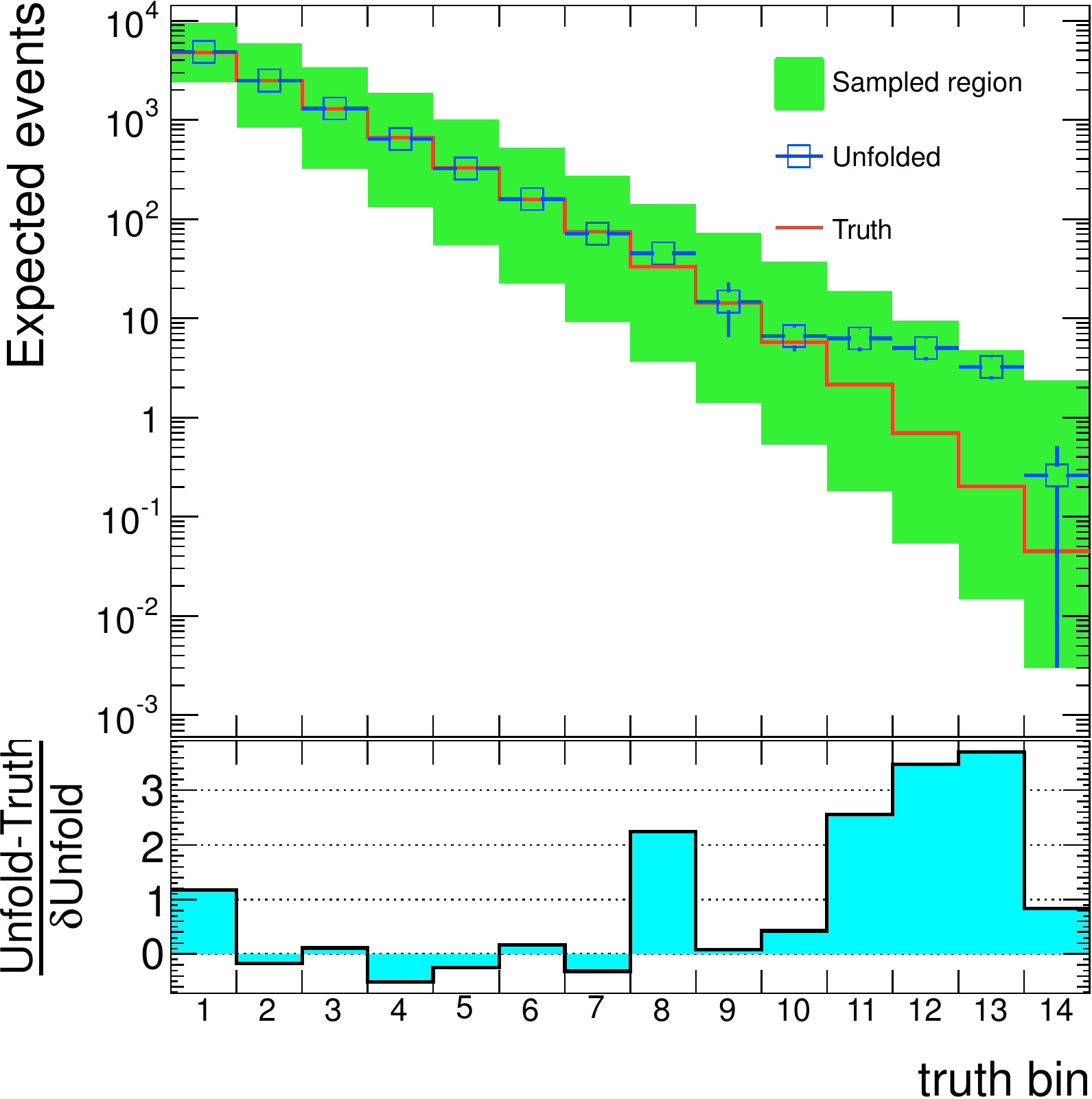}
  }
  \subfigure[$\alpha=40$]{
    \includegraphics[width=0.3\columnwidth]{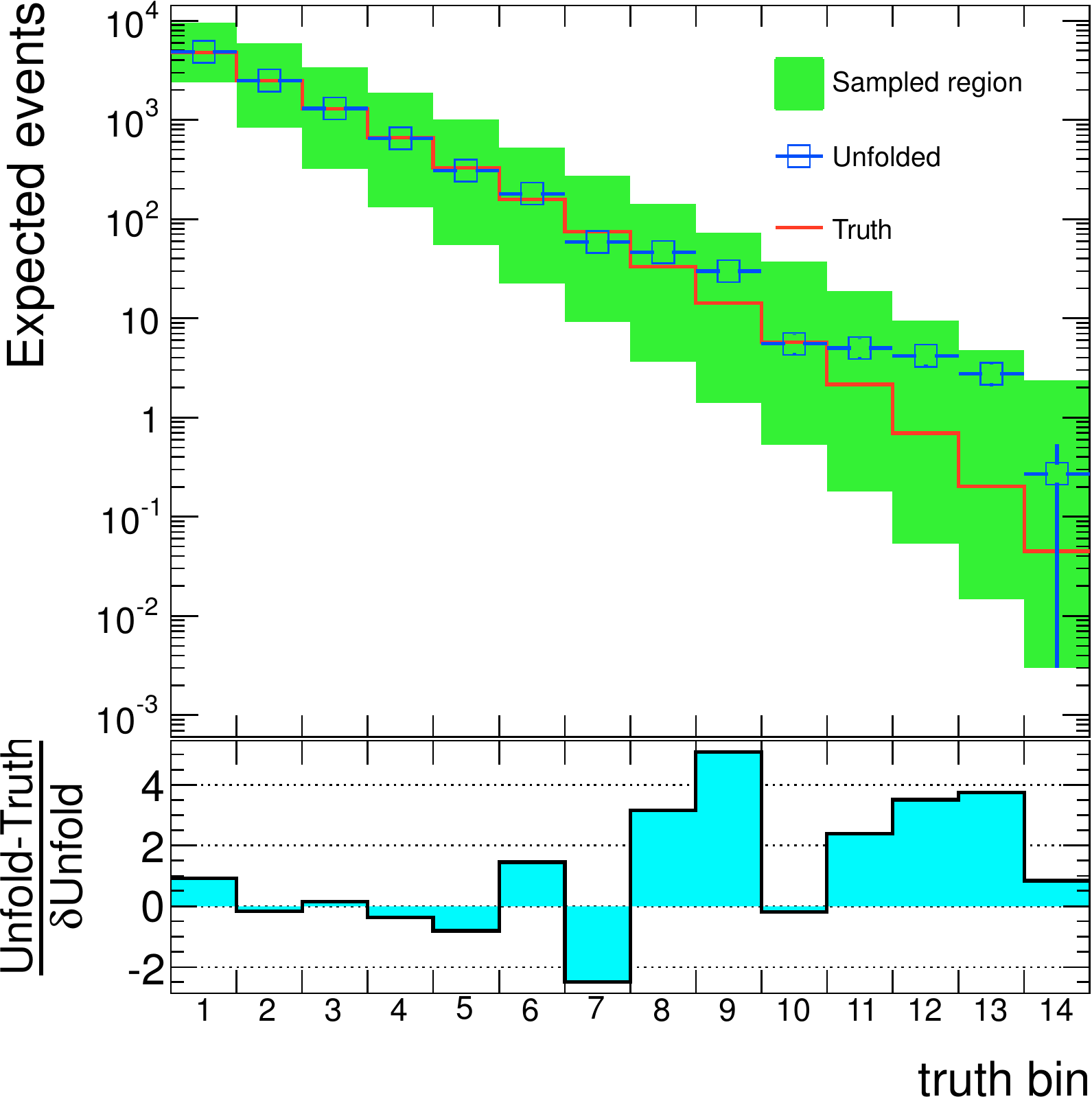}
  }\\
 \subfigure[$\alpha=0$]{
    \includegraphics[width=0.3\columnwidth]{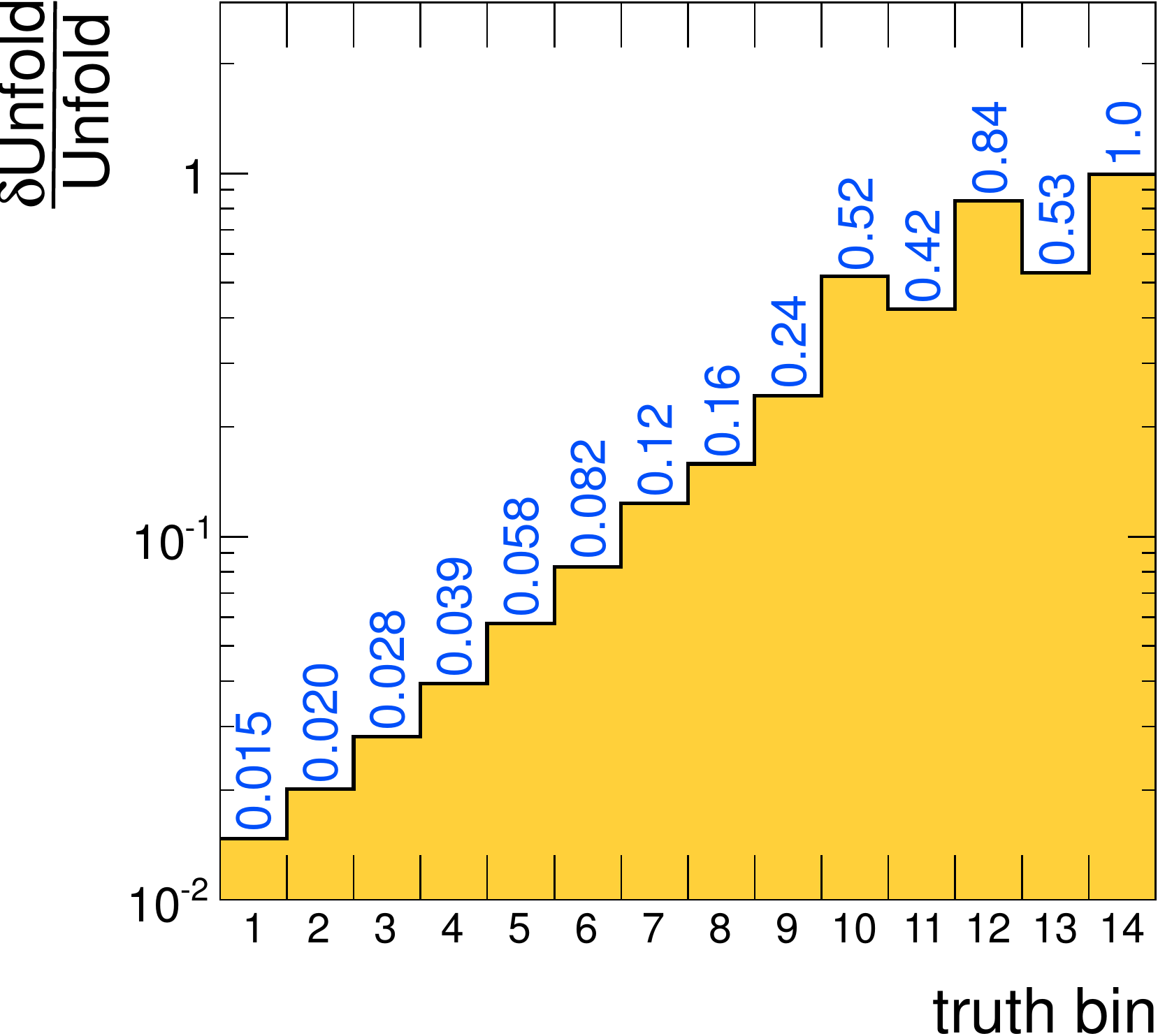}
  }
  \subfigure[$\alpha=20$]{
    \includegraphics[width=0.3\columnwidth]{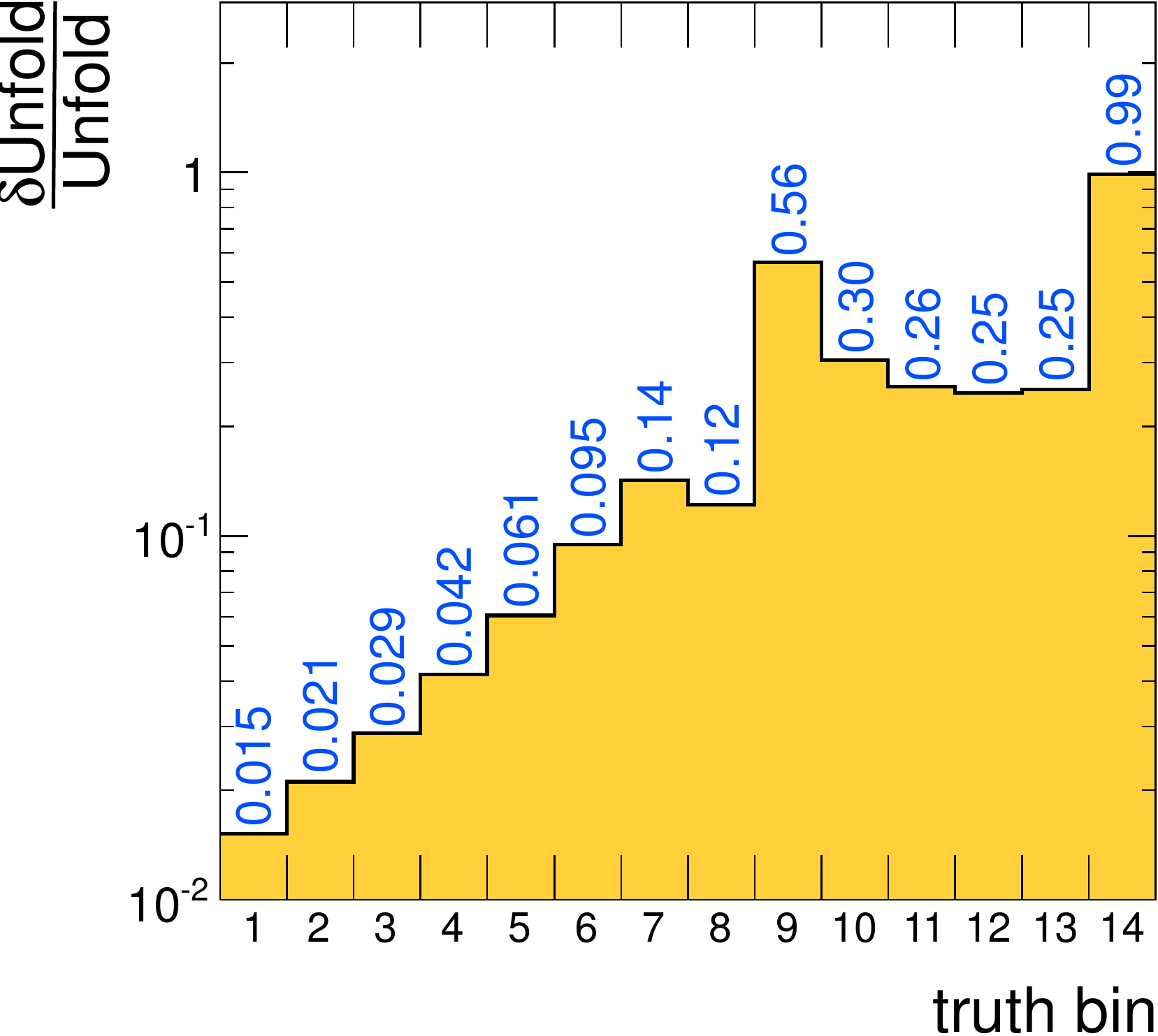}
  }
  \subfigure[$\alpha=40$]{
    \includegraphics[width=0.3\columnwidth]{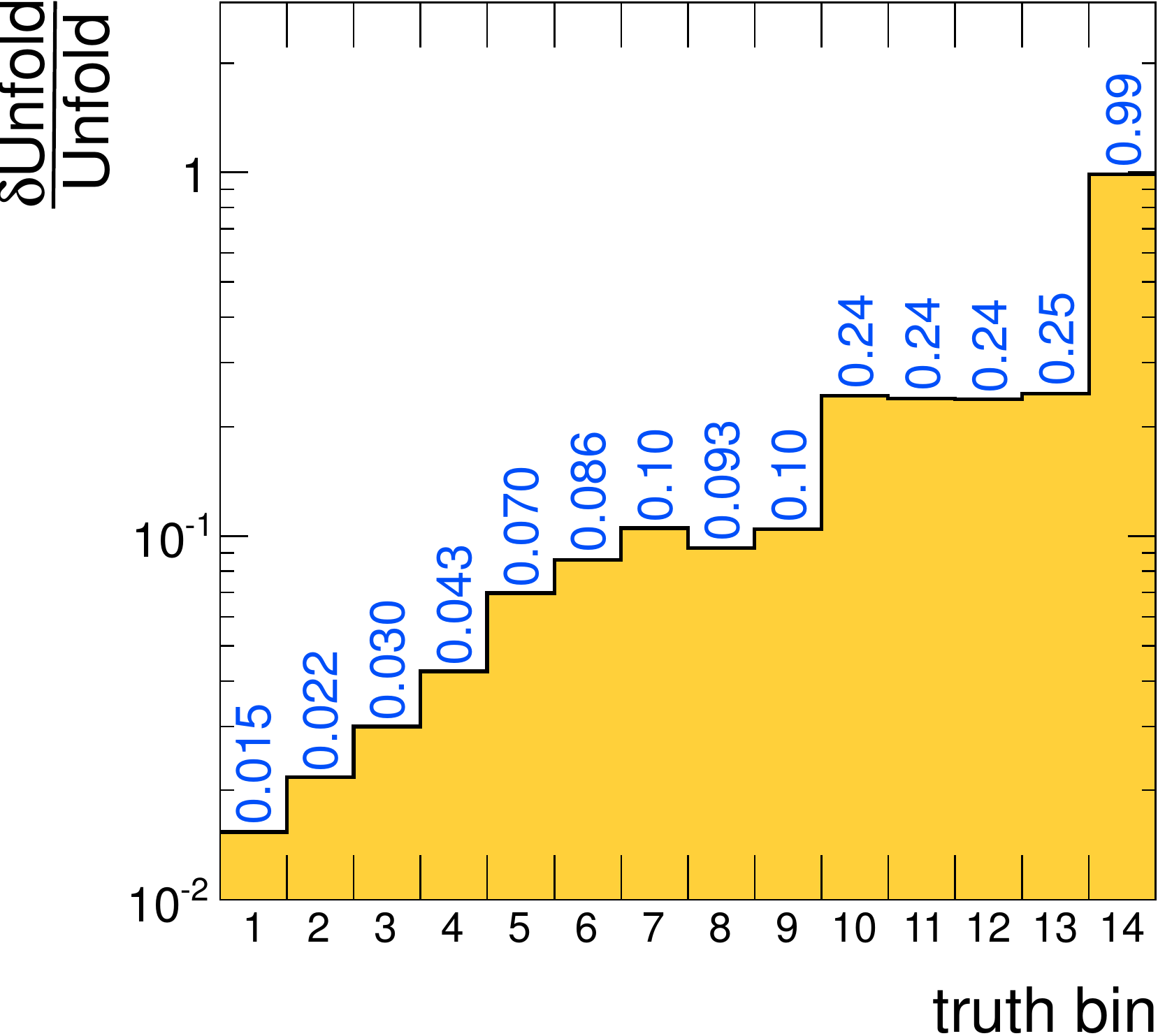}
  }
  \caption{The result of unfolding of Sec.~\ref{sec:regSteepNoSmearing}, with regularization function $S_3$, for three values of $\alpha$.  
\label{fig:unfoldedSteepNoSmearS3}
}
\end{figure}

The attempt with $S_3$ is similarly shown in Fig.~\ref{fig:regFuncSteepNoSmearS3} and \ref{fig:unfoldedSteepNoSmearS3}.  The posterior spread is significantly reduced in the last few truth bins, at the cost of significant bias.



\begin{figure}[H]
  \centering
  \subfigure[$\alpha=0$]{
    \includegraphics[width=0.3\columnwidth]{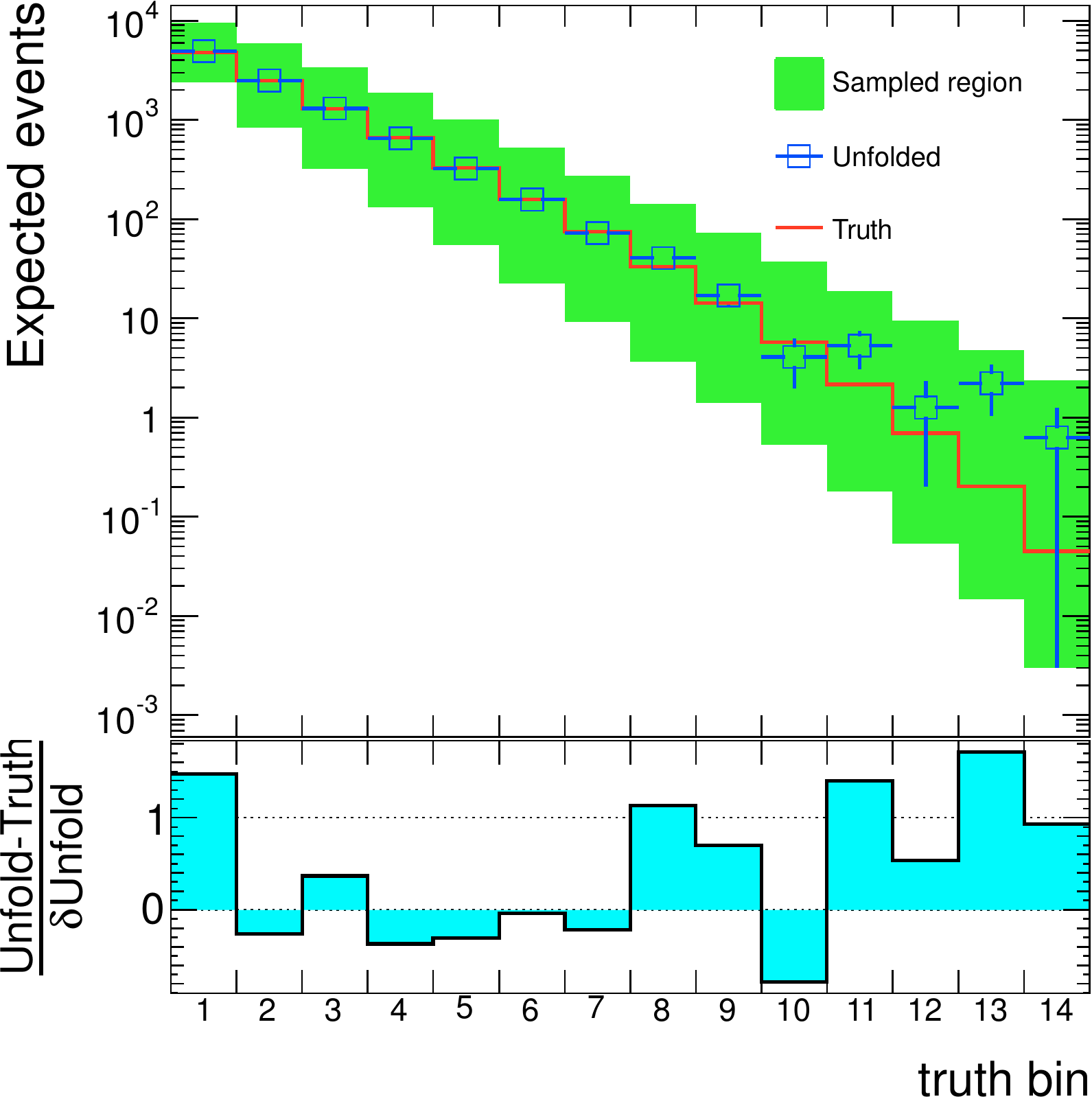}
  }
  \subfigure[$\alpha=1$]{
    \includegraphics[width=0.3\columnwidth]{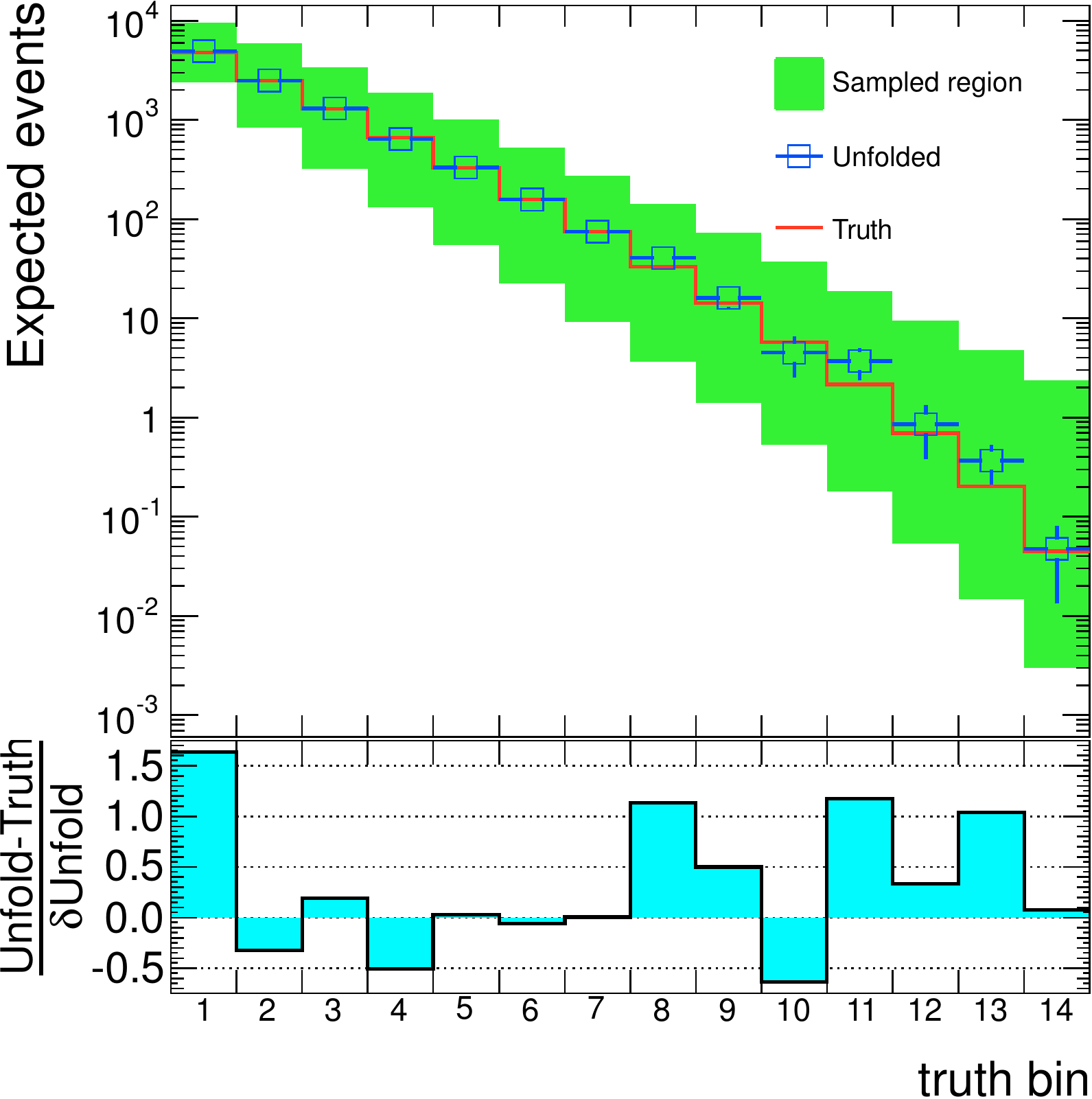}
  }
  \subfigure[$\alpha=10$]{
    \includegraphics[width=0.3\columnwidth]{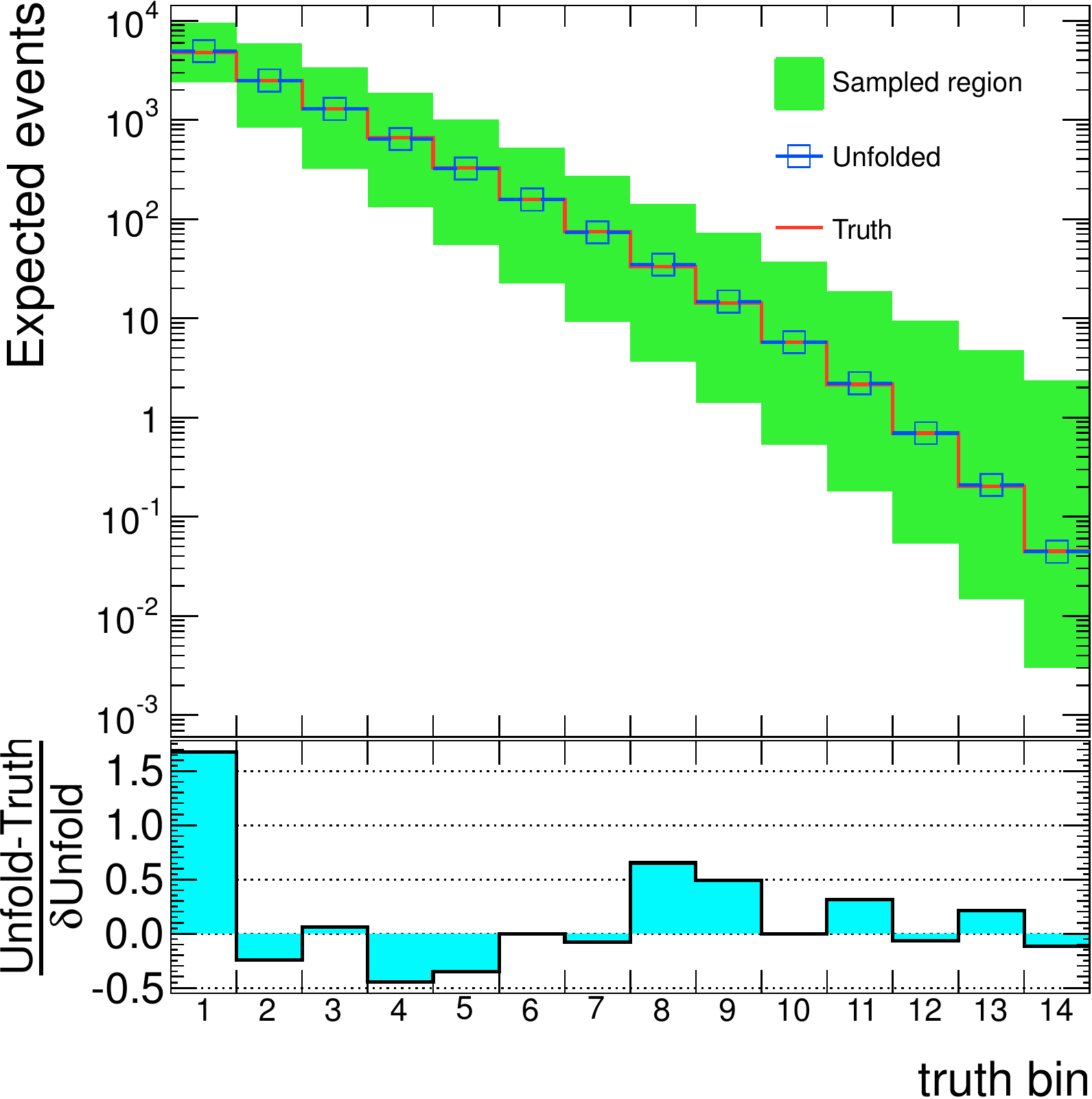}
  }\\
 \subfigure[$\alpha=0$]{
    \includegraphics[width=0.3\columnwidth]{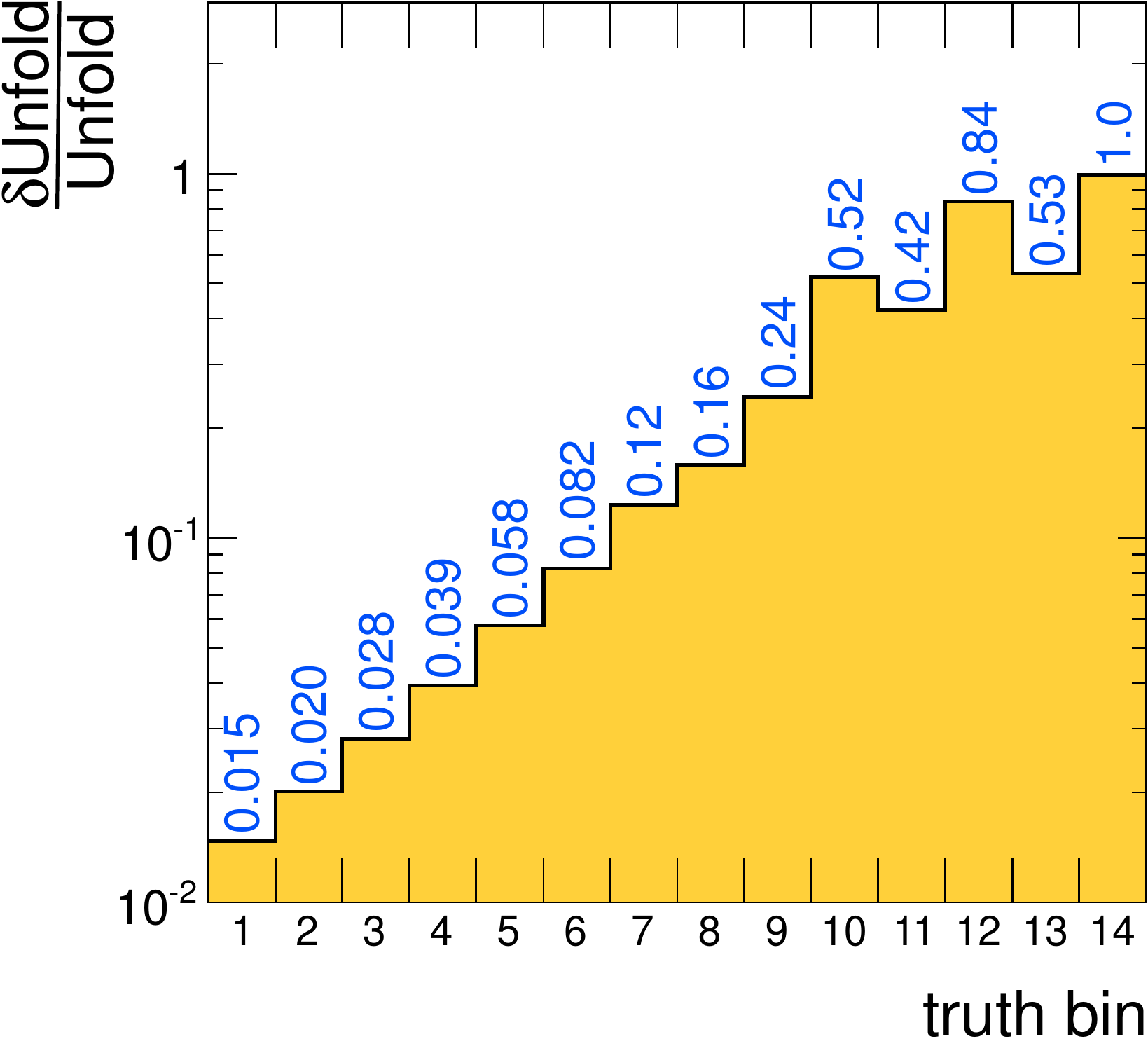}
  }
  \subfigure[$\alpha=1$]{
    \includegraphics[width=0.3\columnwidth]{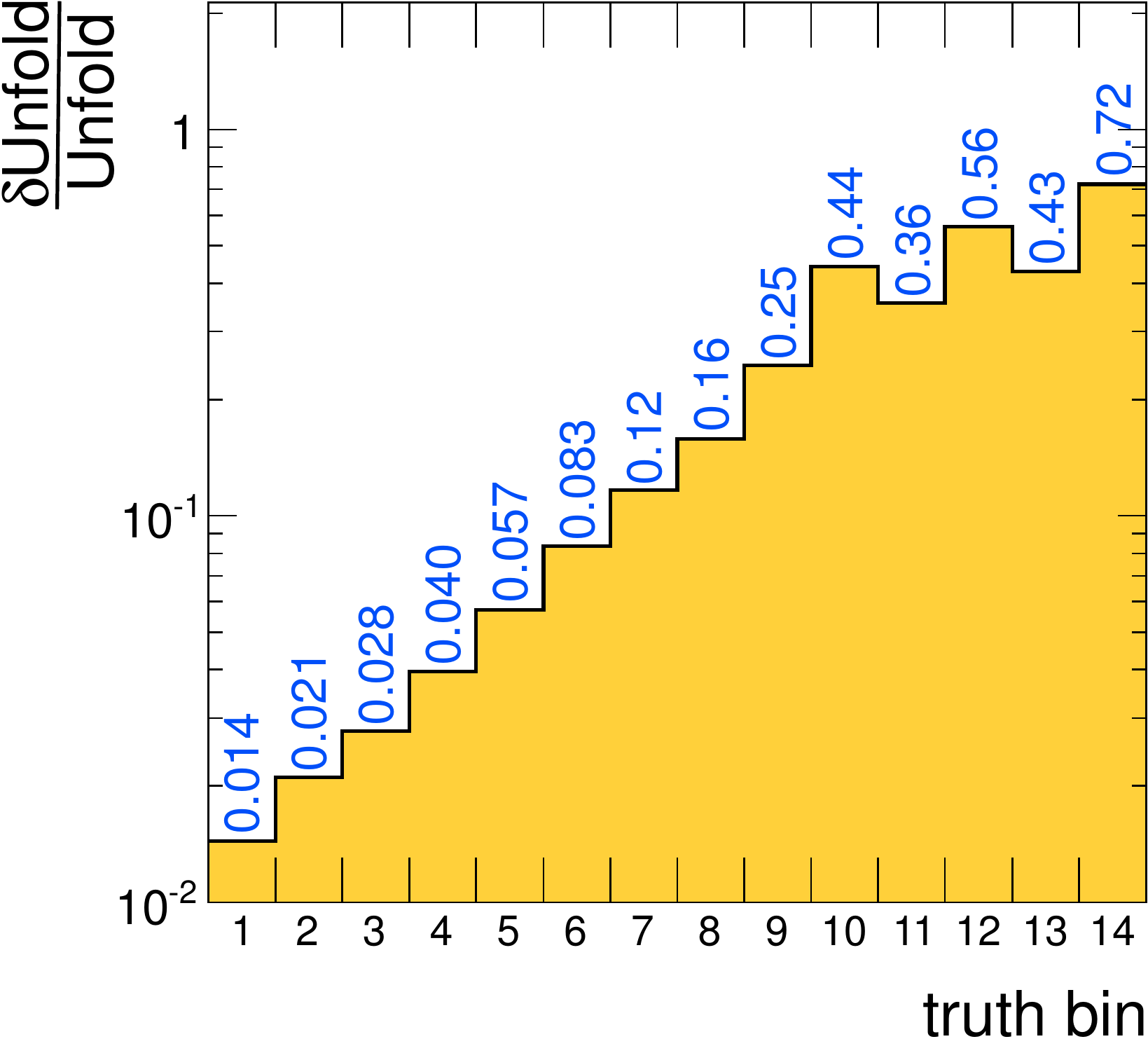}
  }
  \subfigure[$\alpha=10$]{
    \includegraphics[width=0.3\columnwidth]{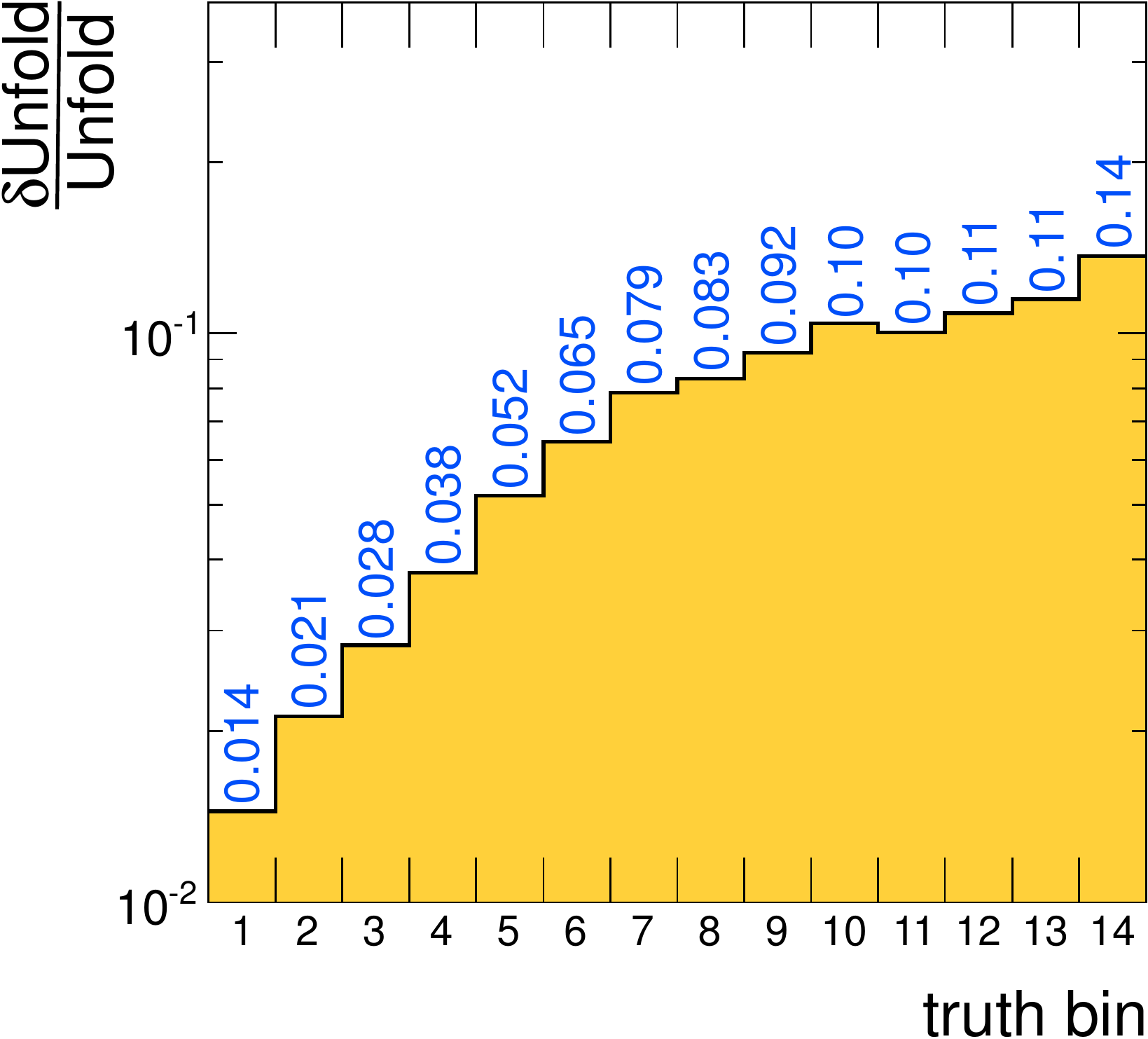}
  }
  \caption{The result of unfolding of Sec.~\ref{sec:regSteepNoSmearing}, with a Gaussian regularization constraint, for three values of $\alpha$ (see Sec.~\ref{sec:regularization}).  
\label{fig:unfoldedSteepNoSmearGaus}
}
\end{figure}

In Fig.~\ref{fig:unfoldedSteepNoSmearGaus} are the results when regularization is made through a Gaussian constraint, of varying RMS, controlled by $\alpha$ as explained in Sec.~\ref{sec:regularization}.  The first row in Fig.~\ref{fig:unfoldedSteepNoSmearGaus} confirms simply that the prior constrains the unfolded spectrum near the truth-level spectrum known from the MC ($\tilde{\T}$) which is used to construct the migrations matrix.  The posterior's spread can be constrained arbitrarily by increasing $\alpha$, and in this idealized case there is no bias cost, because the MC truth-level spectrum is, by construction, the correct one ($\tilde{\T} = \hat{\T}$).  

\begin{figure}[H]
  \centering
  \begin{tabular}{ccccc}
    \includegraphics[width=0.18\columnwidth]{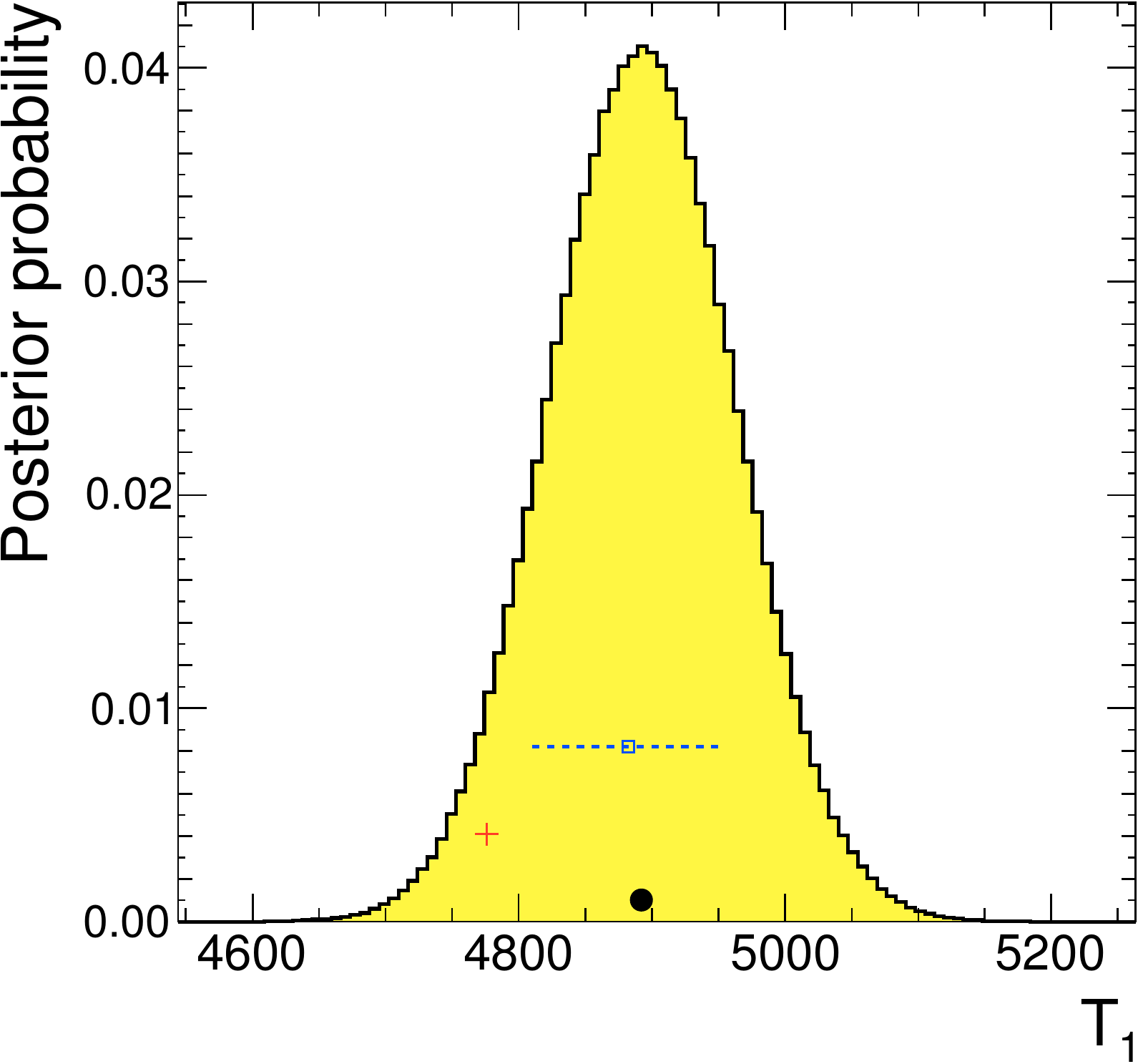} &
    \includegraphics[width=0.18\columnwidth]{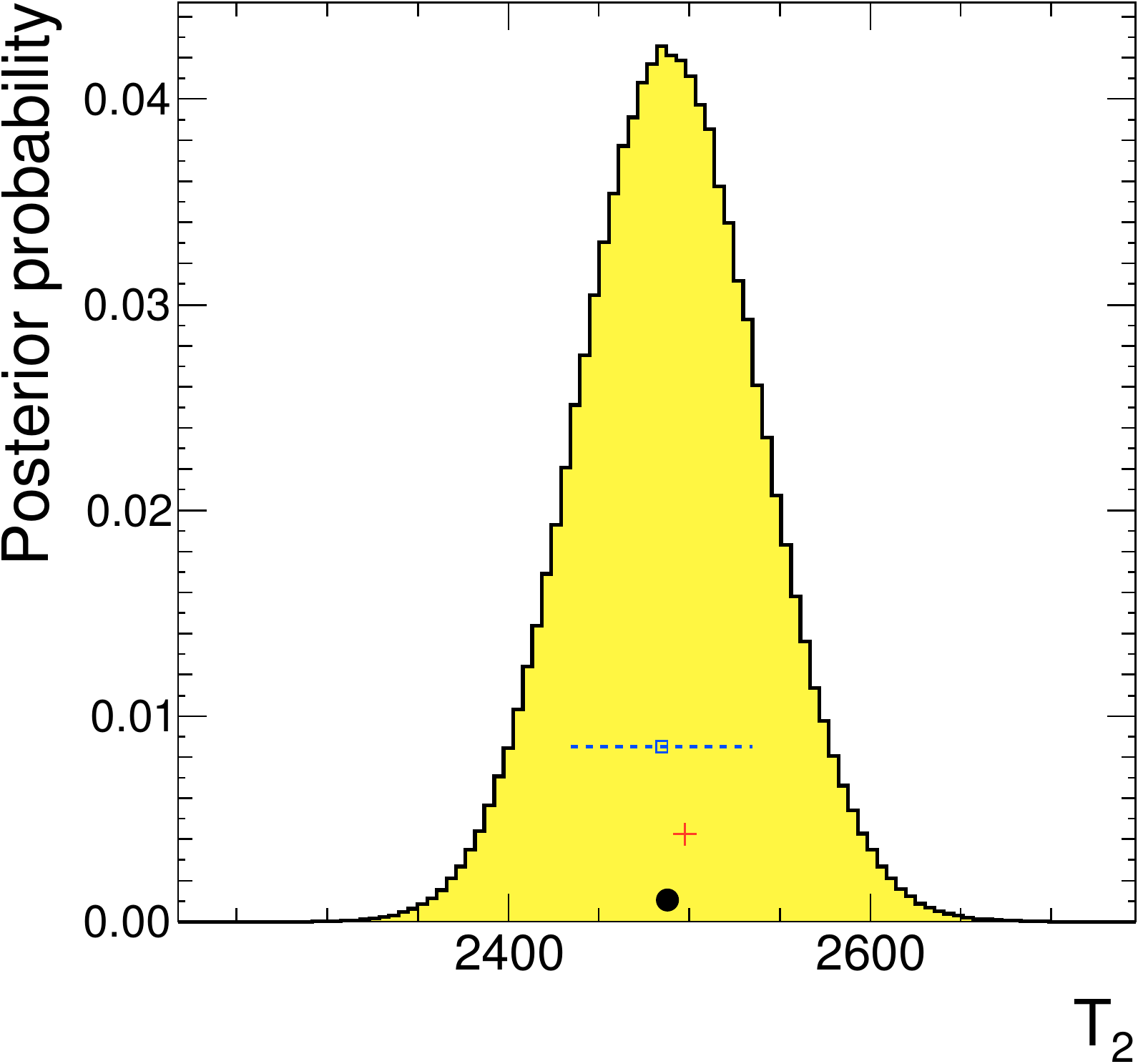} &
    \includegraphics[width=0.18\columnwidth]{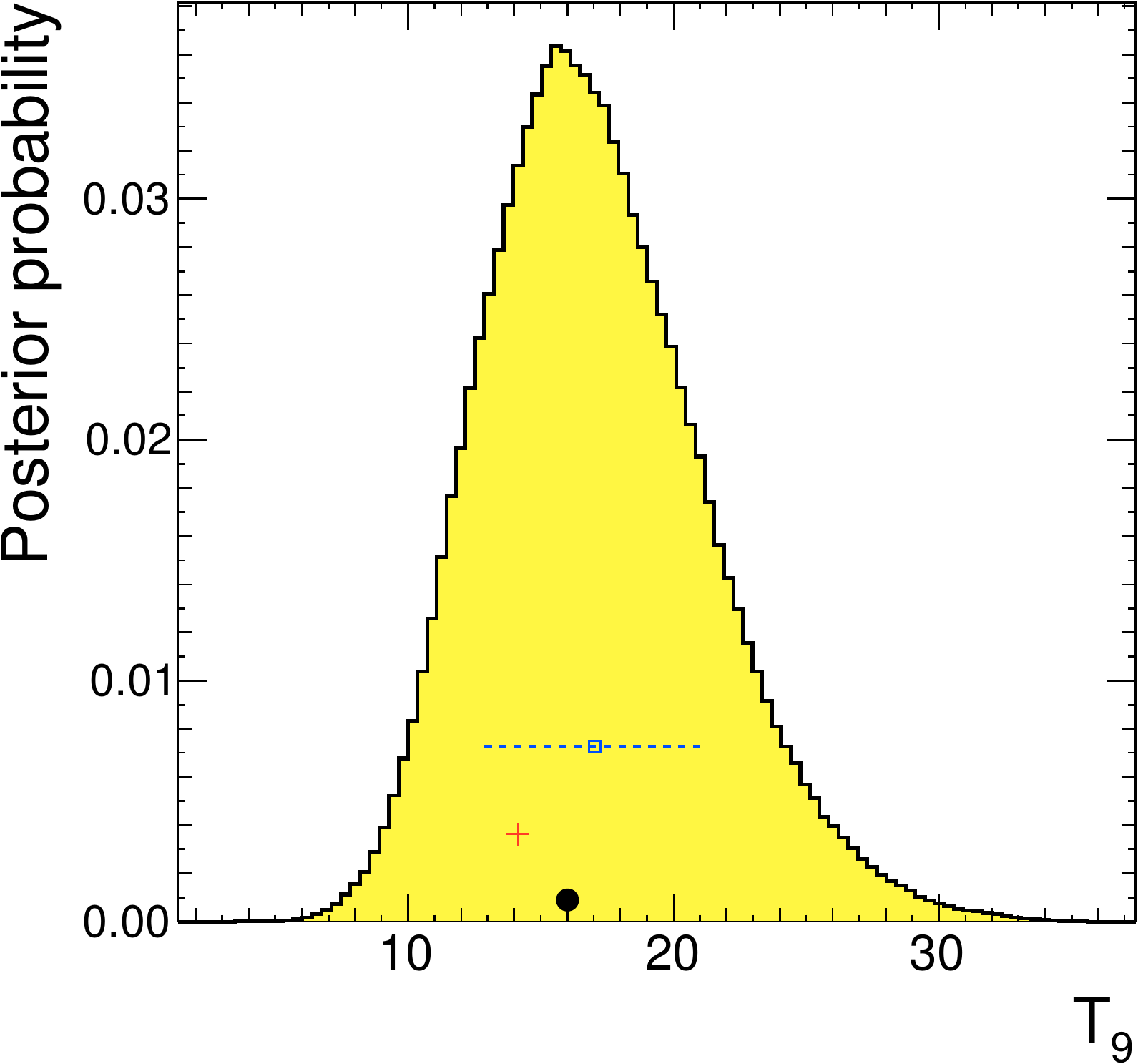} &
    \includegraphics[width=0.18\columnwidth]{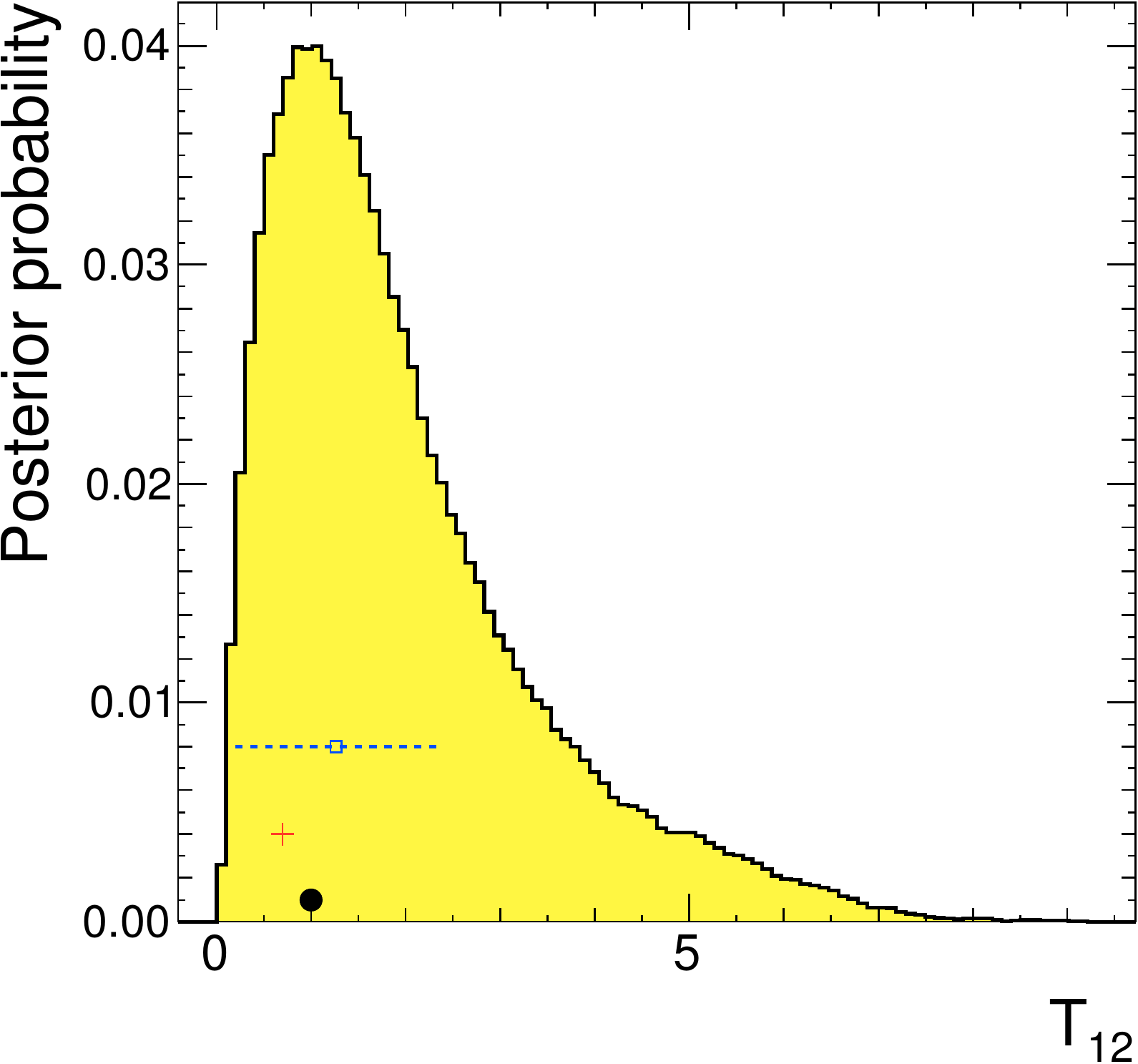} &
    \includegraphics[width=0.18\columnwidth]{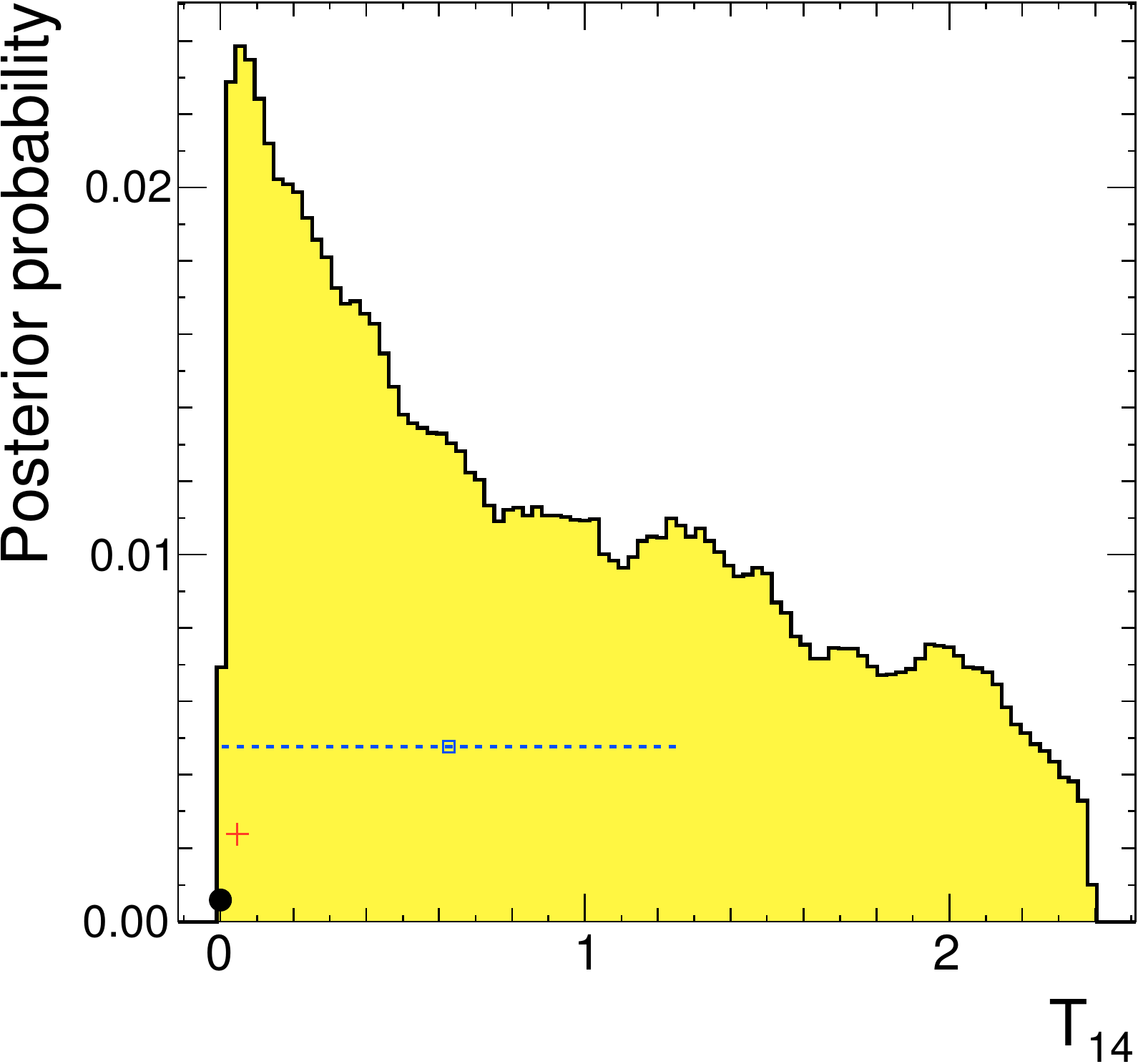} \\

    \includegraphics[width=0.18\columnwidth]{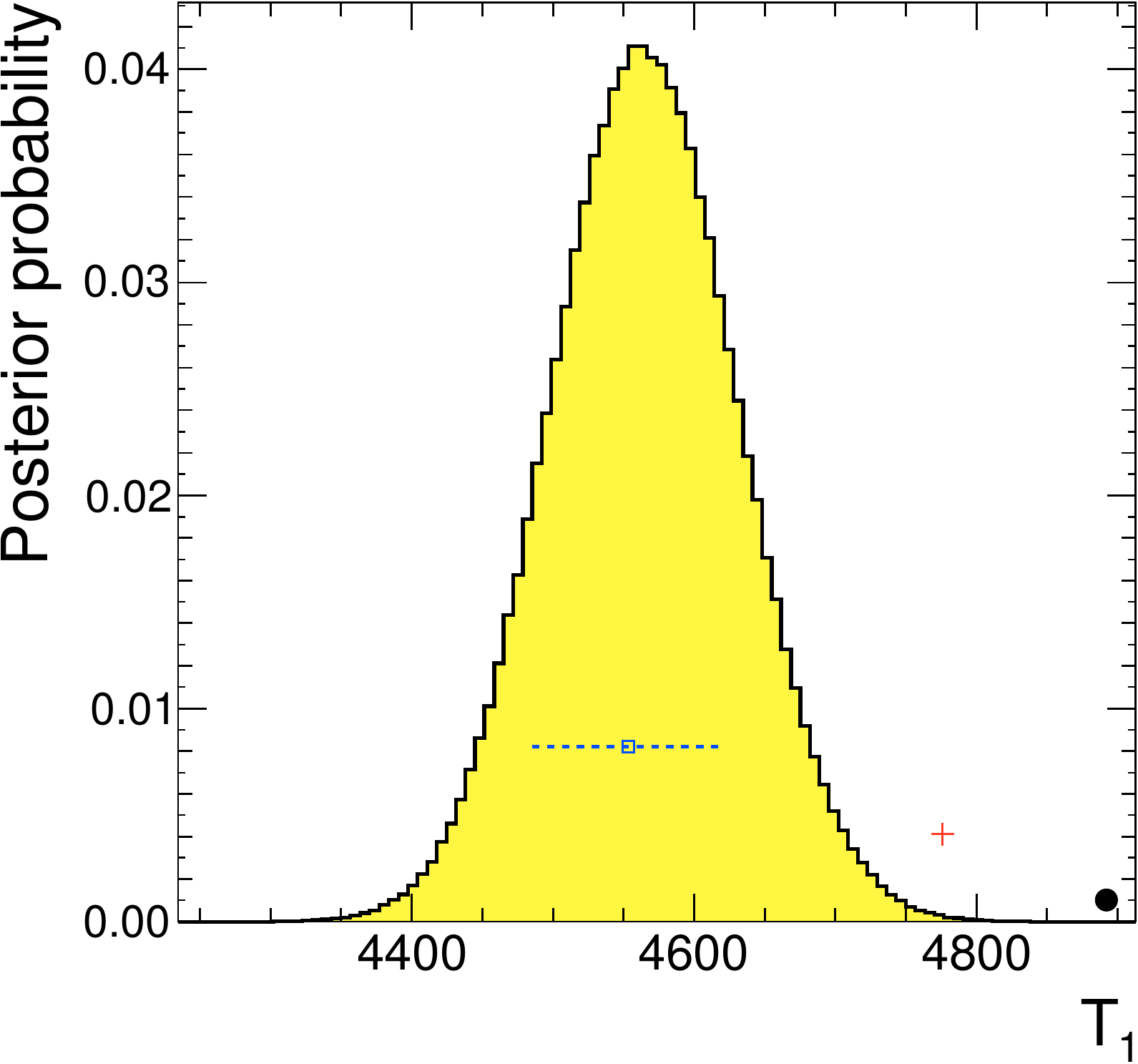} &
    \includegraphics[width=0.18\columnwidth]{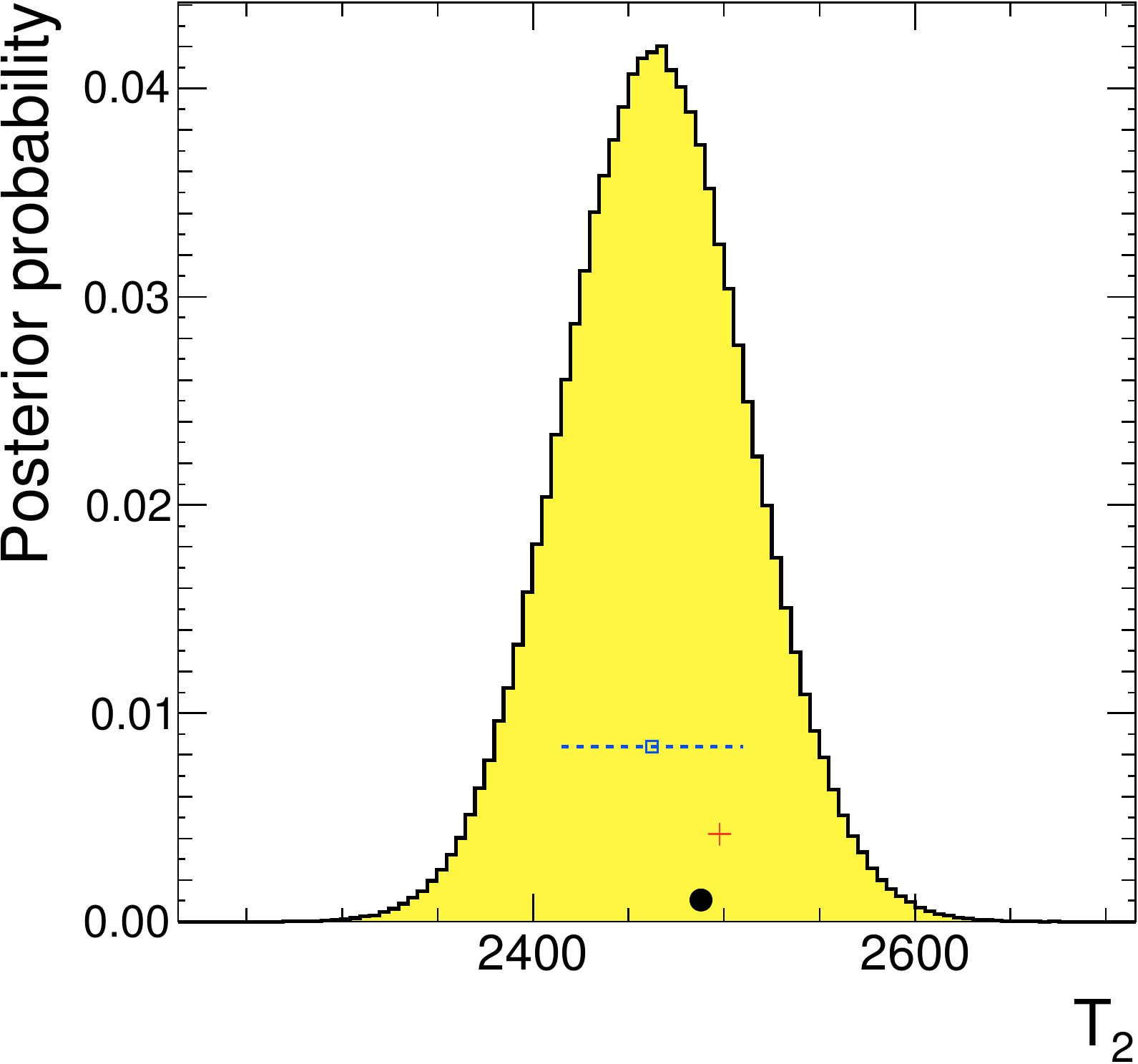} &
    \includegraphics[width=0.18\columnwidth]{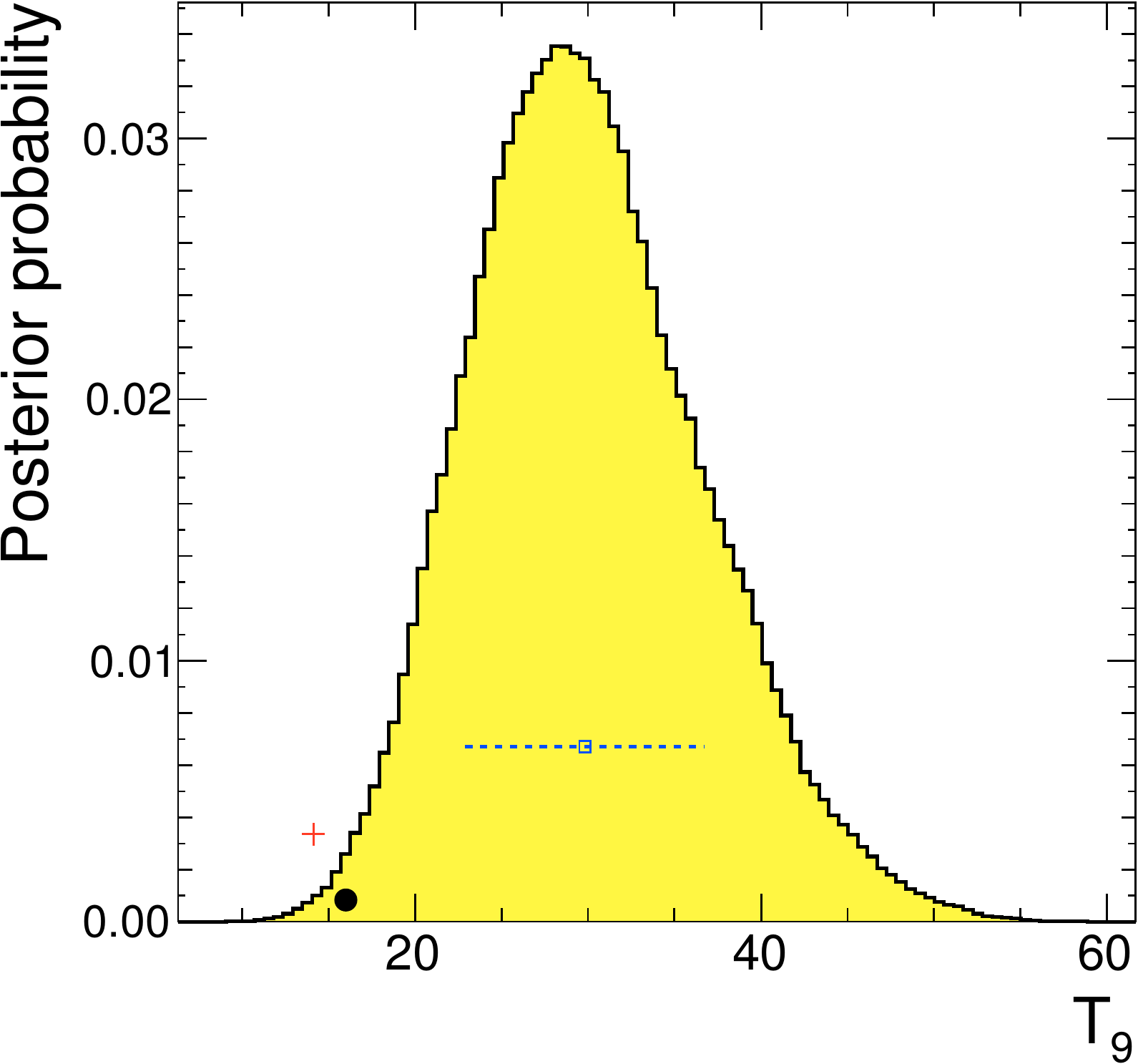} &
    \includegraphics[width=0.18\columnwidth]{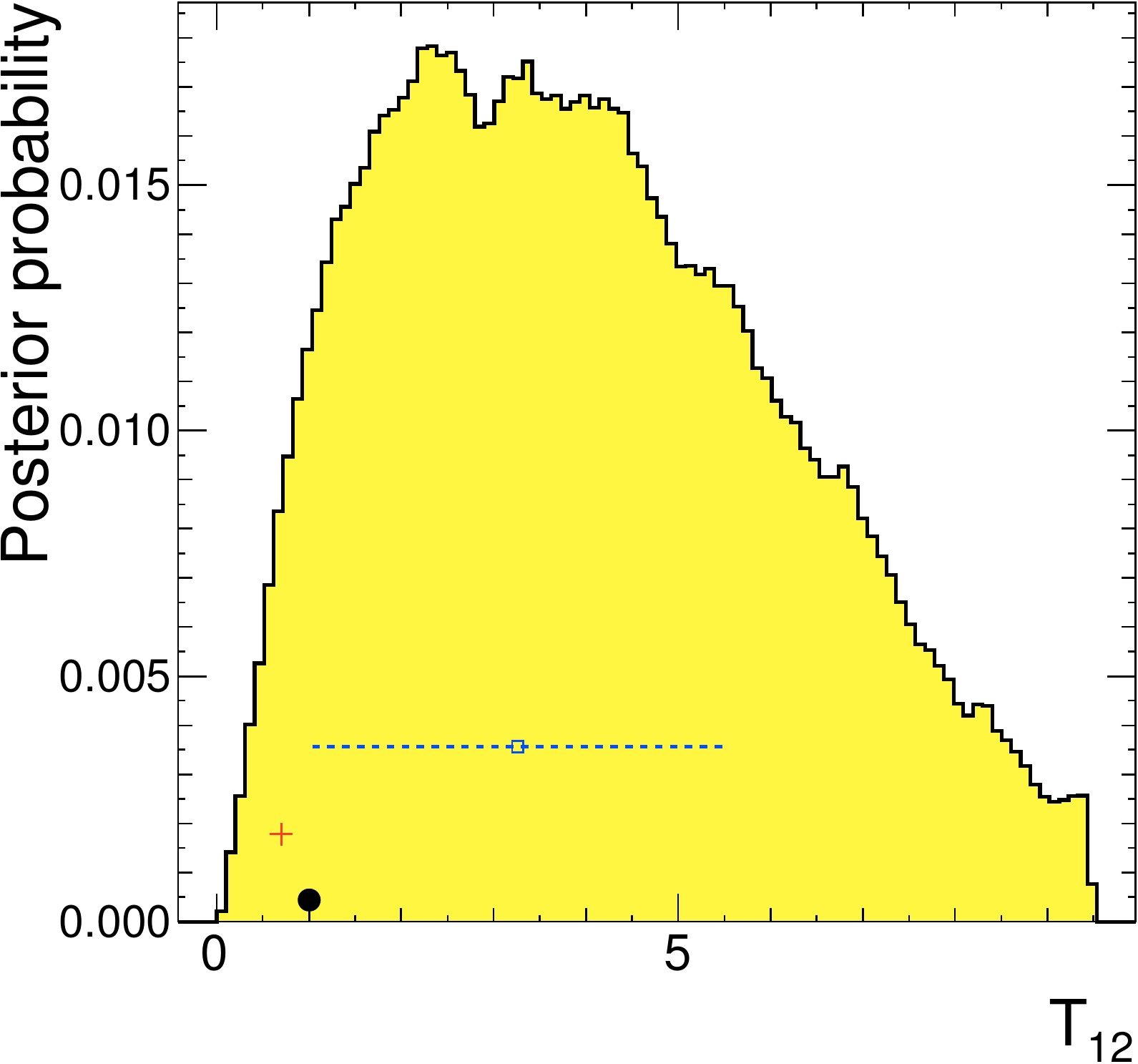} &
    \includegraphics[width=0.18\columnwidth]{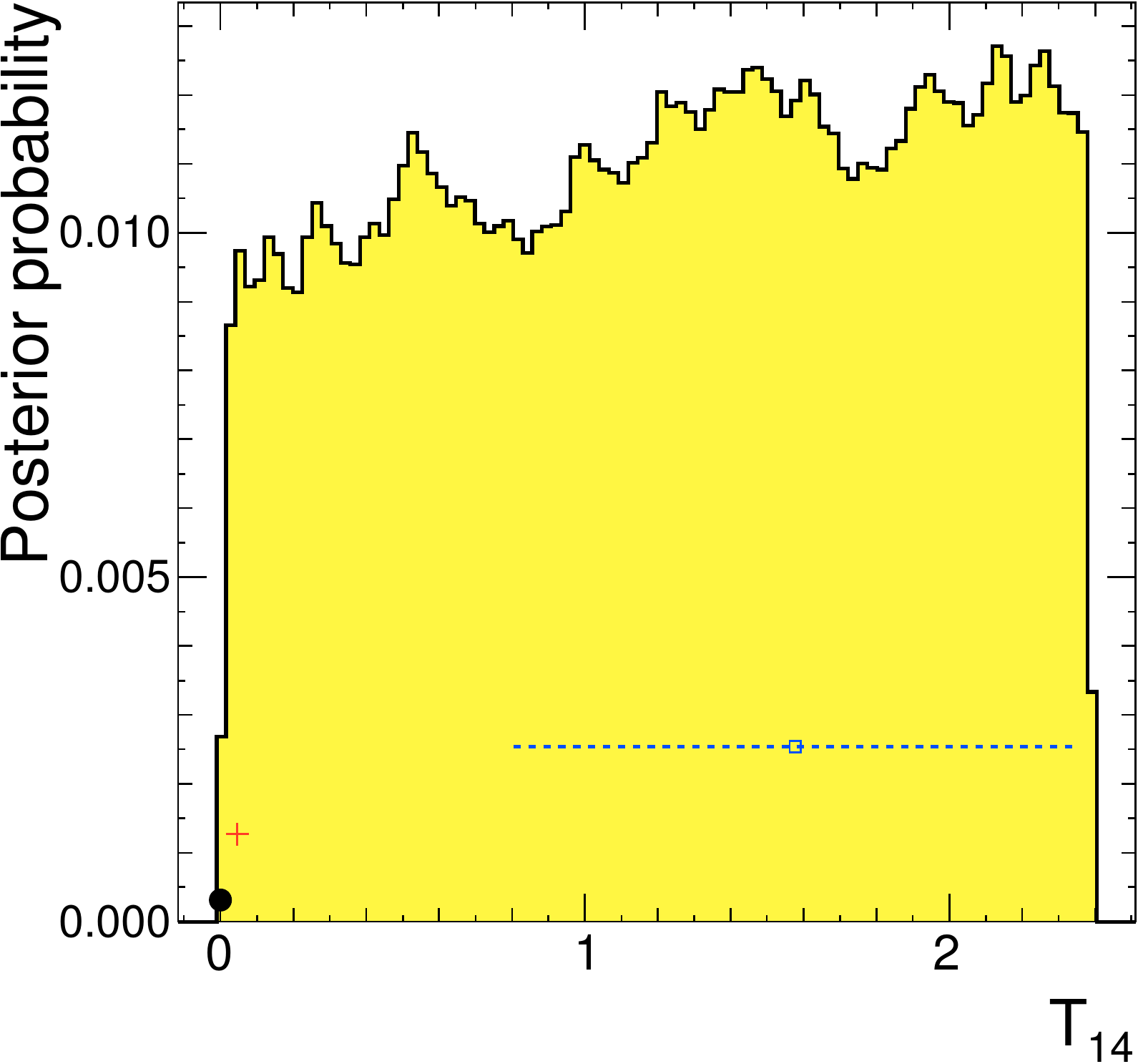} \\

    \includegraphics[width=0.18\columnwidth]{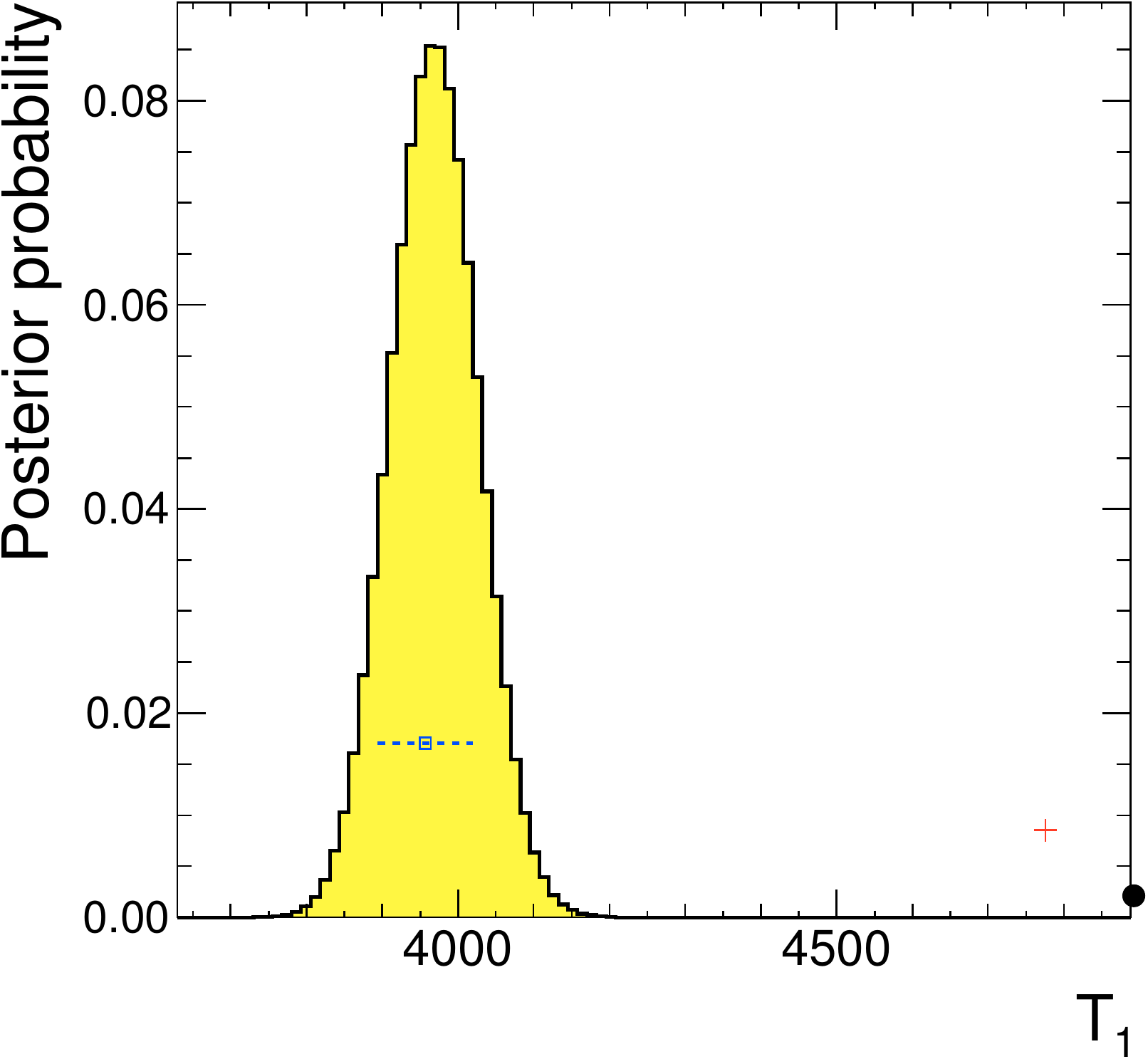} &
    \includegraphics[width=0.18\columnwidth]{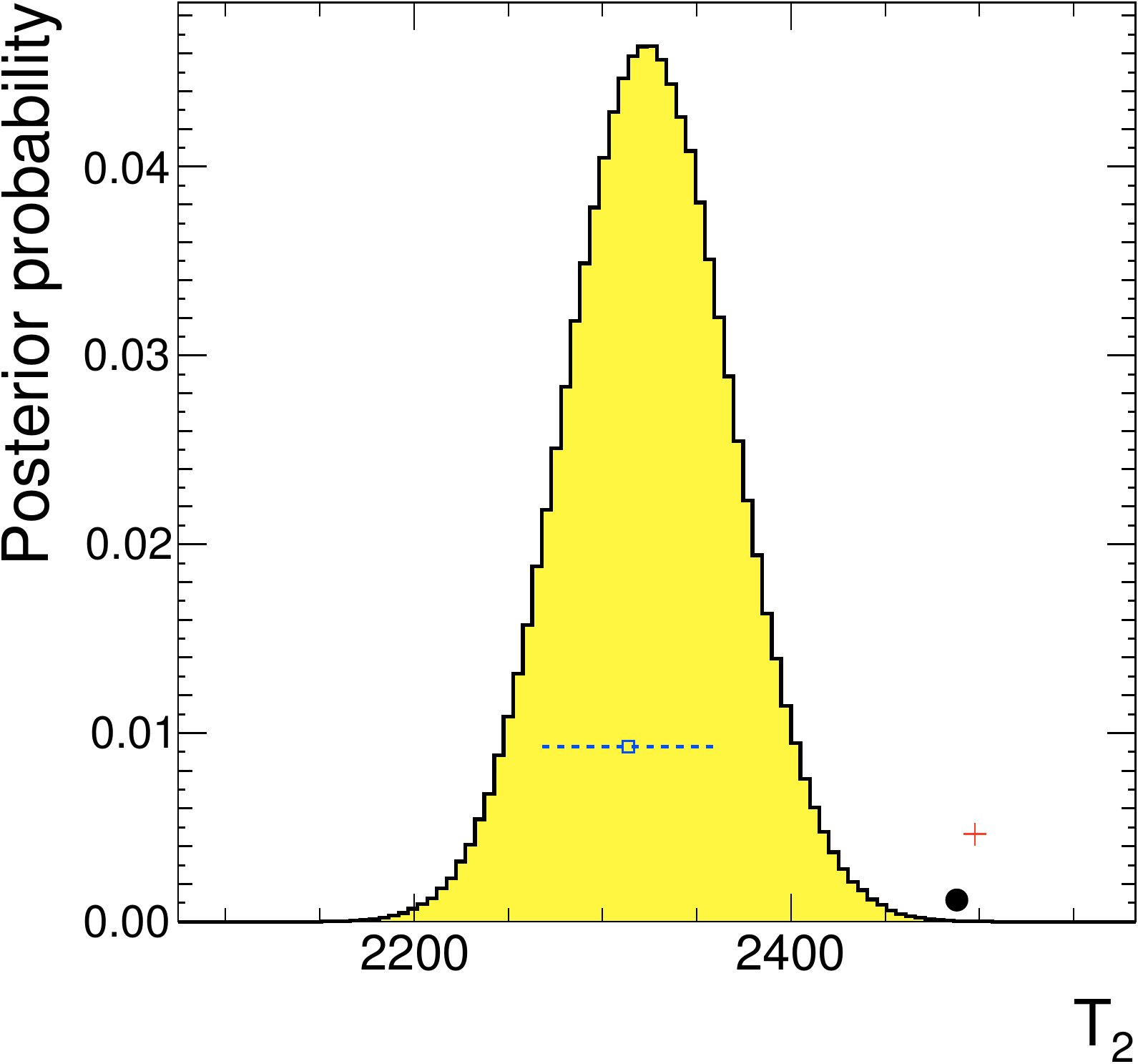} &
    \includegraphics[width=0.18\columnwidth]{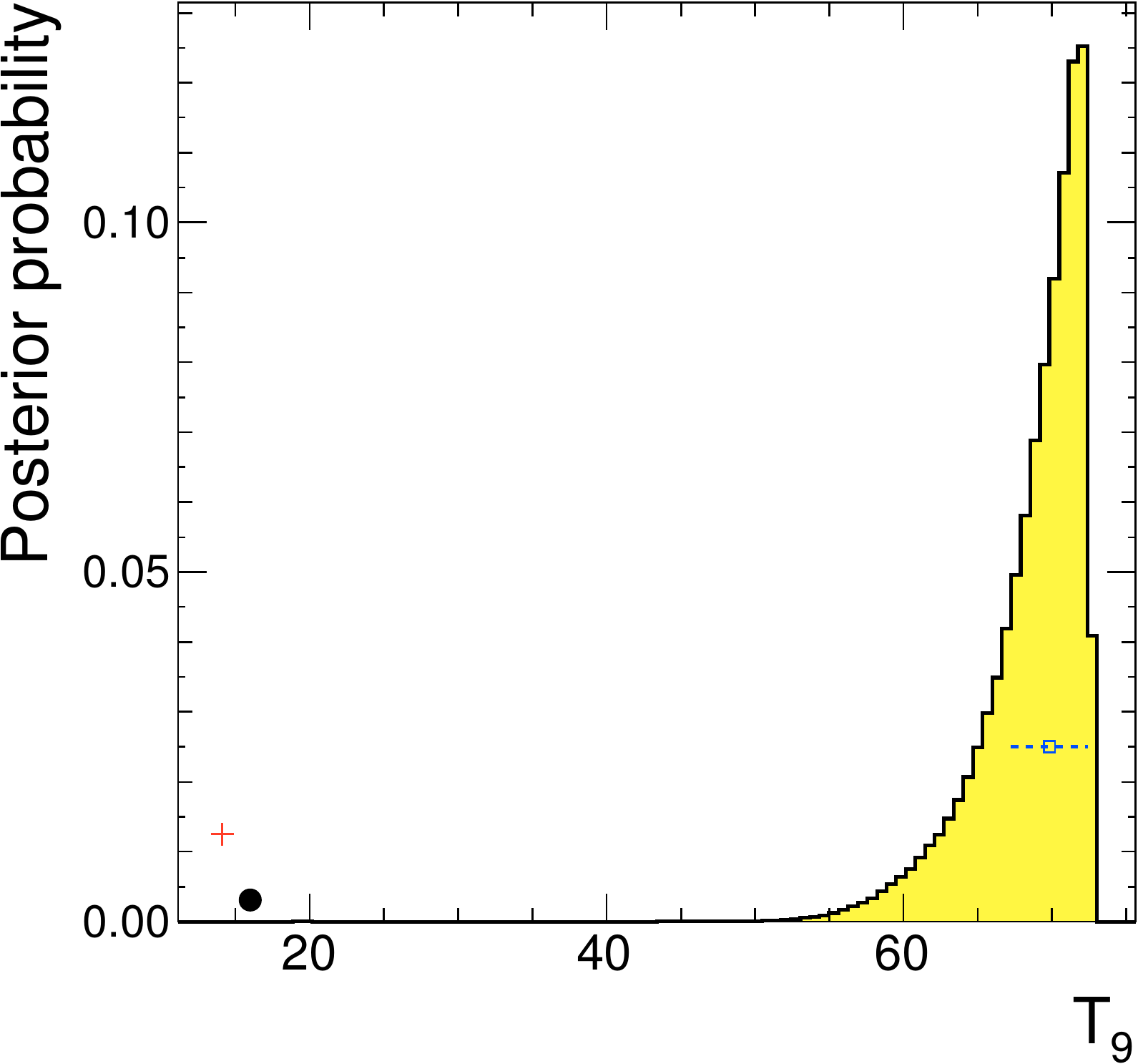} &
    \includegraphics[width=0.18\columnwidth]{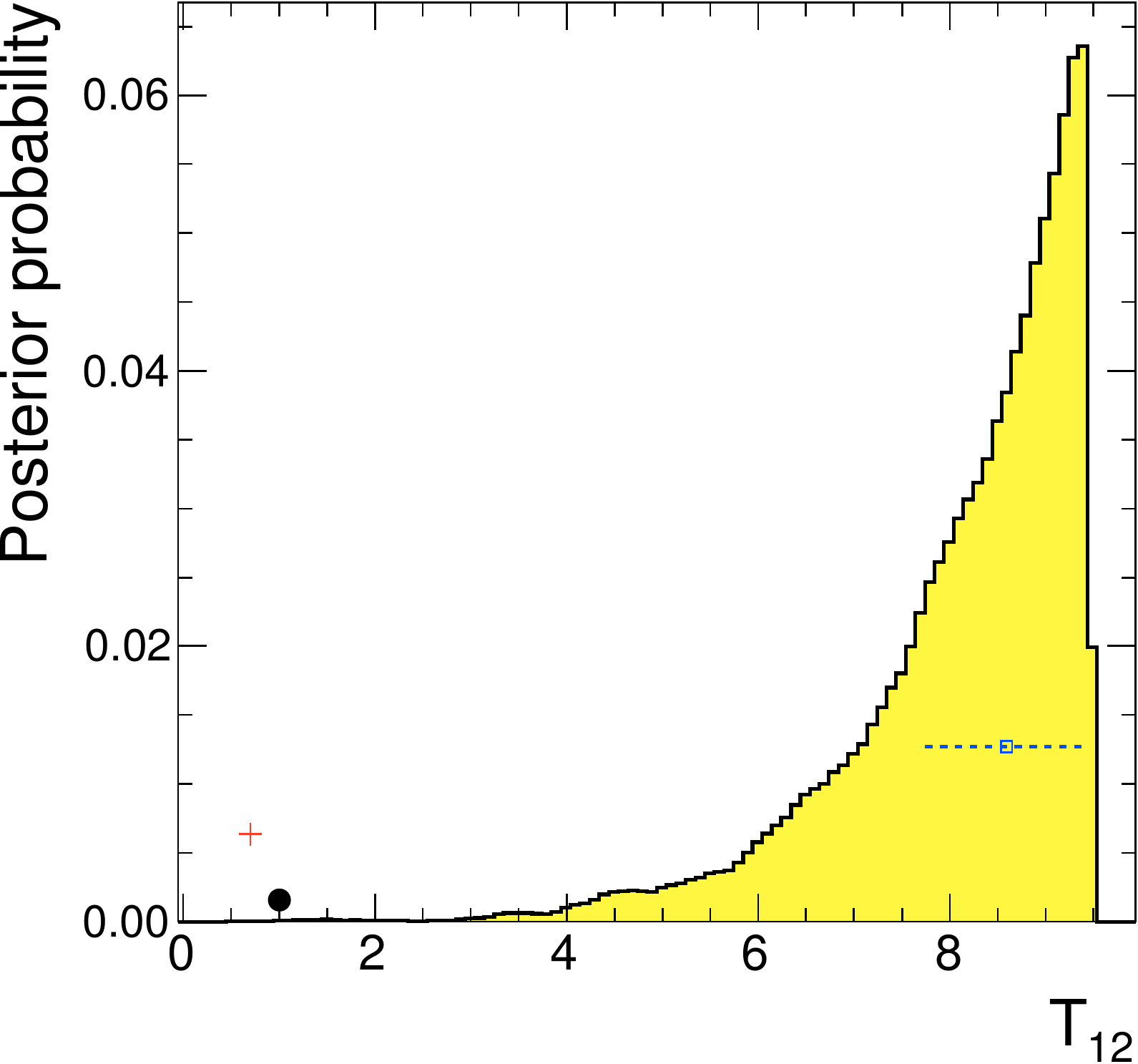} &
    \includegraphics[width=0.18\columnwidth]{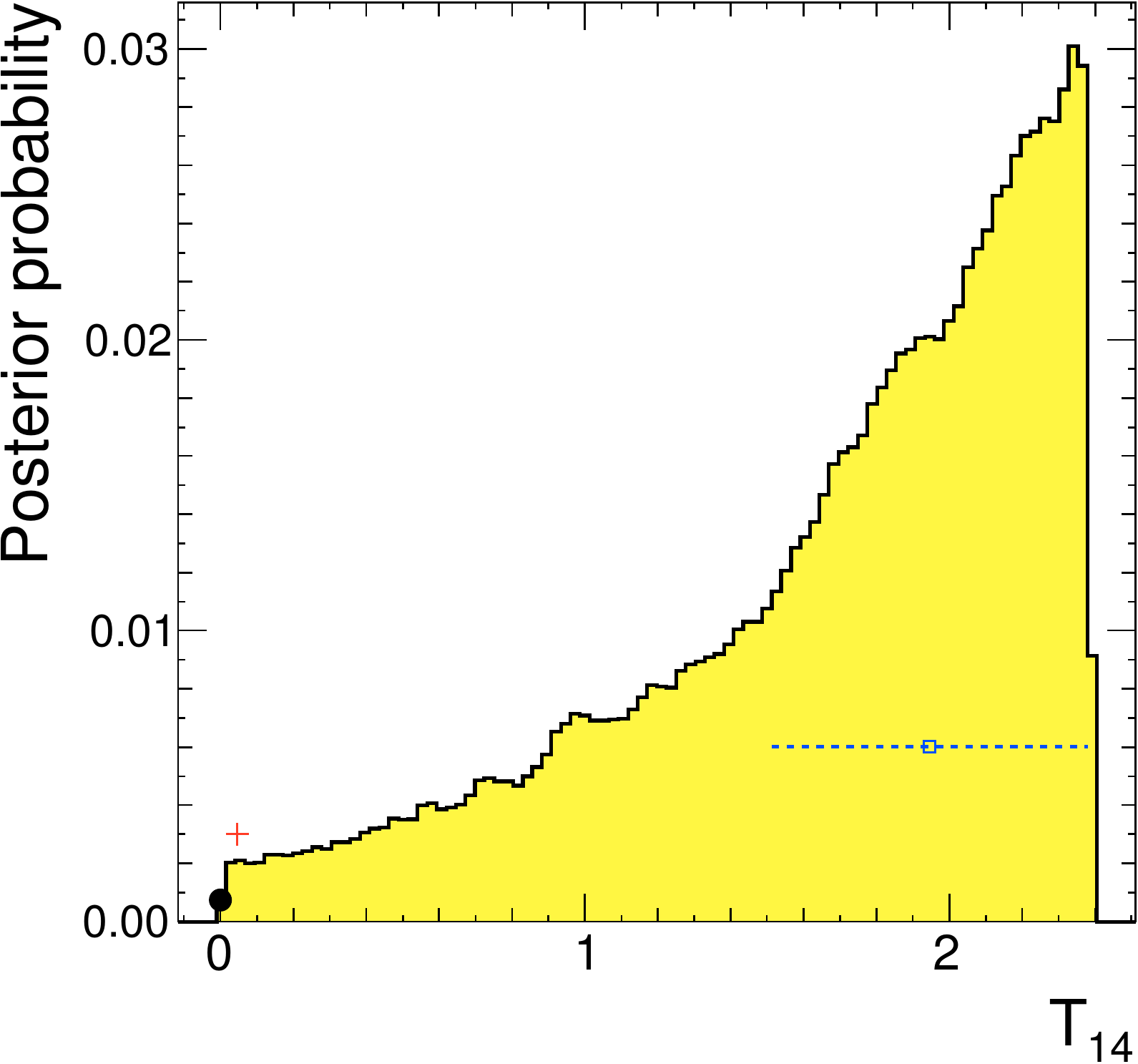} \\

    \includegraphics[width=0.18\columnwidth]{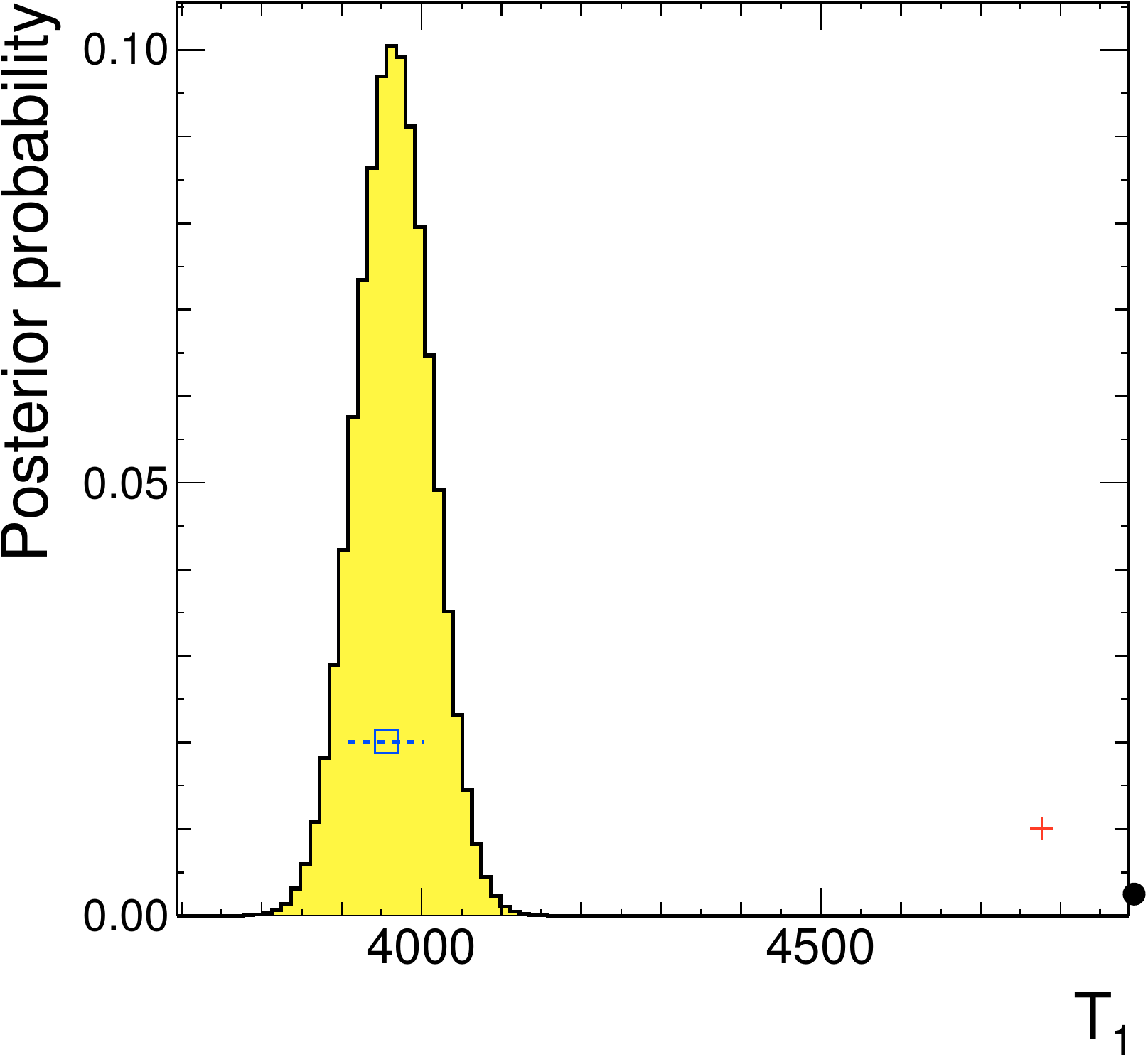} &
    \includegraphics[width=0.18\columnwidth]{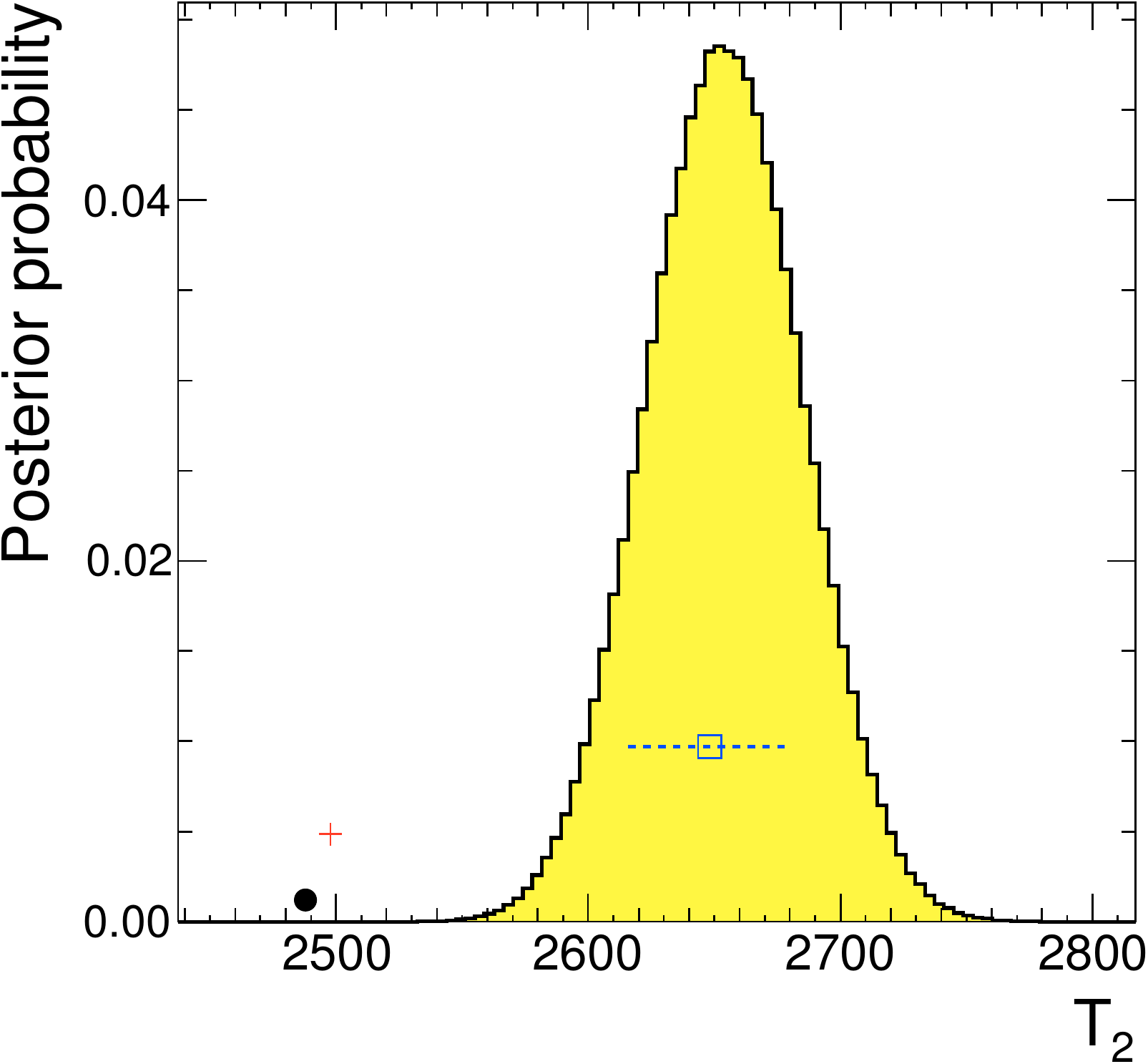} &
    \includegraphics[width=0.18\columnwidth]{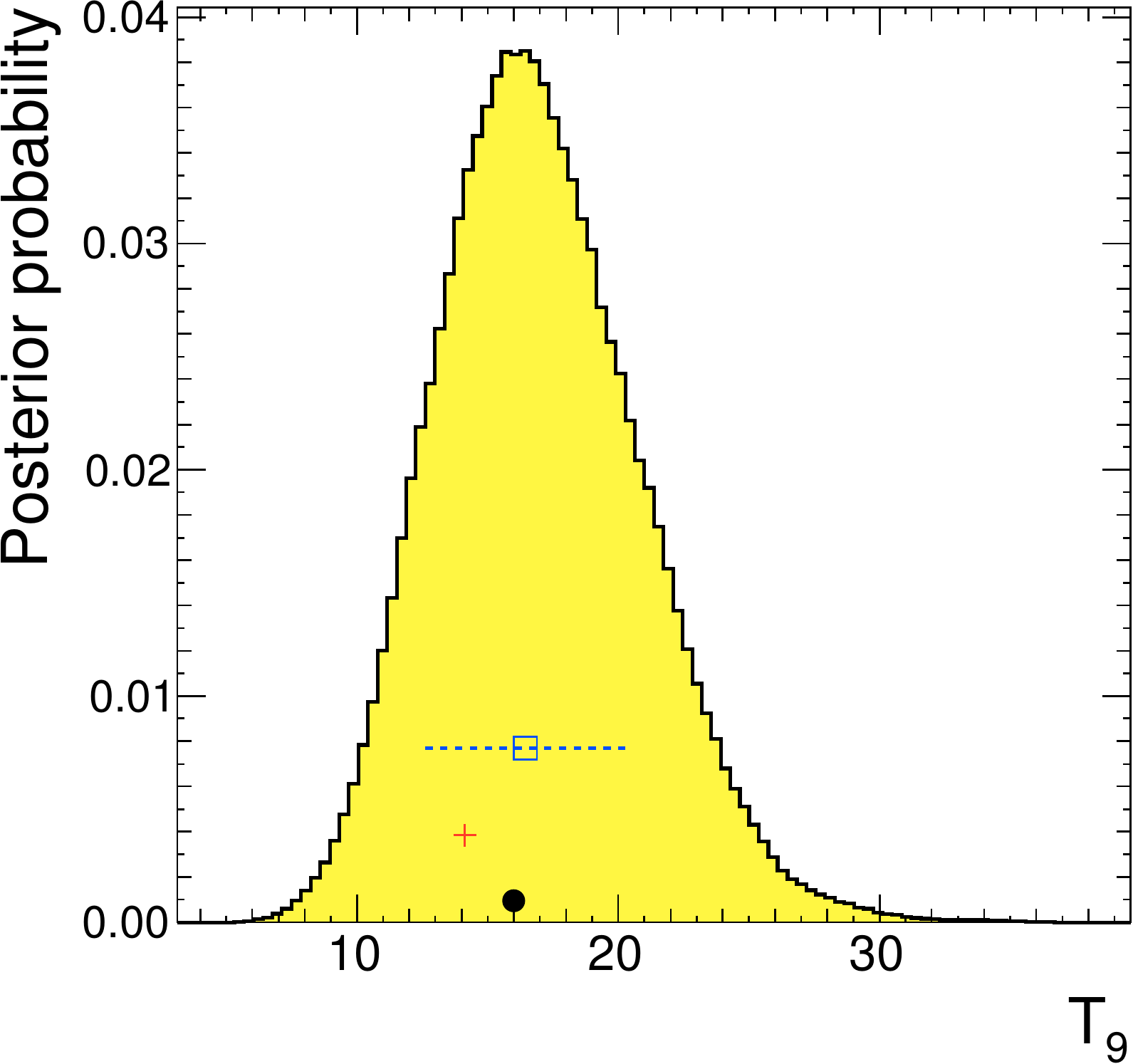} &
    \includegraphics[width=0.18\columnwidth]{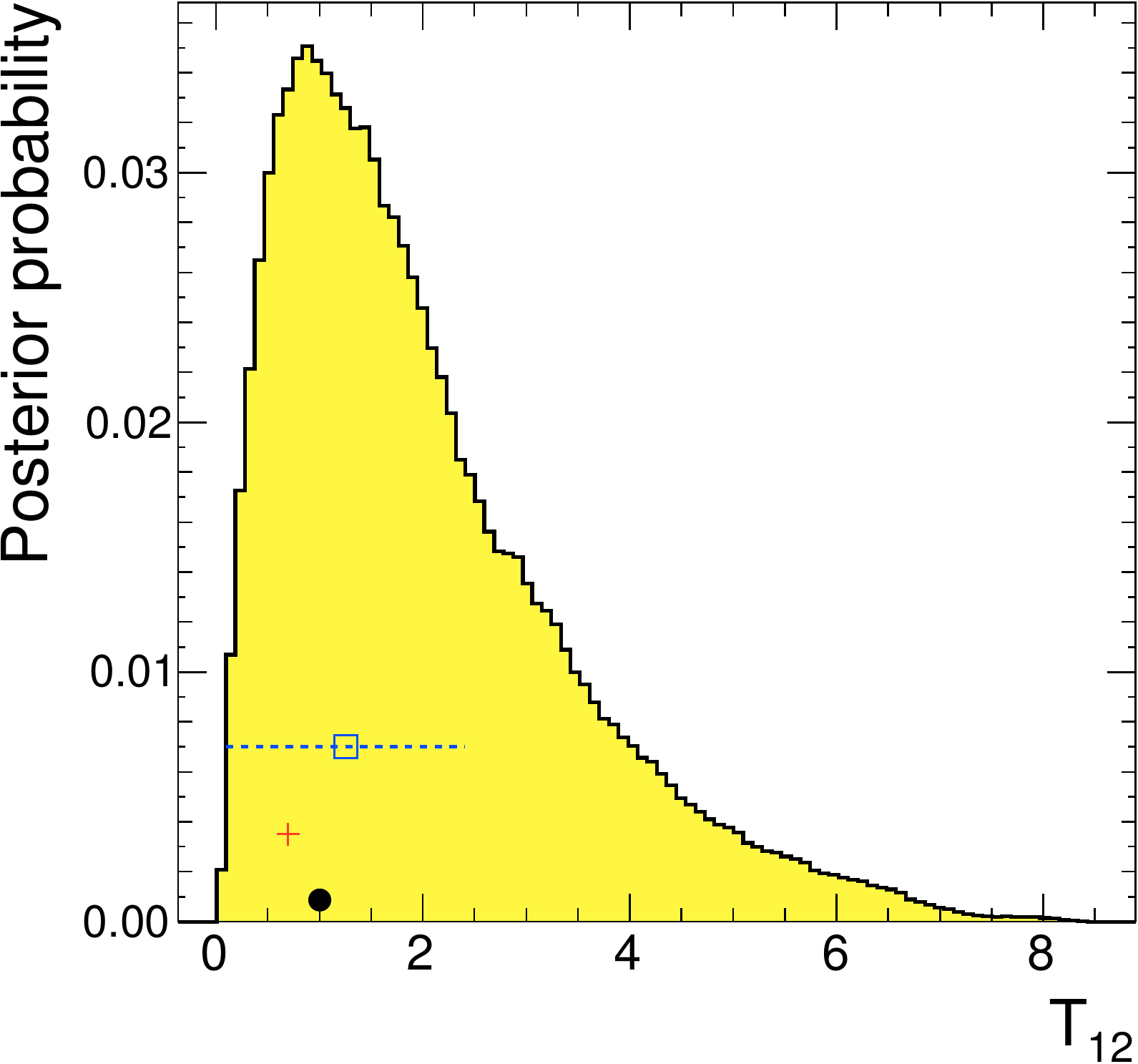} &
    \includegraphics[width=0.18\columnwidth]{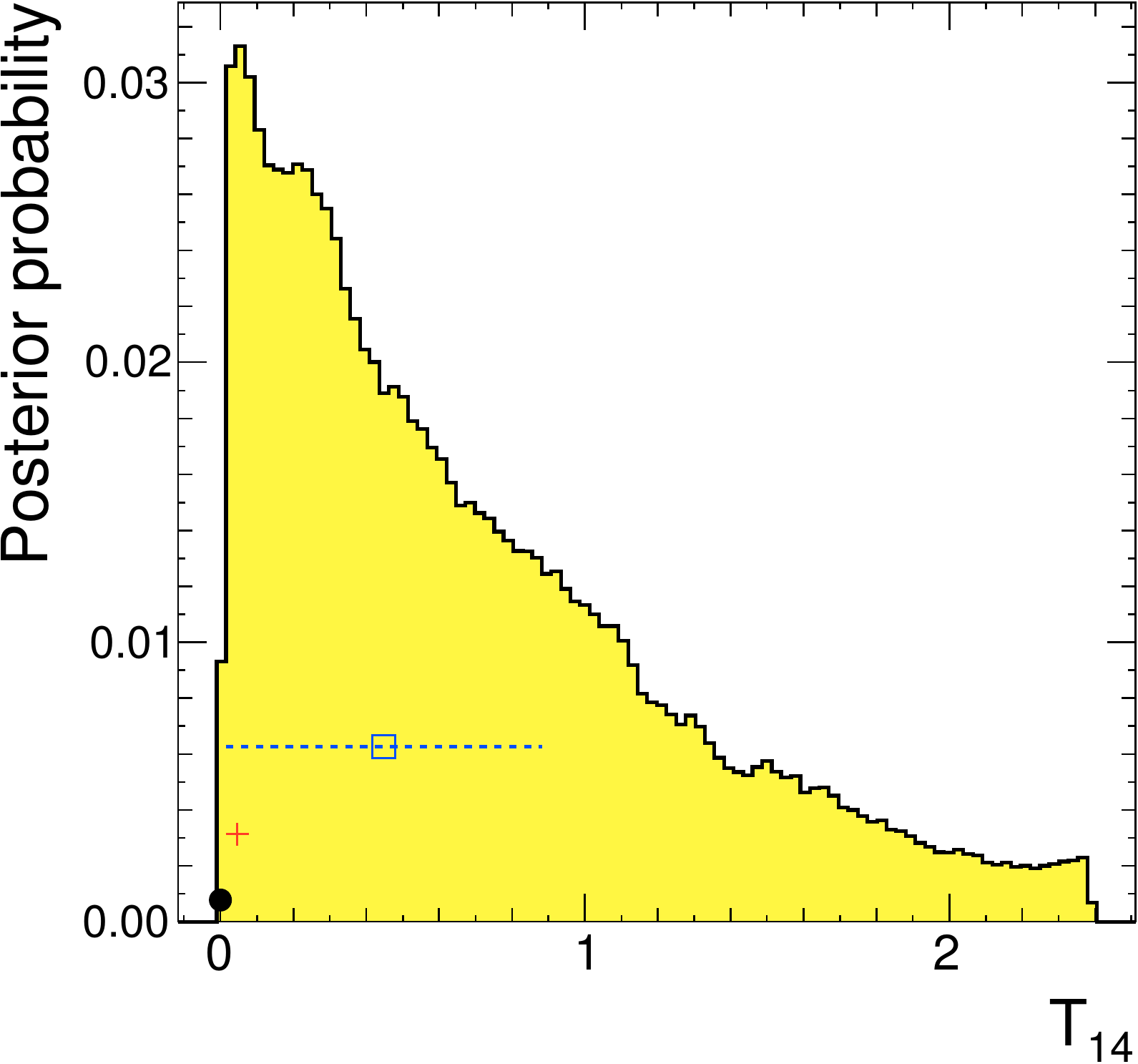} \\

   \includegraphics[width=0.18\columnwidth]{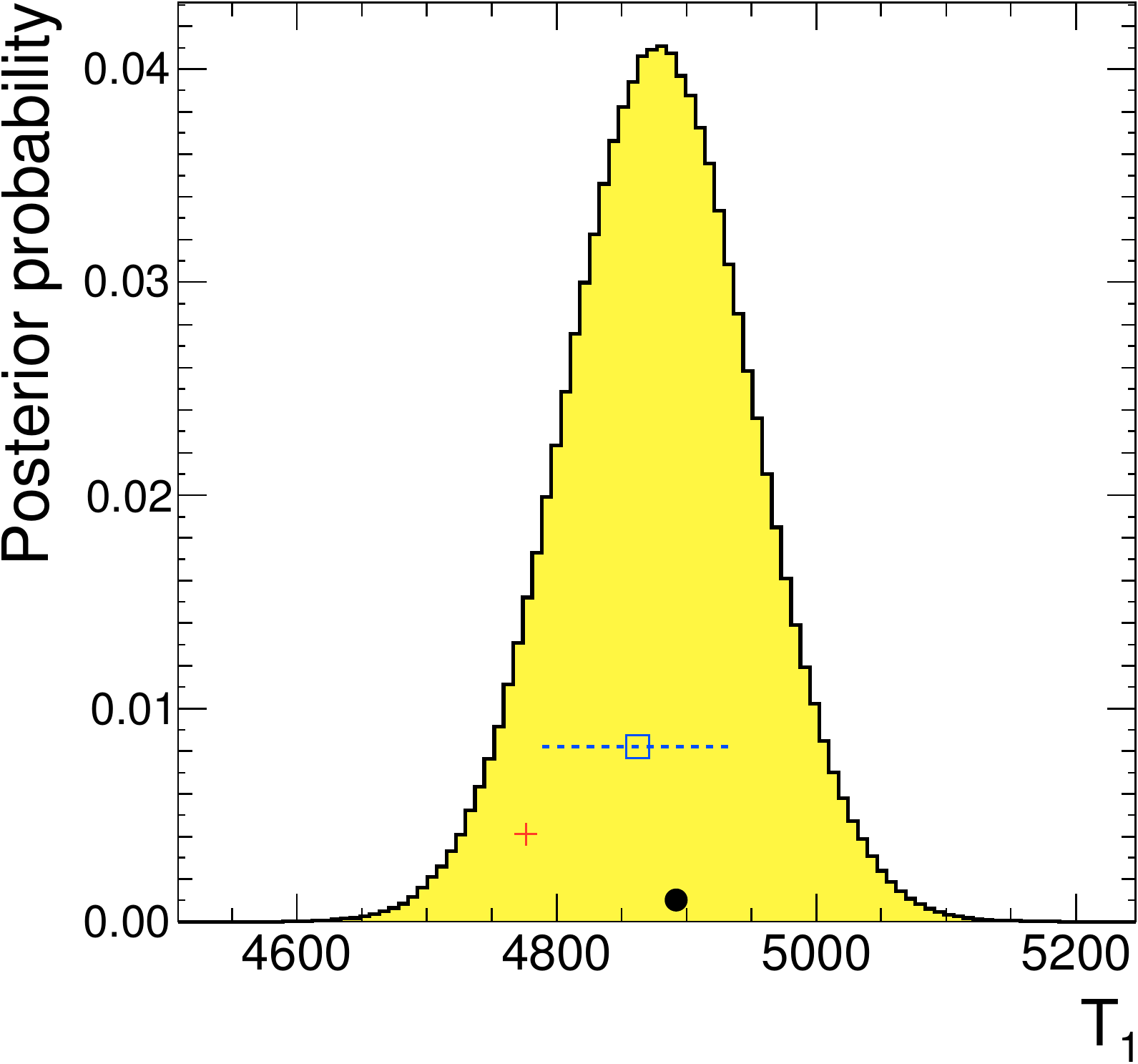} &
    \includegraphics[width=0.18\columnwidth]{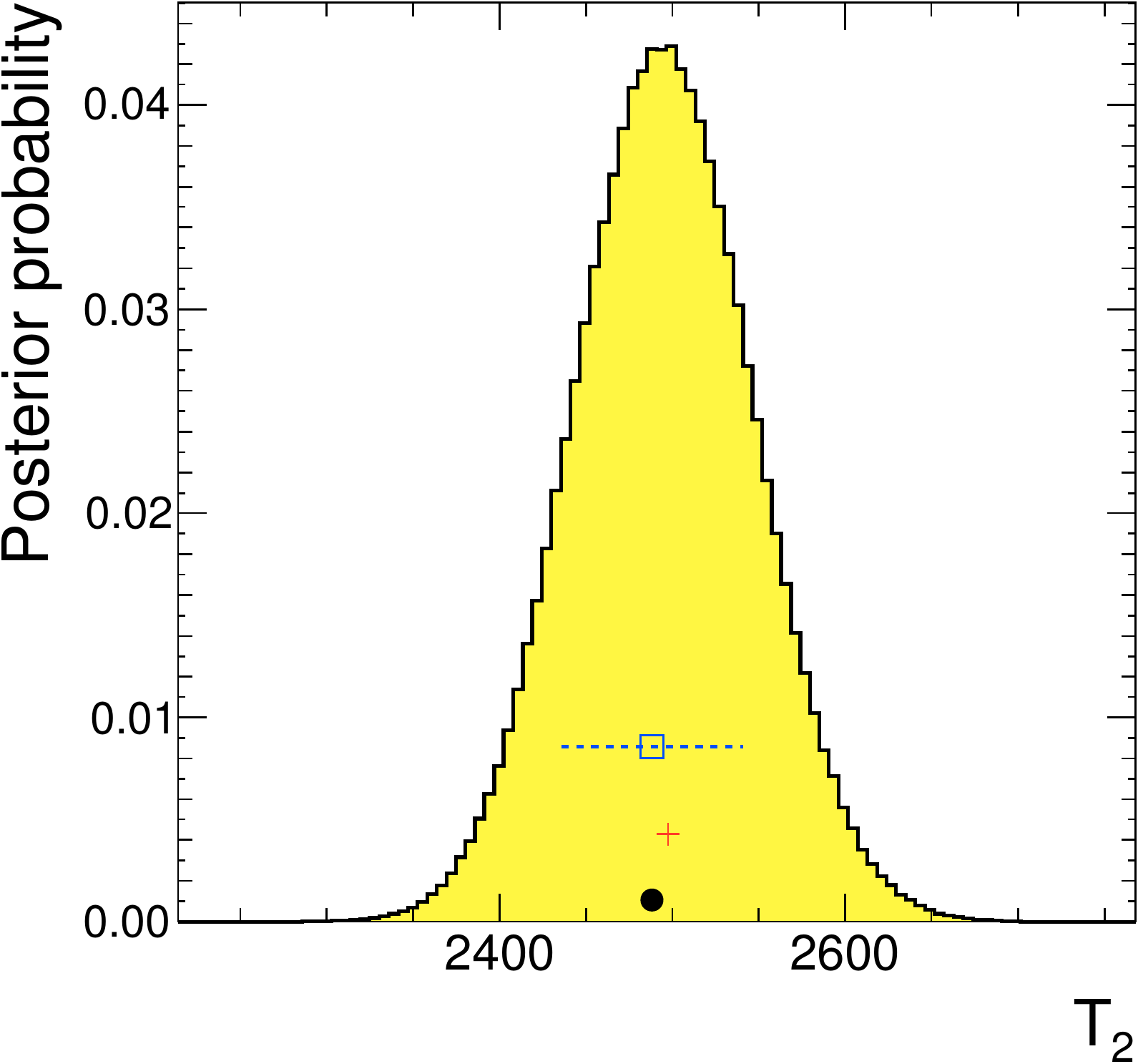} &
    \includegraphics[width=0.18\columnwidth]{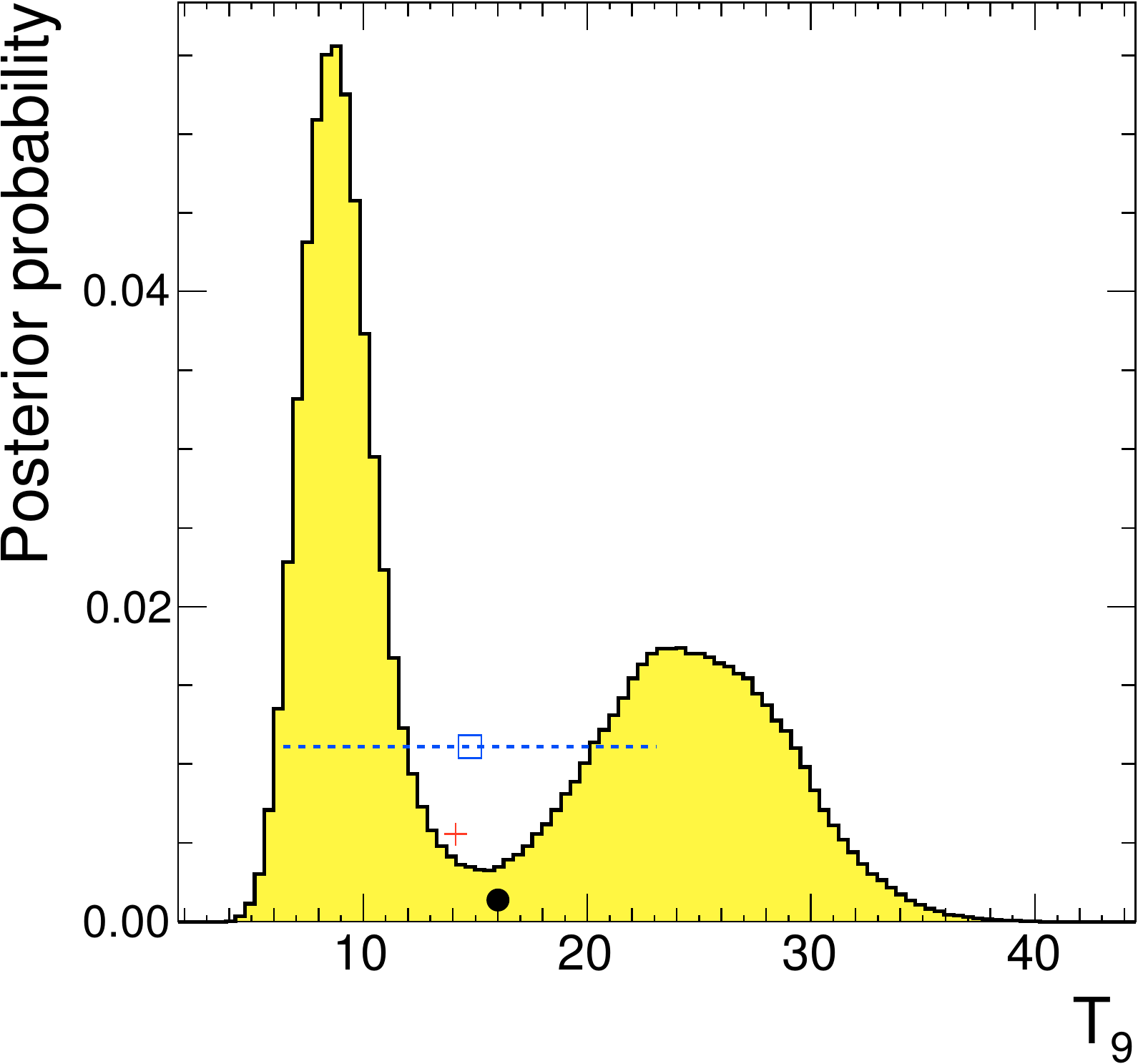} &
    \includegraphics[width=0.18\columnwidth]{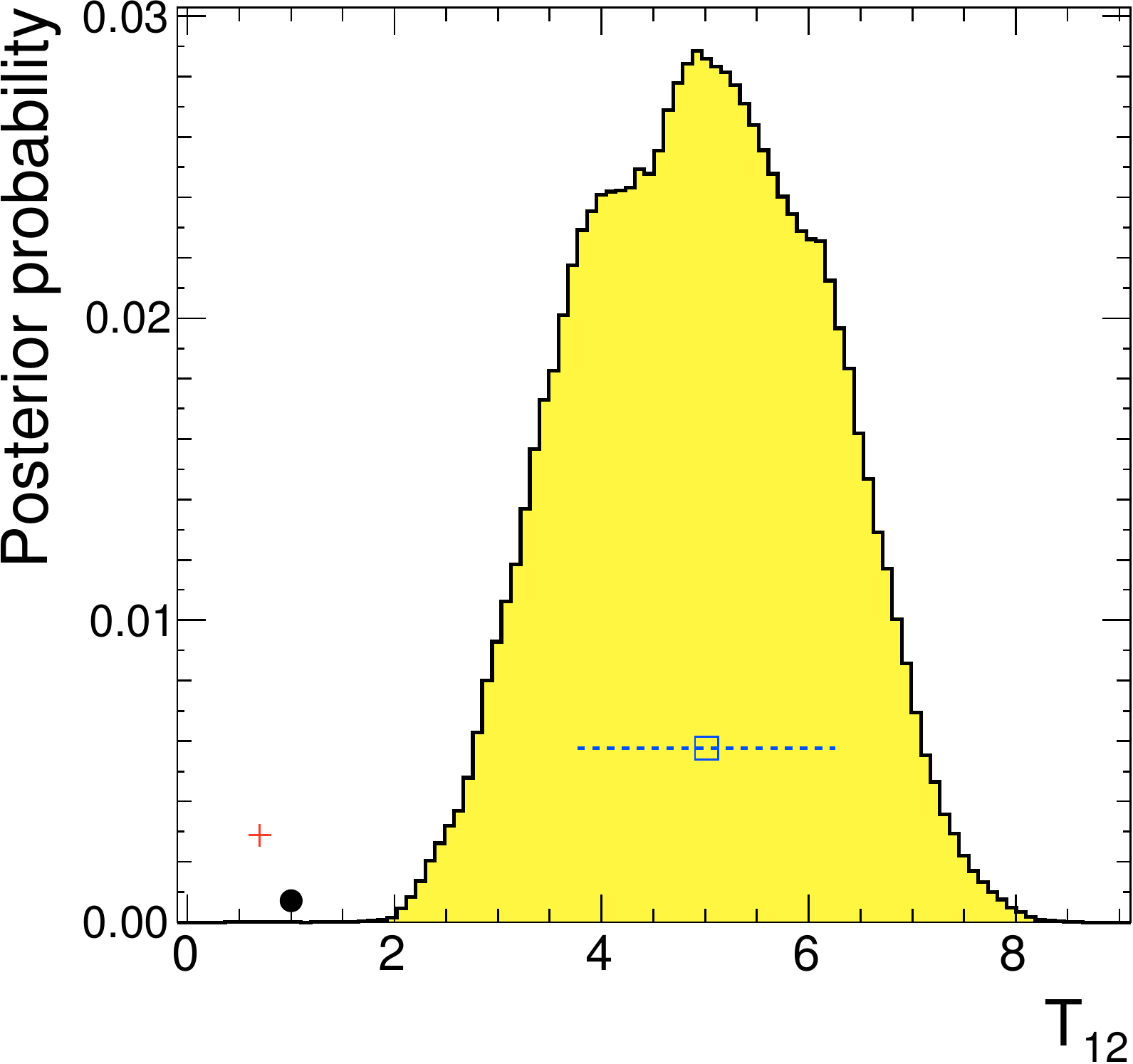} &
    \includegraphics[width=0.18\columnwidth]{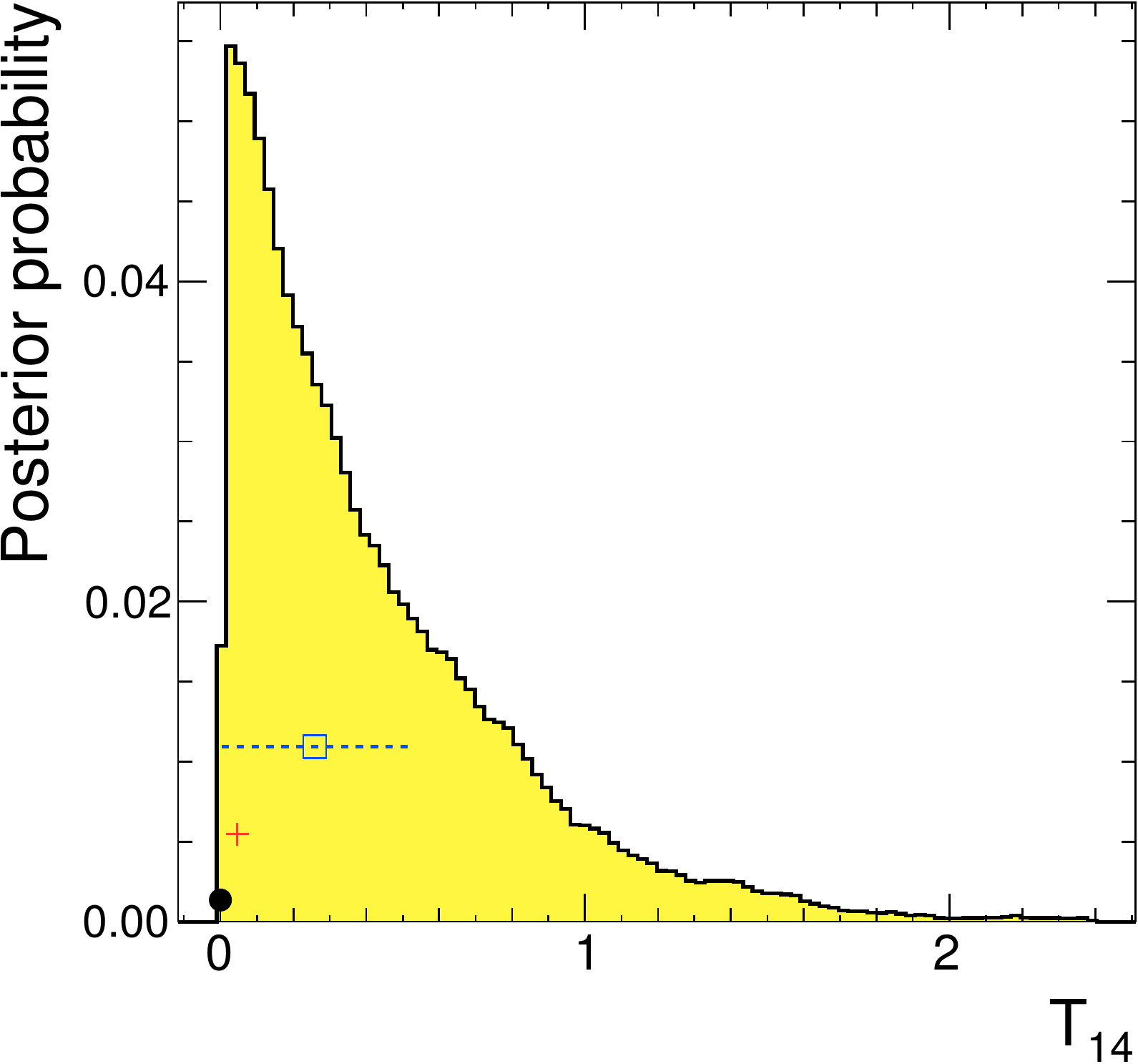} \\

    \includegraphics[width=0.18\columnwidth]{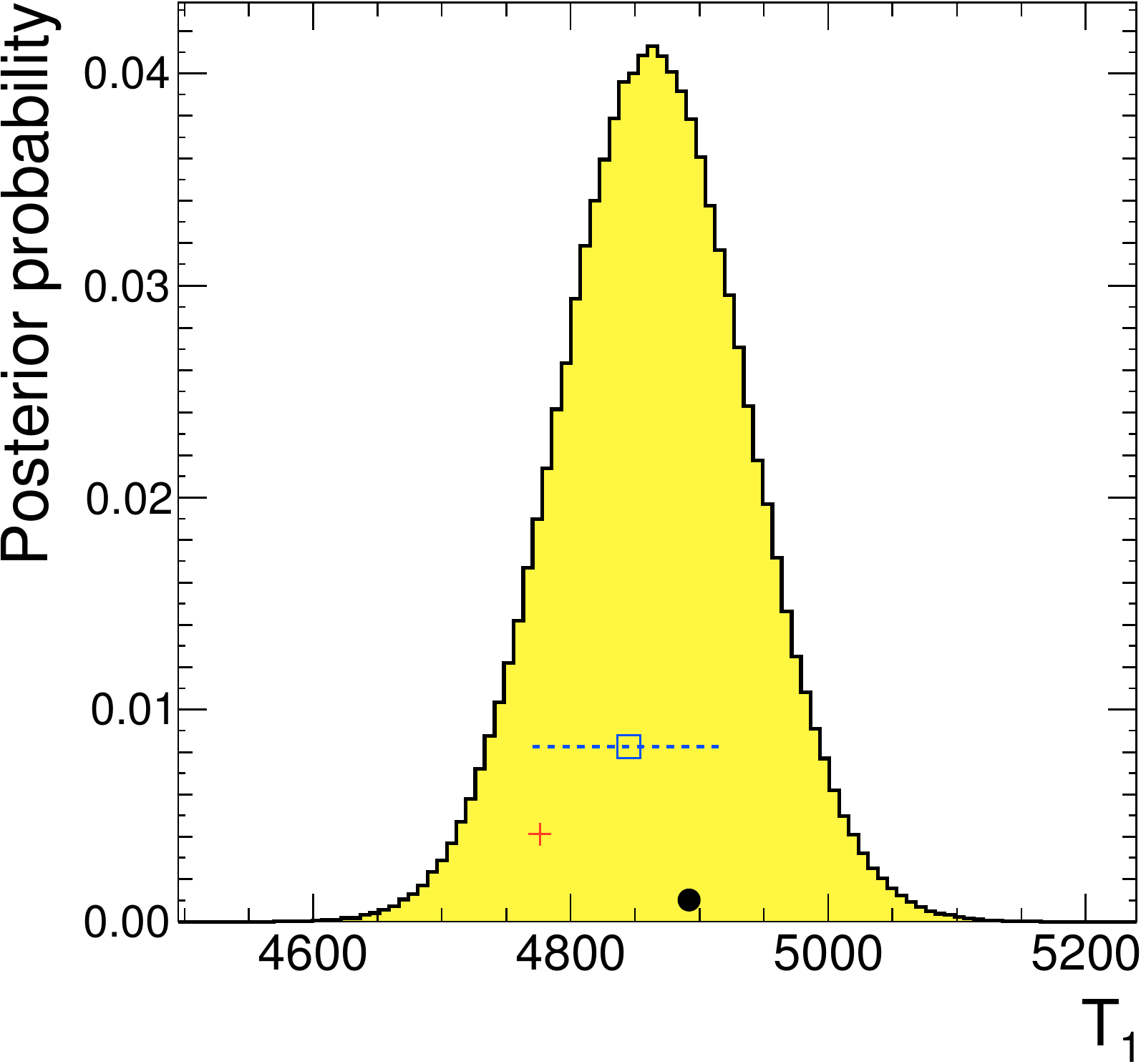} &
    \includegraphics[width=0.18\columnwidth]{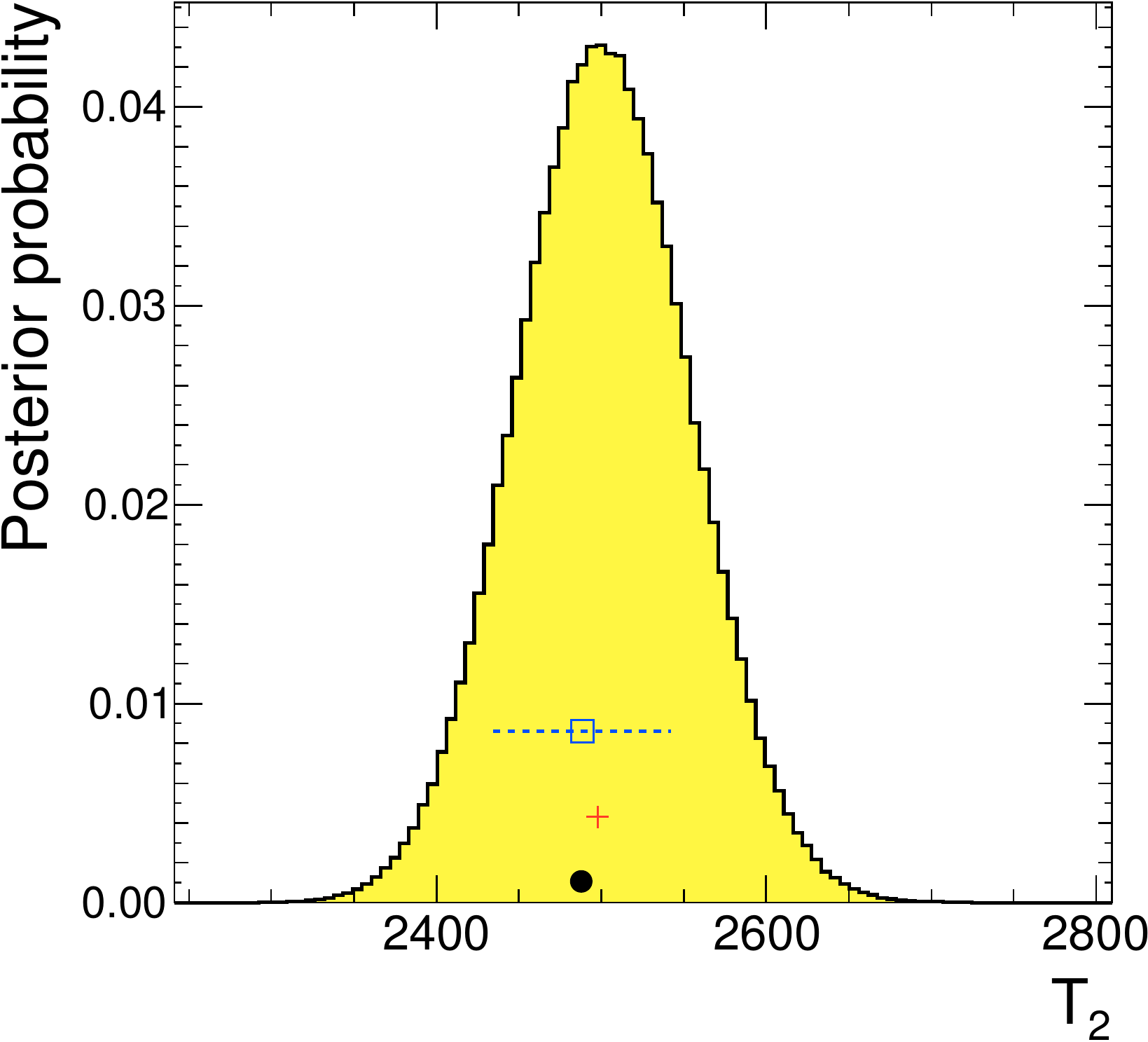} &
    \includegraphics[width=0.18\columnwidth]{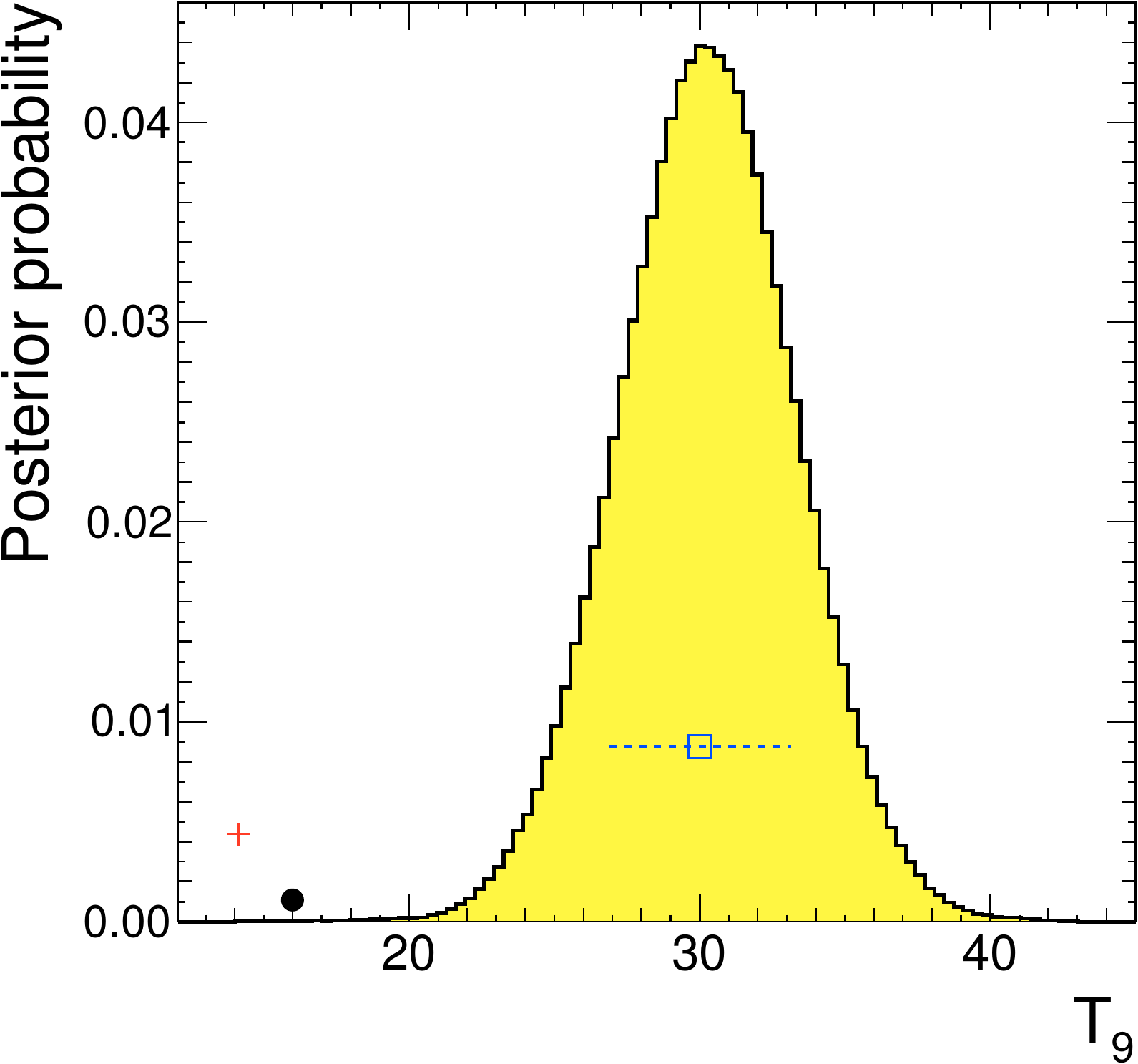} &
    \includegraphics[width=0.18\columnwidth]{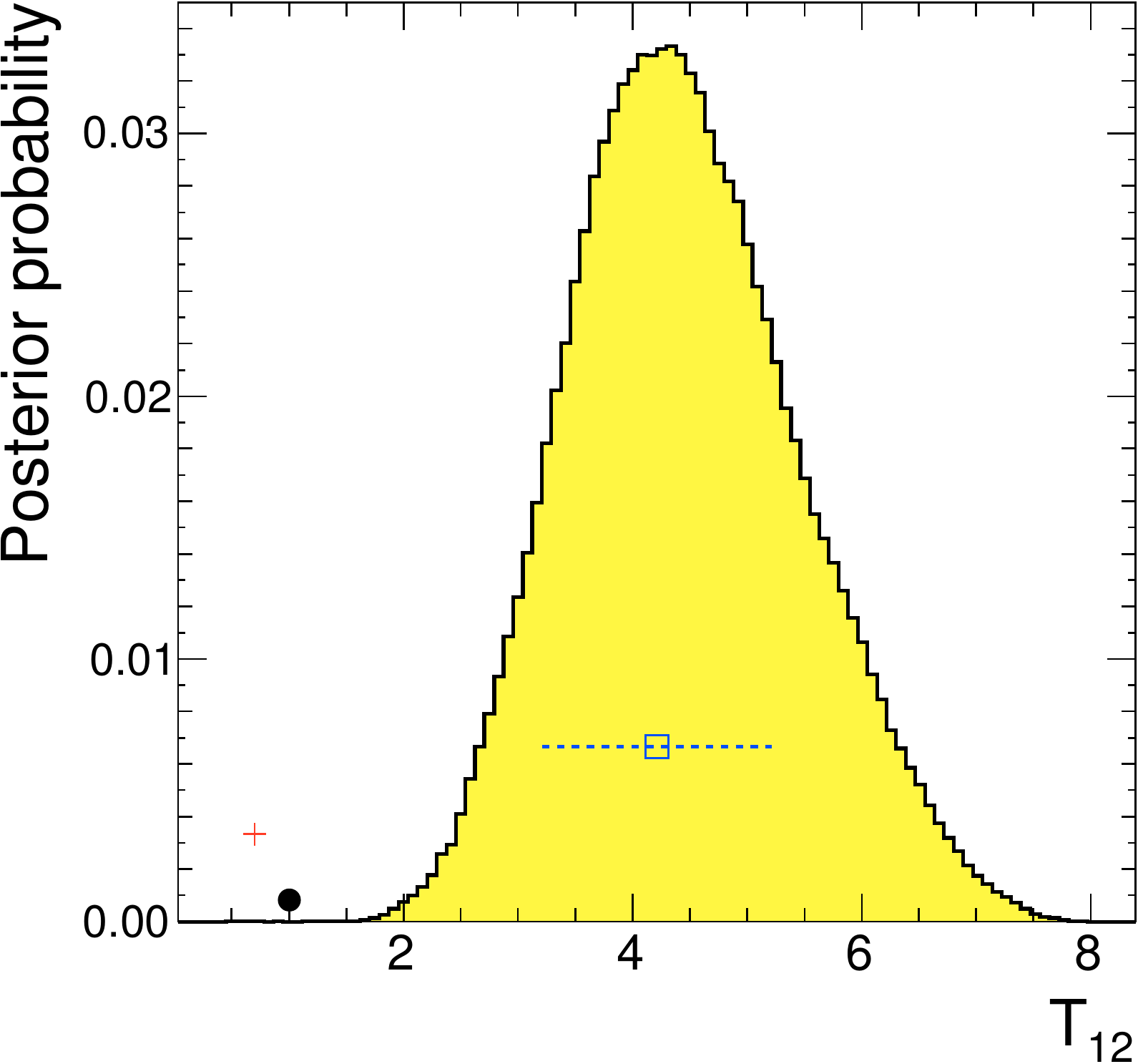} &
    \includegraphics[width=0.18\columnwidth]{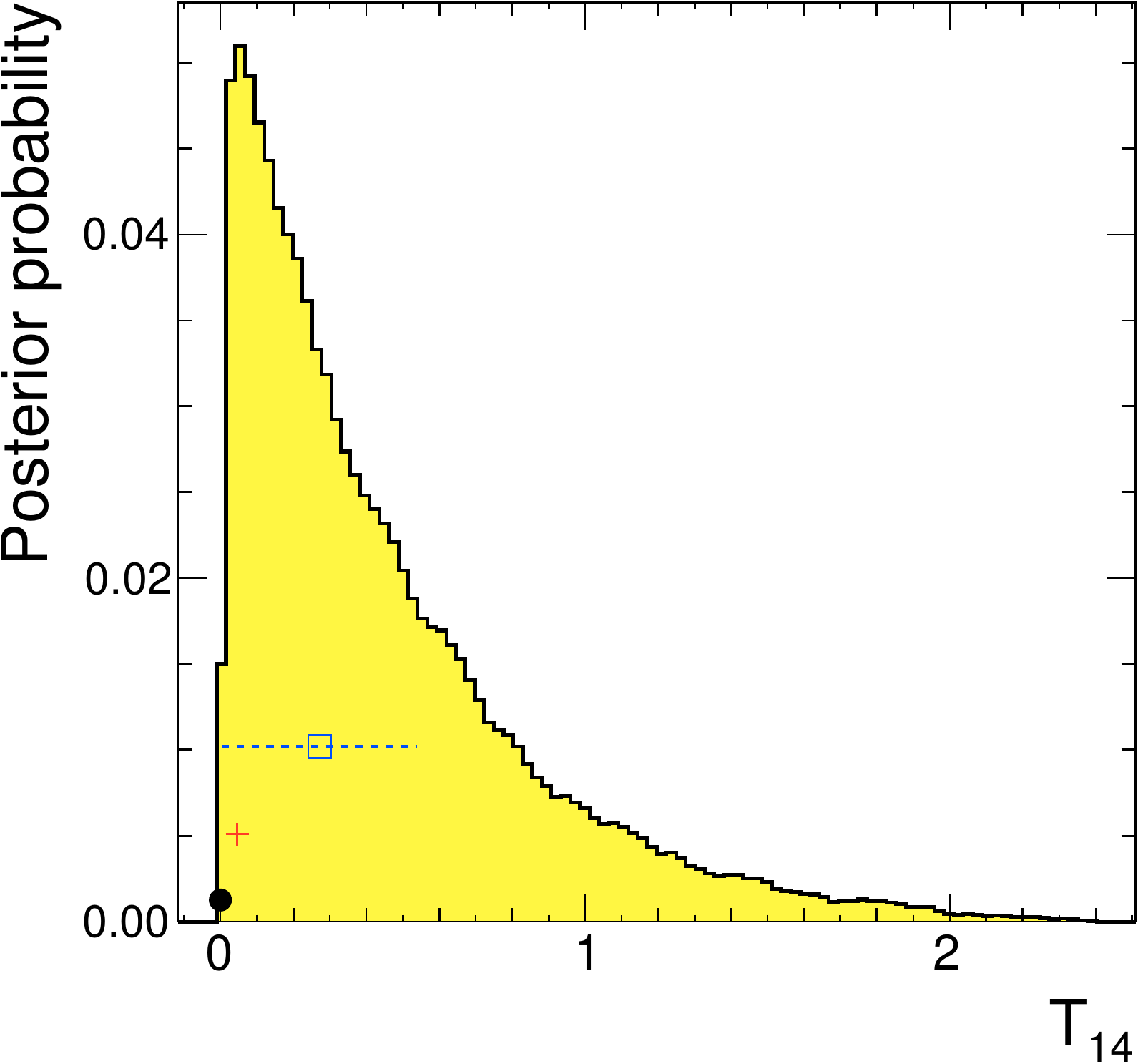} \\

 \end{tabular}
 \caption{Some 1-dimensional distributions from Sec.~\ref{sec:regSteepNoSmearing}.  The columns show $P_{t}(T_t|\tuple{D})$ with $t=\{ 1,2,9,12,14 \}$.  The rows correspond to regularization with $(S,\alpha) = \{ (S_1,0) , (S_1,1\times 10^3) , (S_1, 3\times 10^3) , (S_2,6\times 10^{-4}) , (S_3,20) ,  (S_3,40) \}$, in this order.
\label{fig:1DimSteepNoSmear}
}
\end{figure}

\begin{figure}[H]
  \centering
  \begin{tabular}{ccccc}
    \includegraphics[width=0.18\columnwidth]{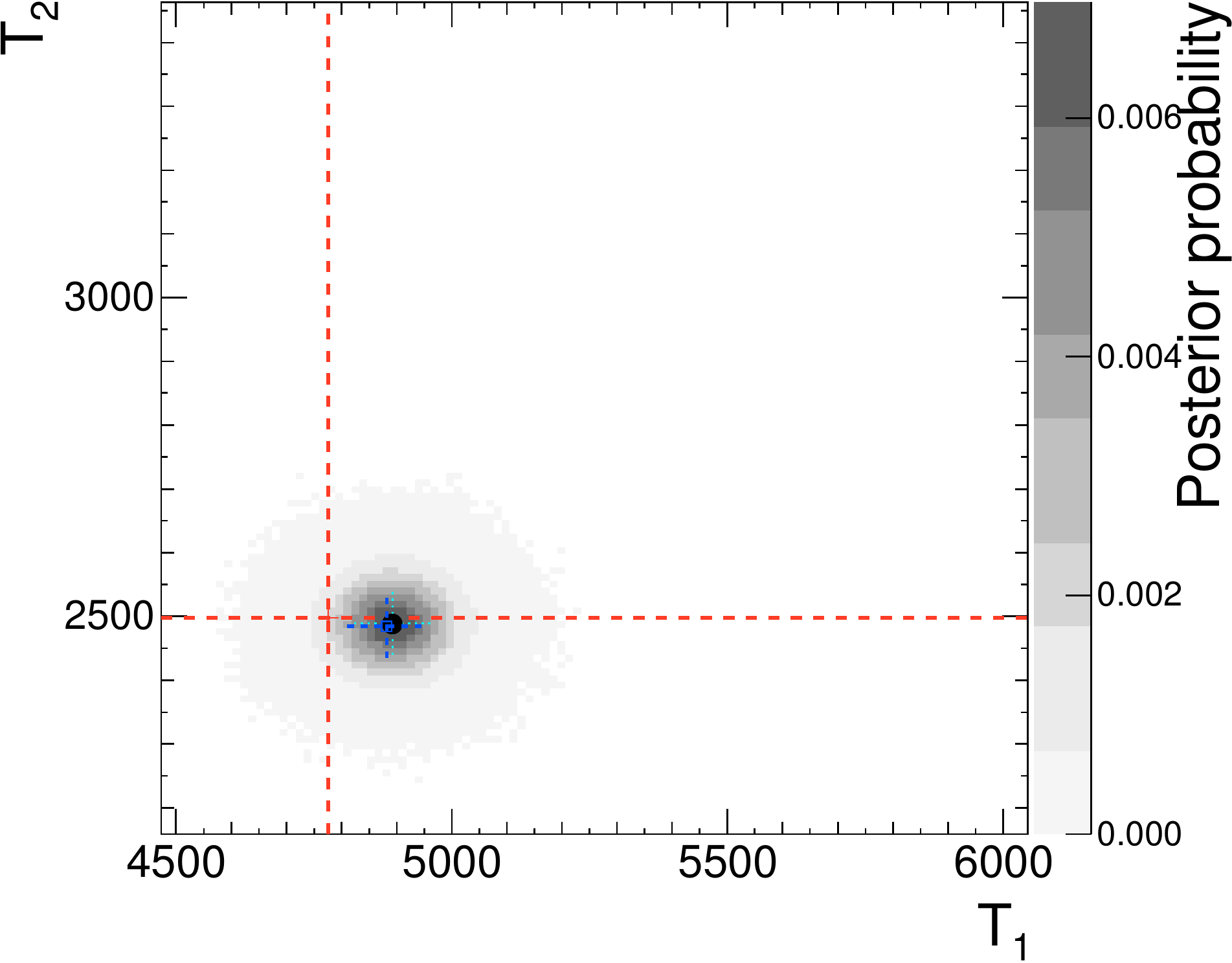} &
    \includegraphics[width=0.18\columnwidth]{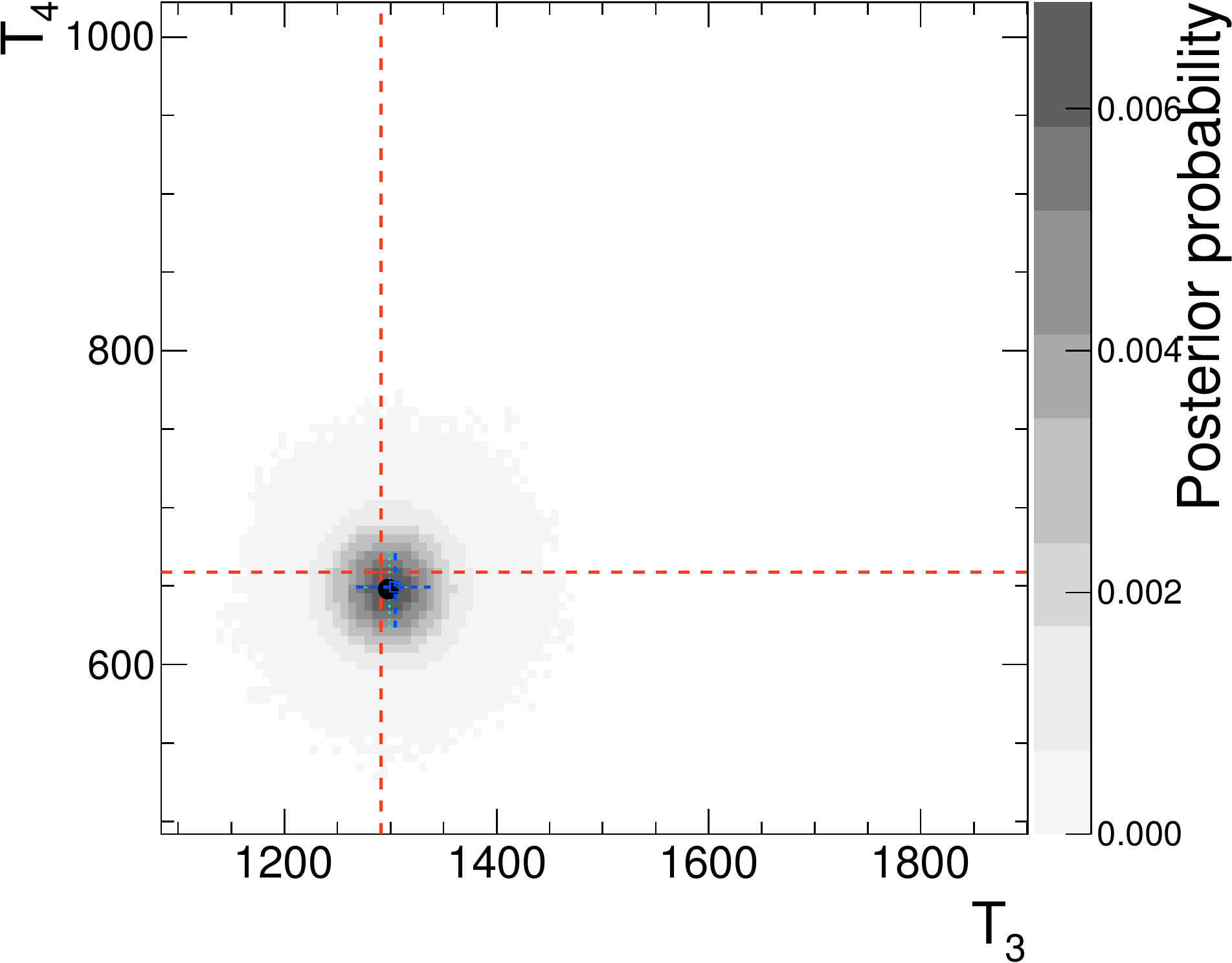} &
    \includegraphics[width=0.18\columnwidth]{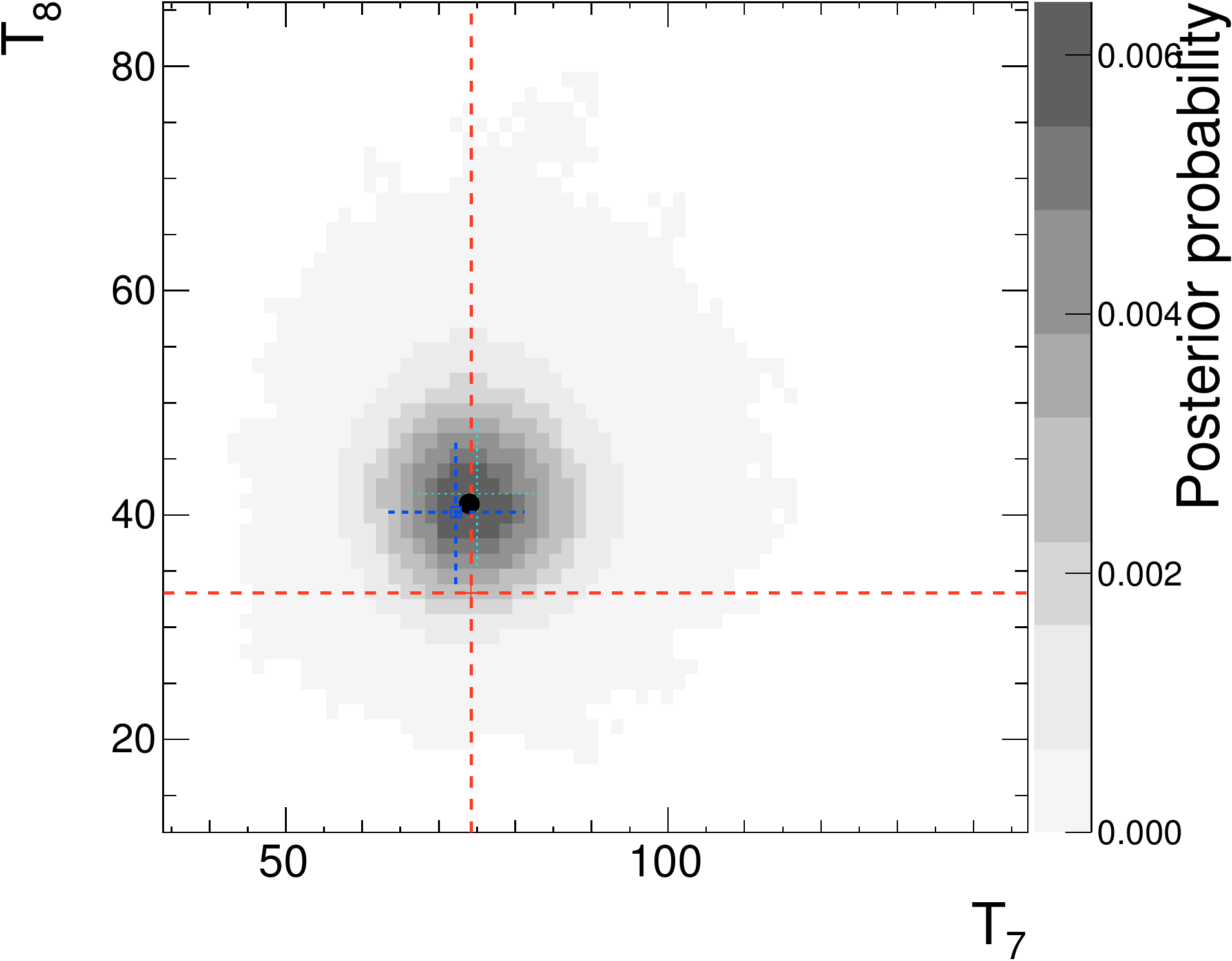} &
    \includegraphics[width=0.18\columnwidth]{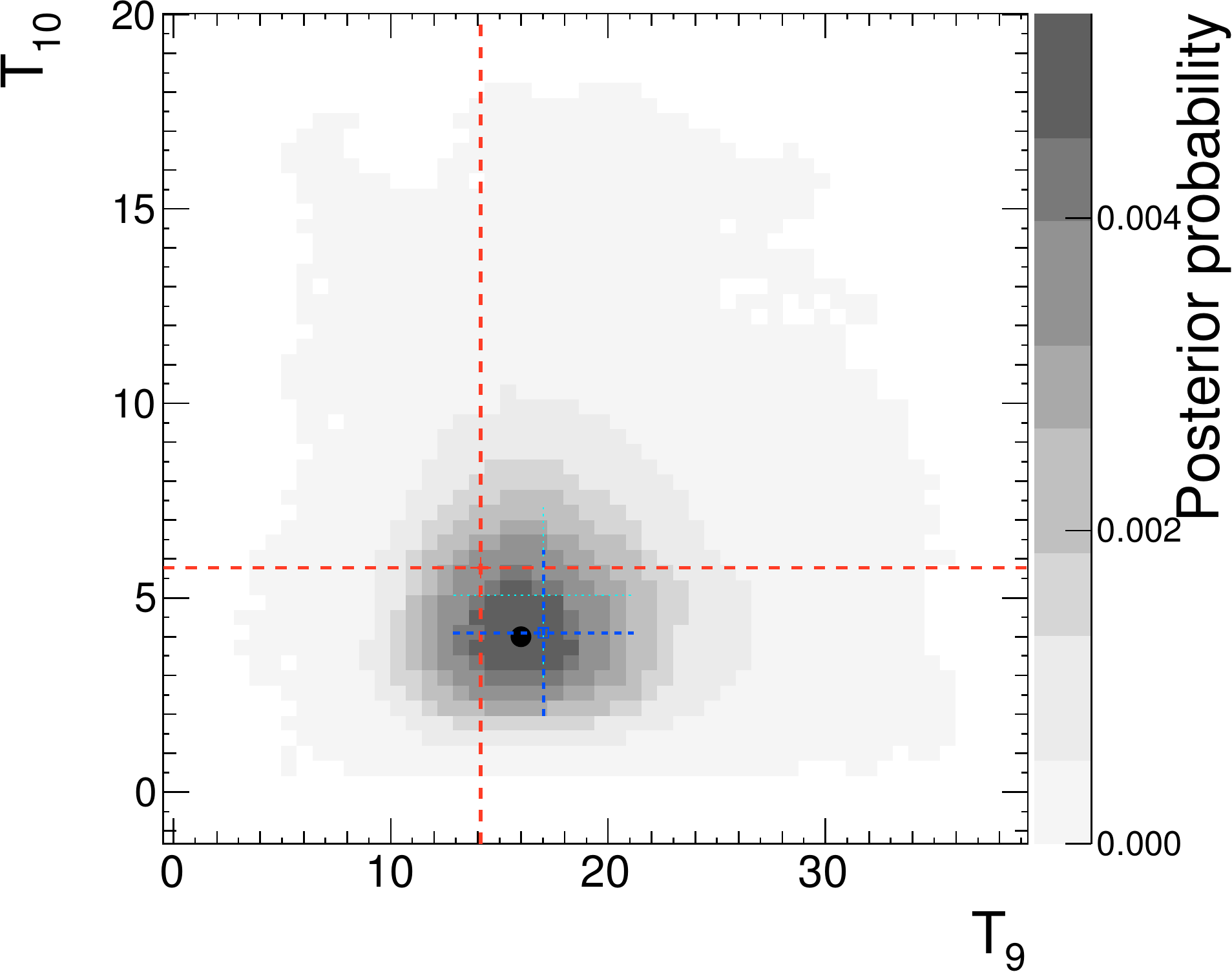} &
    \includegraphics[width=0.18\columnwidth]{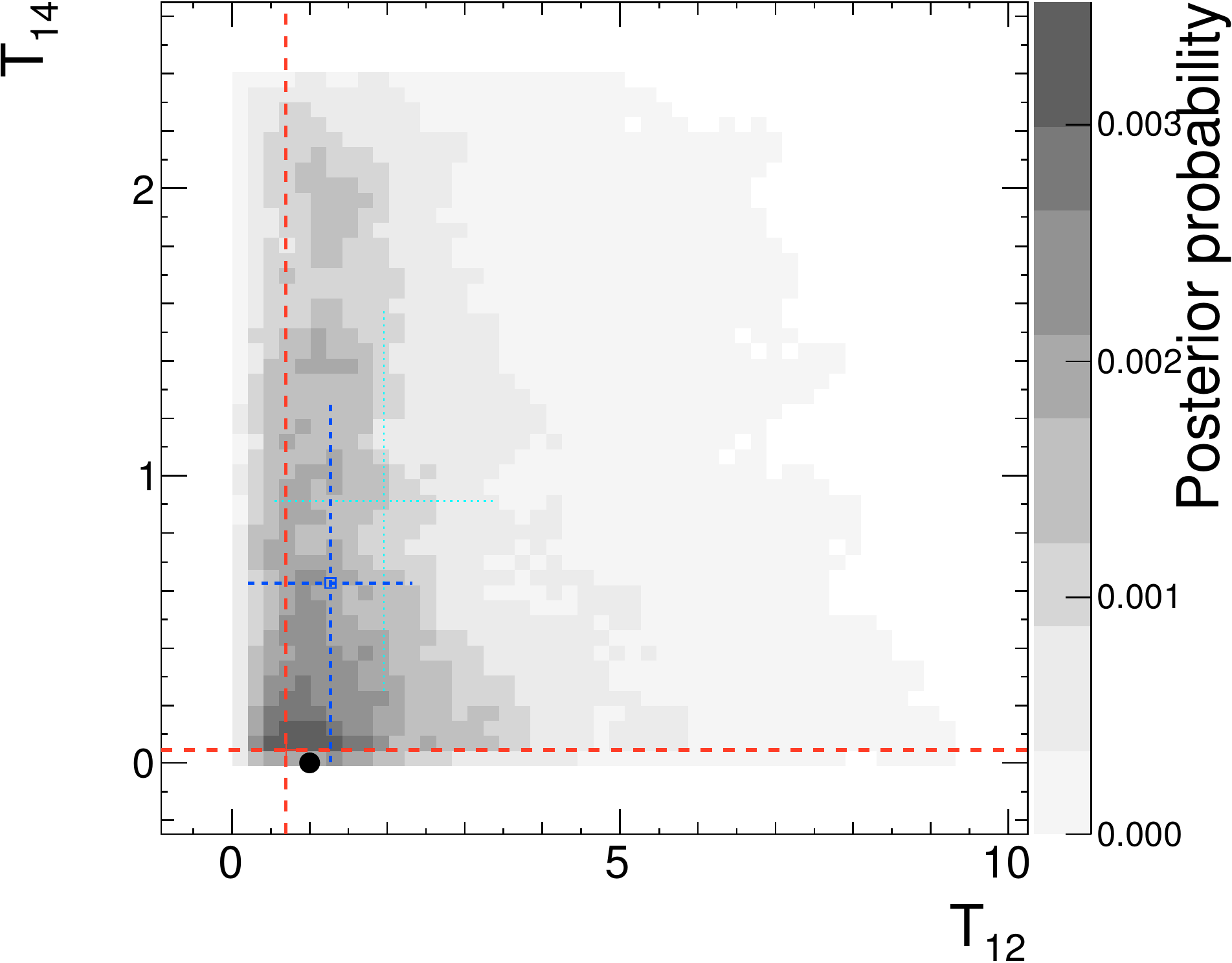} \\

    \includegraphics[width=0.18\columnwidth]{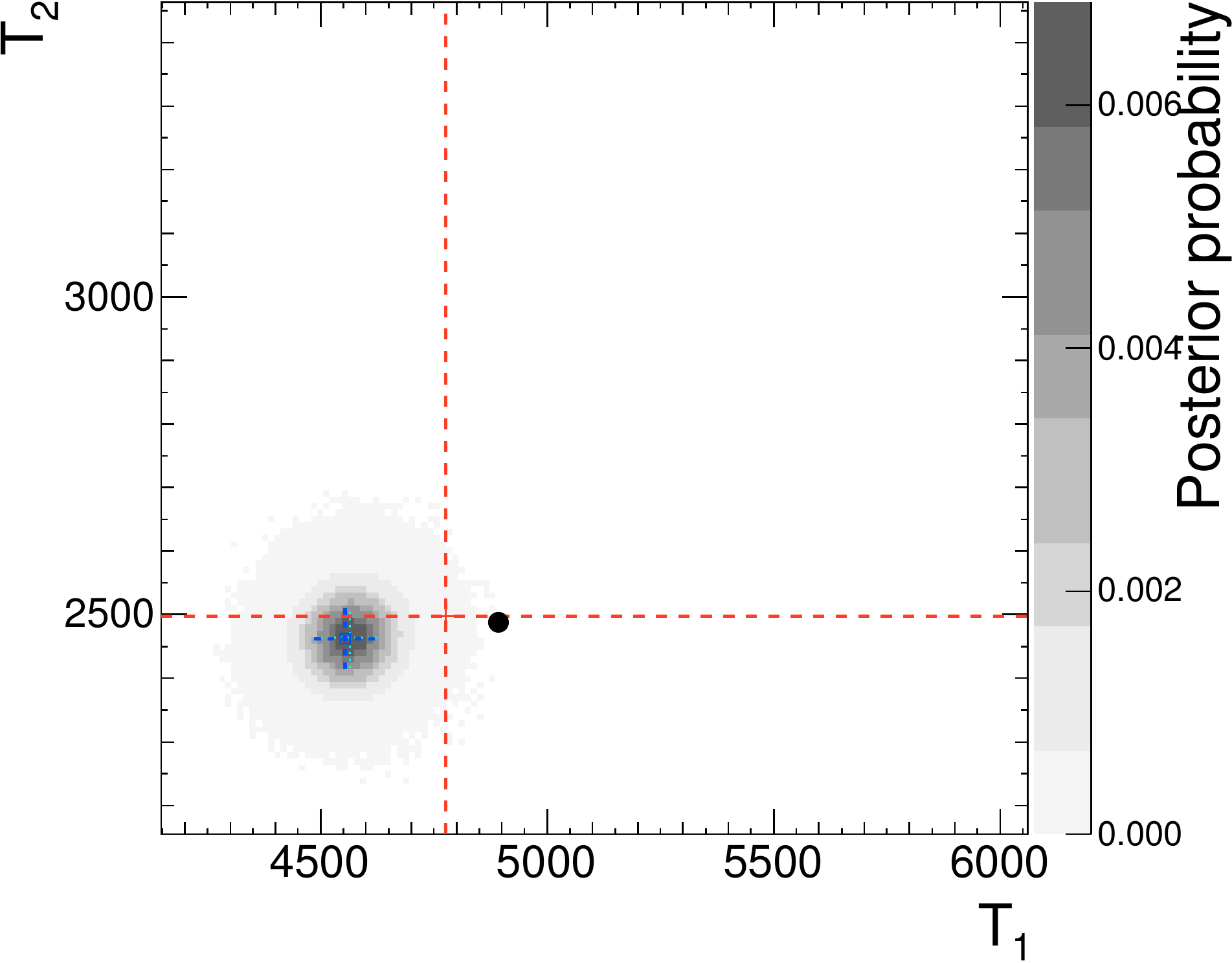} &
    \includegraphics[width=0.18\columnwidth]{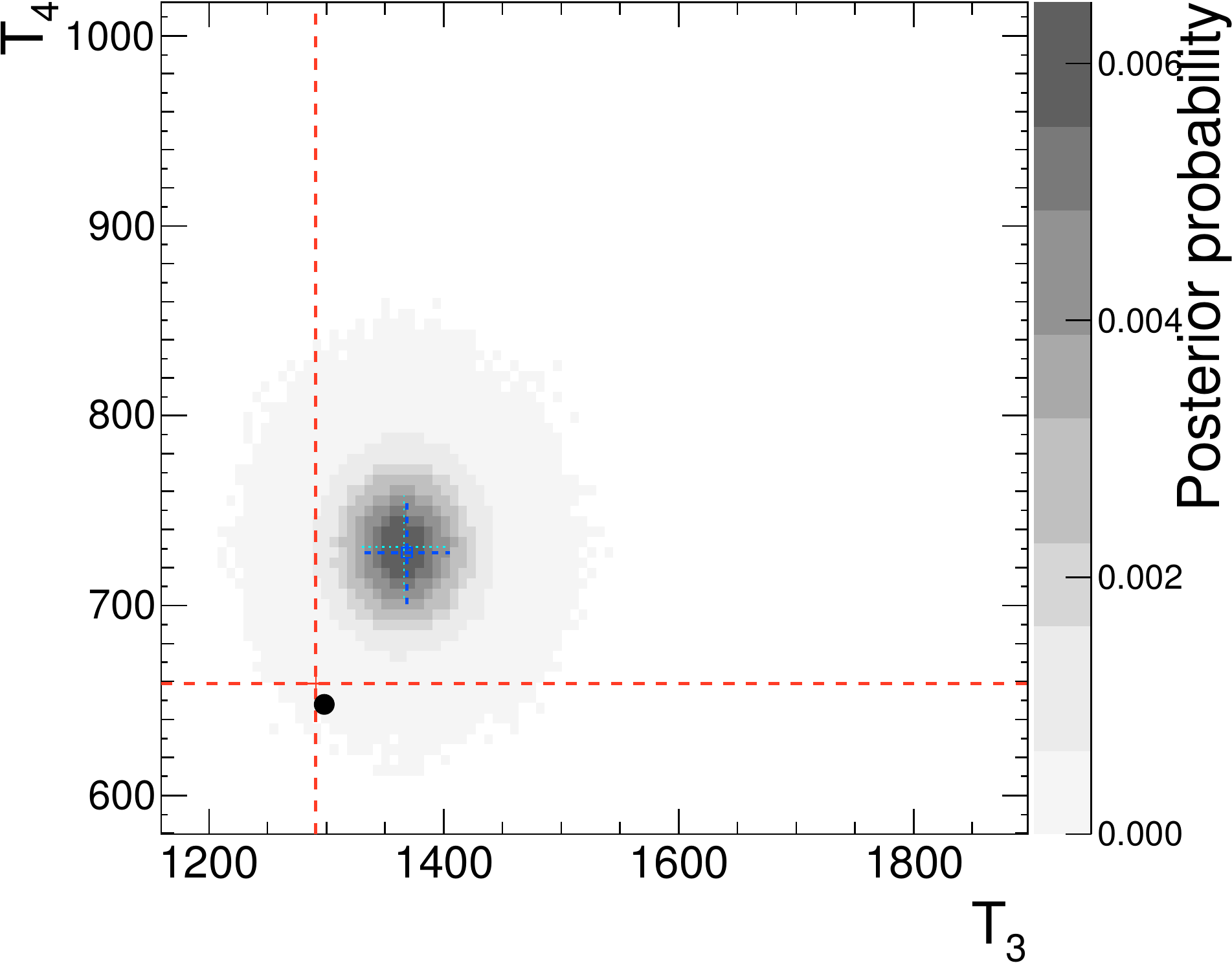} &
    \includegraphics[width=0.18\columnwidth]{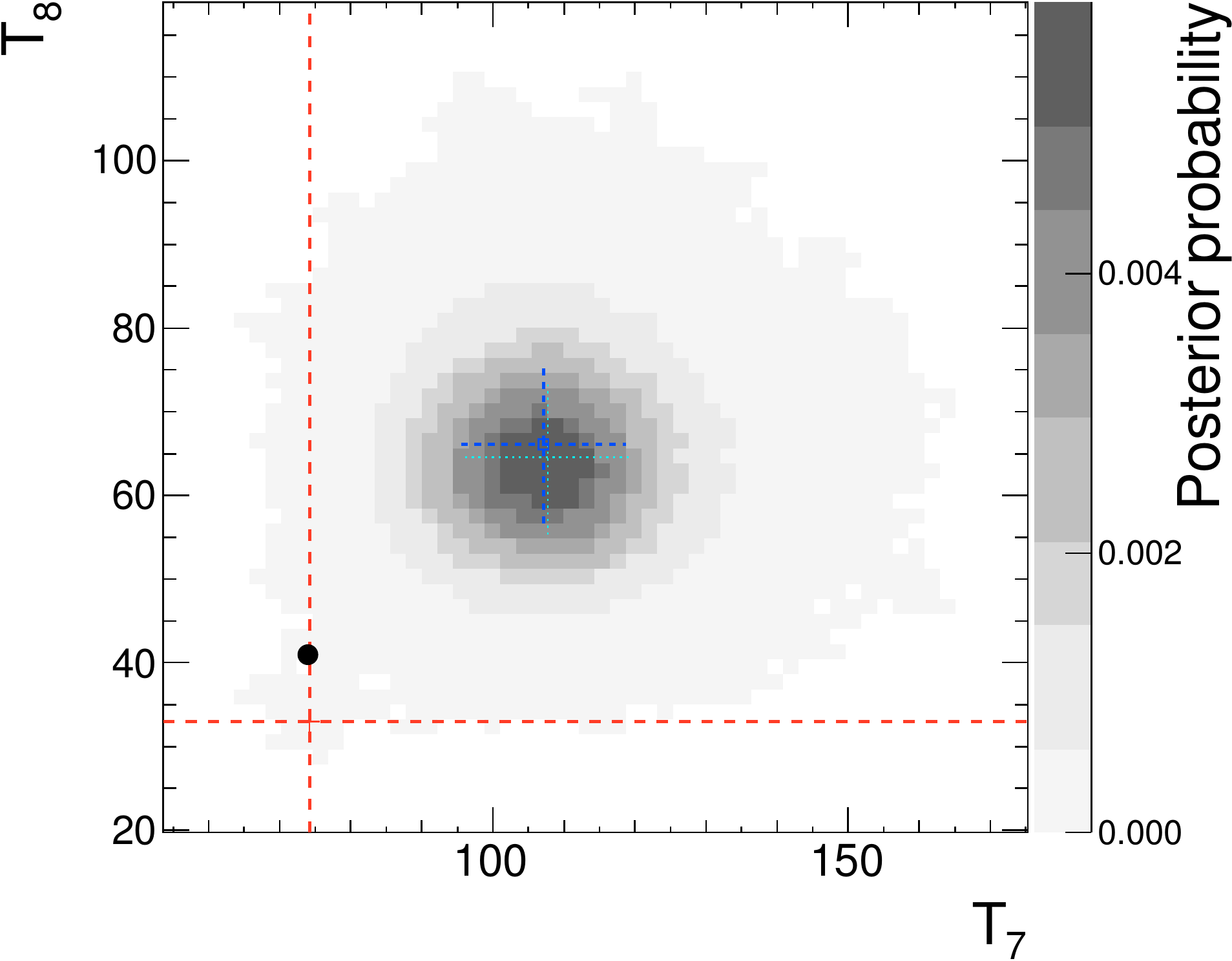} &
    \includegraphics[width=0.18\columnwidth]{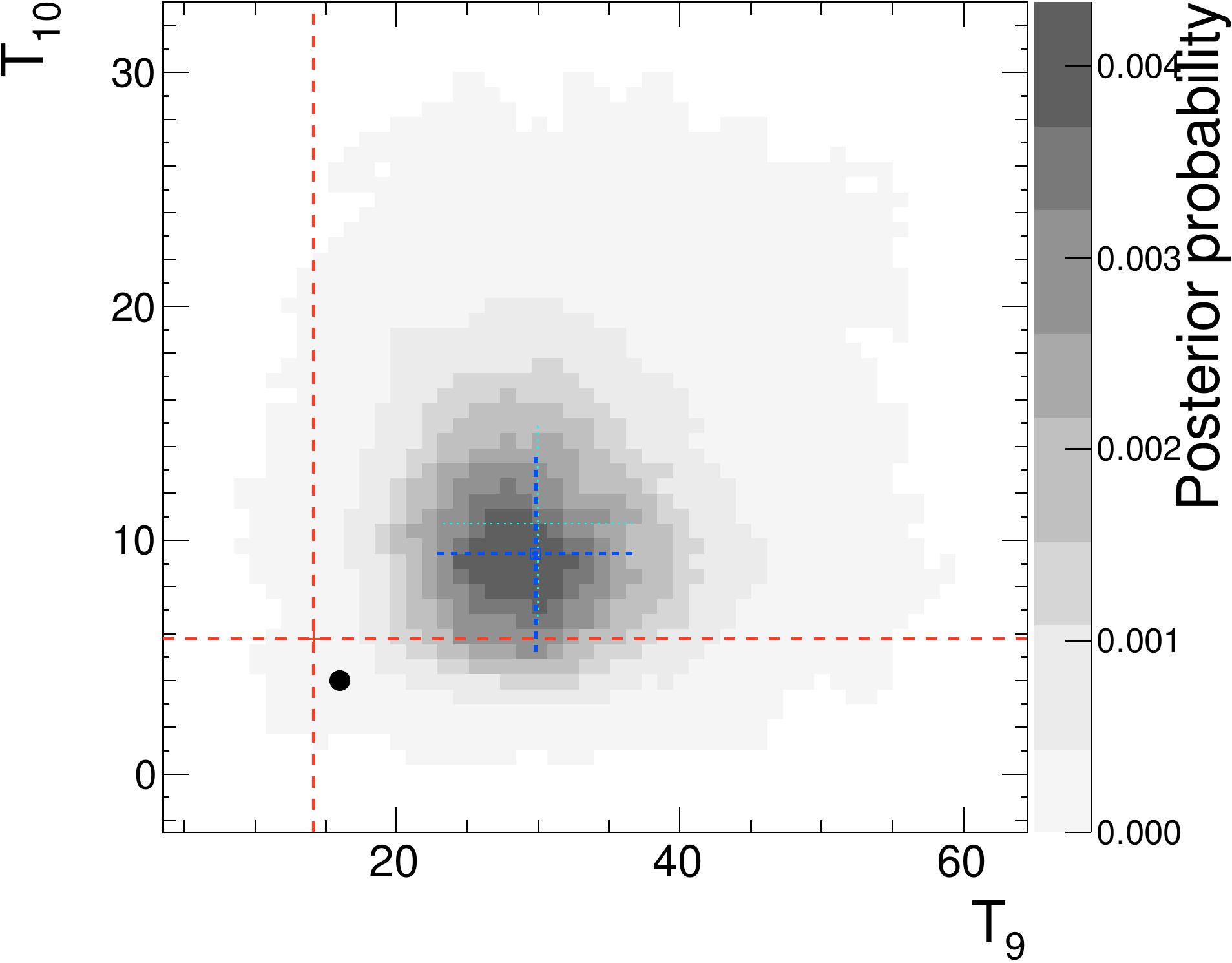} &
    \includegraphics[width=0.18\columnwidth]{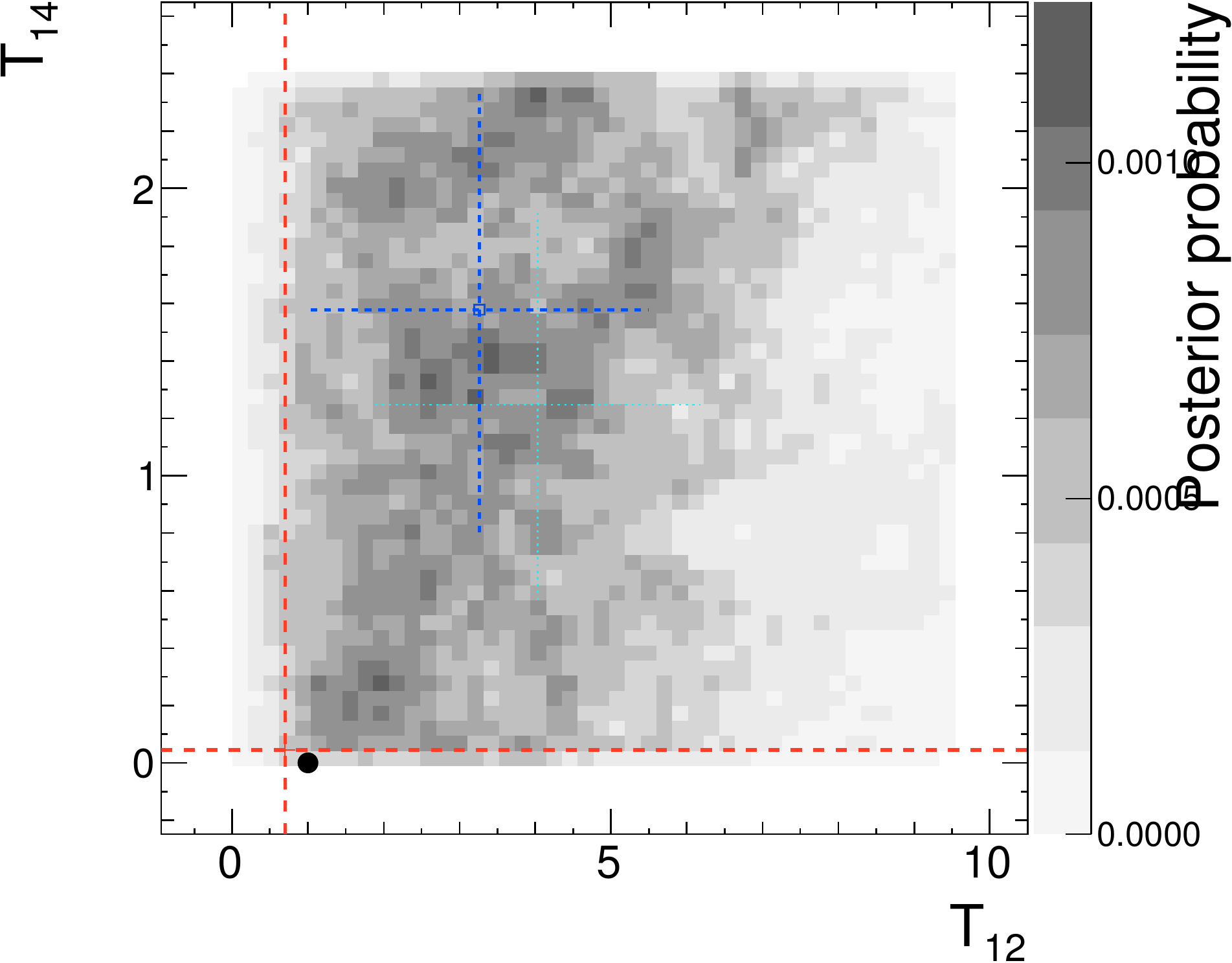} \\

    \includegraphics[width=0.18\columnwidth]{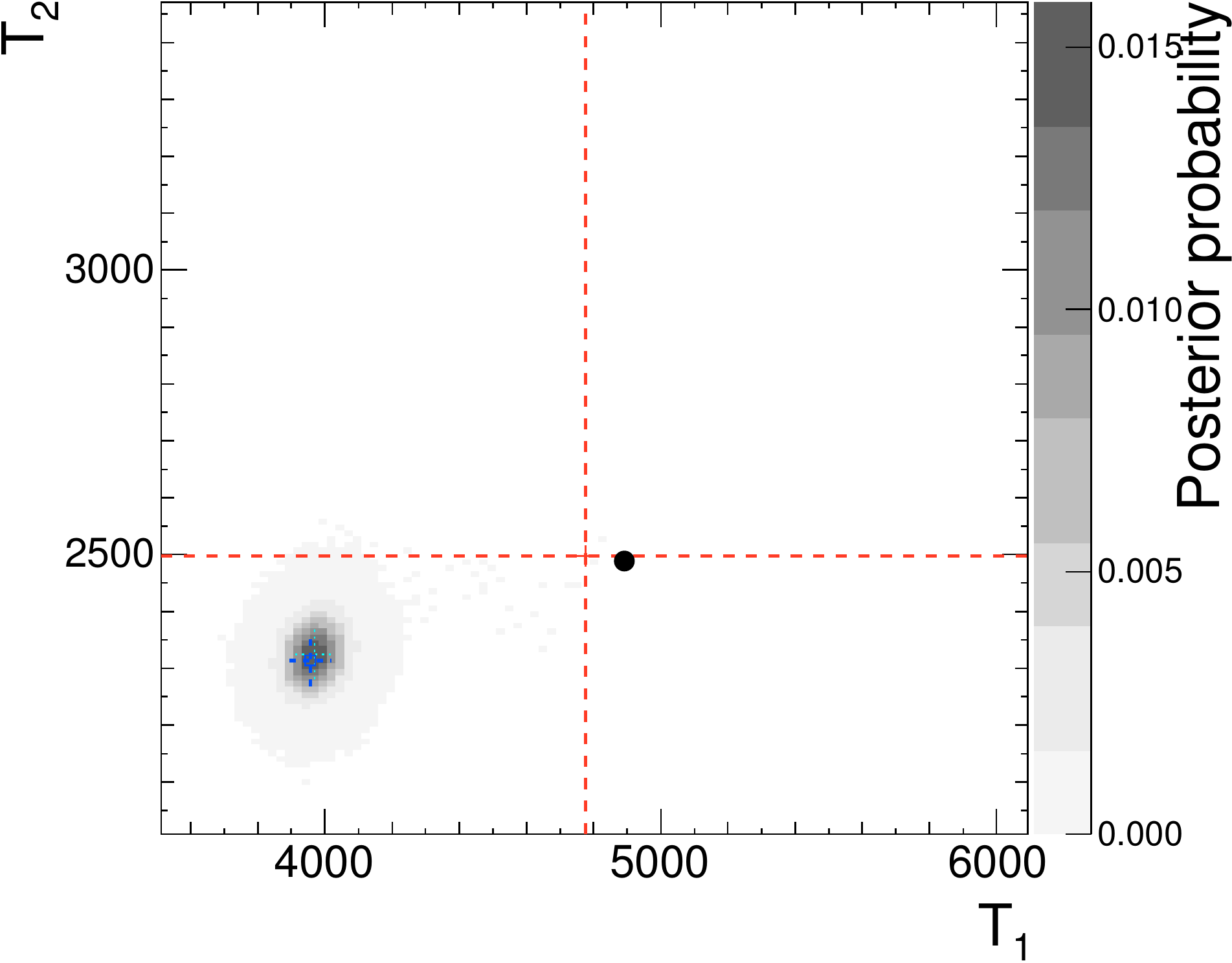} &
    \includegraphics[width=0.18\columnwidth]{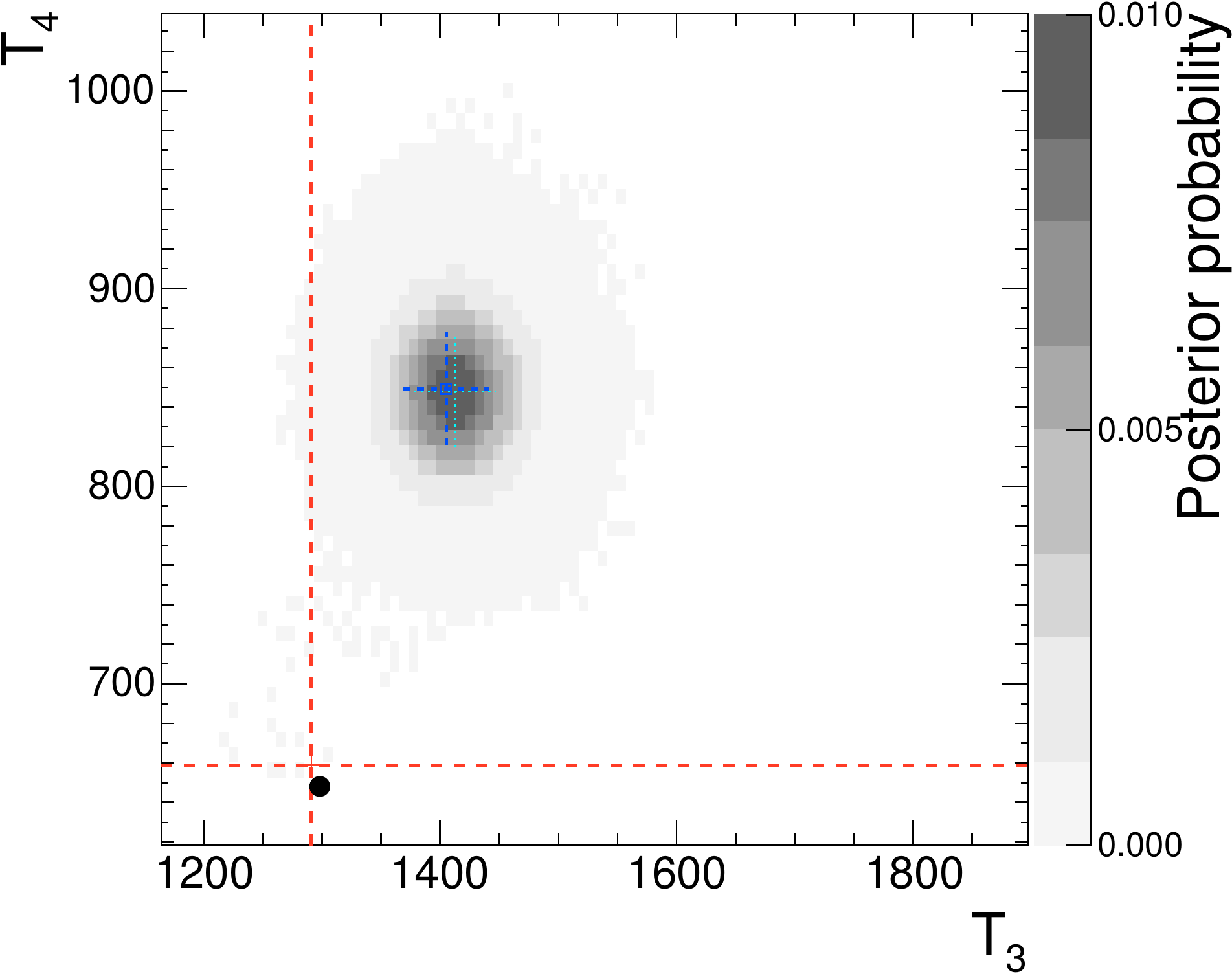} &
    \includegraphics[width=0.18\columnwidth]{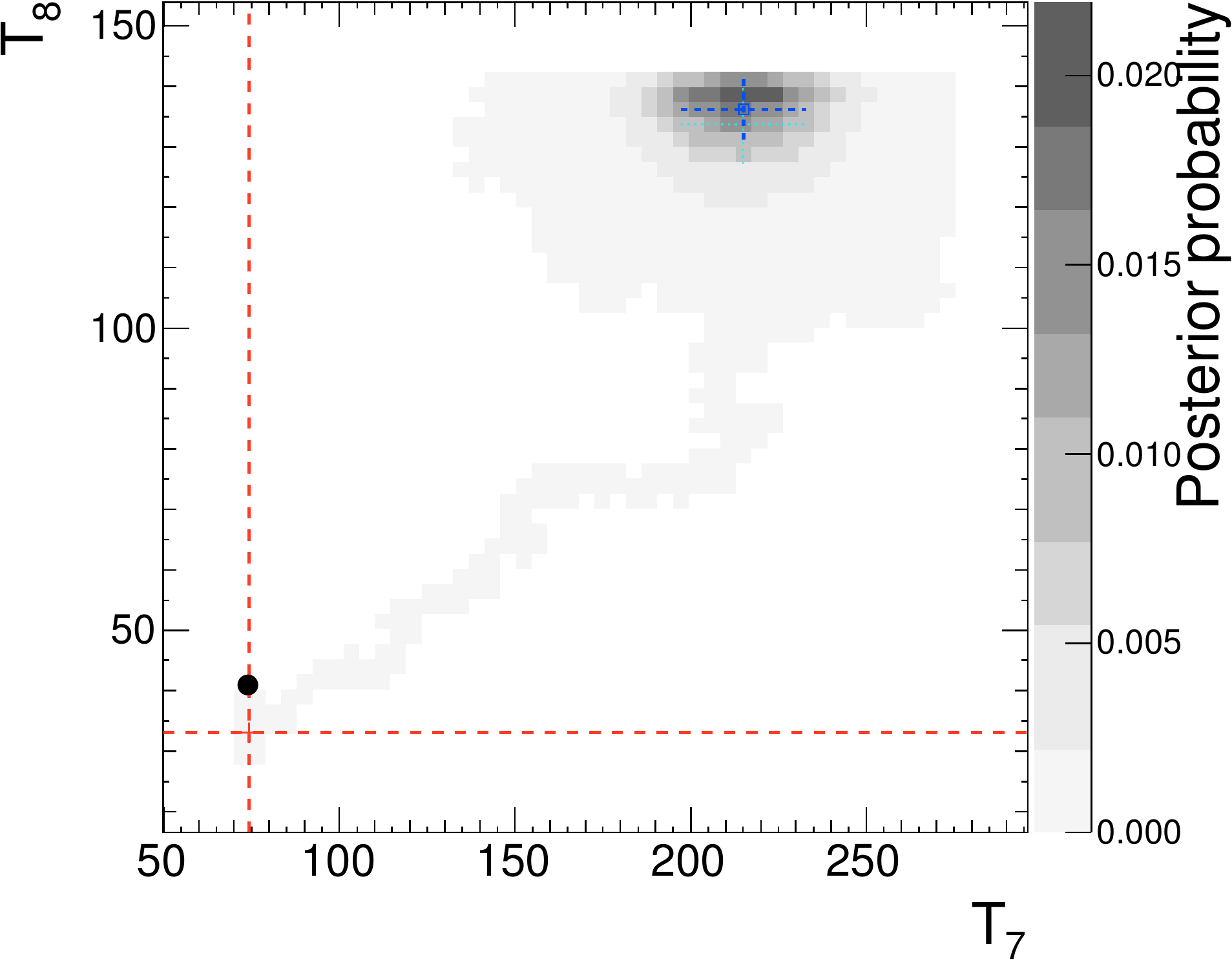} &
    \includegraphics[width=0.18\columnwidth]{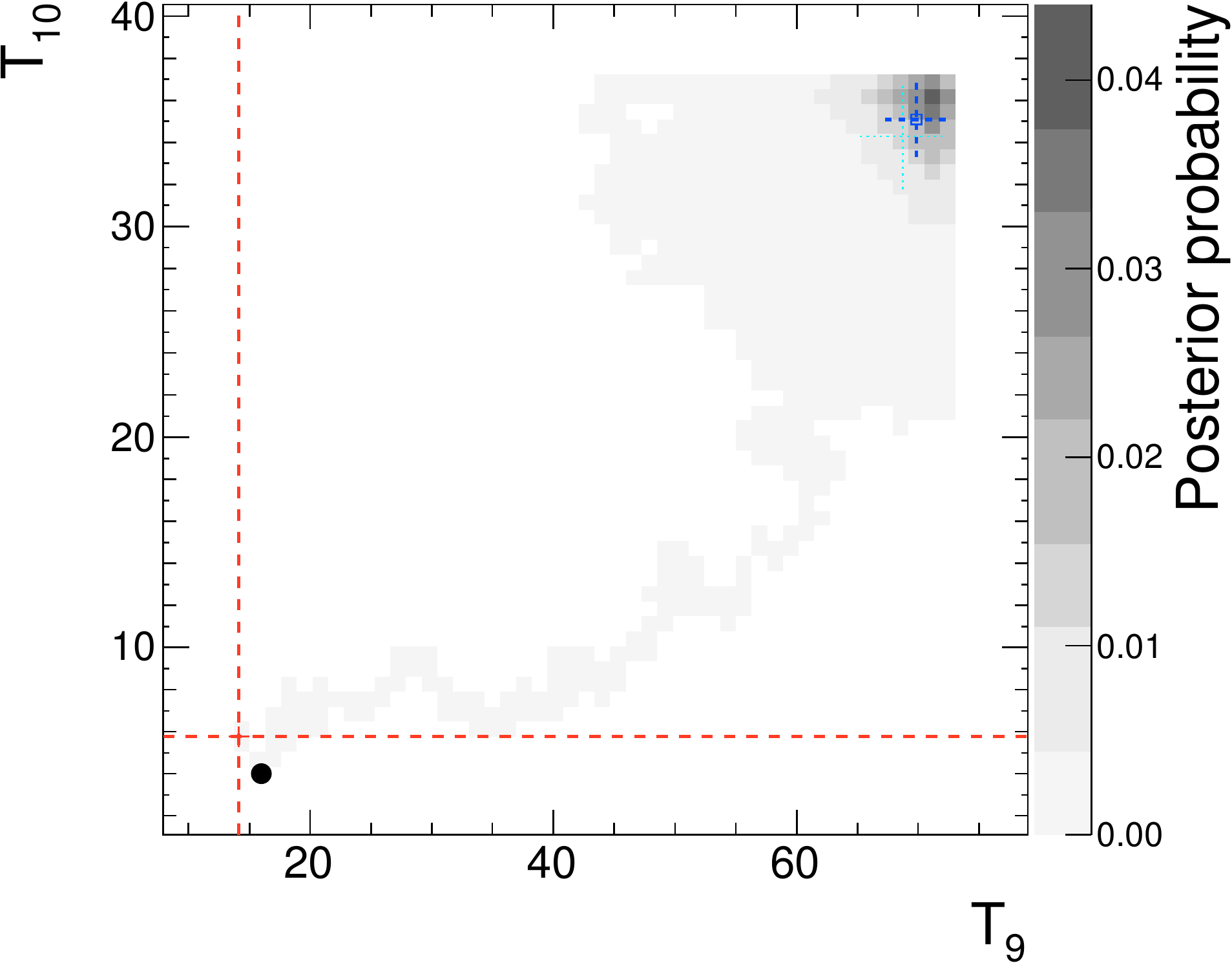} &
    \includegraphics[width=0.18\columnwidth]{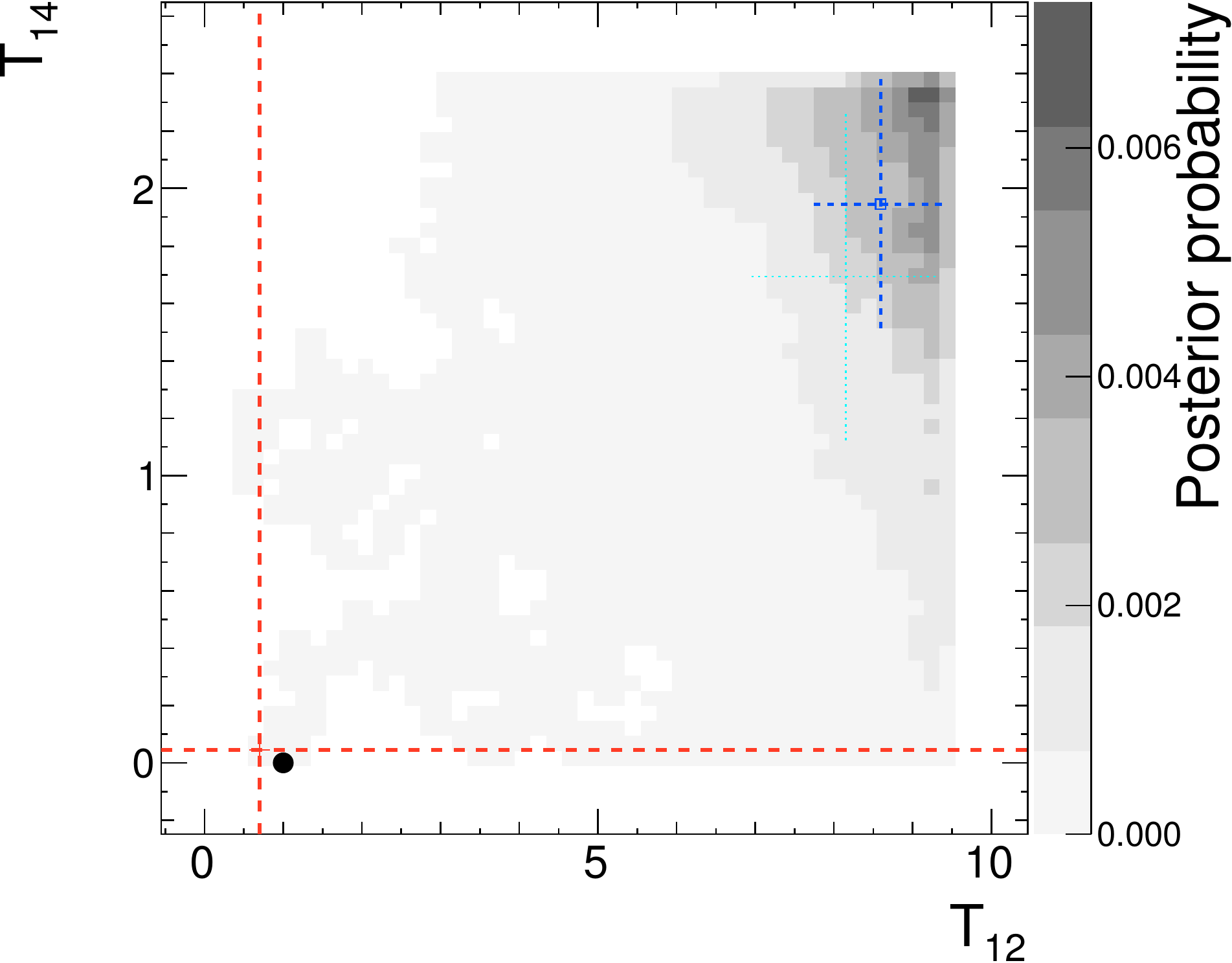} \\

    \includegraphics[width=0.18\columnwidth]{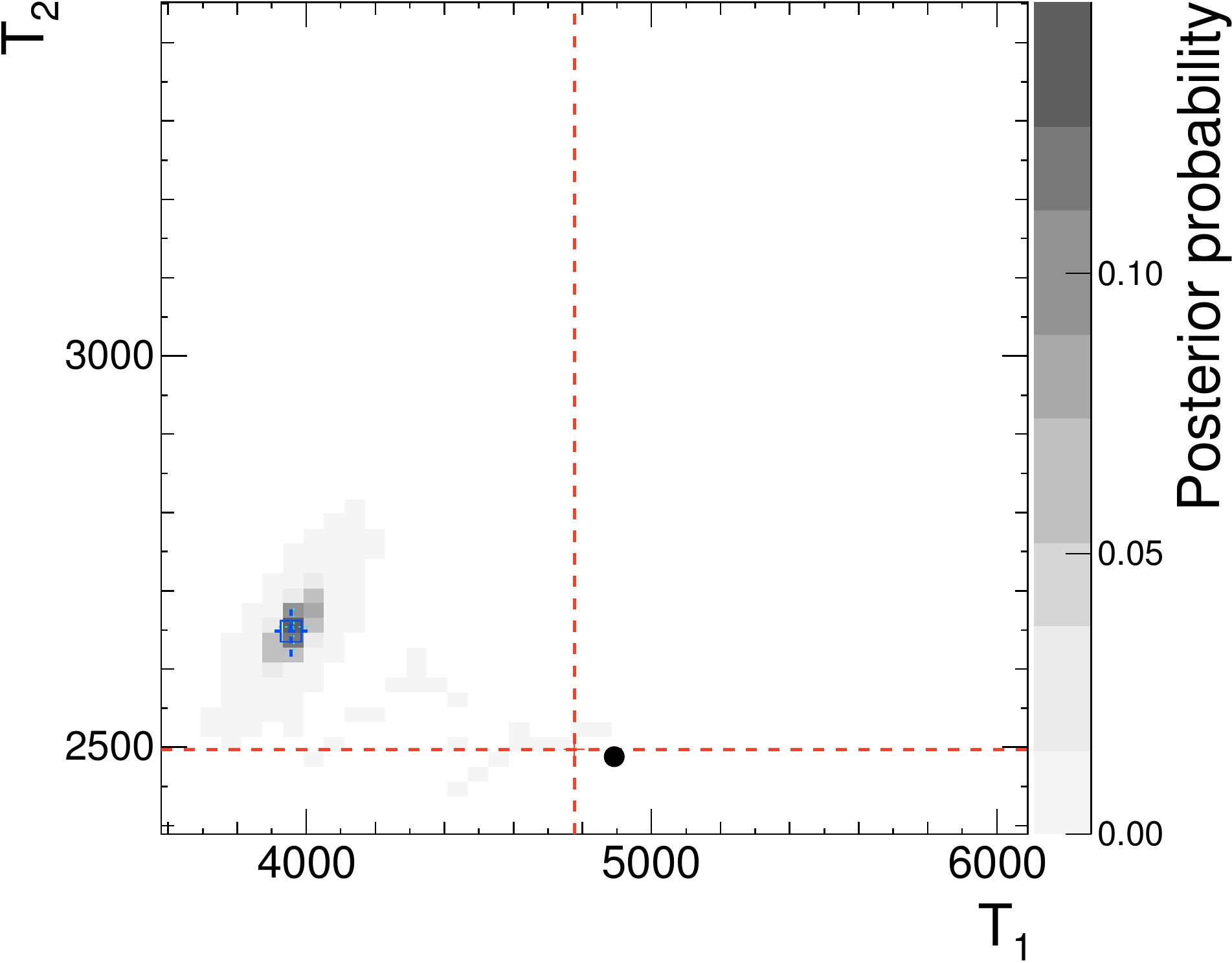} &
    \includegraphics[width=0.18\columnwidth]{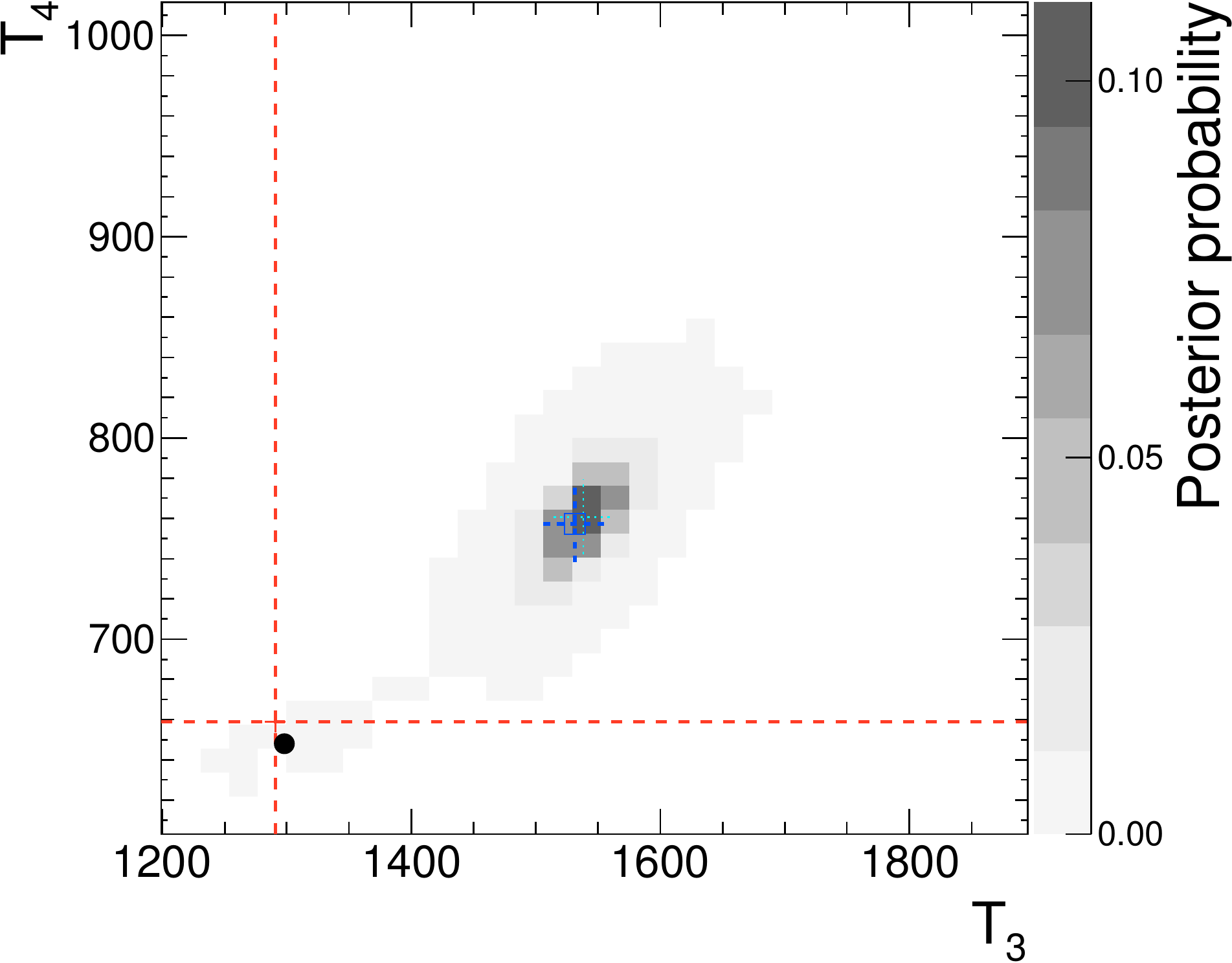} &
    \includegraphics[width=0.18\columnwidth]{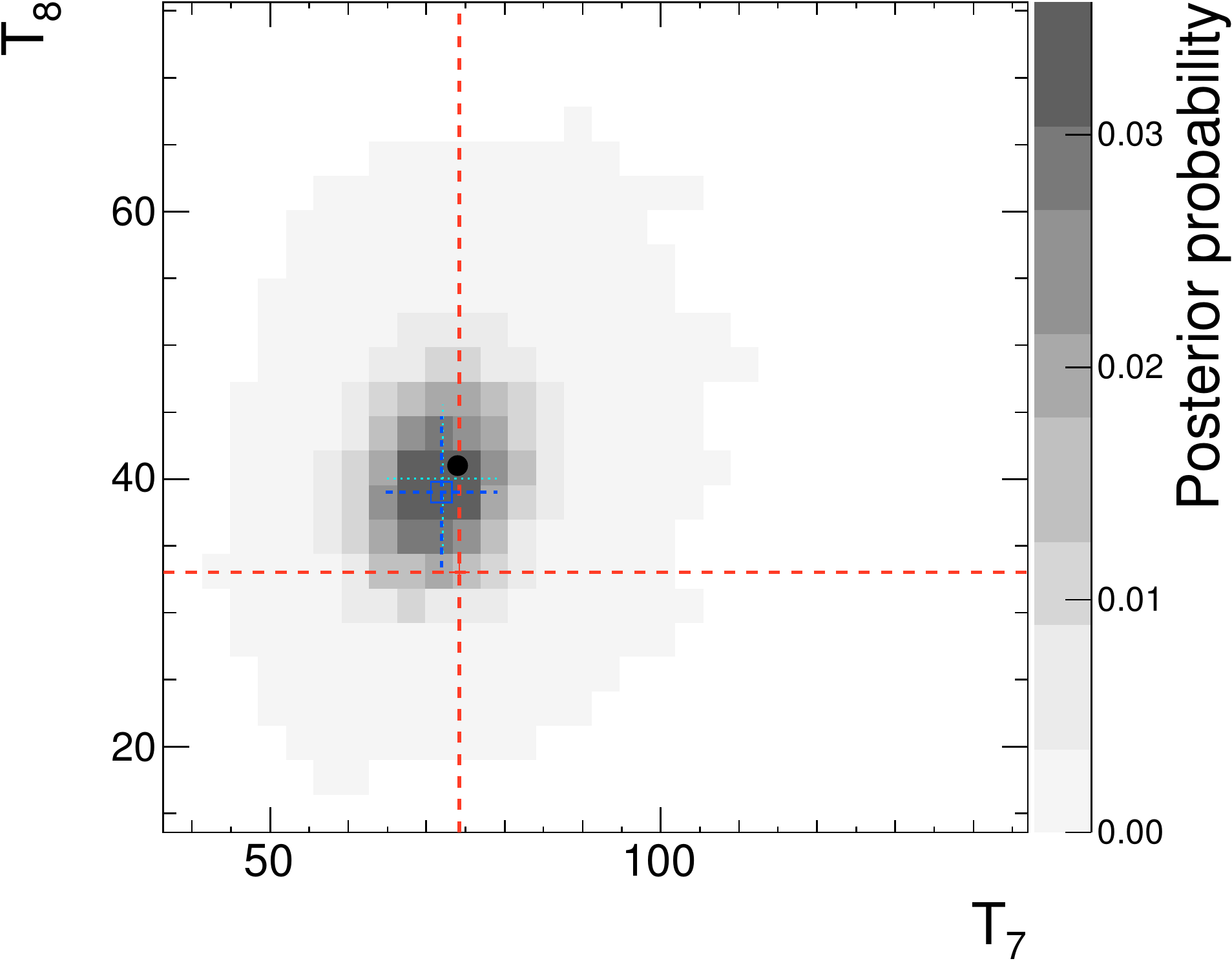} &
    \includegraphics[width=0.18\columnwidth]{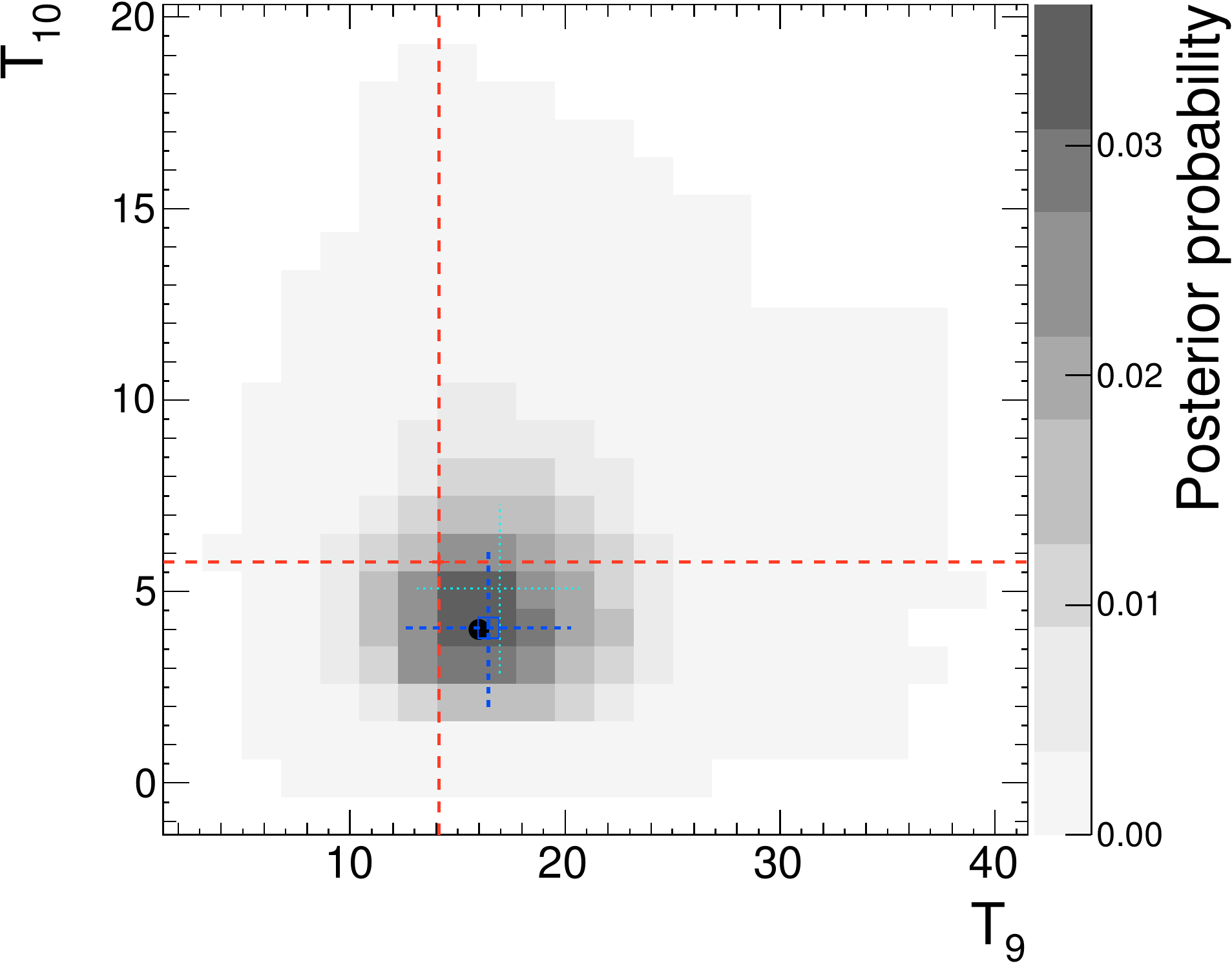} &
    \includegraphics[width=0.18\columnwidth]{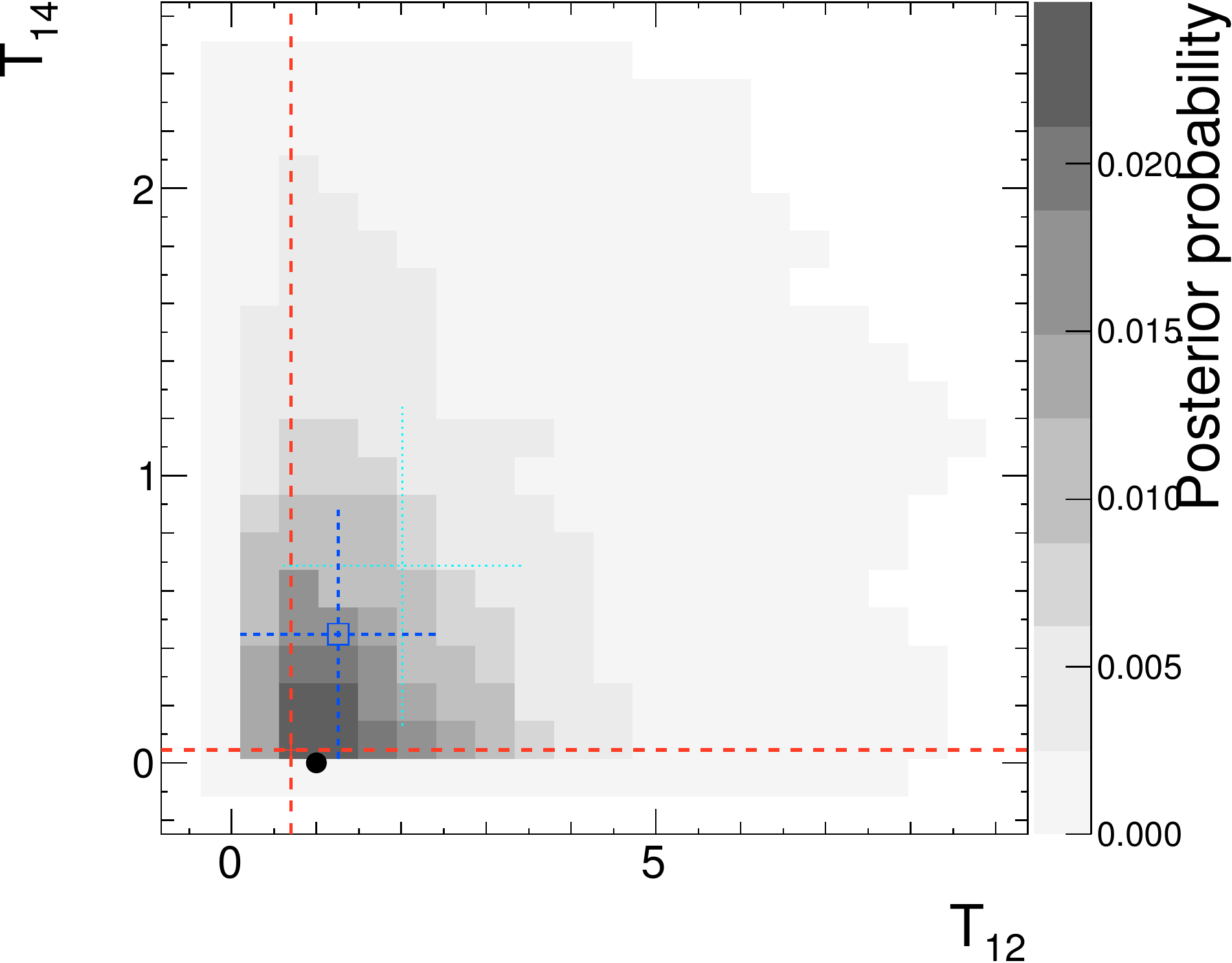} \\

    \includegraphics[width=0.18\columnwidth]{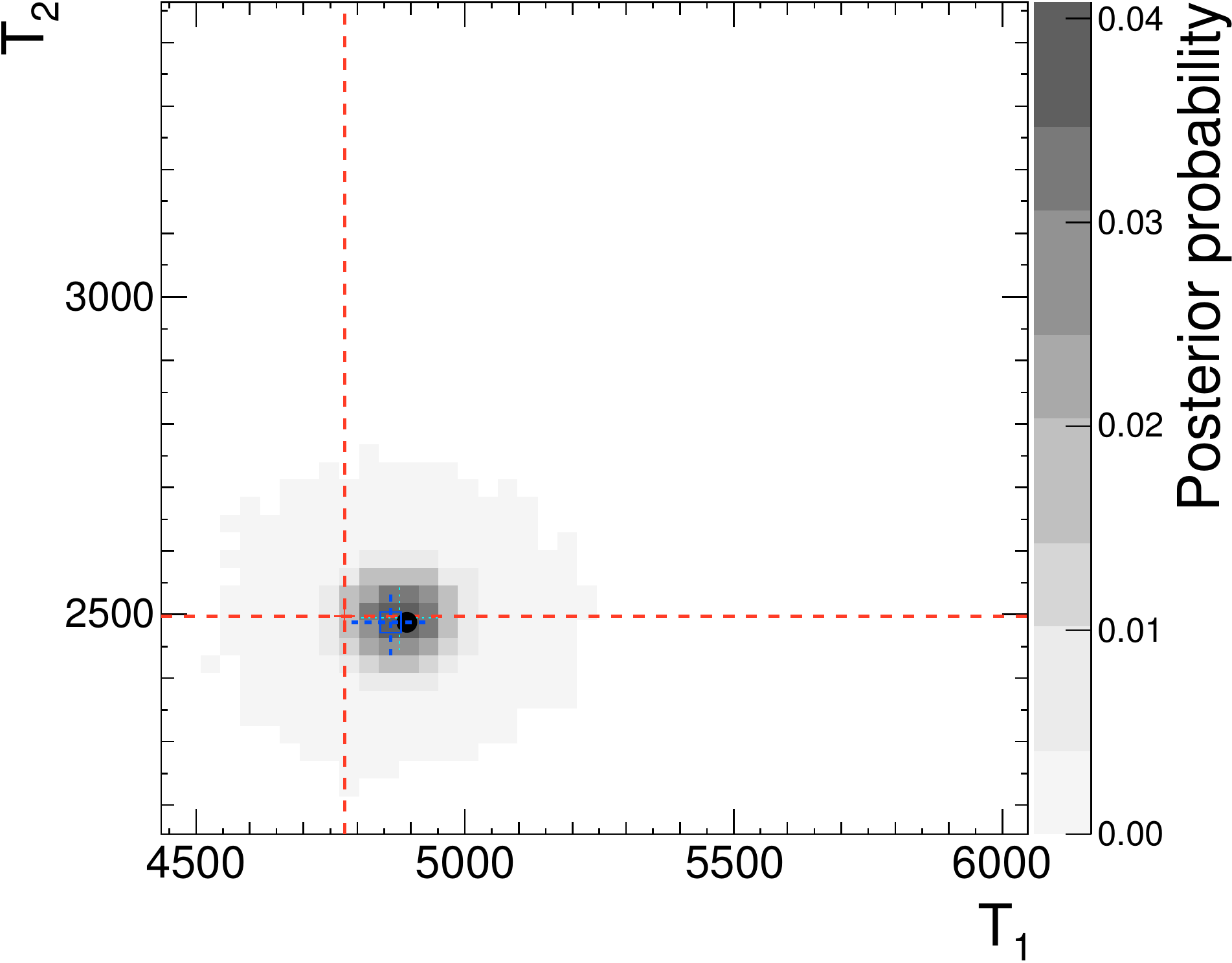} &
    \includegraphics[width=0.18\columnwidth]{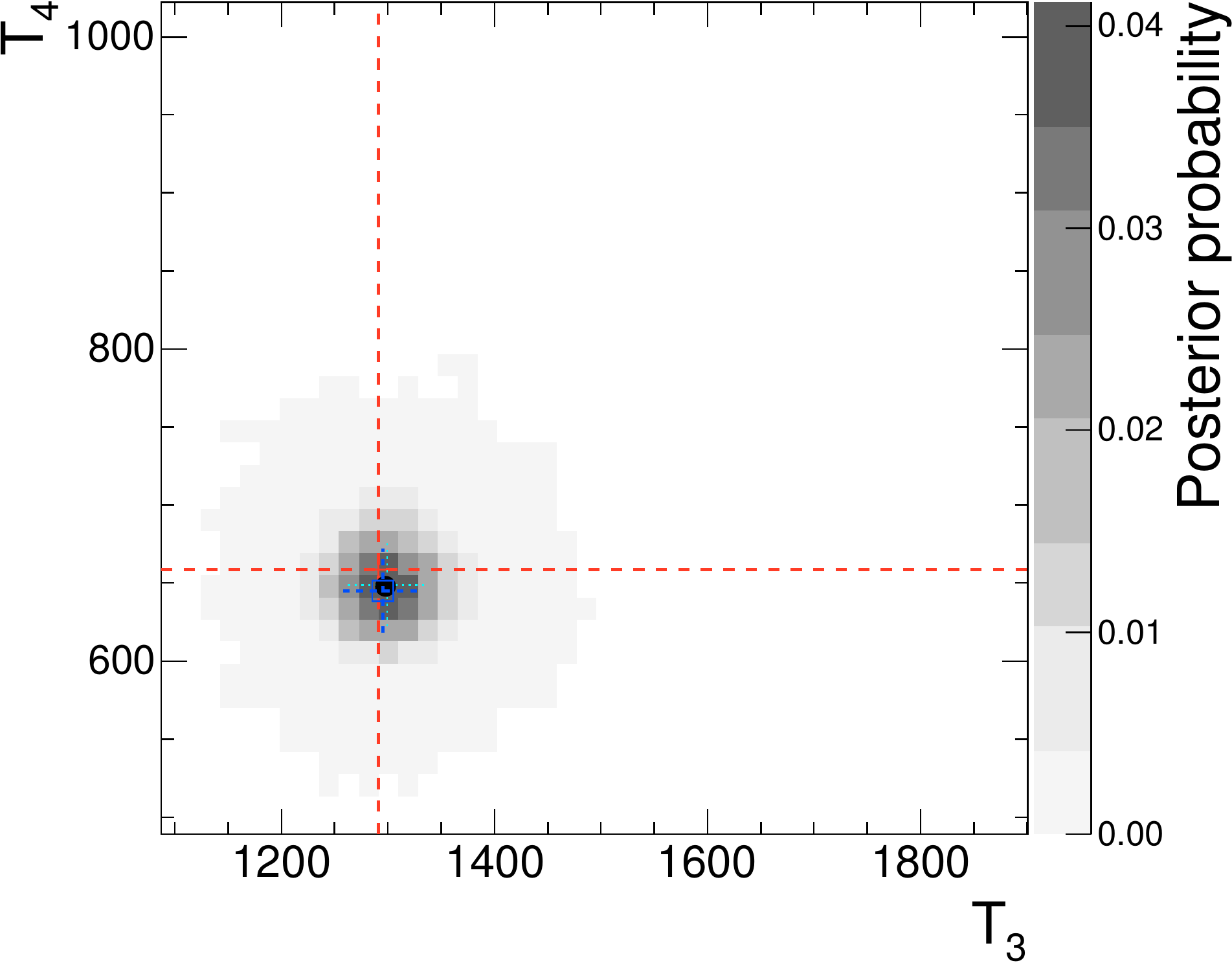} &
    \includegraphics[width=0.18\columnwidth]{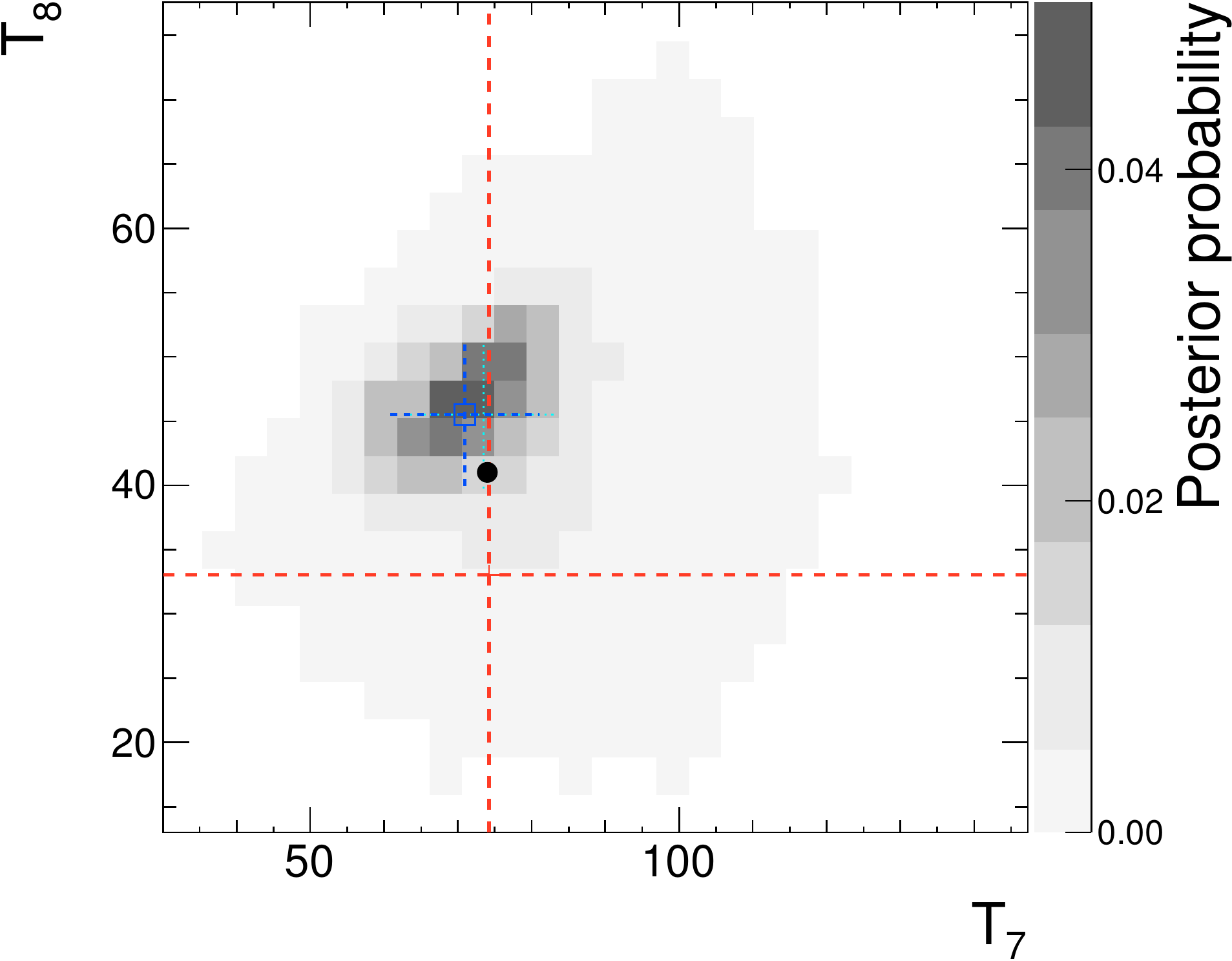} &
    \includegraphics[width=0.18\columnwidth]{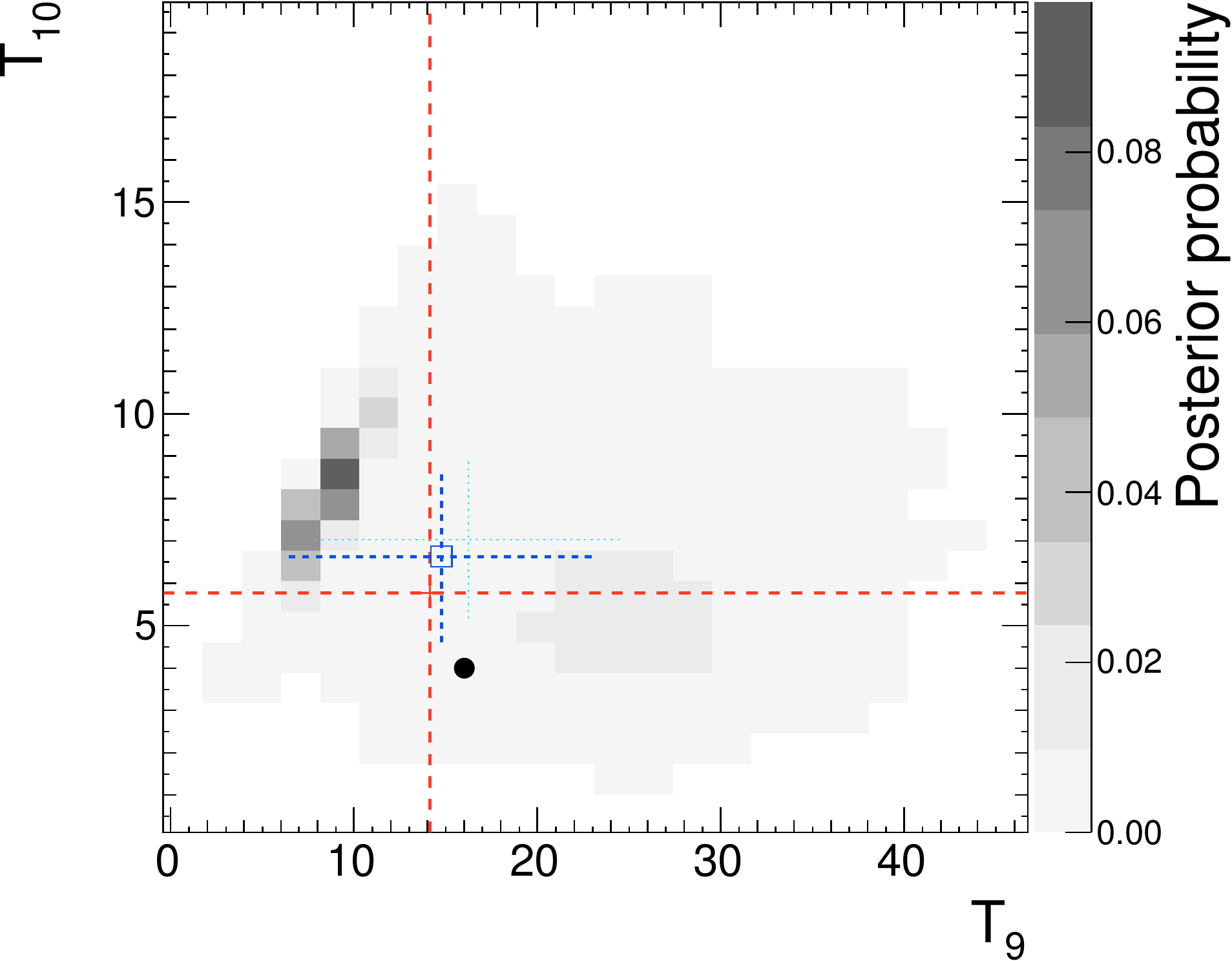} &
    \includegraphics[width=0.18\columnwidth]{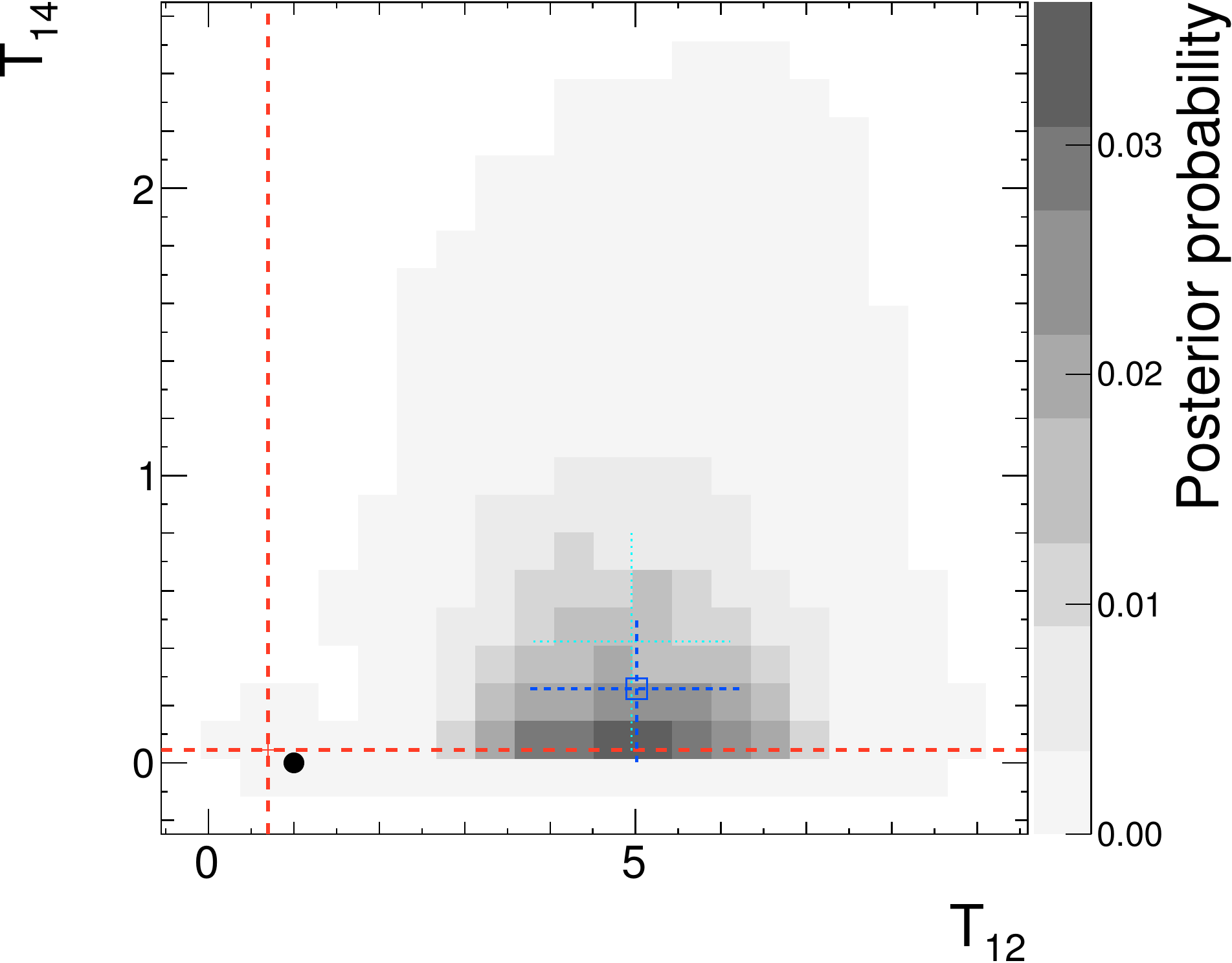} \\

    \includegraphics[width=0.18\columnwidth]{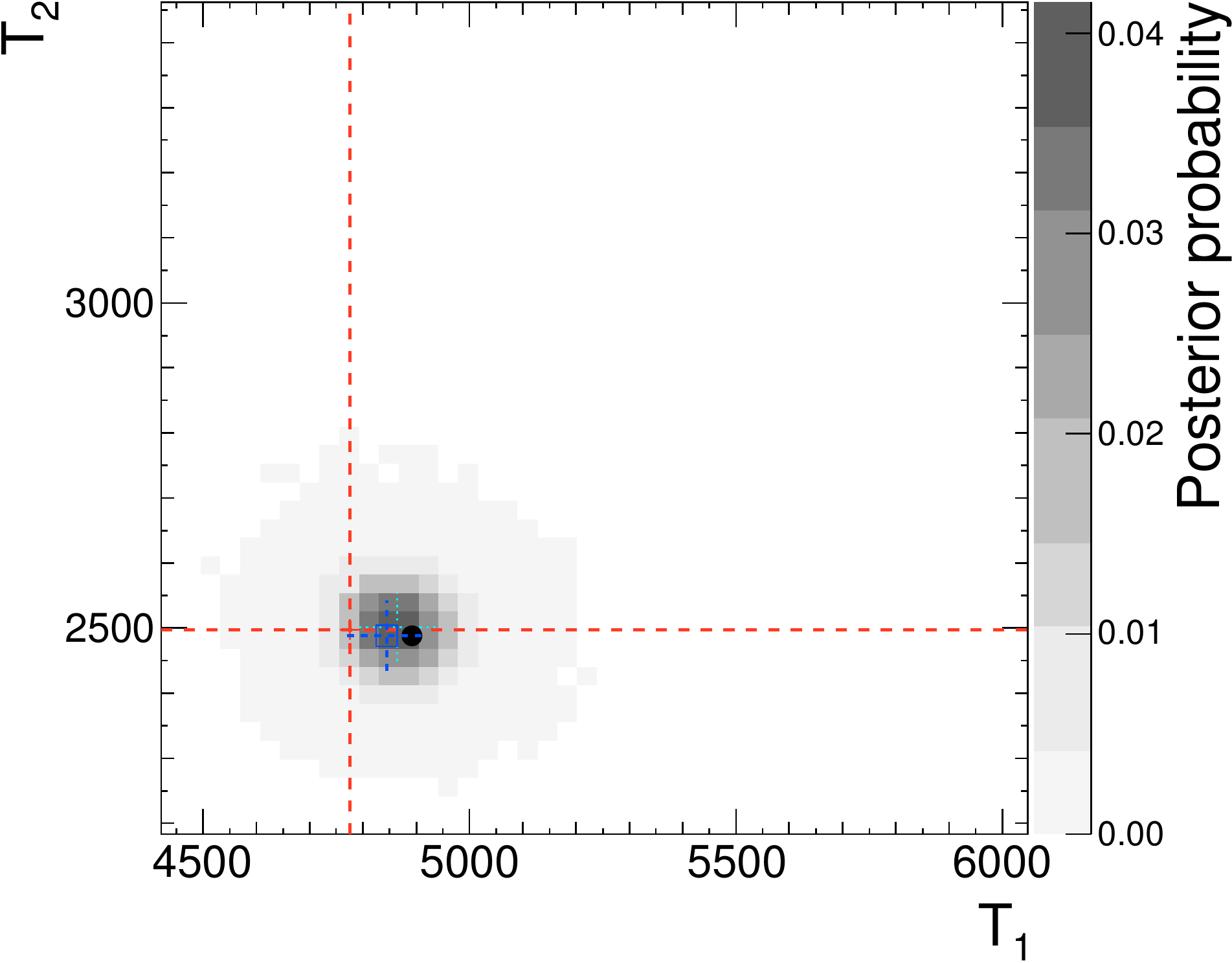} &
    \includegraphics[width=0.18\columnwidth]{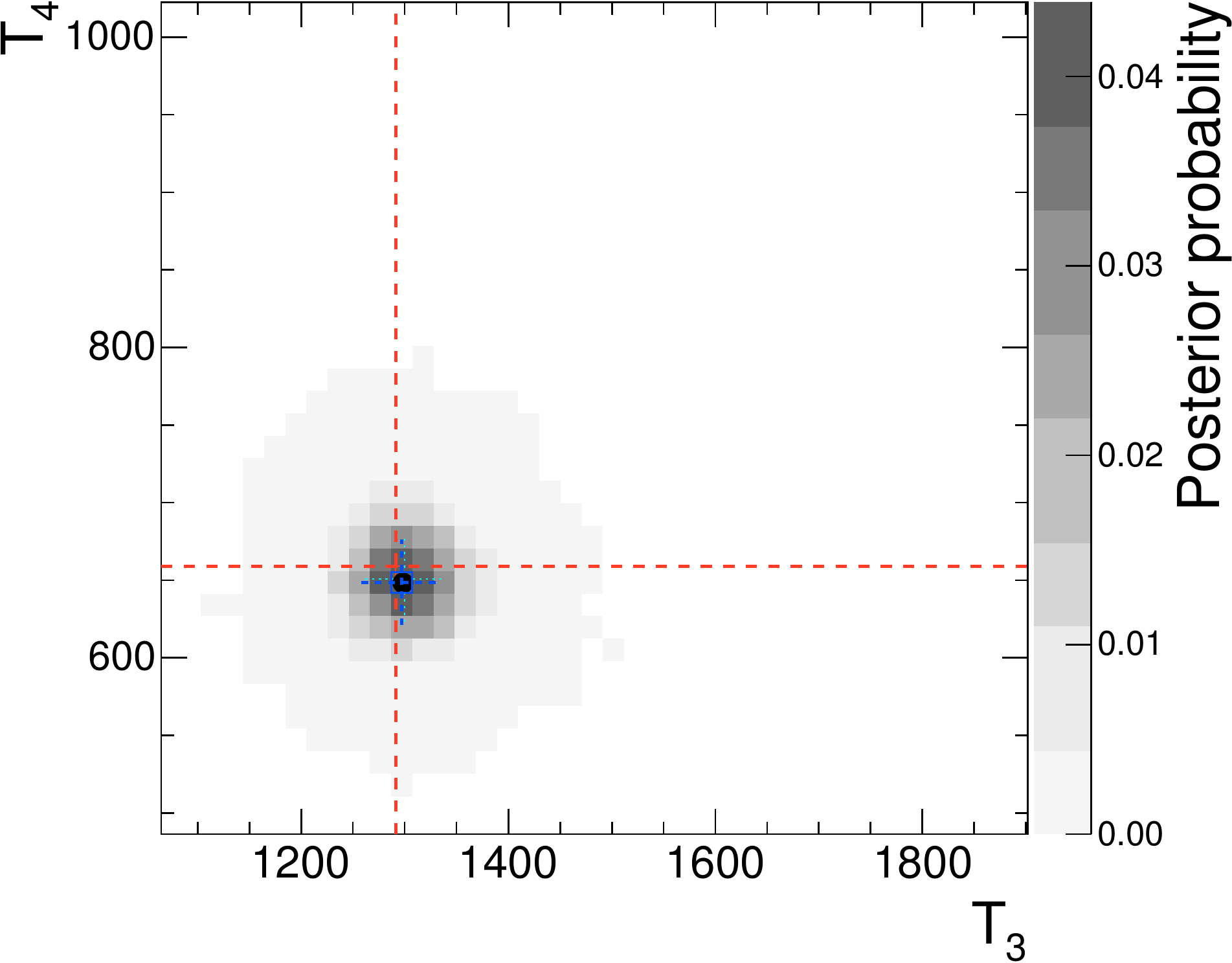} &
    \includegraphics[width=0.18\columnwidth]{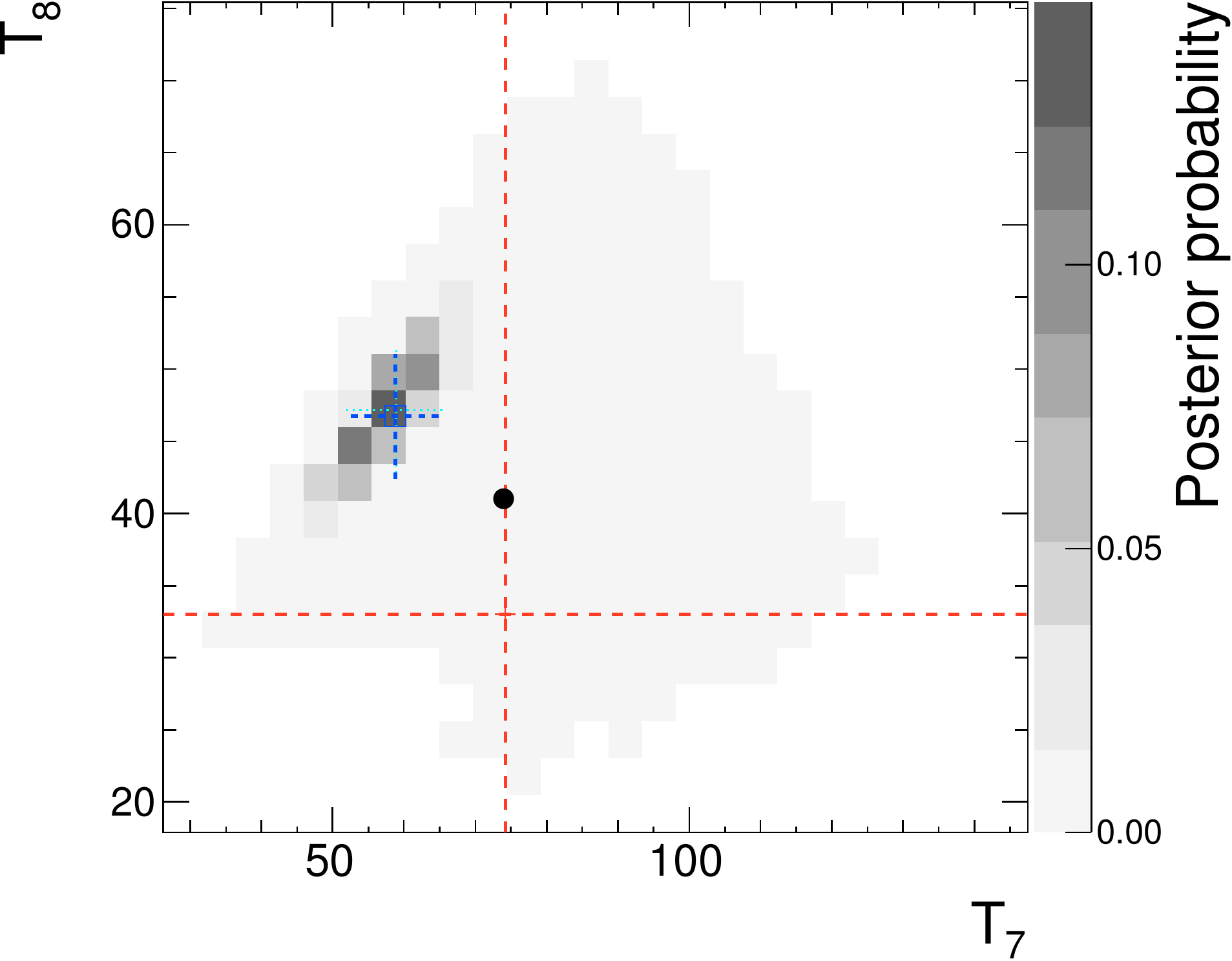} &
    \includegraphics[width=0.18\columnwidth]{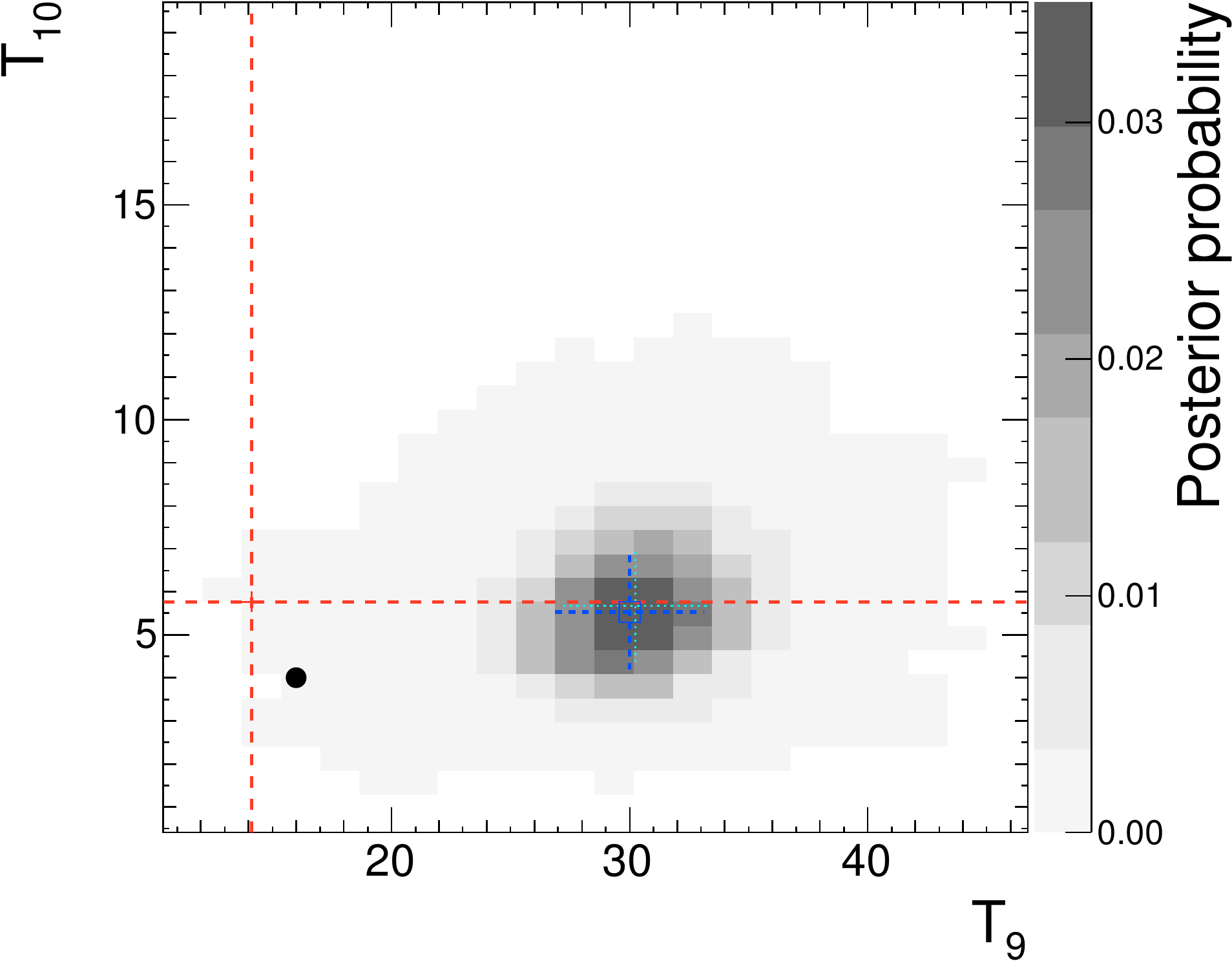} &
    \includegraphics[width=0.18\columnwidth]{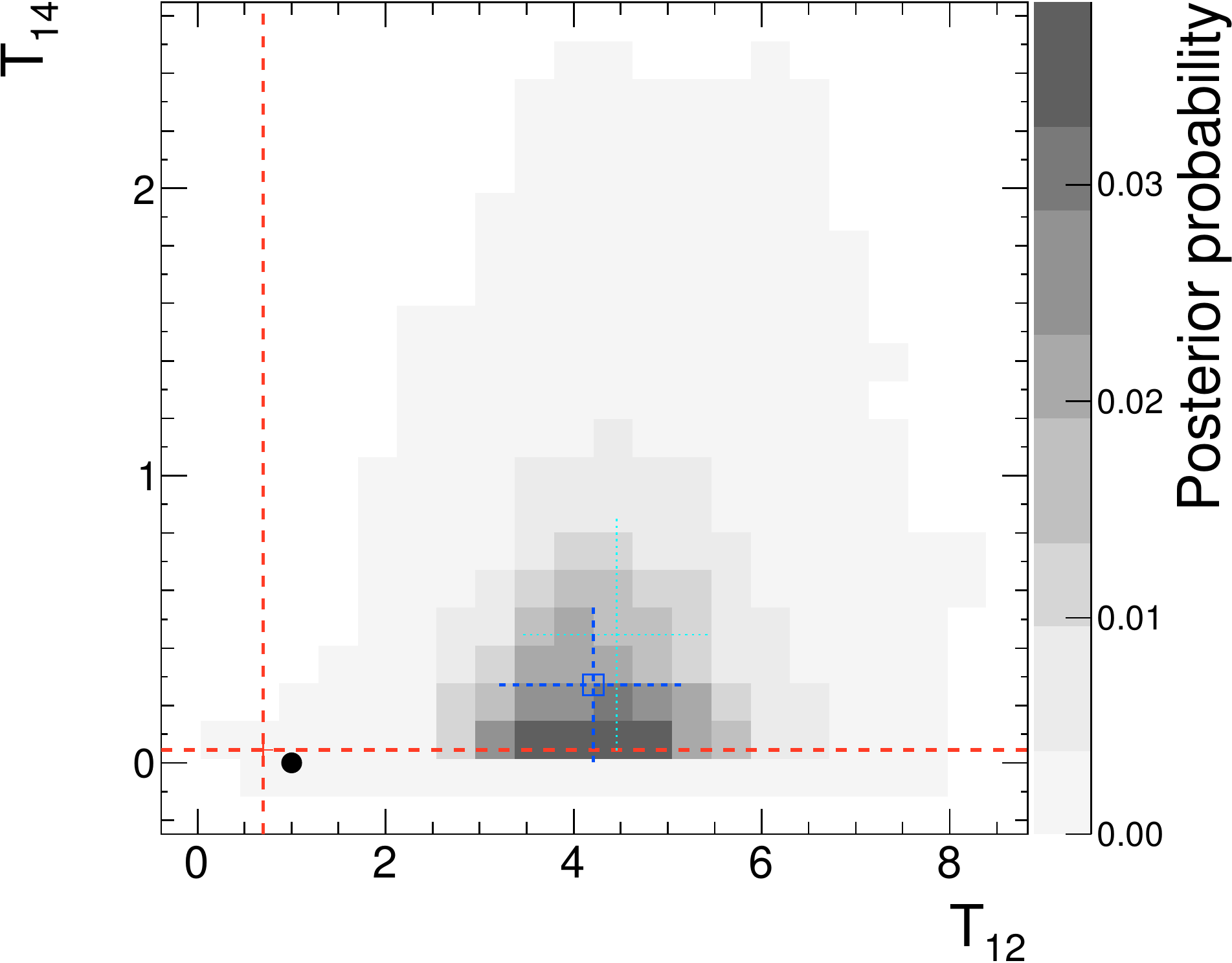} \\

 \end{tabular}
 \caption{Some 2-dimensional distributions from Sec.~\ref{sec:regSteepNoSmearing}.  The columns show $P_{t_1,t_2}(T_{t_1},T_{t_2}|\tuple{D})$ with $(t_1,t_2) = \{ (1,2) , (3,4) , (7,8) , (9,10) , (12,14) \}$.  The rows correspond to regularization with $(S,\alpha) = \{ (S_1,0) , (S_1,1\times 10^3) , (S_1,3\times 10^3) , (S_2,6\times 10^{-4}) , (S_3,20) ,  (S_3,40) \}$, in this order.
\label{fig:2DimSteepNoSmear}
}
\end{figure}

Fig.~\ref{fig:1DimSteepNoSmear} and \ref{fig:2DimSteepNoSmear} show some of the 1-dimensional and 2-dimensional marginal distributions of $p(\T|\D)$, for some of the regularization options tried in this section.  Two effects are notable: (i) Even though there is no smearing, regularization can cause correlations.  (ii) For some regularization functions and parameters (e.g. $(S,\alpha)=(S_3,20)$), it is possible to induce secondary maxima in the posterior, something not observed without regularization.

\subsubsection{With smearing}
\label{sec:regSteepSmearing}

The spectrum used in this section is shown is the same one used in previous sections, and is shown in Fig.~\ref{fig:inputs4}.  The initially sampled hyper-box is the same shown there, and defined in Eq.~\ref{eq:prior4}.  MCMC is used for sampling, without volume reduction.

\begin{figure}[H]
  \centering
  \subfigure[$\alpha=0$]{
    \includegraphics[width=0.3\columnwidth]{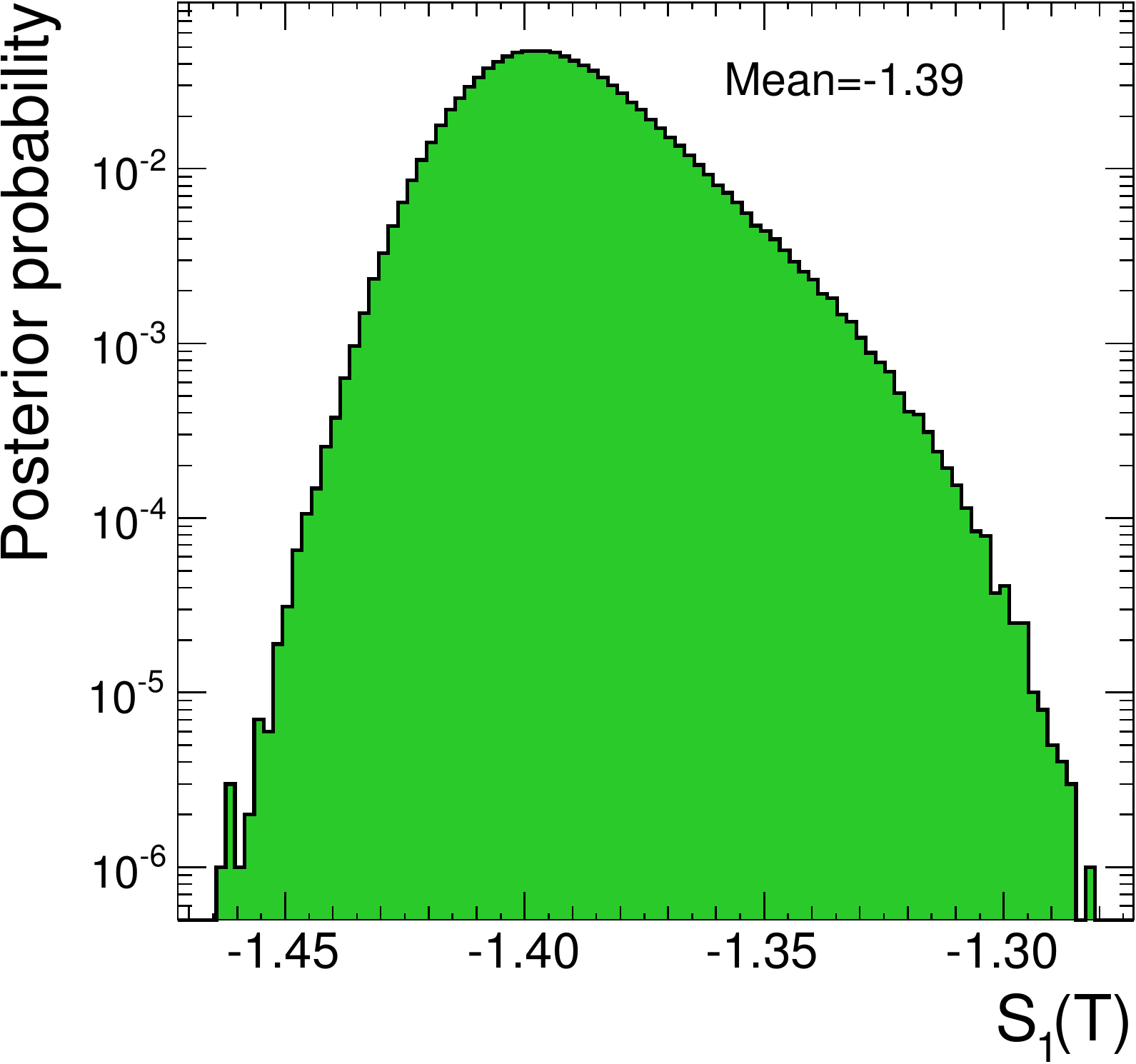}
    \label{fig:regFuncSteepSmearS1a}
  }
  \subfigure[$\alpha=10^3$]{
    \includegraphics[width=0.3\columnwidth]{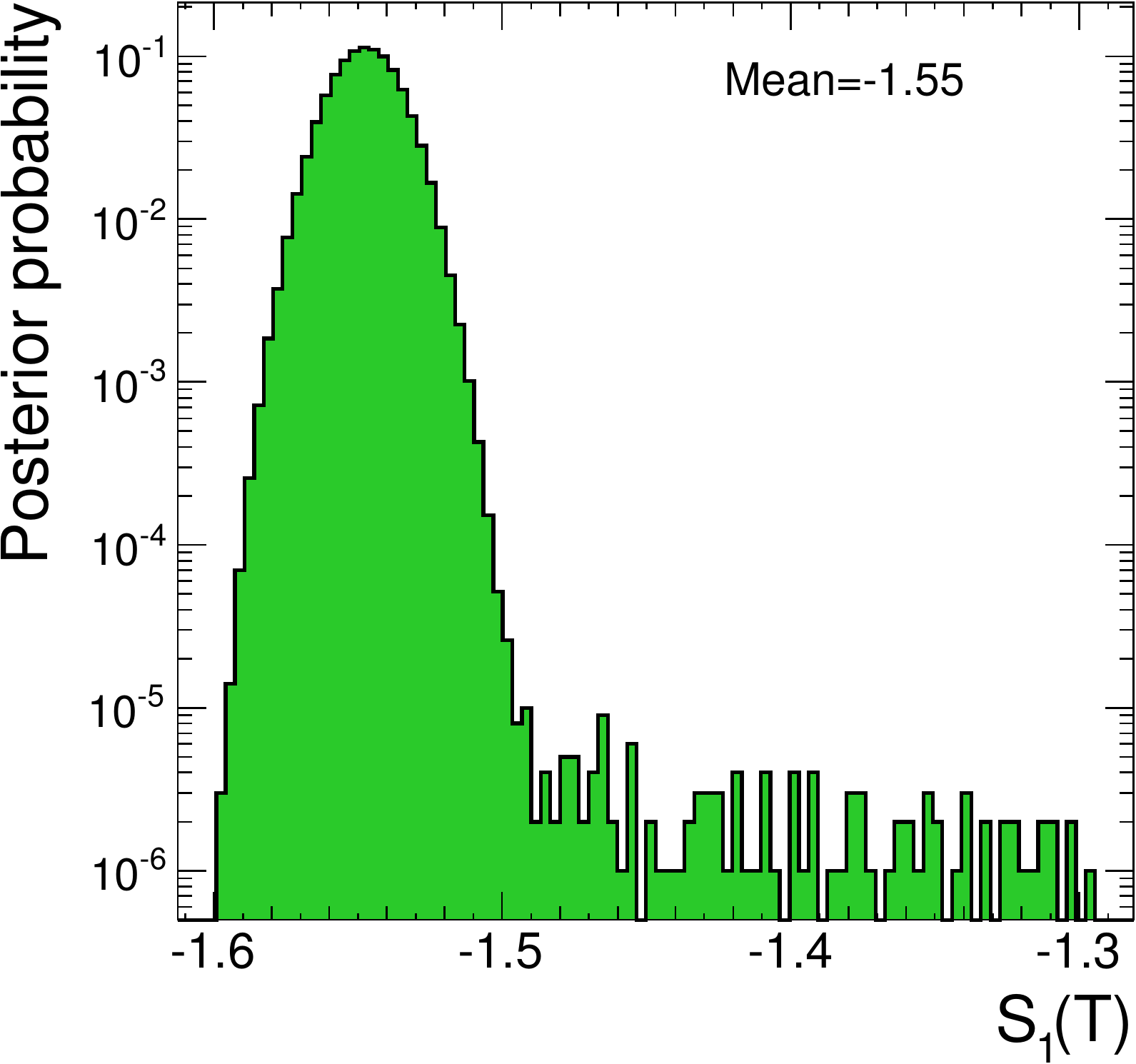}
  }
  \subfigure[$\alpha=3\times 10^3$]{
    \includegraphics[width=0.3\columnwidth]{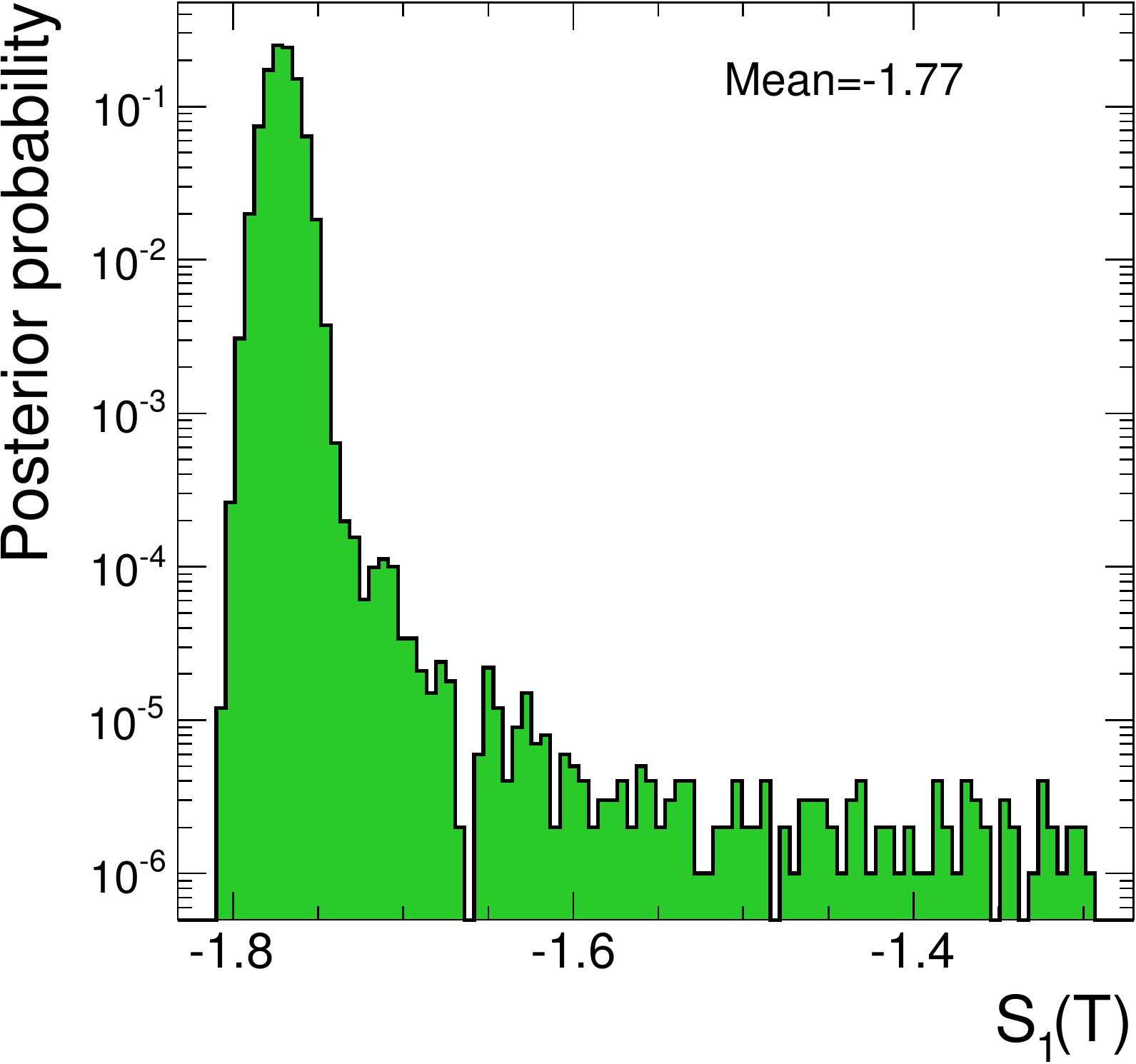}
  }
\caption{The posterior $P(S_1(\tuple{T})|\tuple{D})$, for three different choices of the regularization parameter $\alpha$, corresponding to Sec.~\ref{sec:regSteepSmearing}.
\label{fig:regFuncSteepSmearS1}
}
\end{figure}

\begin{figure}[H]
  \centering
  \subfigure[$\alpha=0$]{
    \includegraphics[width=0.3\columnwidth]{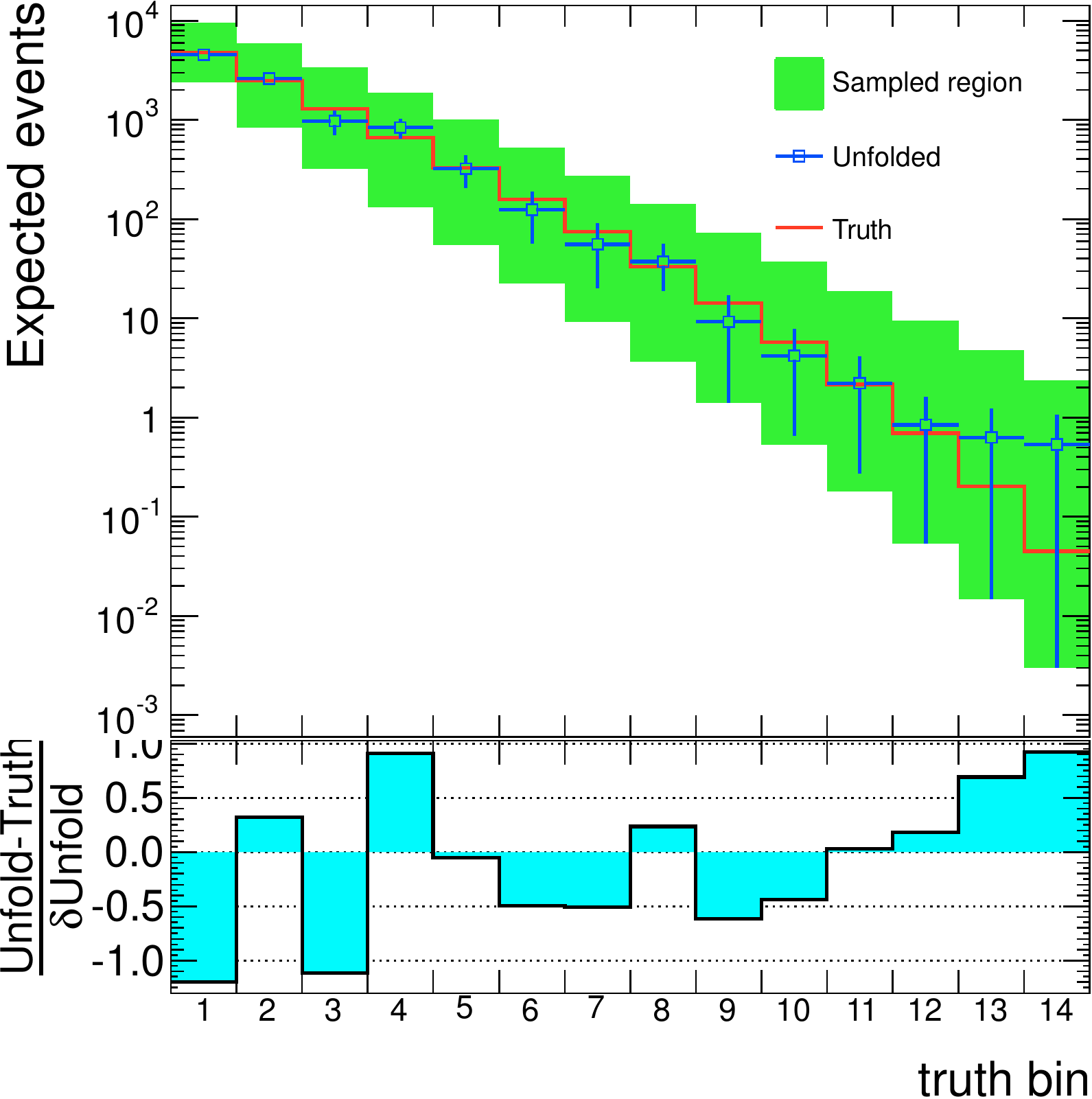}
  }
  \subfigure[$\alpha=10^3$]{
    \includegraphics[width=0.3\columnwidth]{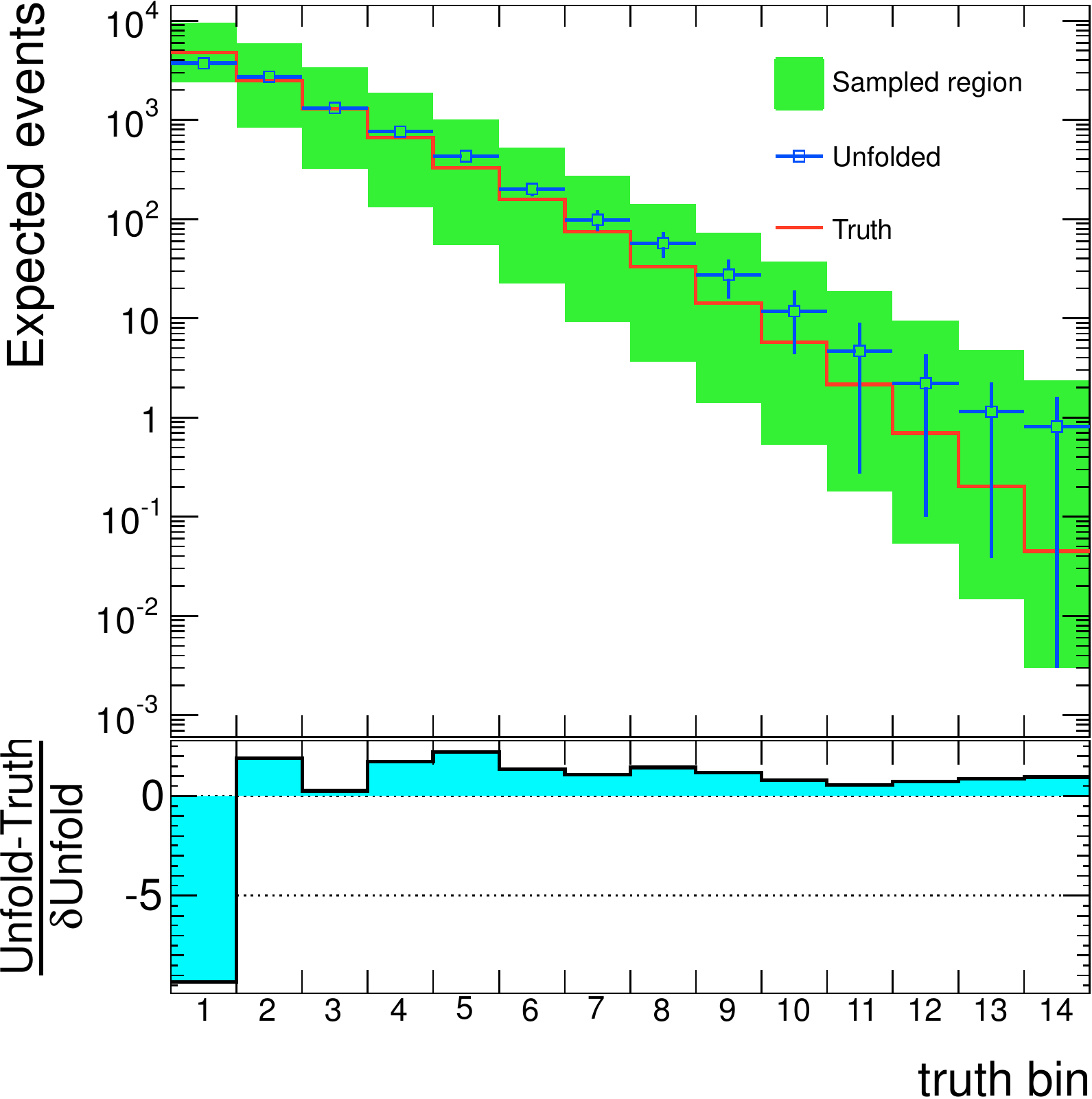}
  }
  \subfigure[$\alpha=3\times 10^3$]{
    \includegraphics[width=0.3\columnwidth]{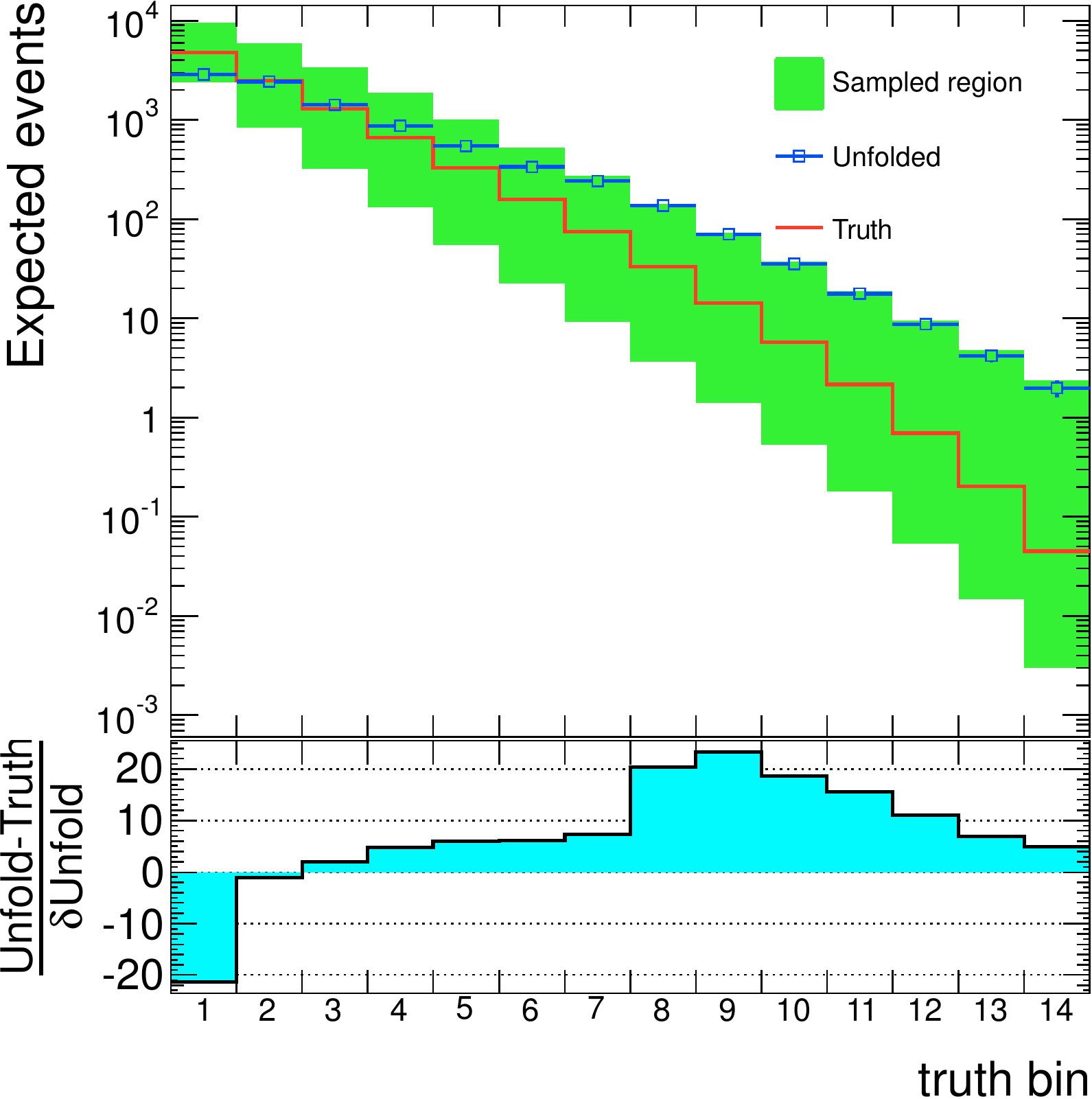}
  }\\
 \subfigure[$\alpha=0$]{
   \includegraphics[width=0.3\columnwidth]{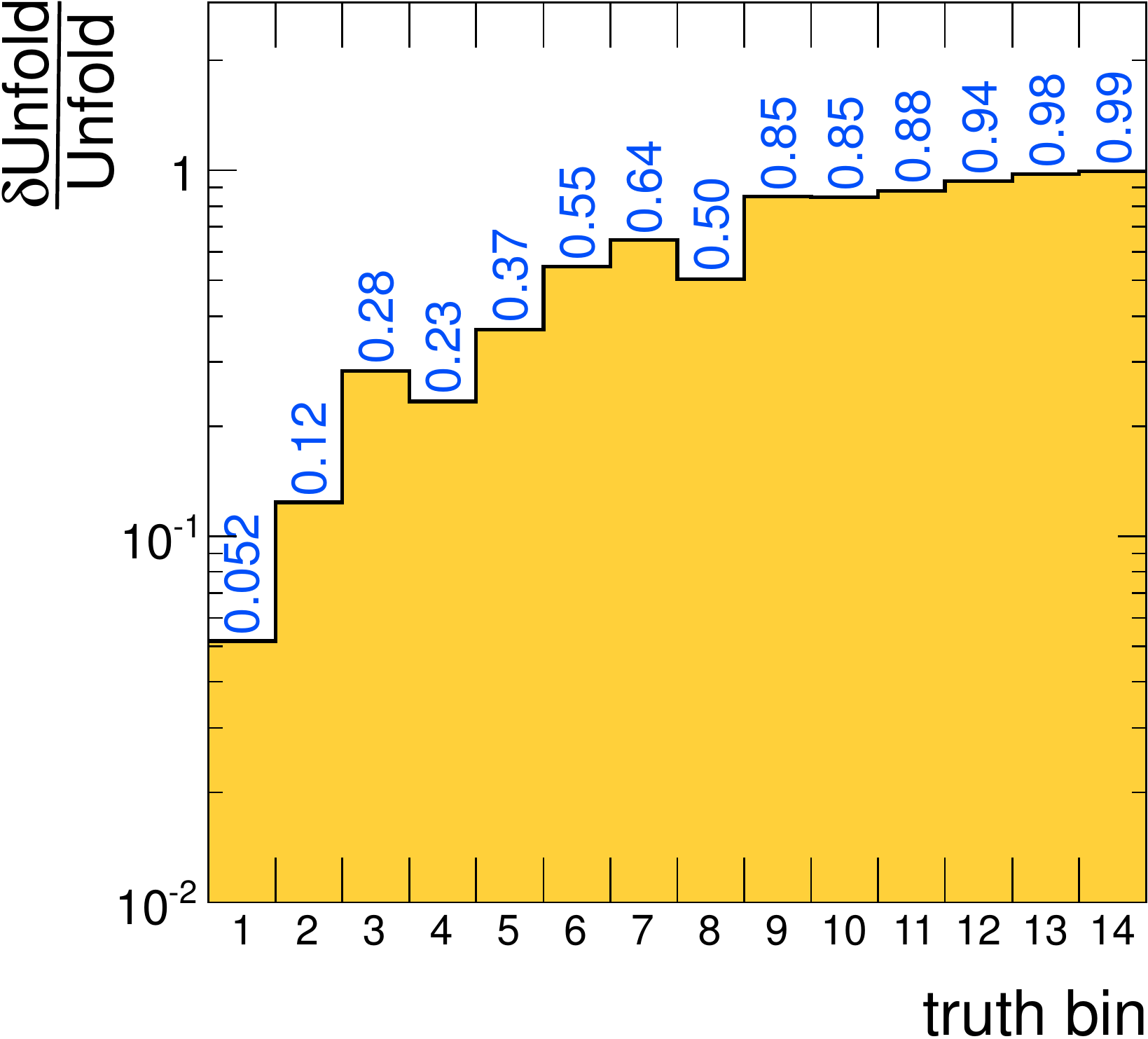}
   \label{fig:unfoldedSteepSmearS1d}
  }
  \subfigure[$\alpha=10^3$]{
    \includegraphics[width=0.3\columnwidth]{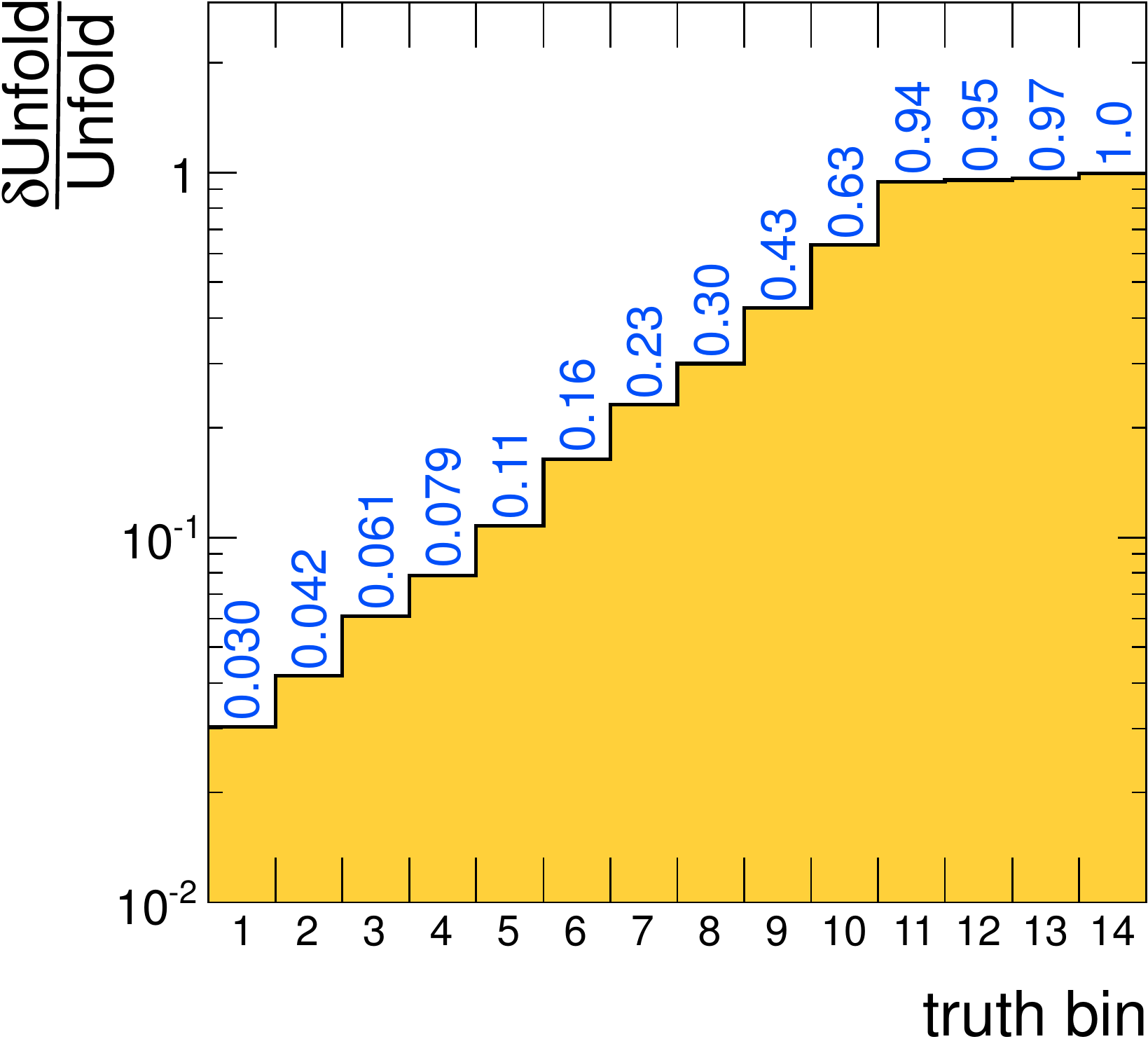}
  }
  \subfigure[$\alpha=3\times 10^3$]{
    \includegraphics[width=0.3\columnwidth]{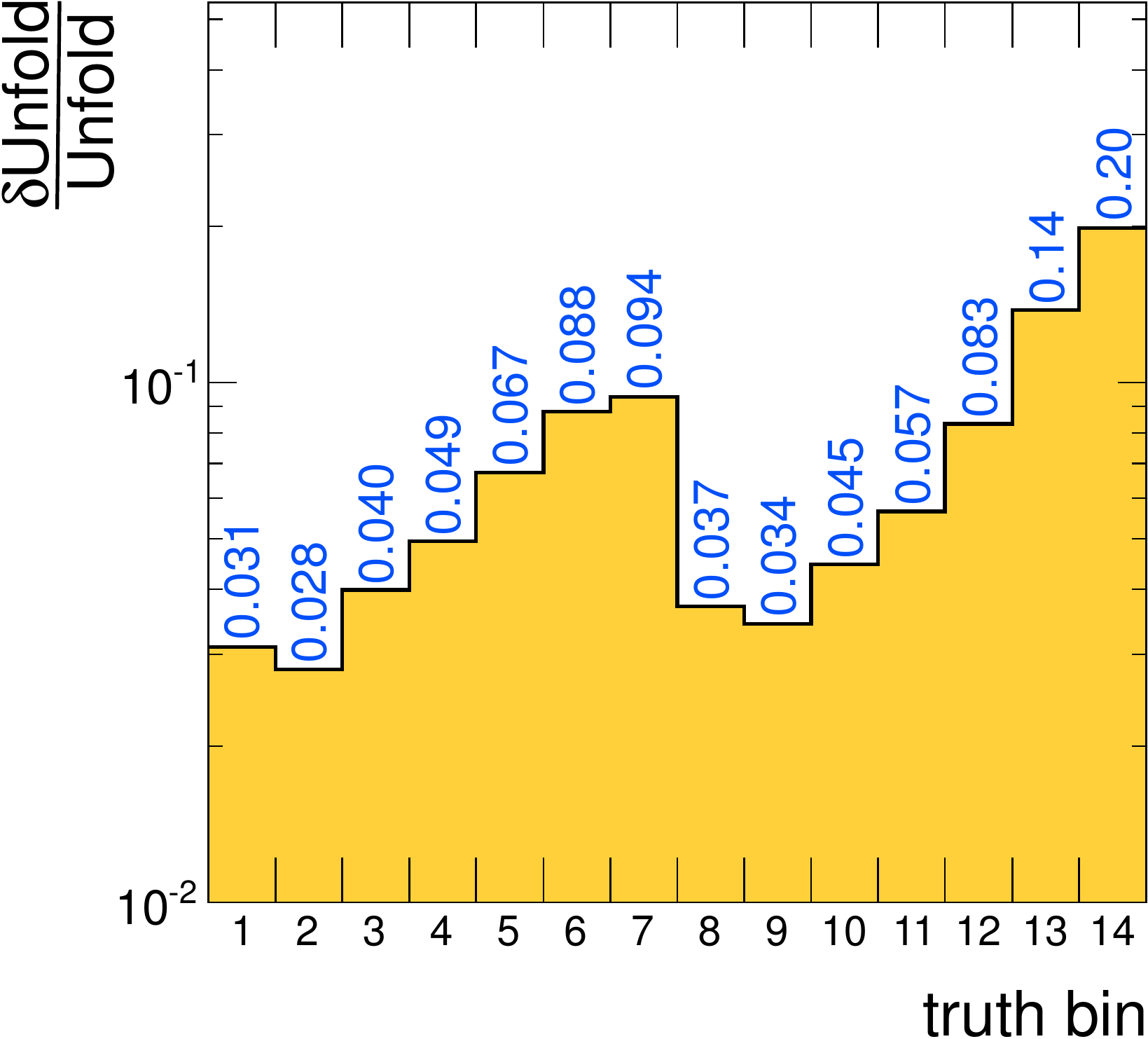}
  }
  \caption{The result of unfolding of Sec.~\ref{sec:regSteepSmearing}, with regularization function $S_1$, for three values of $\alpha$.  
\label{fig:unfoldedSteepSmearS1}
}
\end{figure}

Like in Sec.~\ref{sec:regSteepNoSmearing}, the regularization with negative entropy ($S_1$) is tried first.
The results are shown in Fig.~\ref{fig:regFuncSteepSmearS1} and \ref{fig:unfoldedSteepSmearS1}.  The conclusions are similar to those in Sec.~\ref{sec:regSteepNoSmearing}, with the difference that he posterior spread is large prior to regularization, due to smearing, and it is greatly reduced by regularization.

\begin{figure}[H]
  \centering
  \subfigure[$\alpha=0$]{
    \includegraphics[width=0.3\columnwidth]{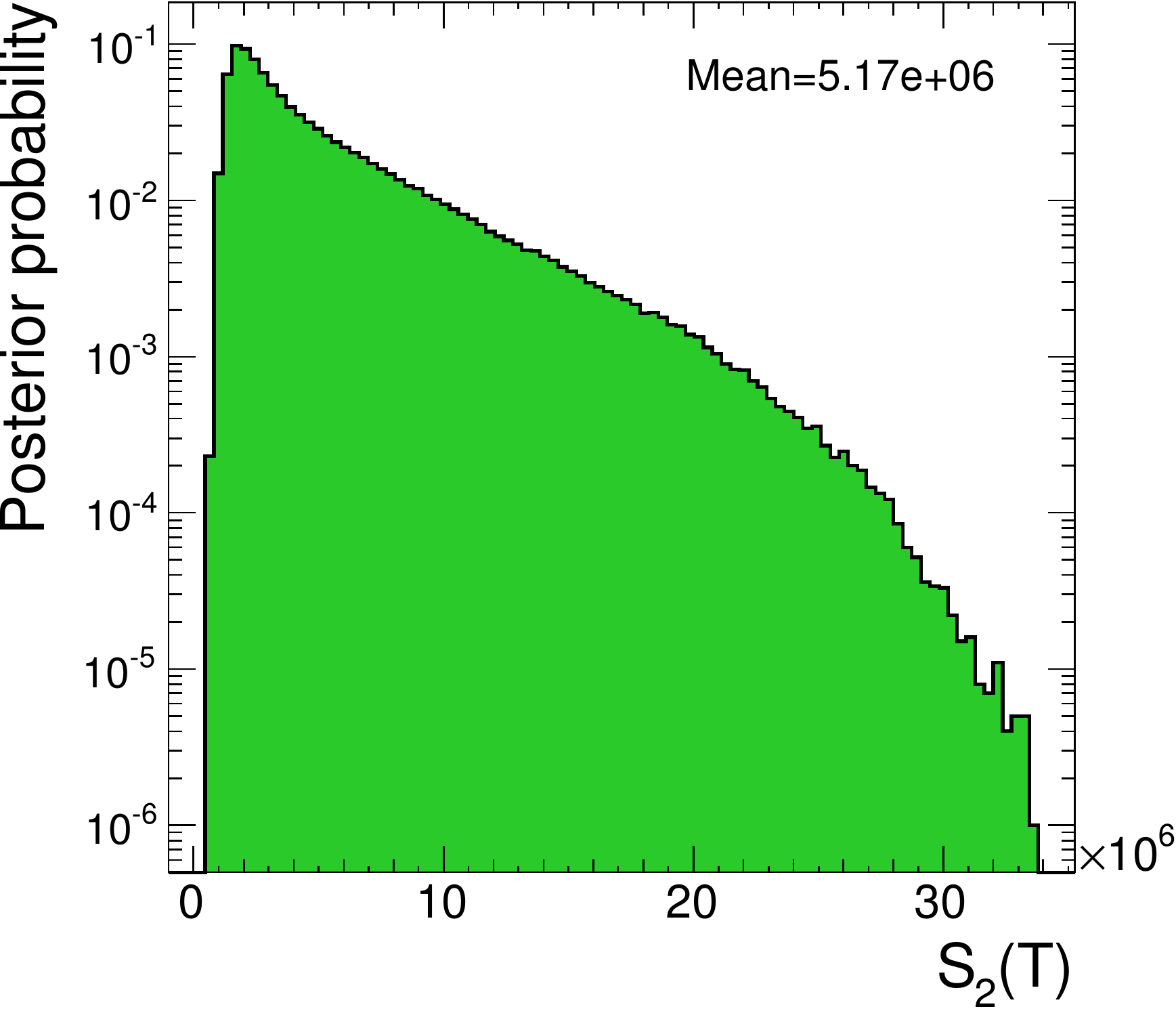}
  }
  \subfigure[$\alpha=3\times 10^{-4}$]{
    \includegraphics[width=0.3\columnwidth]{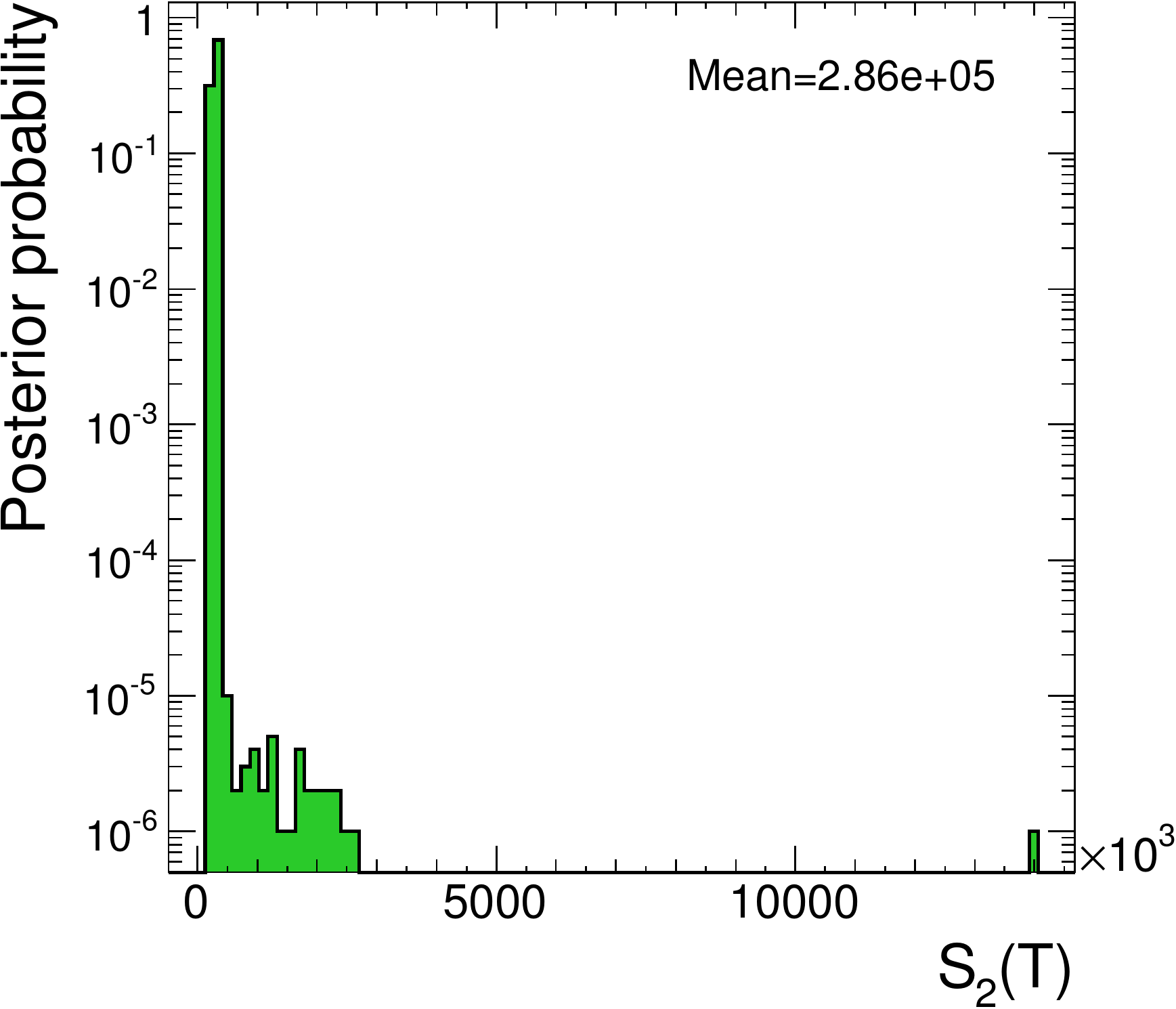}
  }
  \subfigure[$\alpha=6\times 10^{-4}$]{
    \includegraphics[width=0.3\columnwidth]{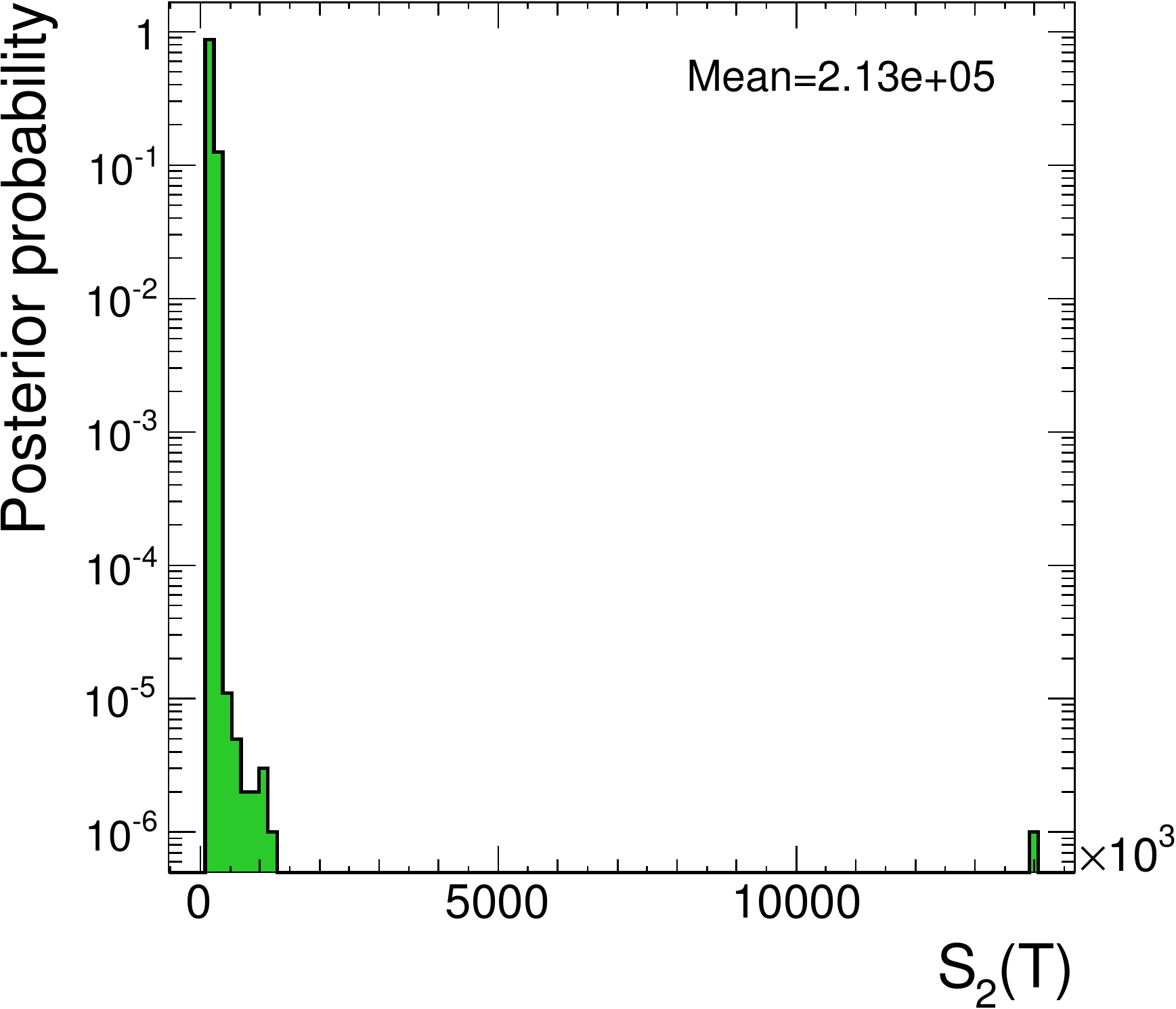}
  }
\caption{The posterior $P(S_2(\tuple{T})|\tuple{D})$, for three different choices of the regularization parameter $\alpha$, corresponding to Sec.~\ref{sec:regSteepSmearing}.
\label{fig:regFuncSteepSmearS2}
}
\end{figure}

\begin{figure}[H]
  \centering
  \subfigure[$\alpha=0$]{
    \includegraphics[width=0.3\columnwidth]{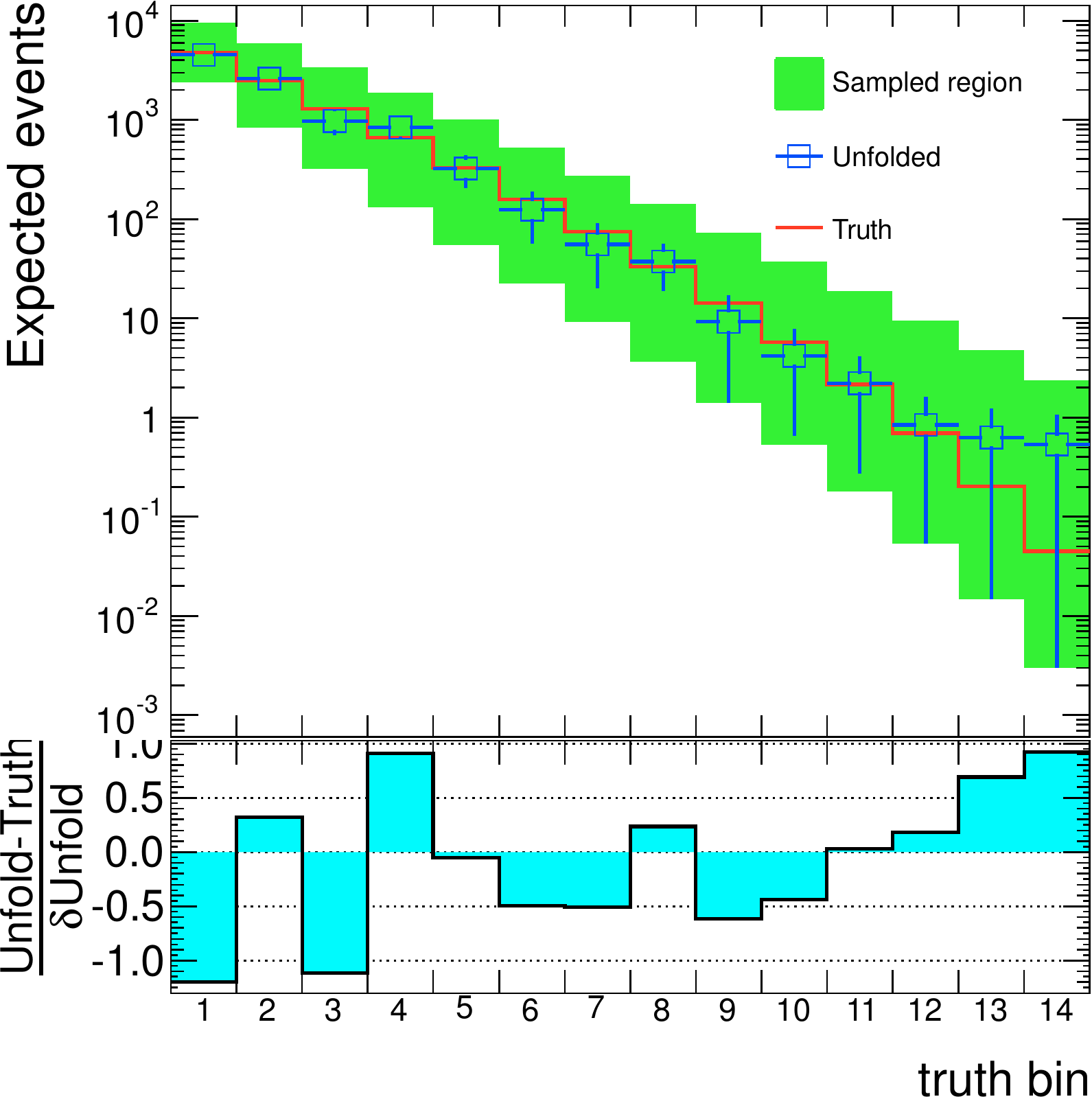}
  }
  \subfigure[$\alpha=3\times 10^{-4}$]{
    \includegraphics[width=0.3\columnwidth]{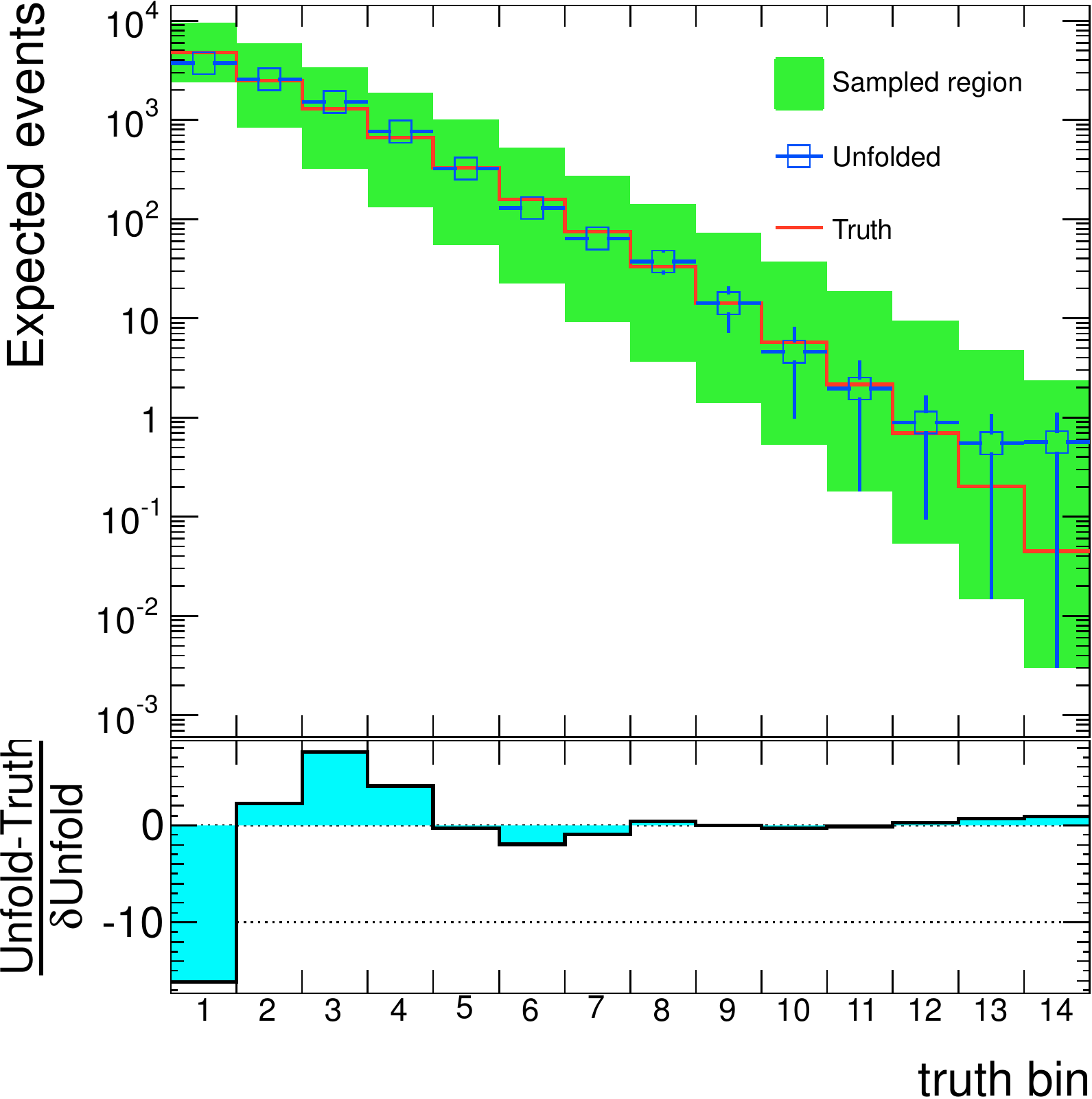}
  }
  \subfigure[$\alpha=6\times 10^{-4}$]{
    \includegraphics[width=0.3\columnwidth]{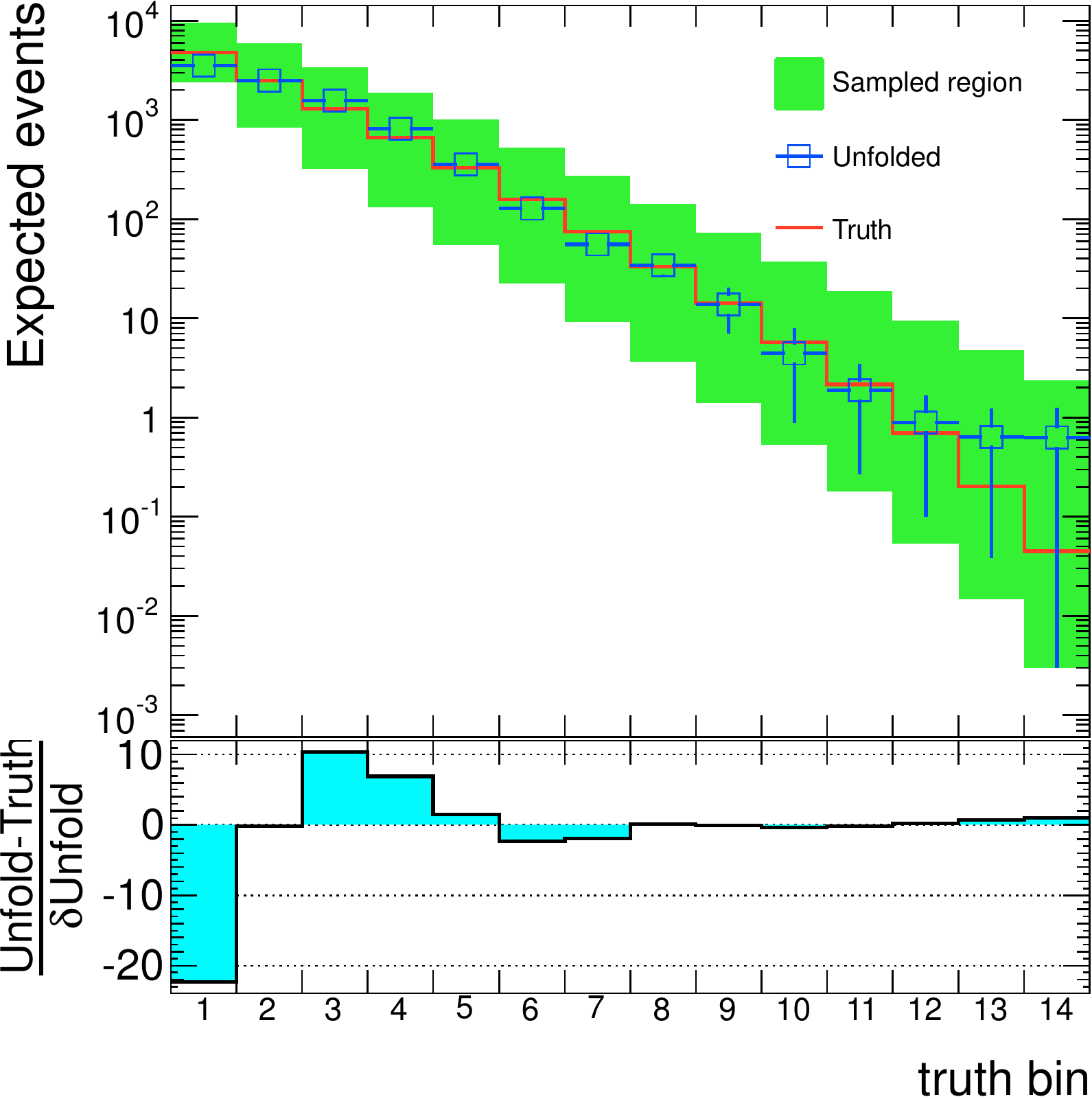}
  }\\
 \subfigure[$\alpha=0$]{
   \includegraphics[width=0.3\columnwidth]{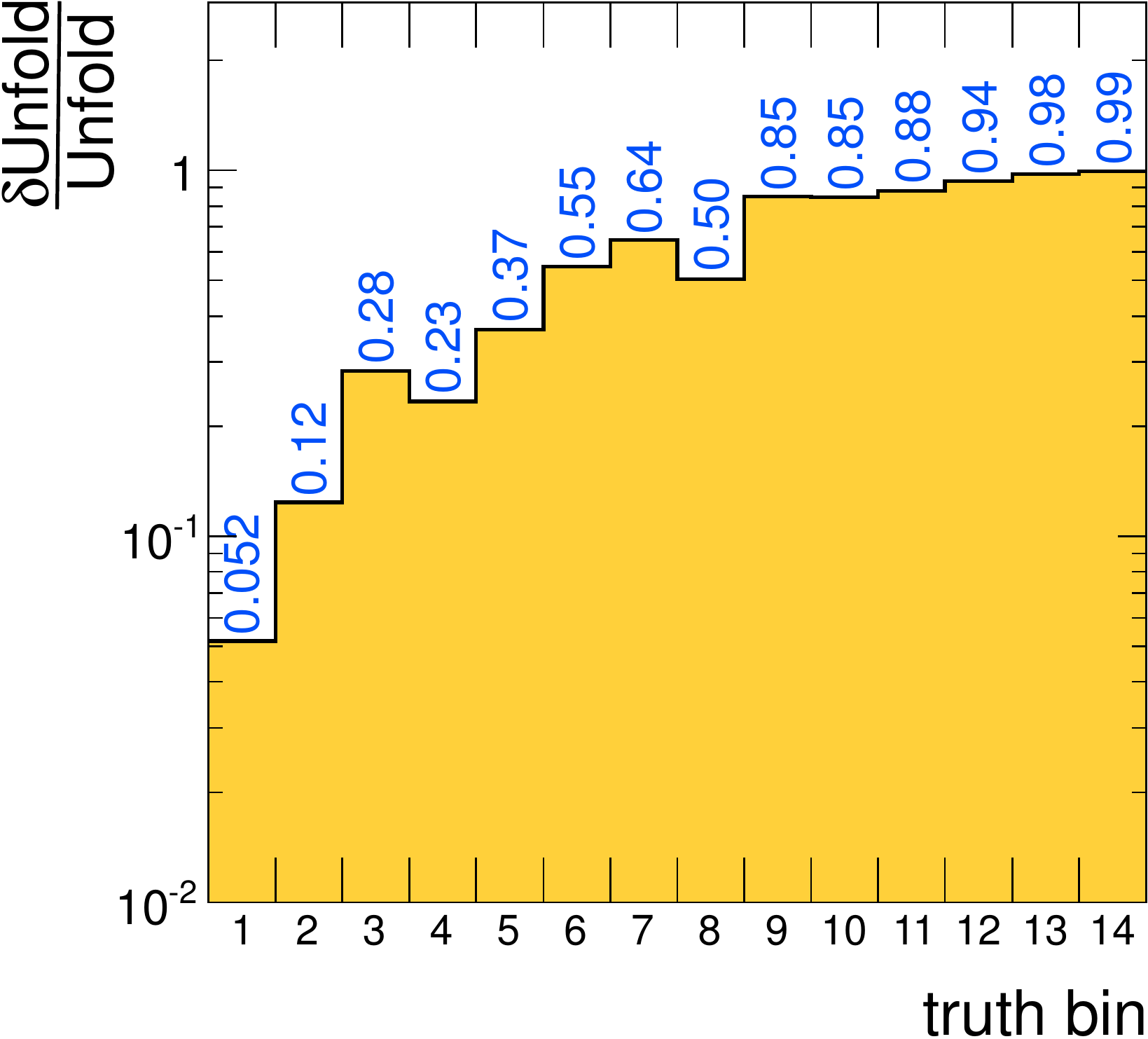}
   \label{fig:unfoldedSteepSmearS2d}
  }
  \subfigure[$\alpha=3\times 10^{-4}$]{
   \includegraphics[width=0.3\columnwidth]{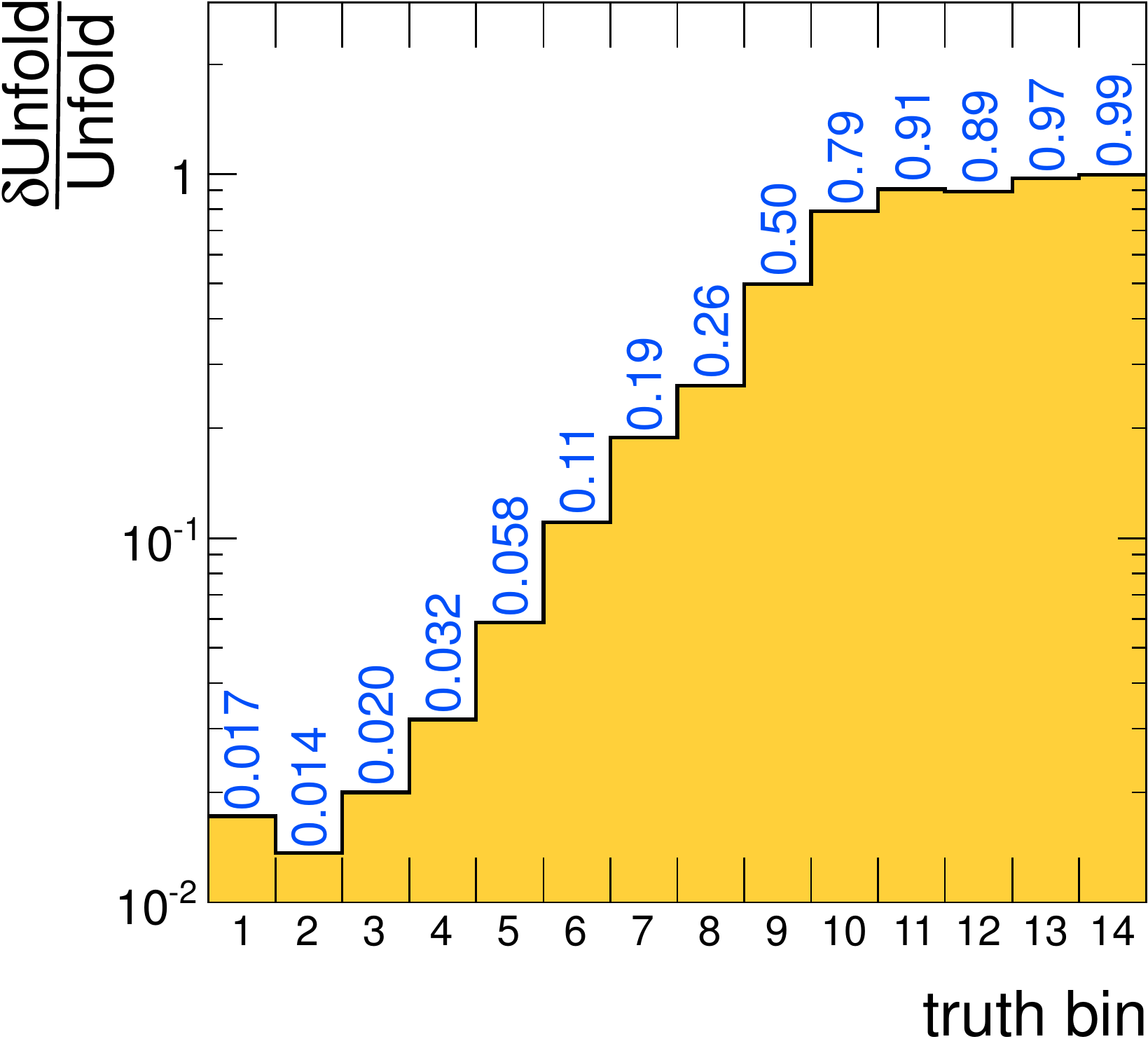}
  }
  \subfigure[$\alpha=6\times 10^{-4}$]{
    \includegraphics[width=0.3\columnwidth]{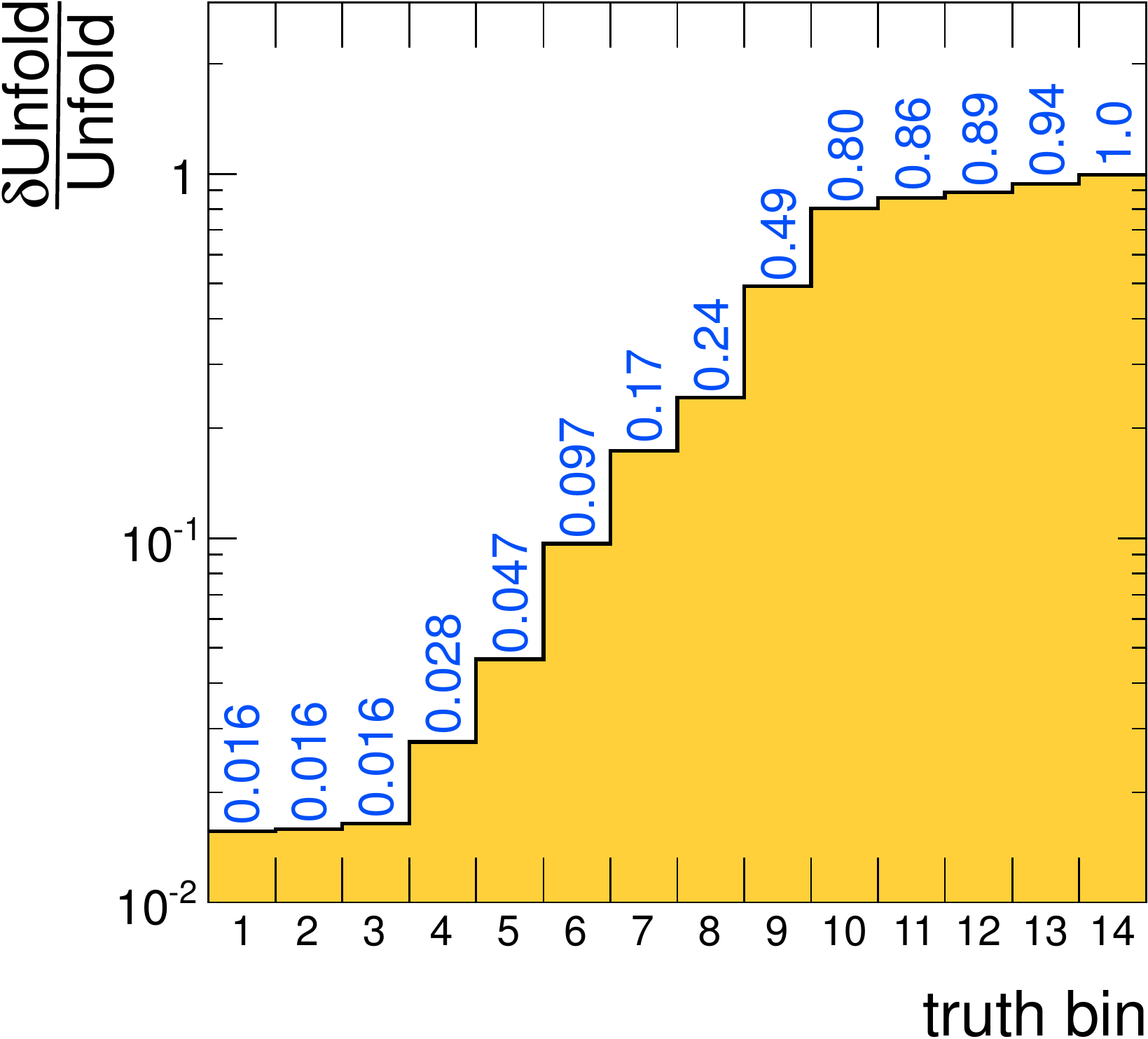}
  }
  \caption{The result of unfolding of Sec.~\ref{sec:regSteepSmearing}, with regularization function $S_2$, for three values of $\alpha$.  
\label{fig:unfoldedSteepSmearS2}
}
\end{figure}

The results of using $S_2$ are shown in Fig.~\ref{fig:regFuncSteepSmearS2}  and Fig.~\ref{fig:unfoldedSteepSmearS2}.  Comparing Fig.~\ref{fig:unfoldedSteepSmearS2d} to Fig.~\ref{fig:unfoldedSteepNoSmearS2d} shows again that smearing increases the spread of the posterior.  That allows regularization, in the presence of smearing (Fig.~\ref{fig:unfoldedSteepSmearS2}), to reduce significantly the posterior spread to levels almost as low as those in Sec.~\ref{sec:regSteepNoSmearing} without smearing, at the cost of significant bias.

\begin{figure}[H]
  \centering
  \subfigure[$\alpha=0$]{
    \includegraphics[width=0.3\columnwidth]{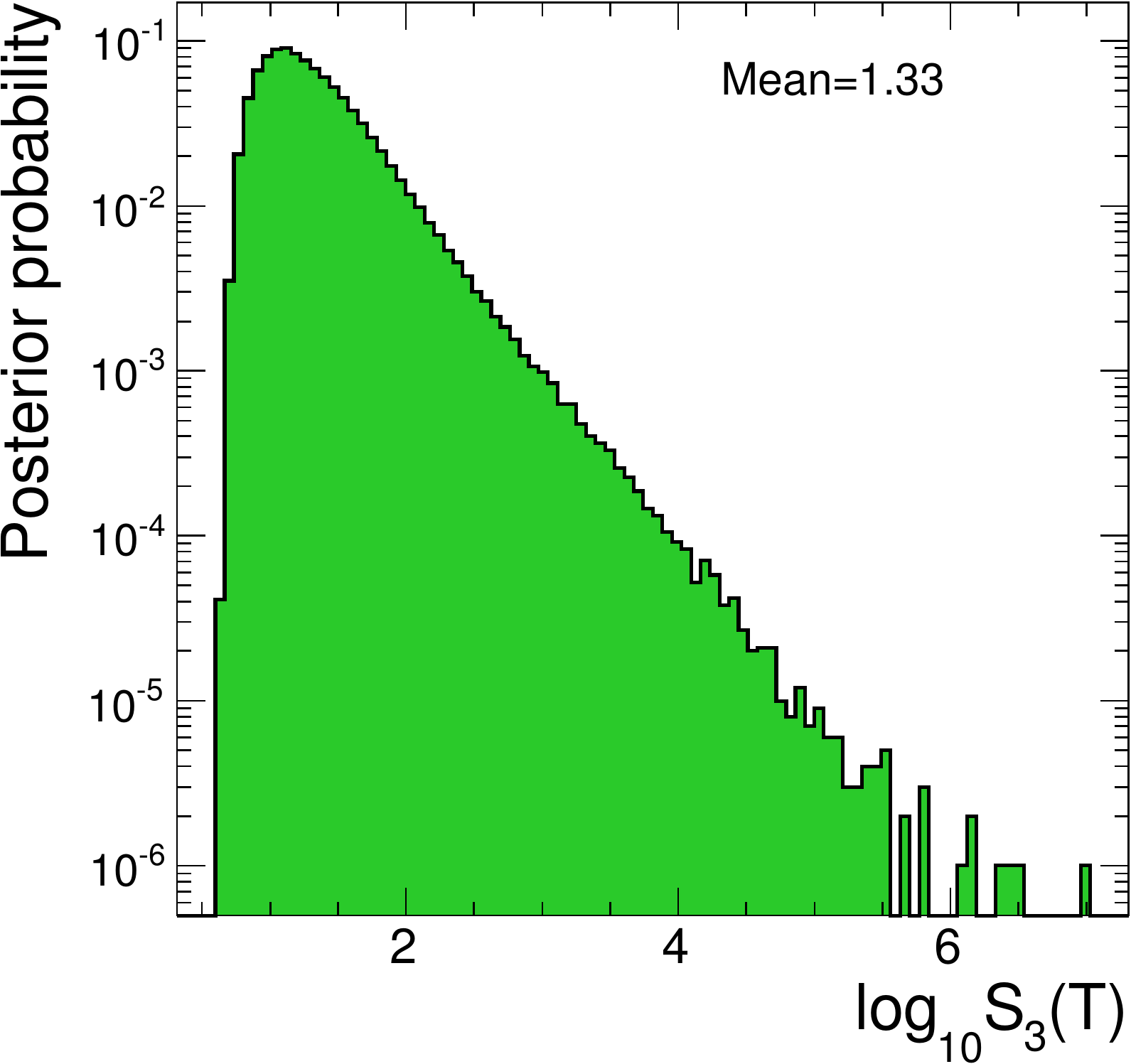}
  }
  \subfigure[$\alpha=20$]{
    \includegraphics[width=0.3\columnwidth]{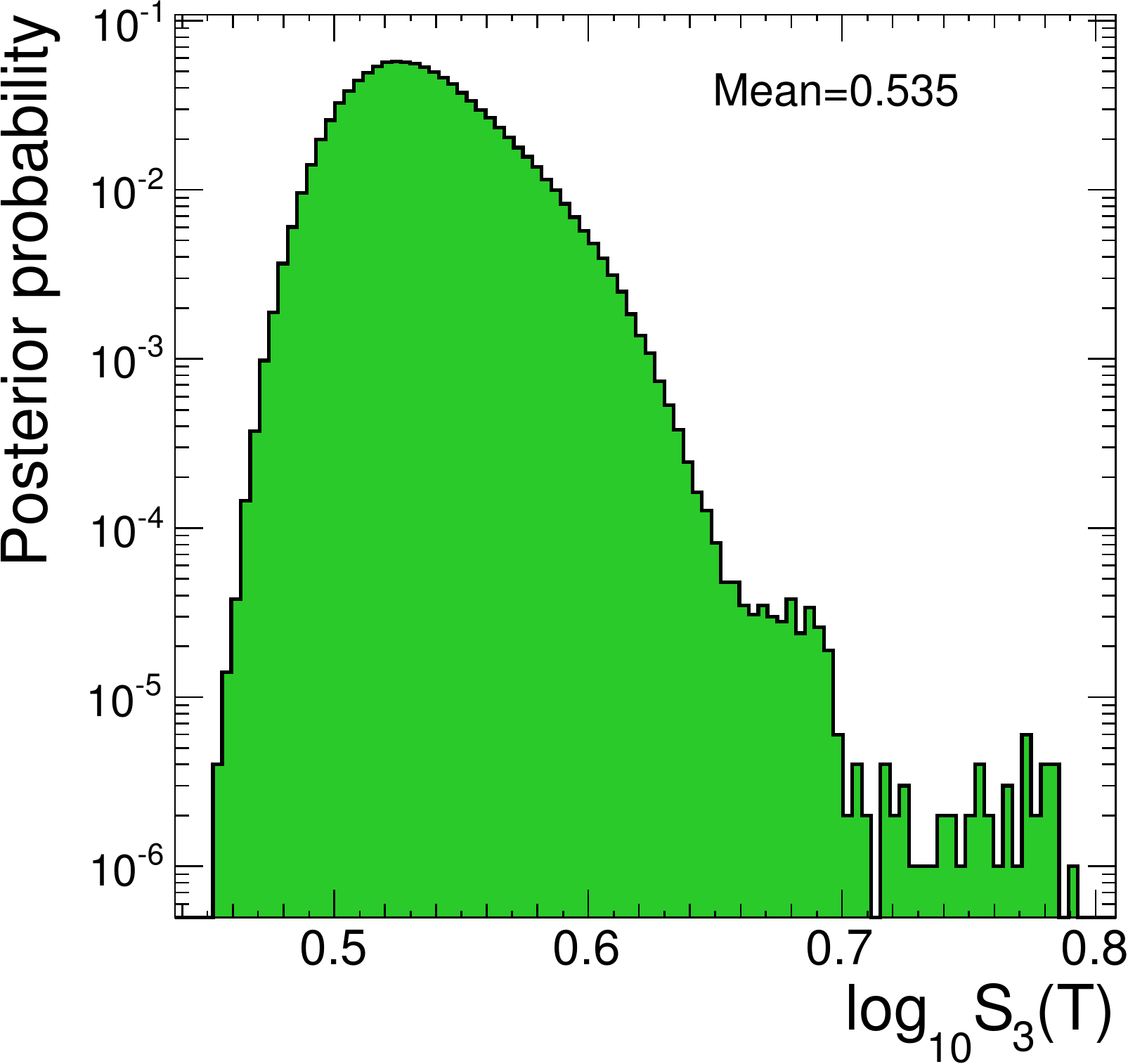}
  }
  \subfigure[$\alpha=40$]{
    \includegraphics[width=0.3\columnwidth]{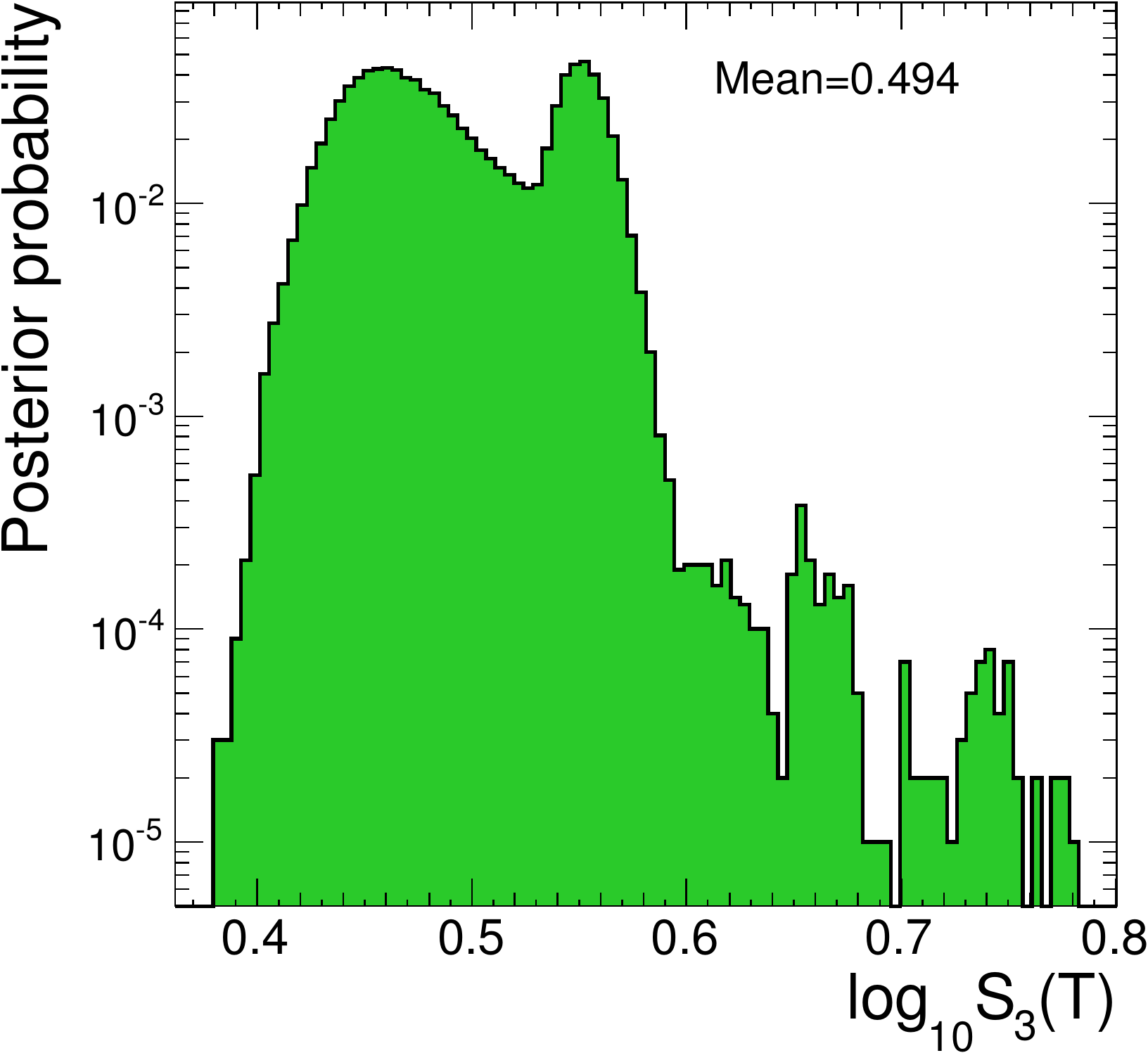}
    \label{fig:regFuncSteepSmearS3c}
  }
\caption{The posterior $P(S_3(\tuple{T})|\tuple{D})$, for three different choices of the regularization parameter $\alpha$, corresponding to Sec.~\ref{sec:regSteepSmearing}.
\label{fig:regFuncSteepSmearS3}
}
\end{figure}

\begin{figure}[H]
  \centering
  \subfigure[$\alpha=0$]{
    \includegraphics[width=0.3\columnwidth]{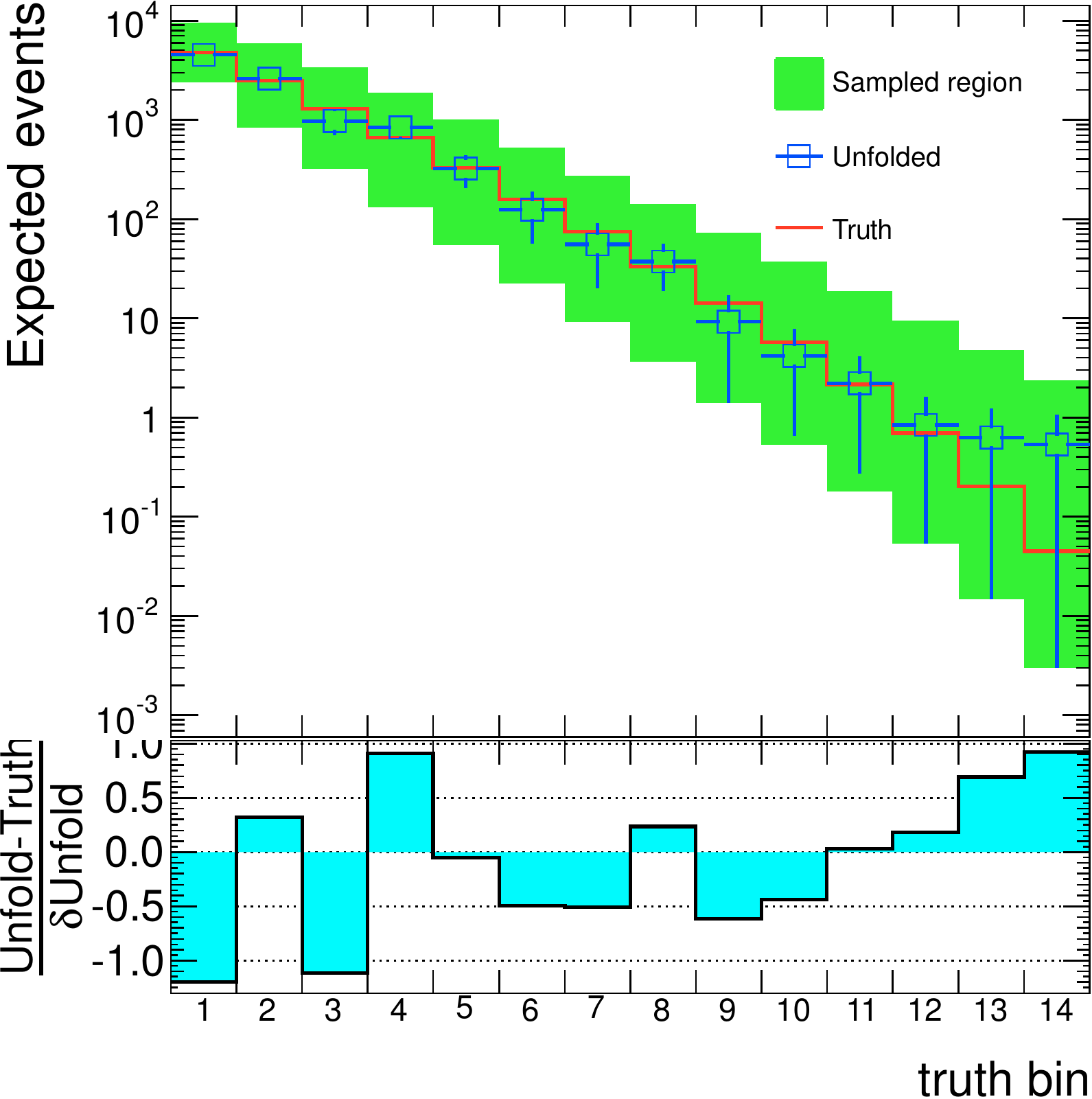}
  }
  \subfigure[$\alpha=20$]{
    \includegraphics[width=0.3\columnwidth]{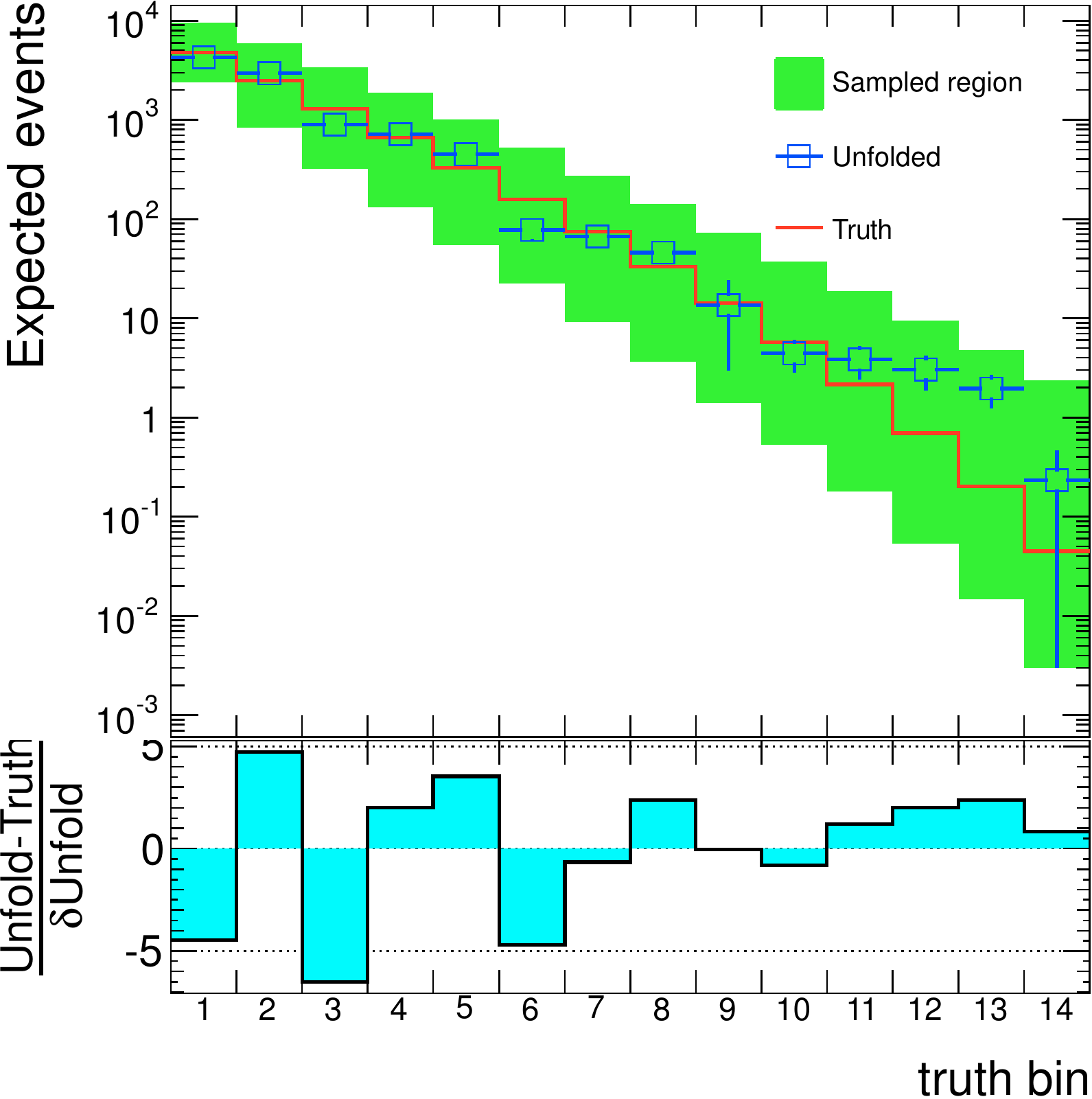}
  }
  \subfigure[$\alpha=40$]{
    \includegraphics[width=0.3\columnwidth]{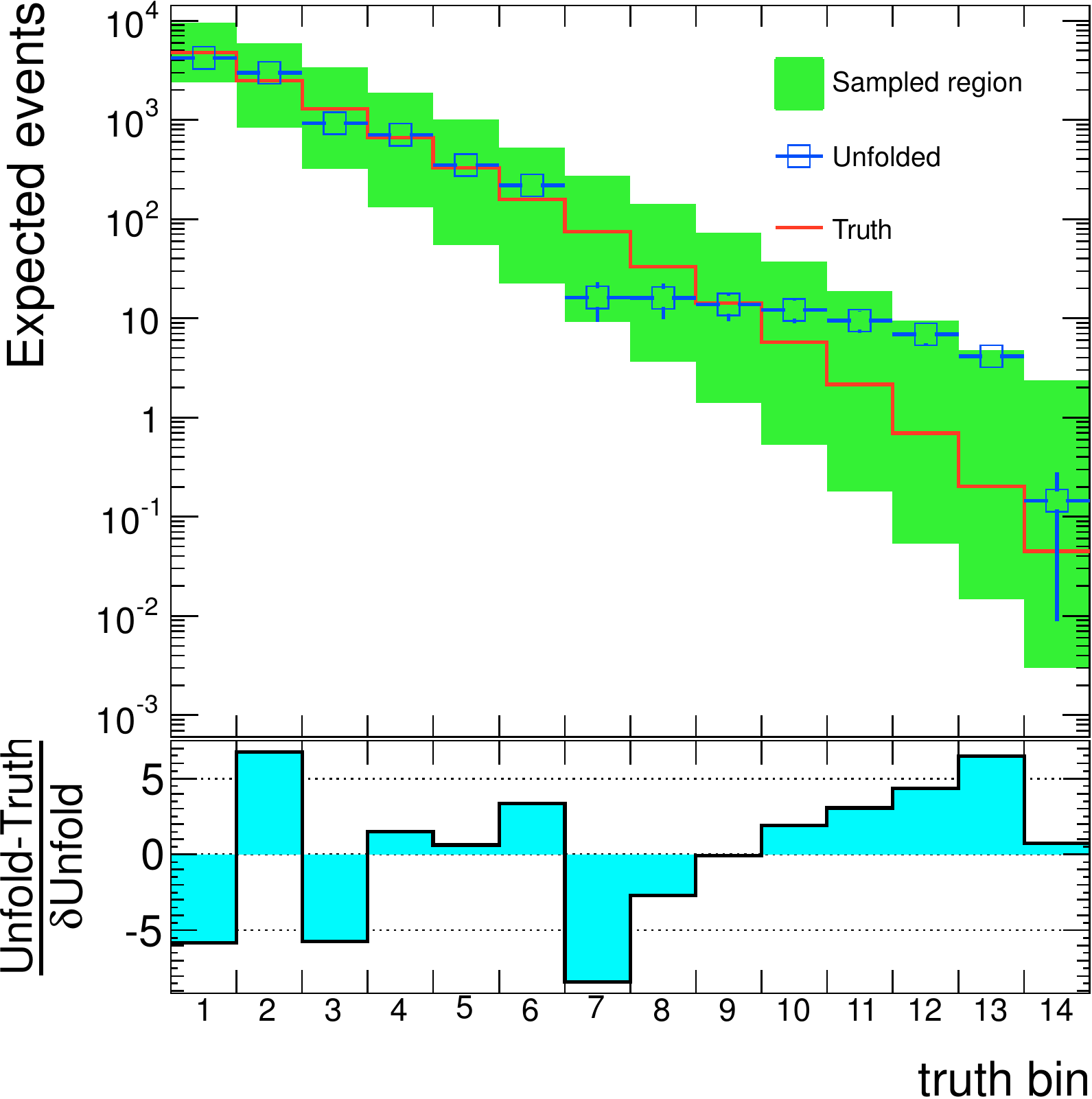}
  }\\
 \subfigure[$\alpha=0$]{
   \includegraphics[width=0.3\columnwidth]{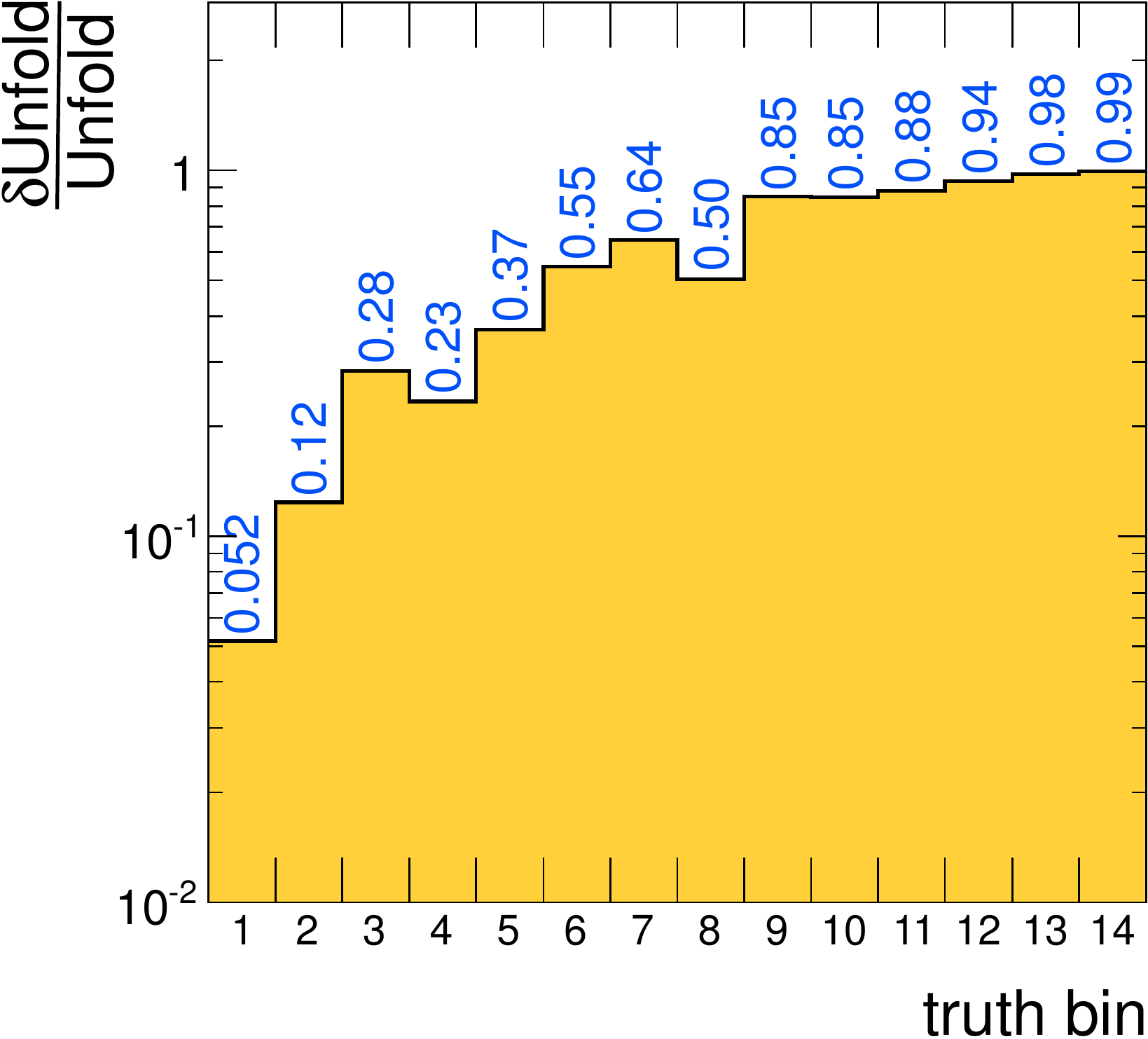}
  }
  \subfigure[$\alpha=20$]{
   \includegraphics[width=0.3\columnwidth]{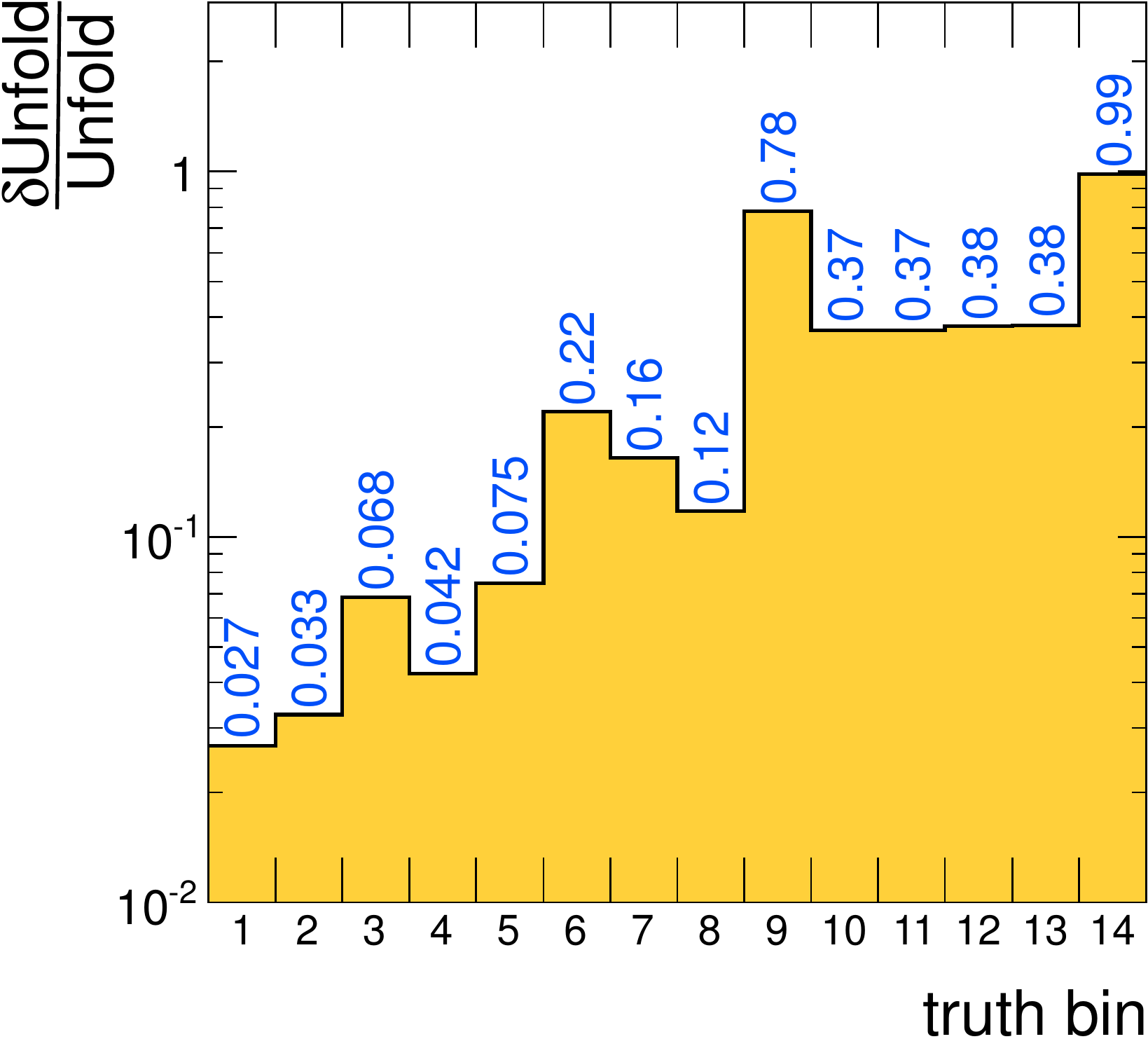}
  }
  \subfigure[$\alpha=40$]{
    \includegraphics[width=0.3\columnwidth]{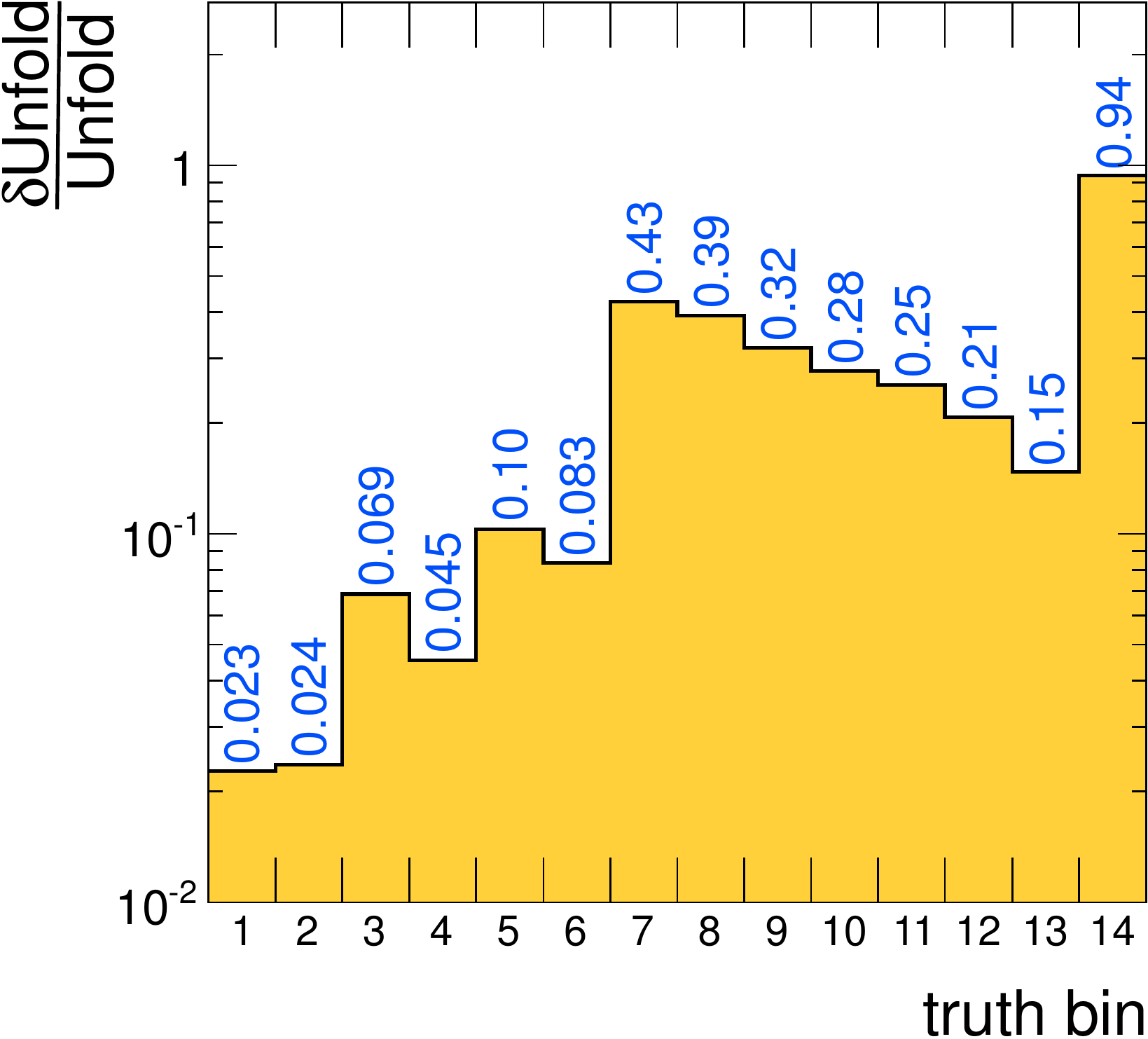}
  }
  \caption{The result of unfolding of Sec.~\ref{sec:regSteepSmearing}, with regularization function $S_3$, for three values of $\alpha$.  
\label{fig:unfoldedSteepSmearS3}
}
\end{figure}

\begin{figure}[H]
  \centering
  \begin{tabular}{ccccc}
   \raisebox{0.09\columnwidth}{$t=1$} &
   \includegraphics[width=0.18\columnwidth]{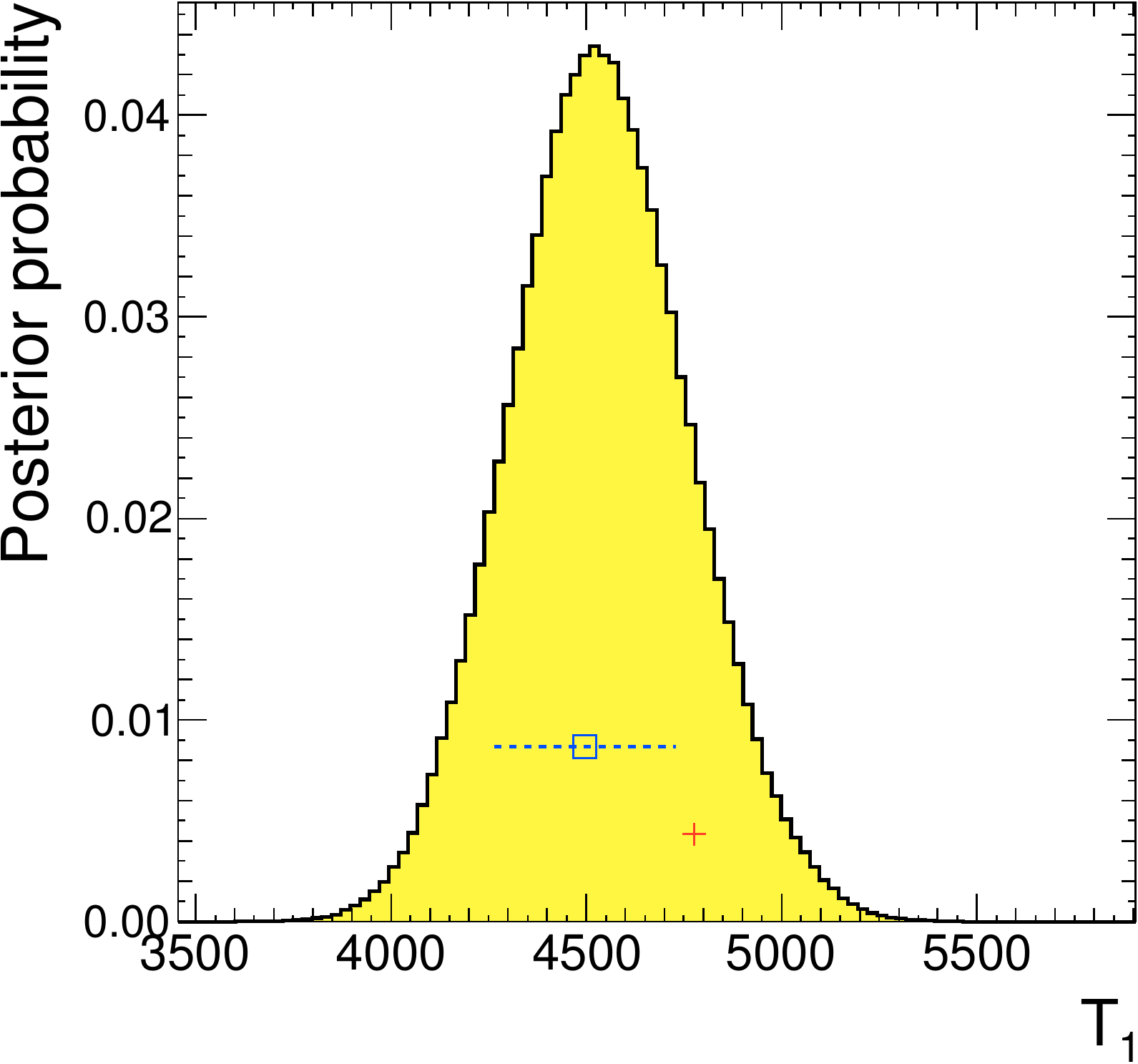} &
   \includegraphics[width=0.18\columnwidth]{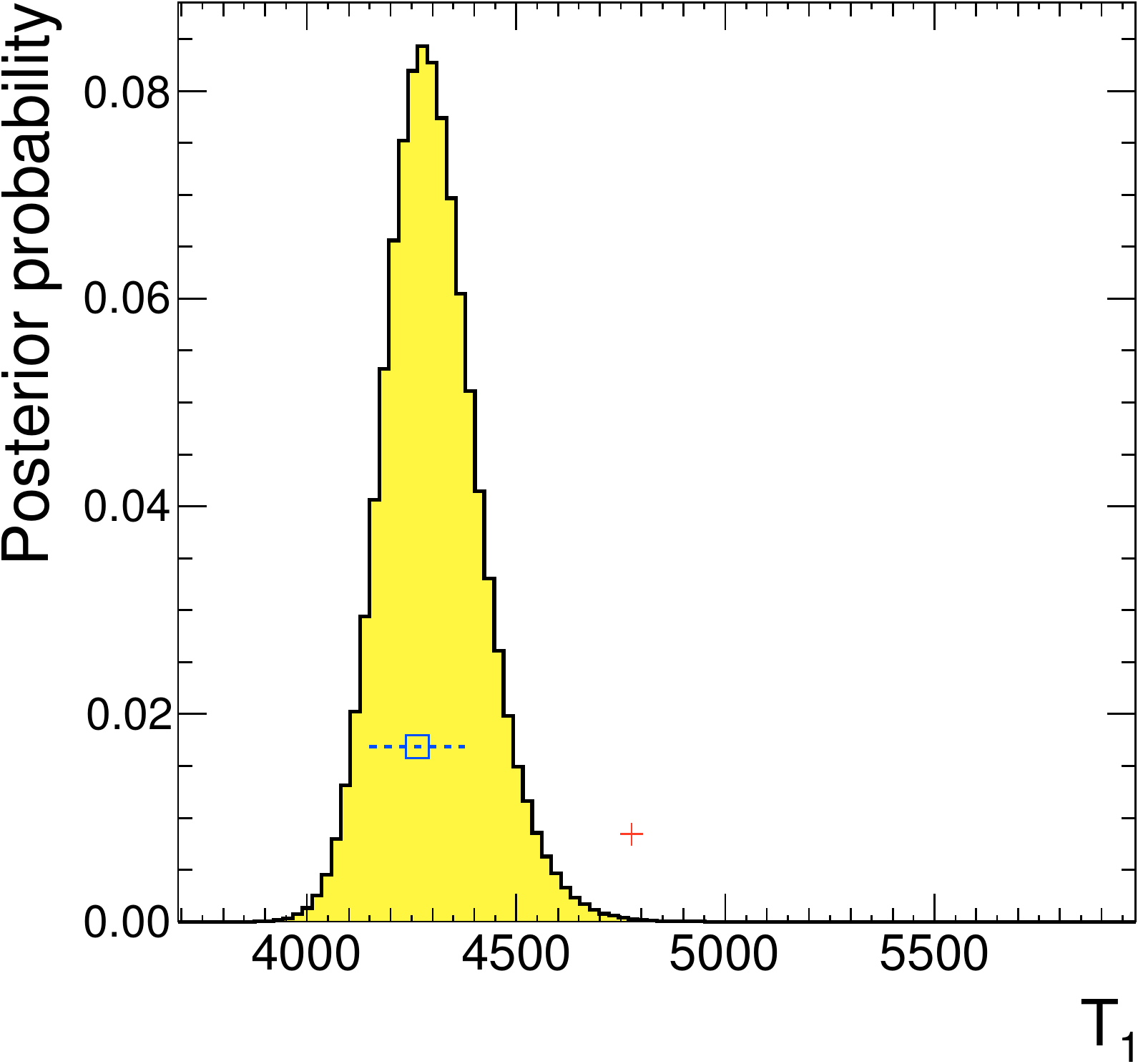} &
   \includegraphics[width=0.18\columnwidth]{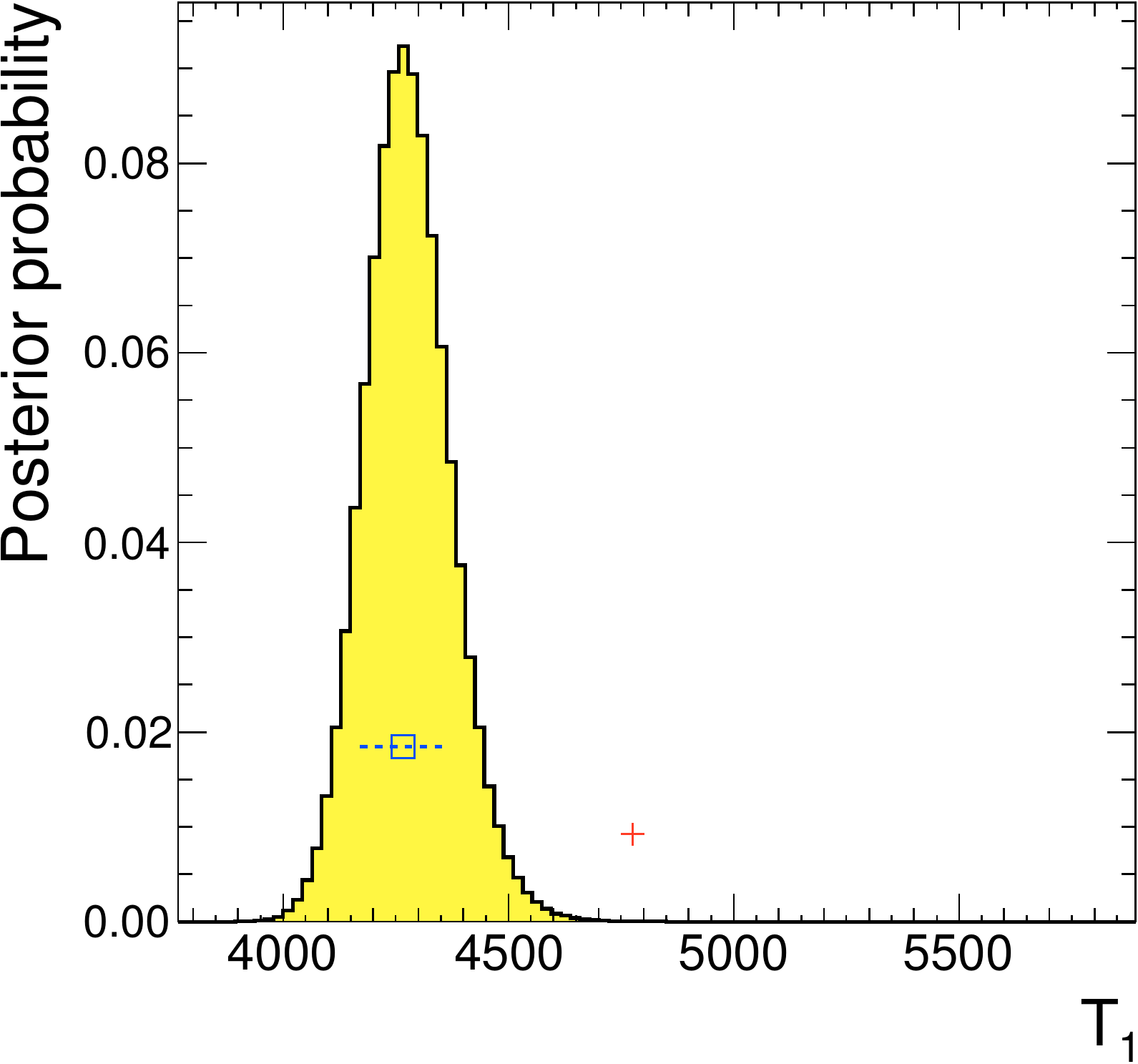} &
   \includegraphics[width=0.18\columnwidth]{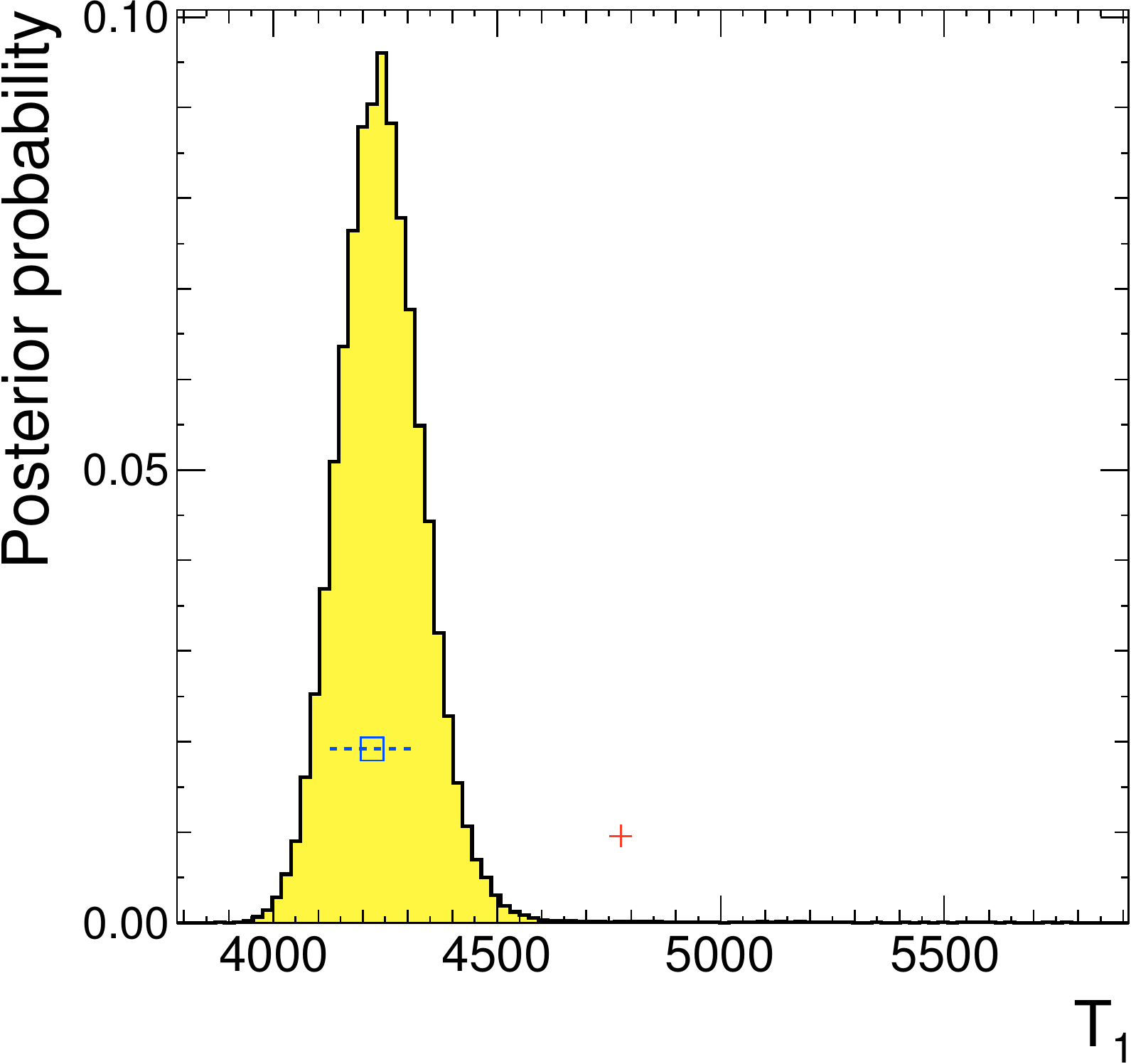} \\

   \raisebox{0.09\columnwidth}{$t=3$} &
   \includegraphics[width=0.18\columnwidth]{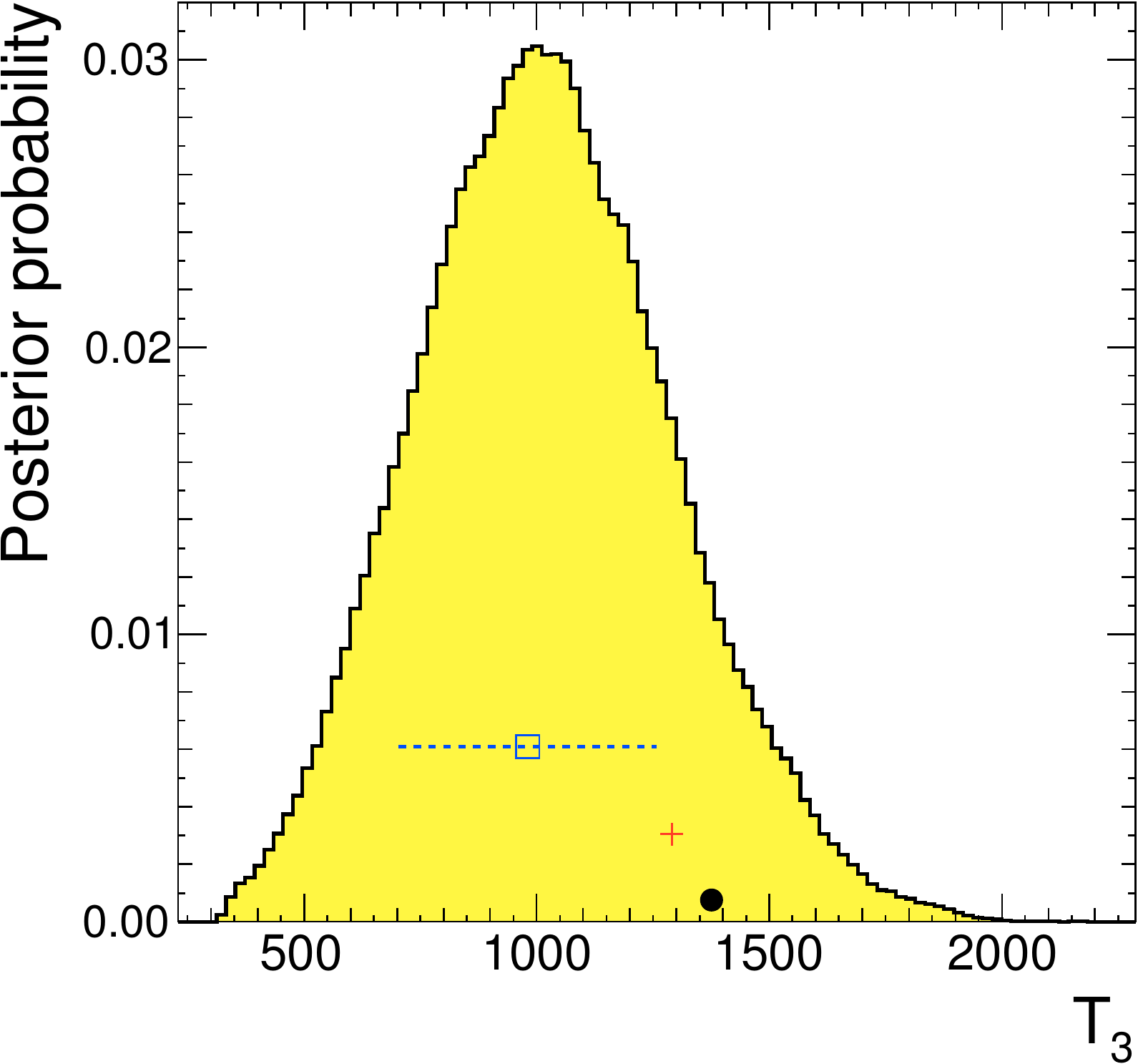} &
   \includegraphics[width=0.18\columnwidth]{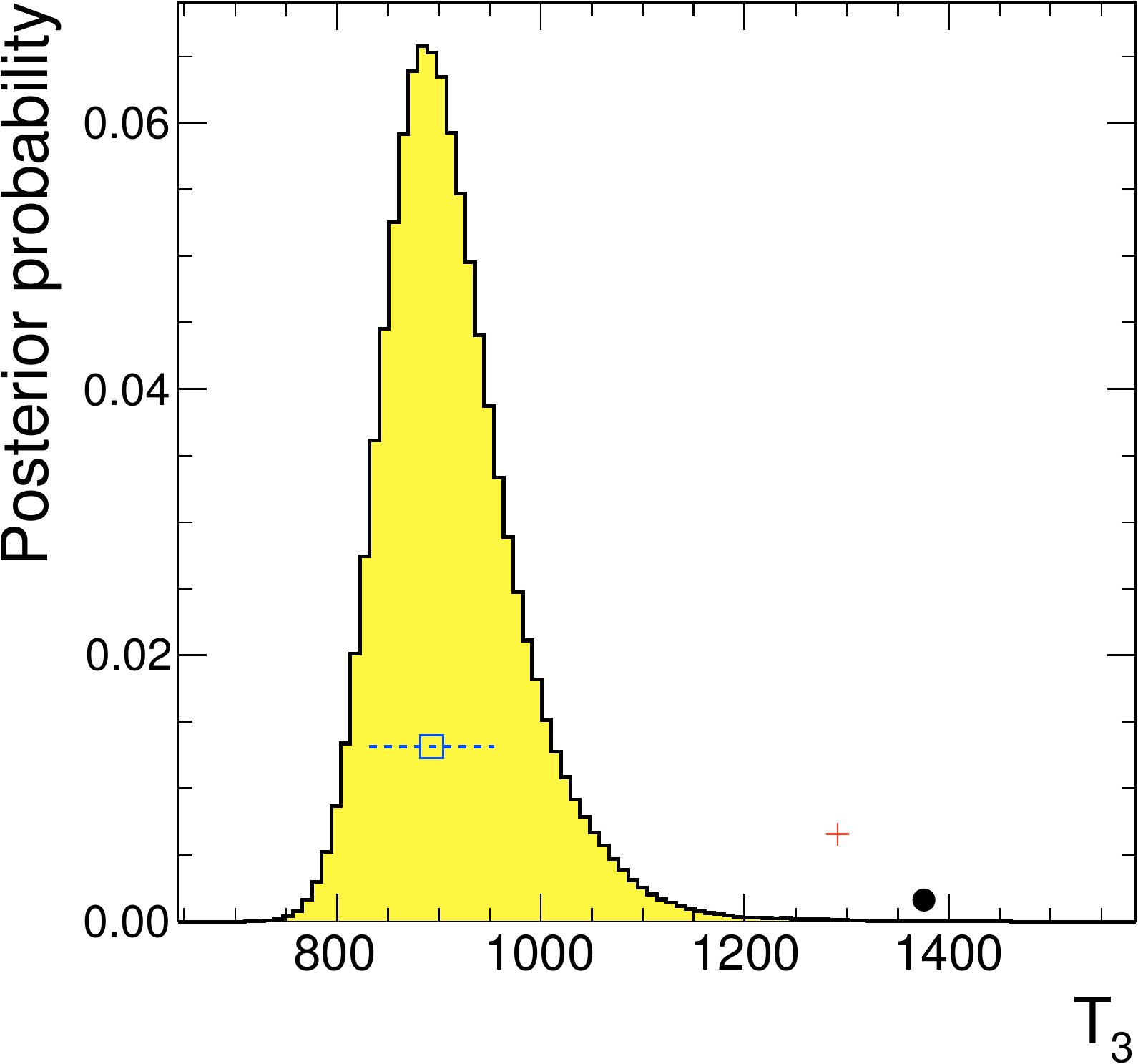} &
   \includegraphics[width=0.18\columnwidth]{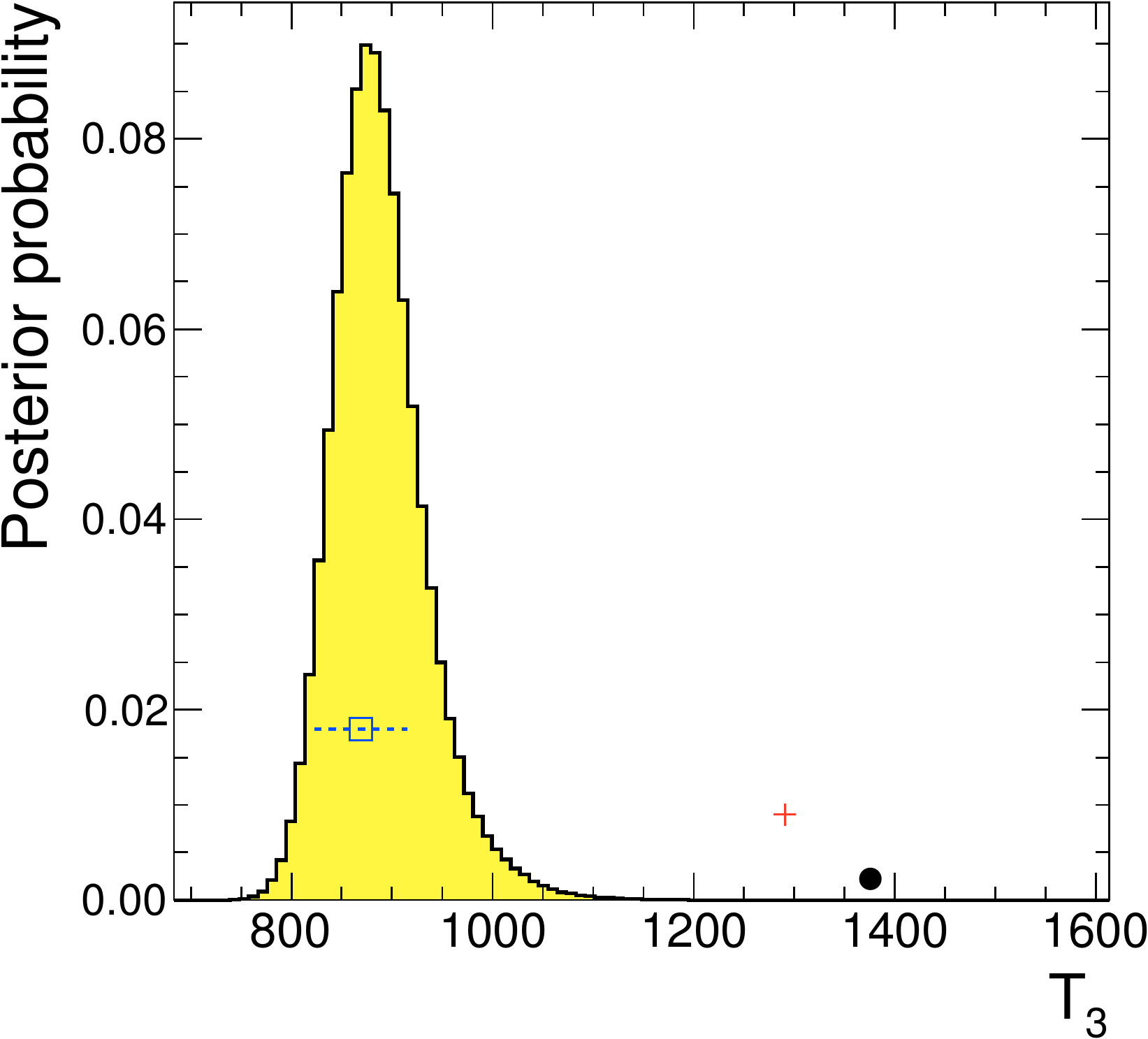} &
   \includegraphics[width=0.18\columnwidth]{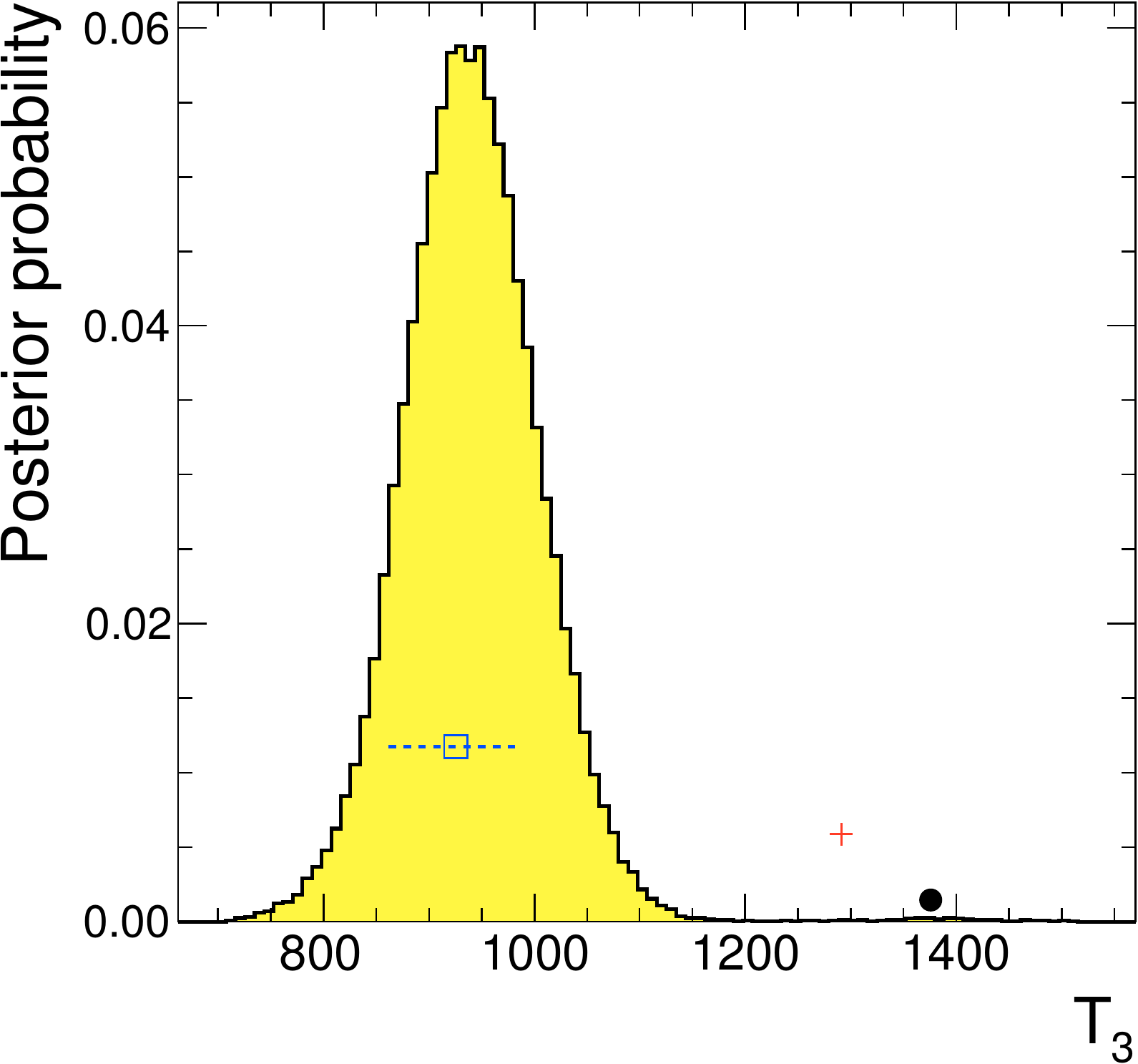} \\

   \raisebox{0.09\columnwidth}{$t=6$} &
   \includegraphics[width=0.18\columnwidth]{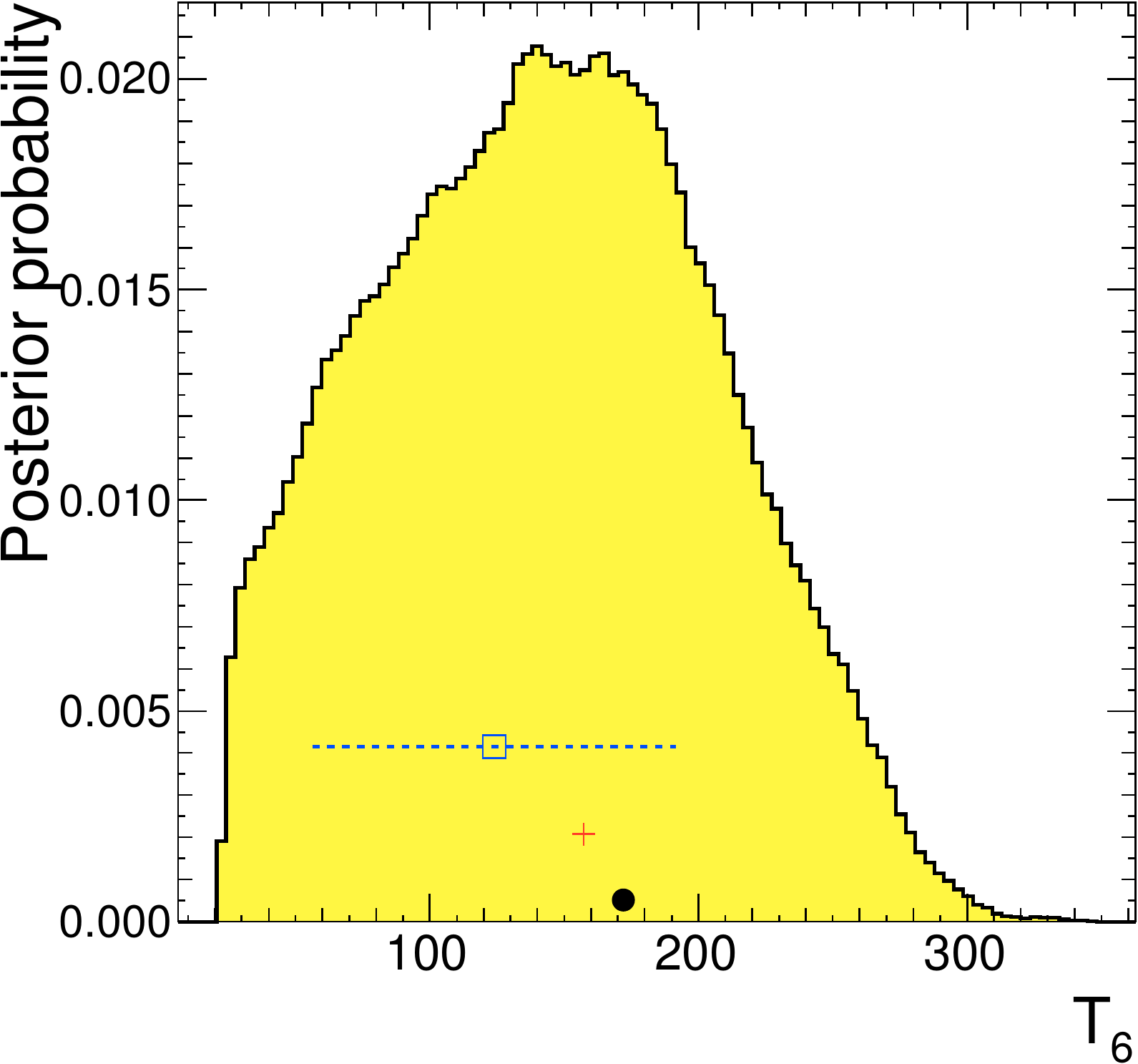} &
   \includegraphics[width=0.18\columnwidth]{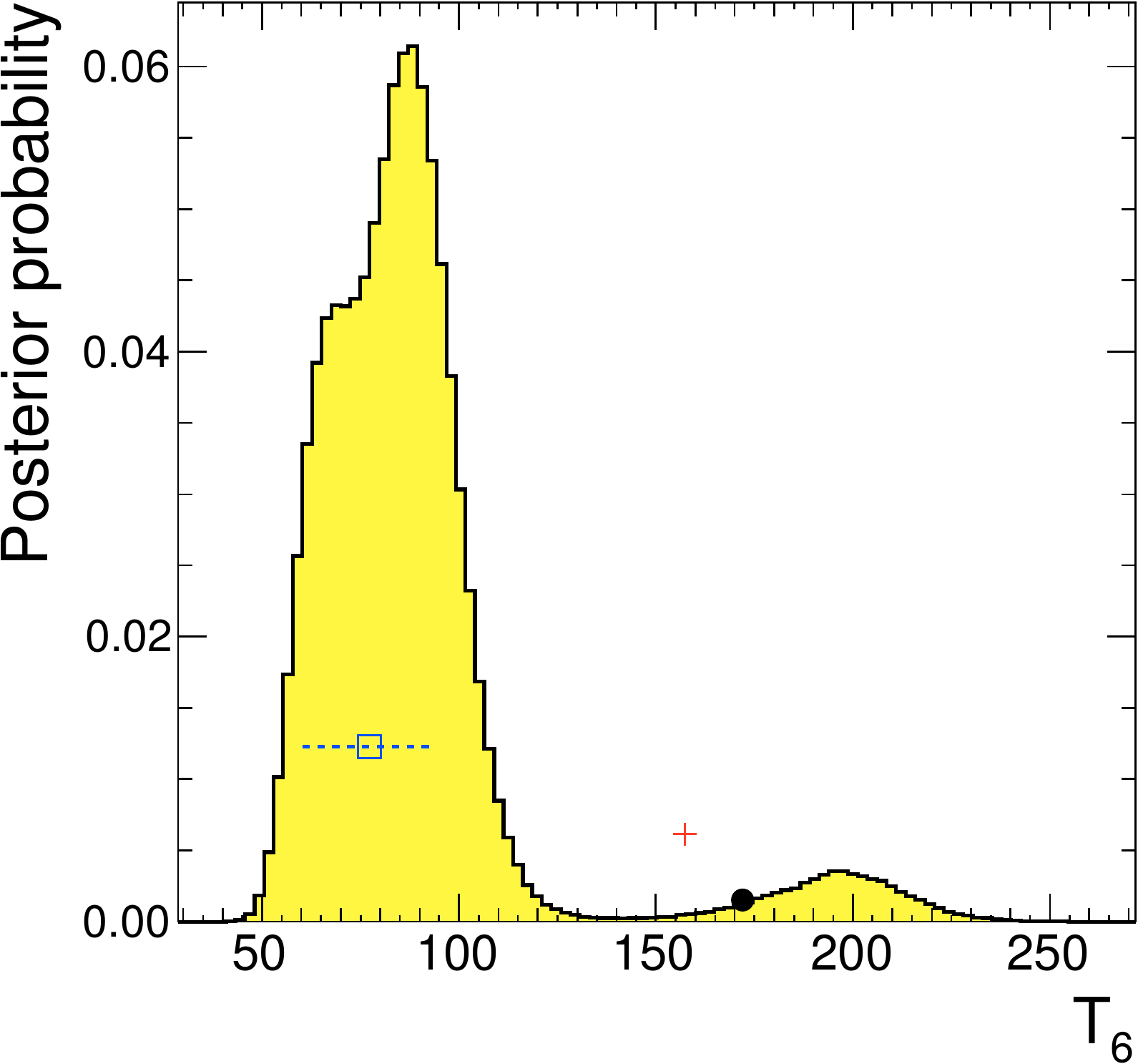} &
   \includegraphics[width=0.18\columnwidth]{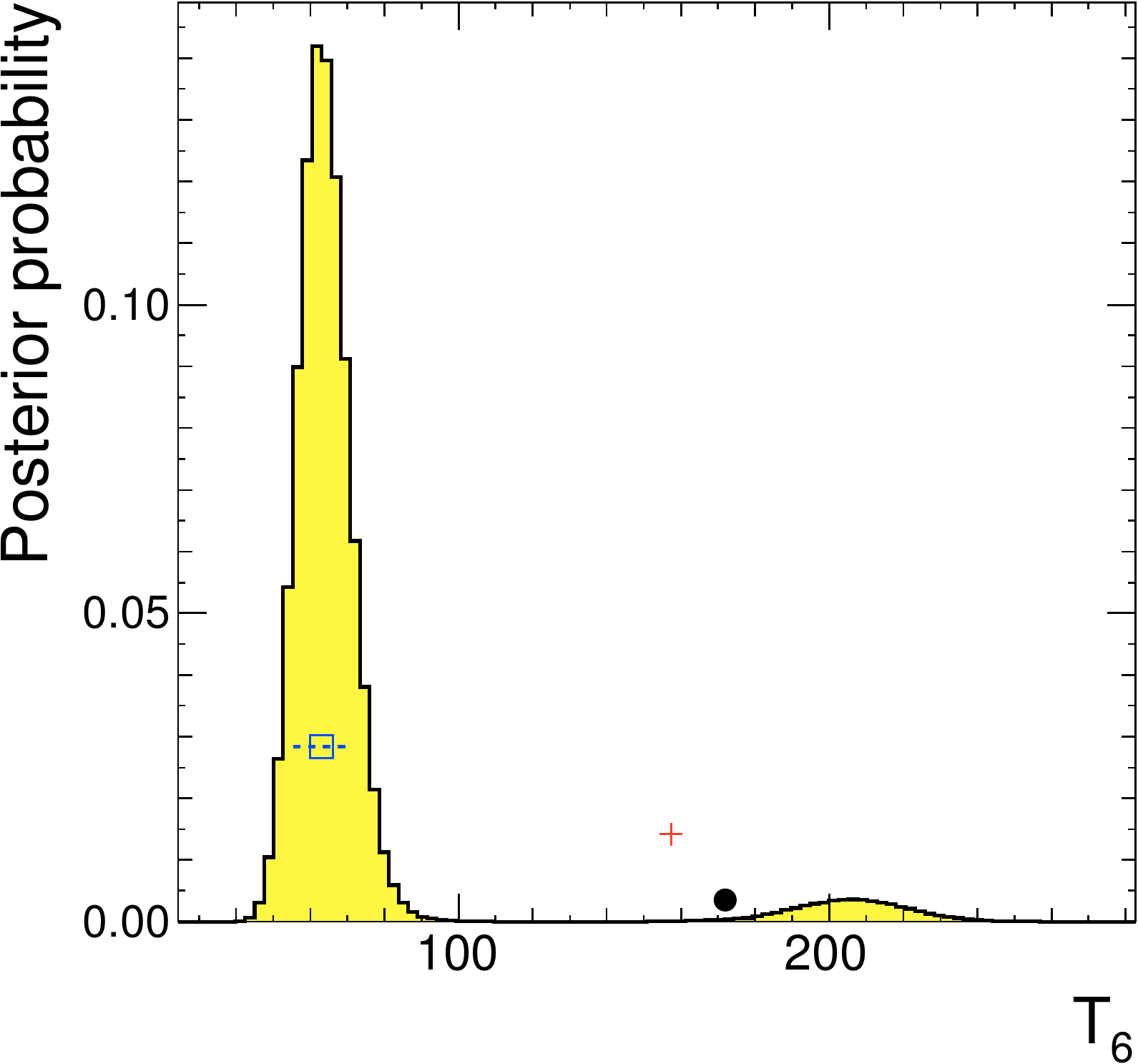} &
   \includegraphics[width=0.18\columnwidth]{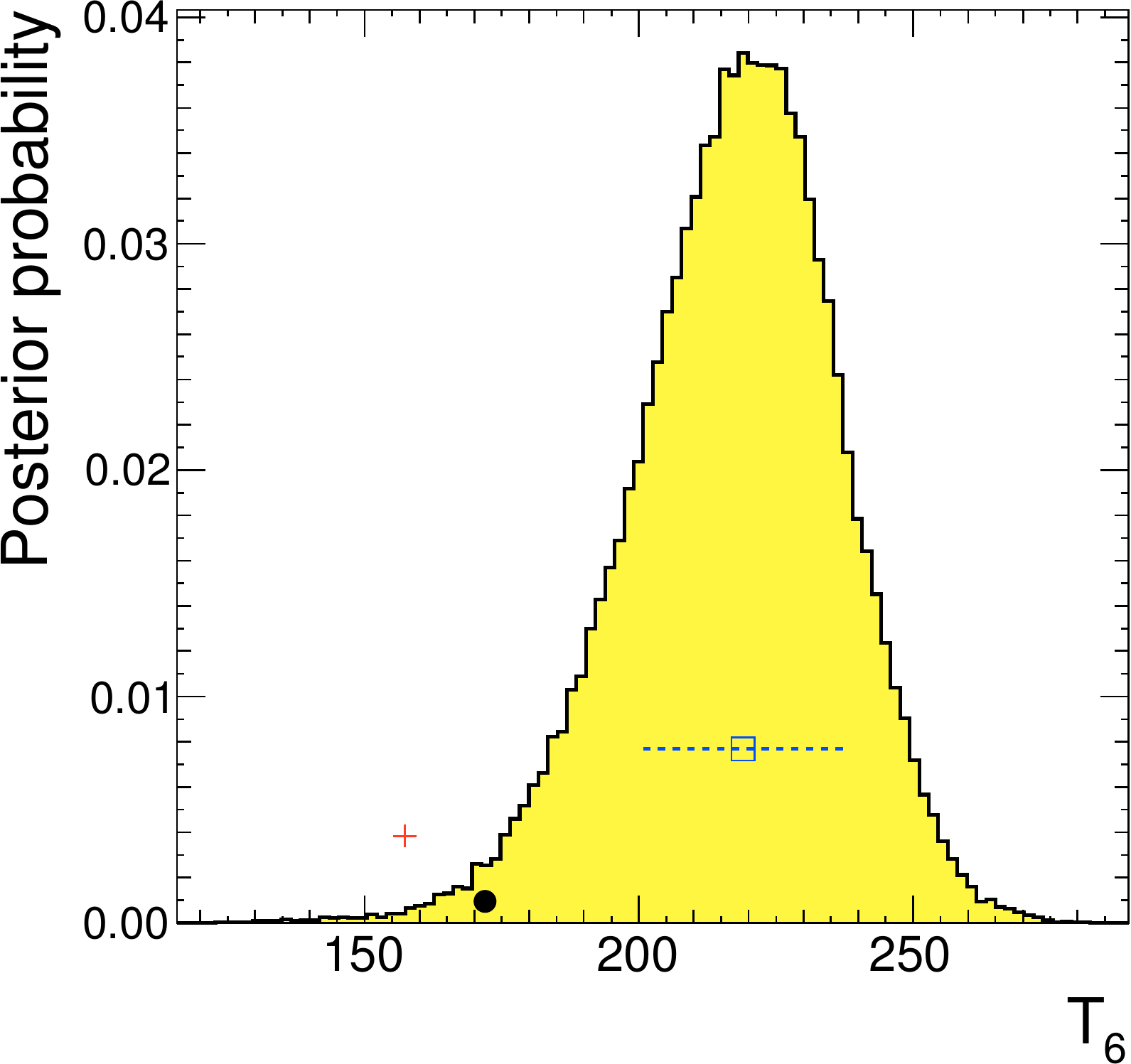} \\

   \raisebox{0.09\columnwidth}{$t=7$} &
   \includegraphics[width=0.18\columnwidth]{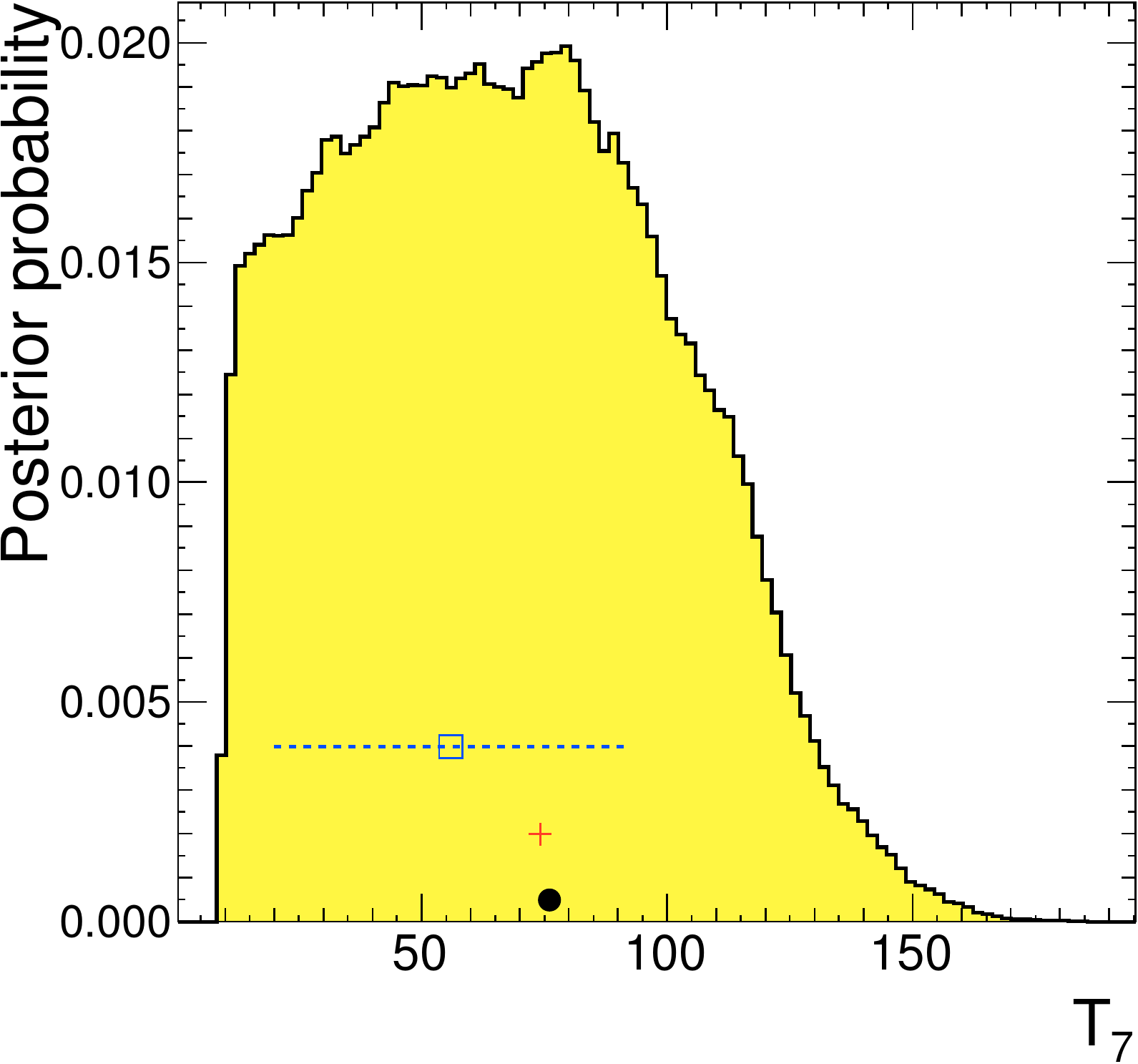} &
   \includegraphics[width=0.18\columnwidth]{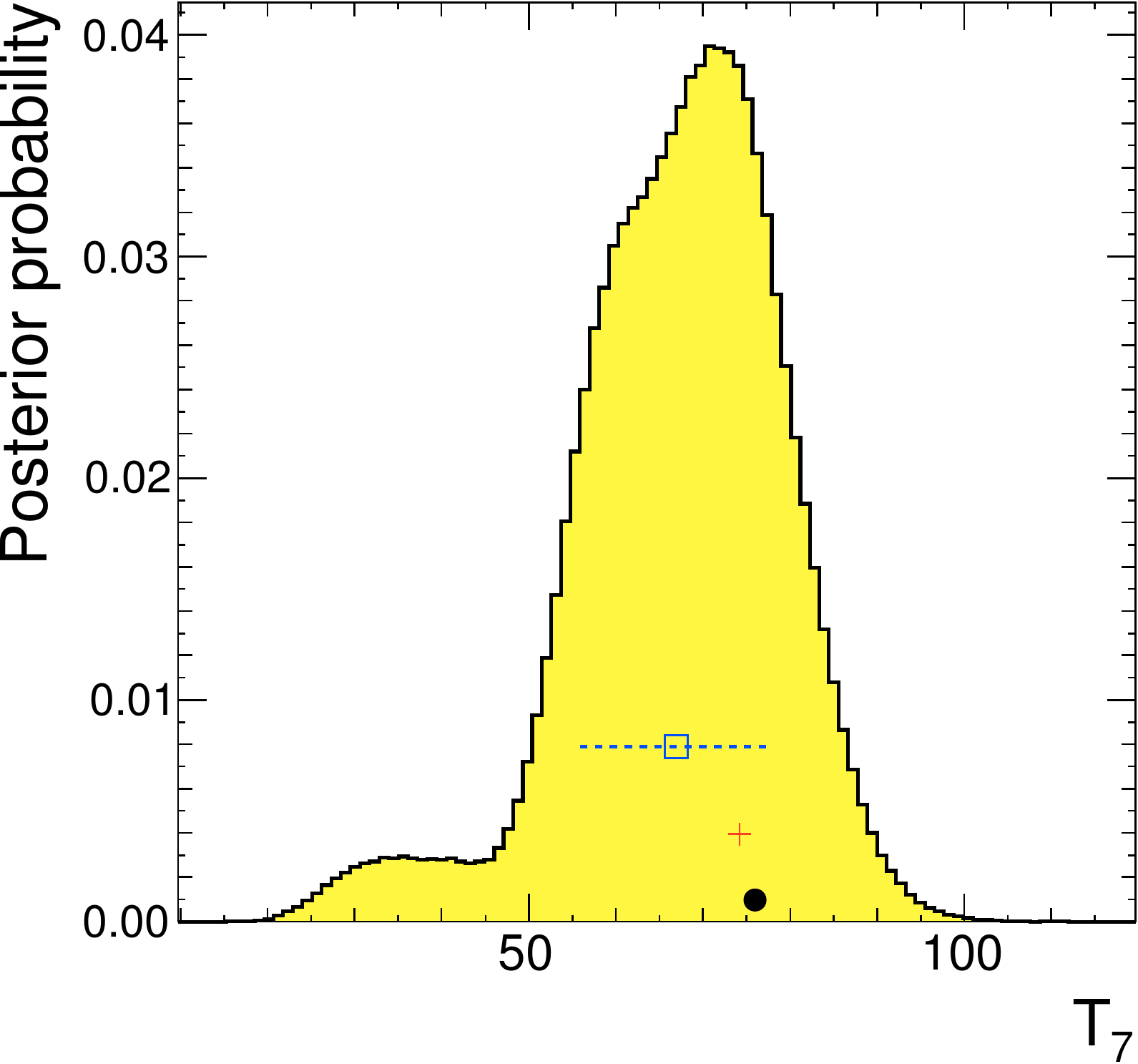} &
   \includegraphics[width=0.18\columnwidth]{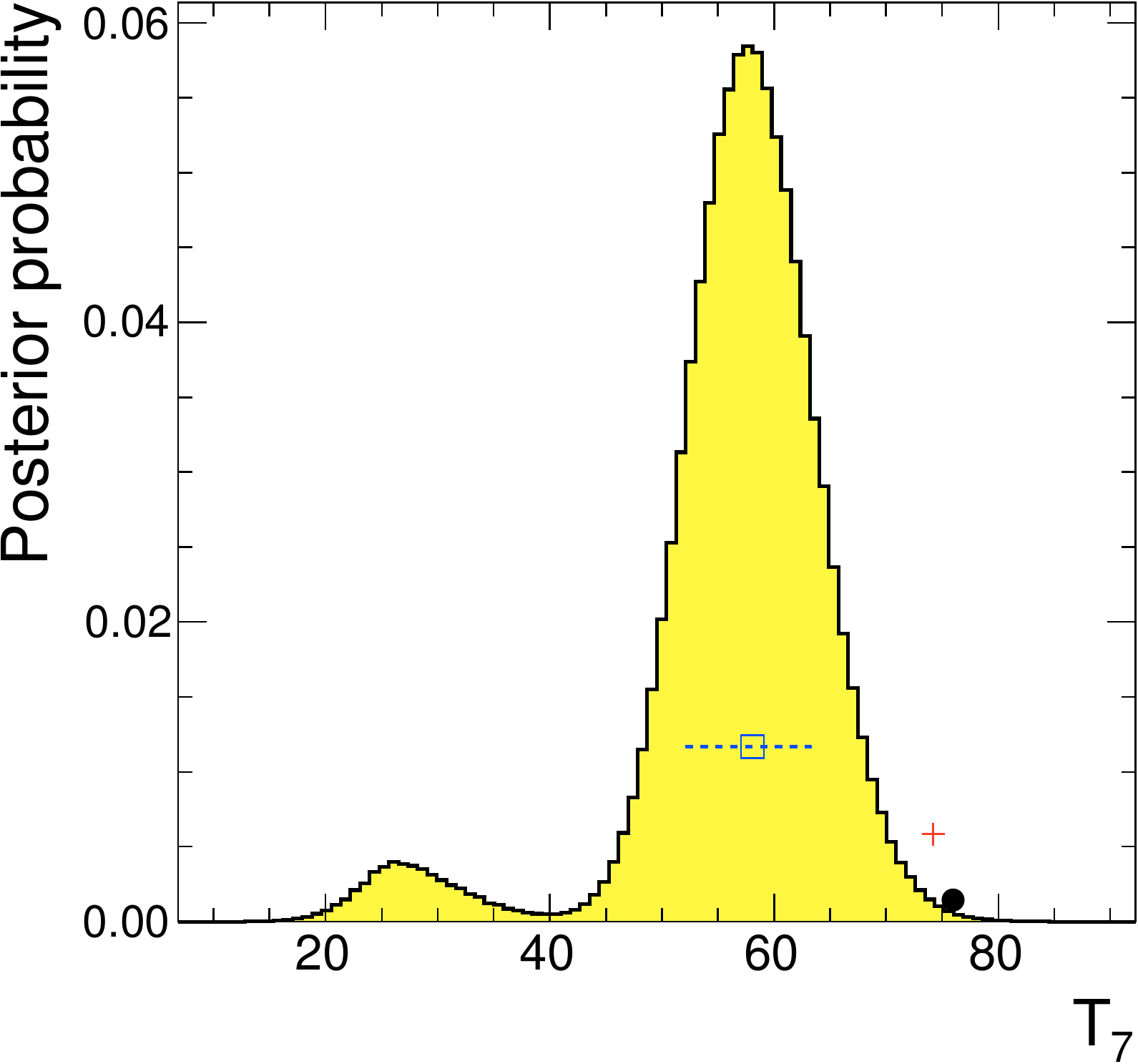} &
   \includegraphics[width=0.18\columnwidth]{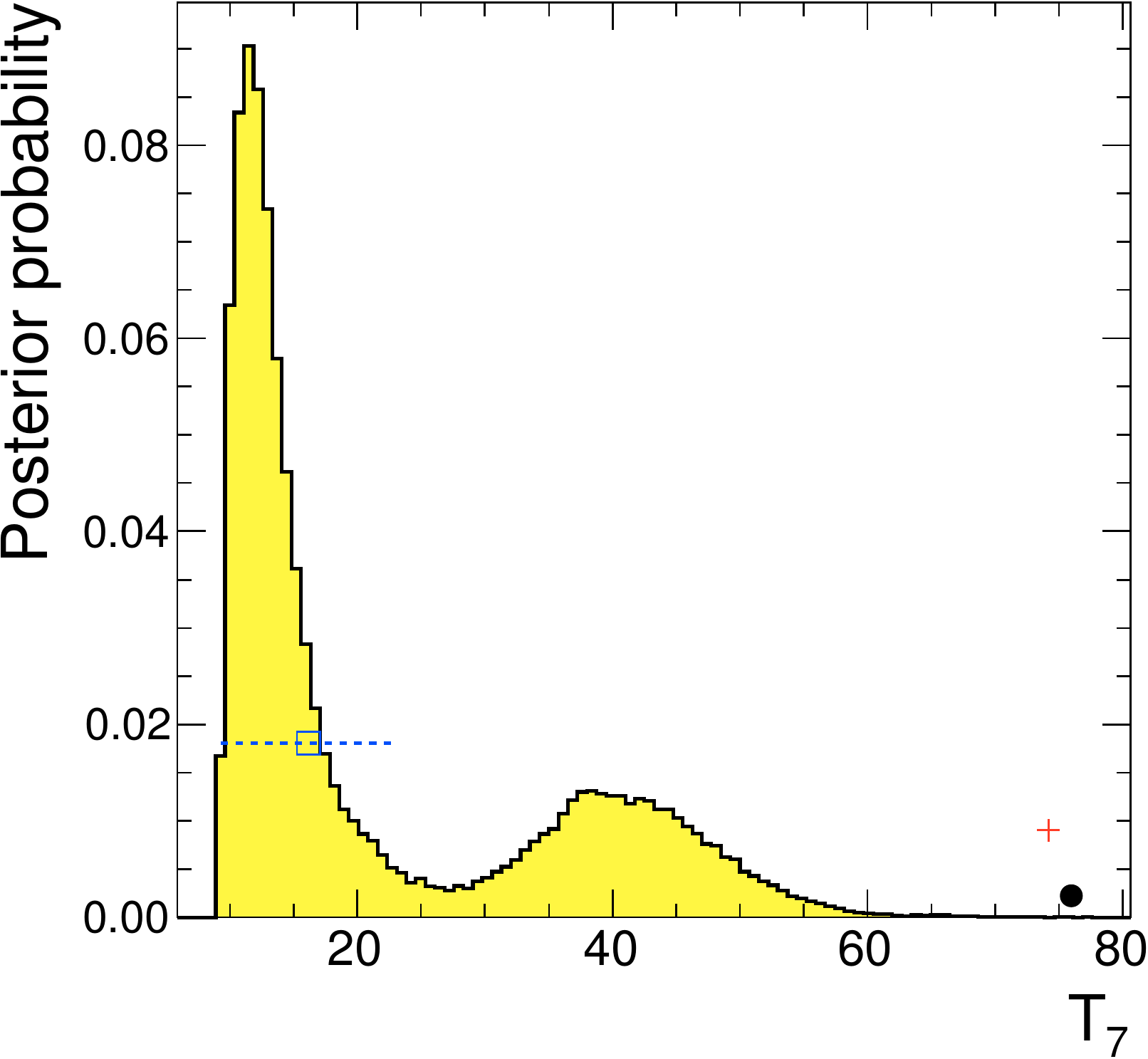} \\

   \raisebox{0.09\columnwidth}{$t=8$} &
   \includegraphics[width=0.18\columnwidth]{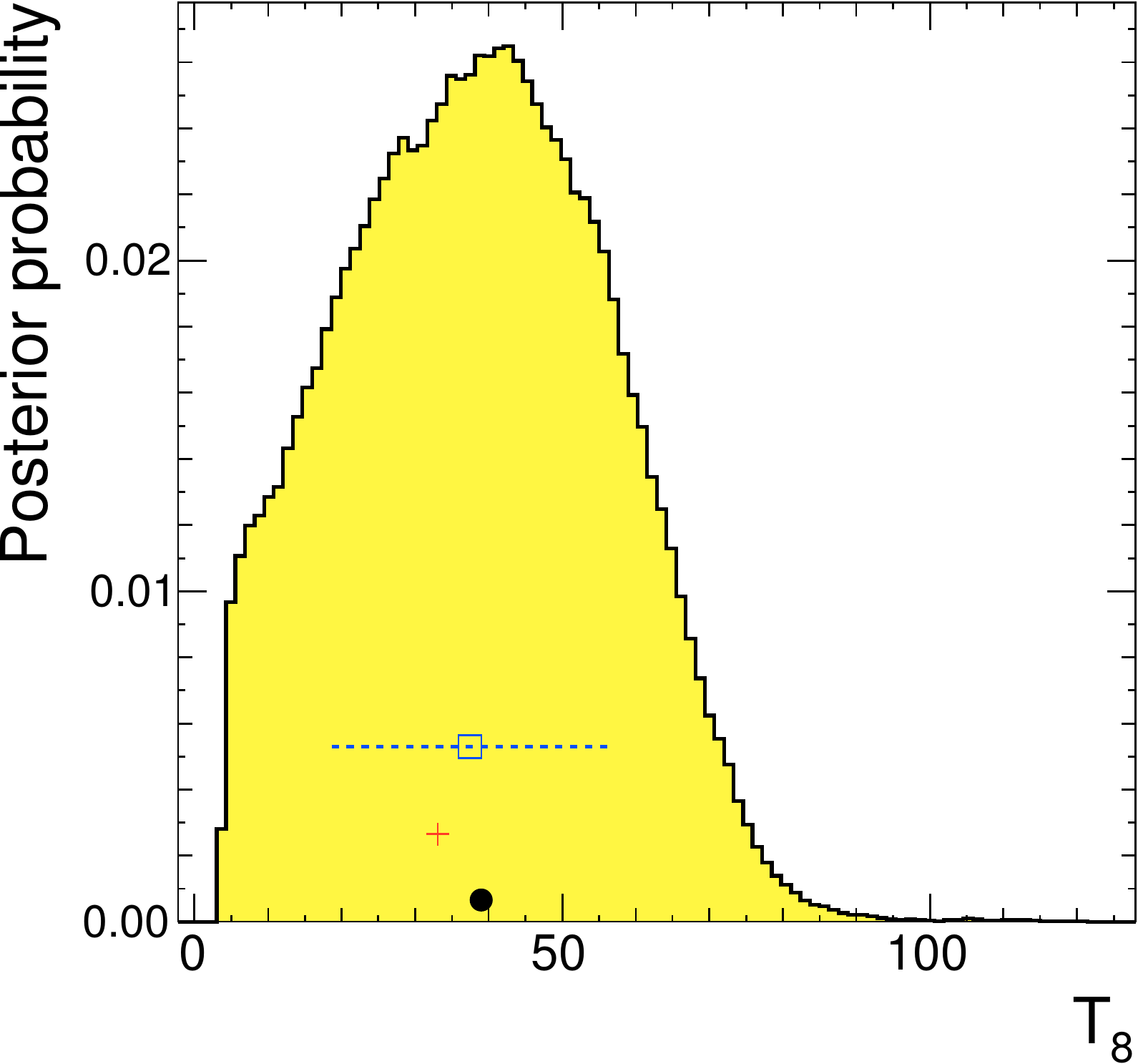} &
   \includegraphics[width=0.18\columnwidth]{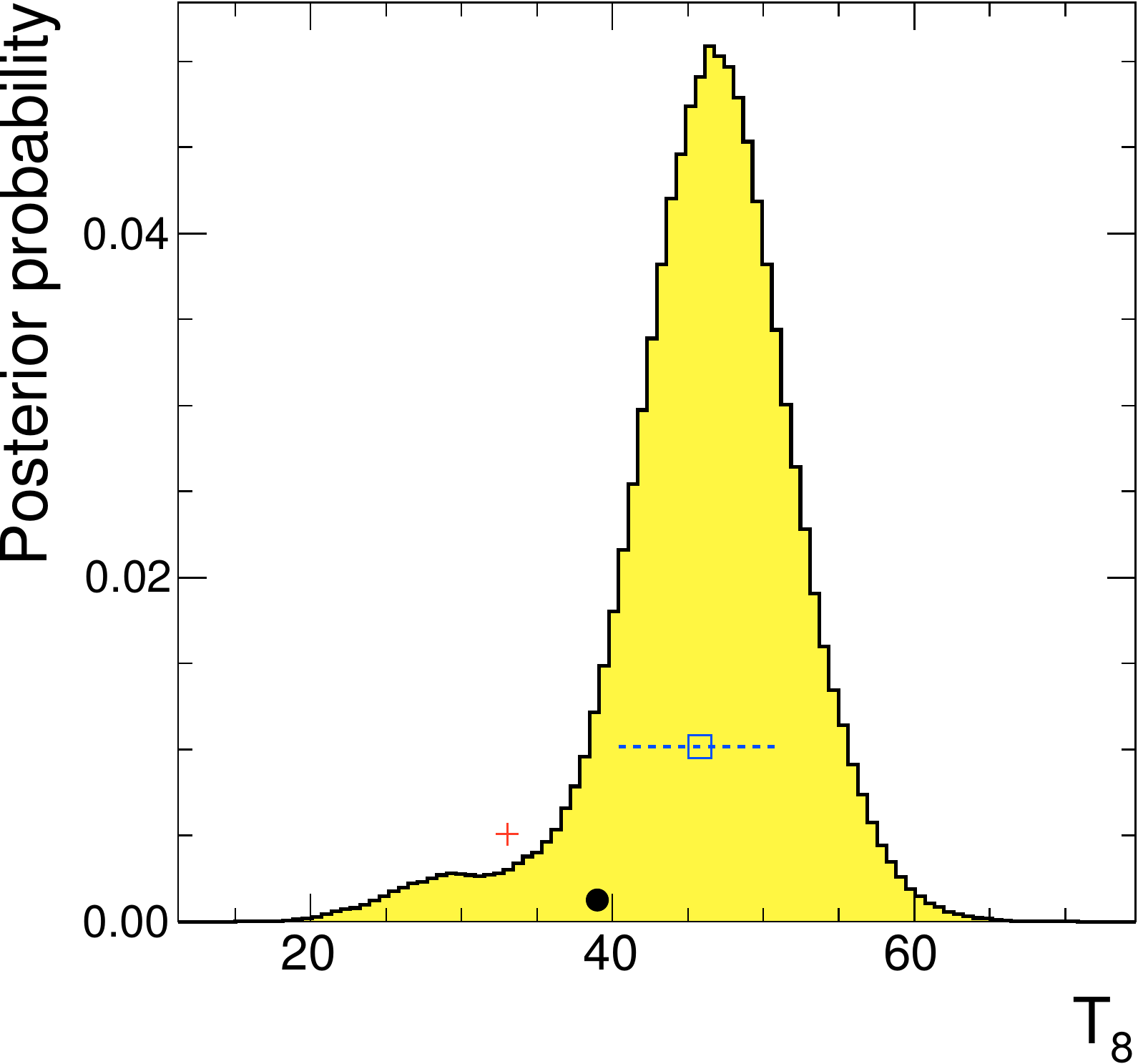} &
   \includegraphics[width=0.18\columnwidth]{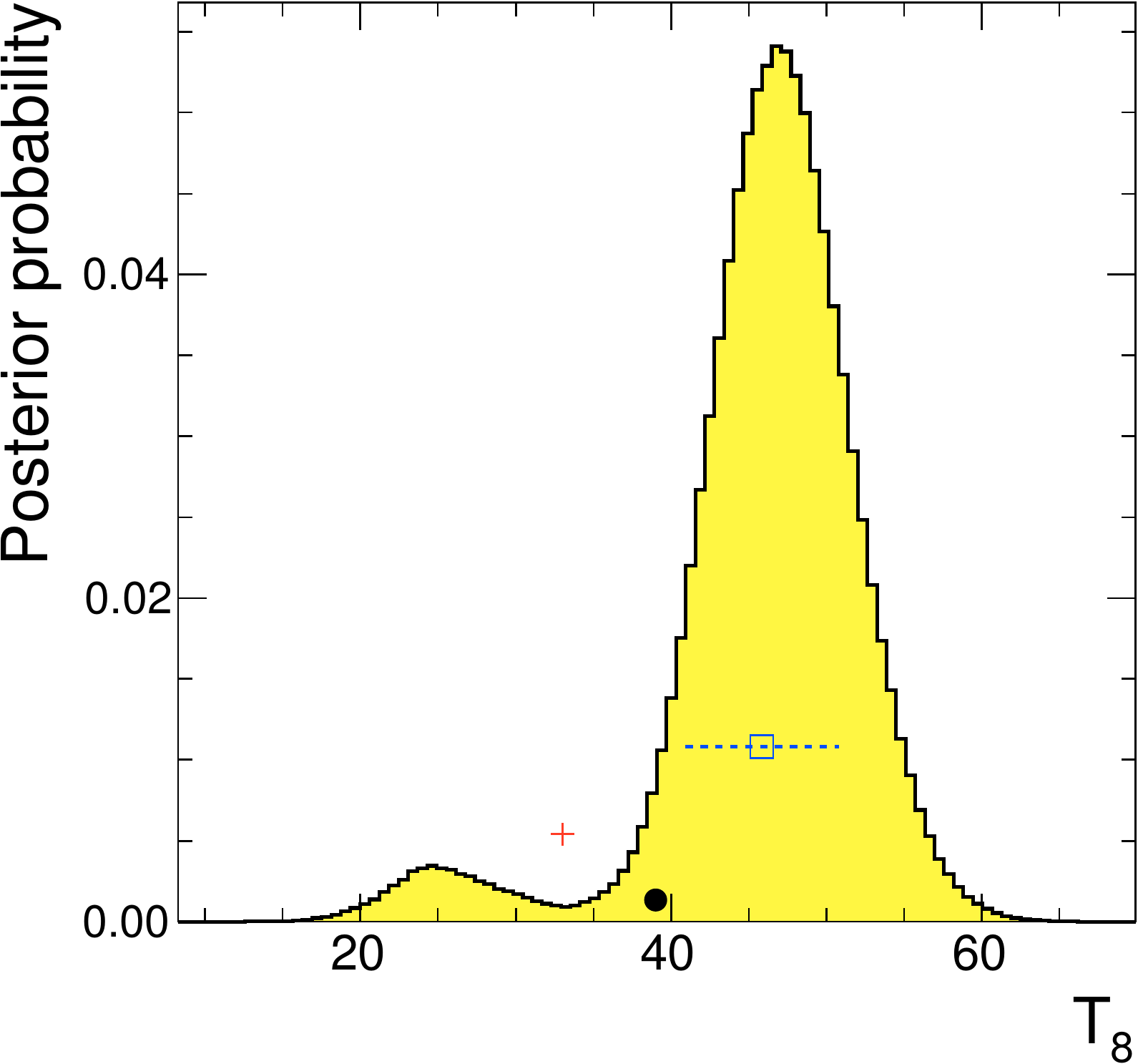} &
   \includegraphics[width=0.18\columnwidth]{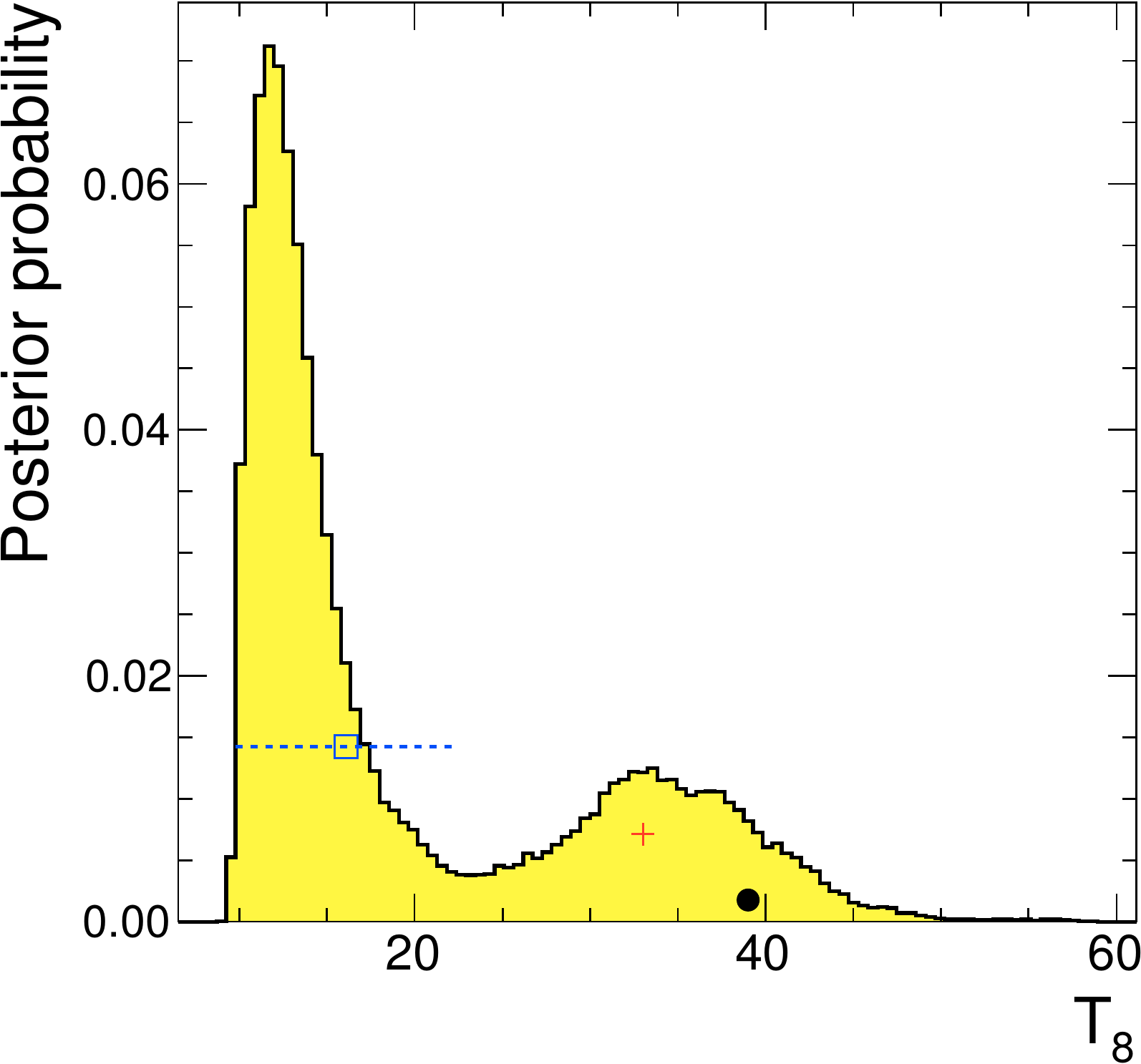} \\

   \raisebox{0.09\columnwidth}{$t=11$} &
   \includegraphics[width=0.18\columnwidth]{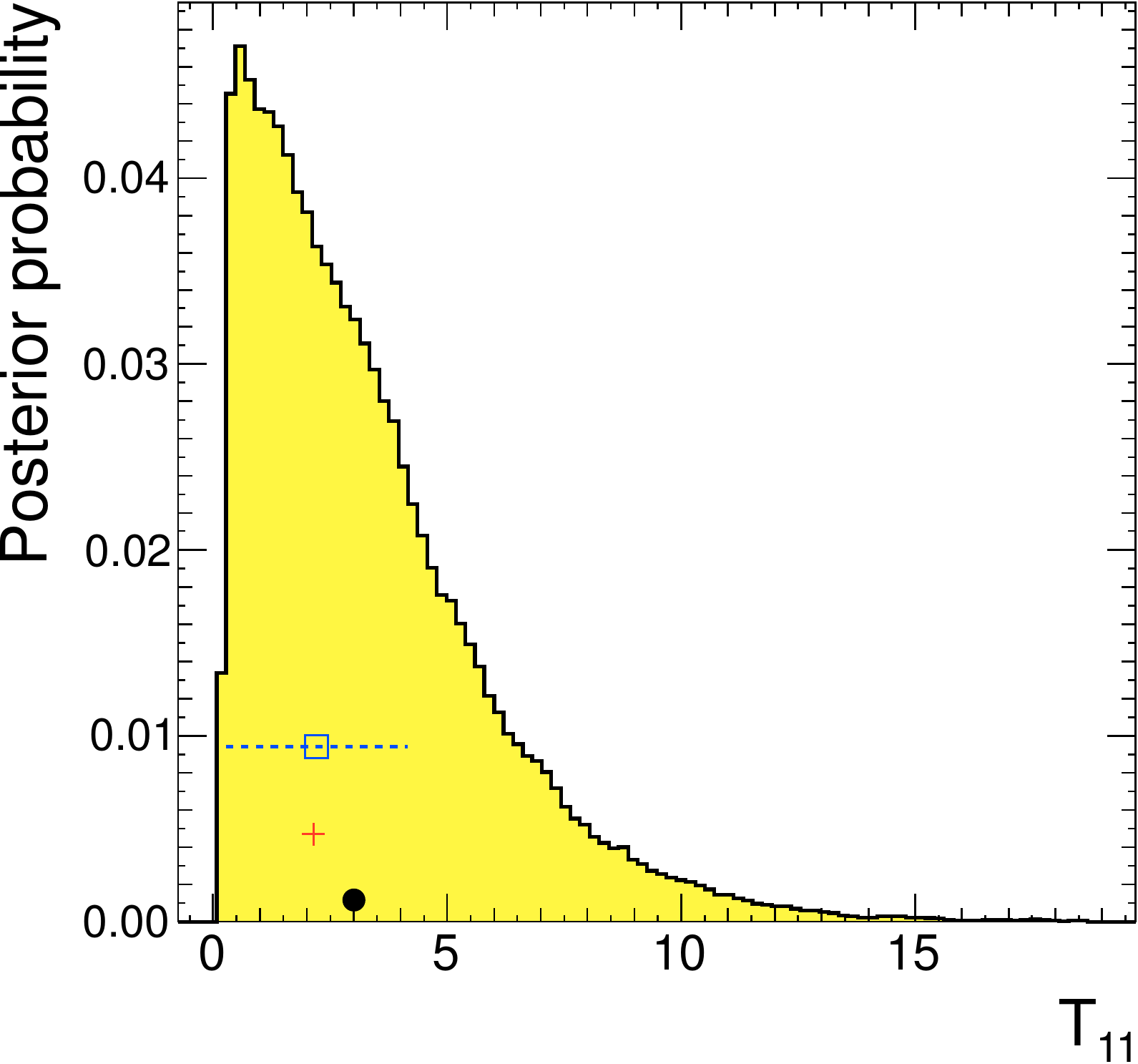} &
   \includegraphics[width=0.18\columnwidth]{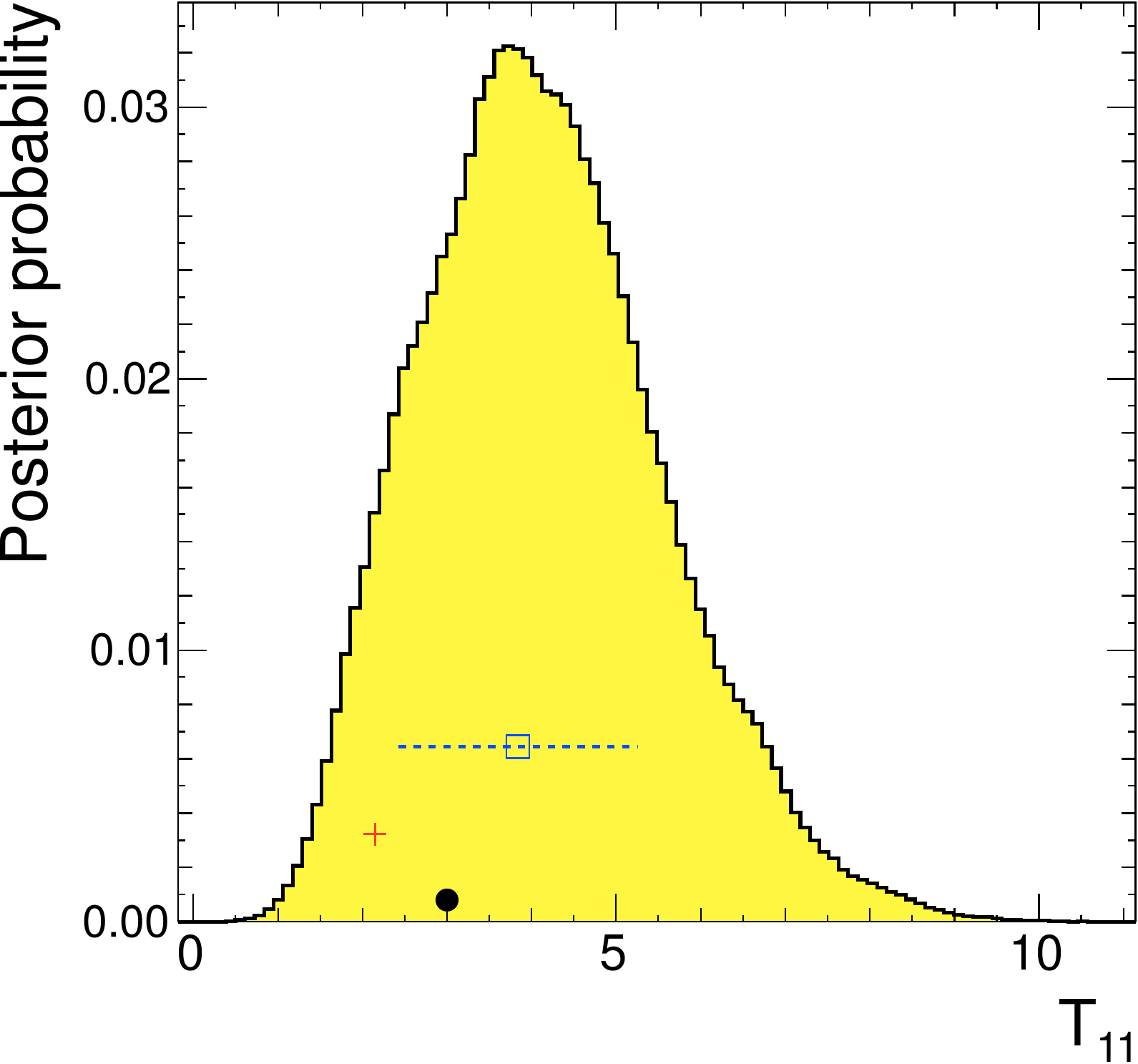} &
   \includegraphics[width=0.18\columnwidth]{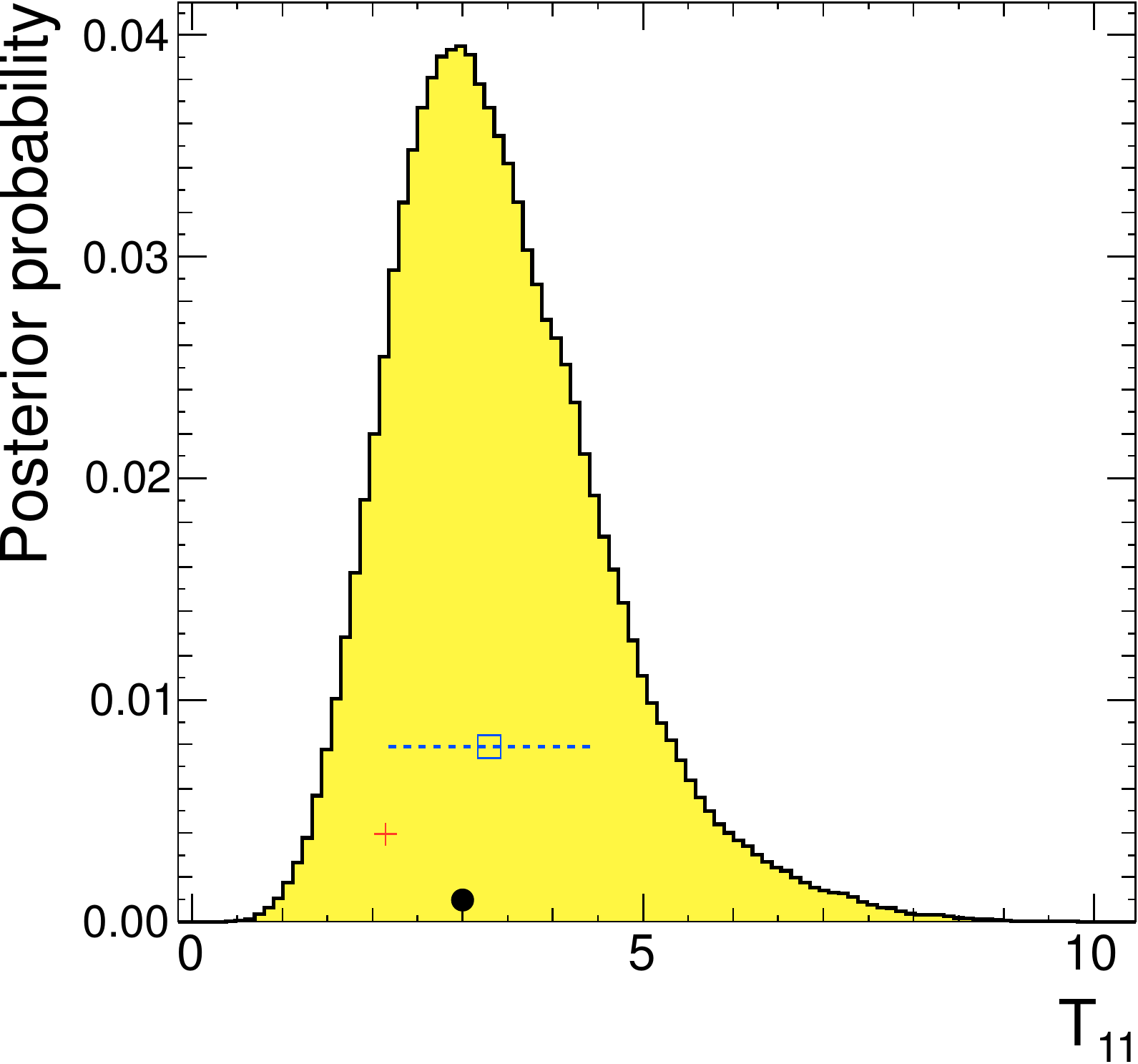} &
   \includegraphics[width=0.18\columnwidth]{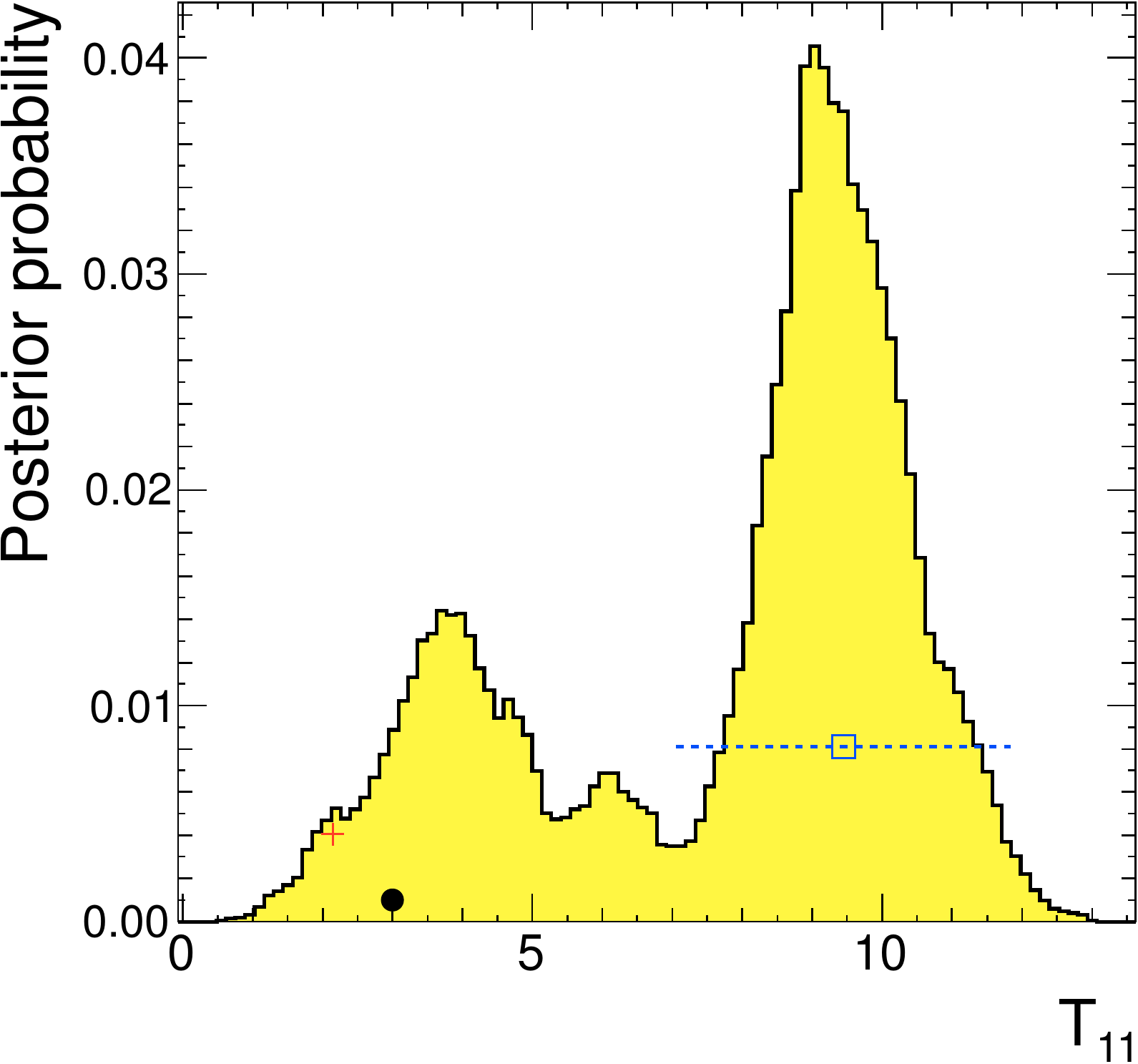} \\

   {}&
   $\alpha = 0$&
   $\alpha = 20$&
   $\alpha = 30$&
   $\alpha = 40$\\

  \end{tabular}
\caption{The 1-dimensional distributions $P_t(T_t|\D)$, for $t=\{1,3,6,7,8,11\}$, computed in Sec.~\ref{sec:regSteepSmearing}, with regularization function $S_3$, and for $\alpha = \{ 0 , 20 , 30 , 40 \}$.
\label{fig:1DimSteepSmearS3}
}
\end{figure}

\begin{figure}[H]
  \centering
  \subfigure[$\alpha=0$]{
    \includegraphics[width=0.3\columnwidth]{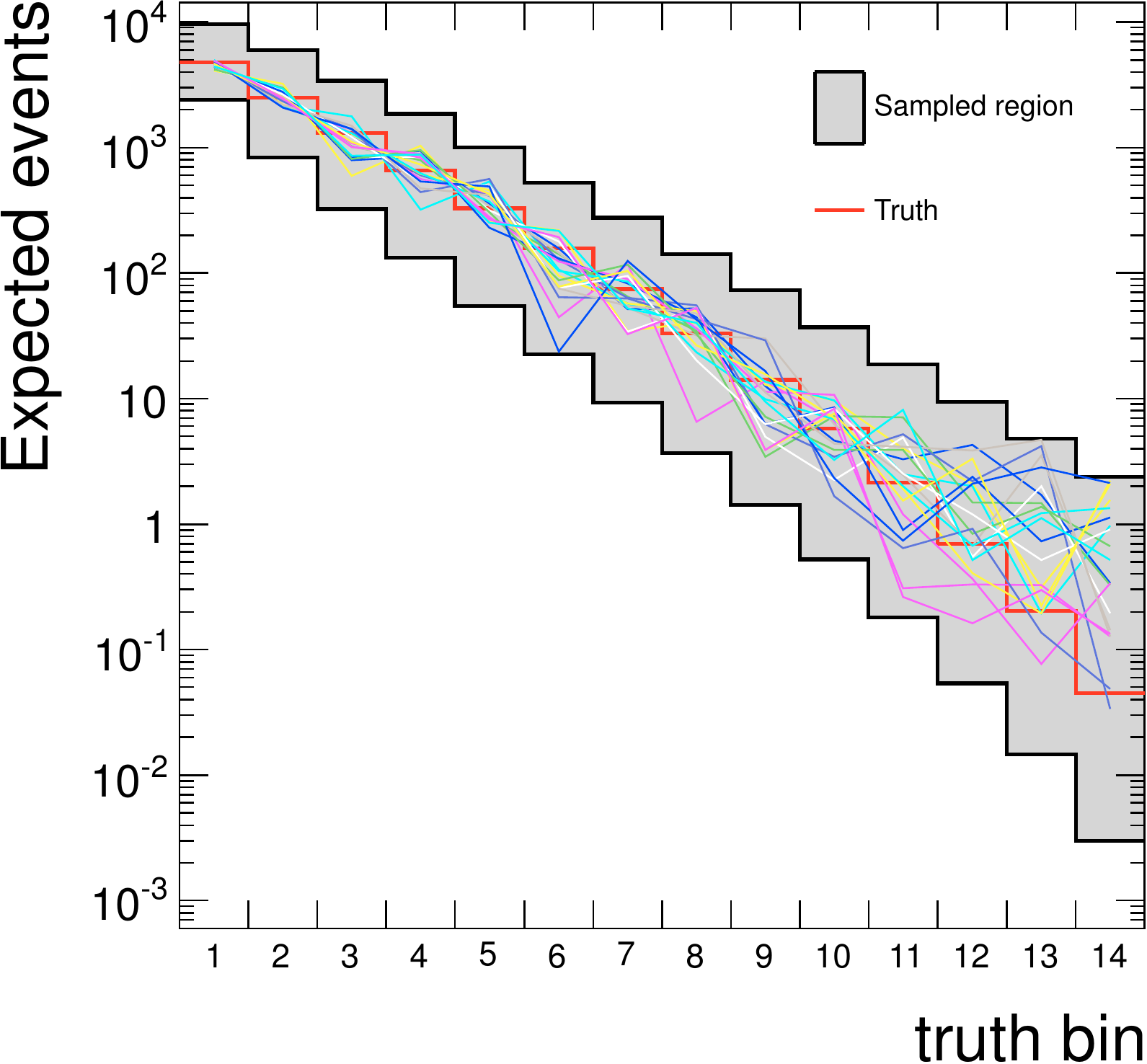}
  }
  \subfigure[$\alpha=20$]{
    \includegraphics[width=0.3\columnwidth]{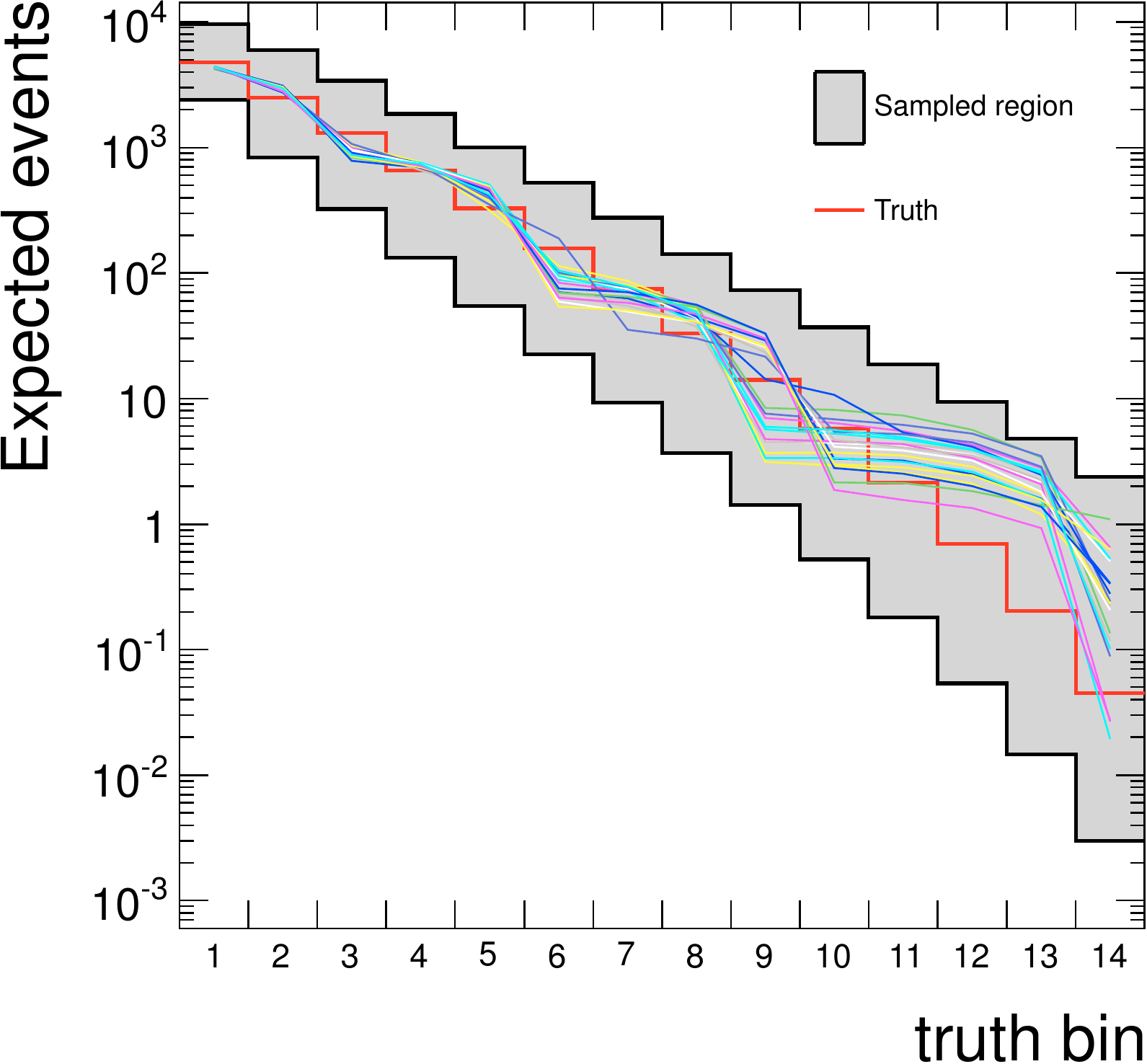}
  }
  \subfigure[$\alpha=40$]{
    \includegraphics[width=0.3\columnwidth]{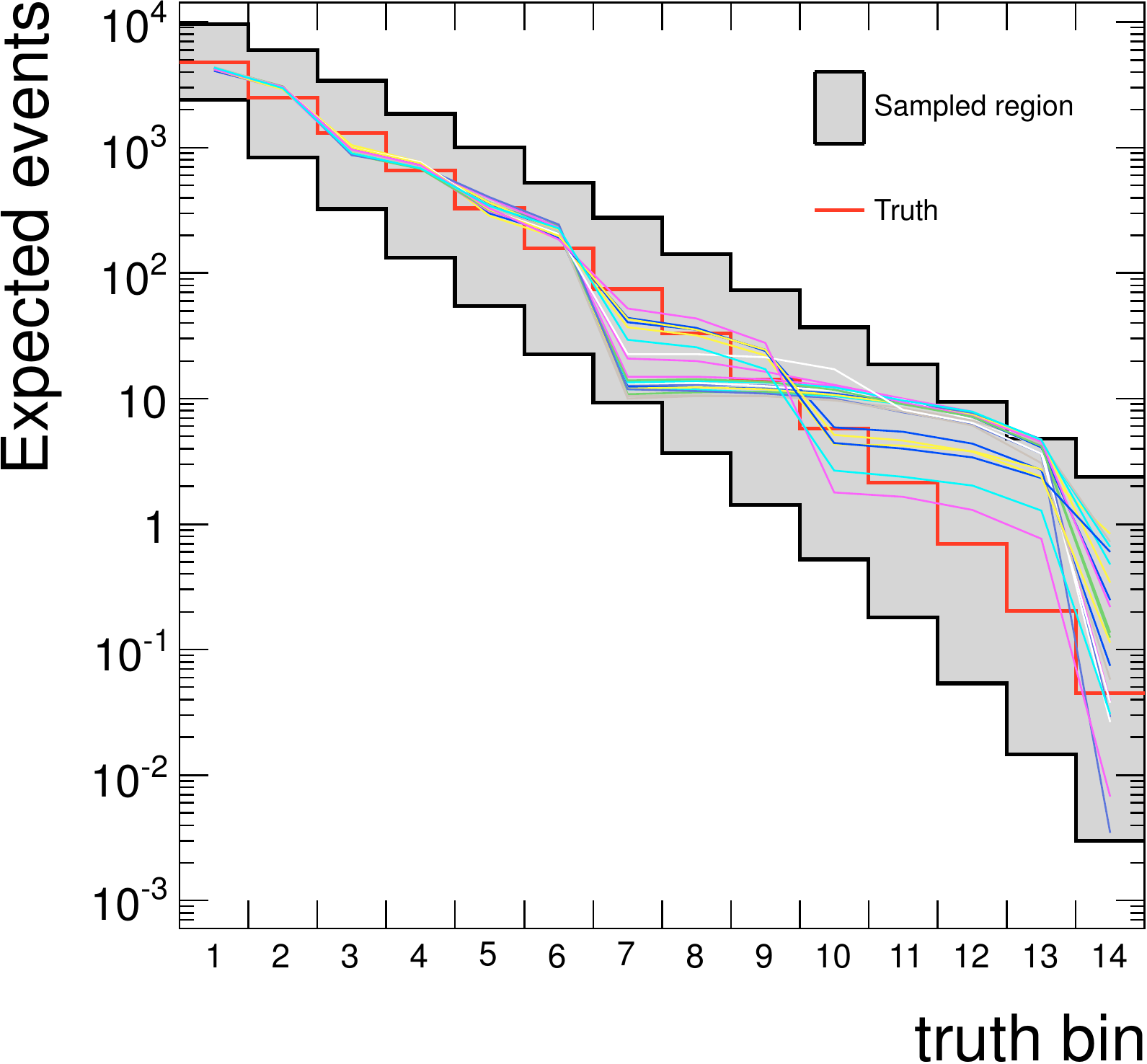}
  }
\caption{From the posteriors computed in Sec.~\ref{sec:regSteepSmearing}, using regularization with $S_3$ and three values of $\alpha$, twenty random $\T$-points are sampled following each posterior, and are overlaid (colored lines) with the sampled hyper-box (gray region) and the actual truth-level spectrum (red histogram).  In (a) the sampled $\T$-points are not regularized, in (b) they are more constant in intervals of bins, and in (c) they tend to be even flatter, and at least two families of spectra seem to be favored, which have their flat regions in different ranges of bins.
\label{fig:pointsSteepSmearS3}
}
\end{figure}

The results with $S_3$ are shown in Fig.~\ref{fig:regFuncSteepSmearS3}  and Fig.~\ref{fig:unfoldedSteepSmearS3}.
It is clear that the tendency is to make the unfolded spectrum more constant.  An interesting effect is observed in Fig.~\ref{fig:regFuncSteepSmearS3c}, where the regularization condition is stronger ($\alpha=40$).  The posterior $p(\T|\D)$ favors two (at least) categories of $\T$-points, some with $\log_{10}S_3(\T)\simeq 5.5$ and some with $\log_{10}S_3(\T) \simeq 4.5$.  This is suggestive that $p(\T|\D)$ has secondary local maxima in its 1-dimensional marginal distributions, which is indeed confirmed in Fig.~\ref{fig:1DimSteepSmearS3}.  The secondary maximum corresponds to $\T$ spectra that have plateaus in different groups of bins, as shown in Fig.~\ref{fig:pointsSteepSmearS3}.

\begin{figure}[H]
  \centering
  \subfigure[$\alpha=0$]{
    \includegraphics[width=0.3\columnwidth]{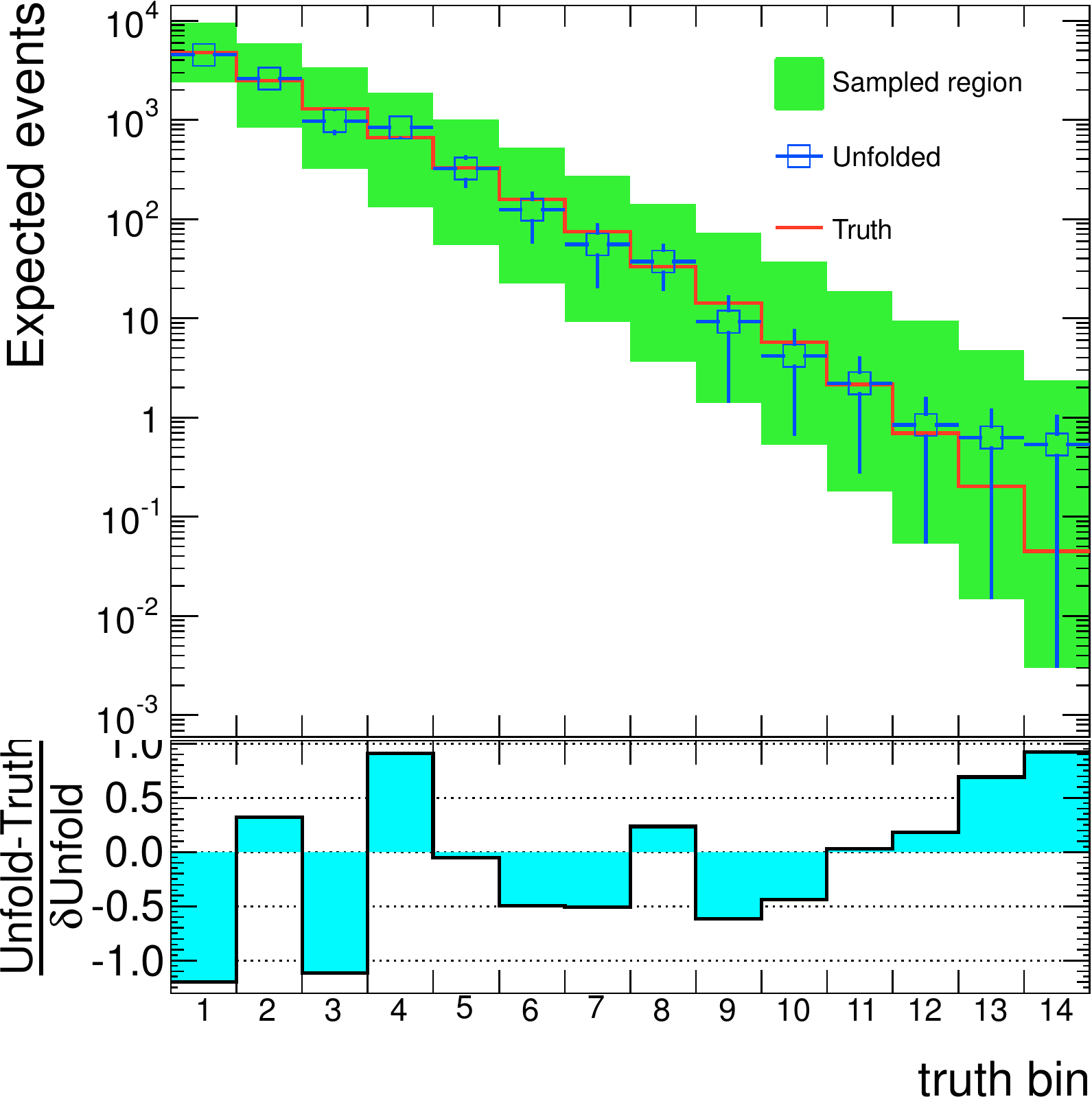}
  }
  \subfigure[$\alpha=1$]{
    \includegraphics[width=0.3\columnwidth]{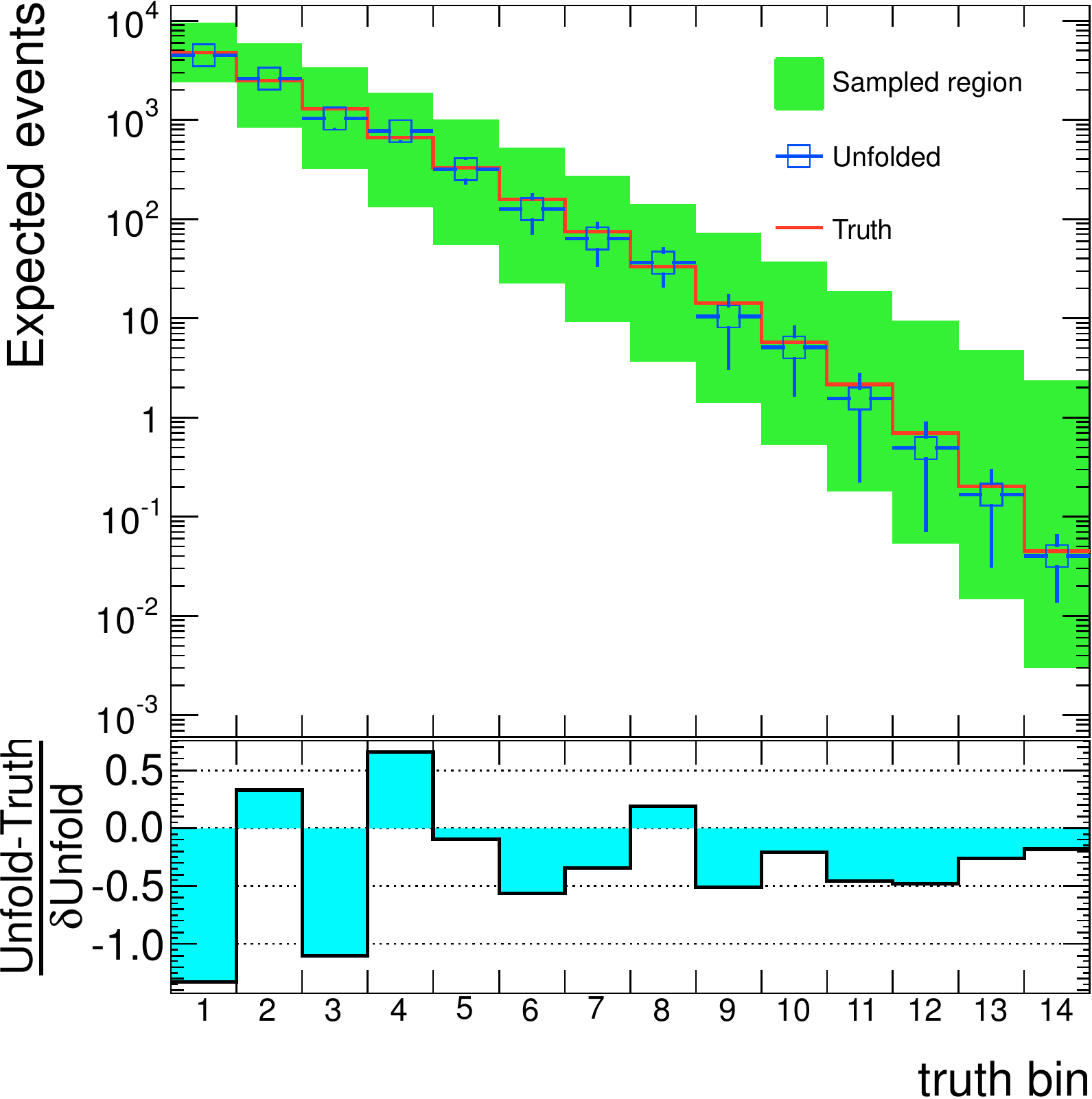}
  }
  \subfigure[$\alpha=10$]{
    \includegraphics[width=0.3\columnwidth]{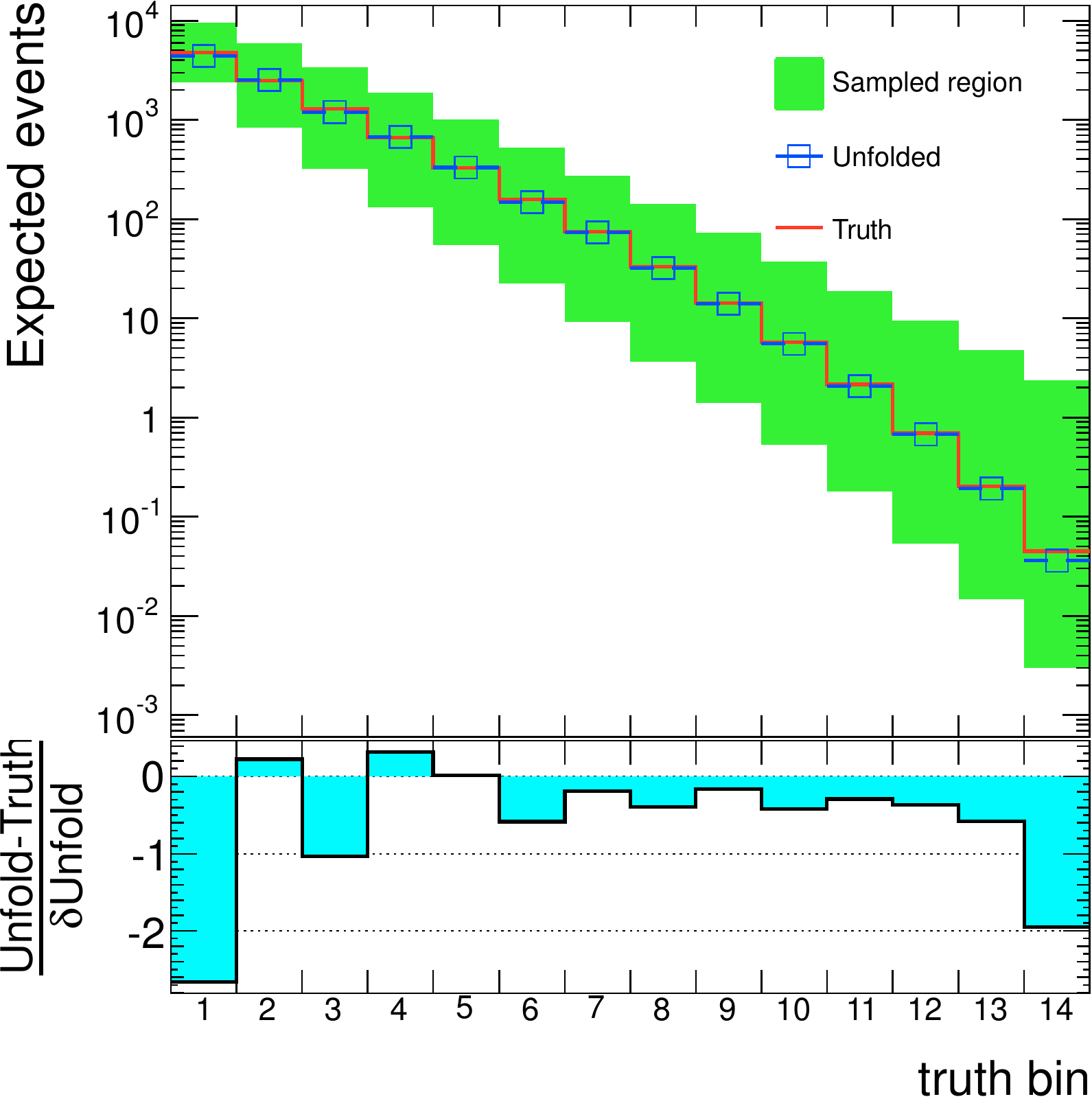}
  }\\
 \subfigure[$\alpha=0$]{
    \includegraphics[width=0.3\columnwidth]{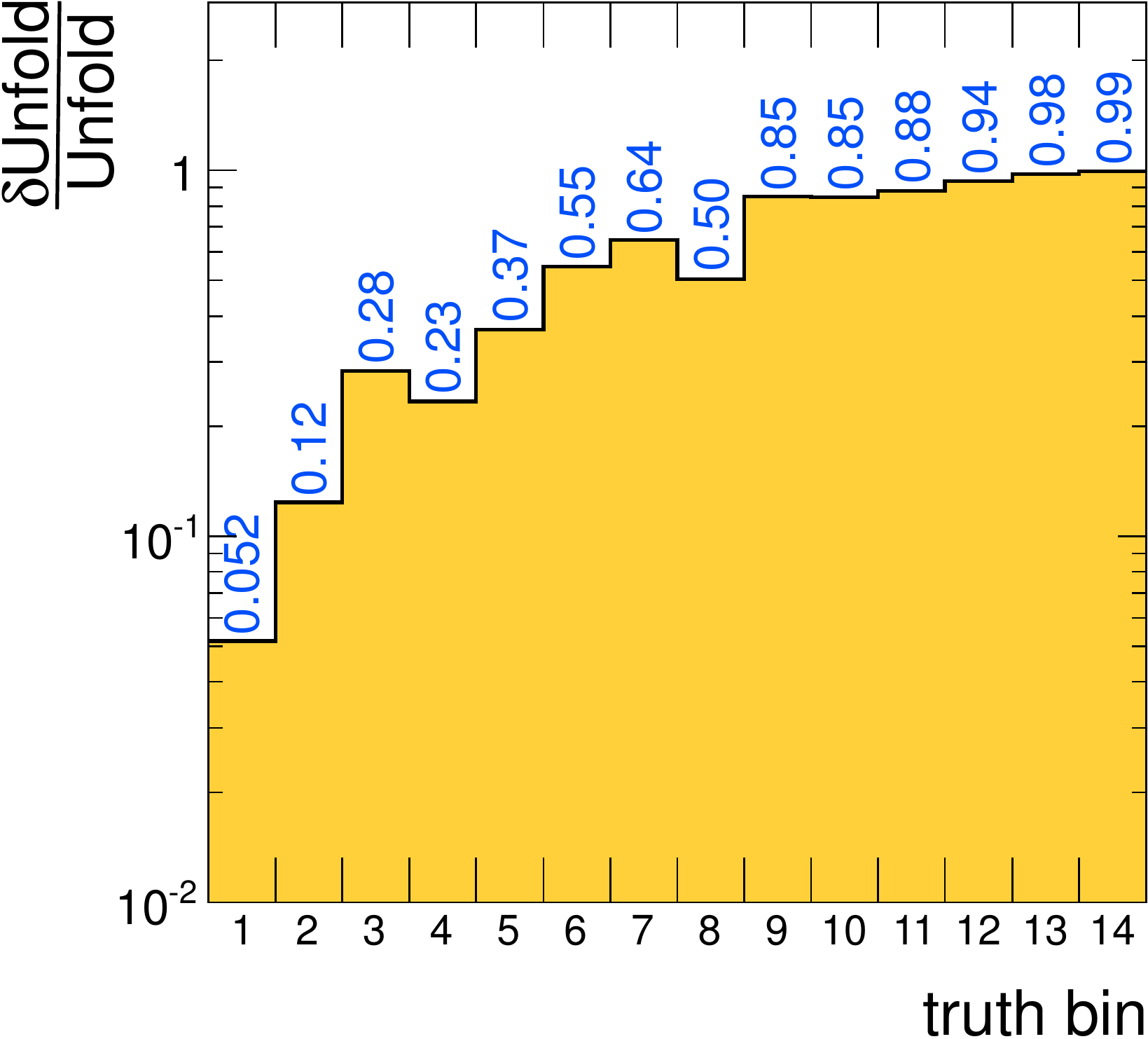}
  }
  \subfigure[$\alpha=1$]{
    \includegraphics[width=0.3\columnwidth]{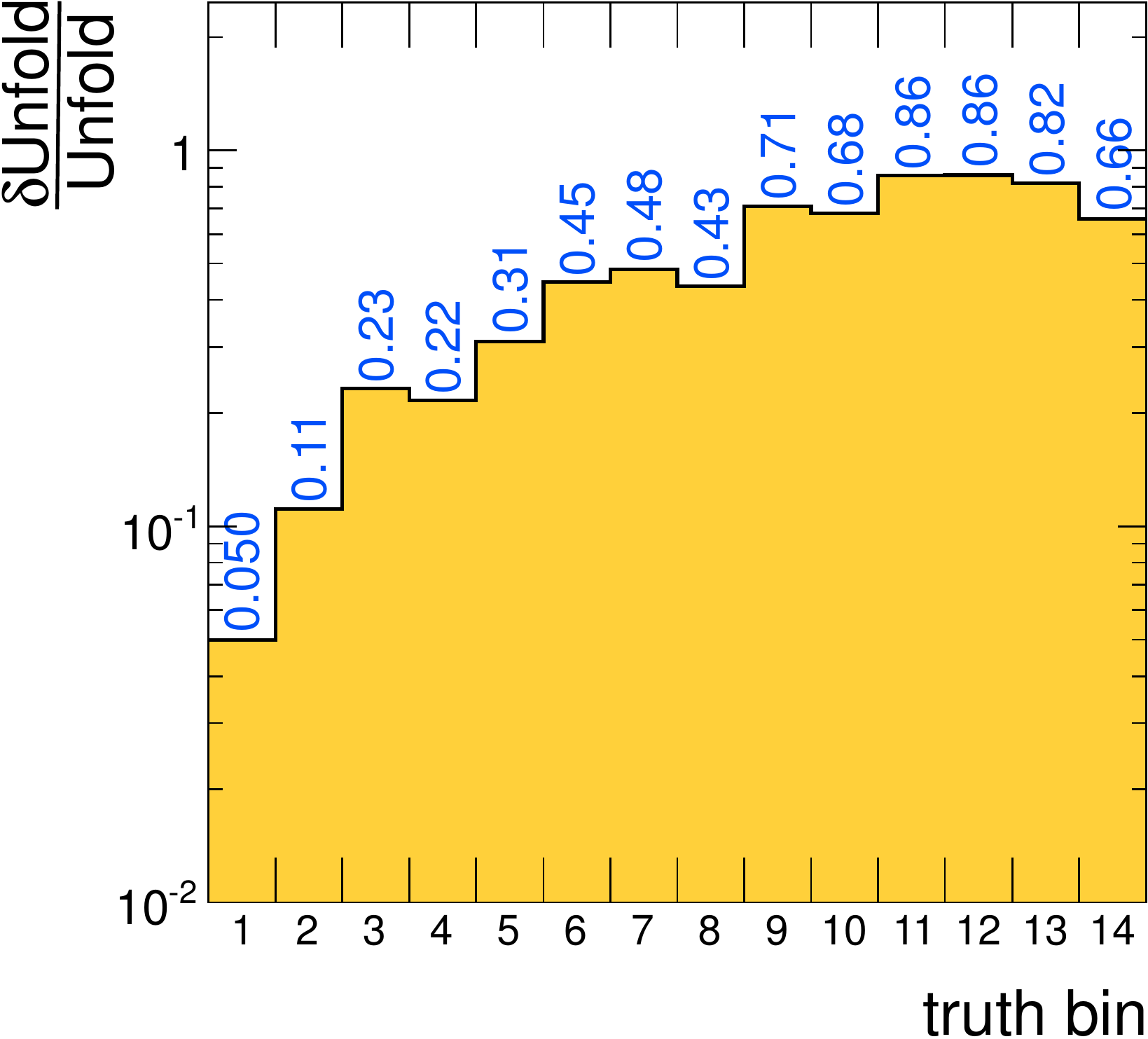}
  }
  \subfigure[$\alpha=10$]{
    \includegraphics[width=0.3\columnwidth]{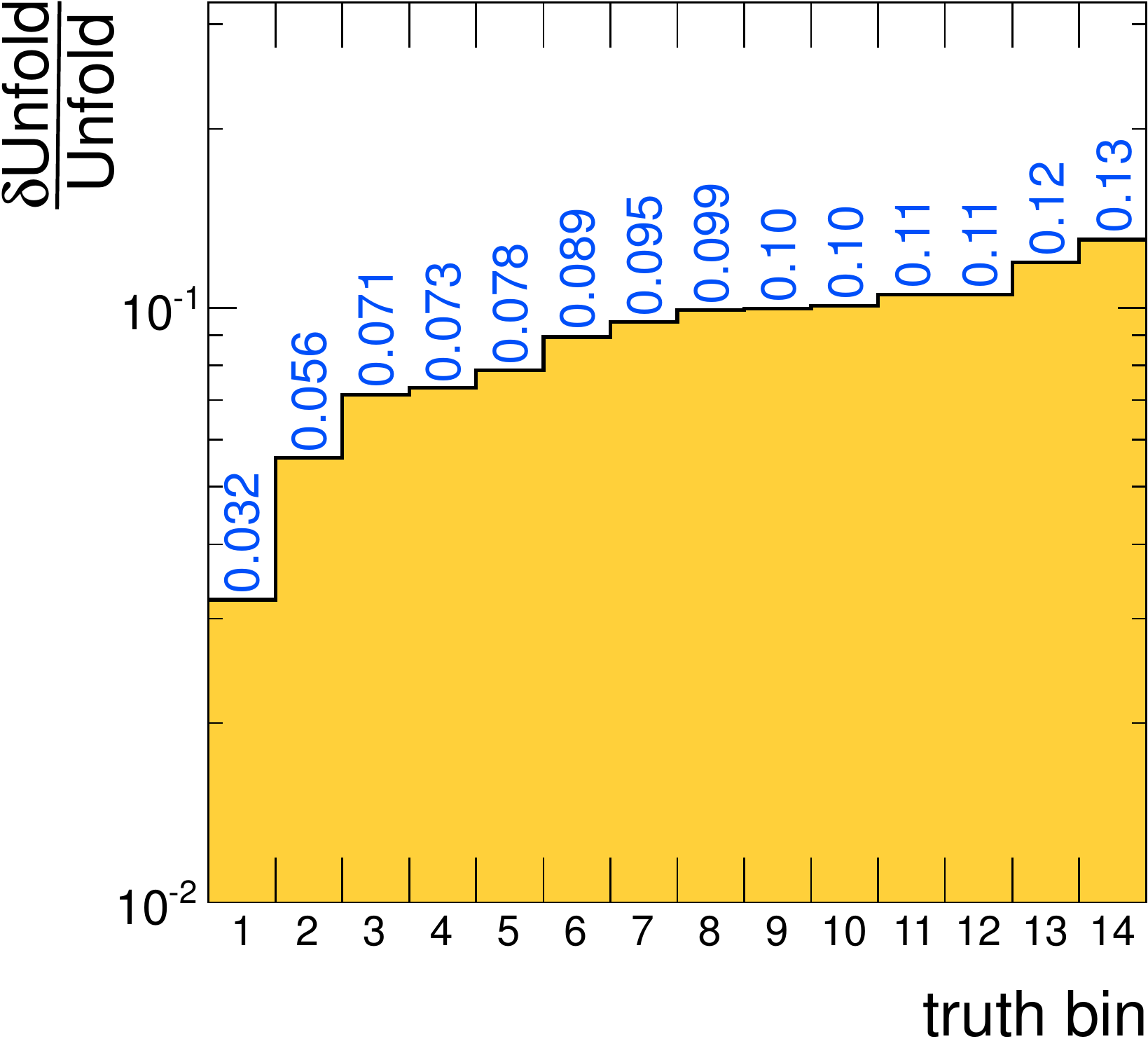}
  }
  \caption{The result of unfolding of Sec.~\ref{sec:regSteepSmearing}, with a Gaussian regularization constraint, for three $\alpha$ values (see Sec.~\ref{sec:regularization}).  
\label{fig:unfoldedSteepSmearGaus}
}
\end{figure}

The results with Gaussian regularization are shown in Fig.~\ref{fig:unfoldedSteepSmearGaus}, with similar behavior as in Sec.~\ref{sec:regSteepNoSmearing}.

It should not be surprising if, with regularization, the unfolded spectrum has smaller uncertainty than the data statistical uncertainty; external information can cause this.  It is obvious, at least in the Gaussian regularization, that the spread of the posterior can become arbitrarily small, regardless of available statistics.
The situation is analogous to fitting a straight line through many points, some of which have small statistical uncertainty, and few of which have large uncertainty.  The straight line is defined by two parameters, which are mostly constrained by the points with small uncertainty.  Even at positions where the data have great uncertainty, the fitted function will have small uncertainty, thanks to the external information that the answer must be a straight line.

\subsection{Spectrum with an unexpected bump}
\label{sec:regSteepBump}

In this section, regularization is tested in the presence of an unknown, smeared bump.  

\begin{figure}[H]
\centering
\subfigure[]{
\includegraphics[height=0.4\columnwidth]{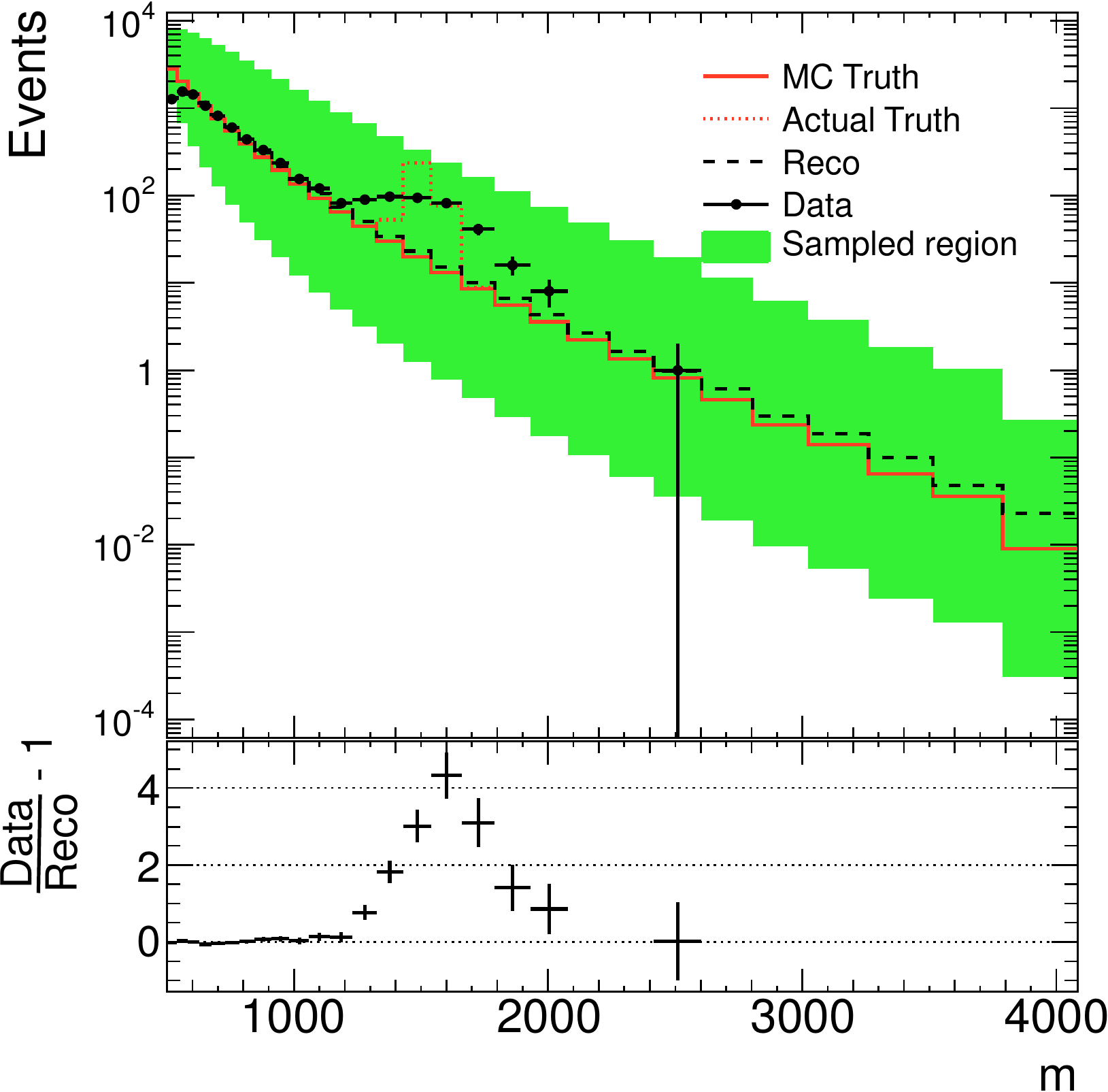}
\label{fig:regGenSteepBumpA}
}
\subfigure[]{
  \includegraphics[height=0.4\columnwidth]{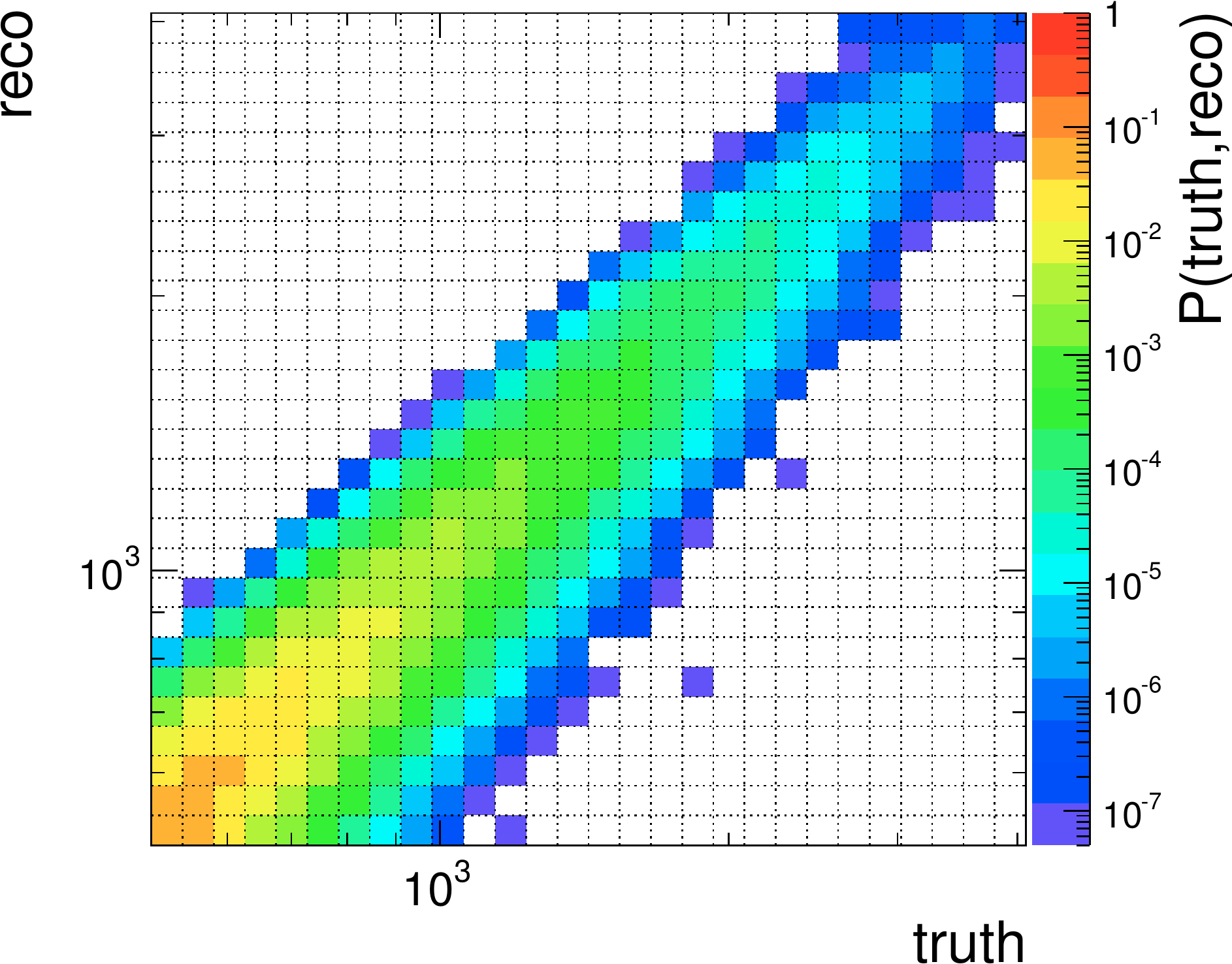}
  \label{fig:regGenSteepBumpB}
}
\caption{(a) The MC truth-level spectrum $\tilde{\T}$ (without the bump), the actual truth-level spectrum $\hat{\T}$ (with the bump), the reconstructed spectrum which corresponds to $\tilde{\T}$ after smearing, the data which follow $\hat{\T}$ after smearing, and the sampled hyper-box used in Sec.~\ref{sec:regSteepBump}.  (b) The migrations matrix, populated with the MC events that compose the MC truth level ($\tilde{\T}$) and the reco spectrum of (a).
\label{fig:regGenSteepBump}
}
\end{figure}

MC events are generated following a steeply falling truth-level spectrum with $N_t=28$ bins, with a prominent Gaussian bump of mean $m=1500$ and RMS=50.  Smearing with parameters $(a,b)=(0.5,0.1)$ applies.  The migrations matrix is built with MC events where the bump is missing, thus, the bump is an unknown feature.  Fig.~\ref{fig:regGenSteepBumpA} shows the truth-level spectrum with and without the bump, the reconstructed spectrum without the bump, which consists of the MC events that populate the migrations matrix (Fig.~\ref{fig:regGenSteepBumpB}), and the data.  Also shown is the sampled hyper-box, which is bigger than in Eq.~\ref{eq:prior4} to accommodate the actual truth-level spectrum (with the bump) $\hat{\T}$, and is defined by:
\begin{equation}
\label{eq:priorBigger}
\begin{split}
T_t^\ulcorner & \equiv \tilde{T}_{t} / (t+1) , t\in[1,N_t] \\
T_t^\urcorner & \equiv  \tilde{T}_t \cdot (t+2) , t\in[1,N_t]
\end{split}
\end{equation}
where $\tilde{\T}$ is the truth level spectrum without the bump, used to populate the migrations matrix.

\begin{figure}[H]
  \centering
  \subfigure[$\alpha=0$]{
    \includegraphics[width=0.3\columnwidth]{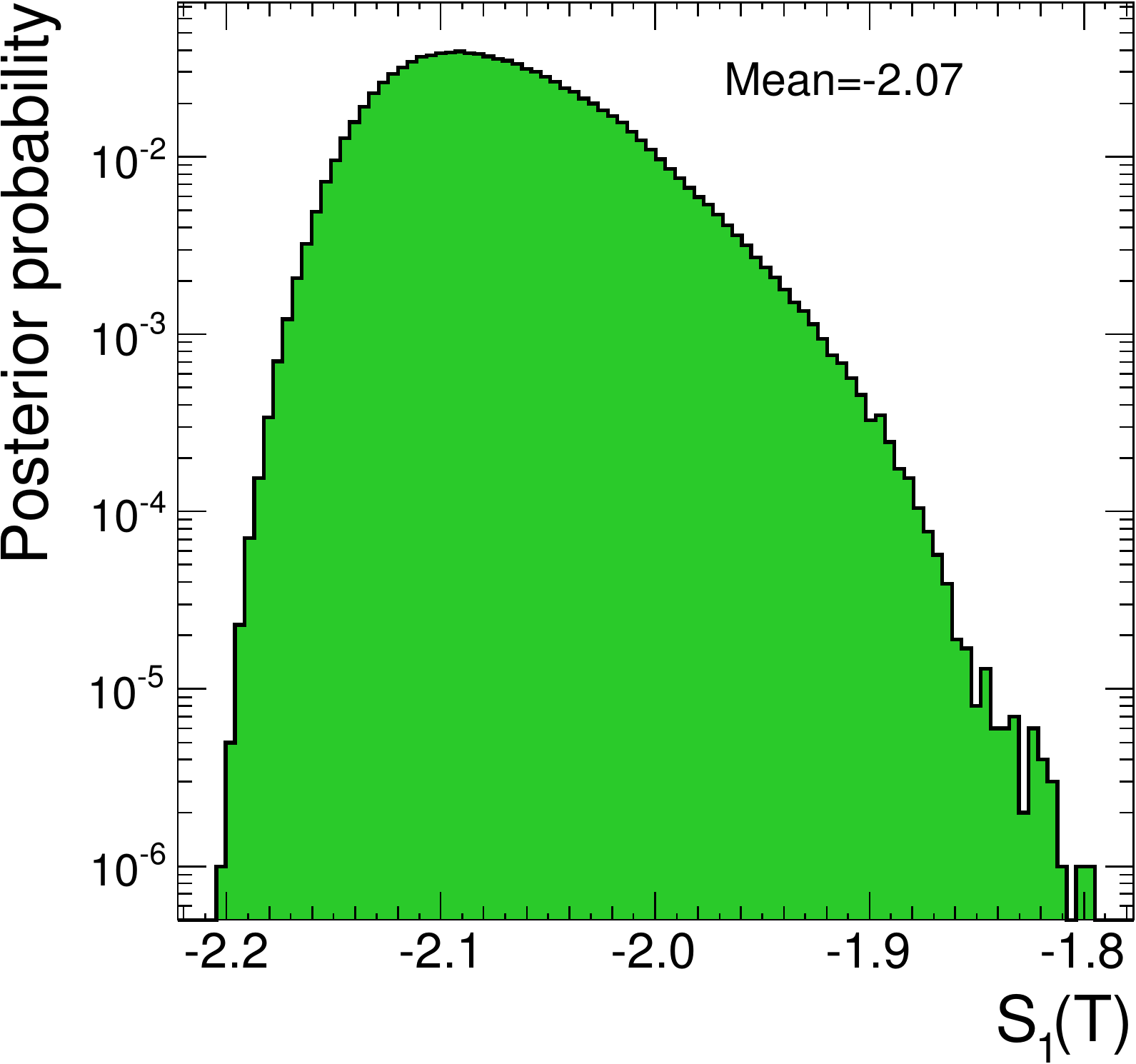}
  }
  \subfigure[$\alpha=10^3$]{
    \includegraphics[width=0.3\columnwidth]{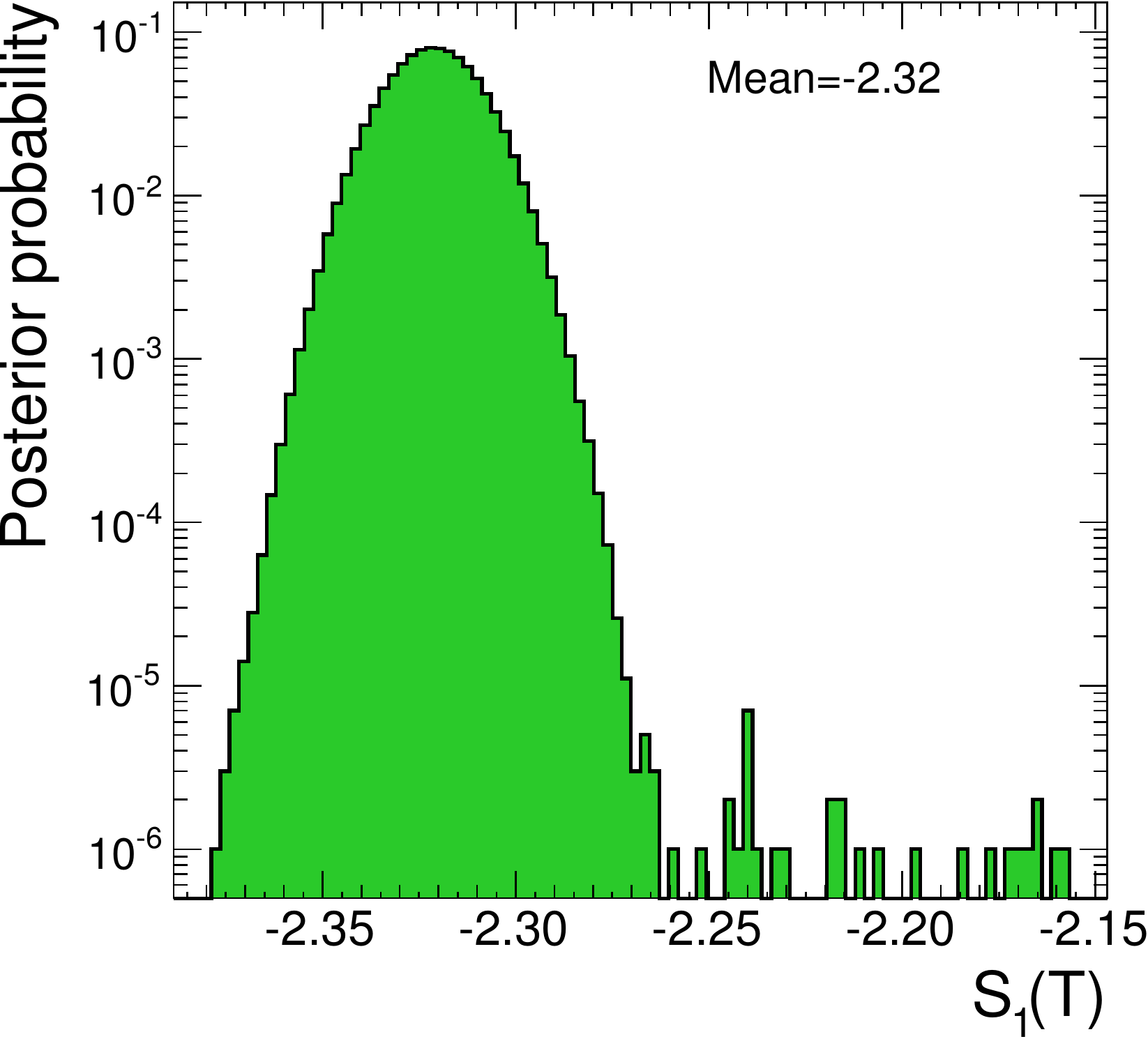}
  }
  \subfigure[$\alpha=3\times 10^3$]{
    \includegraphics[width=0.3\columnwidth]{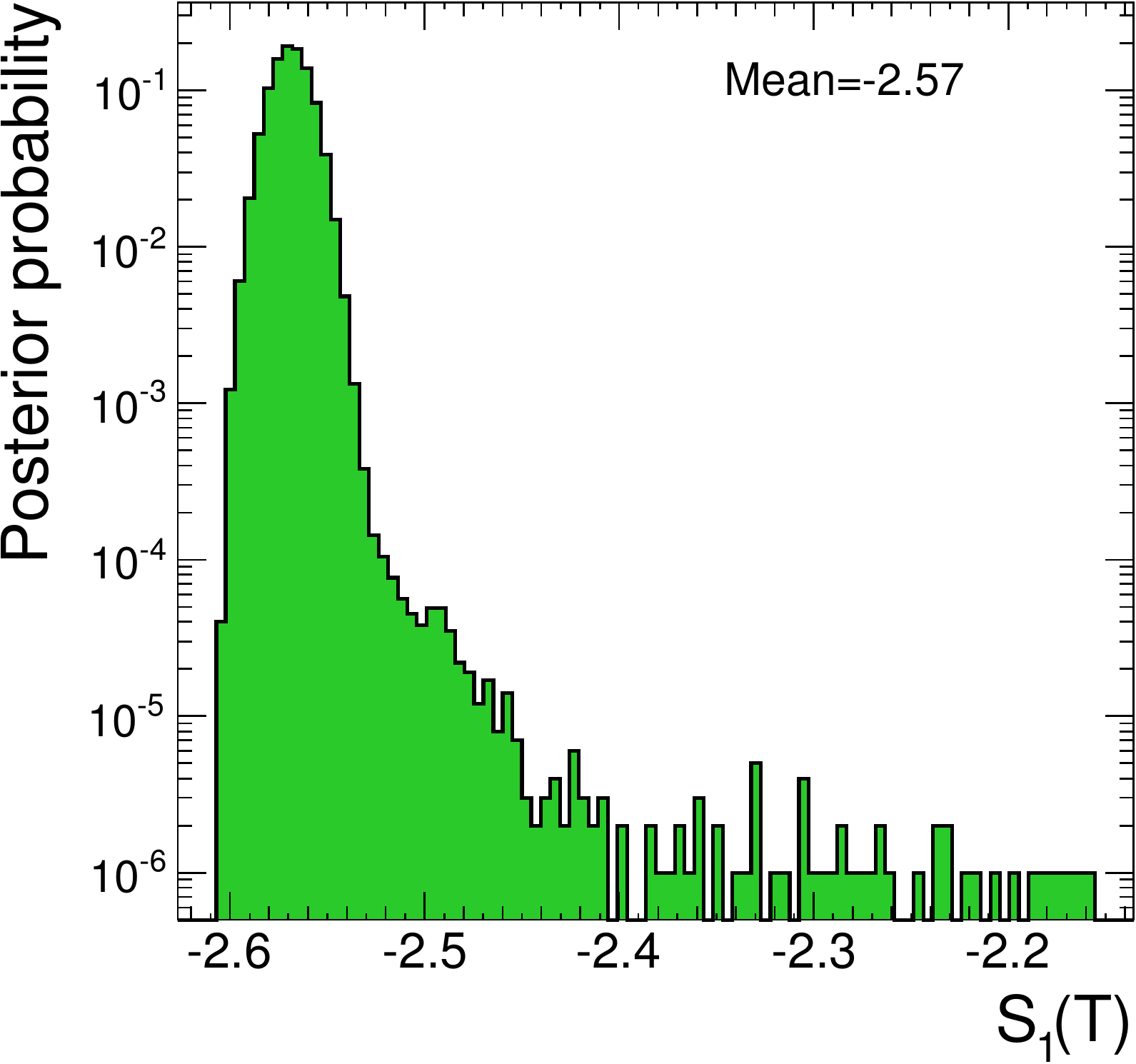}
 }
\caption{The posterior $P(S_1(\tuple{T})|\tuple{D})$, for three different choices of the regularization parameter $\alpha$, corresponding to Sec.~\ref{sec:regSteepBump}.
\label{fig:regFuncSteepBumpS1} 
}
\end{figure}

\begin{figure}[H]
  \centering
  \subfigure[$\alpha=0$]{
    \includegraphics[width=0.3\columnwidth]{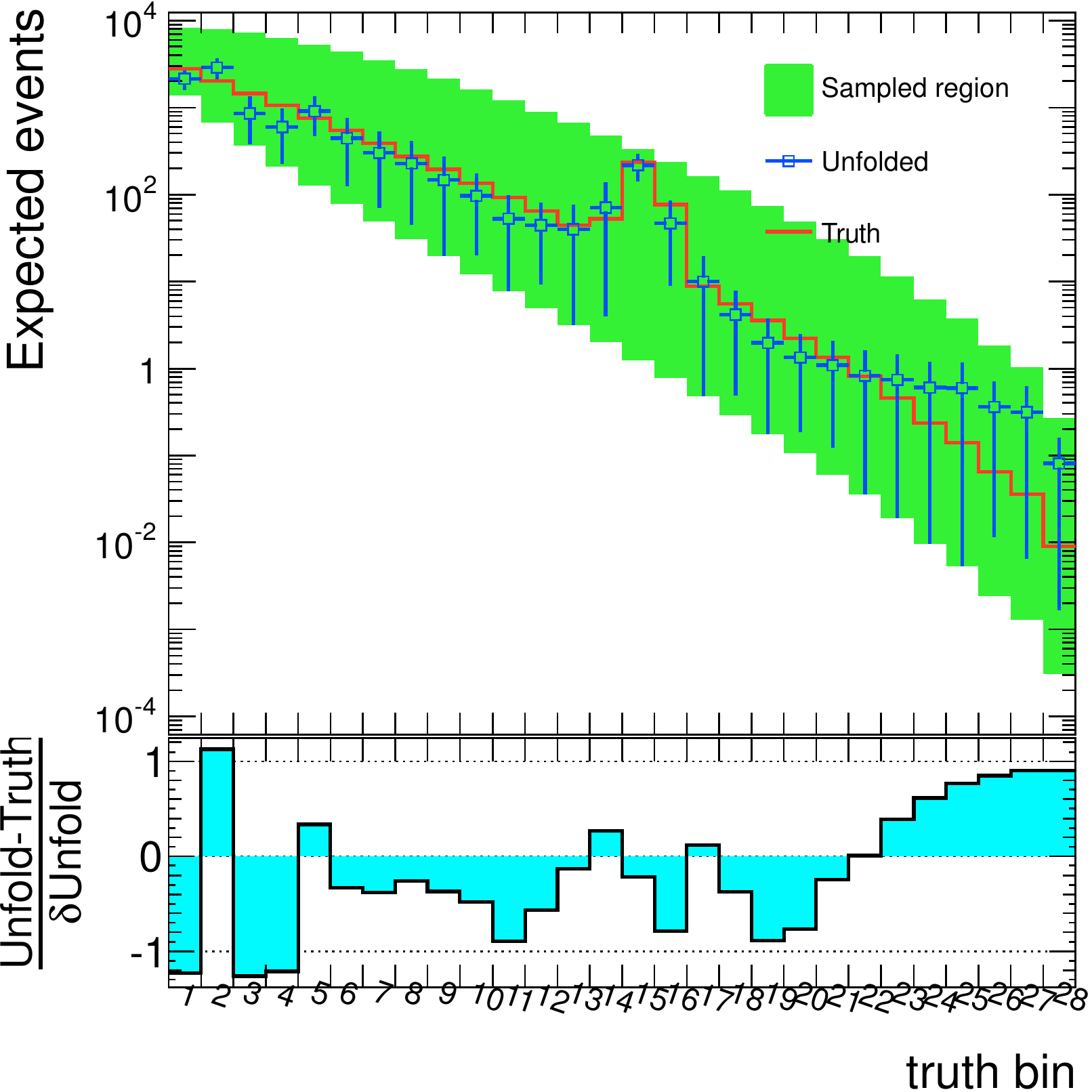}
  }
  \subfigure[$\alpha=10^3$]{
    \includegraphics[width=0.3\columnwidth]{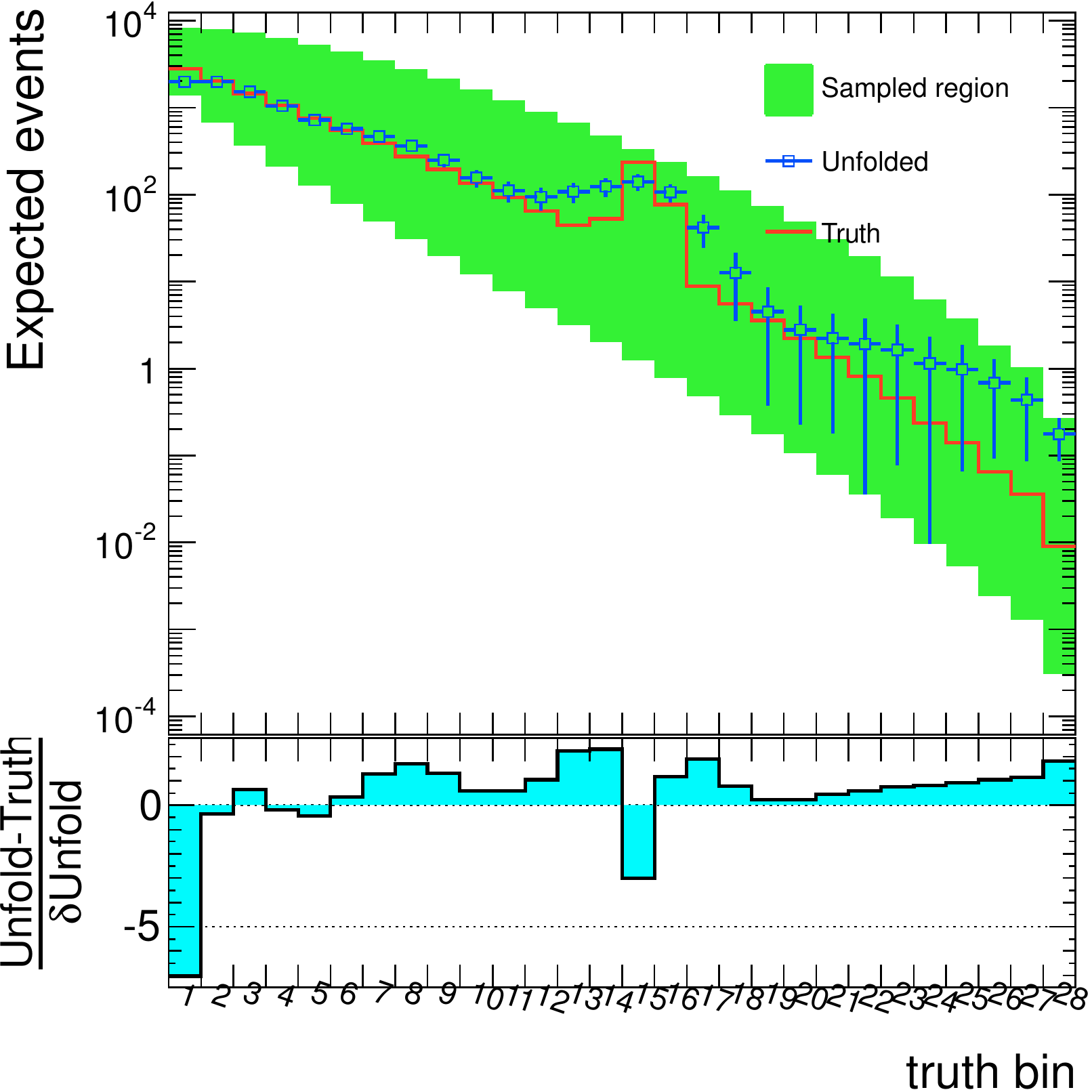}
  }
  \subfigure[$\alpha=3\times 10^3$]{
    \includegraphics[width=0.3\columnwidth]{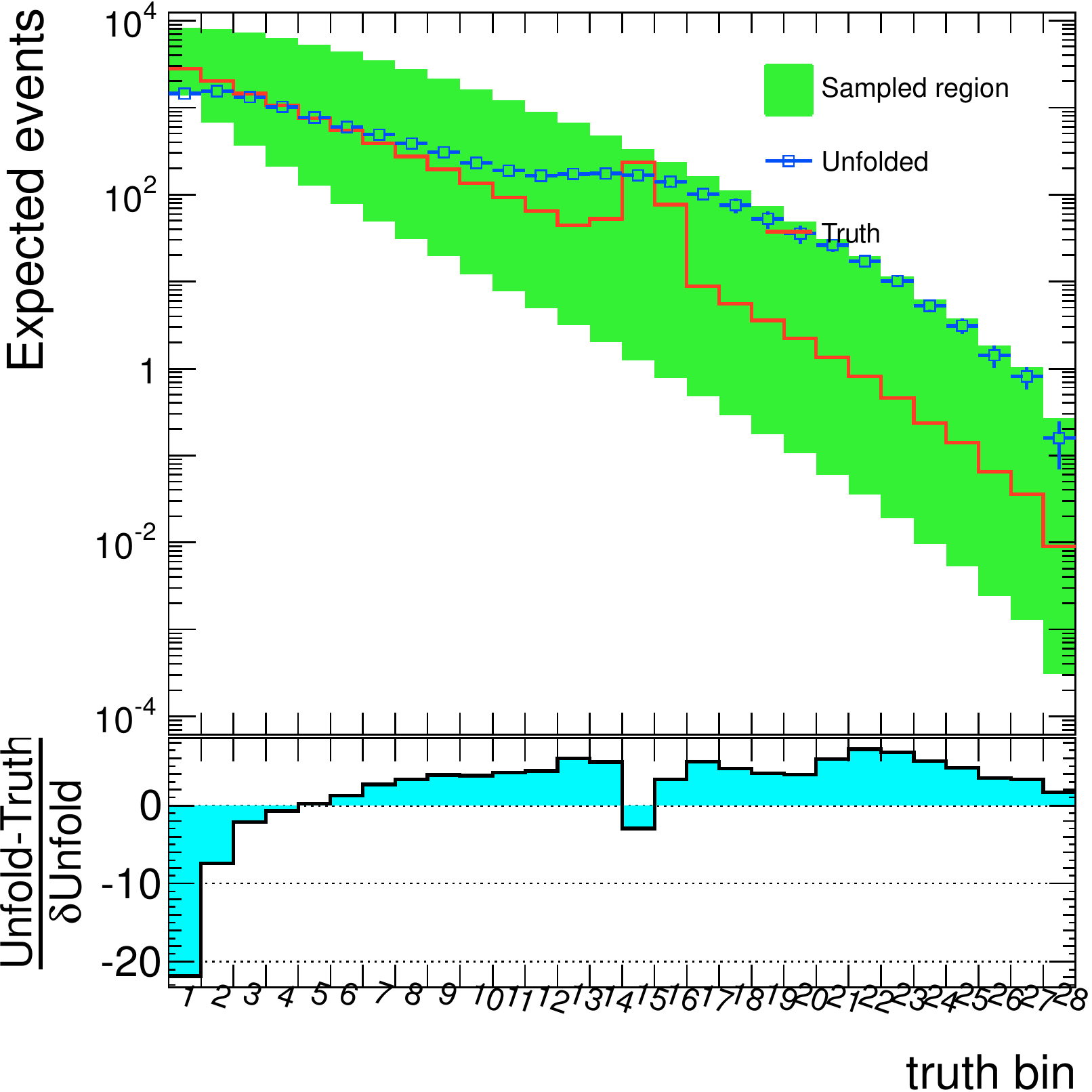}
  }\\
 \subfigure[$\alpha=0$]{
    \includegraphics[width=0.3\columnwidth]{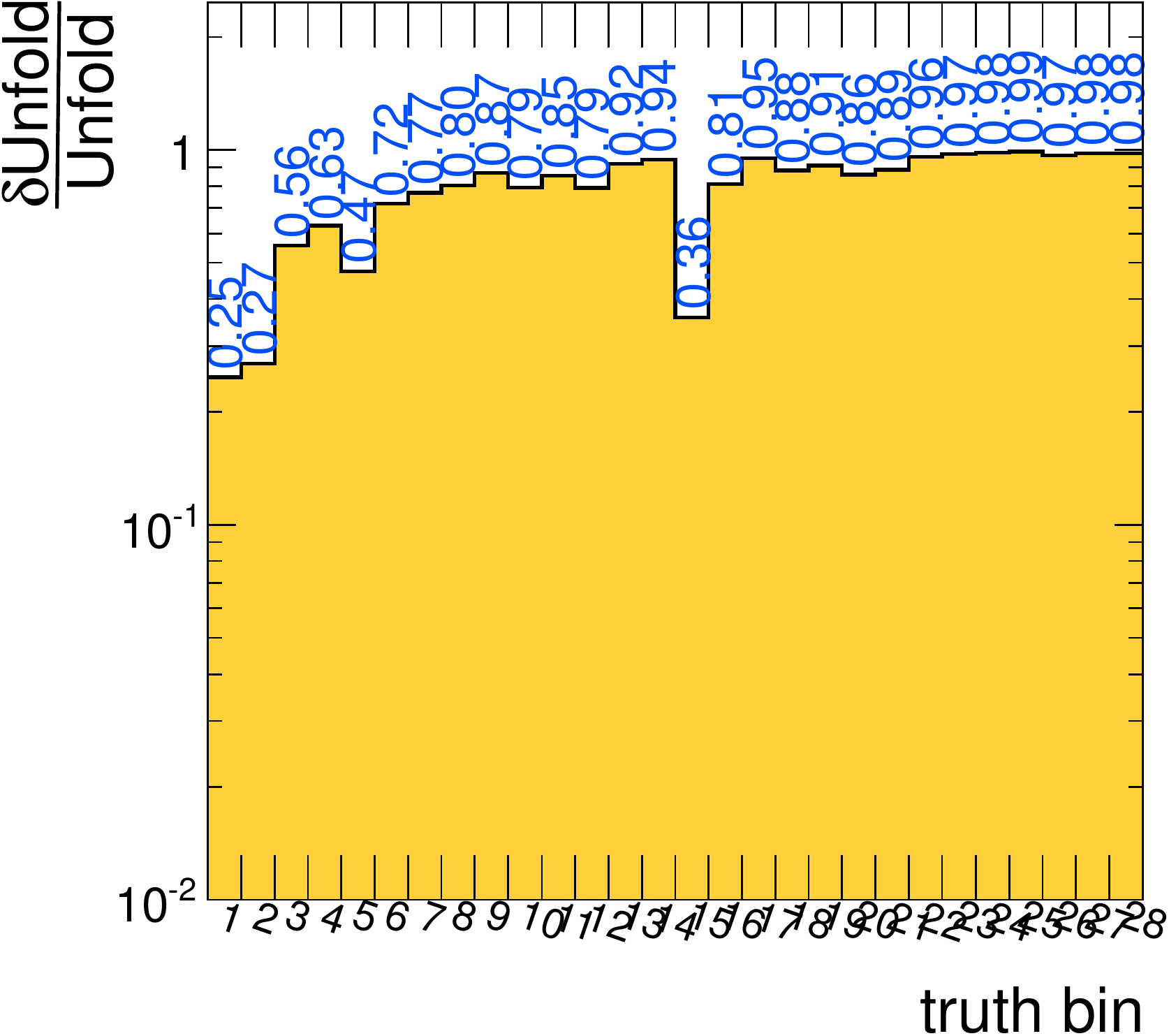}
  }
  \subfigure[$\alpha=10^3$]{
    \includegraphics[width=0.3\columnwidth]{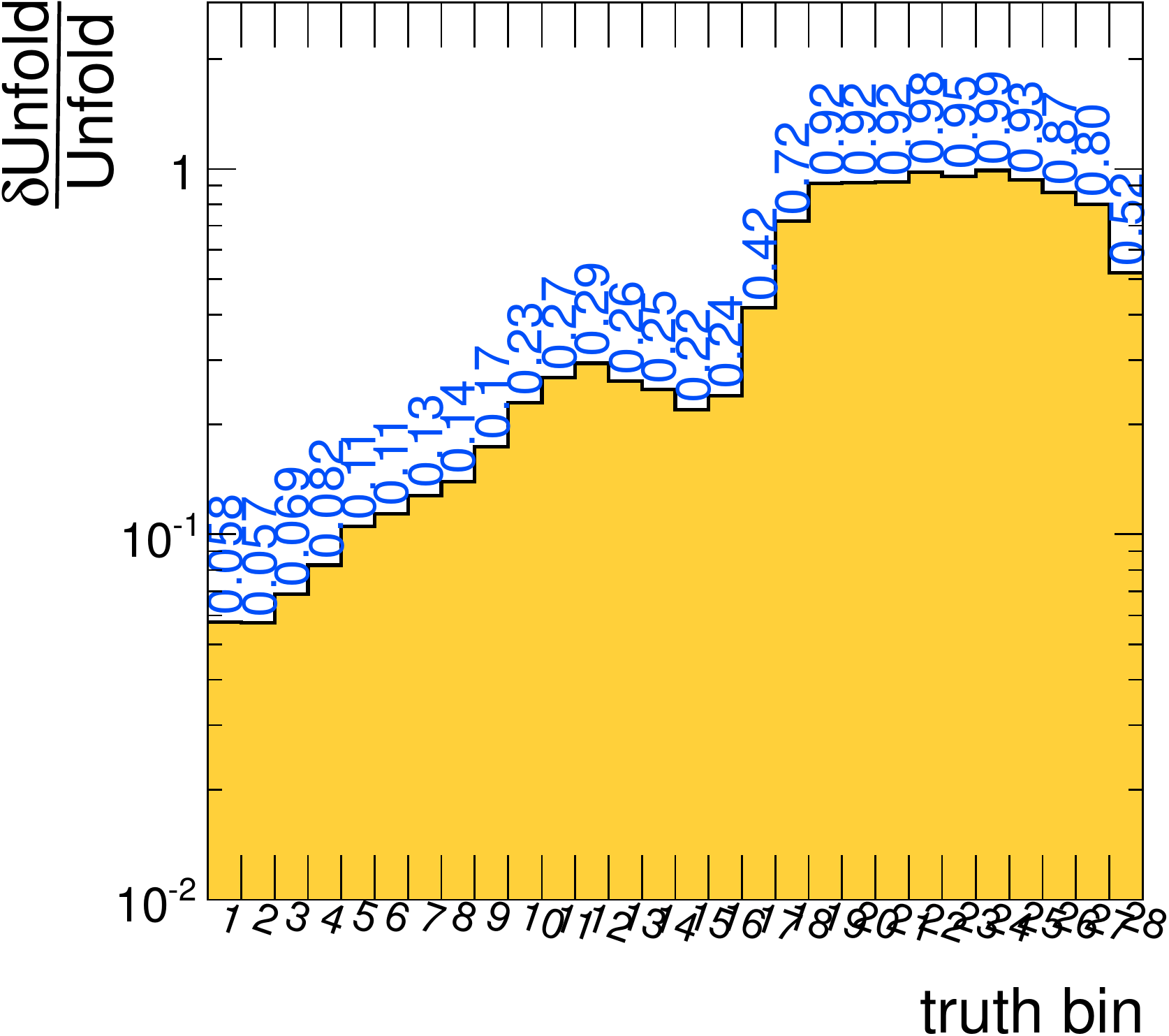}
  }
  \subfigure[$\alpha=3\times 10^3$]{
    \includegraphics[width=0.3\columnwidth]{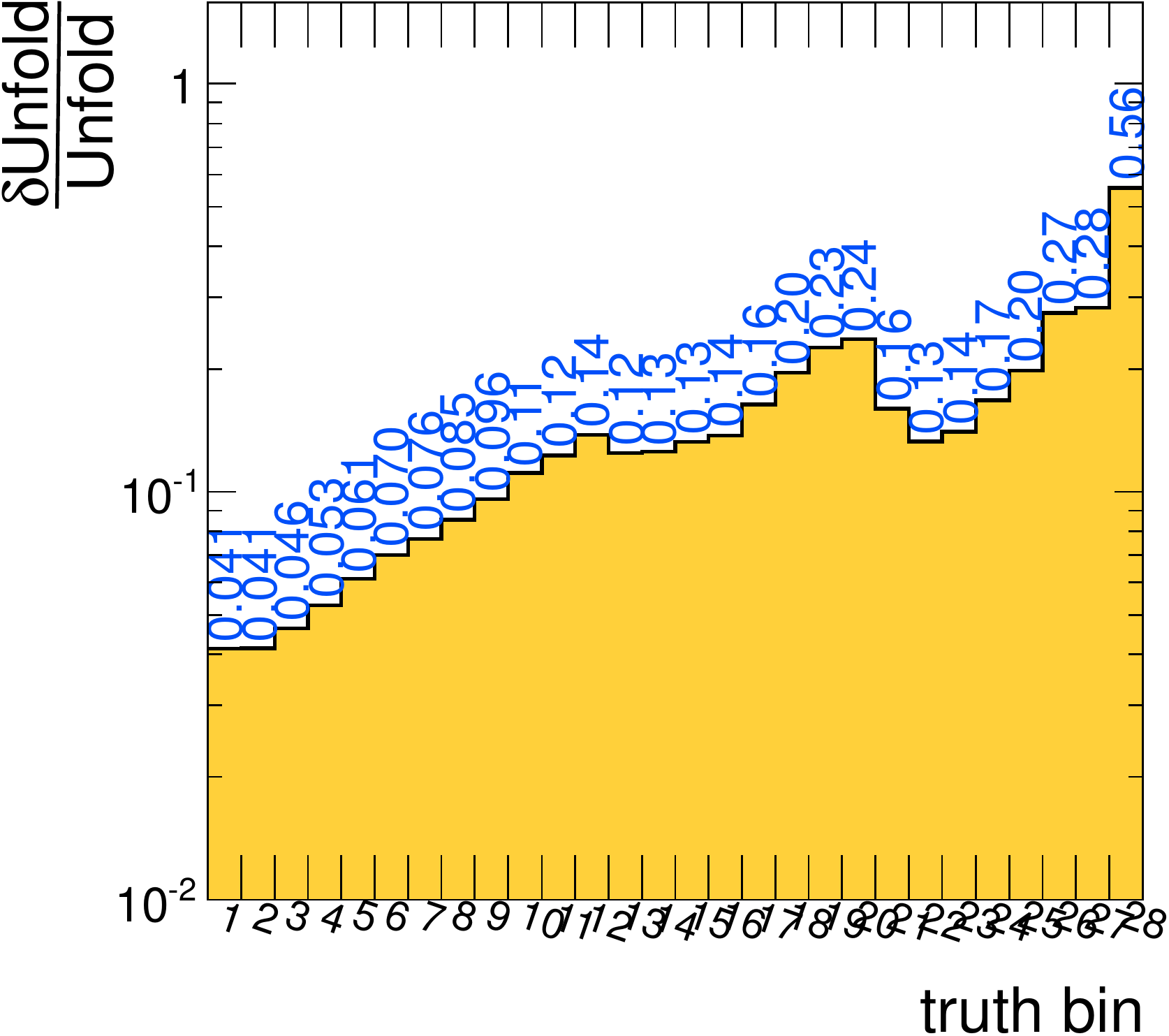}
  }
  \caption{The result of unfolding of Sec.~\ref{sec:regSteepBump}, with regularization function $S_1$, for three $\alpha$ values.  
    \label{fig:unfoldSteepBumpS1}
  }
\end{figure}

Unfolding with negative entropy regularization ($S_1$) is shown in Fig.~\ref{fig:regFuncSteepBumpS1} and \ref{fig:unfoldSteepBumpS1}.  As seen previously (see Sec.~\ref{sec:regSteepSmearing}), regularization with entropy seems to distort the unfolded spectrum significantly.

\begin{figure}[H]
  \centering
  \subfigure[$\alpha=0$]{
    \includegraphics[width=0.3\columnwidth]{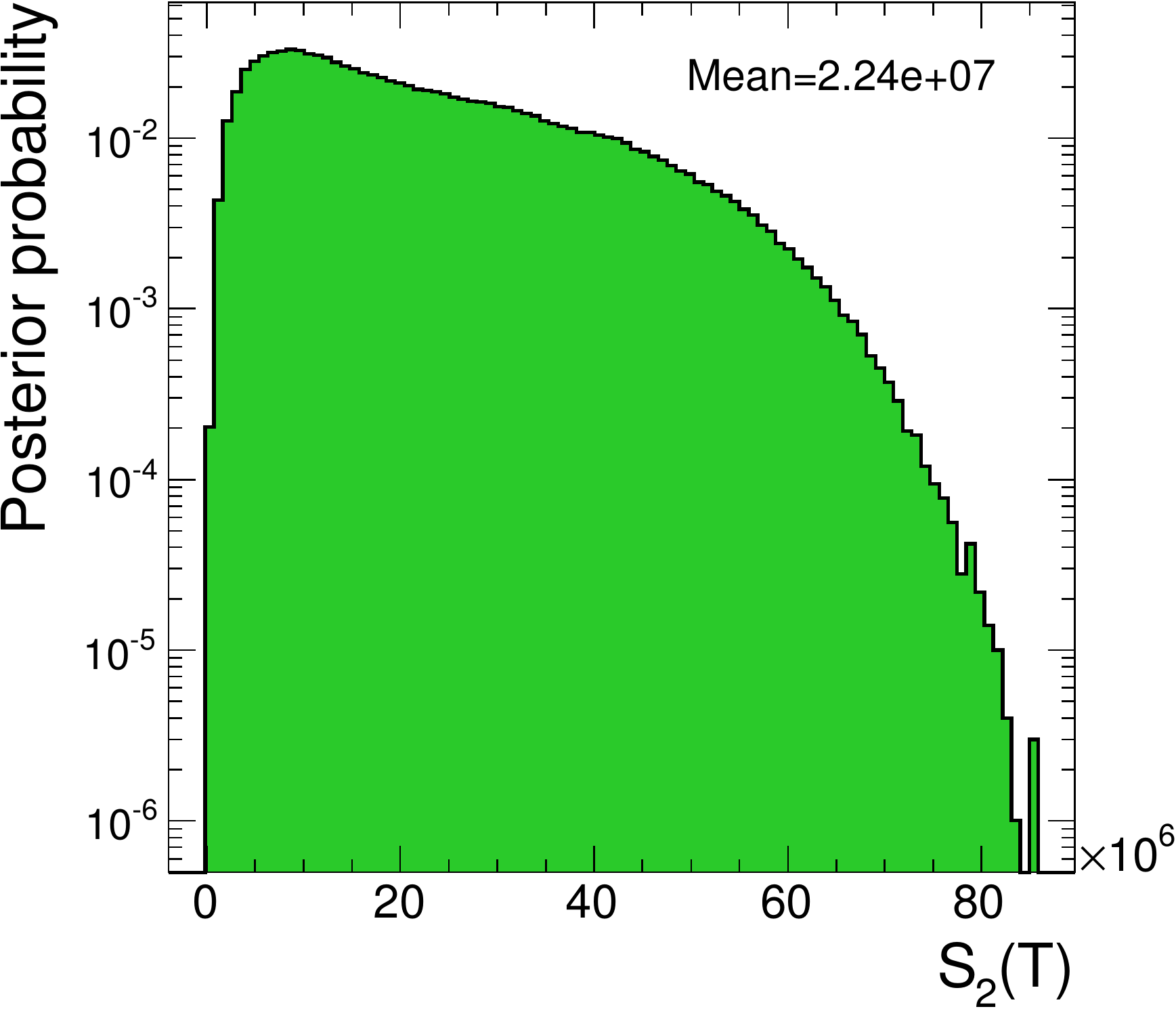}
  }
  \subfigure[$\alpha=3\times 10^{-4}$]{
    \includegraphics[width=0.3\columnwidth]{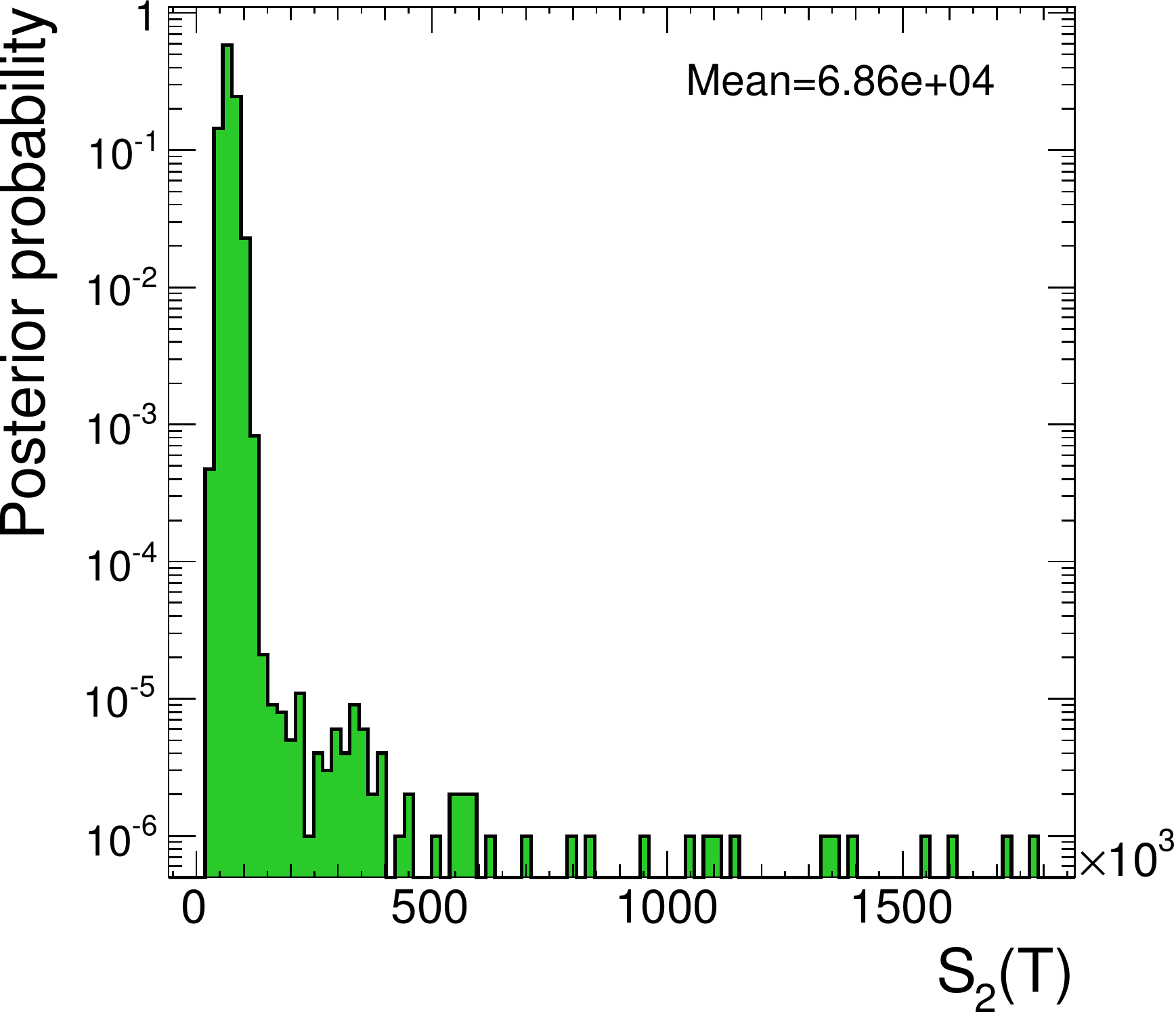}
  }
  \subfigure[$\alpha=6\times 10^{-4}$]{
    \includegraphics[width=0.3\columnwidth]{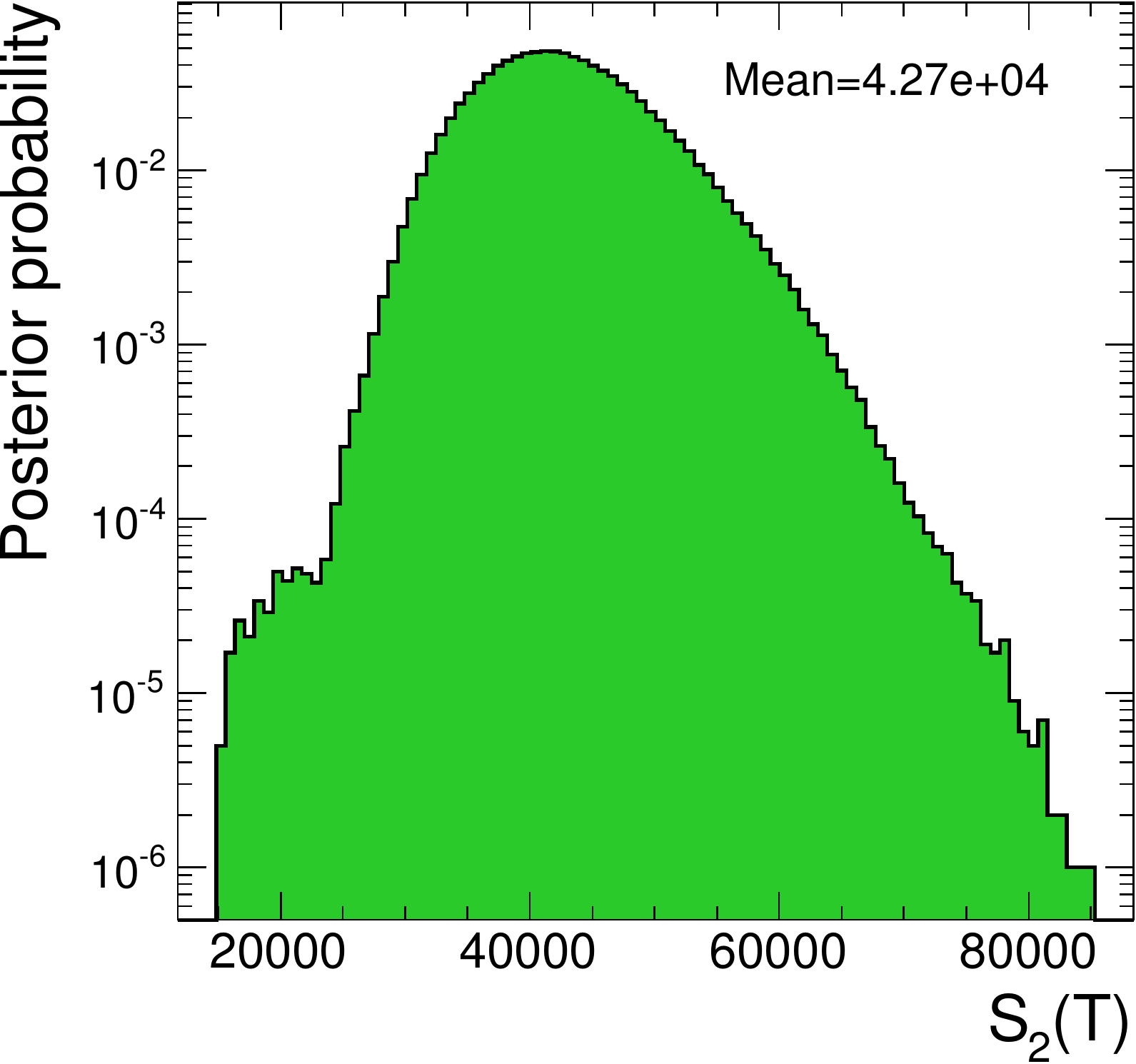}
 }
\caption{The posterior $P(S_2(\tuple{T})|\tuple{D})$, for three different choices of the regularization parameter $\alpha$, corresponding to Sec.~\ref{sec:regSteepBump}.
\label{fig:regFuncSteepBumpS2} 
}
\end{figure}

\begin{figure}[H]
  \centering
  \subfigure[$\alpha=0$]{
    \includegraphics[width=0.3\columnwidth]{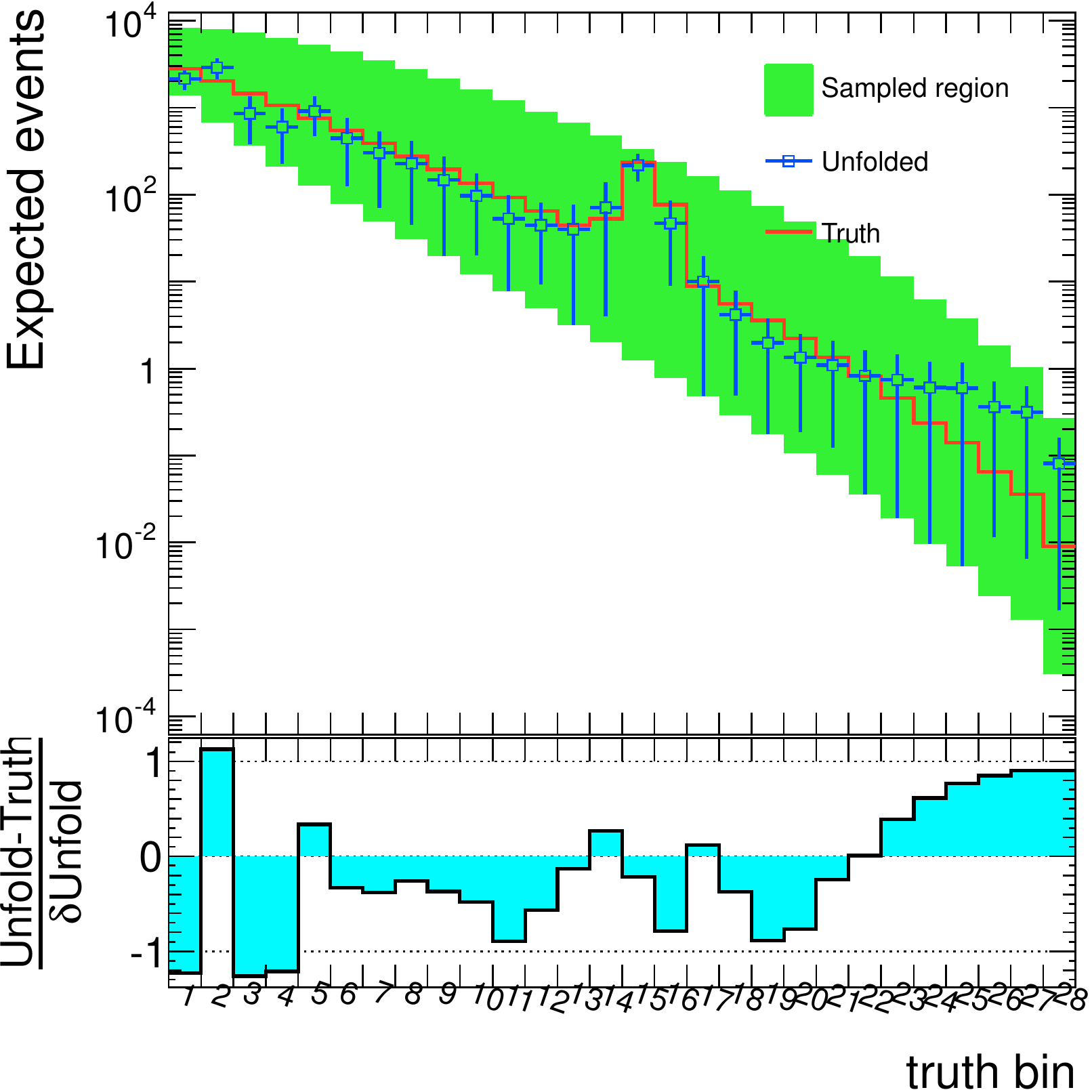}
  }
  \subfigure[$\alpha=3\times 10^{-4}$]{
    \includegraphics[width=0.3\columnwidth]{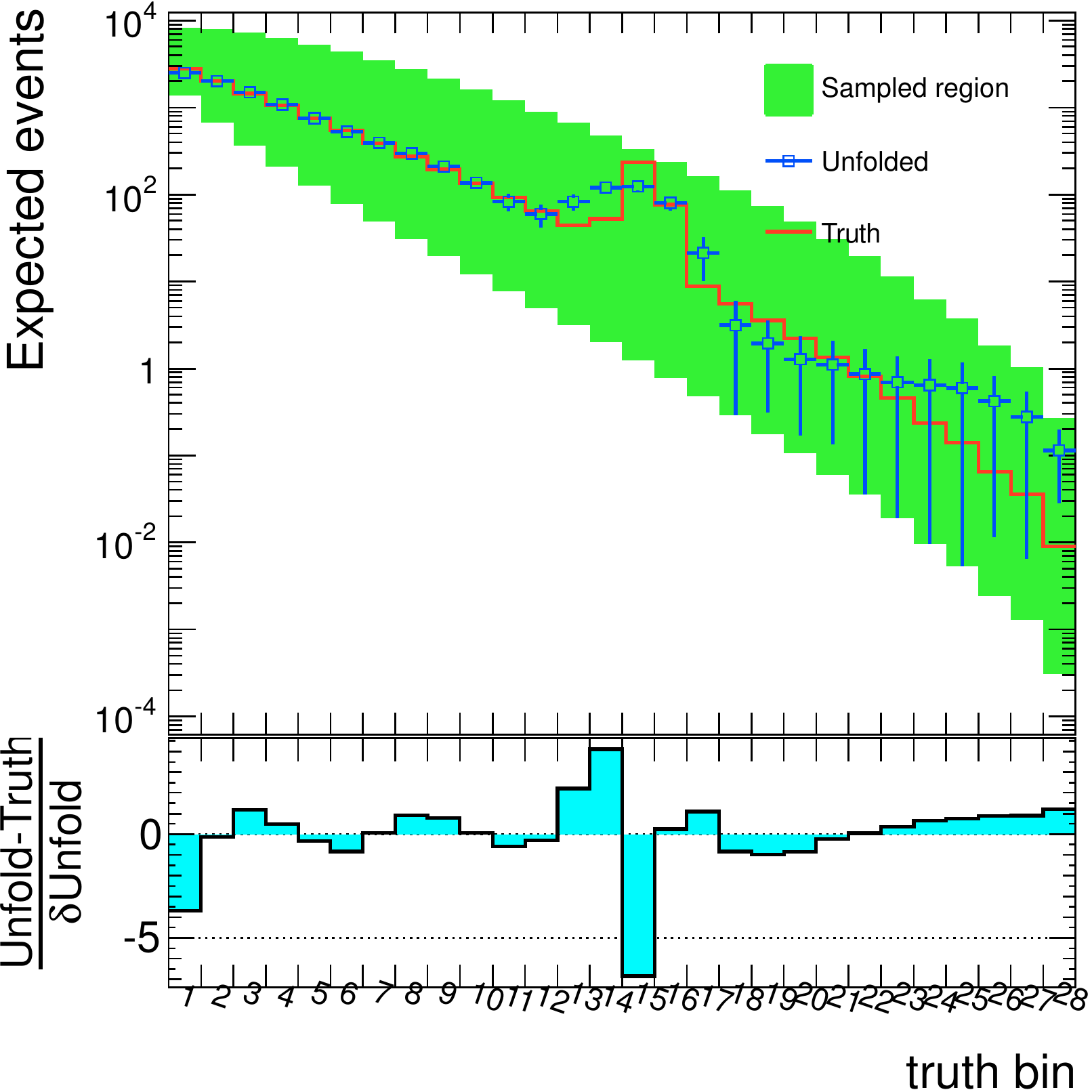}
  }
  \subfigure[$\alpha=6\times 10^{-4}$]{
    \includegraphics[width=0.3\columnwidth]{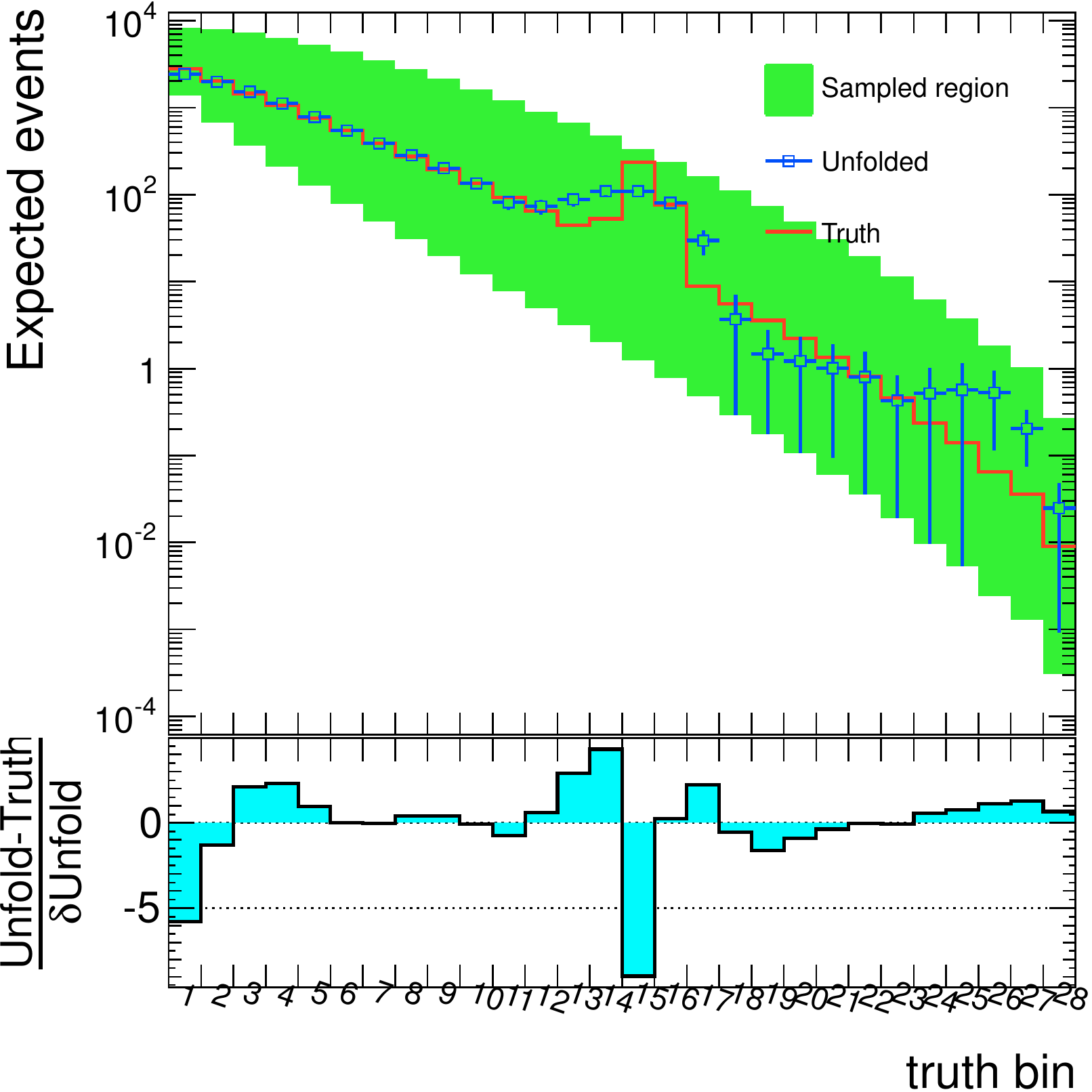}
  }\\
 \subfigure[$\alpha=0$]{
    \includegraphics[width=0.3\columnwidth]{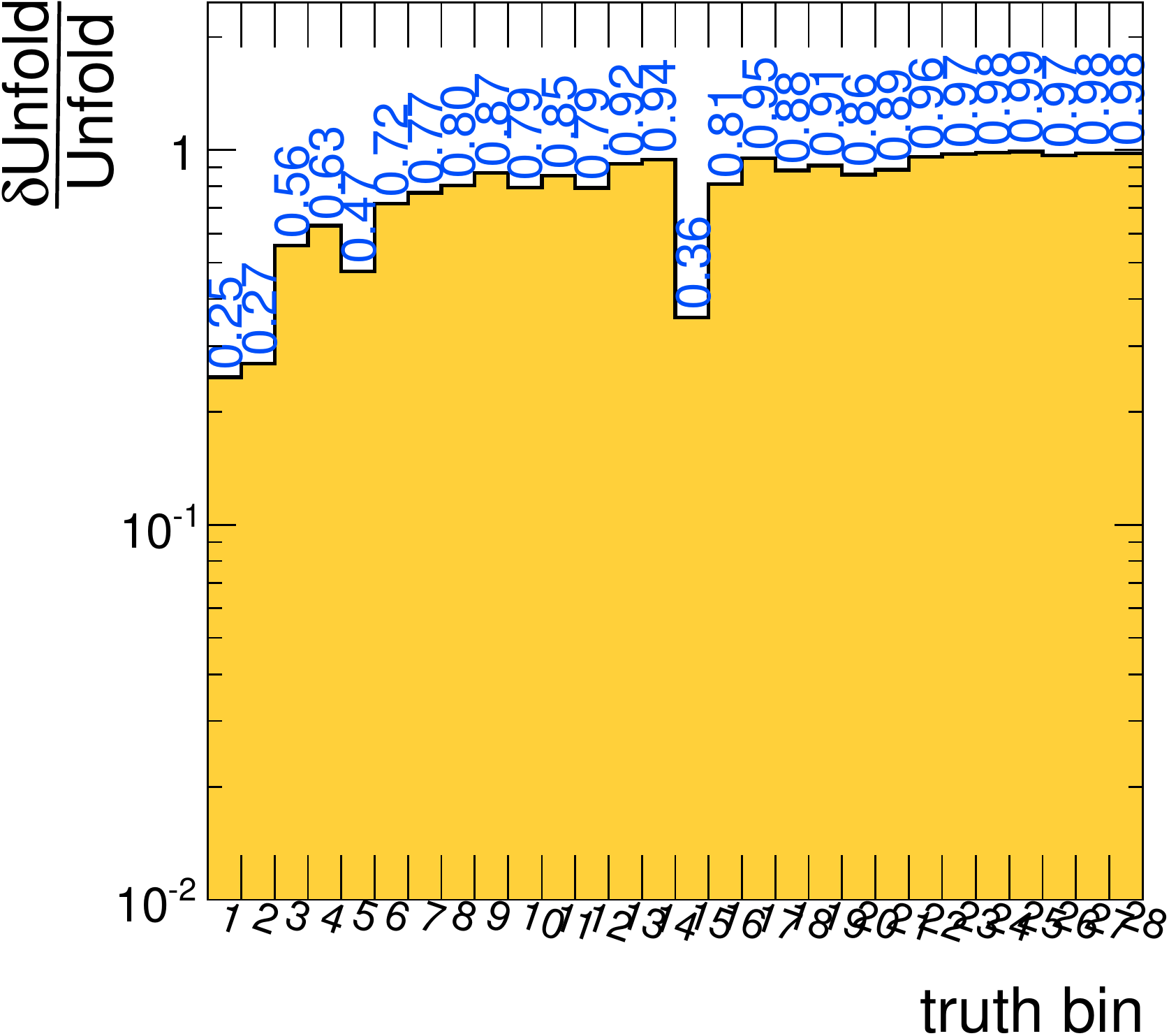}
  }
  \subfigure[$\alpha=3\times 10^{-4}$]{
    \includegraphics[width=0.3\columnwidth]{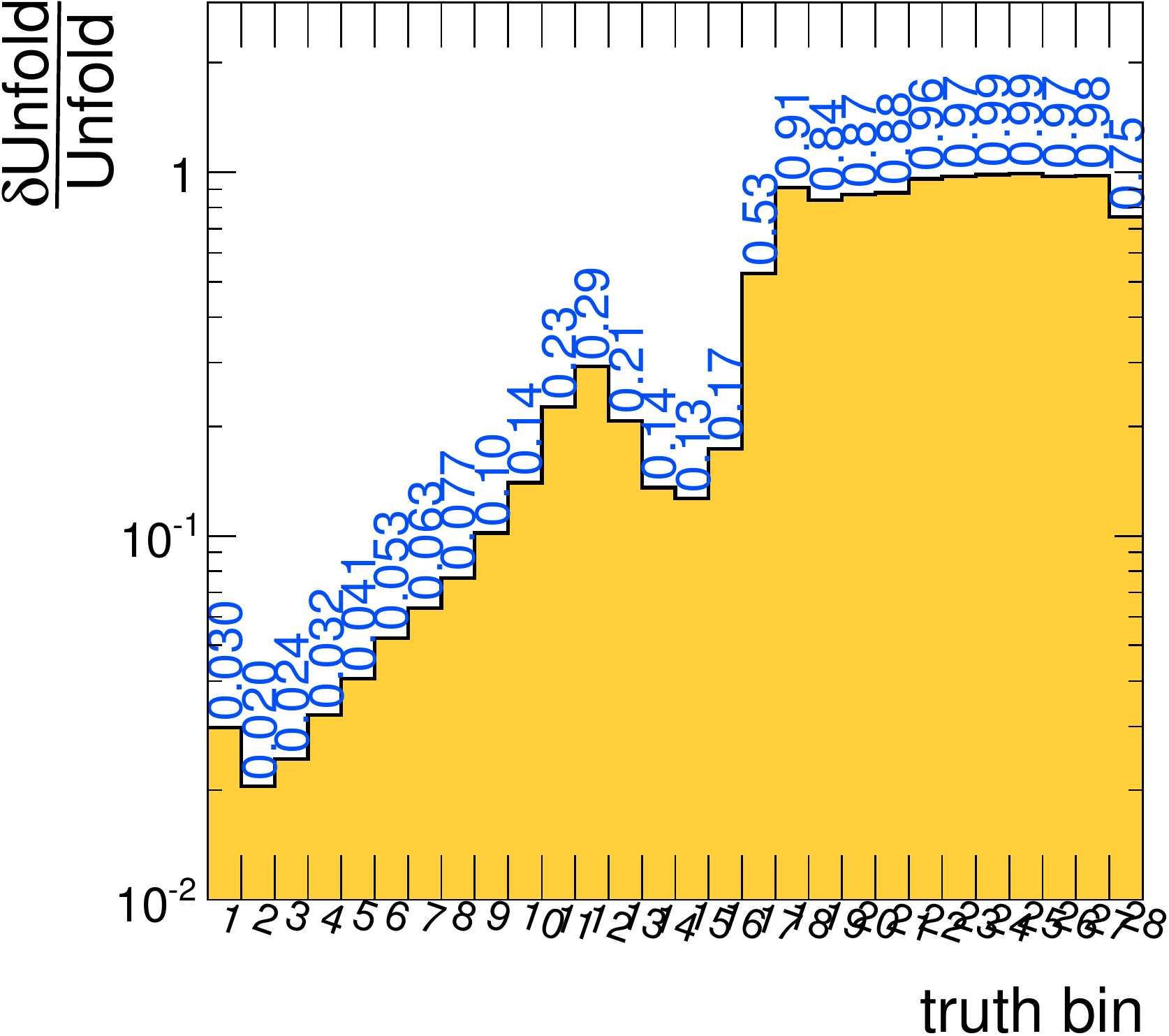}
  }
  \subfigure[$\alpha=6\times 10^{-4}$]{
    \includegraphics[width=0.3\columnwidth]{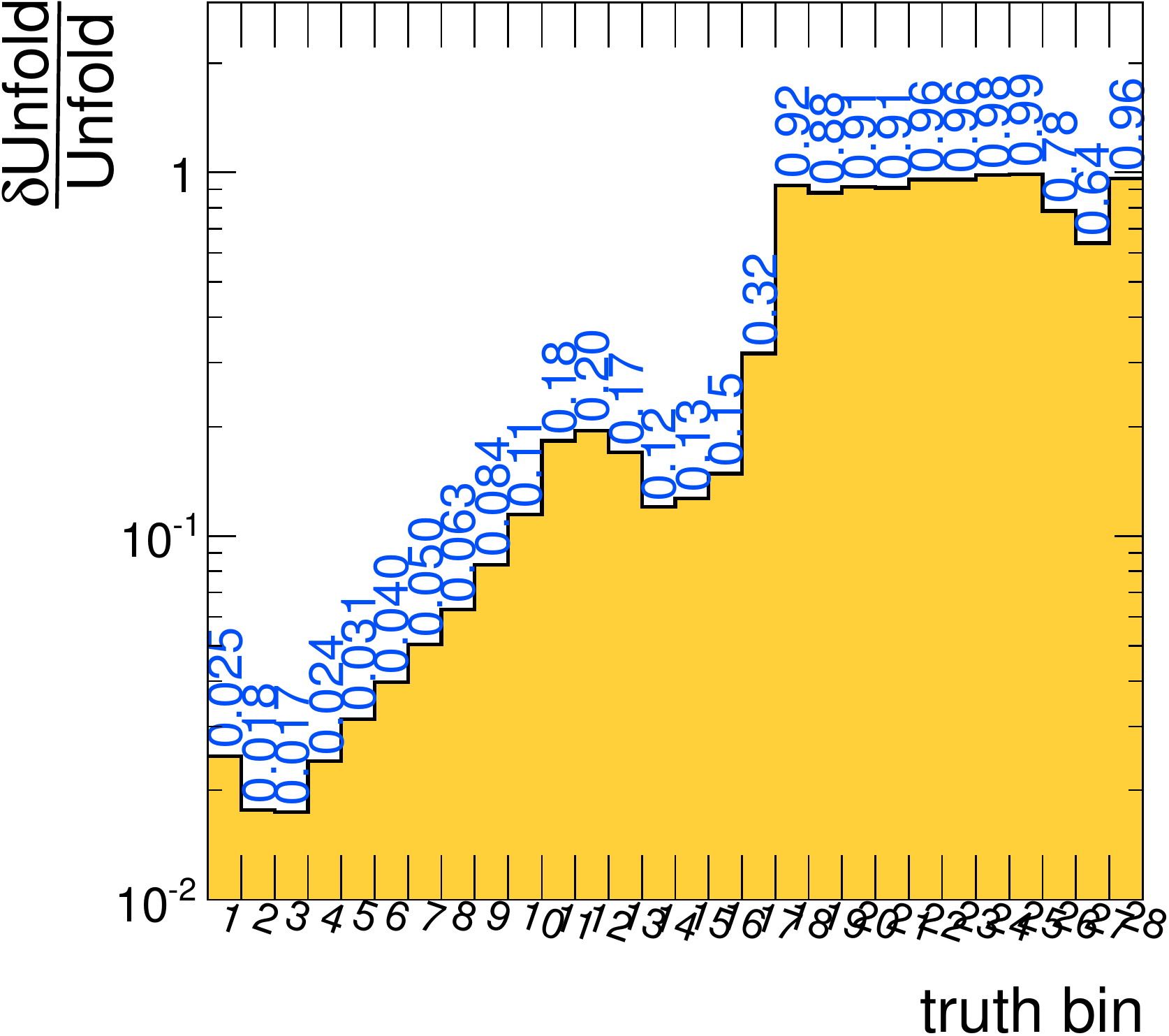}
  }
  \caption{The result of unfolding of Sec.~\ref{sec:regSteepBump}, with regularization function $S_2$, for three $\alpha$ values.  
    \label{fig:unfoldSteepBumpS2}
  }
\end{figure}

Using the regularization function $S_2$ is shown in Fig.~\ref{fig:regFuncSteepBumpS2} and \ref{fig:unfoldSteepBumpS2}.  The spread of the posterior is greatly reduced, especially in the first truth-level bins.  The unfolded spectrum seems to have a bump, but it is significantly wider than it is at truth-level; it has width similar to the data, after smearing.

\begin{figure}[H]
  \centering
  \subfigure[$\alpha=0$]{
    \includegraphics[width=0.3\columnwidth]{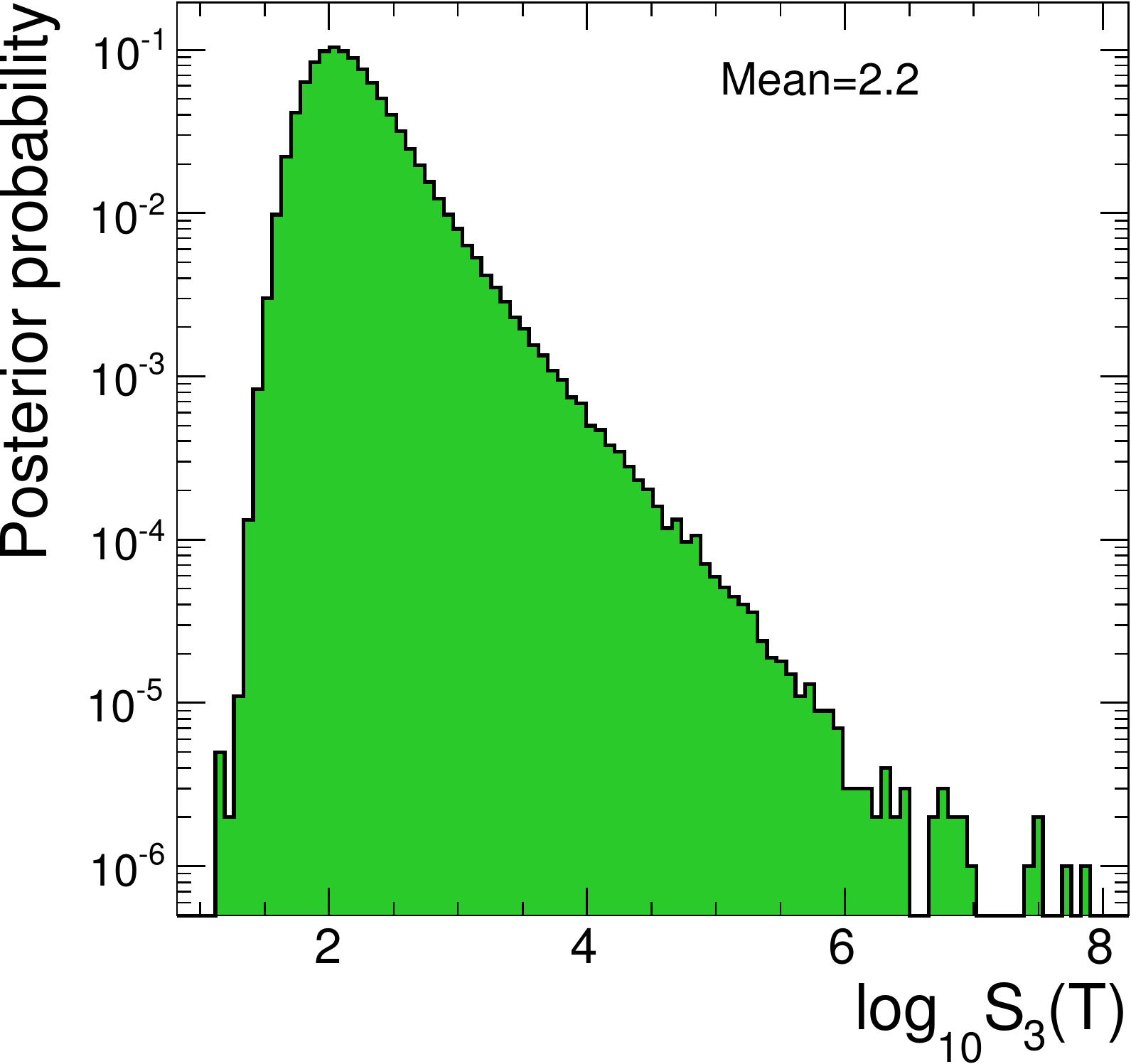}
  }
  \subfigure[$\alpha=10$]{
    \includegraphics[width=0.3\columnwidth]{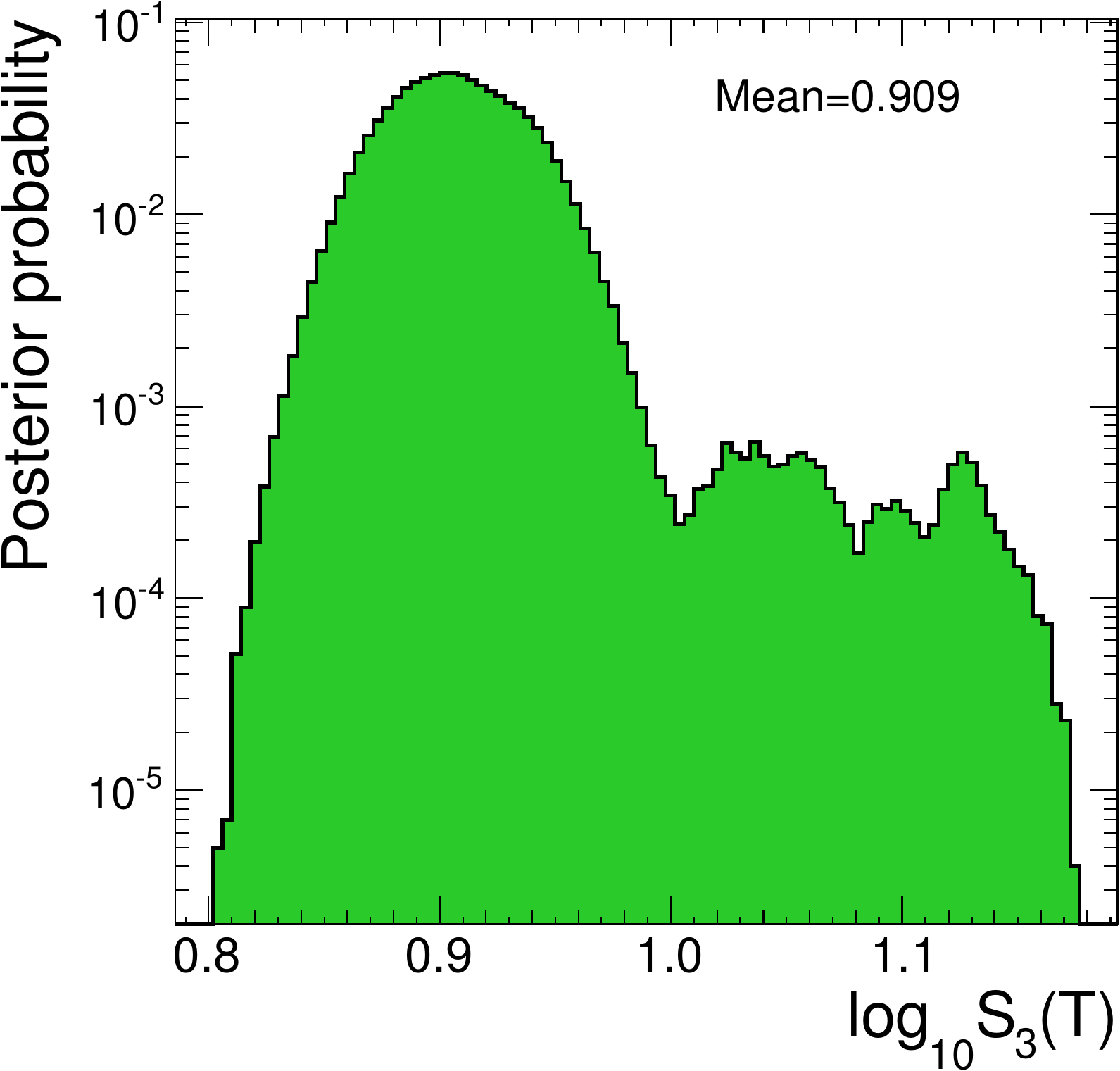}
  }
  \subfigure[$\alpha=20$]{
    \includegraphics[width=0.3\columnwidth]{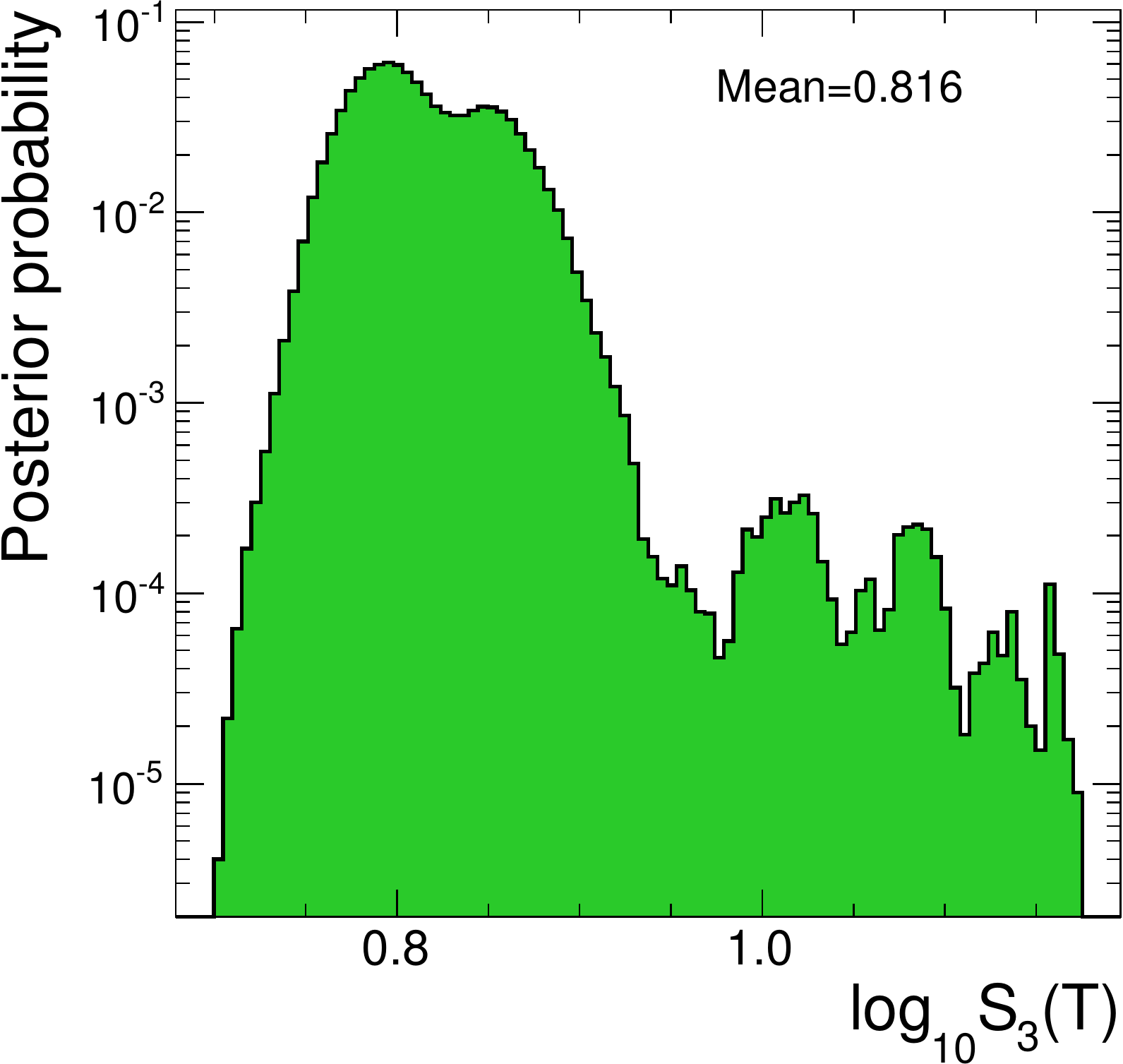}
 }
\caption{The posterior $P(S_3(\tuple{T})|\tuple{D})$, for three different choices of the regularization parameter $\alpha$, corresponding to Sec.~\ref{sec:regSteepBump}.
\label{fig:regFuncSteepBumpS3} 
}
\end{figure}

\begin{figure}[H]
  \centering
  \subfigure[$\alpha=0$]{
    \includegraphics[width=0.3\columnwidth]{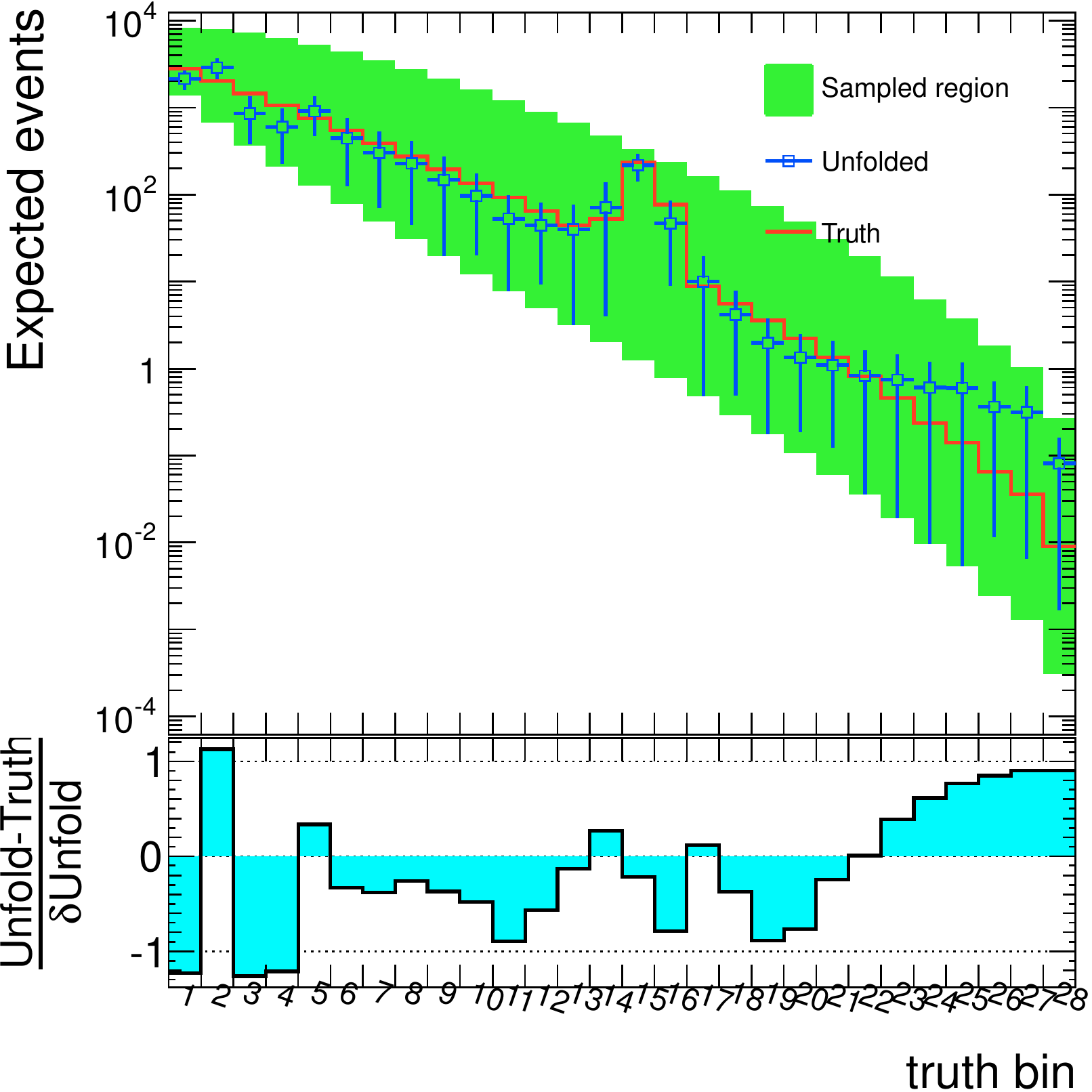}
  }
  \subfigure[$\alpha=10$]{
    \includegraphics[width=0.3\columnwidth]{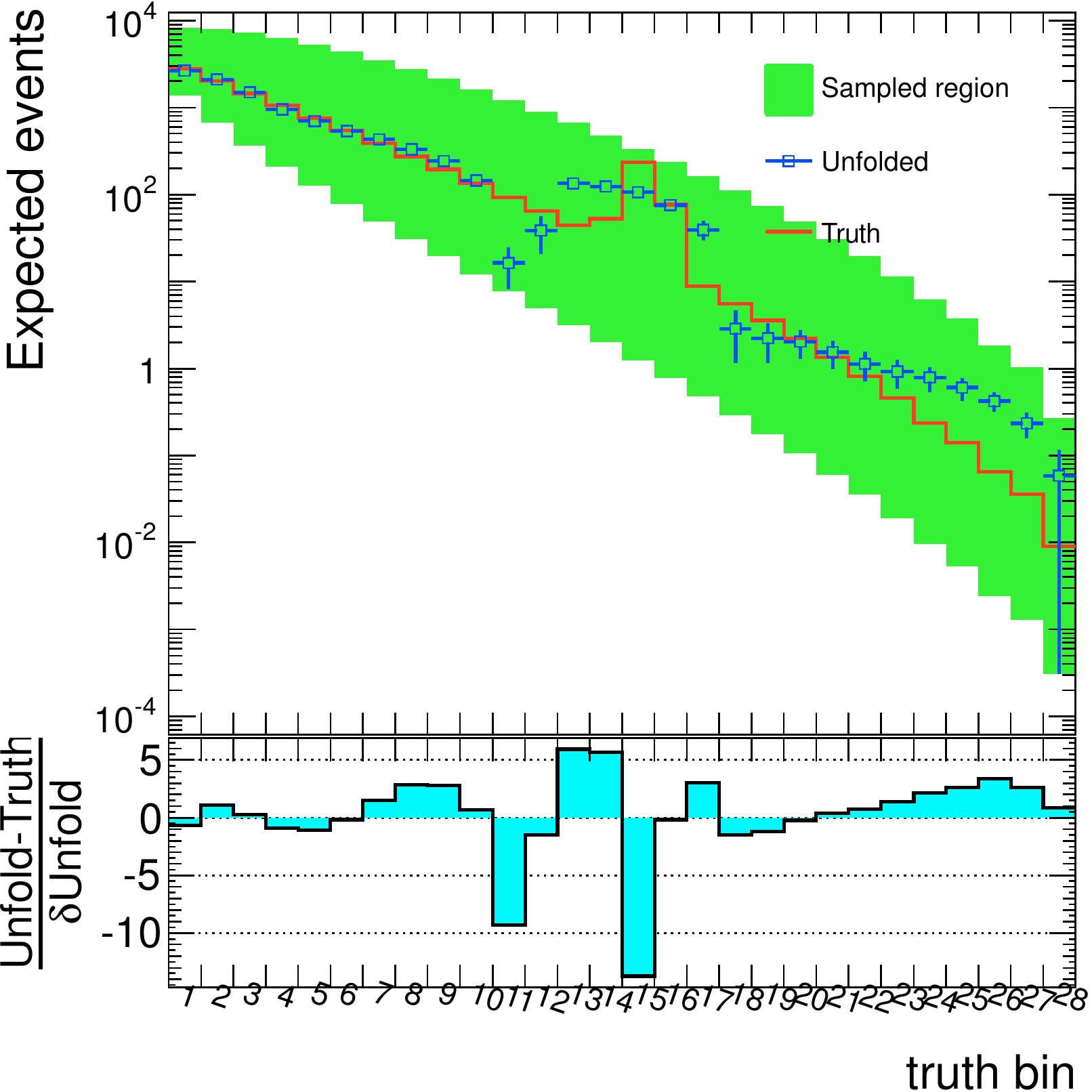}
  }
  \subfigure[$\alpha=20$]{
    \includegraphics[width=0.3\columnwidth]{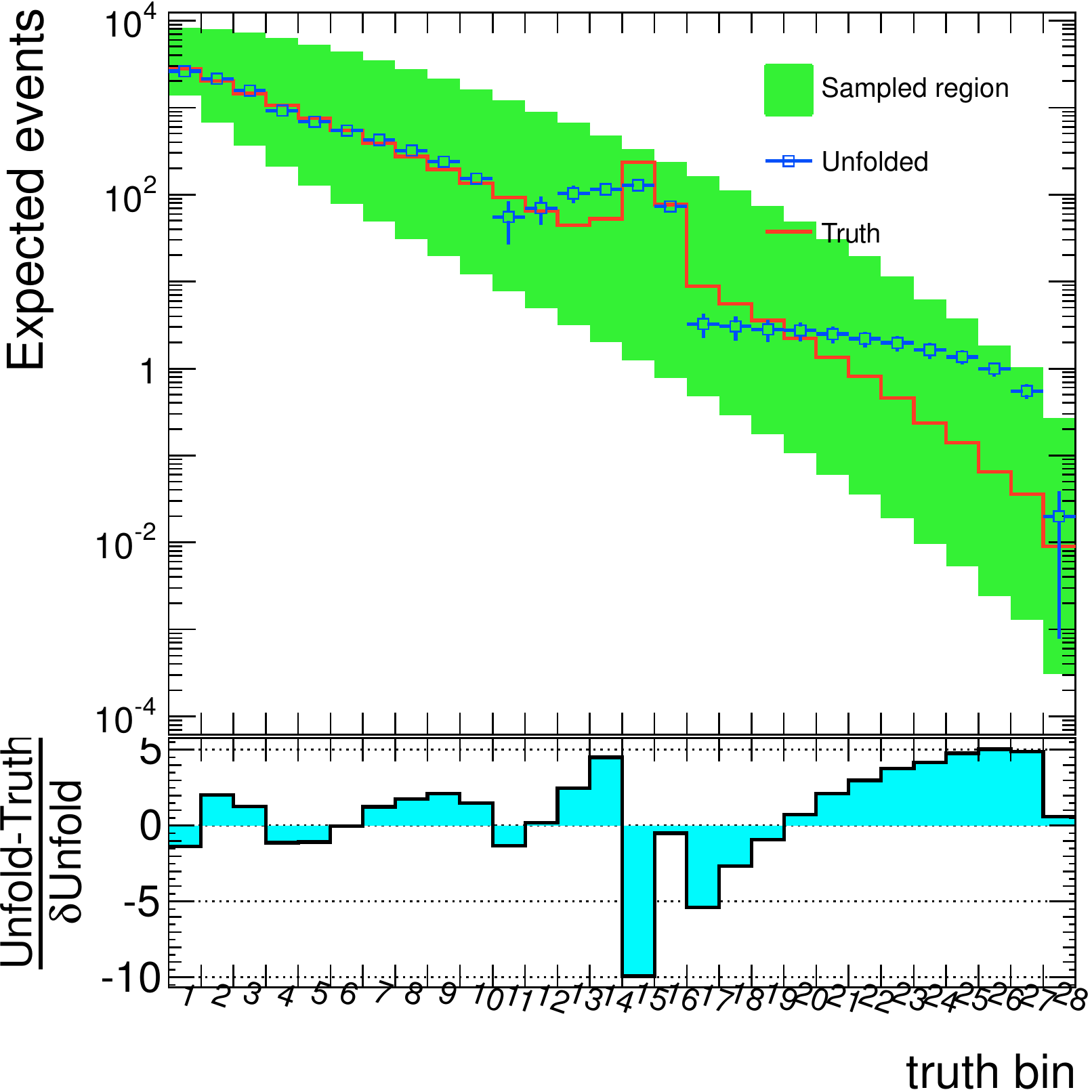}
  }\\
 \subfigure[$\alpha=0$]{
    \includegraphics[width=0.3\columnwidth]{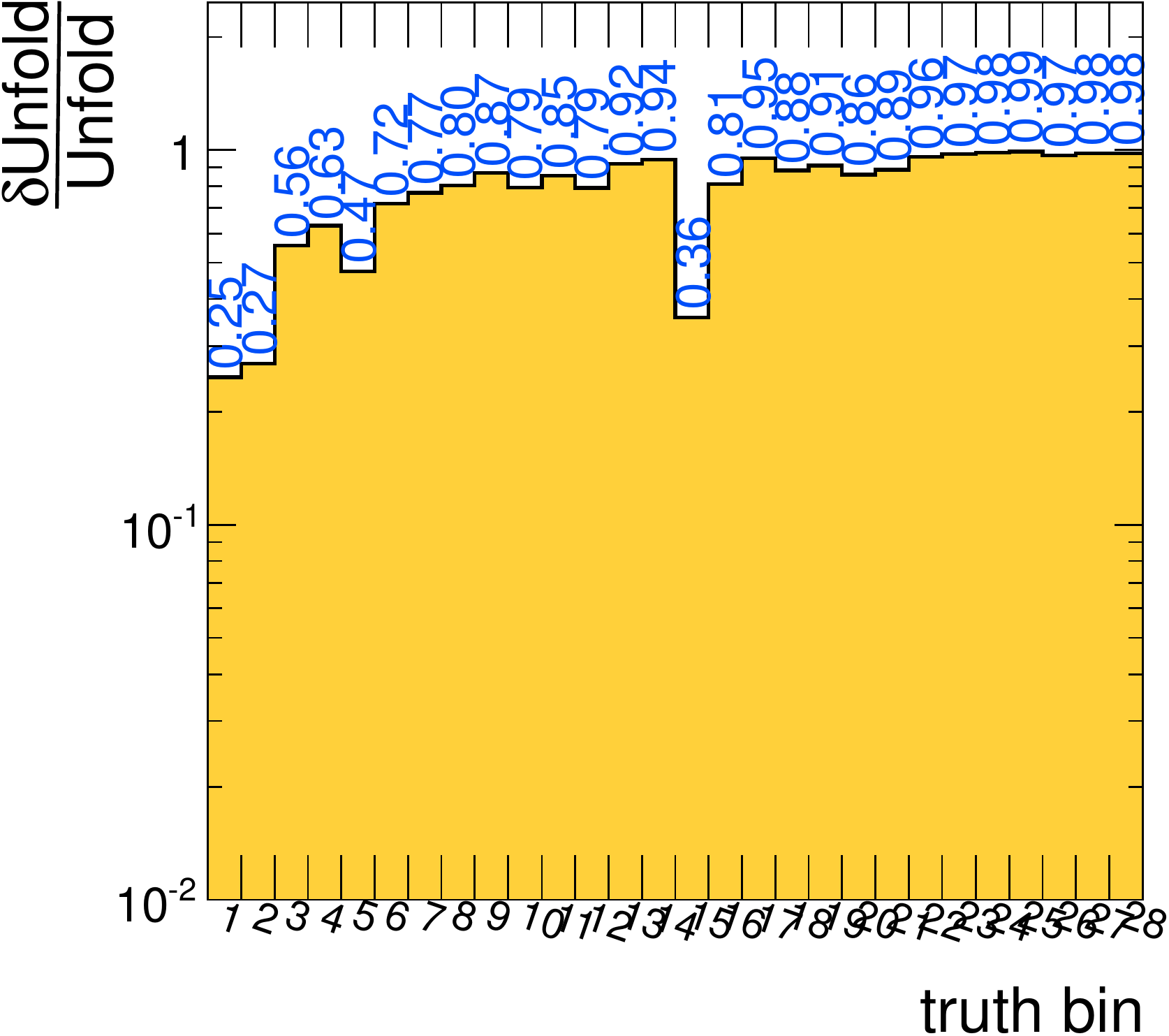}
  }
  \subfigure[$\alpha=10$]{
    \includegraphics[width=0.3\columnwidth]{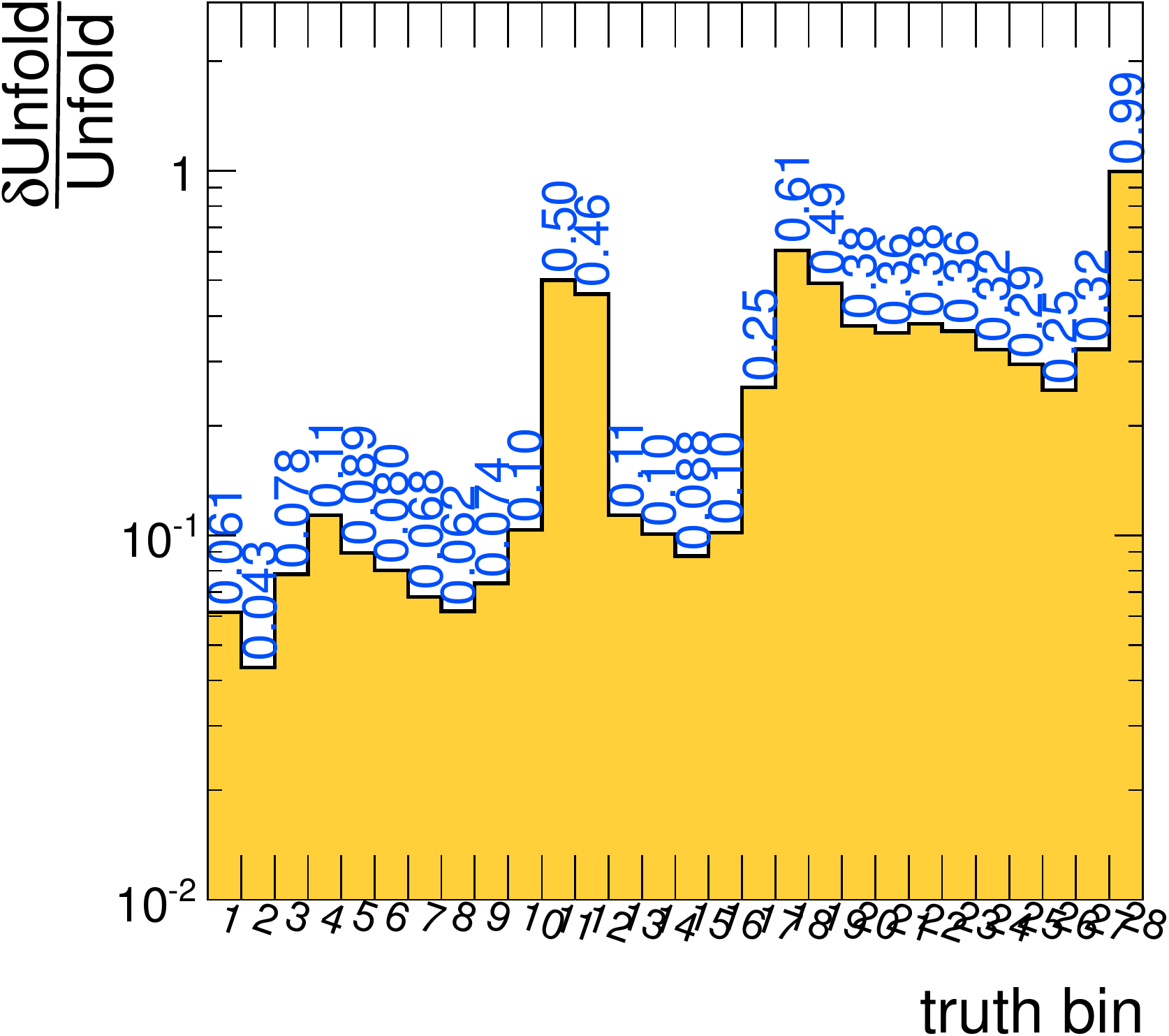}
  }
  \subfigure[$\alpha=20$]{
    \includegraphics[width=0.3\columnwidth]{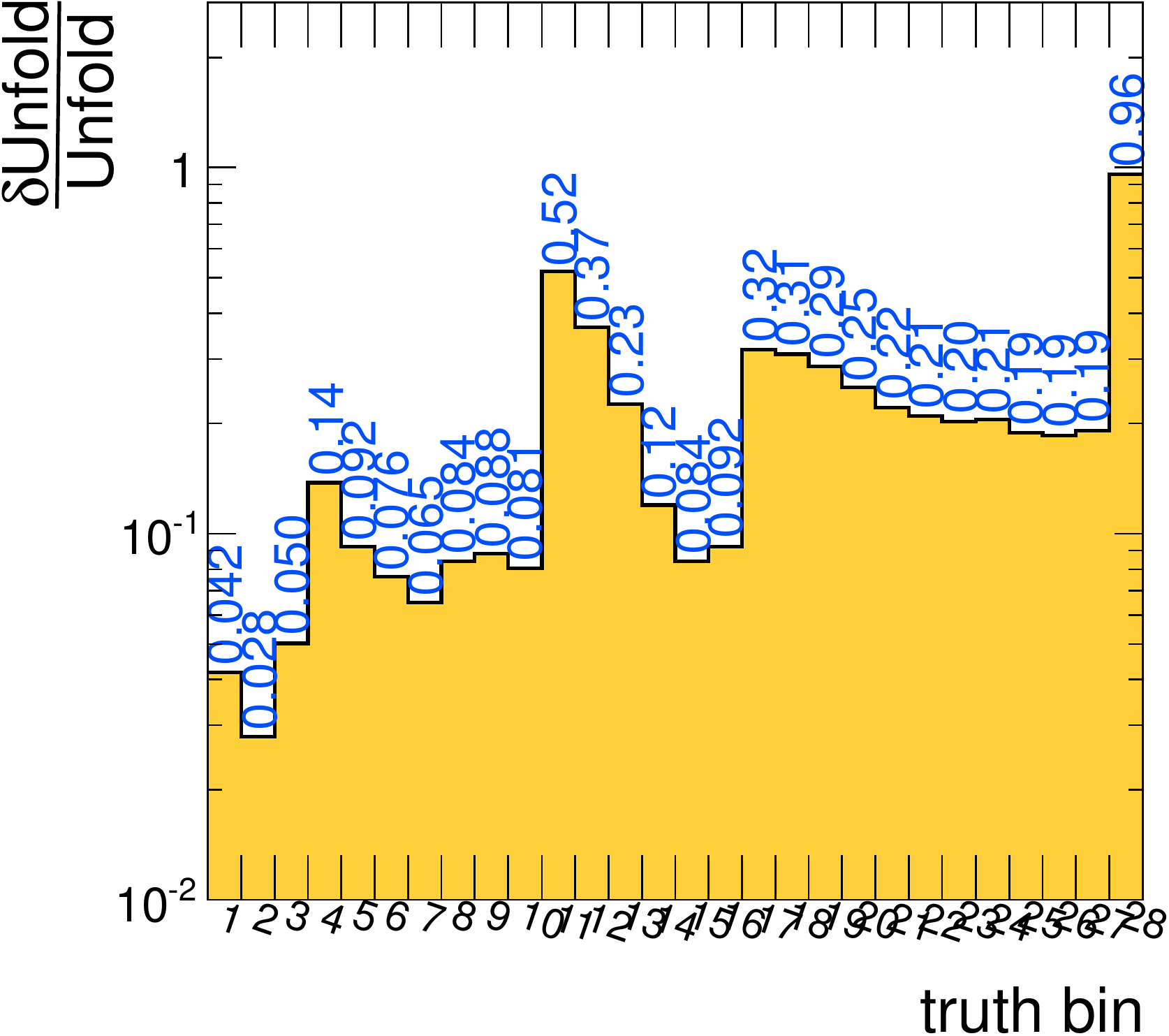}
  }
  \caption{The result of unfolding of Sec.~\ref{sec:regSteepBump}, with regularization function $S_3$, for three $\alpha$ values.  
    \label{fig:unfoldSteepBumpS3}
  }
\end{figure}

Using the regularization function $S_3$ is shown in Fig.~\ref{fig:regFuncSteepBumpS3} and \ref{fig:unfoldSteepBumpS3}. The unfolded spectrum tends to be flat in long intervals, thus obscuring the shape of the bump.

\begin{figure}[H]
  \centering
  \subfigure[$\alpha=0$]{
    \includegraphics[width=0.3\columnwidth]{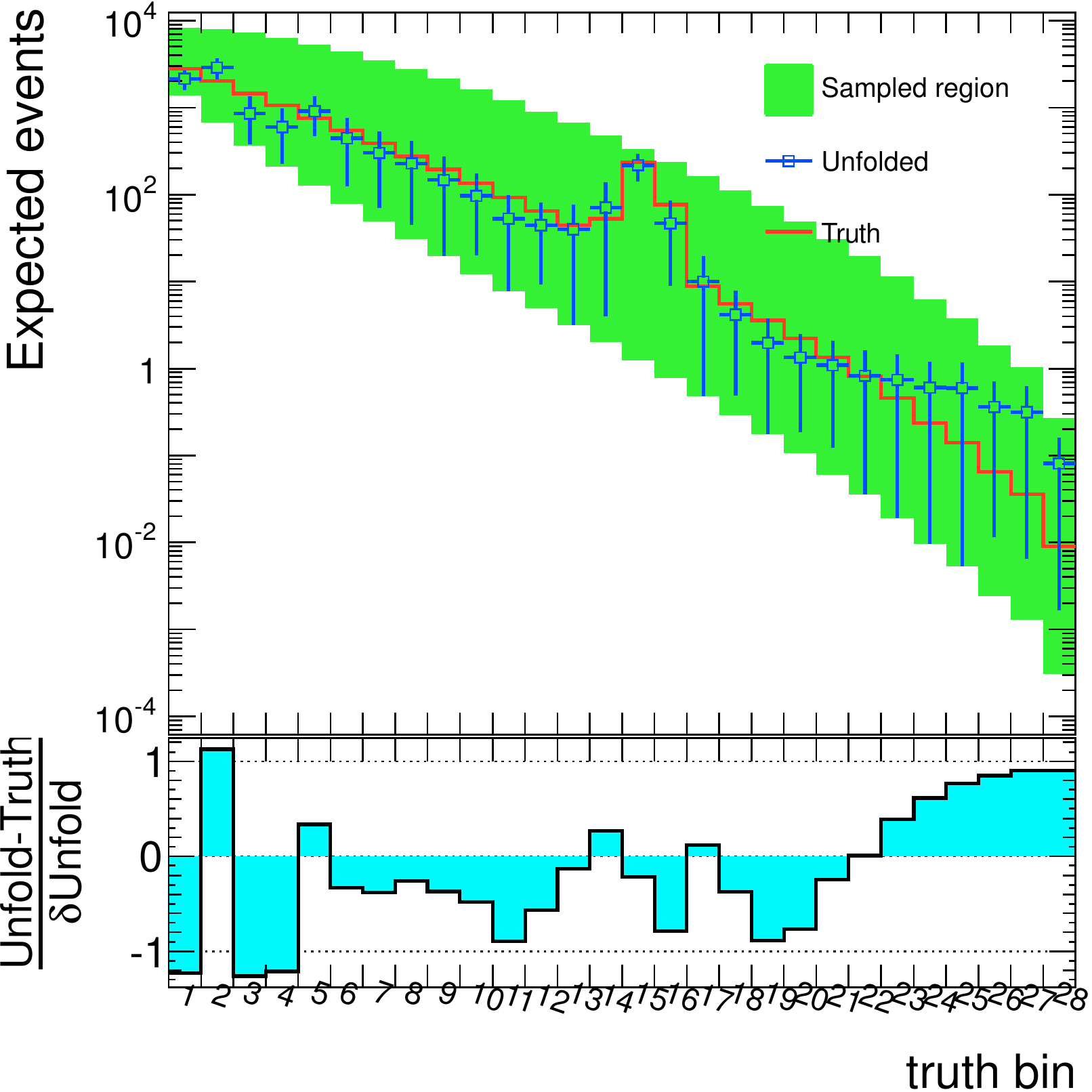}
  }
  \subfigure[$\alpha=1$]{
    \includegraphics[width=0.3\columnwidth]{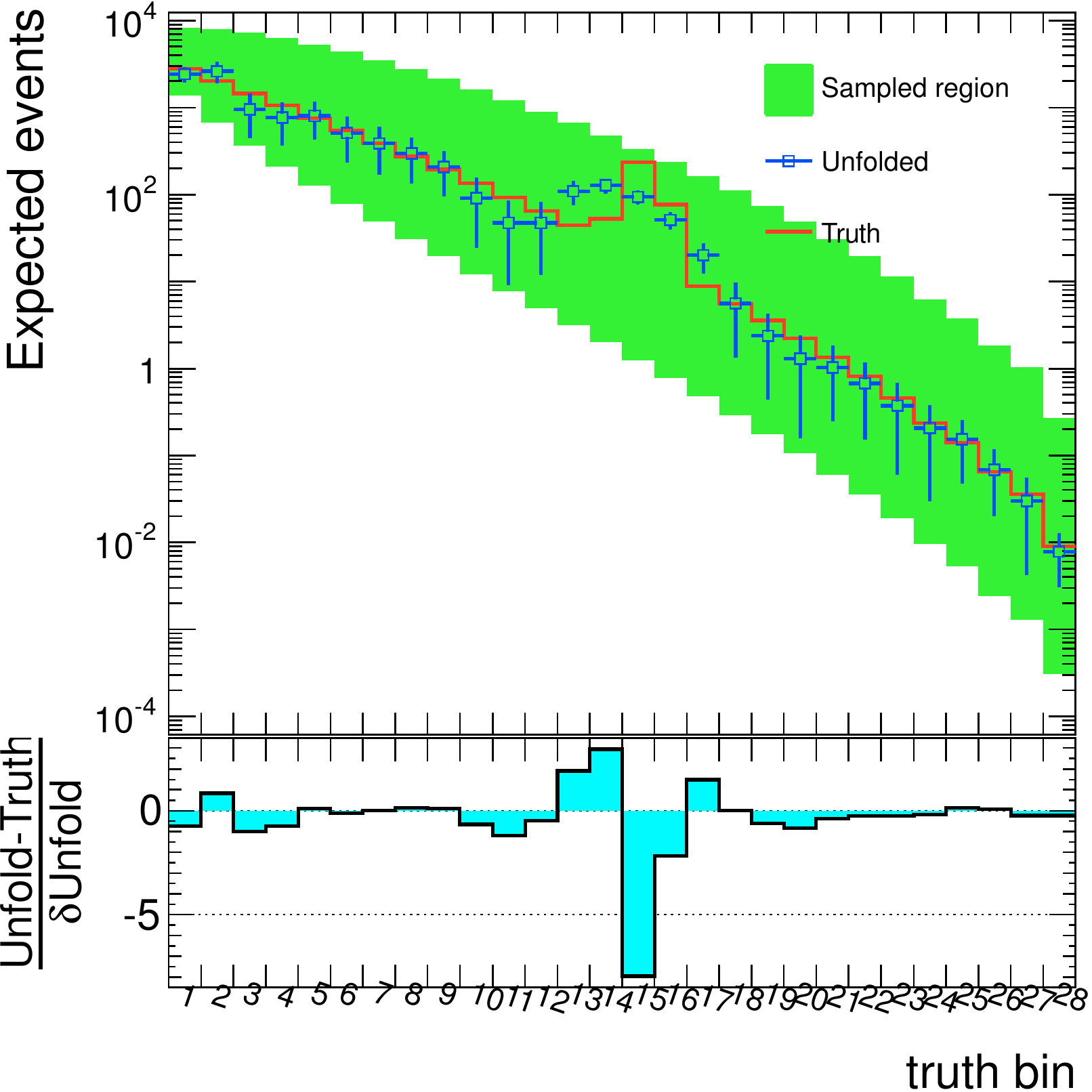}
  }
  \subfigure[$\alpha=10$]{
    \includegraphics[width=0.3\columnwidth]{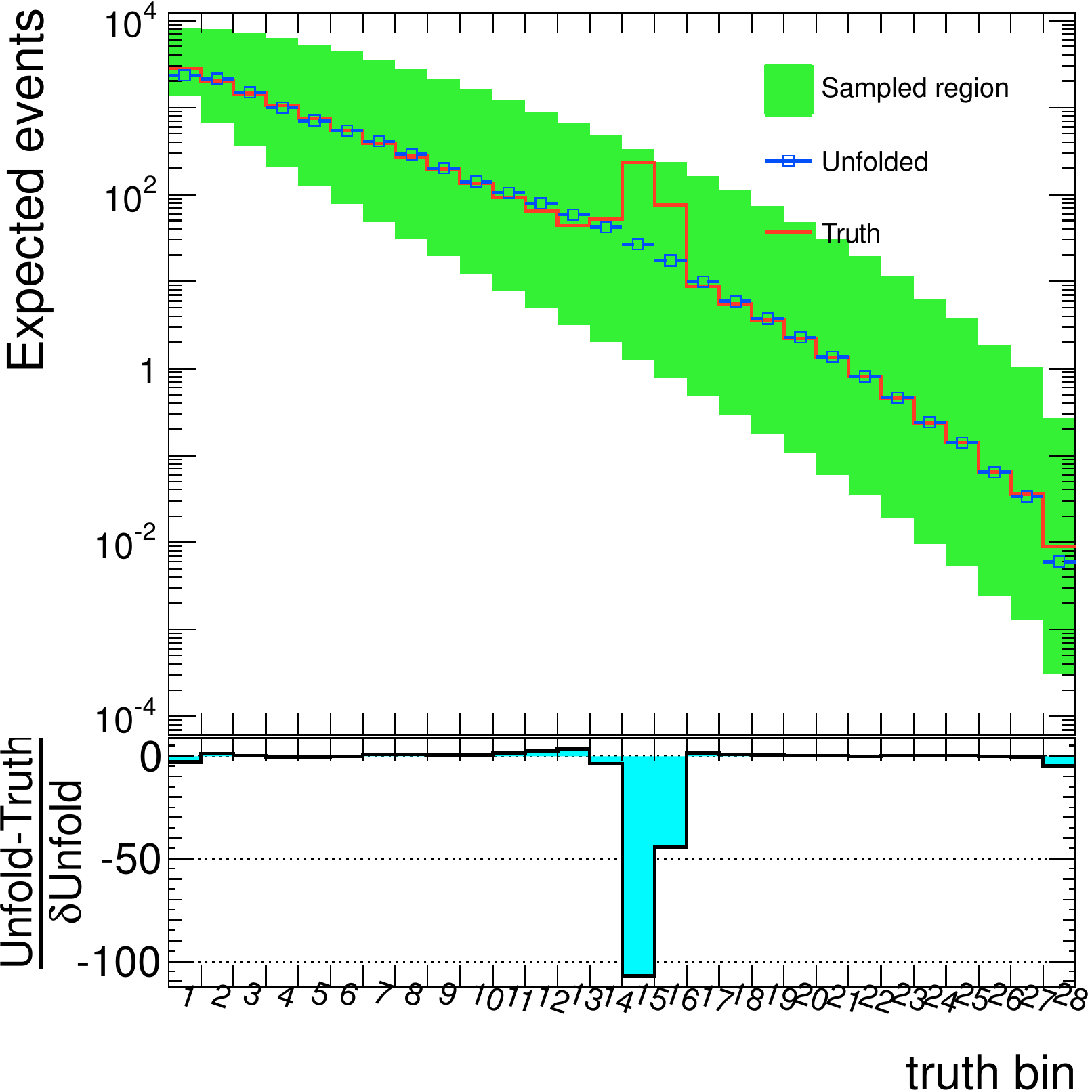}
  }\\
 \subfigure[$\alpha=0$]{
    \includegraphics[width=0.3\columnwidth]{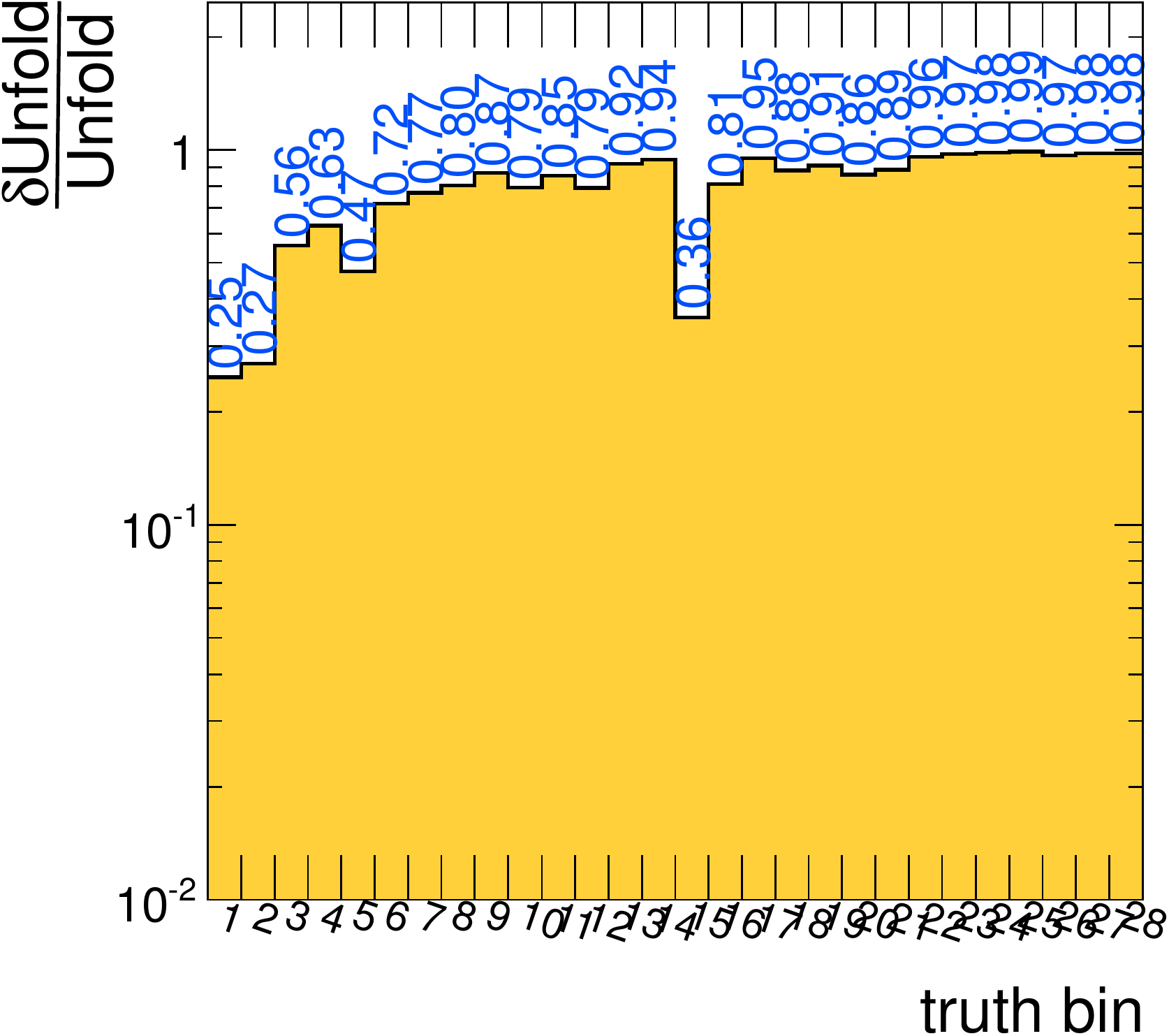}
  }
  \subfigure[$\alpha=1$]{
    \includegraphics[width=0.3\columnwidth]{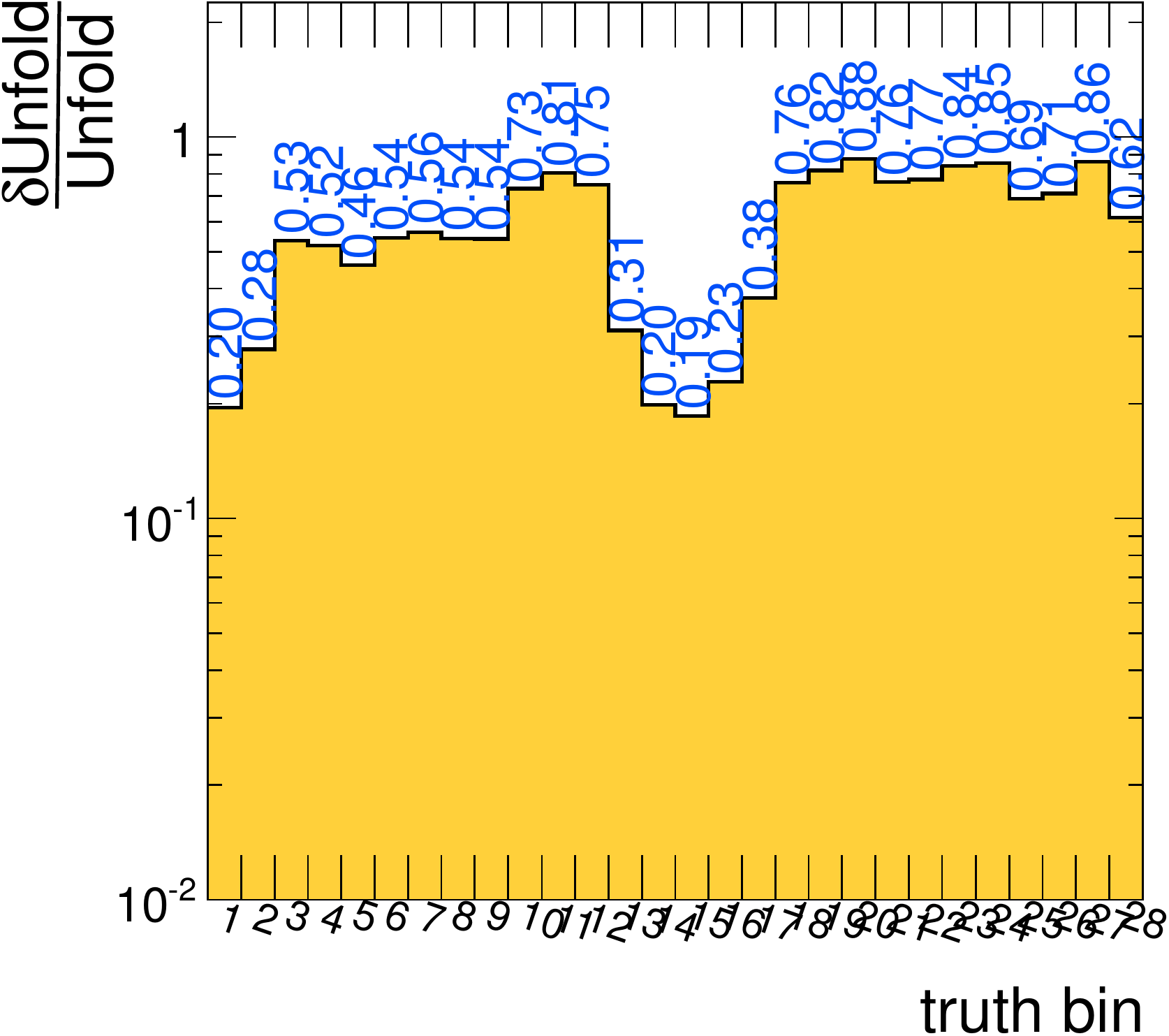}
  }
  \subfigure[$\alpha=10$]{
    \includegraphics[width=0.3\columnwidth]{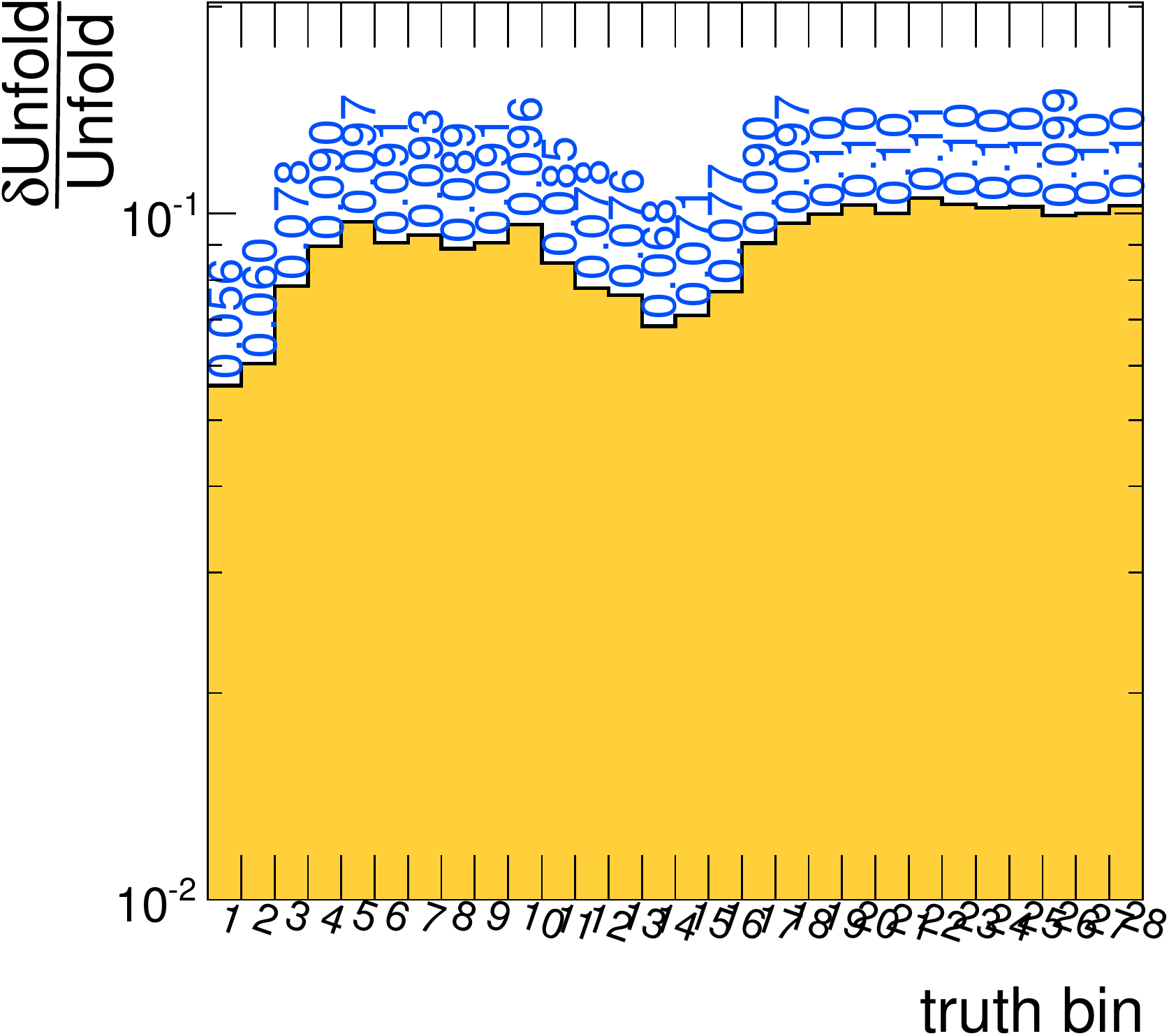}
  }
  \caption{The result of unfolding of Sec.~\ref{sec:regSteepBump}, with Gaussian regularization, for three $\alpha$ values.  
    \label{fig:unfoldSteepBumpGaus}
  }
\end{figure}

Finally, FBU with a Gaussian constraint is shown in Fig.~\ref{fig:unfoldSteepBumpGaus}.  As expected, a stronger constraint leads to a narrow posterior which concentrates around $\tilde{\T}$ instead of $\hat{\T}$.  Mild values of $\alpha$, such as $\alpha=1$, don't tend to make the bump in the unfolded spectrum as narrow as it is at truth-level; the bump maintains its reco-level width.

None of the regularization methods tried seems to improve how the bump appears in the unfolded spectrum.  Its shape is either obscured, or it remains as wide as it is in the data, after smearing.  When having no regularization, nevertheless, the unfolded spectrum peaks in a narrow region, similar to how narrow the bump is in $\hat{\T}$ (see examples with $\alpha=0$ in any of Fig.~\ref{fig:unfoldSteepBumpGaus}, \ref{fig:unfoldSteepBumpS3}, \ref{fig:unfoldSteepBumpS2}, or \ref{fig:unfoldSteepBumpS1}).  The problem is that the posterior is spread wide without regularization, resulting in large error bars in the unfolded spectrum.

\subsection{Spectrum with an expected bump}
\label{sec:regSteepBumpExpected}

In this section, regularization is tested in the presence of an known, smeared bump.  

\begin{figure}[H]
\centering
\subfigure[]{
\includegraphics[height=0.4\columnwidth]{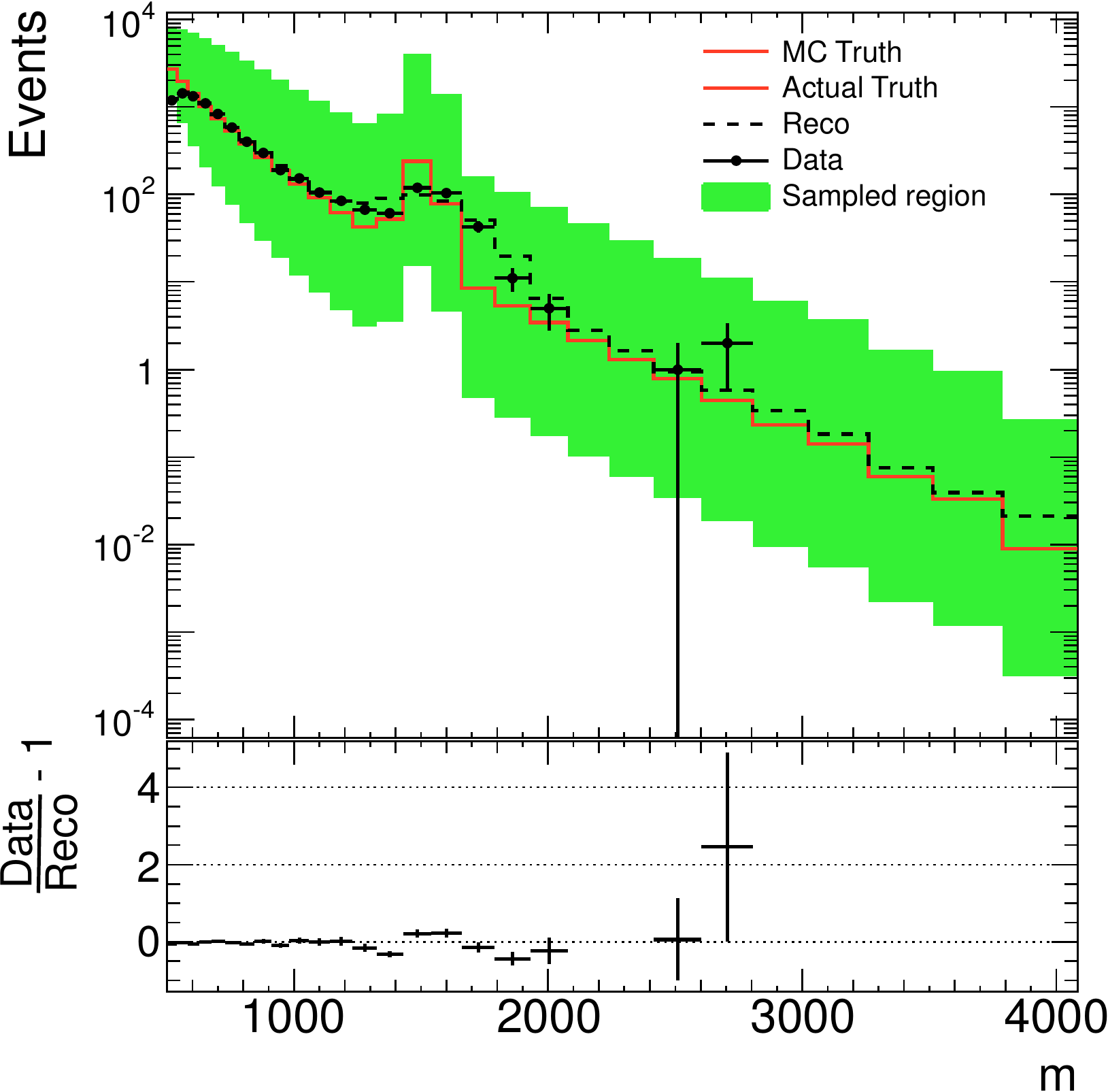}
}
\subfigure[]{
  \includegraphics[height=0.4\columnwidth]{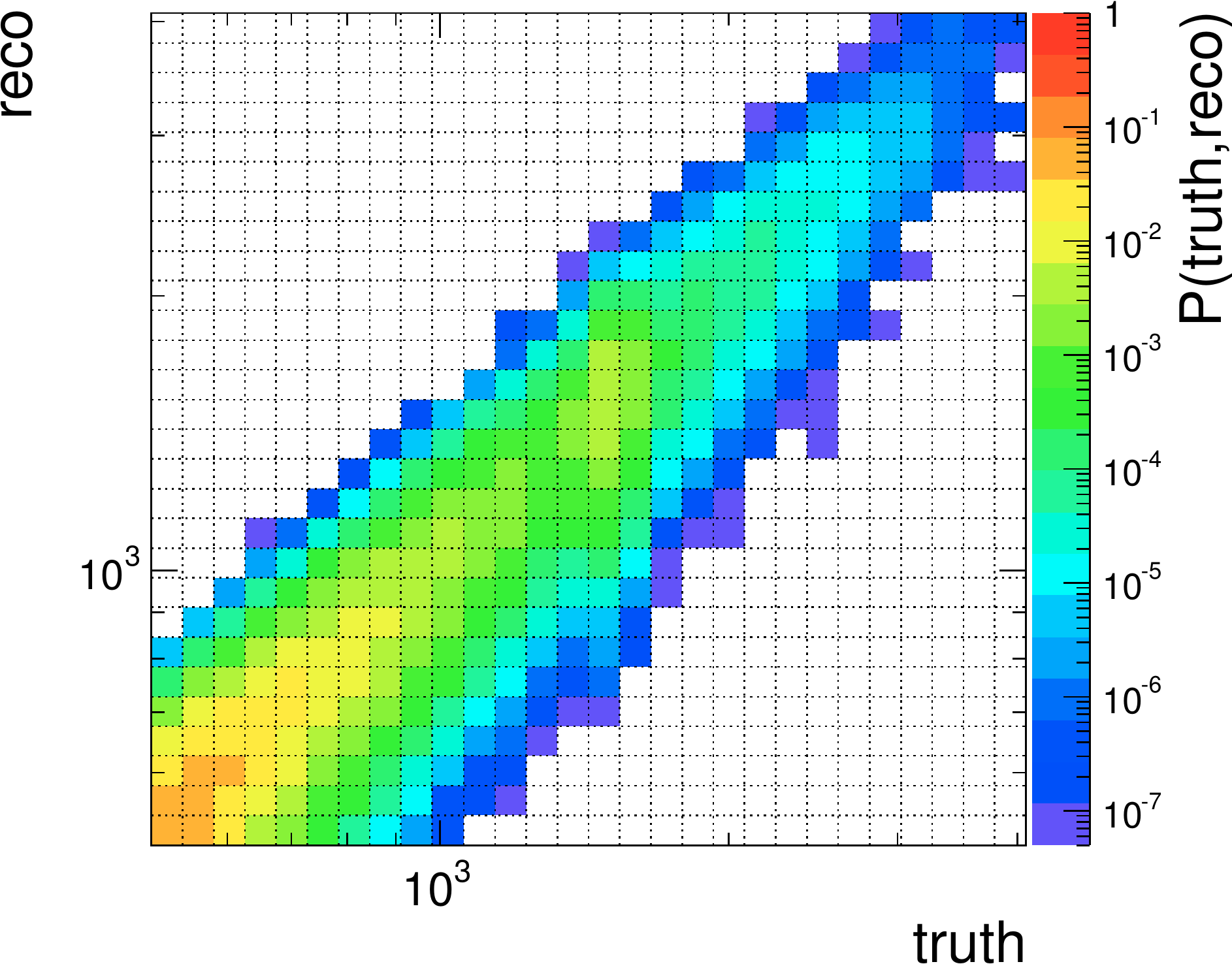}
}
\caption{(a) The MC truth-level spectrum $\tilde{\T}$ and the actual truth-level spectrum $\hat{\T}$, where $\tilde{\T} = \hat{\T}$, the reconstructed spectrum which corresponds to $\tilde{\T}$ after smearing, the data which follow $\hat{\T}$ after smearing, and the sampled hyper-box used in Sec.~\ref{sec:regSteepBumpExpected}.  (b) The migrations matrix, populated with the MC events that compose the MC truth level ($\tilde{\T} = \hat{\T}$) and the reco spectrum of (a).
\label{fig:regGenSteepBumpExpected}
}
\end{figure}

The data contain the same feature as in Sec.~\ref{sec:regSteepBump}, except that here the MC contains the same feature at truth-level ($\tilde{\T} = \hat{\T}$), so, the migrations matrix is aware of it.
The input spectra, the migrations matrix, and the sampled region are shown in Fig.~\ref{fig:regGenSteepBumpExpected}.  The sampled region is defined, as before, by Eq.~\ref{eq:priorBigger}.

\begin{figure}[H]
  \centering
  \subfigure[$\alpha=0$]{
    \includegraphics[width=0.3\columnwidth]{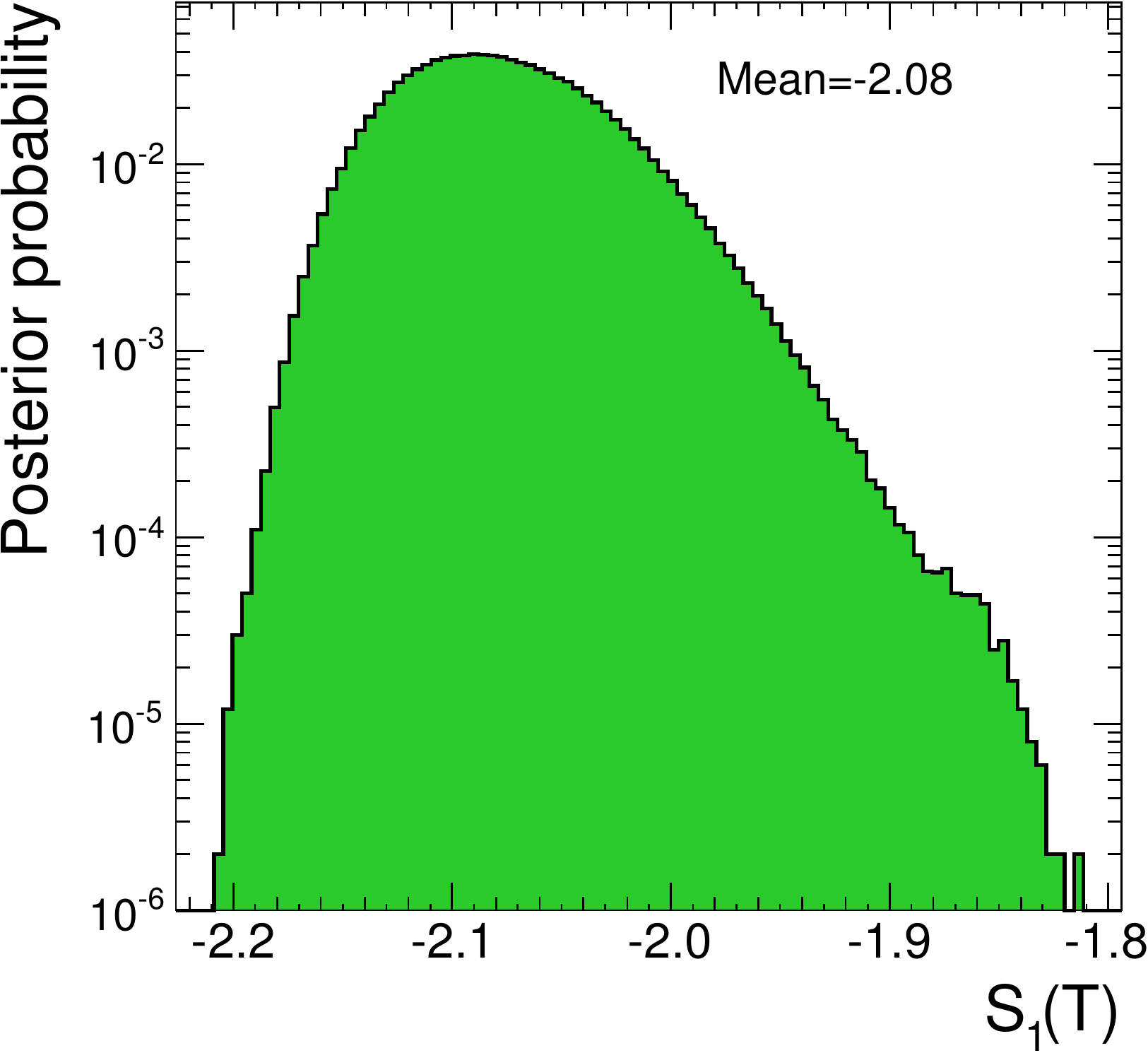}
  }
  \subfigure[$\alpha=10^3$]{
    \includegraphics[width=0.3\columnwidth]{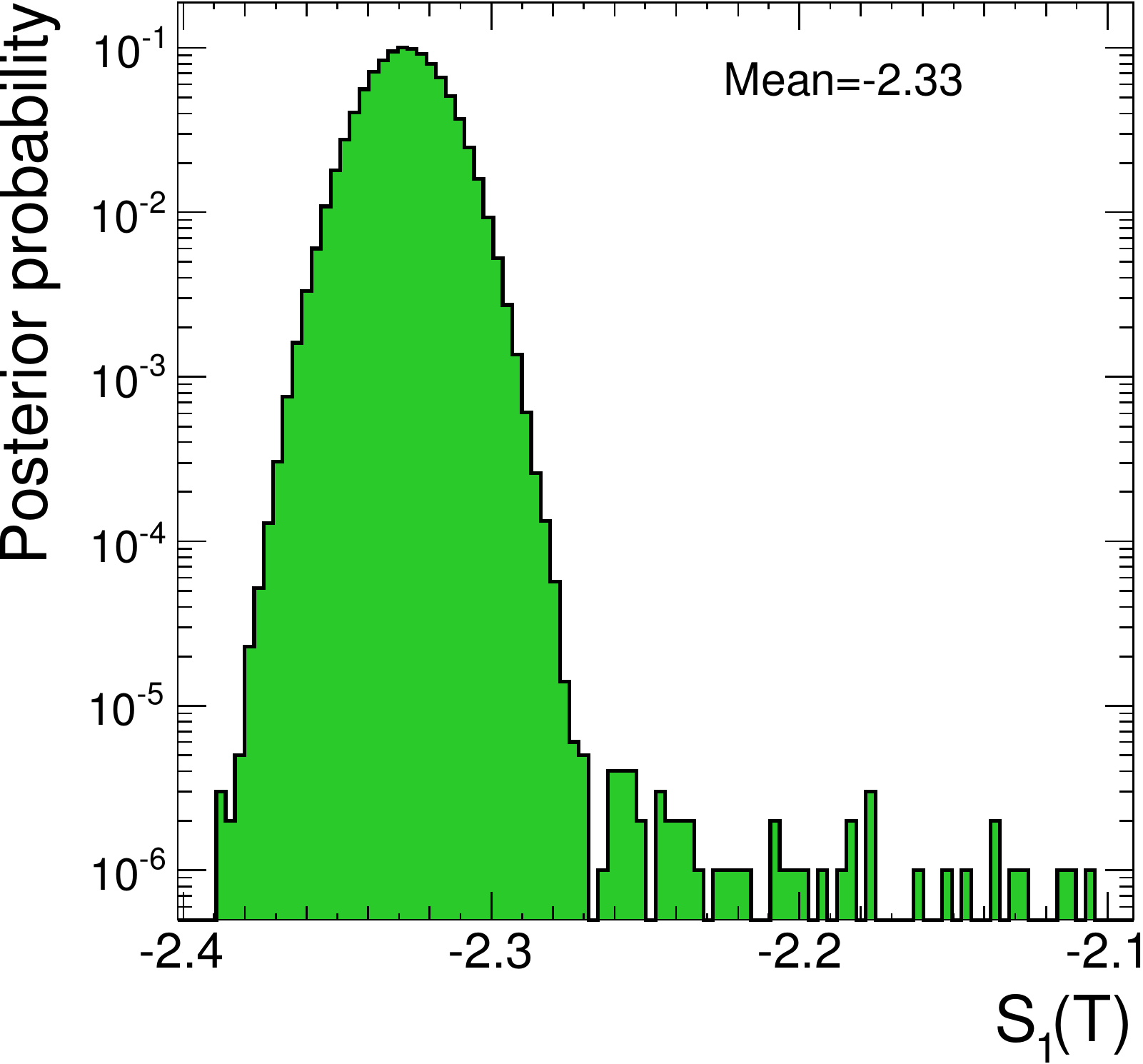}
  }
  \subfigure[$\alpha=3\times 10^3$]{
    \includegraphics[width=0.3\columnwidth]{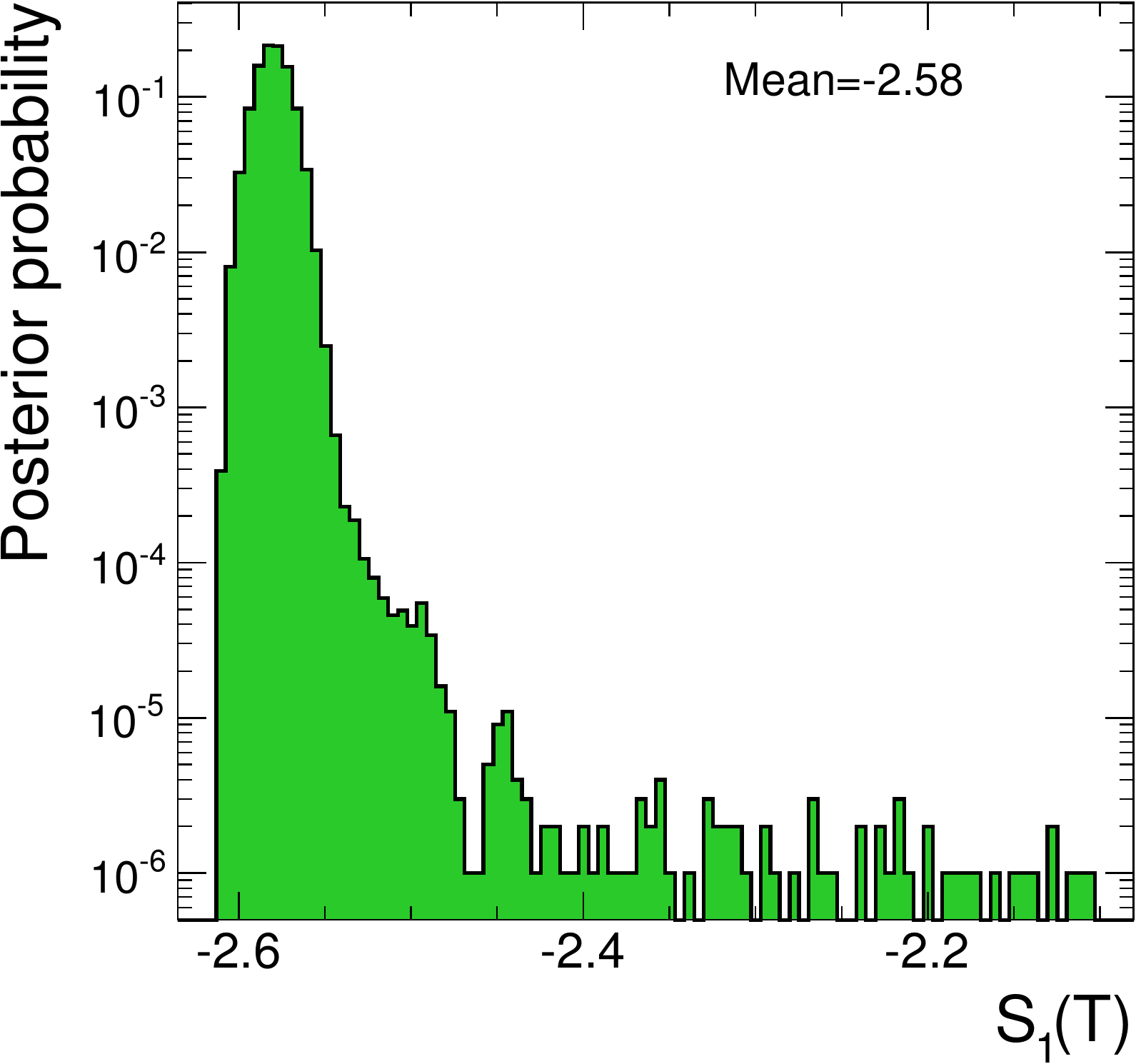}
 }
\caption{The posterior $P(S_1(\tuple{T})|\tuple{D})$, for three different choices of the regularization parameter $\alpha$, corresponding to Sec.~\ref{sec:regSteepBumpExpected}.
\label{fig:regFuncSteepBumpExpectedS1} 
}
\end{figure}

\begin{figure}[H]
  \centering
  \subfigure[$\alpha=0$]{
    \includegraphics[width=0.3\columnwidth]{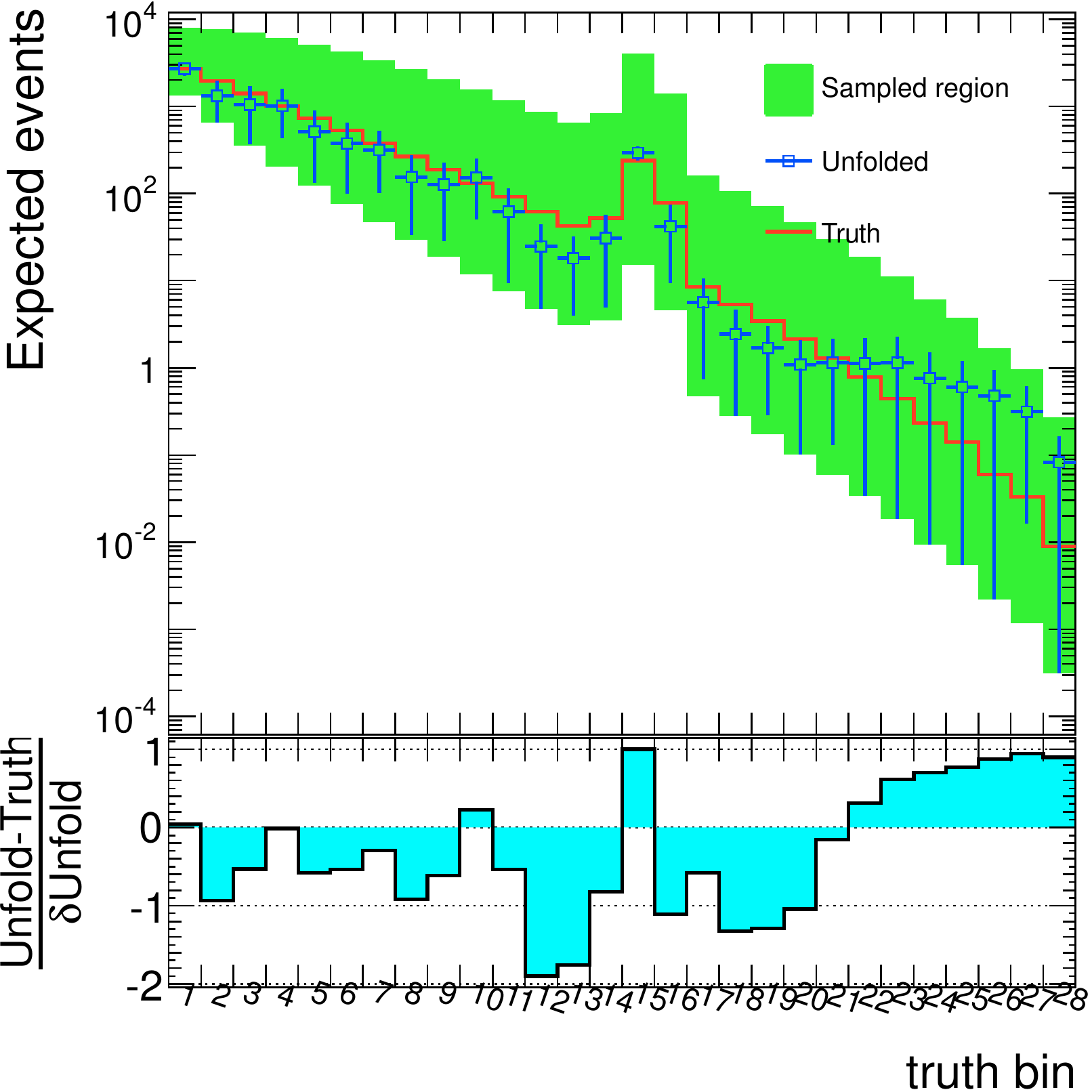}
  }
  \subfigure[$\alpha=10^3$]{
    \includegraphics[width=0.3\columnwidth]{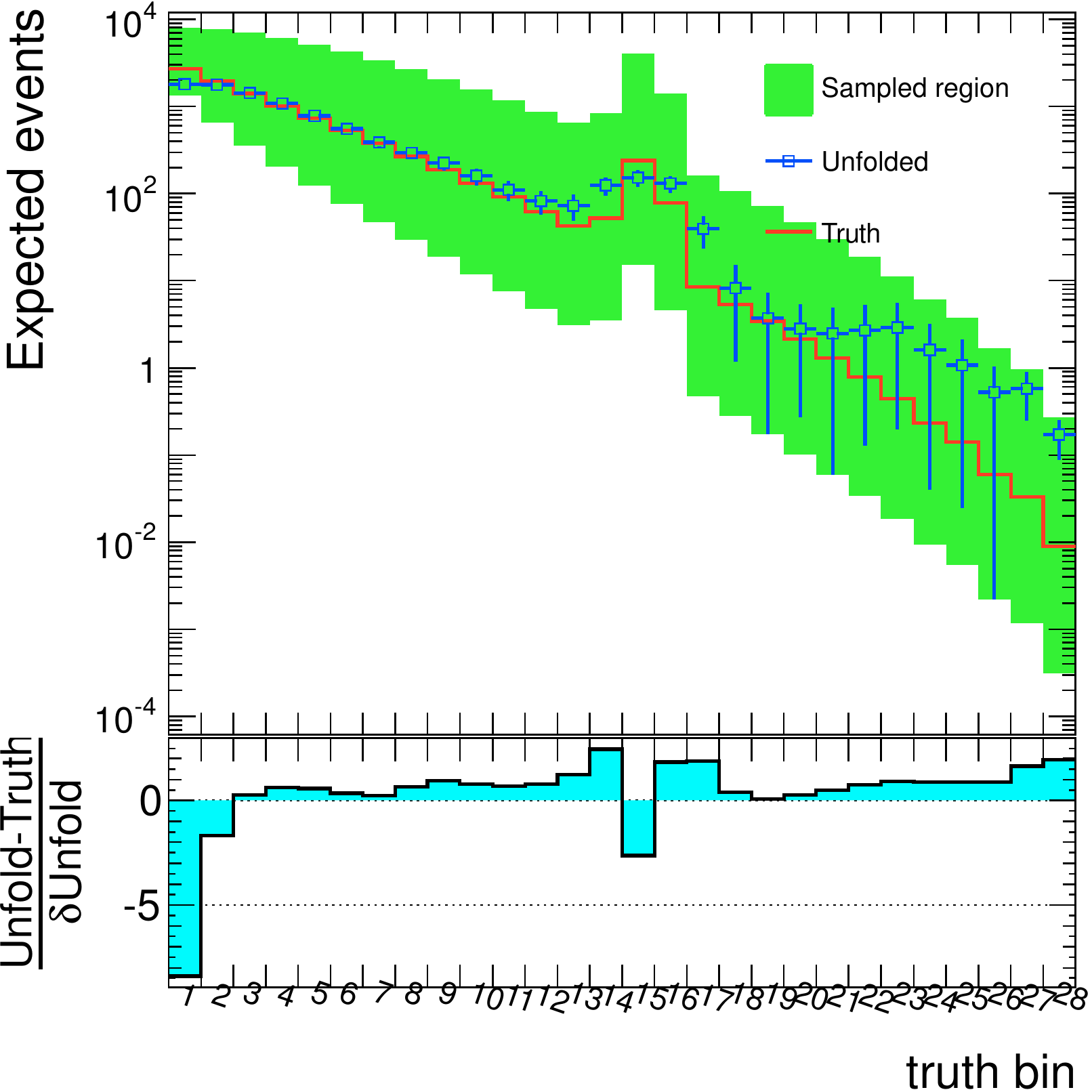}
  }
  \subfigure[$\alpha=3\times 10^3$]{
    \includegraphics[width=0.3\columnwidth]{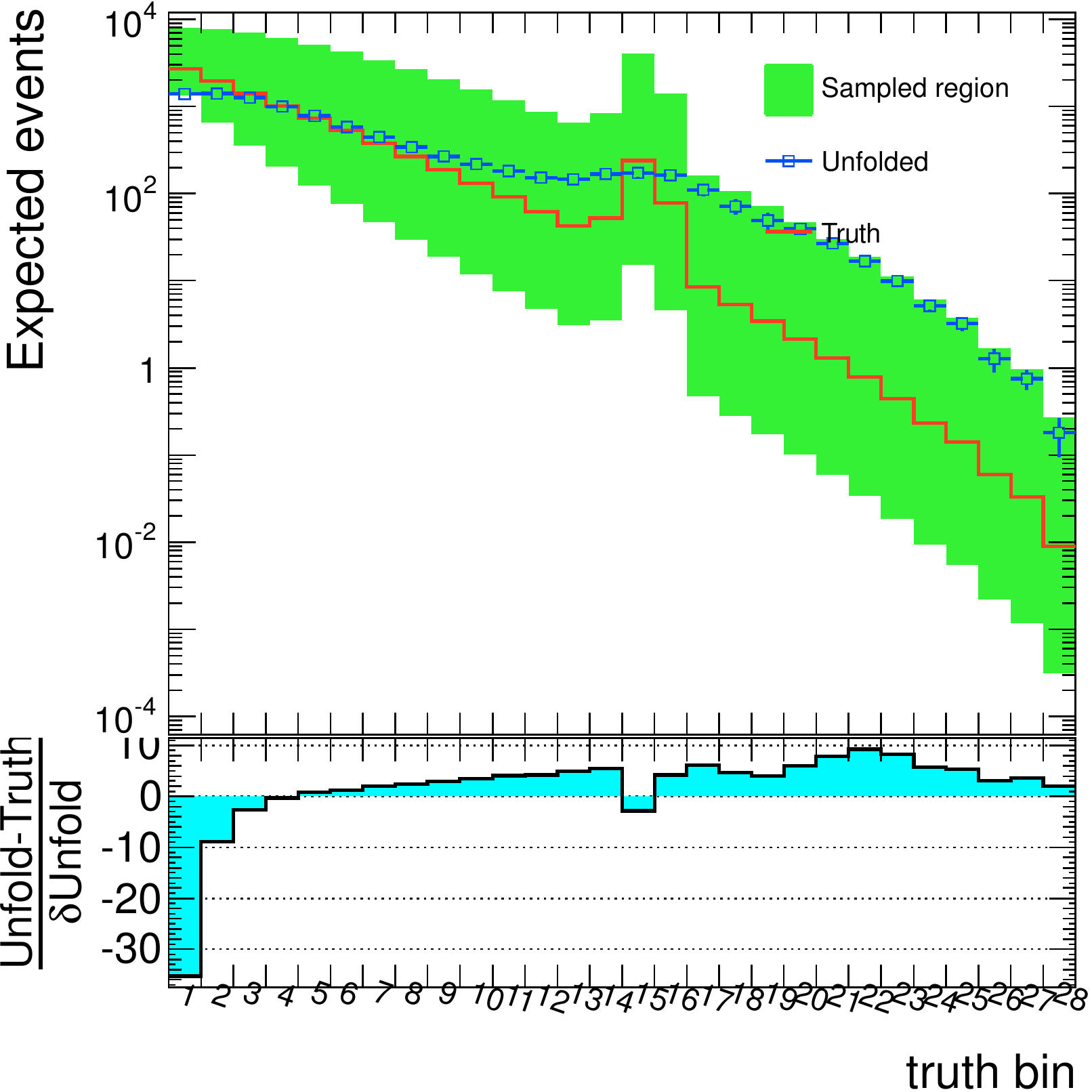}
  }\\
 \subfigure[$\alpha=0$]{
    \includegraphics[width=0.3\columnwidth]{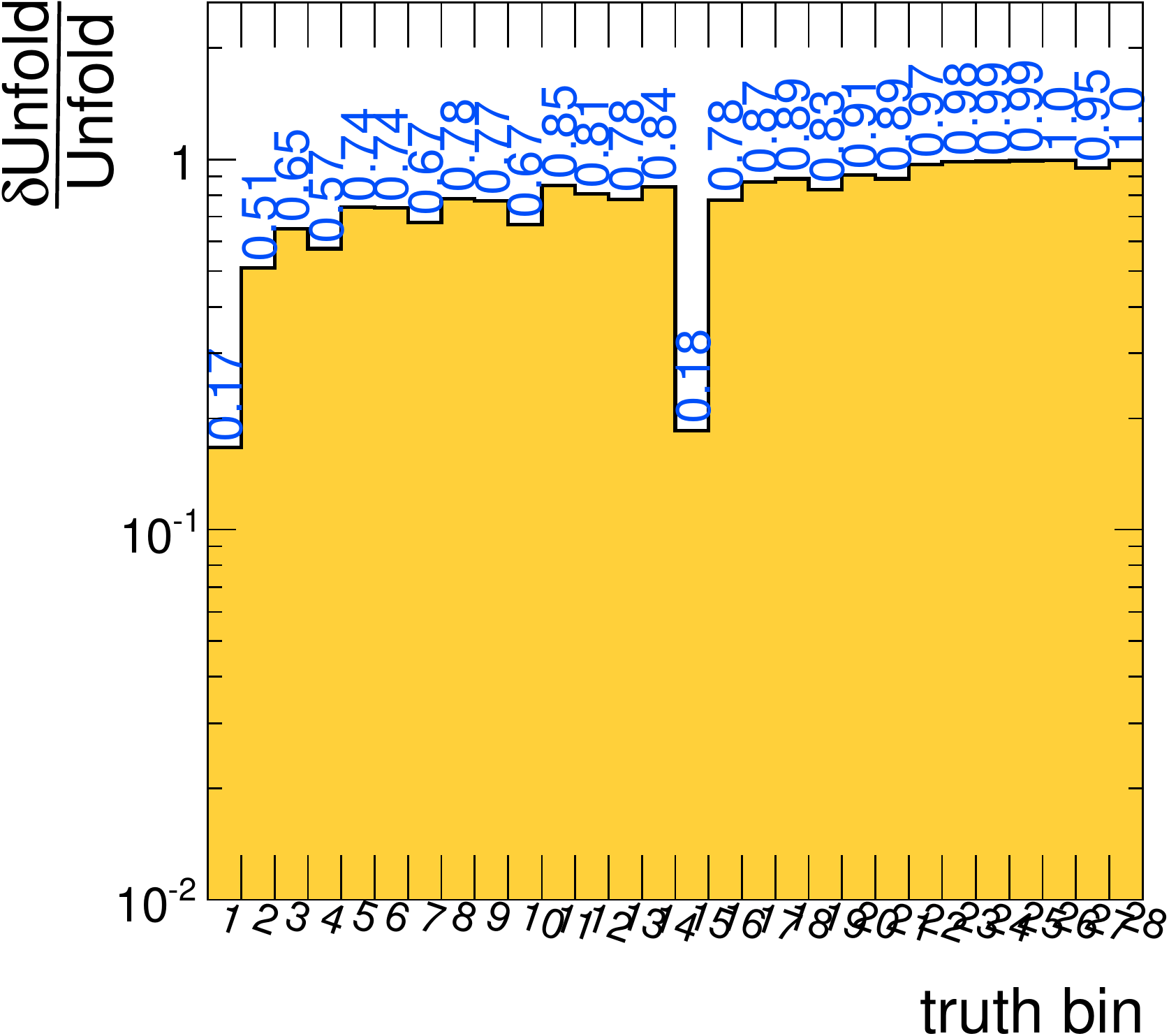}
  }
  \subfigure[$\alpha=10^3$]{
    \includegraphics[width=0.3\columnwidth]{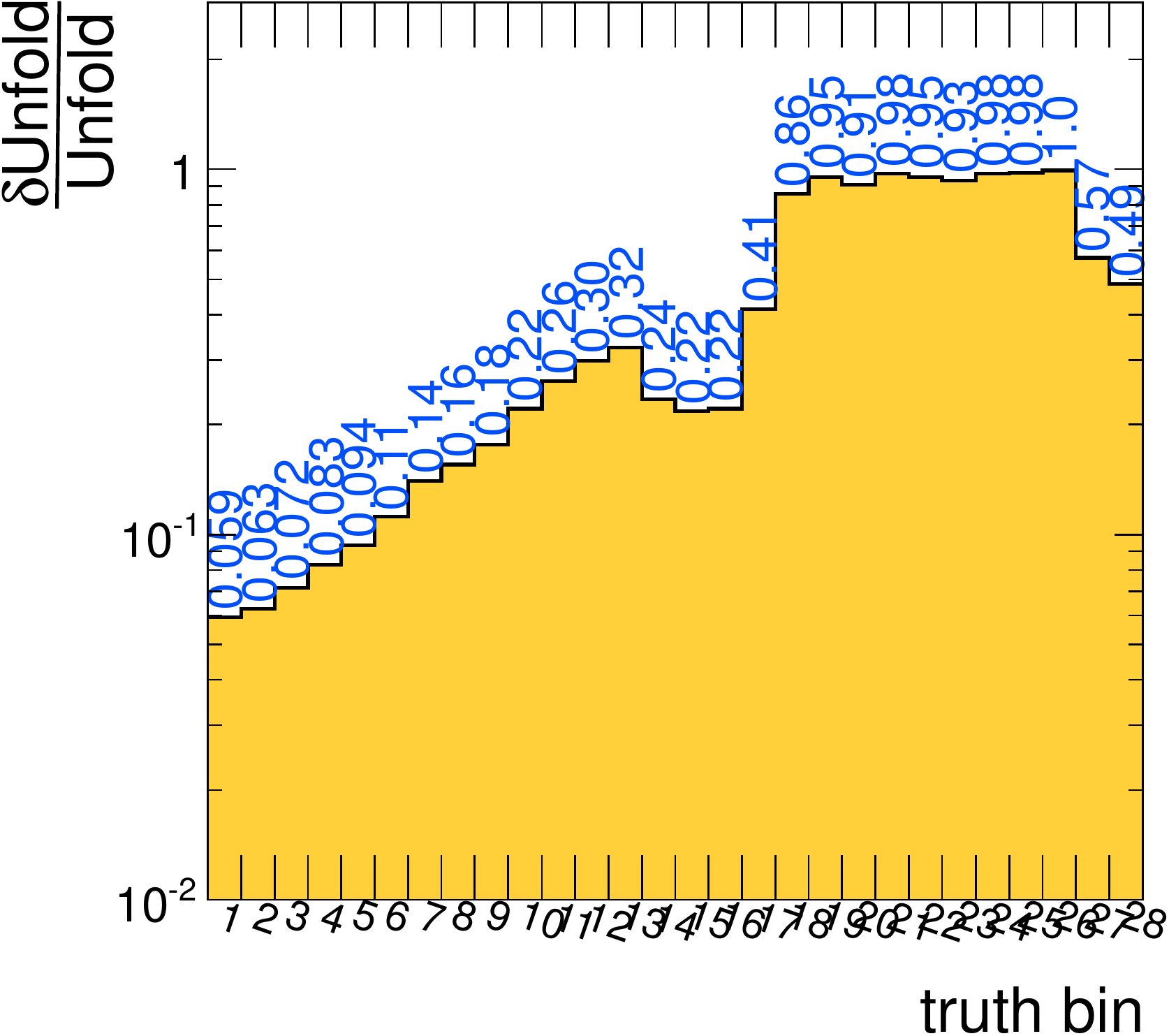}
  }
  \subfigure[$\alpha=3\times 10^3$]{
    \includegraphics[width=0.3\columnwidth]{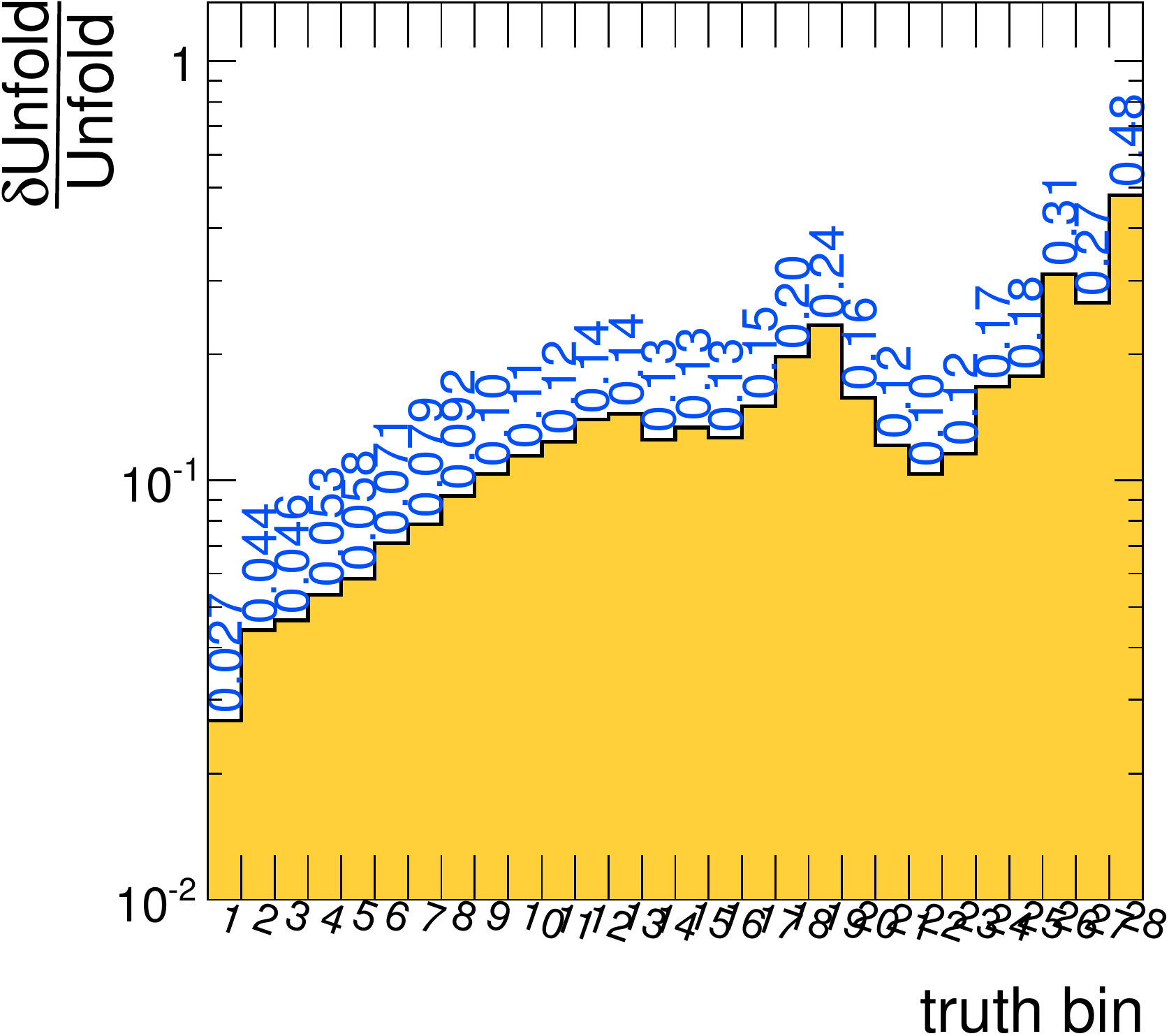}
  }
  \caption{The result of unfolding of Sec.~\ref{sec:regSteepBumpExpected}, with regularization function $S_1$, for three $\alpha$ values.  
    \label{fig:unfoldSteepBumpExpectedS1}
  }
\end{figure}

Unfolding with $S_1$ is shown in Fig.~\ref{fig:regFuncSteepBumpExpectedS1} and \ref{fig:unfoldSteepBumpExpectedS1}.  The results are not much different from Sec.~\ref{sec:regSteepBump}, Fig.~\ref{fig:unfoldSteepBumpS1}, where the bump was expected.  Entropy demands the unfolded histogram to be closer to horizontal, and the truth-level width of the bump is not resolved with this regularized unfolding.  It seems better resolved without regularization ($\alpha=0$), but the spread of the posterior is then larger.

\begin{figure}[H]
  \centering
  \subfigure[$\alpha=0$]{
    \includegraphics[width=0.3\columnwidth]{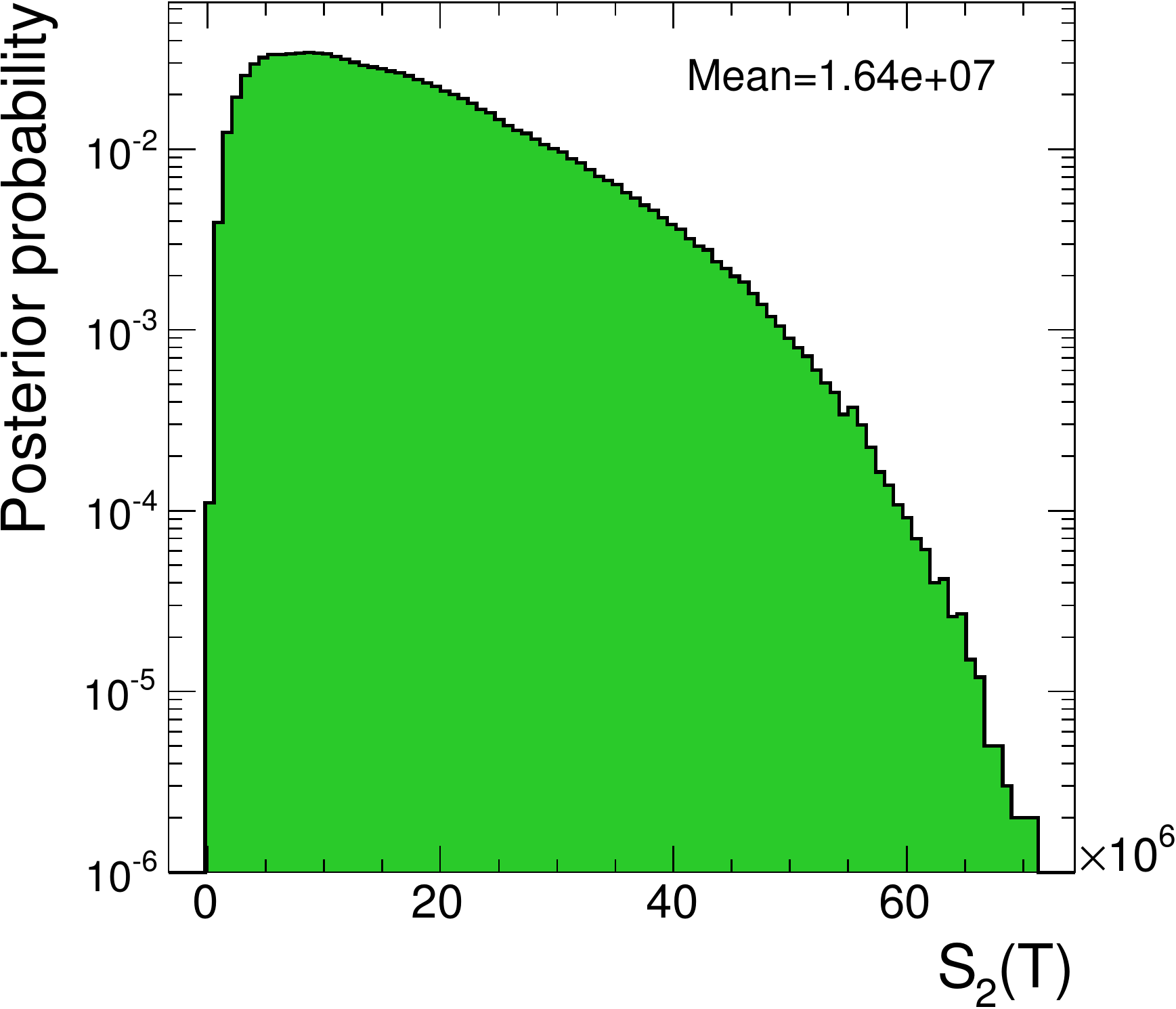}
  }
  \subfigure[$\alpha=3\times 10^{-4}$]{
    \includegraphics[width=0.3\columnwidth]{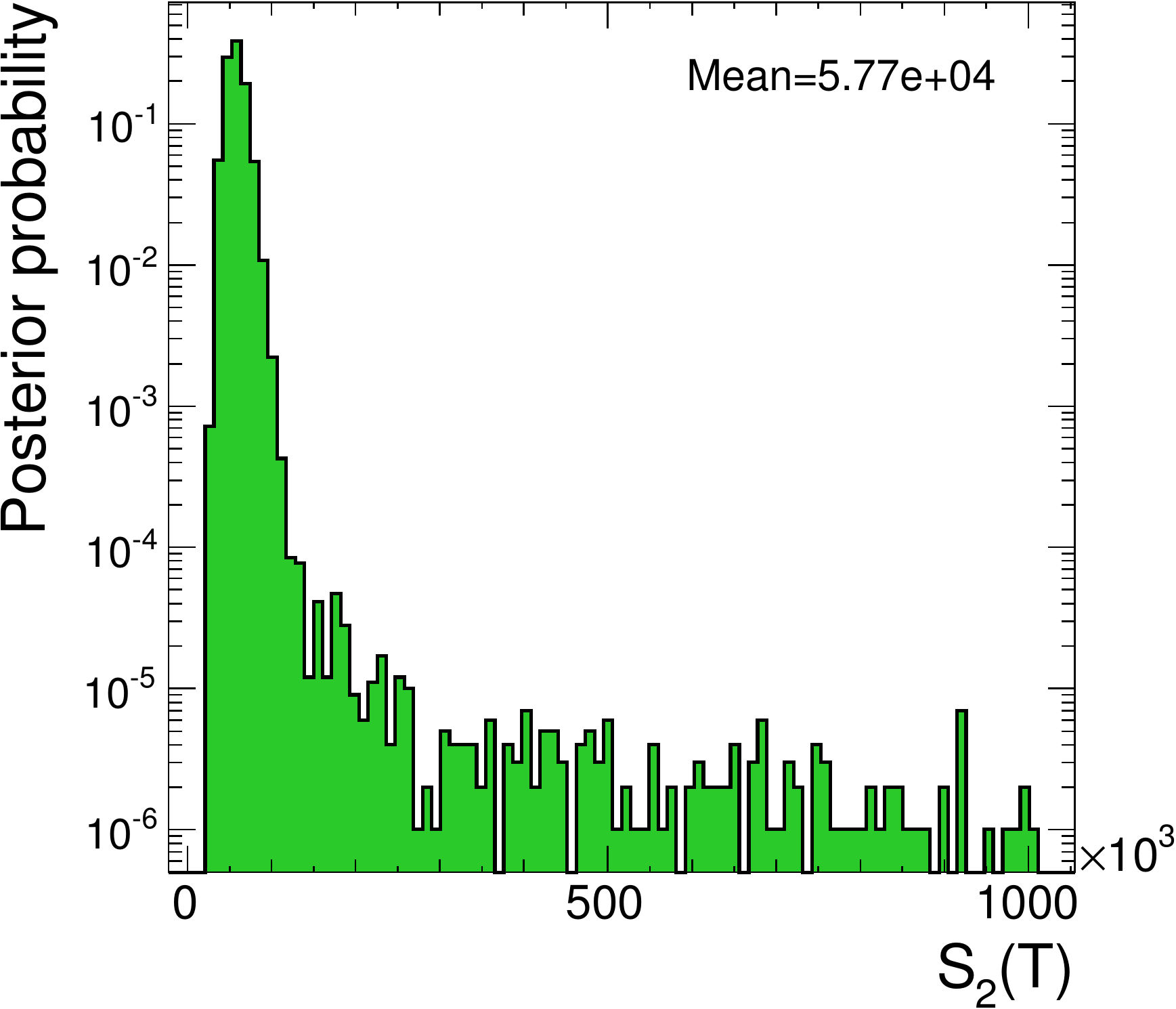}
  }
  \subfigure[$\alpha=6\times 10^{-4}$]{
    \includegraphics[width=0.3\columnwidth]{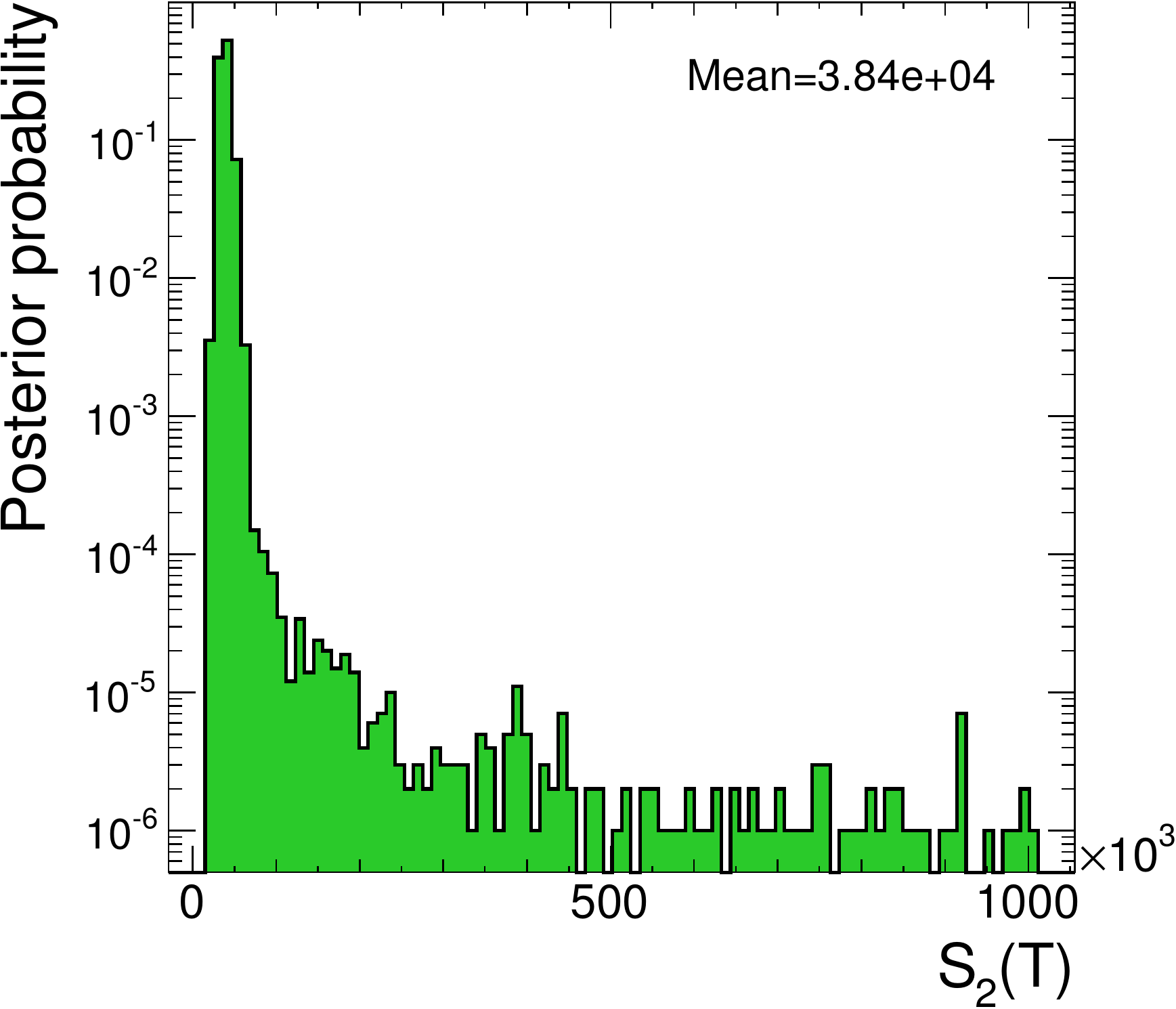}
 }
\caption{The posterior $P(S_2(\tuple{T})|\tuple{D})$, for three different choices of the regularization parameter $\alpha$, corresponding to Sec.~\ref{sec:regSteepBumpExpected}.
\label{fig:regFuncSteepBumpExpectedS2} 
}
\end{figure}

\begin{figure}[H]
  \centering
  \subfigure[$\alpha=0$]{
    \includegraphics[width=0.3\columnwidth]{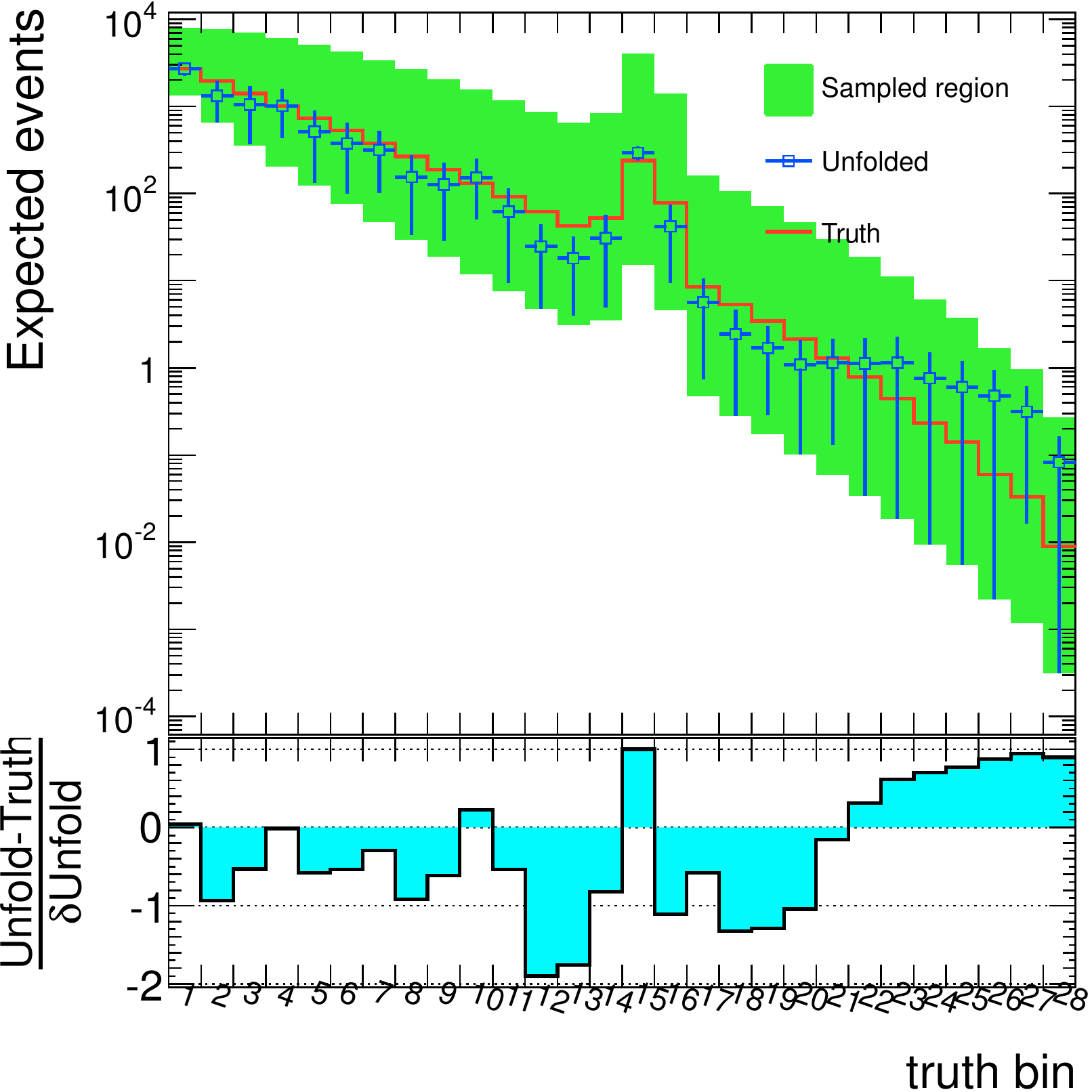}
  }
  \subfigure[$\alpha=3\times 10^{-4}$]{
    \includegraphics[width=0.3\columnwidth]{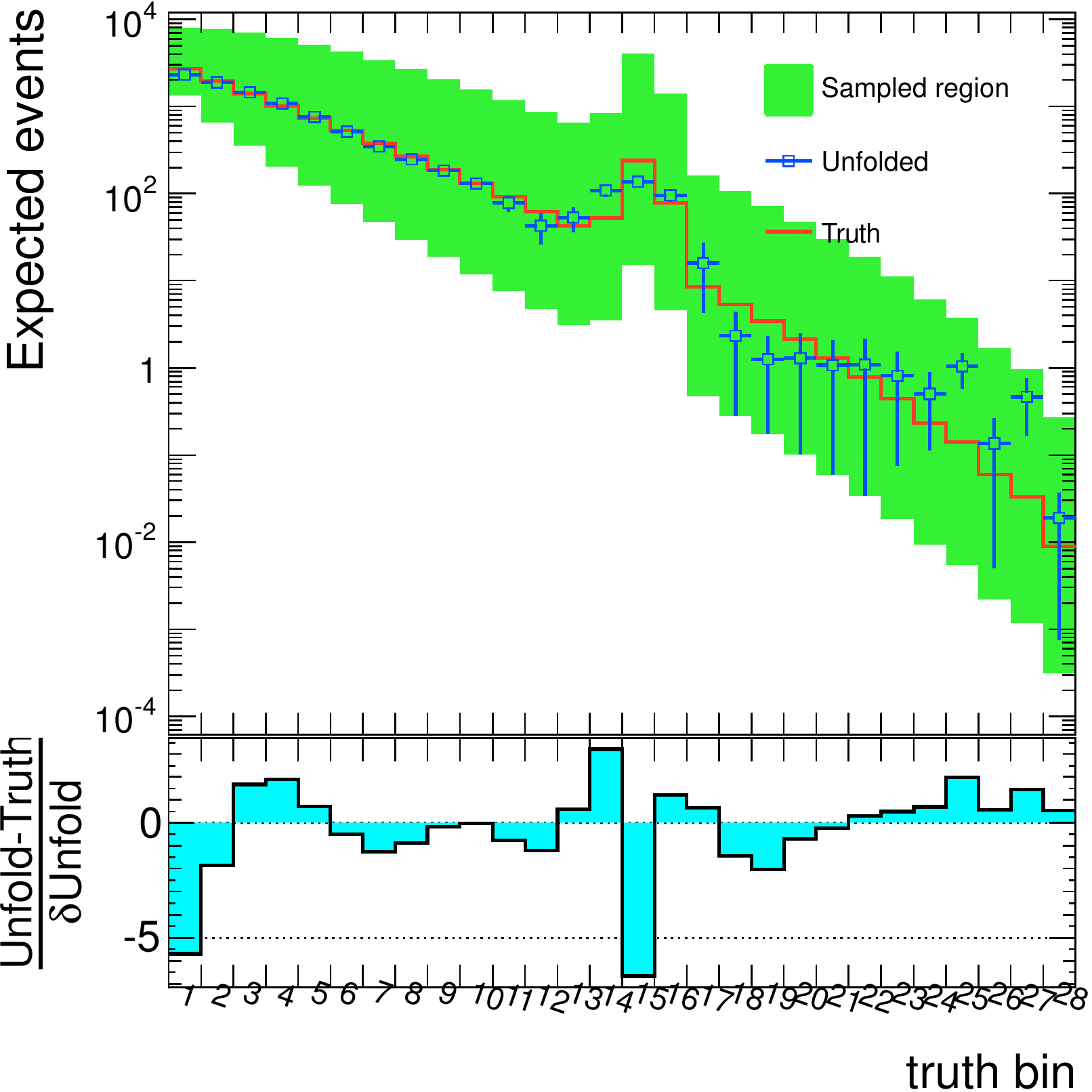}
  }
  \subfigure[$\alpha=6\times 10^{-4}$]{
    \includegraphics[width=0.3\columnwidth]{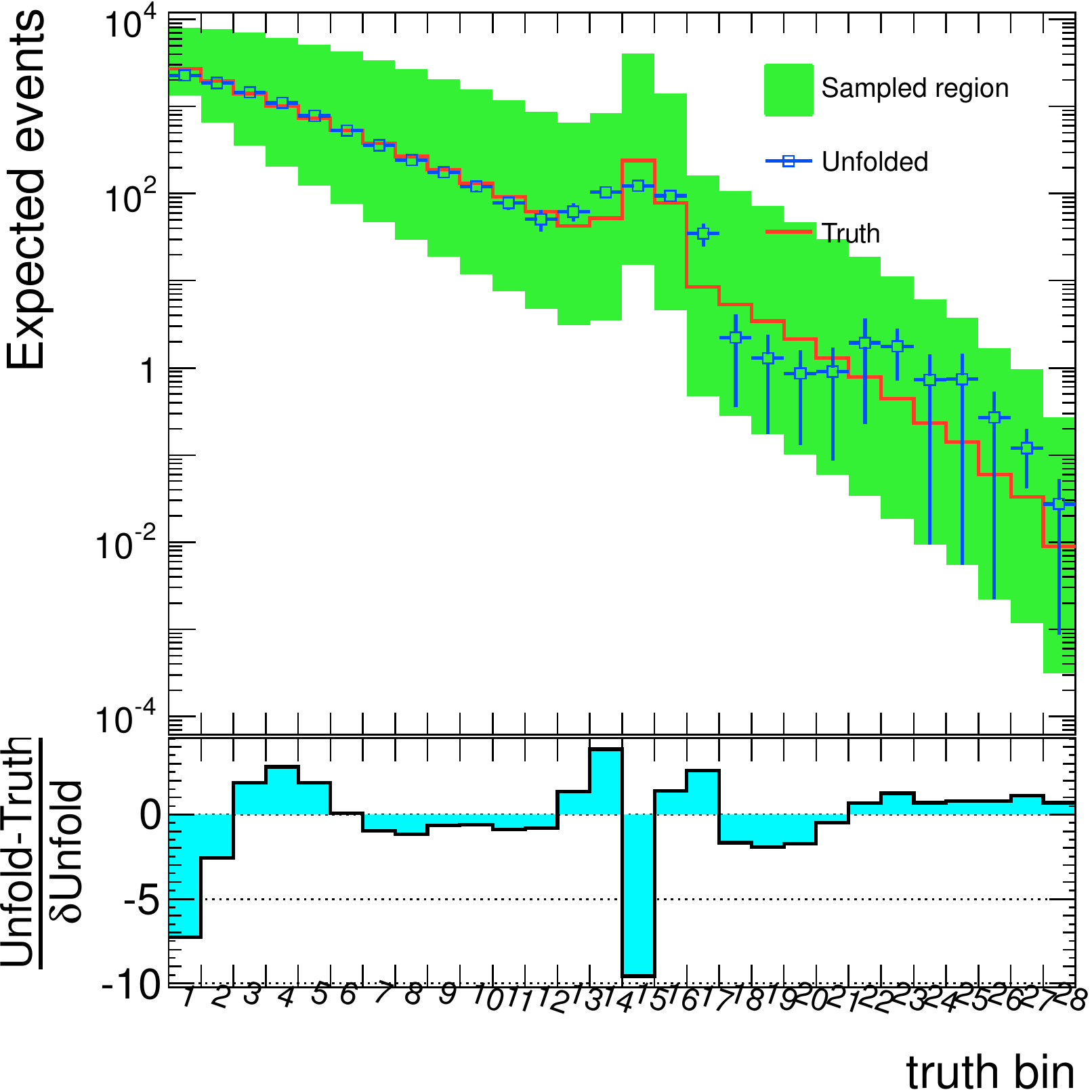}
  }\\
 \subfigure[$\alpha=0$]{
    \includegraphics[width=0.3\columnwidth]{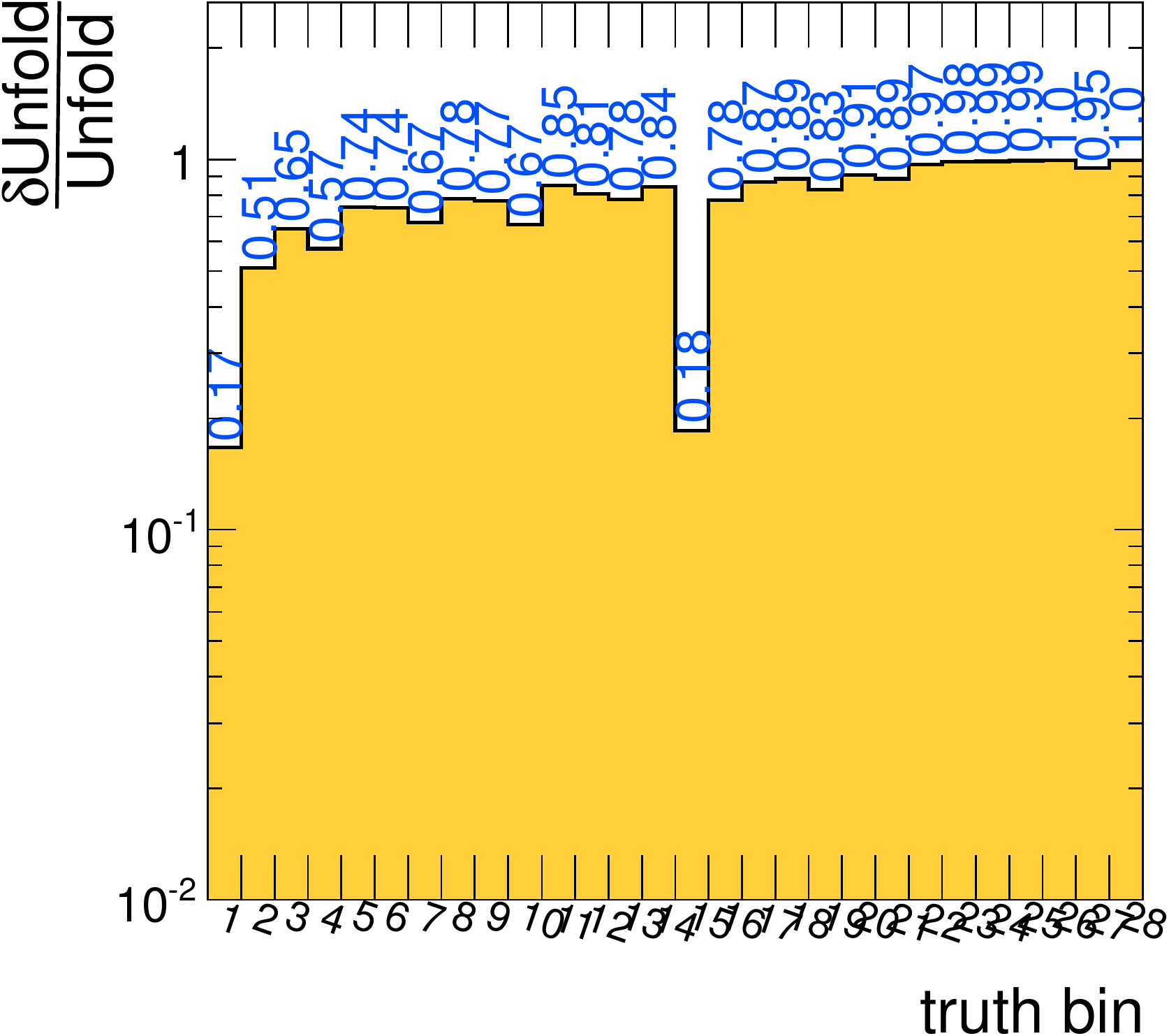}
  }
  \subfigure[$\alpha=3\times 10^{-4}$]{
    \includegraphics[width=0.3\columnwidth]{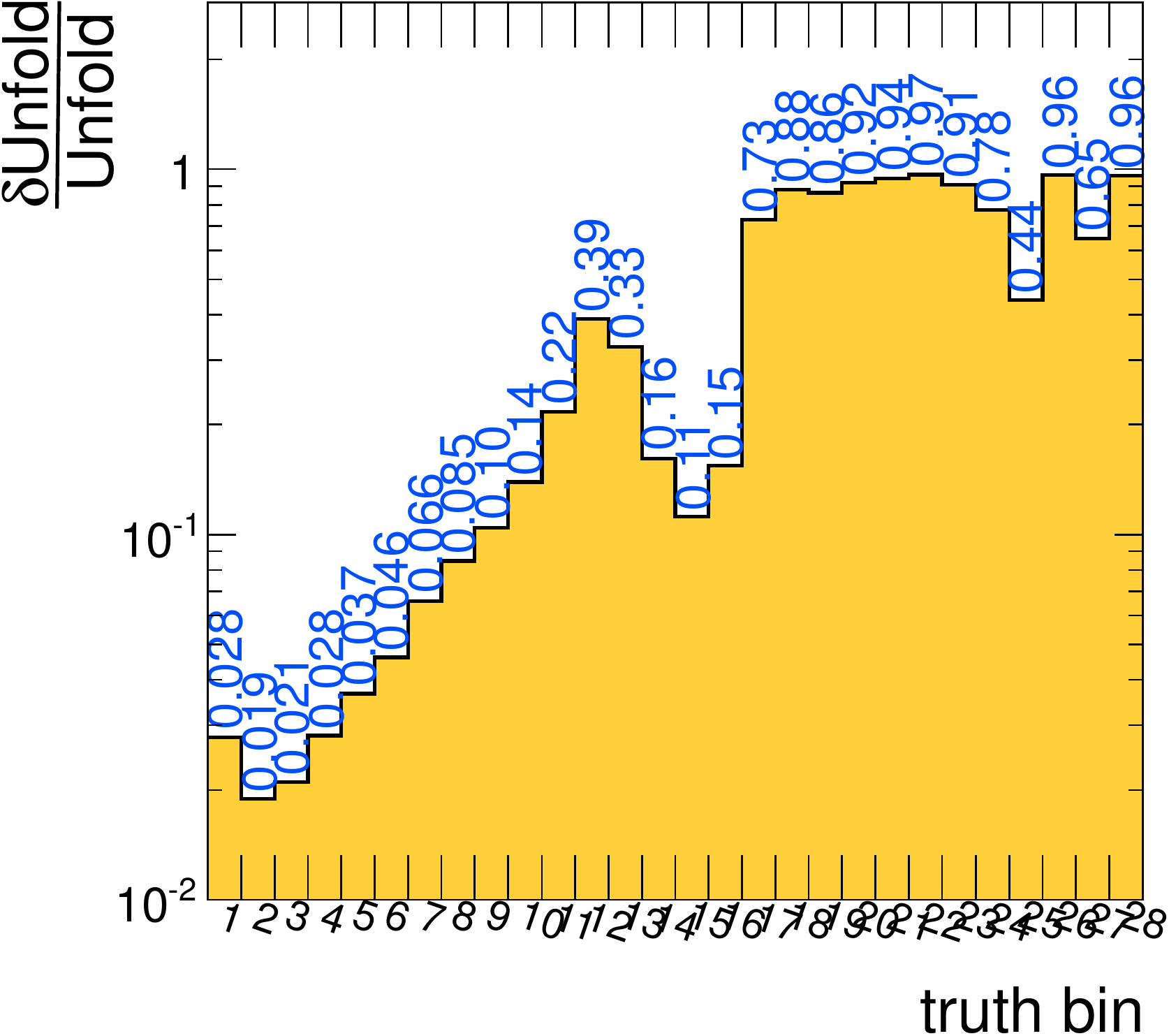}
  }
  \subfigure[$\alpha=6\times 10^{-4}$]{
    \includegraphics[width=0.3\columnwidth]{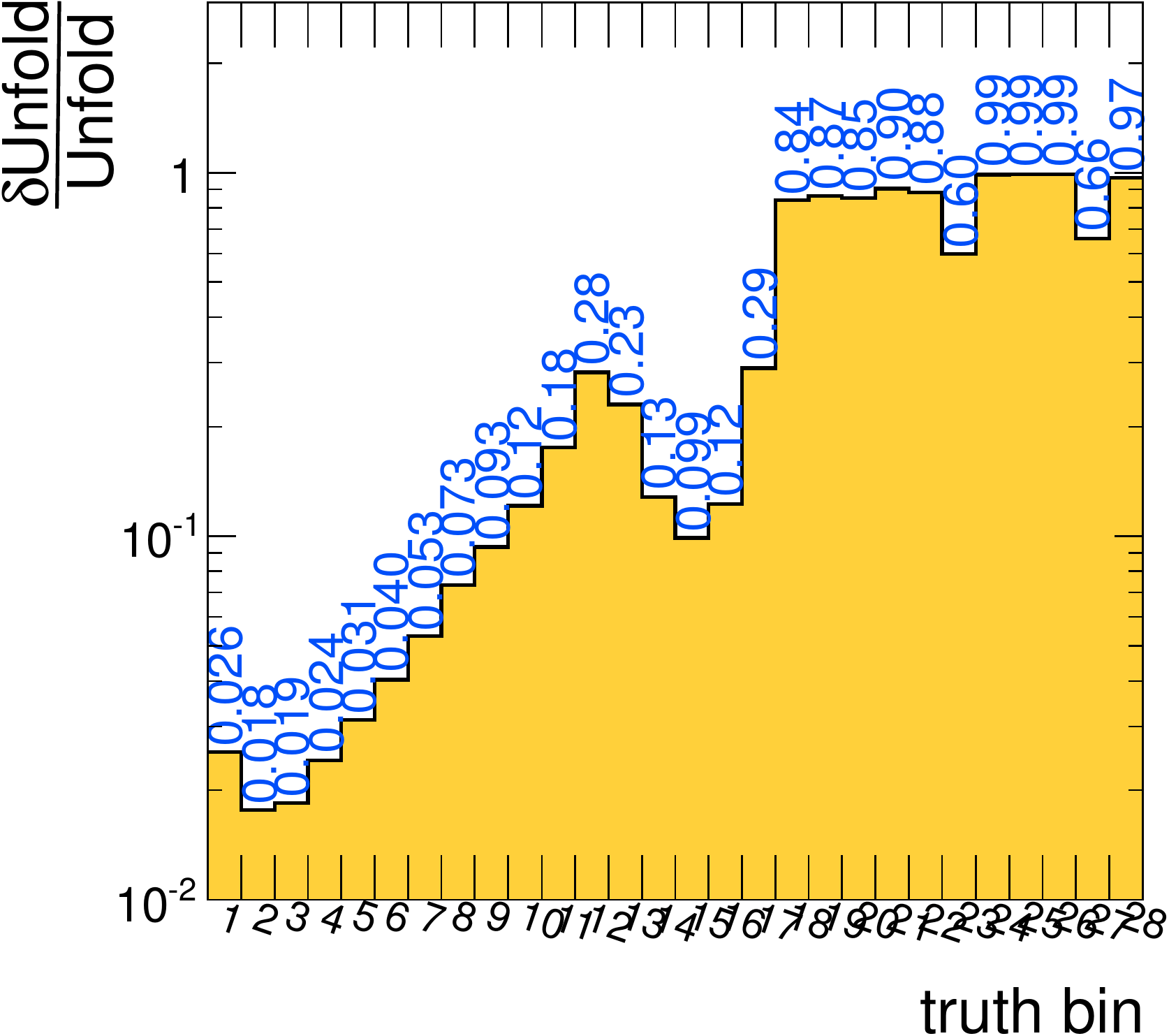}
  }
  \caption{The result of unfolding of Sec.~\ref{sec:regSteepBumpExpected}, with regularization function $S_2$, for three $\alpha$ values.  
    \label{fig:unfoldSteepBumpExpectedS2}
  }
\end{figure}

Using the regularization function $S_2$ is shown in Fig.~\ref{fig:regFuncSteepBumpExpectedS2} and \ref{fig:unfoldSteepBumpExpectedS2}.    The results are similar to Sec.~\ref{sec:regSteepBump}, Fig.~\ref{fig:unfoldSteepBumpS2}, where the bump was expected.

\begin{figure}[H]
  \centering
  \subfigure[$\alpha=0$]{
    \includegraphics[width=0.3\columnwidth]{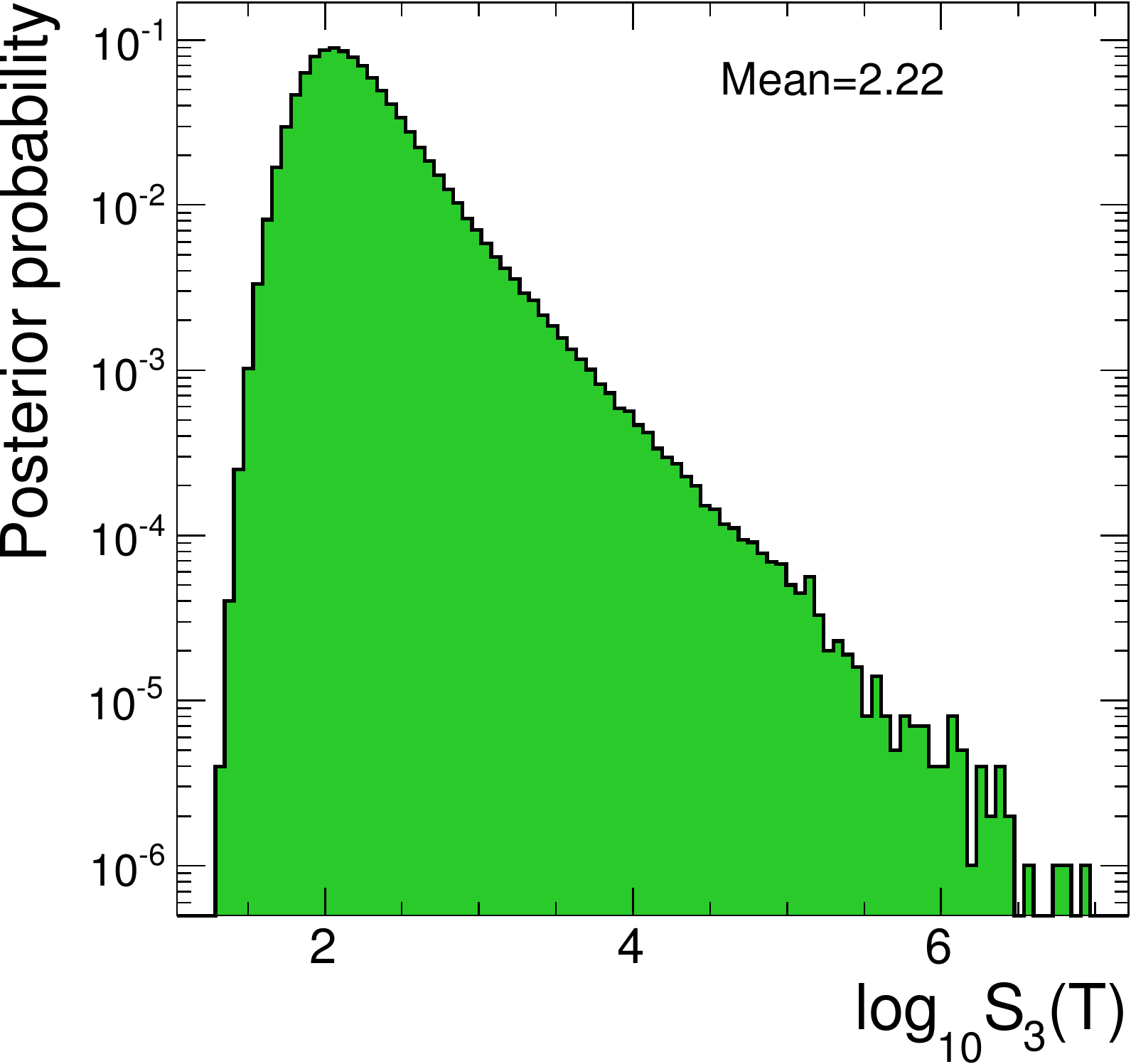}
  }
  \subfigure[$\alpha=10$]{
    \includegraphics[width=0.3\columnwidth]{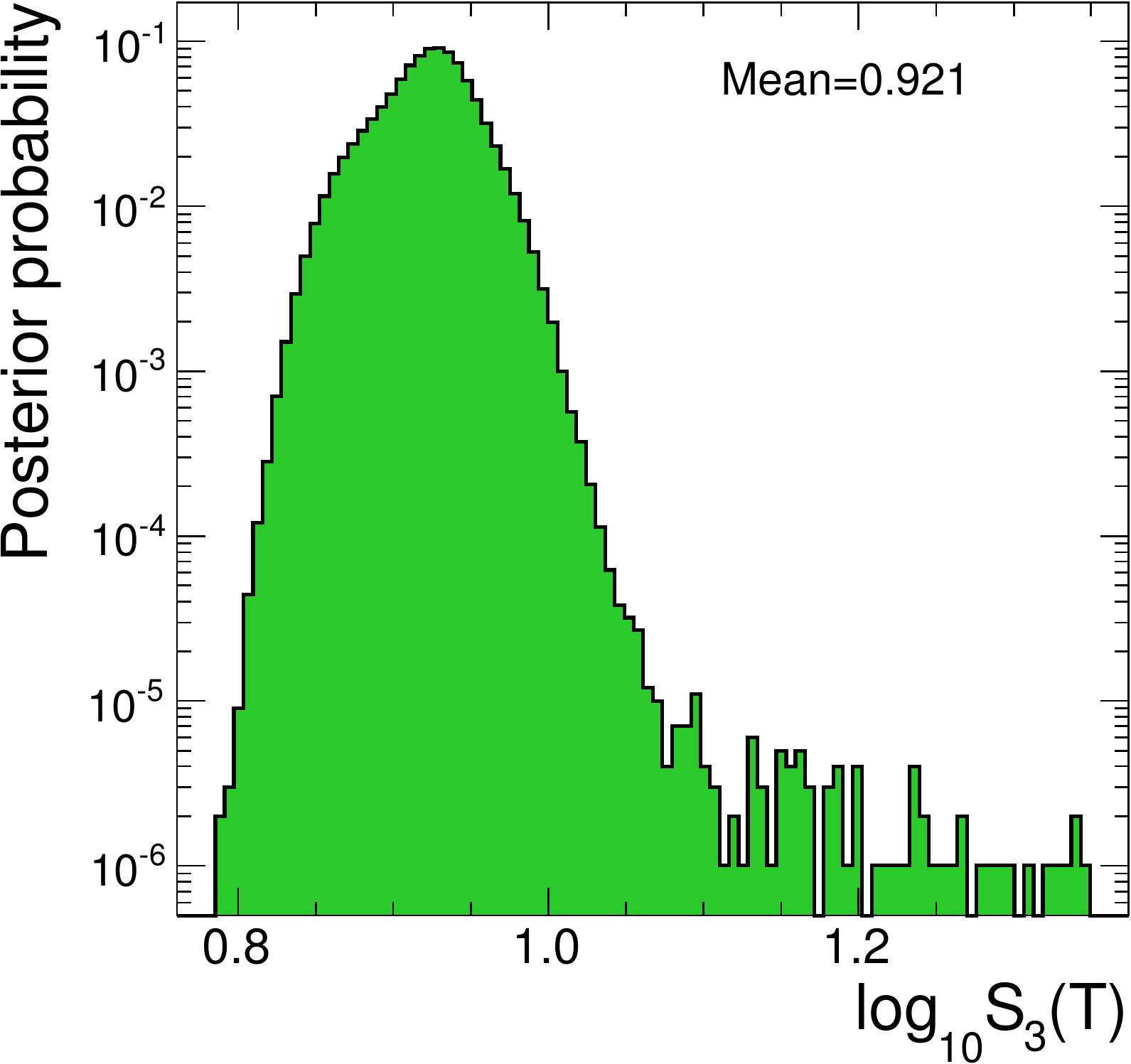}
  }
  \subfigure[$\alpha=20$]{
    \includegraphics[width=0.3\columnwidth]{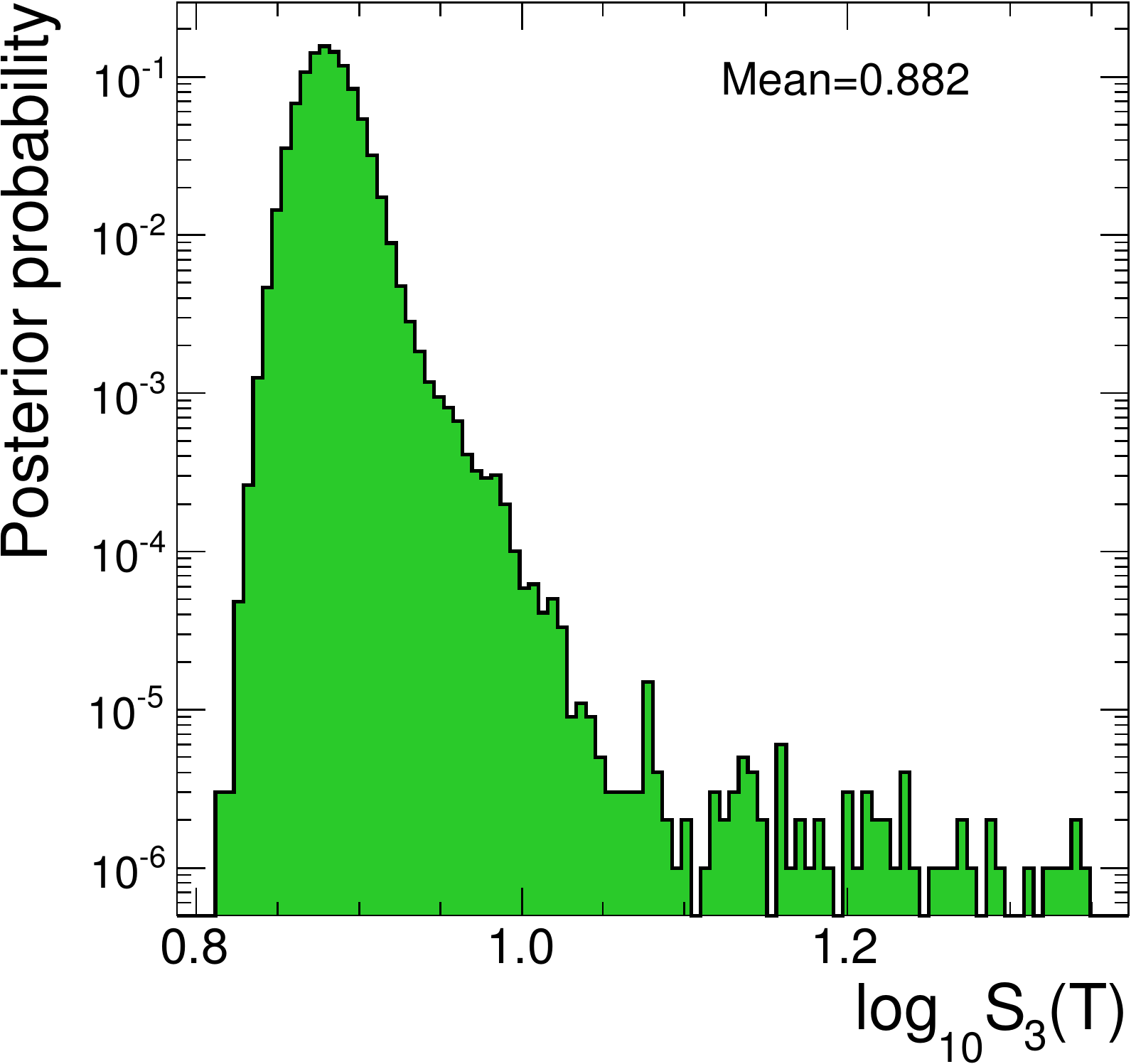}
 }
\caption{The posterior $P(S_3(\tuple{T})|\tuple{D})$, for three different choices of the regularization parameter $\alpha$, corresponding to Sec.~\ref{sec:regSteepBumpExpected}.
\label{fig:regFuncSteepBumpExpectedS3} 
}
\end{figure}

\begin{figure}[H]
  \centering
  \subfigure[$\alpha=0$]{
    \includegraphics[width=0.3\columnwidth]{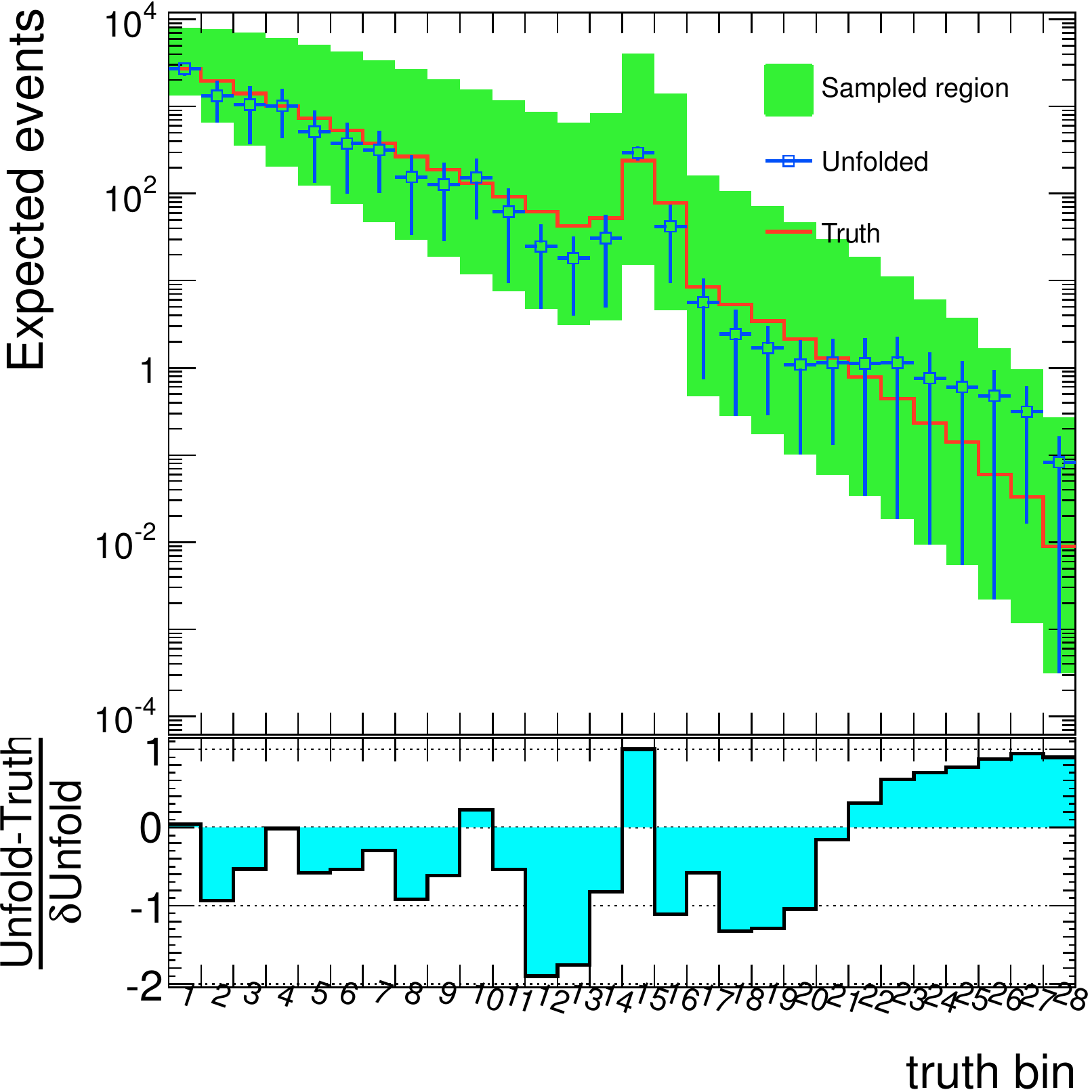}
  }
  \subfigure[$\alpha=10$]{
    \includegraphics[width=0.3\columnwidth]{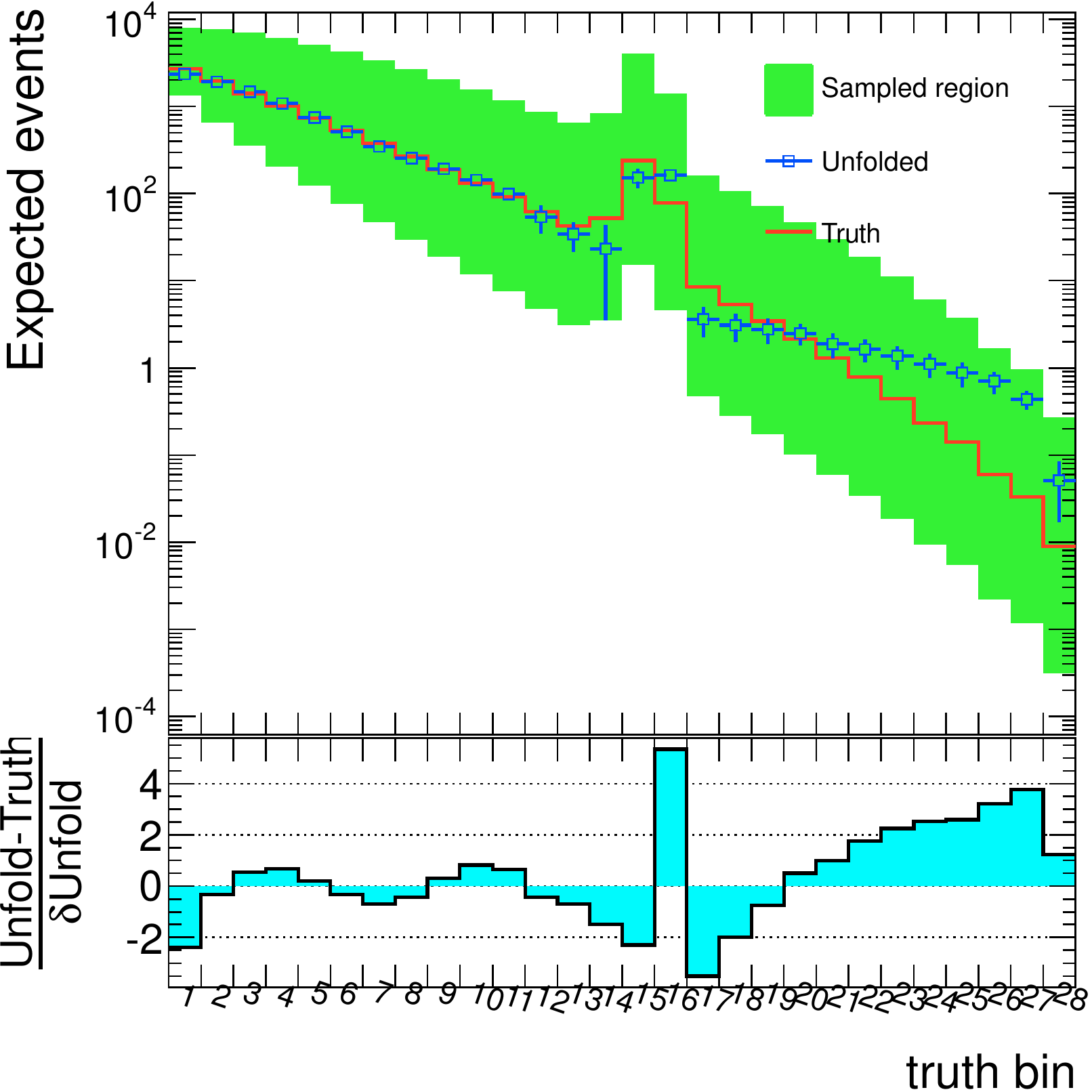}
  }
  \subfigure[$\alpha=20$]{
    \includegraphics[width=0.3\columnwidth]{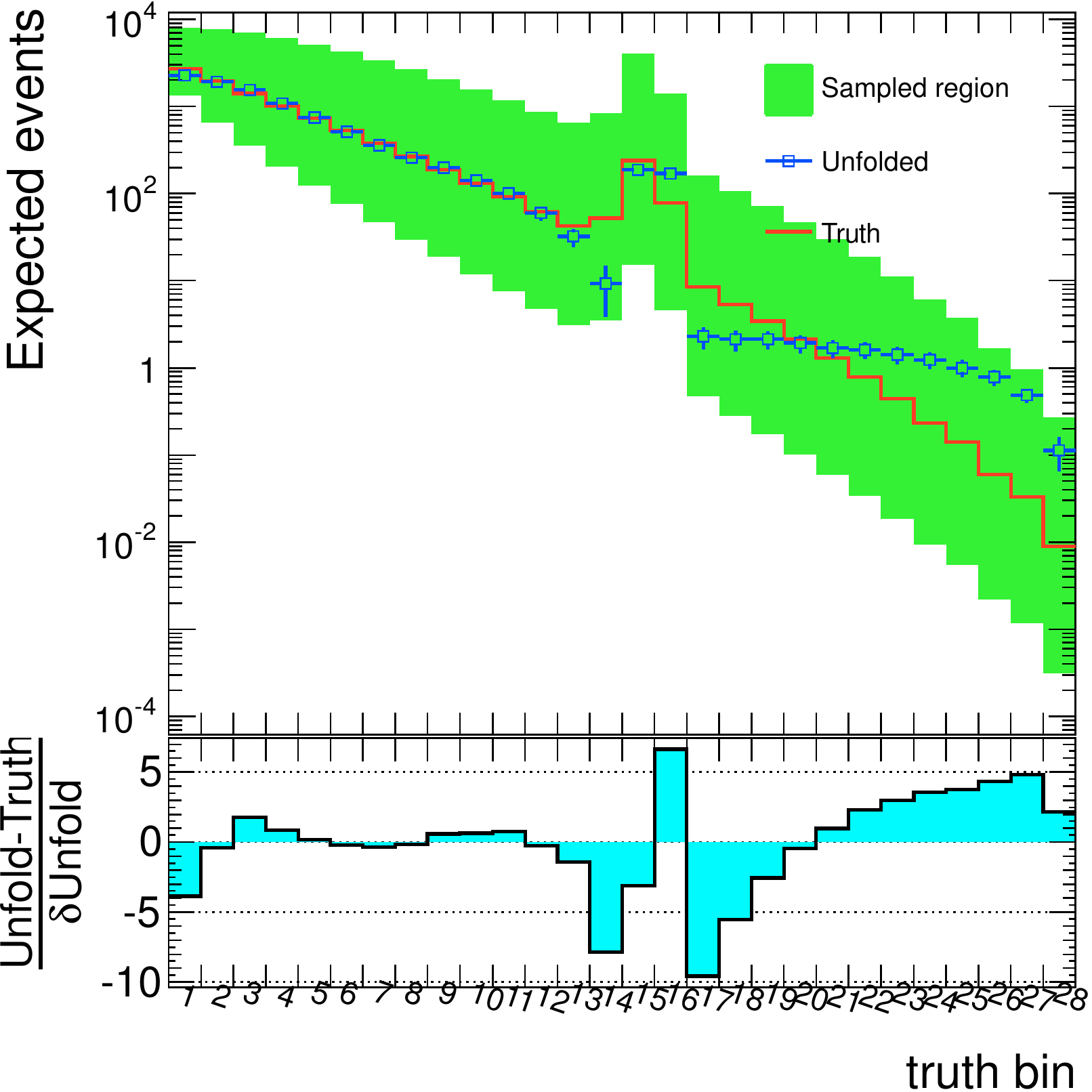}
  }\\
 \subfigure[$\alpha=0$]{
    \includegraphics[width=0.3\columnwidth]{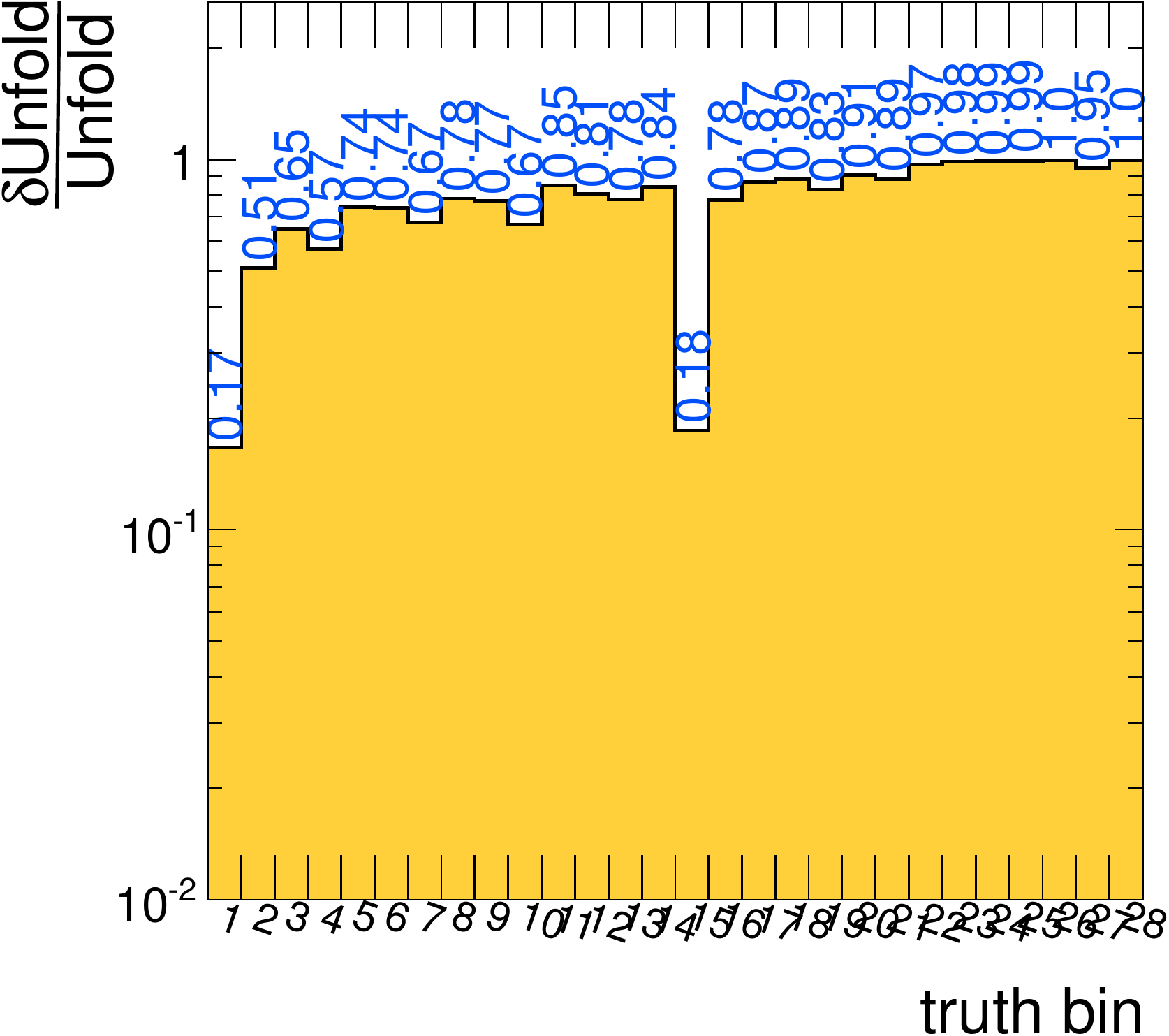}
  }
  \subfigure[$\alpha=10$]{
    \includegraphics[width=0.3\columnwidth]{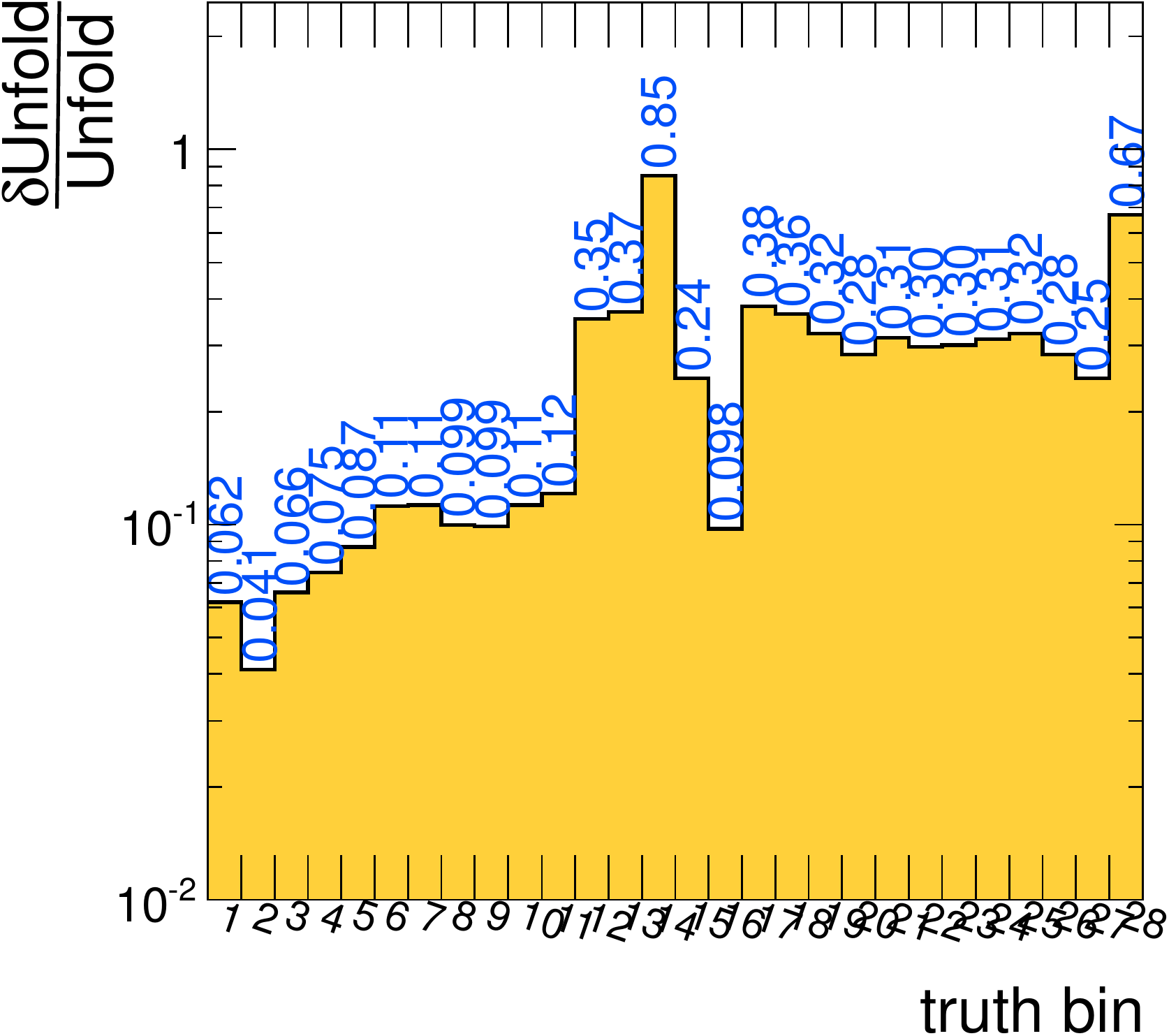}
  }
  \subfigure[$\alpha=20$]{
    \includegraphics[width=0.3\columnwidth]{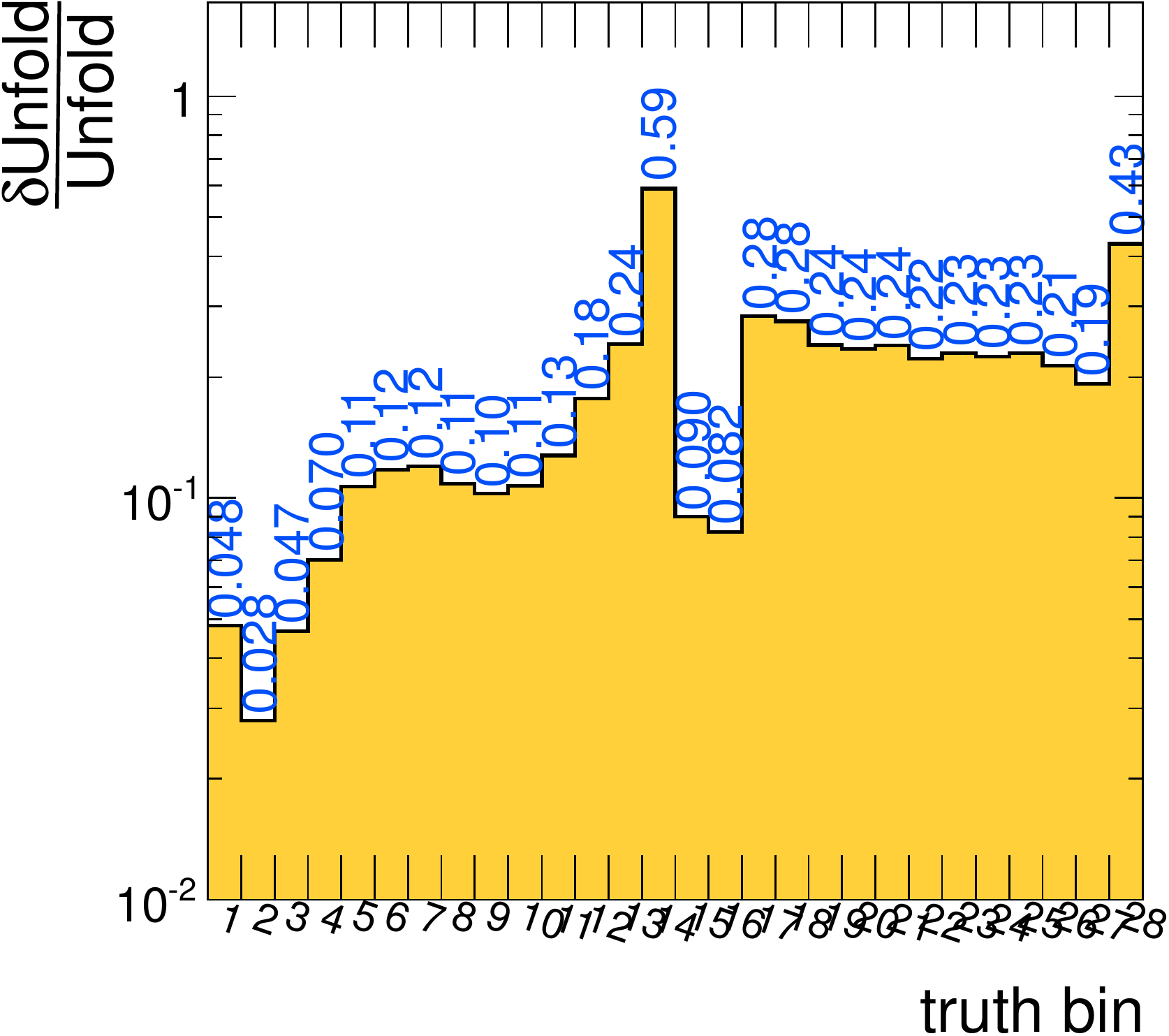}
  }
  \caption{The result of unfolding of Sec.~\ref{sec:regSteepBumpExpected}, with regularization function $S_3$, for three $\alpha$ values.  
    \label{fig:unfoldSteepBumpExpectedS3}
  }
\end{figure}

Using the regularization function $S_3$ is shown in Fig.~\ref{fig:regFuncSteepBumpExpectedS3} and \ref{fig:unfoldSteepBumpExpectedS3}.    The results are similar to Sec.~\ref{sec:regSteepBump}, Fig.~\ref{fig:unfoldSteepBumpS3}.

\begin{figure}[H]
  \centering
  \subfigure[$\alpha=0$]{
    \includegraphics[width=0.3\columnwidth]{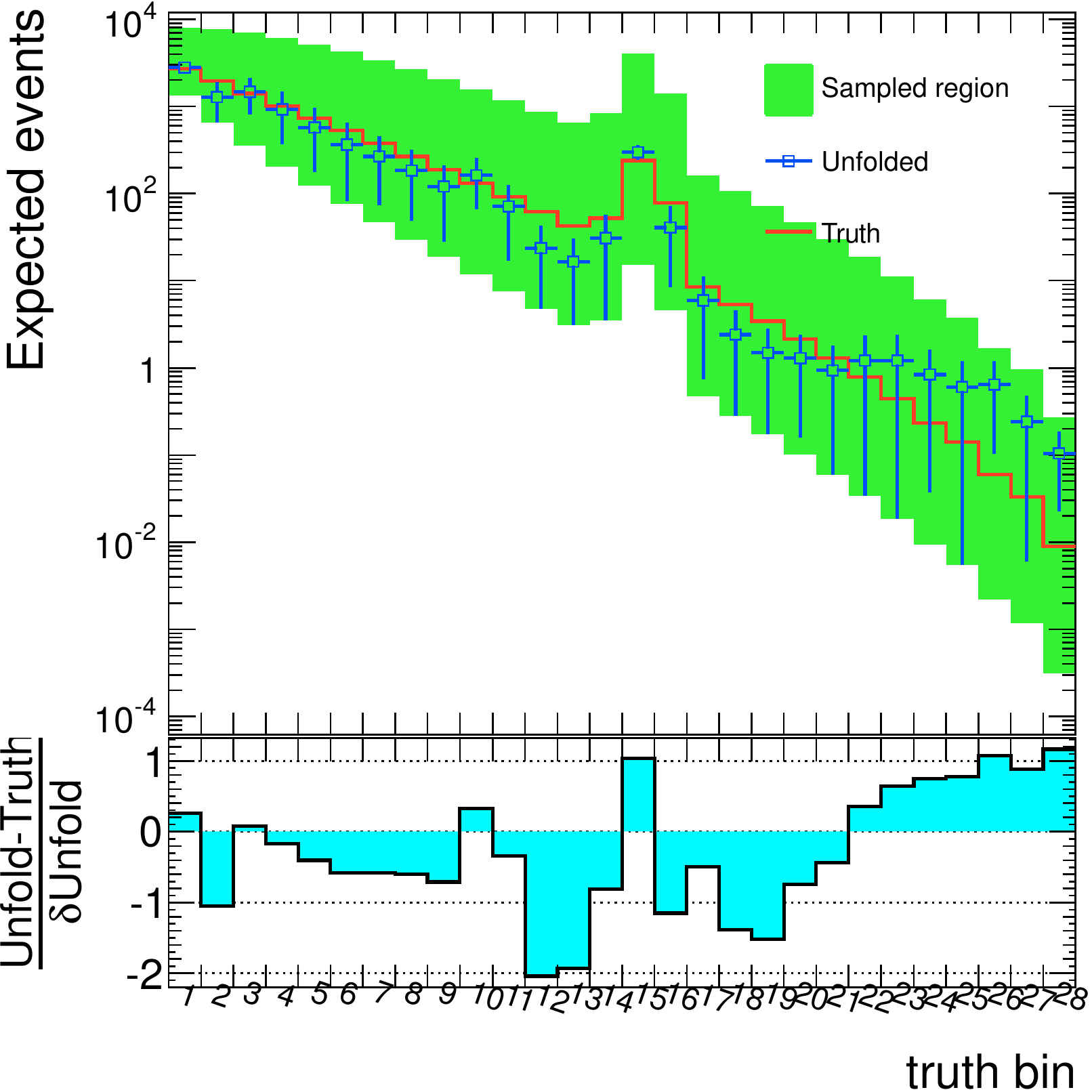}
  }
  \subfigure[$\alpha=1$]{
    \includegraphics[width=0.3\columnwidth]{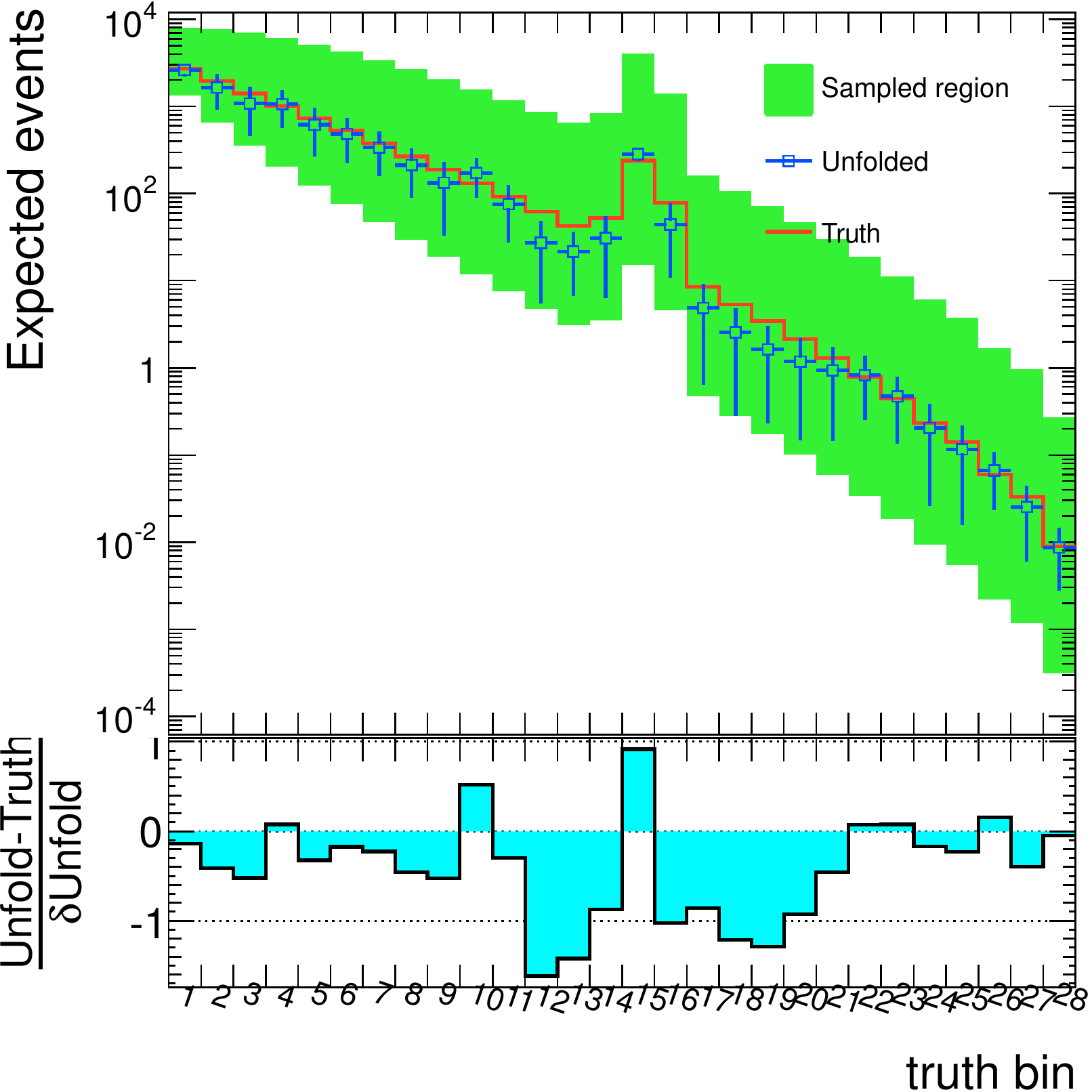}
  }
  \subfigure[$\alpha=10$]{
    \includegraphics[width=0.3\columnwidth]{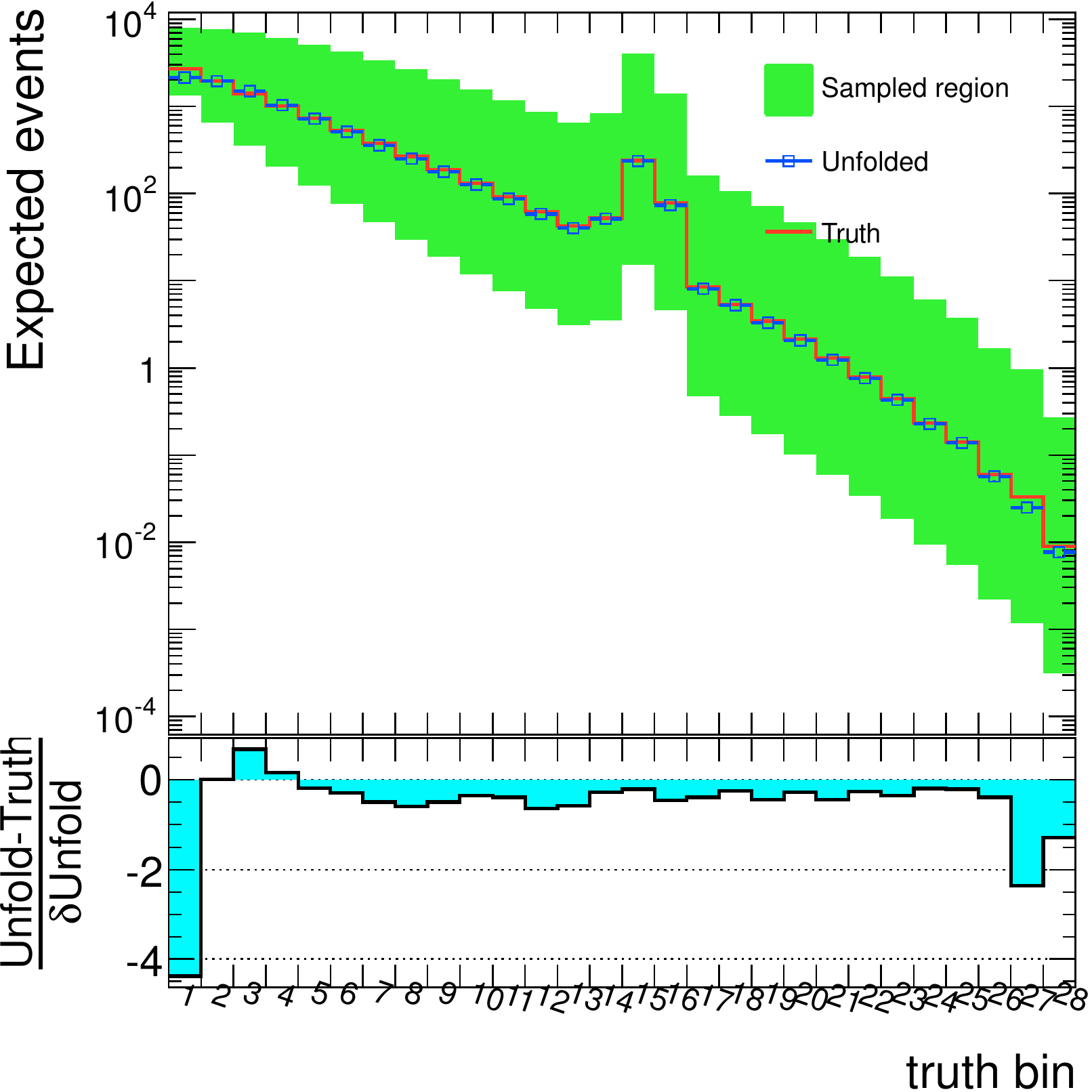}
  }\\
 \subfigure[$\alpha=0$]{
    \includegraphics[width=0.3\columnwidth]{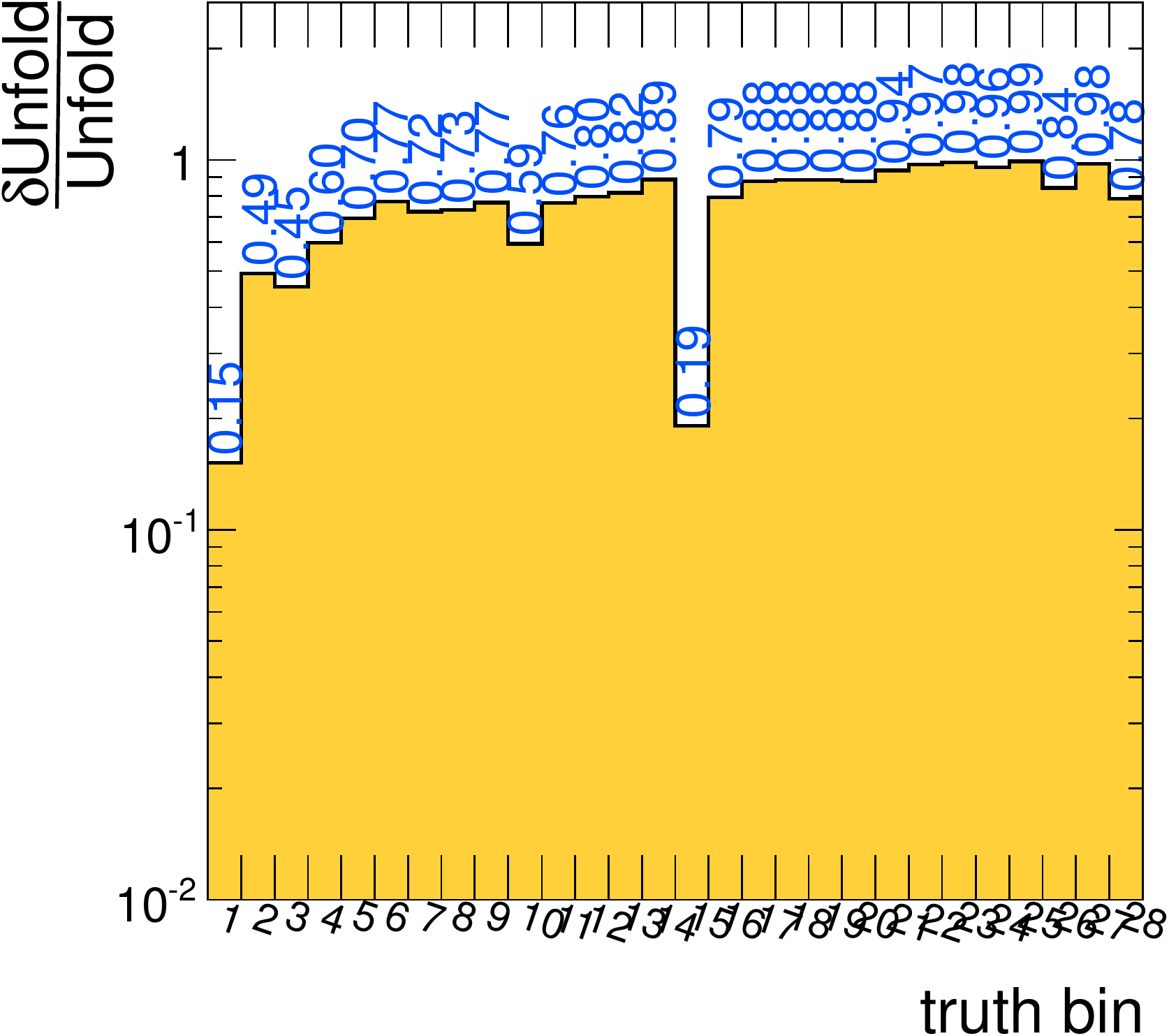}
  }
  \subfigure[$\alpha=1$]{
    \includegraphics[width=0.3\columnwidth]{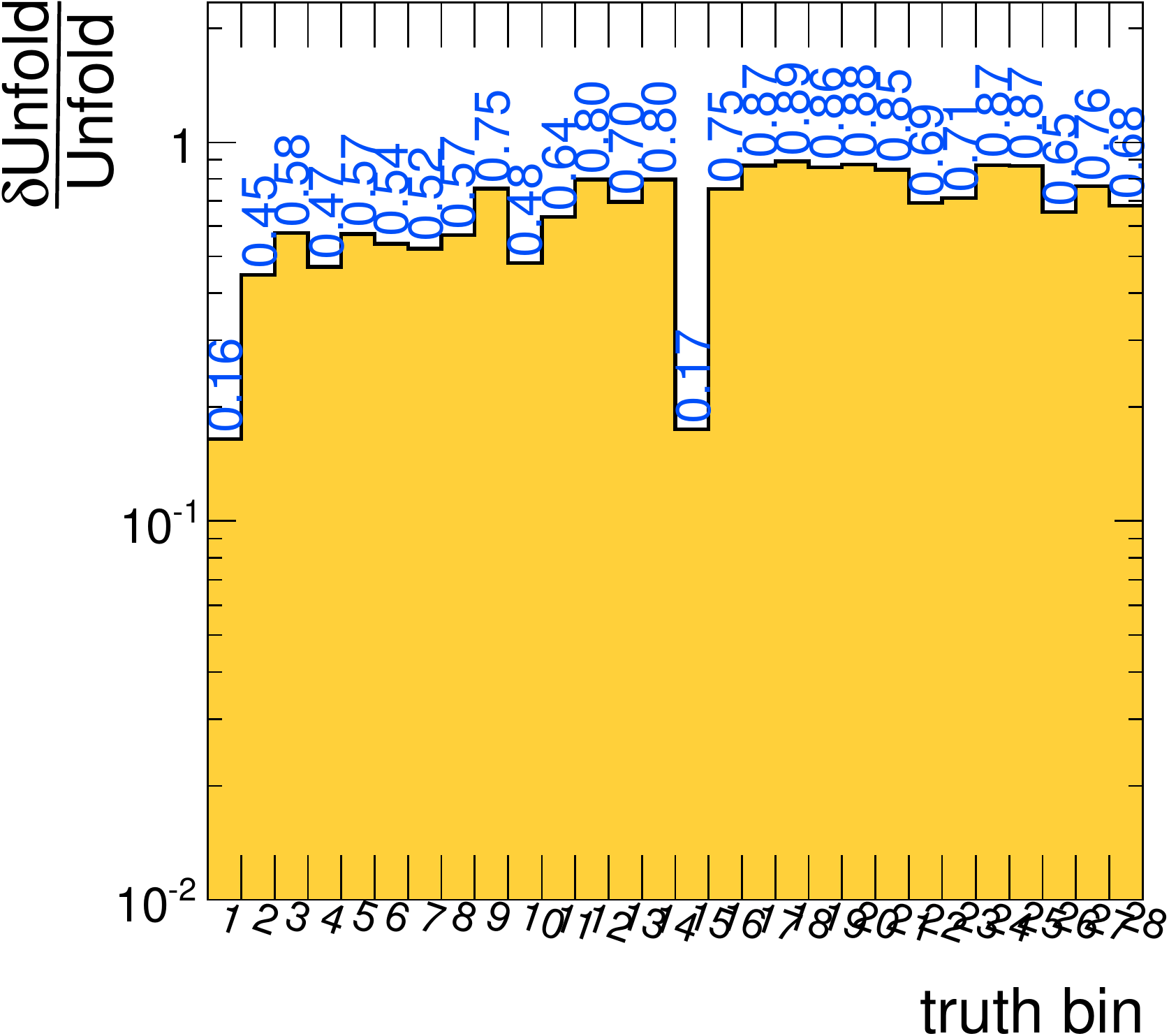}
  }
  \subfigure[$\alpha=10$]{
    \includegraphics[width=0.3\columnwidth]{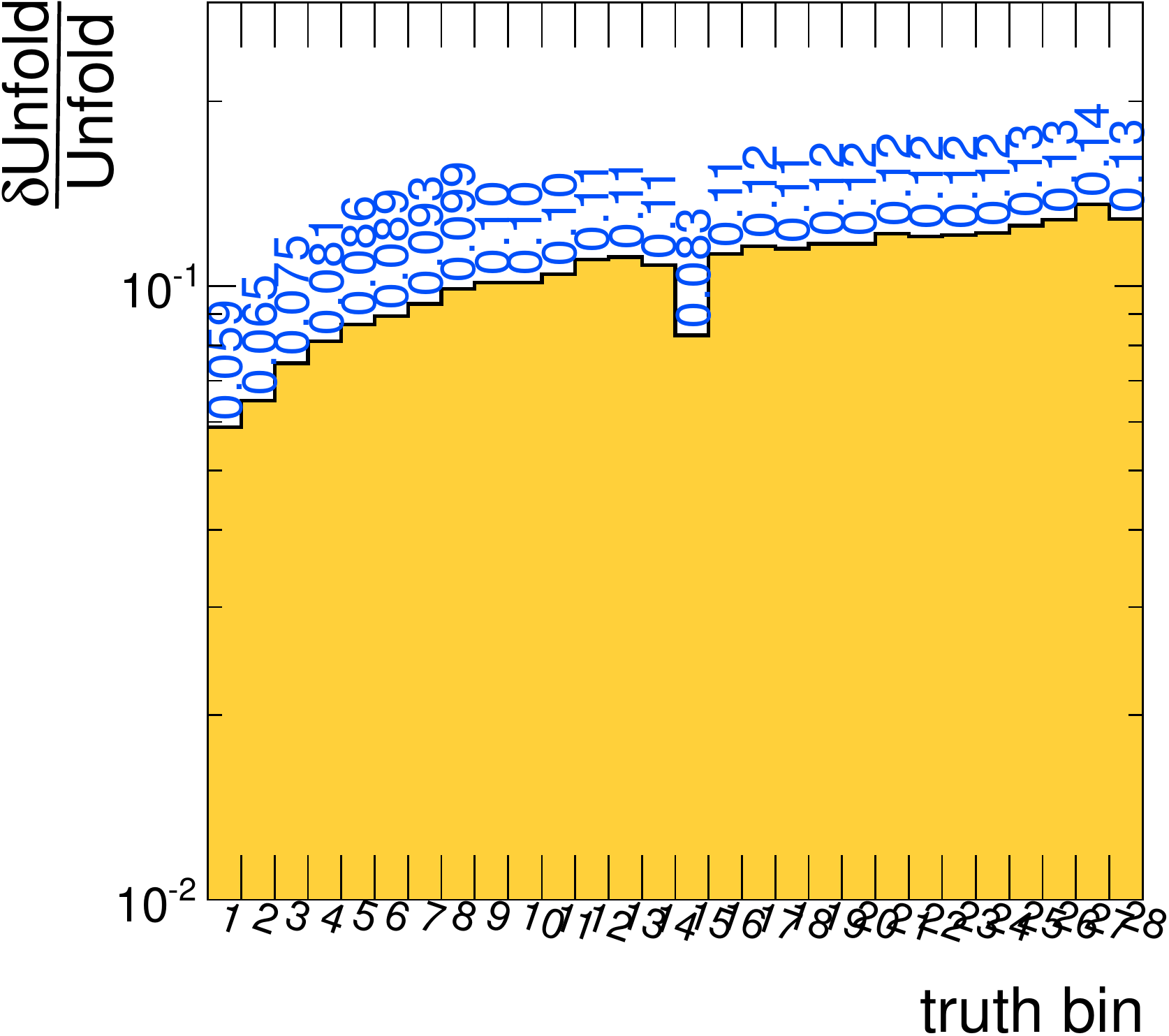}
  }
  \caption{The result of unfolding of Sec.~\ref{sec:regSteepBumpExpected}, with Gaussian regularization, for three $\alpha$ values.  
    \label{fig:unfoldSteepBumpExpectedGaus}
  }
\end{figure}

Finally, FBU with a Gaussian constraint is shown in Fig.~\ref{fig:unfoldSteepBumpExpectedGaus}.  As expected, a stronger constraint leads to a narrow posterior which concentrates around $\tilde{\T}$, which is by construction the same as $\hat{\T}$, which means that the posterior narrows down to $\hat{\T}$ with increasing values of $\alpha$.  This, unfortunately, is too ideal to be realistic.  If one knows already $\hat{\T}$, then there is no need for unfolding, or even for any data.  In reality $\tilde{\T}$ will not be known to be exactly equal to $\hat{\T}$, even for expected bumps (e.g. the $Z$ boson mass peak).

None of the regularization methods tried seems to improve how the bump appears in the unfolded spectrum, even when the bump is known at the MC level.  Only exception is the Gaussian constraint, which, as discussed, is not a realistic scenario; how close $\tilde{\T}$ is to $\hat{\T}$ would have to be considered as a systematic uncertainty.  

From the examples in Sec.~\ref{sec:regSteepBump} and \ref{sec:regSteepBumpExpected}, it seems that regularization indeed reduces the spread of the posterior, but it increases the bias significantly, distorting a feature such as a bump.  The only regularization condition that could help resolve a feature more finely than it appears after smearing, is with an $S(\T)$ tailored to the actual truth-level spectrum $\hat{\T}$ (e.g. a Gaussian constraint towards $\tilde{\T}$, which is necessarily assumed to be $\simeq \hat{\T}$).  If an unexpected feature is present, this is not possible (since $\tilde{\T} \neq \hat{\T}$), so, it is difficult to ensure that the regularization condition will reflect an actual property of $\hat{\T}$.  If it doesn't reflect an actual property of $\hat{\T}$ (e.g., if $S(\T) = S_1(\T)$ in the presence of a bump), then the feature may be very distorted by regularization, even if the feature was expected.  It would, maybe, be preferable to apply no regularization, i.e., to assume a constant prior, in which case the feature will not be distorted, and the truth-level (narrow) bump will be estimated without bias, even if the bump is unexpected.  But the spread of the posterior will be so big, that probably the feature will not be any more clear than before unfolding.

\section{Interpretation of unfolding}

It is a common misconception that the result of unfolding is ``corrected data'', which can be used to set limits and test hypotheses as if it was data, provided only that some attention is paid to the covariance between bins.  SVD and iterative unfolding, which provide an estimator (i.e., a single \T) and a covariance matrix, make this misconception easier, when $N_t = N_r$, because the user inputs a histogram $\D$ and gets another similar histogram as output.

The origin of this misconception is that the data (\D) are often thought to have uncertainty in each bin, which is uncorrelated between independent bins.  From this one can easily be led to think that unfolding may introduce bin correlations but, apart from that complication, its answer can be used as a ``corrected'' replacement of the original \D.

The impression that data (\D) have uncertainty is wrong to start with.  When one observes $D_r$ events in the $r^{\rm th}$ reco-level bin, there is no statistical uncertainty about how many events were counted, and, assuming we know how to count correctly, there is no systematic uncertainty either.  It is customary to plot $\D$ with error bars equal to $\sqrt{D_r}$ in bin $r$, and this is what makes people think that $\D$ has an uncertainty. Variance, however, is only a property of probability distributions; not of actual observations.

How it became customary to draw error bars of $\sqrt{D_r}$ around $\D$ is a question for the historian of science, but the interpretation of these error bars should be the following.  When we observe $D_r$ independent random events in bin $r$, this number is assumed to be pulled from a Poisson distribution with mean $\hat{R}_r$.  One doesn't know $\hat{R}_r$, but can try to infer it.  To do so one can classically construct the maximum likelihood estimator (MLE), $\bar{R}_r$, which maximizes the likelihood
\begin{equation}
L(D_r|R_r) = \frac{R_r^{D_r}}{D_r!}e^{-R_r}.
\end{equation}
Maximizing this results in $\bar{R}_r = D_r$.
This MLE is an unbiased estimator of $\hat{R}_r$, because its expectation value is 
\begin{equation}
E(\bar{R}_r) = E(D_r) = \hat{R}_r.
\end{equation}
The standard deviation of a maximum likelihood estimator is estimated as explained in \cite{cowanBook} (Fig.\ 6.4), and it reflects the width of $L(D_r|R_r)$ near its maximum.
This is totally equivalent to solving Bayes' equation for a constant prior $\pi(R_r)$:
\begin{equation}
p(R_r|D_r) \propto L(D_r|R_r).
\end{equation}
As mentioned in Sec.~\ref{sec:conceptualization}, the classical MLE is nothing but the mode of the Bayesian posterior, if the prior is assumed constant, and the variance of the MLE reflects the width of this posterior.
If $D_r$ is a large enough number, then the above $p(R_r|D_r)$ is approximated well by a Gaussian of mean $D_r$ and standard deviation $\sqrt{D_r}$. Similarly, the MLE of $\hat{R}_r$ is $\bar{R}_r = D_r$, which suggests\footnote{It is understood that $\bar{R}_r \neq \hat{R}_r$, but only $E(\bar{R}_r) = \hat{R}_r$. So, loosely speaking, this suggestion would be valid on average, in a frequentist sense.} that, \emph{if} indeed $\hat{R}_r = D_r$, then if the experiment was repeated infinite times these hypothetical data would be distributed like a Gaussian of mean $D_r$ and standard deviation $\sqrt{D_r}$.
So, when the data $D_r$ are drawn with error bars of size $\sqrt{D_r}$ it should be understood that these error bars don't refer to the standard deviation of $D_r$, since such a thing is not defined, but they refer to the standard deviation of $p(R_r|D_r)$, or of the distribution suggested by the estimator $\bar{R}_r$.  The mode of $p(R_r|D_r)$ happens to be numerically equal to $D_r$, so the error bars are centered at $D_r$, but that's about all they have to do with $D_r$ itself.  They reflect an inference about $\hat{R}_r$.
If one wants to test how compatible $D_r$ is with a theoretical hypothesis which predicts an expected number of events $\tilde{R}_r$, then the customary error bar of $\sqrt{D_r}$ is irrelevant.  The actual likelihood of $D_r$, under the hypothesis of $\tilde{R}_r$, is given by
\begin{equation}
\frac{\tilde{R}_r^{D_r}}{D_r!} e^{-\tilde{R}_r} \simeq \frac{1}{\sqrt{2\pi \tilde{R}_r}} e^{-\frac{(D_r-\tilde{R}_r)^2}{2\tilde{R}_r}}
\end{equation}
and \emph{not} by
\begin{equation}
\frac{1}{\sqrt{2\pi D_r}} e^{-\frac{(D_r-\tilde{R}_r)^2}{2D_r}}.
\end{equation}
The correct way to visualize this comparison would be to plot $D_r$ without any error bars, and then superimpose the value $\tilde{R}_r$ surrounded by error bars of size $\sqrt{\tilde{R}_r}$, if $\tilde{R}_r$ is large enough to justify the Gaussian approximation.  More on this in \cite{plottingSignificance, errorsOnBkg}.

So, the data (\D) never had uncertainty to begin with, and it is misleading to think that the result of unfolding is just ``corrected data'' with bin correlations.

\subsection{Using $p(\T|\D)$ to estimate parameters at truth-level}
\label{sec:limits}

To set limits on the expected number of reconstructed events of a hypothetical signal, $s$, one needs to compute
\begin{equation}
p(s|D) \propto L(D|s) \cdot \pi(s).
\end{equation}
This is very simple to do at reco-level, if it is known how the signal is reconstructed, namely, if model of the detector response is available.  Details can be found in Ref.~\cite{choudalakisLimits}.

A reason unfolding is used in HEP is to make it easy for theorists to test their theories without having a model of the detector.  Here it will be examined how this can be done with FBU, and under what conditions it is correct.

Let's assume that the Standard Model (SM), at truth-level, at the luminosity corresponding to the analyzed data, predicts a spectrum $\T^{SM}$, and the assumed new physics (NP) adds on top of that a spectrum $\T^{NP}$.  For simplicity, possible destructive interference is ignored.  For example, $\T^{SM}$ could be a steeply falling spectrum, and $\T^{NP}$ be a bump.  Let's define as parameter of interest the signal cross-section in units of pb, denoted by $\sigma$.  To make $\sigma$ explicitly appear in the equations, it is convenient to write $T^{NP}$ as $\sigma \cdot \T^{SNP}$, where $\T^{SNP}$ has the same shape as $\T^{NP}$, but is scaled\footnote{SNP stands here for ``scaled new physics''.} to the integrated luminosity that corresponds to 1 \ipb.
Let's further assume that the response matrix for SM is $\P^{SM}$ and for NP it is $\P^{NP}$.  The elements of a response matrix (see Sec.~\ref{sec:nomenclature}) reflect both the probability of being smeared from one bin into another, and the probability to be reconstructed in \emph{any} of the considered reco-level bins.

To infer the $\sigma$ of this hypothetical signal one needs to compute 
\begin{equation}
p(\sigma|\D\land\P^{SM}\land\P^{NP}\land\T^{SNP}\land\T^{SM}).
\end{equation}
For brevity, let's denote with $K$ the condition $\P^{SM}\land\P^{NP}\land\T^{SNP}\land\T^{SM}$.
Then,
\begin{equation}
p(\sigma|\D\land K) \propto L(\D|\sigma\land K) \cdot \pi(\sigma\land K).
\end{equation}
Assuming that all components of $K$ are not uncertain, the joint prior $\pi(\sigma\land K)$ can be written as a product of $\pi(\sigma)$ and $\delta$-functions that pinpoint each component of $K$ to its known value.  For brevity of notation, we can omit $K$, just like $\M$ was omitted from Eq.~\ref{eq:bayesWithM}.  It is, of course, remembered that $K$ is silently assumed.
\begin{equation}
\label{eq:posteriorSigma}
p(\sigma|\D) \propto L(\D|\sigma) \cdot \pi(\sigma).
\end{equation}

How can this be computed using the output of FBU?
The information given to the theorist is $p(\T|\D)$, from Eq.~\ref{eq:bayes}.
From this, and the stated regularization ($\pi(\T)$), the theorist can extract
\begin{equation}
\label{eq:likelihoodInverted}
L(\D|\T) \propto \frac{p(\T|\D)}{\pi(\T)}.
\end{equation}

Assuming $K$ and a value of $\sigma$, the expected reconstructed events are, in analogy to Eq.~\ref{eq:RwithBmatrix}, given by
\begin{equation}
\R = \B + (\P^{SM})^T \T^{SM} + \sigma (\P^{NP})^T \T^{SNP}.
\end{equation}
The likelihood of \D\ under this assumption ($L(\D|\sigma)$, in abbreviated notation) can now be computed by the theorist, by evaluating Eq.~\ref{eq:likelihoodInverted} at the \T\ which corresponds to the same \R\ as the hypothesized NP.  Namely,
\begin{eqnarray}
  \nonumber \R &=& \B + \P^T \T \\
  \nonumber    &=& \B + (\P^{SM})^T \T^{SM} + \sigma (\P^{NP})^T \T^{SNP},
\end{eqnarray}
which gives
\begin{eqnarray}
  \label{eq:theTneeded0}
\nonumber \T =&{}& (\P^T)^{-1}(\P^{SM})^T \T^{SM} \\
              &+& \sigma (\P^T)^{-1}(\P^{NP})^T \T^{SNP},
\end{eqnarray}
where $\P$ is the response matrix used in the provided unfolding.

It is reasonable for a theorist to assume that $\P = \P^{SM}$, since usually SM MC is used to construct \P\ for the unfolding.  As long as the theorist doesn't want to assume his own detector model, $(\P^T)^{-1}(\P^{SM})^T = (\P^T)^{-1}\P^T = \mathbb{I}$, so,
\begin{equation}
  \label{eq:theTneeded}
  \T = \T^{SM} + \sigma (\P^T)^{-1}(\P^{NP})^T \T^{SNP}.
\end{equation}
Regarding $\P^{NP}$, it is reasonable to assume that the signal and the SM are subject to the same detector energy resolution and other hardware sources of smearing which are represented in \P.  However, depending on the character of the signal, the efficiency of being reconstructed may not be the same as for SM.  The difference could be an overall multiplicative factor, e.g., some branching ratio, but it could also vary from bin to bin.
The theorist needs to check, or at least argue, whether the following is true
\begin{equation}
  P(r|t) = P^{NP}(r|t)\ \ \ , \forall (r, t).
\end{equation}
To answer, he needs to have a MC generator and detector simulation for his NP, from which he can estimate $P^{NP}(r|t)$, and he needs to be supplied with the elements of the \P\ used in the unfolding, to compare the elements of the two matrices.  So, one needs to be aware that the plain answer from unfolding is not enough to be sure it is interpreted correctly.  Unfortunately, some kind of detector model is still needed by the theorist to obtain $\P^{NP}$, and the experimentalists need to publish the elements of \P.

If one is convinced that $\P = \P^{NP}$, then Eq.~\ref{eq:theTneeded0} is further simplified to
\begin{equation}
  \label{eq:theTneeded2}
\T = \T^{SM} + \sigma \T^{SNP}.
\end{equation}

The procedure to arrive at $p(\sigma|\D)$ is to assume an array of $\sigma$ values, then for each $\sigma$ compute the $\T$ of Eq.~\ref{eq:theTneeded0} (or \ref{eq:theTneeded} or \ref{eq:theTneeded2}, if allowed), then insert this \T\ into Eq.~\ref{eq:likelihoodInverted} to calculate $L(\D|\T)$ up to a constant, and finally insert this $L(\D|\T)$ into Eq.~\ref{eq:posteriorSigma} to compute $p(\sigma|\D)$ up to a constant, assuming the prior $\pi(\sigma)$.  When this is done for all $\sigma$ values, the function $p(\sigma|\D)$ will not be normalized to 1 yet, so, this needs to be done eventually.

\subsection{Hypothesis testing}
\label{sec:hypo}

In Sec.~\ref{sec:limits} it was shown how to compute the probability of any hypothetical NP cross-section $\sigma$, using the output of FBU.  Bayesian hypothesis comparison is then straight forward, and consists in comparing the posterior probabilities of two competing hypotheses, of which one could be just the SM hypothesis.  More discussion in Ref.~\cite{DAgostiniHypo}.

For a frequentist hypothesis test\footnote{For a detailed introduction in frequentist hypothesis testing, see Ref.~\cite{BH}, and \cite{DAgostiniHypo} for some fair criticism.} at truth-level, it is possible to define the following null hypothesis: The given truth-level spectrum $\check{\T}$ is pulled from the posterior of FBU, $p(\T|\D)$.  This $\check{\T}$ could be the prediction of a theorist who is curious if it is consistent with the posterior about $\hat{\T}$.

A test statistic can easily be defined using truth-level quantities:
\begin{equation}
\chi^2(\T) \equiv -\log p(\T|\D).
\end{equation}
The corresponding \pval is the probability that a \T\ sampled from $p(\T|\D)$ would be at least as unlikely, according to $p(\T|\D)$, as $\check{\T}$.  Namely, 
\begin{equation}
\pval = P(\chi^2(\T) > \chi^2(\check{\T}) | \T \sim p(\T|\D)),
\end{equation}
where $\T \sim p(\T|\D)$ means ``\T\ is randomly sampled from $p(\T|\D)$''.

\pvals are notoriously easy to misinterpret.  The following is an attempt to explain how this \pval should be understood (adapted from \cite{BH}).

First, let's imagine a robot, Rejectron, which can utter only this sentence that rejects the null hypothesis: ``THIS $\check{\T}$ IS NOT PULLED FROM $p(\T|\D)$''.  It says this mechanically whenever a $\check{\T}$ with $\pval \le \alpha$ is presented to it.  A knob on its chest adjusts the value of $\alpha \in [0,1]$.  If infinite $\check{\T}$'s are presented to it, and they follow $p(\T|\D)$, then the robot is guaranteed to make a false rejection of the null hypothesis with frequency $\alpha$.

Now, let's say that the specific $\check{\T}$ a theorist proposes has $\pval = \gamma$.  If the robot's $\alpha$ is less than $\gamma$, then the robot will stay silent.  If $\alpha = \gamma$ then the robot will reject the null hypothesis.  If $\alpha > \gamma$, the robot will still reject the null hypothesis, but its false rejection probability in an infinite ensemble of spectra pulled from $p(\T|\D)$ would be larger than $\gamma$. So, the minimum value one could set $\alpha$ to, and still reject the null hypothesis when $\check{\T}$ is presented to the robot, is $\gamma$.  So, the $\pval$ of $\check{\T}$ is the minimum possible false-rejection frequency (i.e., Type-I error rate) of a robot (i.e., a decision algorithm) which rejects the null hypothesis when $\check{\T}$ is presented to it.

So, is this \pval\ saying how likely $\check{\T}$ is to originate from $p(\T|\D)$?  No, although this is a common misinterpretation of \pvals.  Is it the probability that $\check{\T} = \hat{\T}$?  Obviously not.  This \pval\ says more about Rejectron than about $\check{\T}$.  

If the theorist wants to know how likely his theory is, according to the result of FBU, he can define a volume in the vicinity of $\check{\T}$ that he considers representative of his theory, and integrate $p(\T|\D)$ in that volume.  An integration is necessary, since individual \T\ points are not assigned a probability, but a probability \emph{density}.  As explained in Sec.~\ref{sec:limits}, this is only correct as long as his theory does not involve a different migration model ($\P^{NP}$) from the migration model used to derive the unfolding (\P).  Interestingly, this should be the case when $\check{T}$ is the SM, and the \P\ used in unfolding was also derived from SM MC.  So, it is straight forward for a theorist to compute the probability that SM is true by integrating $p(\T|\D)$ in a volume around $\T^{SM}$ which represents the theoretical uncertainty in the SM prediction, i.e., uncertainty from higher order terms, parton distribution uncertainty, etc.

\section{Conclusion}

A fully bayesian unfolding (FBU) method is formulated, and presented in numerous examples.

To conclude, some observations made in Sec.~\ref{sec:applications} and \ref{sec:regularization} will be summarized concisely.
Then, some final recommendations will be given in Sec.~\ref{sec:recommendations}.

\subsection{Summary of observations}
\begin{itemize}
\item The asymmetric, non-Gaussian shape of $p(\T|\D)$ is evident in truth-level bins with low statistics.
\item No smearing leads to no correlations, and $p(\T|\D)$ is maximized at $\T = \D$.
\item Smearing increases the spread of $p(\T|\D)$, and introduces correlations.
\item Large smearing, even when event counts are large, can cause  $p(\T|\D)$ to be non-Gaussian, due to the $T_t > 0$ boundary.
\item By reducing the volume of the sampled hyper-box, uniform sampling is fast even in FBU with large $N_t$.
\item MCMC is much more efficient, and qualitatively it gives very similar results, but it is not as accurate, due to anomalies in the marginal distributions of $p(\T|\D)$.
\item It is possible to infer $T_t$ in bins where there are no data, provided that migrations are possible from these bins into regions where data exist.
\item Expected bumps can become more sharp with FBU, without regularization, provided that the smearing is not big enough to make the uncertainties comparable to the bump itself.

\item Regularization can affect correlations.
\item Regularization can cause secondary maxima in the posterior.
\item In the presence of smearing, there is more potential for regularization to reduce the posterior's spread.
\item Regularization doesn't help make an unexpected bump more sharp.  It either obscures it, or it maintains the width it has in the data after smearing.  
\item Unfolding without regularization reconstructs unexpected bumps with the right width, but the posterior is too spread out, which may obscure the bump unless it is enormous.
\item Regularization doesn't necessarily make an expected bump more sharp, unless $\pi(\T)$ is tailored to $\tilde{\T}$ (e.g., with a Gaussian constraint), and $\tilde{\T} \simeq \hat{\T}$.
\end{itemize}

\subsection{Final Recommendations}
\label{sec:recommendations}

\begin{itemize}
\item An unfolded spectrum is not ``corrected data'', and should not be thought of in such terms.
\item Unfolding is a non-parametric inference procedure, and it should be avoided unless it, \emph{per se}, is the end goal of the analysis.
\item If the goal is to set a limit or estimate an unknown parameter, or to test a hypothesis, it is much simpler to use the actual data ($\D$) for this.
\item If unfolding is used, all choices and details must be presented transparently.  It is not enough to say ``an unfolding technique was used.''  The regularization condition (i.e.\ the prior) needs to be published, as well as the elements of $\P$.
\item A result with no regularization (constant prior) has special properties and should be shown by default, even if some regularization is used eventually.  Alternative regularization choices are encouraged if believed to be reasonable.  The prior, thus the regularization, is assumed subjective and informative, although special ``non-informative'' priors can be considered as well.
\item Reporting the result of some unfolding procedure \emph{does not justify} omitting the data ($\D$).  This is hard to emphasize enough.  Unfolding does not replace the data.
\end{itemize}

\section*{Acknowledgements}

I thank Diego Casadei for his remarks on the first version of this document, Giulio D' Agostini and Glen Cowan for their interest, Davide Gerbaudo and Francesco Rubbo for being my first collaborators to try FBU, Kerim Suruliz who was the first to implement FBU in BAT \cite{bat} thus providing a valuable cross-check and making FBU accessible to BAT users, and the HEP graduate students at the University of Chicago for their insightful questions.

\end{multicols}


\begin{multicols}{2}
  

\end{multicols}

\begin{figure}[H]
\centering
\begin{tabular}{ccccc}
   \includegraphics[width=0.18\columnwidth]{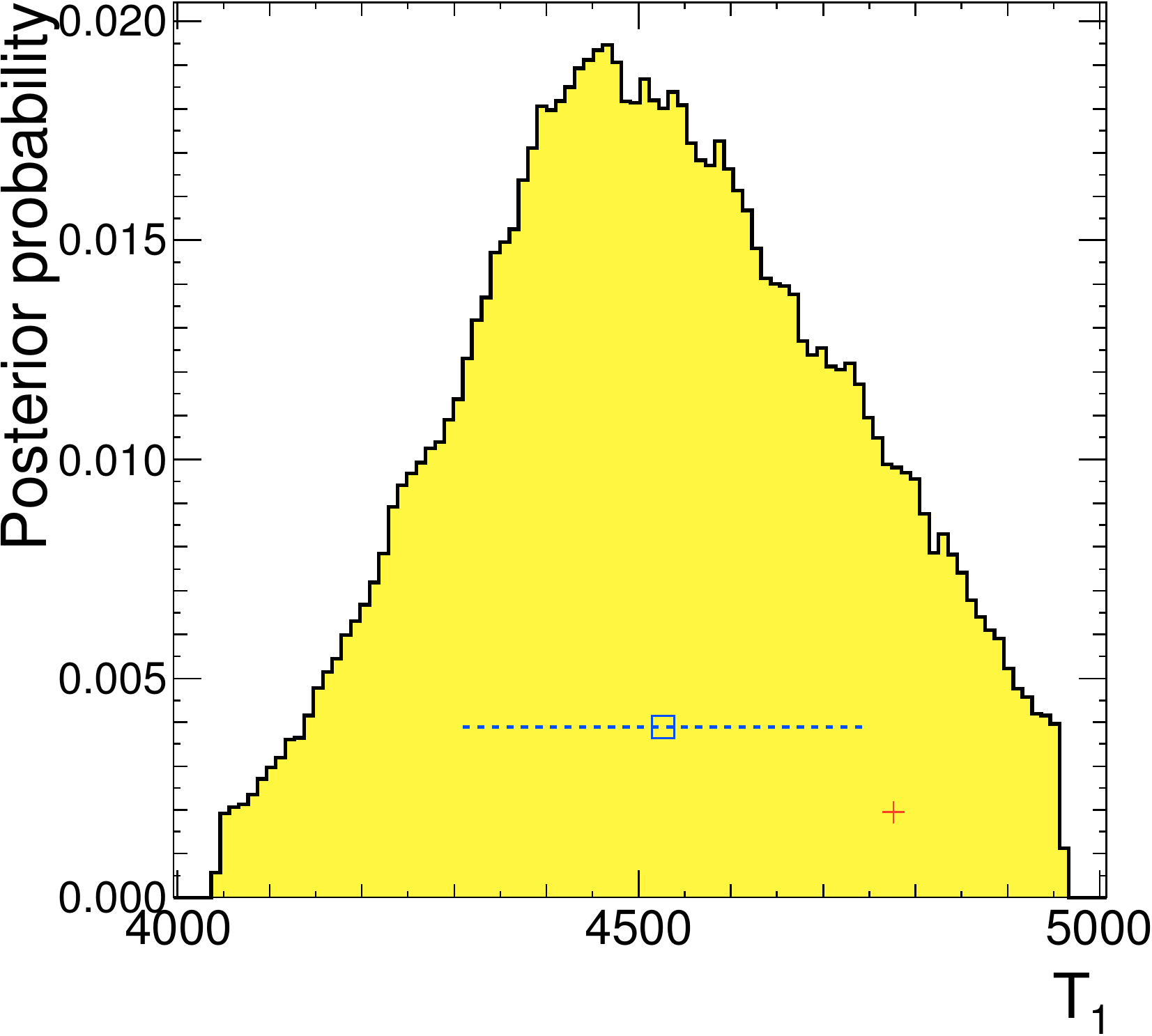} &
   \includegraphics[width=0.18\columnwidth]{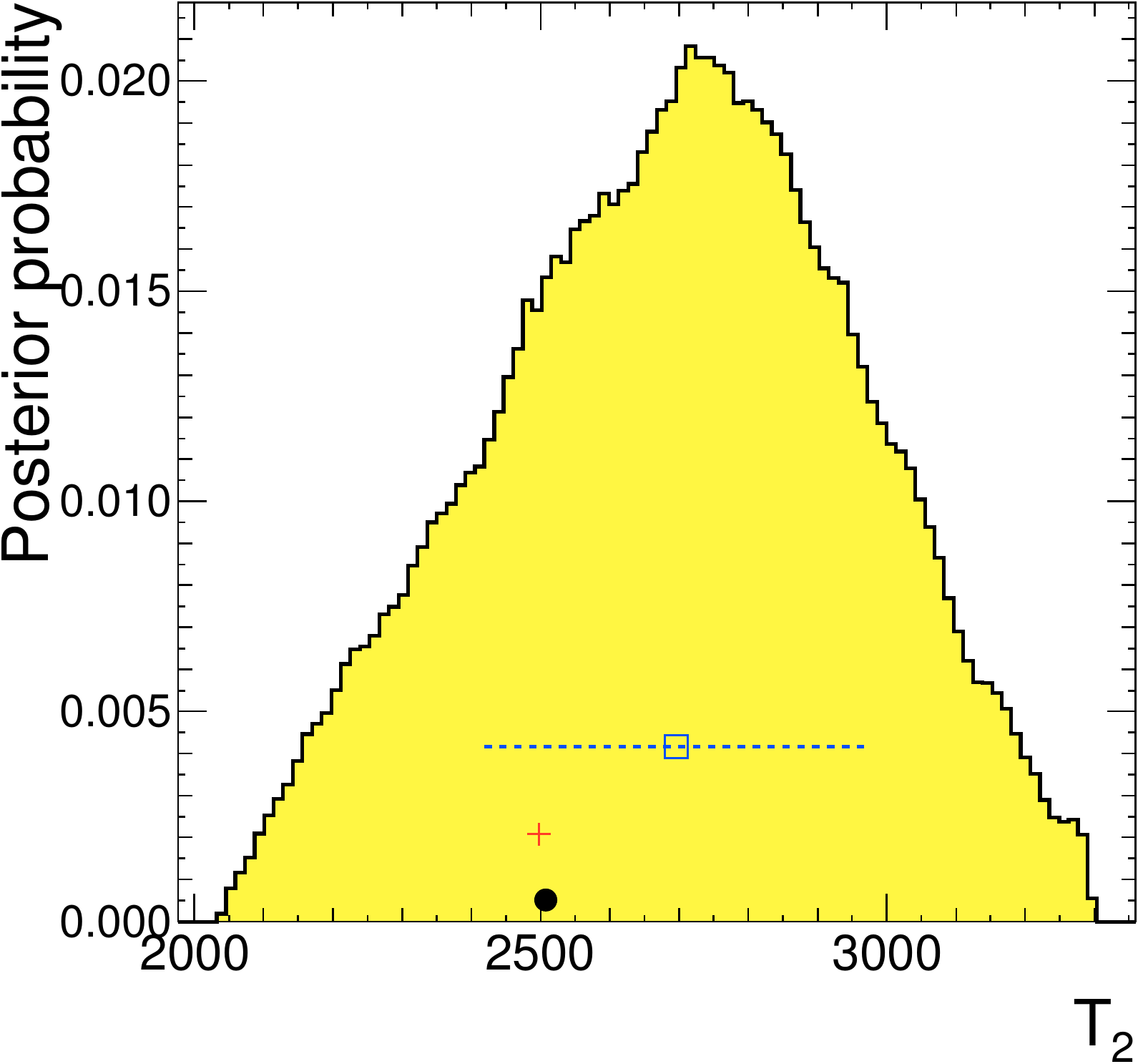} &
   \includegraphics[width=0.18\columnwidth]{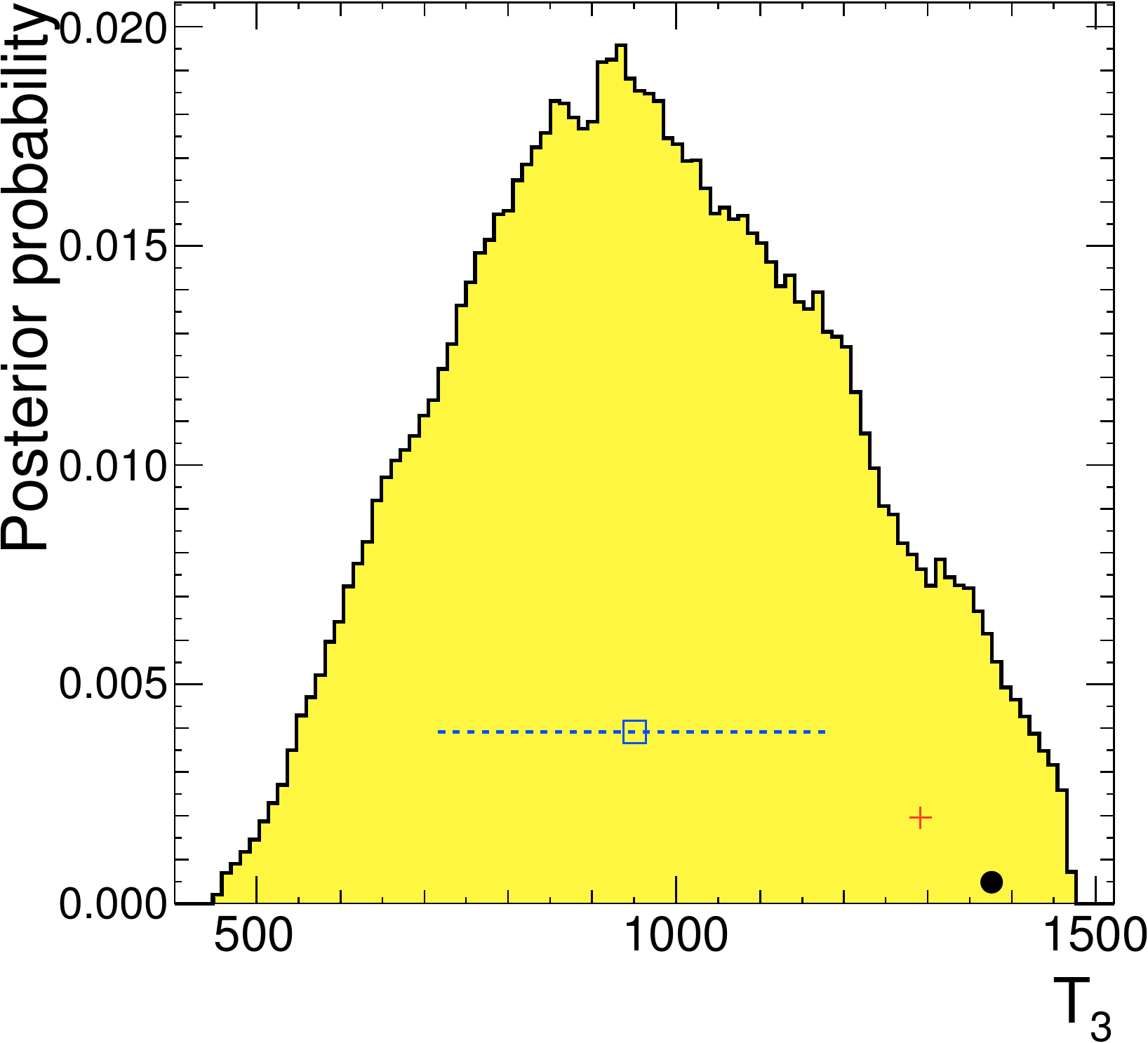} &
   \includegraphics[width=0.18\columnwidth]{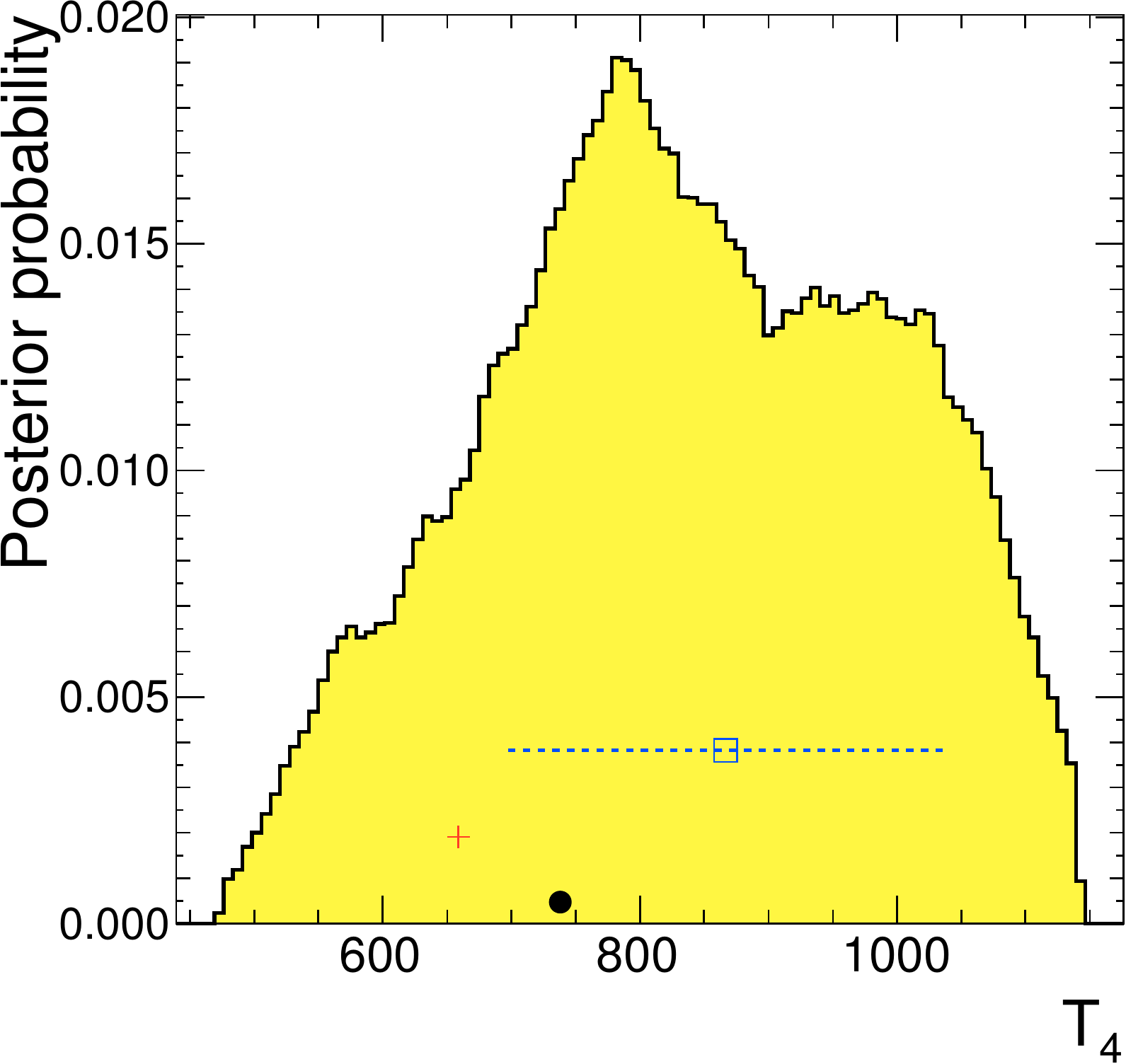} &
   \includegraphics[width=0.18\columnwidth]{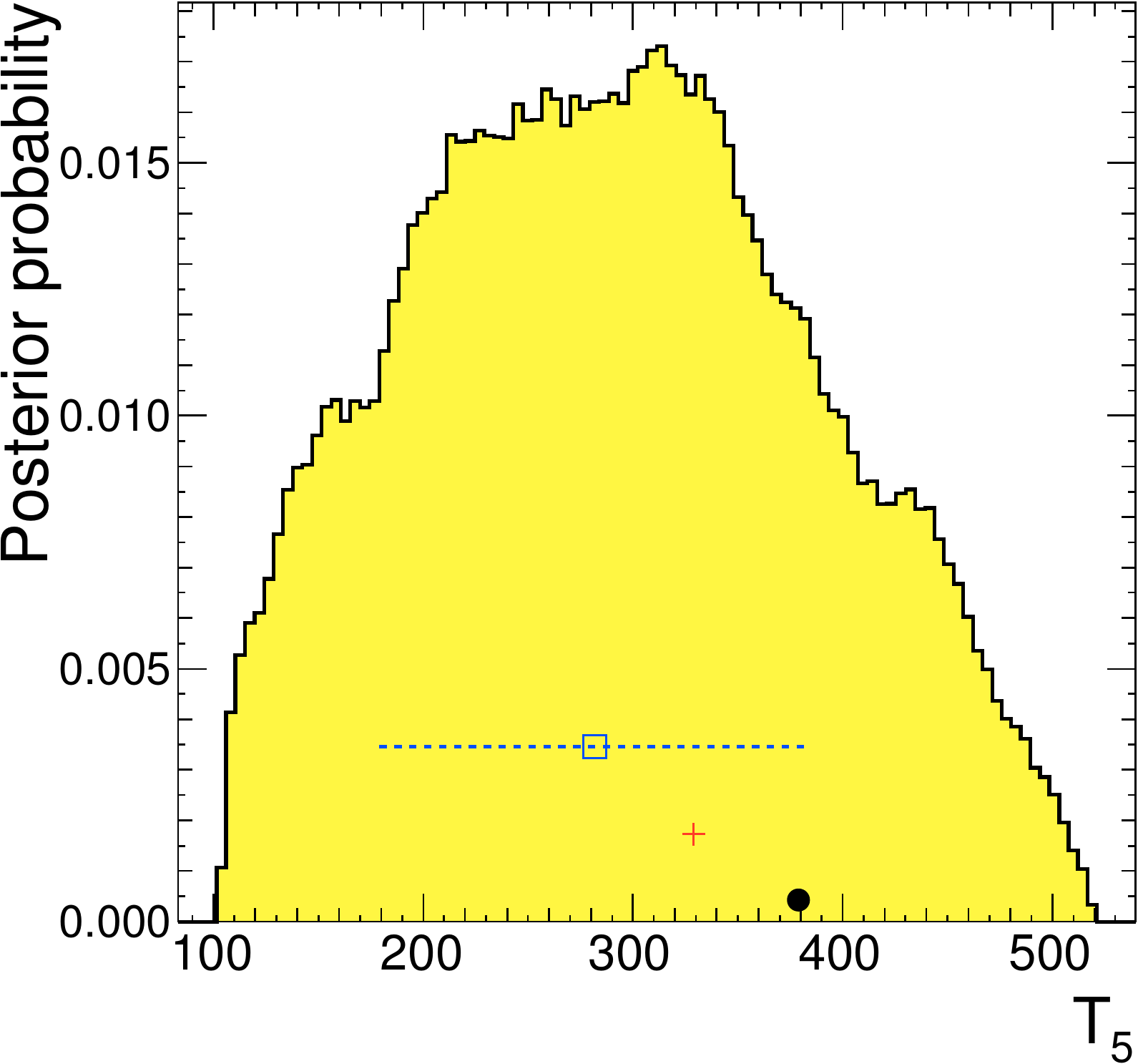} \\

   \includegraphics[width=0.18\columnwidth]{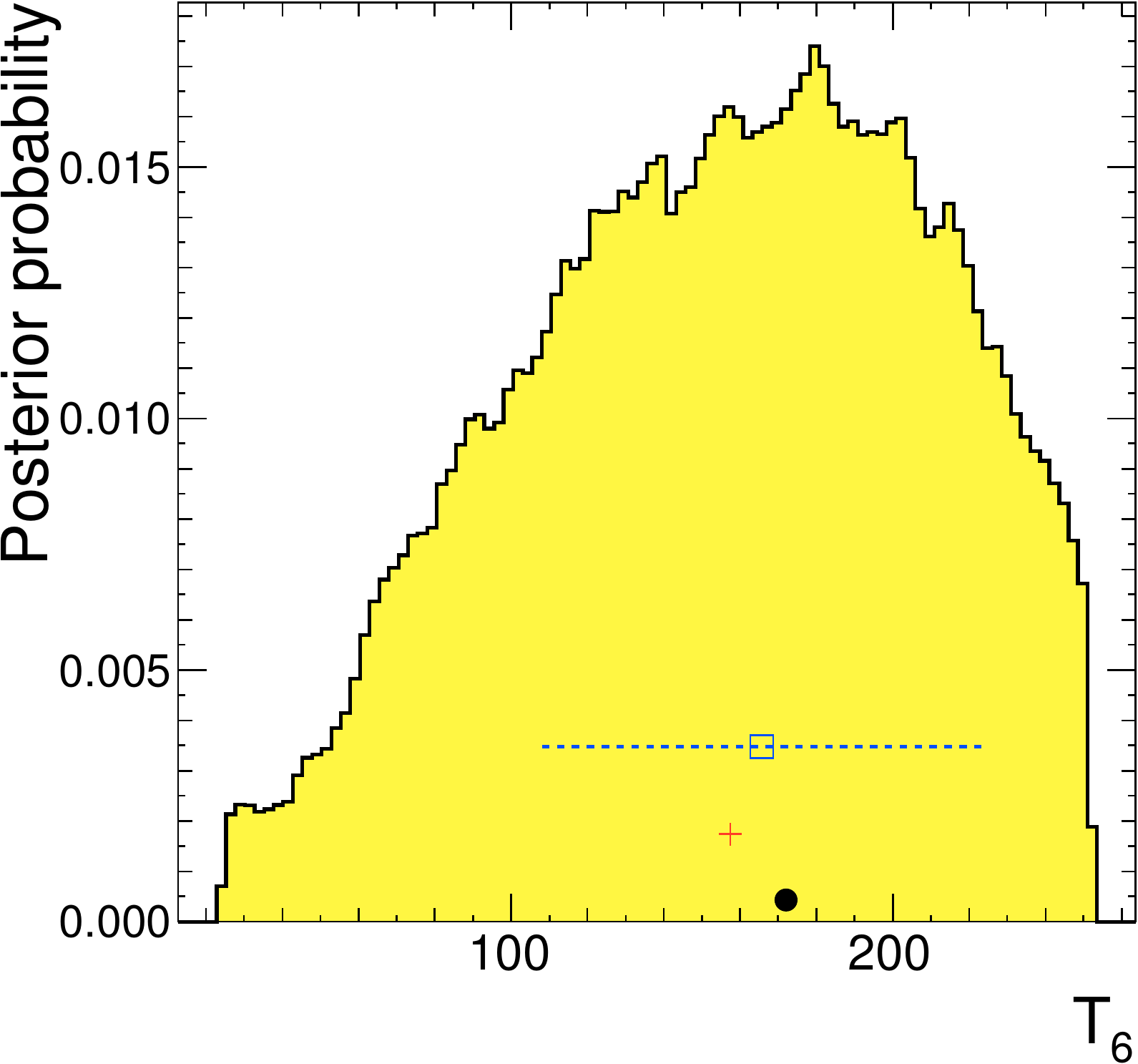} &
   \includegraphics[width=0.18\columnwidth]{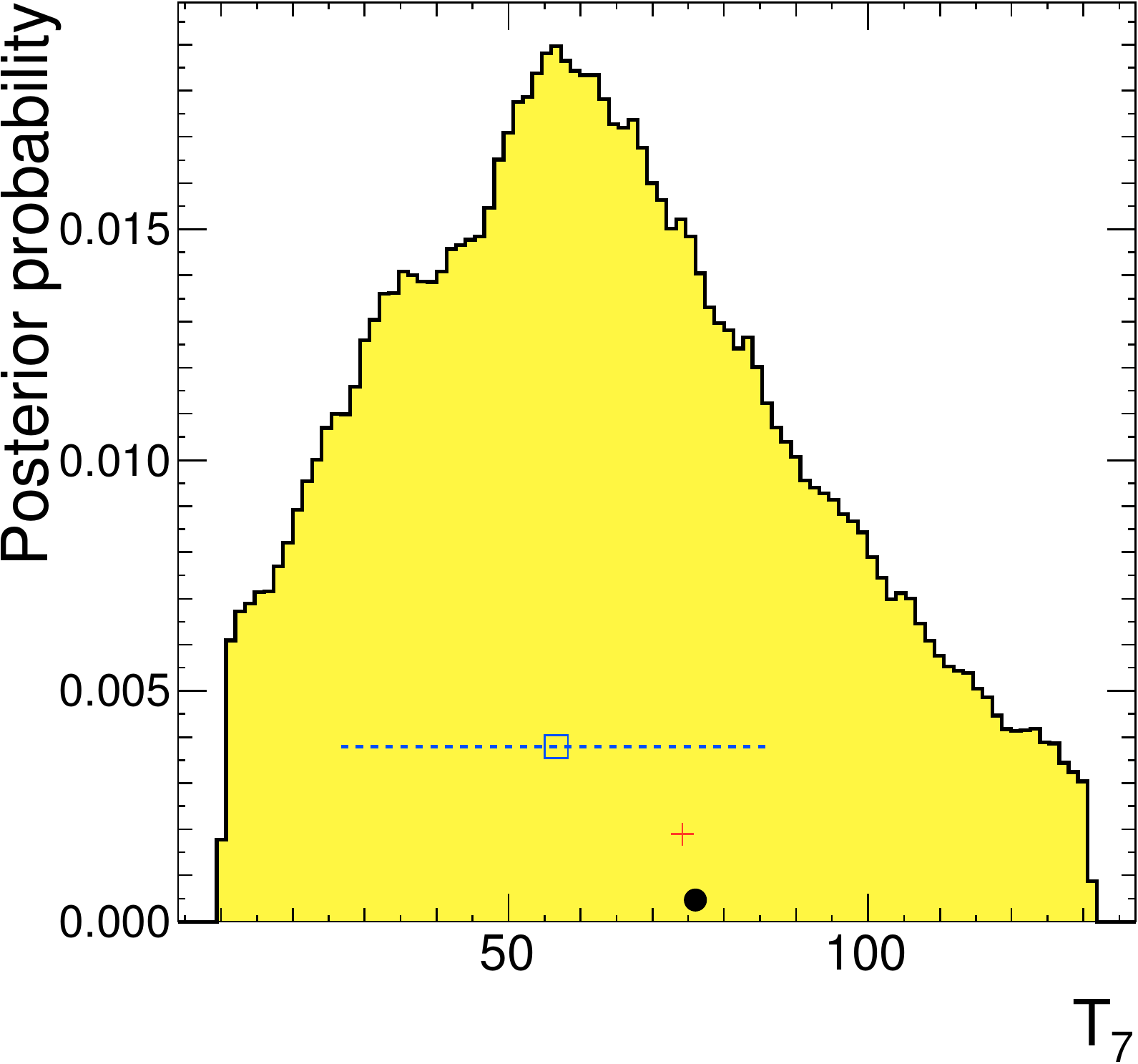} &
   \includegraphics[width=0.18\columnwidth]{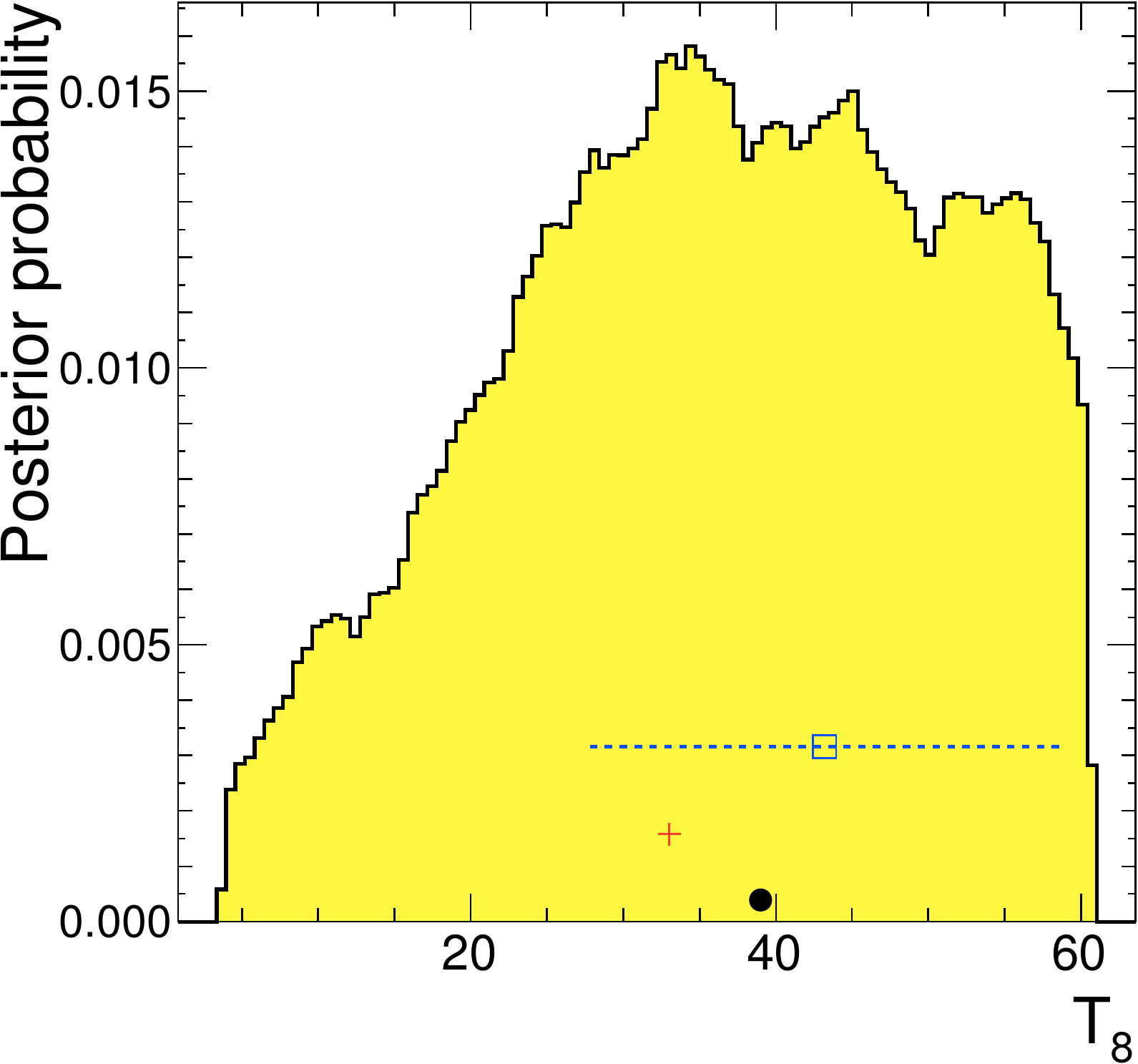} &
   \includegraphics[width=0.18\columnwidth]{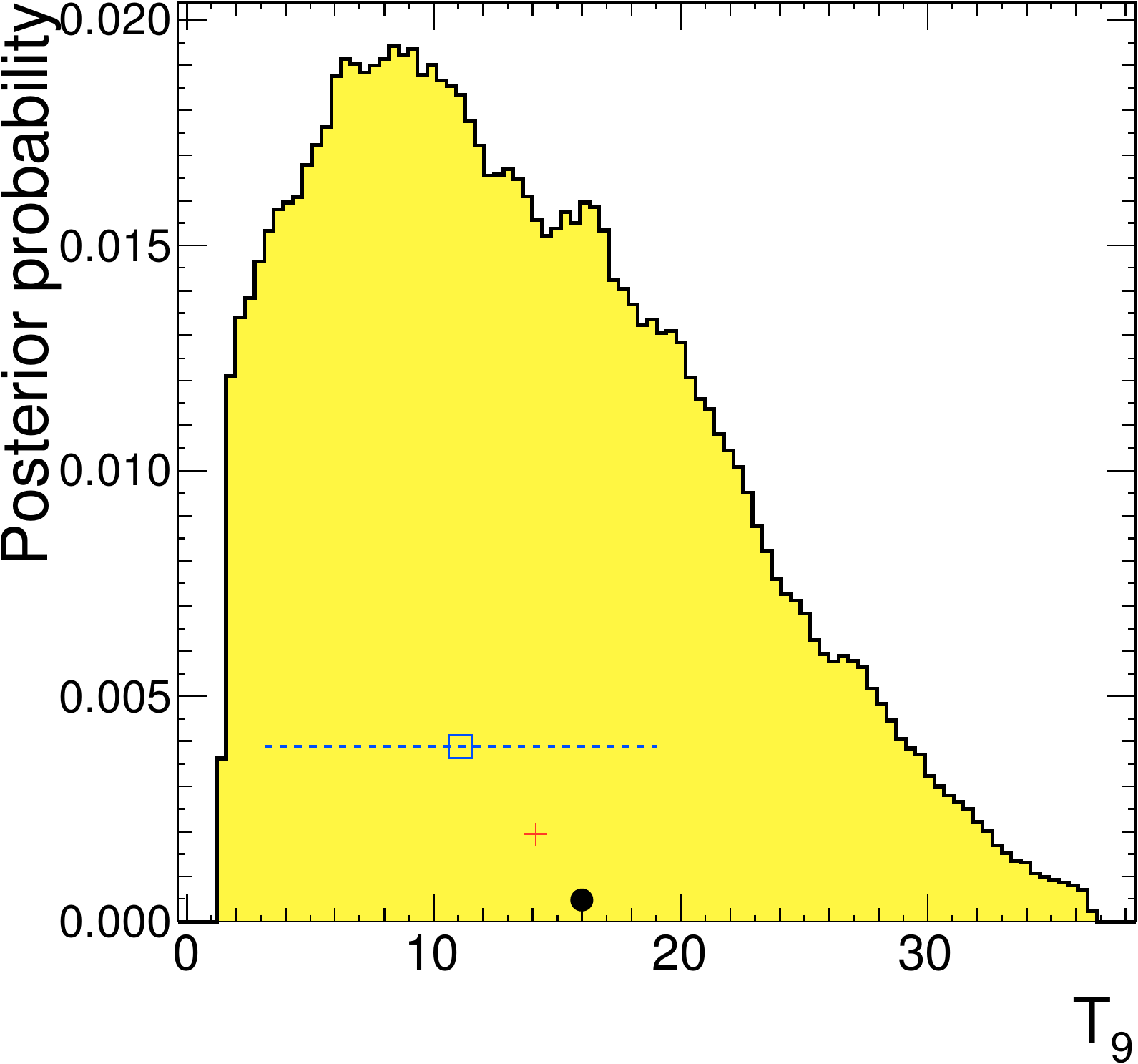} &
   \includegraphics[width=0.18\columnwidth]{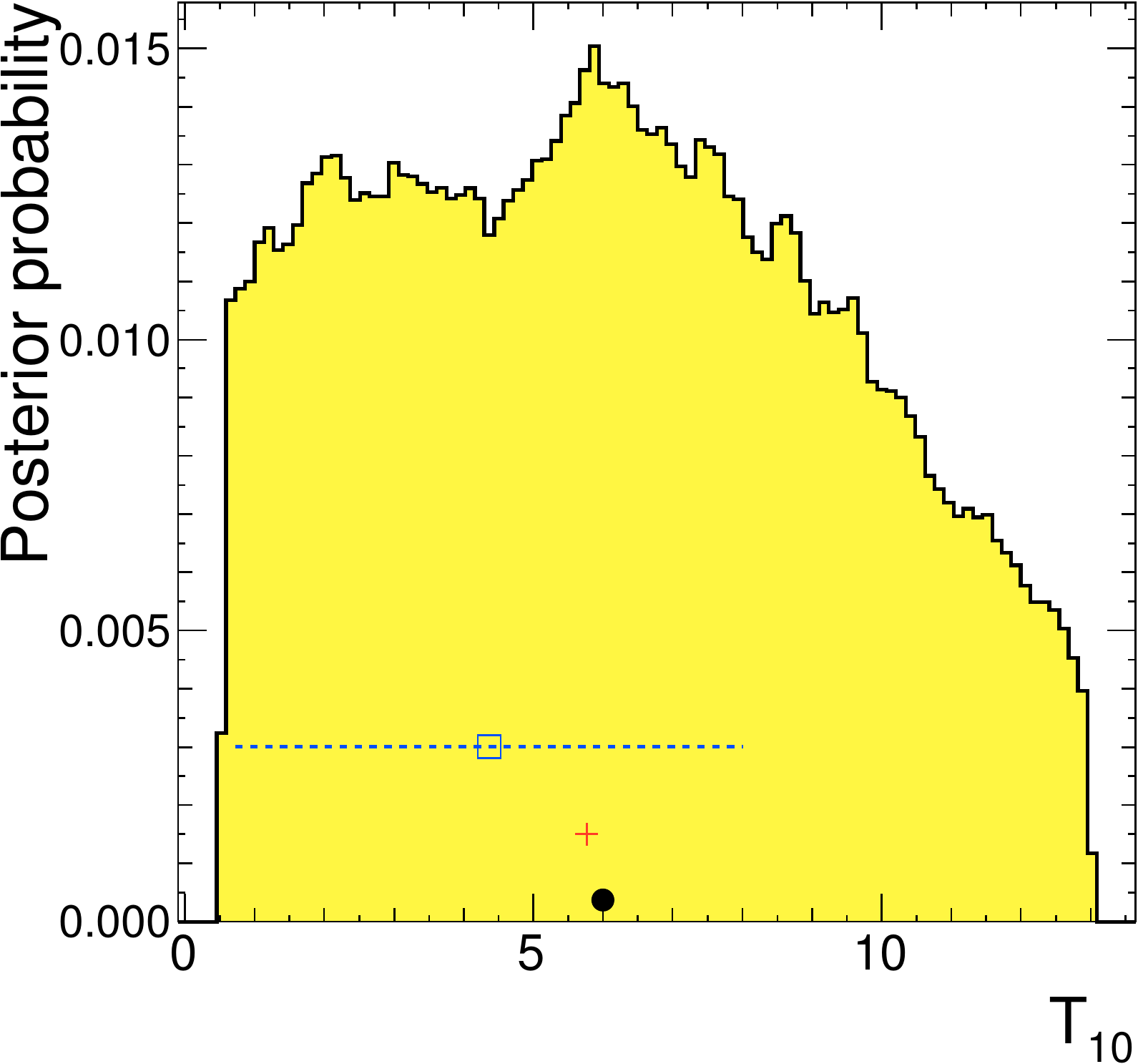} \\

   \includegraphics[width=0.18\columnwidth]{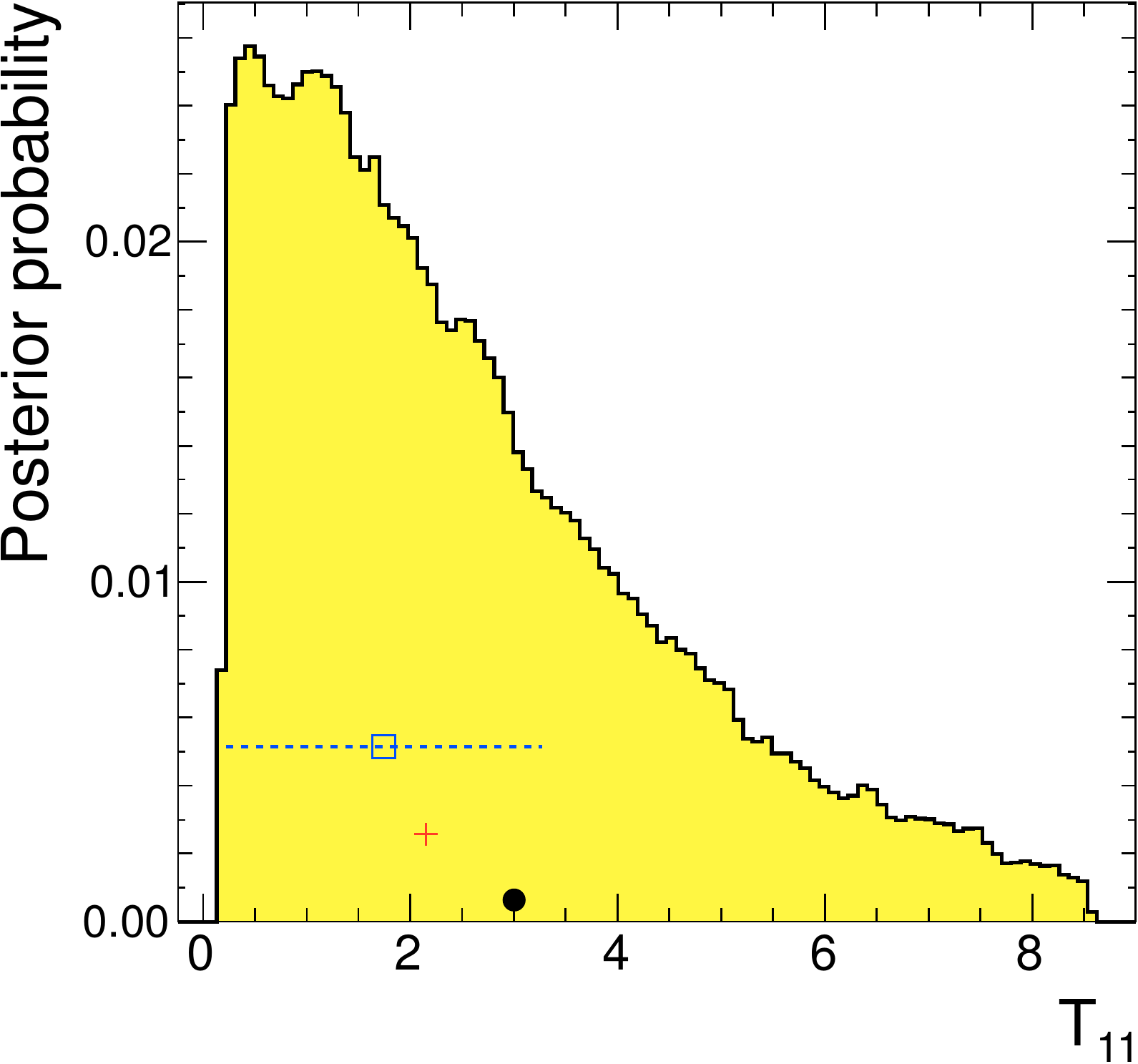} &
   \includegraphics[width=0.18\columnwidth]{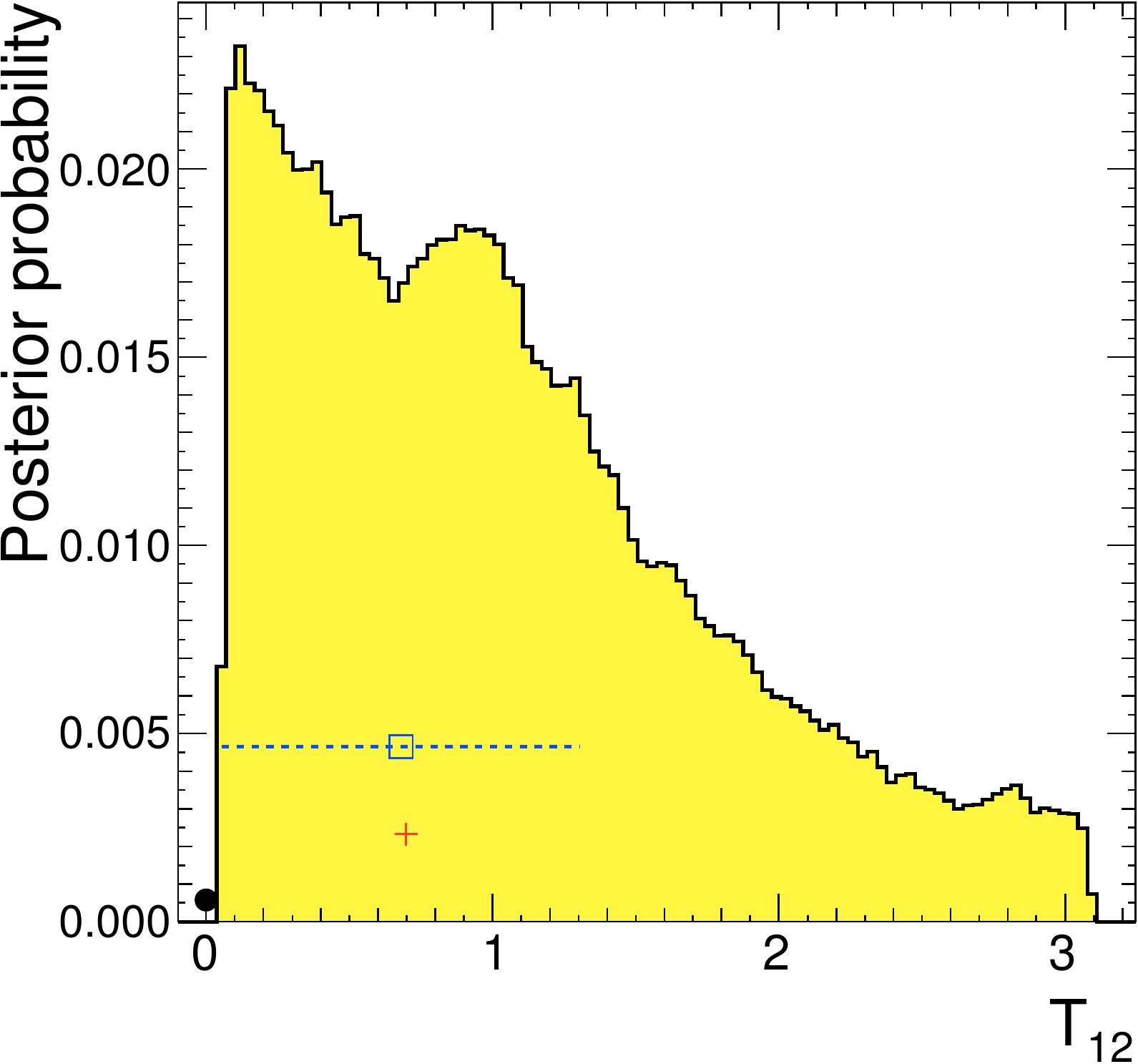} &
   \includegraphics[width=0.18\columnwidth]{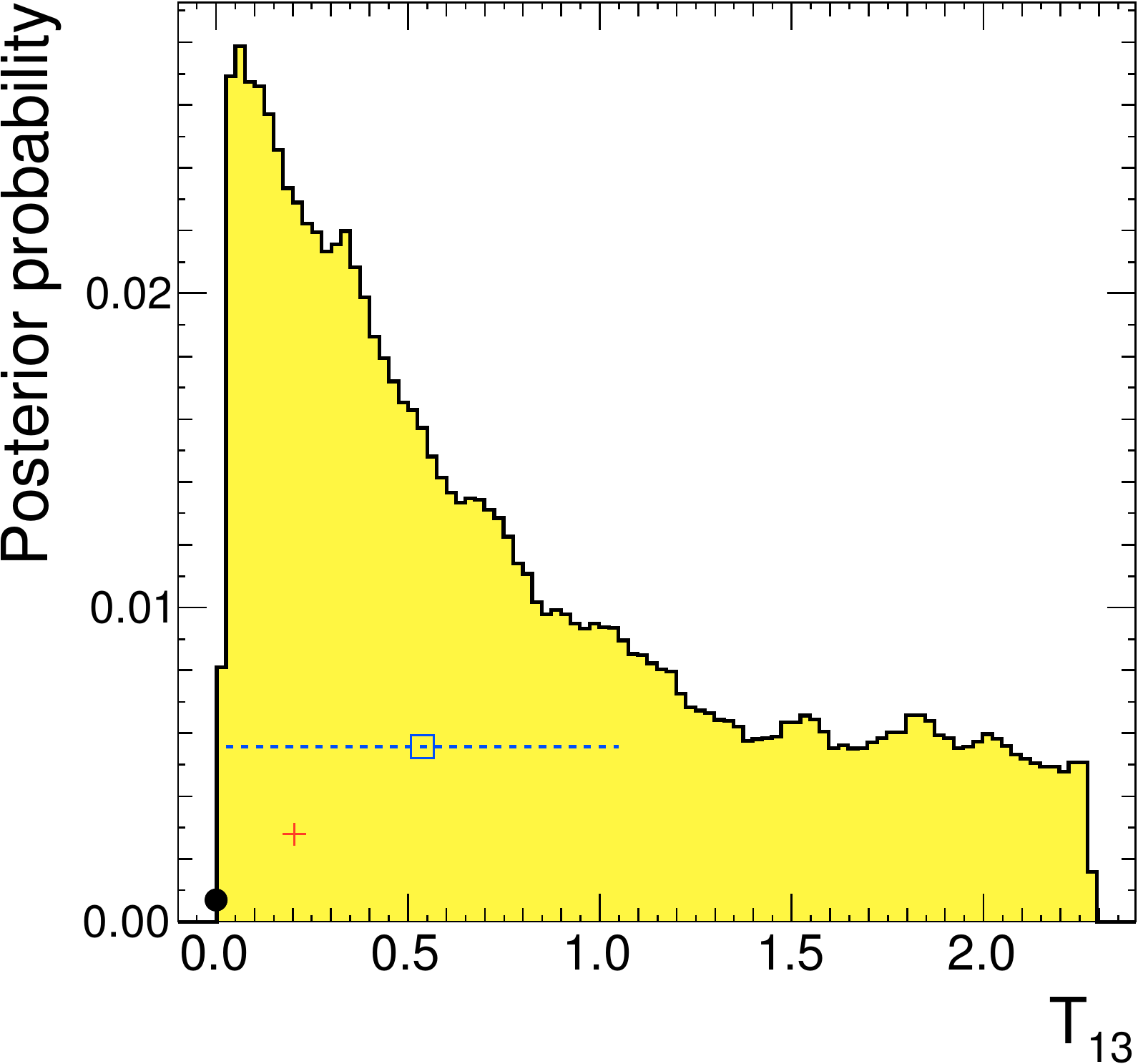} &
   \includegraphics[width=0.18\columnwidth]{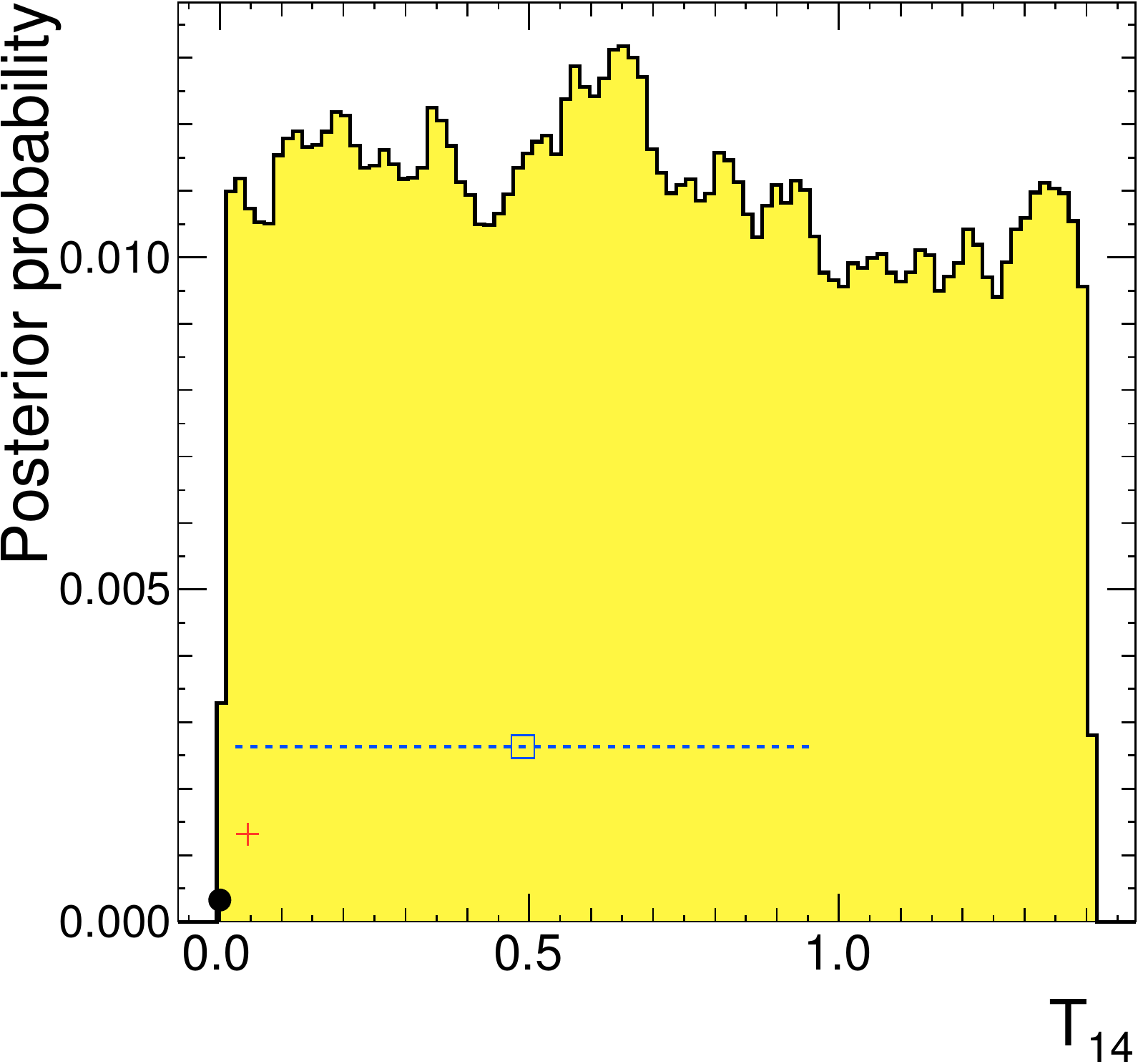} 
 \end{tabular}
 \caption{The 1-dimensional marginal distributions of $p(\tuple{T}|\tuple{D})$ in the example of Sec.~\ref{sec:example5}, sampling the reduced (``new'') volume in Fig.~\ref{fig:priorRedefined4}.  More anomalies are observed than in Fig.~\ref{fig:1Dim5}.
\label{fig:1Dim5reduced}}
\end{figure}

\begin{figure}[H]
  \centering
  \subfigure[]{
    \includegraphics[width=0.3\columnwidth]{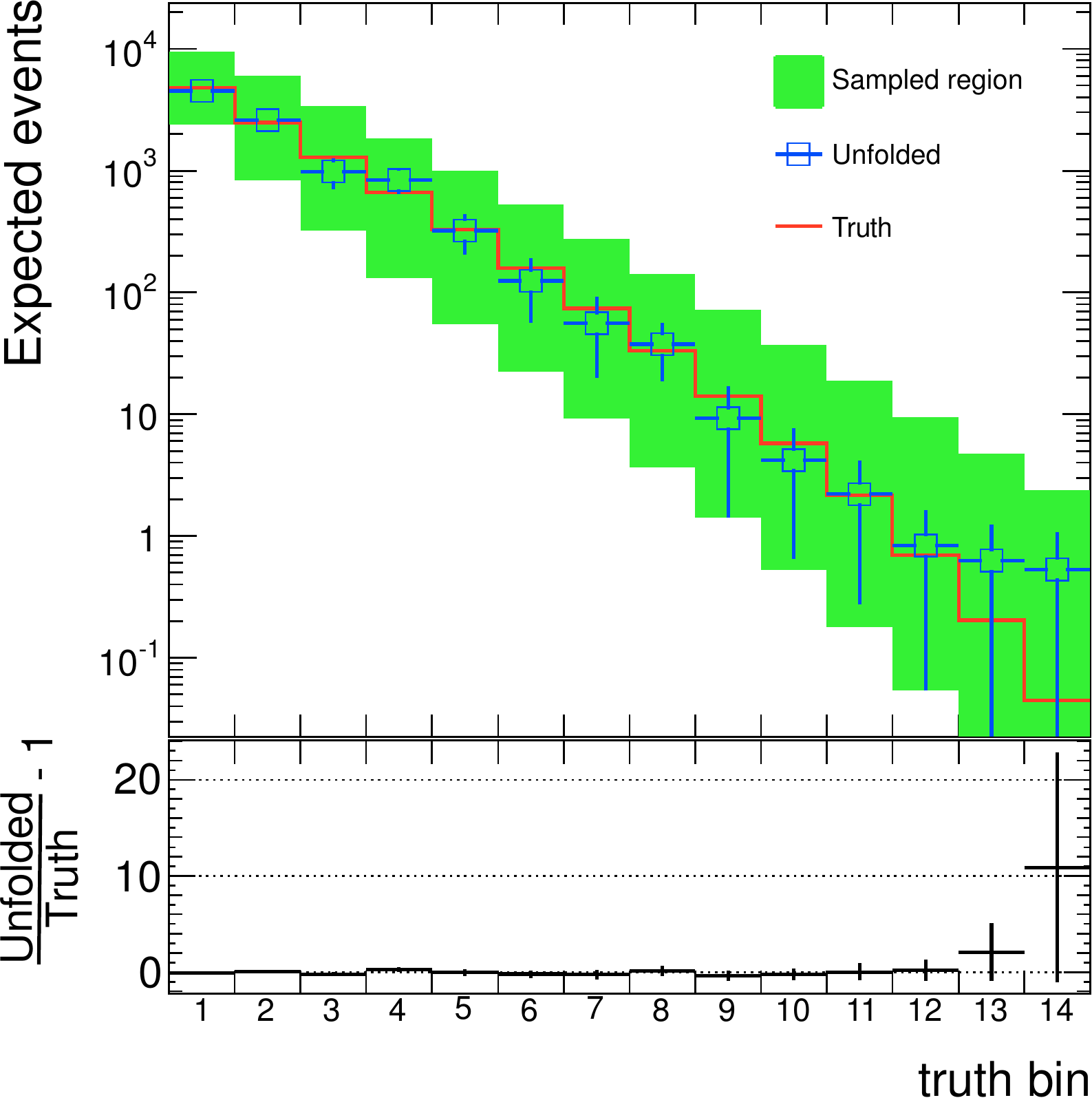}
  }
  \subfigure[]{
    \includegraphics[width=0.3\columnwidth]{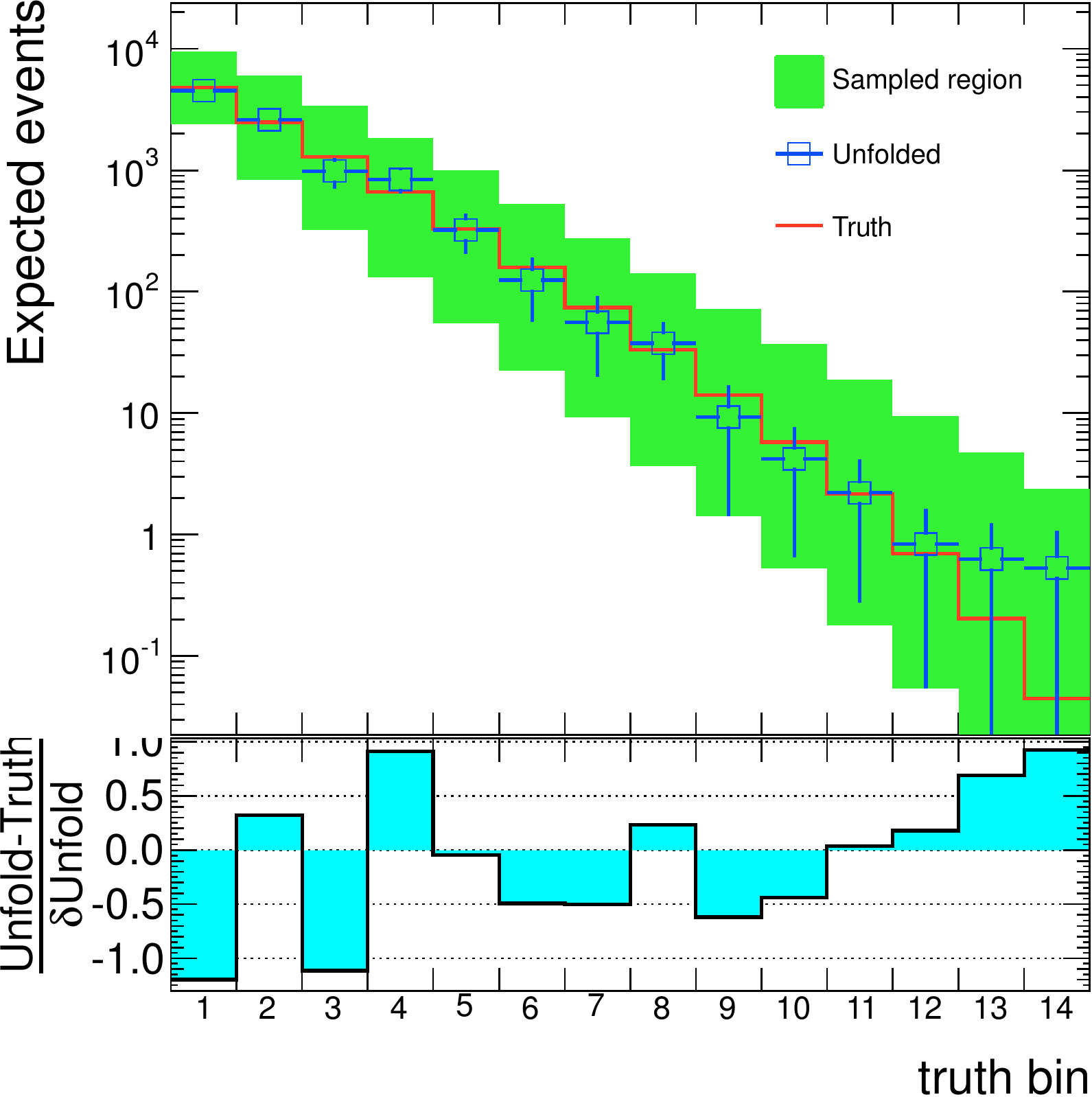}
  }
   \subfigure[]{
    \includegraphics[width=0.3\columnwidth]{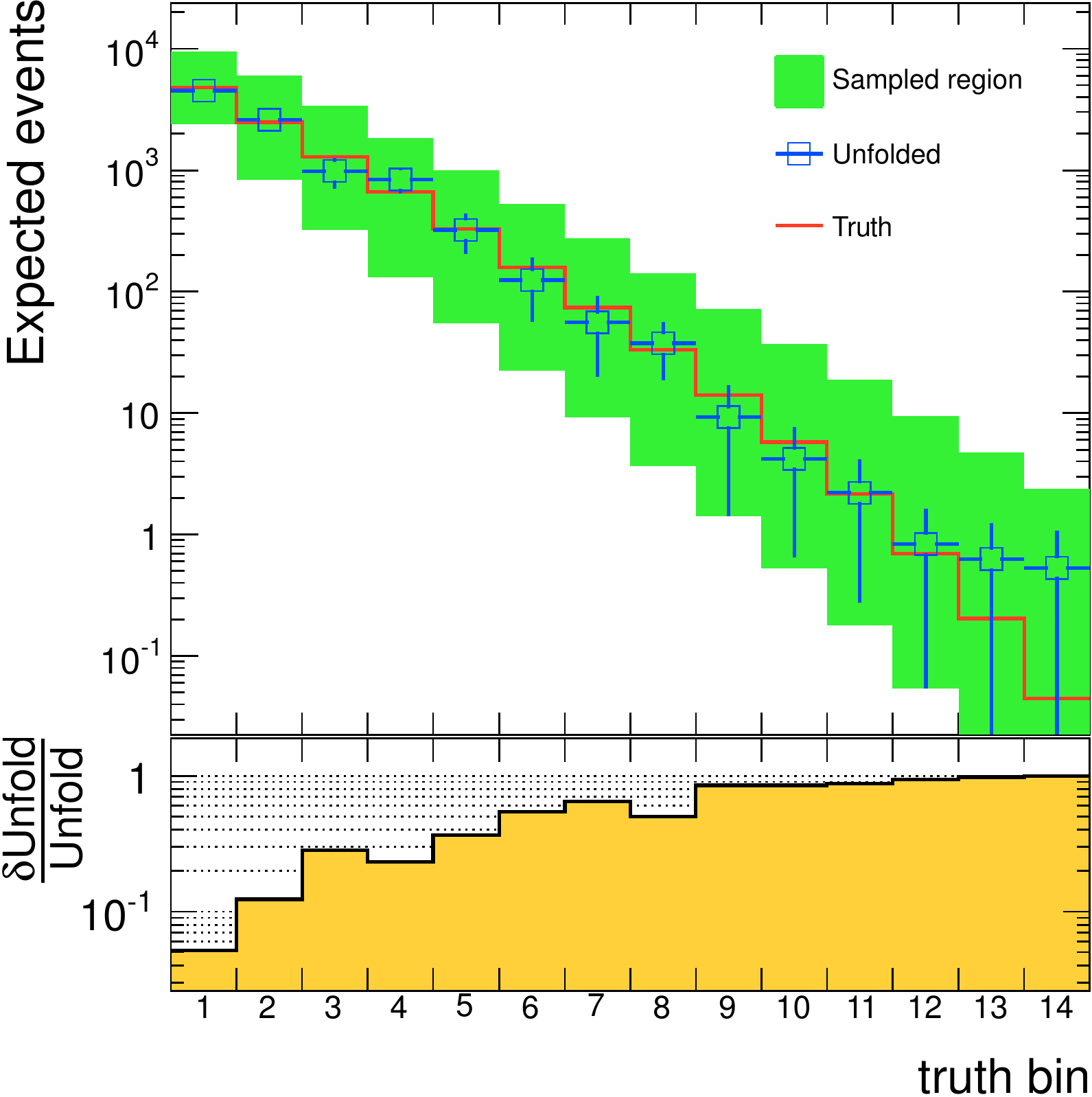}
  } \\
  \subfigure[]{
    \includegraphics[width=0.3\columnwidth]{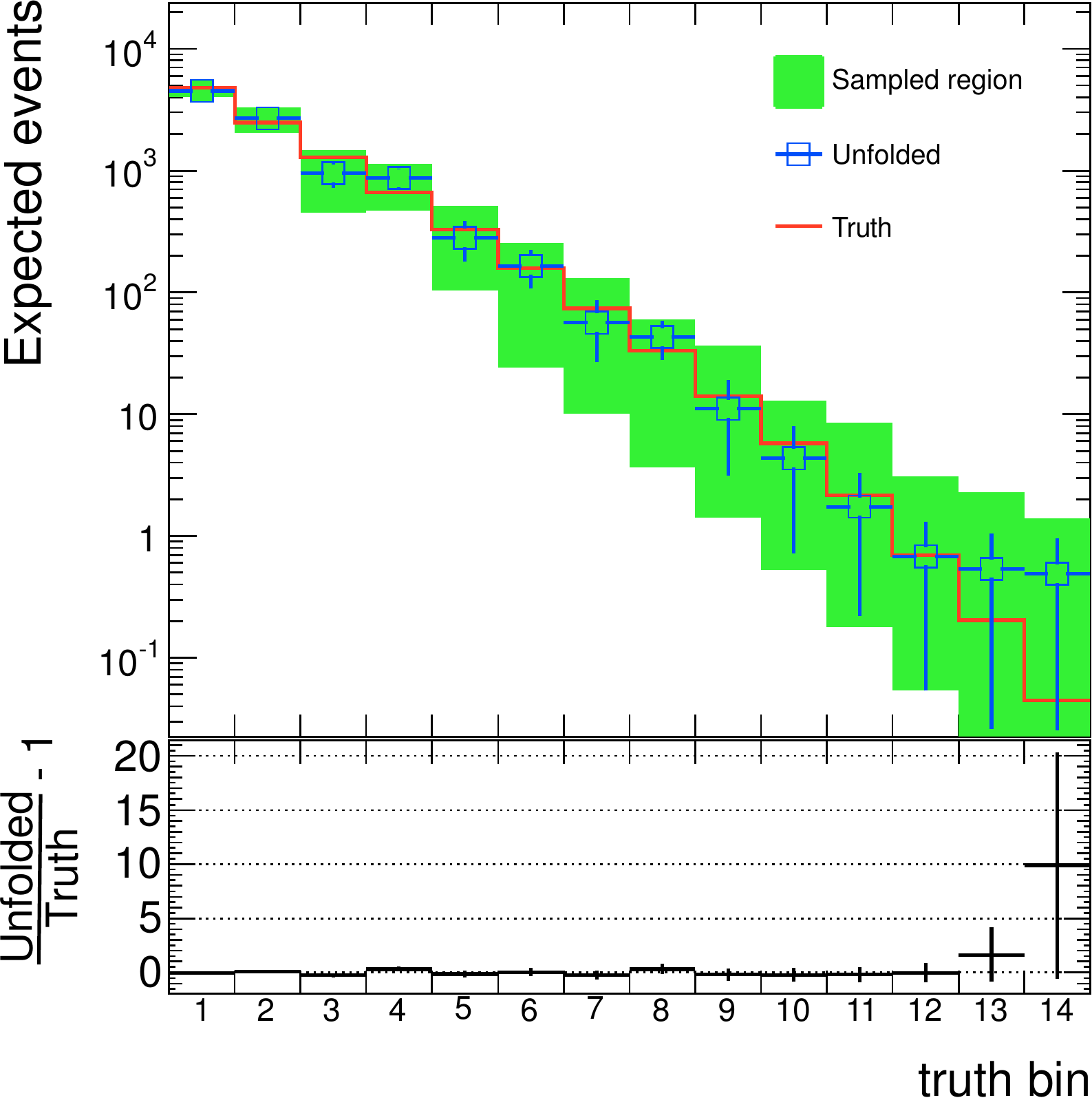}
  }
  \subfigure[]{
    \includegraphics[width=0.3\columnwidth]{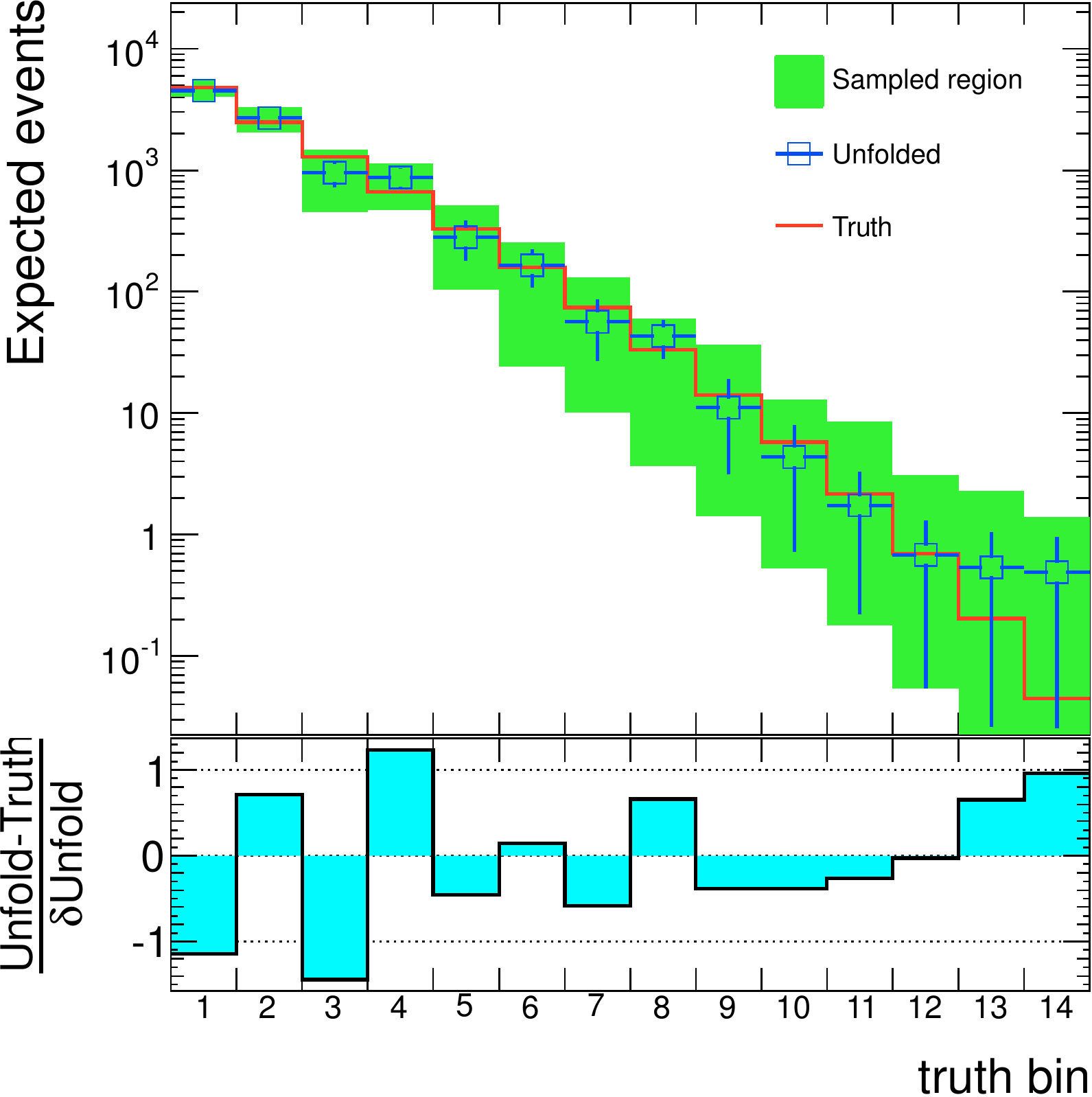}
  }
   \subfigure[]{
    \includegraphics[width=0.3\columnwidth]{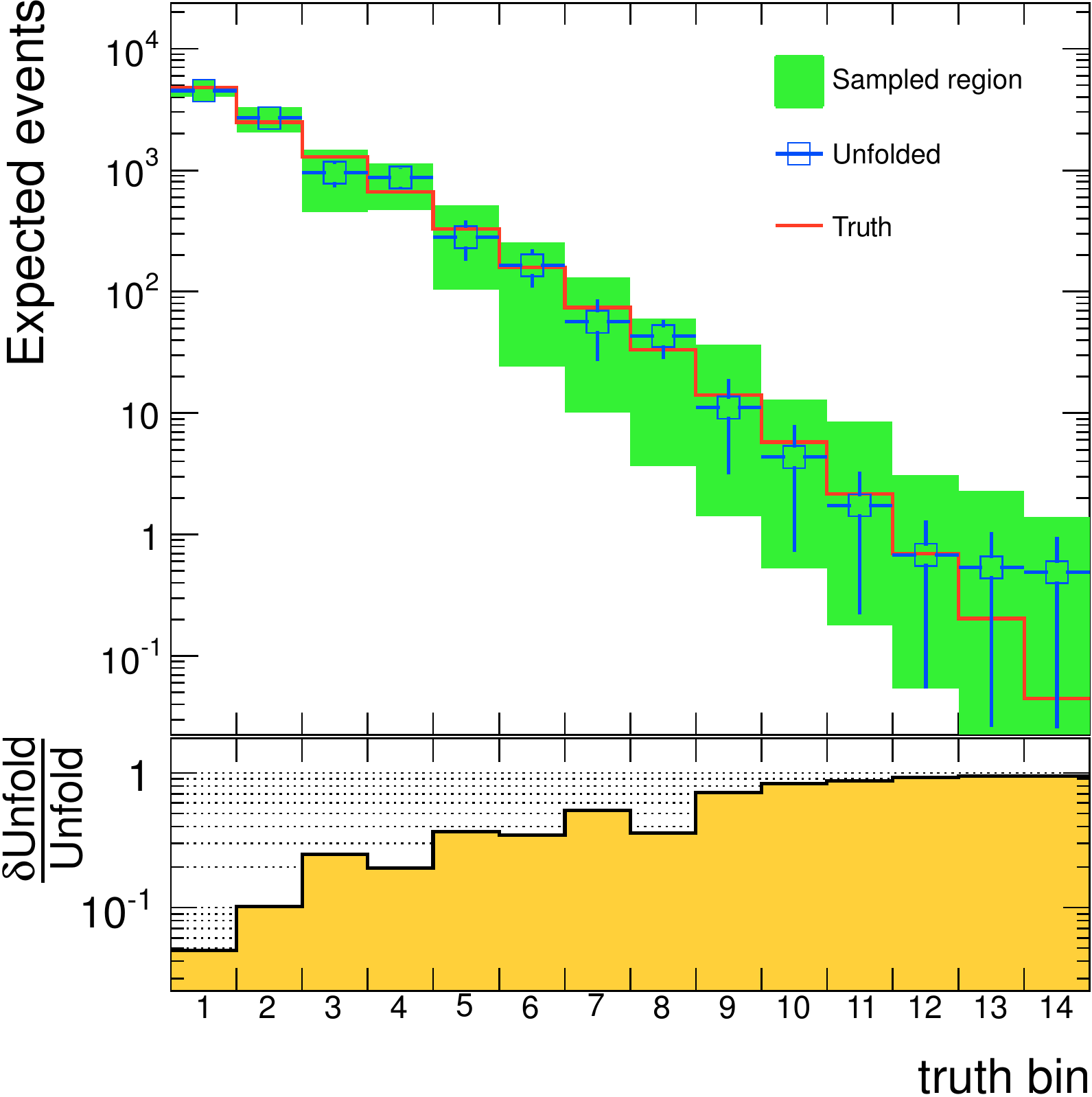}
  } 
  \caption{The unfolded spectrum of the example in Sec.~\ref{sec:example5}.  In (a), (b) and (c) the result is obtained without reducing the volume of the initial sampled hyper-box, and in (d), (e) and (f) the initial hyper-box has been reduced. The unfolded spectrum doesn't change much with volume reduction, even though in Fig.~\ref{fig:1Dim5reduced} it is seen that this introduces some anomalies.  
\label{fig:unfolded5}}
\end{figure}


\end{document}